%% file: Thesis.tex
\pdfoutput=1
\documentclass[paper=a4,fontsize=12pt,toc=bibliography,british,final]{scrbook}		

\usepackage[ngerman,british]{babel}
\usepackage{amsmath}
\usepackage[utf8]{inputenc}
\usepackage{lmodern}			
\usepackage[T1]{fontenc}
\usepackage{amsmath}
\usepackage{amssymb}
\usepackage{mathtools}
\usepackage{graphicx}
\usepackage{calc}
\usepackage{multicol}
\usepackage{paralist}
\usepackage[section]{placeins}
\usepackage{varioref}
\usepackage{url}
\usepackage{hyperref}
\usepackage[hypcap]{caption}		
\usepackage{tabularx}

\usepackage[babel,kerning=true,spacing=true,protrusion=true,expansion=true]{microtype}

\usepackage{subfigure}

\allowdisplaybreaks[1]

\usepackage[numbers,sort&compress]{natbib}

\input{Thesis_Abkuerzungen}

\newlength{\mathnumlength}
\newlength{\mathnumlengthlong}
\newenvironment{mydedication}
{\vspace{6ex}\vspace*{\fill}\begin{quotation}\begin{center}\begin{em}\large}
{\par\end{em}\end{center}\end{quotation}\vspace*{\fill}}

\begin{document}

\selectlanguage{british}
\input{Thesis_Titlepage}

\tableofcontents
\input{Thesis_Introduction}

\part{Theoretical framework}
\label{part:TheoFramework}
\input{Thesis_fRG}

\input{Thesis_VertexParametrization}

\input{Thesis_ChannelDecomposition}

\input{Thesis_WICharge}

\part{Applications}
\label{part:Applications}
\input{Thesis_RedPairFWModel}

\input{Thesis_AttractiveHubbard}
\input{Thesis_RepulsiveHubbard}

\part{Summary}
\input{Thesis_Conclusion}

\input{Thesis_DeutscheZusammenfassung}

\part{Appendix}
\appendix
\input{Thesis_Appendix}

\bibliographystyle{aipnumunsrt4-1}

\input{Thesis.bbl}

\input{Thesis_Backmatter}
\end{document}

%% file: Thesis_Abkuerzungen.tex
\newcommand{\e}[1]{\text{e}^{#1}}
\newcommand{\Glam}{\mathcal G^\Lambda}
\newcommand{\Gamlam}{\Gamma^\Lambda}
\newcommand{\Gamlamtilde}{\tilde\Gamma^\Lambda}
\newcommand{\Gamlamcoeff}[2]{\Gamma^{(#1)\,\Lambda}_{#2}}
\newcommand{\Vertex}[1]{\Gamma^{(4)\,\Lambda}_{#1}}
\newcommand{\VertexNull}[1]{\Gamma^{(4)\,\Lambda_0}_{#1}}
\newcommand{\tr}{\operatorname{tr}}
\newcommand{\dfunc}[2][]{\frac{\delta #1}{\delta #2}}
\newcommand{\secdfunc}[3]{\frac{\delta^2 #1}{\delta {#2}\delta {#3}}}

\newcommand{\conj}{\operatorname{conj.}}

\newcommand{\G}[1]{G^\Lambda_{#1}}

\newcommand{\SL}[1]{S^\Lambda_{#1}}

\newcommand{\mtau}[2]{\tau^{(#1)}_{#2}}

\newcommand{\intdrei}[1]{\int\negthickspace\negthinspace\frac{d^3 #1}{(2\pi)^3}}
\newcommand{\intzwei}[1]{\int\negthickspace\negthinspace\frac{d^2 #1}{(2\pi)^2}}
\newcommand{\inteins}[1]{\int\negthickspace\negthinspace\frac{d #1}{2\pi}}

\newcommand{\R}[1]{R^\Lambda_{#1}}
\newcommand{\tR}[1]{\tilde R^\Lambda_{#1}}
\newcommand{\cc}[1]{c^\dagger_{#1}}
\newcommand{\ca}[1]{c_{#1}}
\newcommand{\ccup}[1]{c^\dagger_{#1\uparrow}}
\newcommand{\caup}[1]{c_{#1\uparrow}}
\newcommand{\ccdown}[1]{c^\dagger_{#1\downarrow}}
\newcommand{\cadown}[1]{c_{#1\downarrow}}
\newcommand{\psic}[1]{\bar\psi_{#1}}
\newcommand{\psia}[1]{\psi_{#1}}
\newcommand{\psicup}[1]{\bar\psi_{#1\uparrow}}
\newcommand{\psiaup}[1]{\psi_{#1\uparrow}}
\newcommand{\psicdown}[1]{\bar\psi_{#1\downarrow}}
\newcommand{\psiadown}[1]{\psi_{#1\downarrow}}
\newcommand{\phic}[1]{\bar\phi_{#1}}
\newcommand{\phia}[1]{\phi_{#1}}
\newcommand{\phicp}[1]{\bar\phi_{#1+}}
\newcommand{\phiap}[1]{\phi_{#1+}}
\newcommand{\phicm}[1]{\bar\phi_{#1-}}
\newcommand{\phiam}[1]{\phi_{#1-}}
\DeclareMathOperator \re {Re}
\DeclareMathOperator \im {Im}
\DeclareMathOperator \sgn {sgn}

\newcommand{\hFB}[4]{h^{#1,\Lambda}_{#2}(#3,#4)}
\newcommand{\VPH}[2]{V^{\text{PH},\Lambda}_{#1}(#2)}

\newcommand{\etal}{\textit{et~al.}}
\newcommand{\ie}{\textit{i.\,e.}}

%% file: Thesis_Titlepage.tex

\begin{titlepage}
\begin{center}
\vspace*{3cm}
{\huge\usekomafont{title} Functional renormalization group study of fluctuation effects in fermionic superfluids}

\vspace{3cm} {\small Von der Fakultät Mathematik und Physik der Universität Stuttgart zur\\
Erlangung der Würde eines Doktors der Naturwissenschaften\\
(Dr.\ rer.\ nat.) genehmigte Abhandlung}

\vspace{2cm}

\small vorgelegt von
\bigskip
\large \\ \textbf{Andreas Eberlein}
\bigskip

\small aus Lichtenfels

\vspace{2cm}

\centering
\begin{tabular}{ll}
	Hauptberichter: &		Prof. Dr. Walter Metzner\\
	Mitberichter:		&		Prof. Dr. Alejandro Muramatsu\\
									&		Prof. Dr. Manfred Salmhofer\\[7.5mm]
\end{tabular}

Tag der mündlichen Prüfung: 22. März 2013\\

\vspace{\fill} \textsc{Max-Planck-Institut für Festkörperforschung\\
Stuttgart 2013}

\end{center}
\end{titlepage}

\cleardoublepage

\begin{mydedication}
	For my parents
\end{mydedication}

%% file: Thesis_Introduction.tex
\chapter{Introduction}
\label{chap:Introduction}

\section{Context and motivation}
This thesis is concerned with ground state properties of two-dimensional correlated superconductors and fermionic superfluids. The former have attracted a lot of research interest since the discovery of high-temperature superconductivity in cuprate materials by Bednorz and Müller~\cite{Bednorz1986} more than a quarter of a century ago. Somewhat later, successes in experiments with cold atomic gases of fermions (for reviews see~\cite{Bloch2008,Esslinger2010}) opened a new chapter not only in atomic physics but also in condensed matter physics~\cite{Lewenstein2007} and renewed the interest in low-dimensional systems of correlated fermions. In such systems, fluctuation effects are particularly strong due to the low dimensionality. This often leads to competing instabilities and rich phase diagrams. 
The behaviour of low-dimensional fermionic systems at finite temperatures is also of interest, because it may deviate from Landau-Fermi liquid theory. The cuprates for example exhibit anomalous transport properties in their normal state (see~\cite{Hussey2008} for a review), where in particular the so-called pseudogap is observed. The latter characterizes a phase without apparent symmetry breaking, but with gaps in the one-particle and magnetic excitation spectra, whose origin is not understood to date. A similar effect was predicted for attractively interacting fermions~\cite{Randeria1992} and has recently been observed in experiments with cold atoms~\cite{Feld2011}.

Shortly after the discovery of high-temperature superconductivity in the cuprates, Anderson~\cite{Anderson1987} suggested that the Hubbard model on the two-dimensional square lattice should contain the essential ingredients to describe these materials. Zhang and Rice~\cite{Zhang1988} emphasized that the interesting physics at low energies should even be captured in the single-band Hubbard model. 
The latter was originally proposed by Hubbard~\cite{Hubbard1963} for the description of ferromagnetism and correlation driven metal-insulator transitions in systems with partially occupied atomic $d$-shells. Besides, it was independently suggested for the study of ferromagnetism in such systems by Kanamori~\cite{Kanamori1963} and Gutzwiller~\cite{Gutzwiller1963}. Since then, it has been used to describe very different phenomena in condensed matter physics. The progress in the field of cold atom physics made it even possible to simulate this model using optical lattices. 
The three-dimensional fermionic case was first realized by Köhl~\etal~\cite{Koehl2005}, allowing for the observation of a metal to band insulator transition. Recently, first indications of short-range antiferromagnetic correlations were observed~\cite{Greif2012-arXiv}.

\begin{figure}
	\centering
	\includegraphics[width=0.25\linewidth]{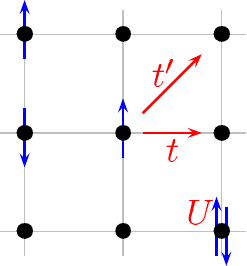}
	\caption{Visualization of the parameters in the Hubbard model.}
	\label{fig:Intro:Lattice}
\end{figure}

The Hubbard model is described by the Hamiltonian
\begin{equation}
	\mathcal H = \sum_{i,j, \sigma} t_{ij} \cc{i\sigma} \ca{j \sigma} - \mu \sum_{i,\sigma} n_{i\sigma} + U \sum_i n_{i\uparrow} n_{i\downarrow}
	\label{eq:Intro:HubbardHamiltonian}
\end{equation}
where $t_{ij}$ describes the hopping amplitudes between the lattice sites labelled by $i$ and $j$, $c^{(\dagger)}_{i\sigma}$ annihilates (creates) a fermion with spin projection $\sigma$ on site $i$ and $n_{i\sigma} = \cc{i\sigma} \ca{i\sigma}$ is the density of fermions with spin projection $\sigma$ on site $i$. The chemical potential $\mu$ determines the fermionic density and is included for convenience.
In this work, the hopping amplitudes are restricted to
\begin{equation}
	t_{ij} = \left\{\begin{array}{r@{\quad} l}
	                	-t	&	\text{in case sites $i$, $j$ are nearest neighbours and} \\
										-t'	&	\text{in case sites $i$, $j$ are next-nearest neighbours.}
	                \end{array}
\right.
\end{equation}
The fermions interact via an on-site interaction of strength $U$. This short-range interaction can be seen as arising from the Coulomb interaction between charged particles that is screened by degrees of freedom which are not contained in the (effective) model any more. In this sense one can speak about superconductivity in the Hubbard model, while it is more appropriate to speak about superfluidity in case the microscopic interaction is short ranged. Note however that the spectrum of collective excitations is different for charged or neutral particles~\cite{Anderson1958,Belkhir1994,Altland2006}: 
In a superfluid, the phase mode of the gap is a massless Goldstone mode with a linear dispersion. In a superconductor, it becomes an excitation with the dispersion of the plasma mode due to the Coulomb interaction via the Anderson-Higgs mechanism. The model parameters are visualized in figure~\ref{fig:Intro:Lattice}. After Fourier transformation to momentum space, the fermionic dispersion reads
\begin{equation}
	\xi(\boldsymbol k) = -2 t (\cos k_x + \cos k_y) - 4 t' \cos k_x \cos k_y - \mu,
	\label{eq:Intro:Dispersion}
\end{equation}
where momenta are measured in units of the inverse lattice constant. In the following, the nearest neighbour hopping $t$ is used as unit of energy and is set to one.

The seemingly simple model in~\eqref{eq:Intro:HubbardHamiltonian} can lead to complex behaviour through the competition between the kinetic energy that delocalizes the fermions and the interaction energy that localizes them. In the cuprates the kinetic and interaction energies seem to be of similar size~\cite{Basov2011}. The resulting competition between delocalization and localization tendencies gives rise to enhanced charge as well as spin fluctuations and a strong renormalization of fermionic quasiparticles. The intermediate correlation regime is also of interest in systems of attractively interacting fermions, where the so-called BCS-BEC crossover occurs~\cite{Eagles1969,Leggett1980,Nozieres1985}. 
It describes a state that is neither a BCS superfluid of weakly bound Cooper pairs nor an interacting (hard-core) Bose gas of small molecules. The former is found for weak and the latter for strong attraction between the fermions. The intermediate coupling regime is characterized by strong fluctuations.

The regime where the kinetic and interaction energies are comparable is difficult to describe theoretically. Most methods approach it either from weak or from strong coupling. Strong-coupling methods include Quantum Monte Carlo (QMC) simulations, dynamical mean-field theory (DMFT)~\cite{Metzner1989,Georges1992b} (see~\cite{Georges1996} for a review), slave-particle gauge theories (see~\cite{Lee2006} for a review) and large-$N$~\cite{Sachdev1991} or gradient~\cite{Falb2008} expansions for the $t$-$J$ model. QMC is exact up to numerical and statistical errors and was extensively applied to the Hubbard model (see for example~\cite{Hirsch1985,White1989,Loh1990,Aimi2007,Yanagisawa2010,Furukawa1992,Scalettar1989,Singer1996}, or~\cite{Bulut2002,Scalapino2007} for reviews). 
However, for the repulsive Hubbard model away from half-filling, the so-called sign-problem imposes severe limitations on the accessible temperatures and system sizes~\cite{Loh1990}, making the detection of $d$-wave superfluidity a formidable task~\cite{White1989,Loh1990,Aimi2007,Yanagisawa2010}. 
DMFT solves the atomic limit exactly and can be seen as a large-$N$ expansion in the coordination number of the lattice. One milestone achieved by DMFT was to provide an accurate description of the Mott-Hubbard metal-insulator transition in high dimensions~\cite{Jarrell1992,Rozenberg1992,Georges1992}.
The method or its cluster extensions can describe long-range order also for non-local order parameters~\cite{Maier2000}, but the accessible correlation lengths for fluctuations are small (restricted by the size of the cluster, see for example~\cite{Maier2005}). This prevents from describing the effects of spatially long-range collective mode fluctuations in correlated superfluids or superconductors.

Weak-coupling methods usually depart from a Fermi liquid or mean-field state and treat the fluctuations around it. Such methods include self-consistent resummations of perturbation theory like the fluctuation exchange approximation (FLEX)~\cite{Bickers1989} (see~\cite{Scalapino1995} for a review), the parquet approach~\cite{Bickers1991a}, self-consistent perturbation theory around an ordered state~\cite{Georges1991} or renormalization group (RG) methods~\cite{Solyom1979,Salmhofer2007book,Metzner2012}. 
The FLEX approximation allows to describe spin-fluctuation mediated pairing in the repulsive Hubbard model and yielded a lot of insight into the properties of the normal and superfluid phases (see for example the book by Manske~\cite{Manske2004} for an overview). However, it resums only a small set of diagrams of perturbation theory and its application cannot be justified in the interesting case of competing instabilities. This is different for the parquet approach, which takes all interaction channels into account on equal footing. The price to pay is that the parquet self-consistency equations are very complex and were solved numerically only in a few cases (see for example~\cite{Zheleznyak1997}). (Functional) Renormalization group methods proved as a vital alternative, as they resum perturbation theory in a scale-separated way and treat all interaction channels on equal footing. Besides, they are physically transparent and able to treat critical phenomena or fluctuations with long correlation lengths.

\begin{figure}
	\centering
	\includegraphics[width=0.45\linewidth]{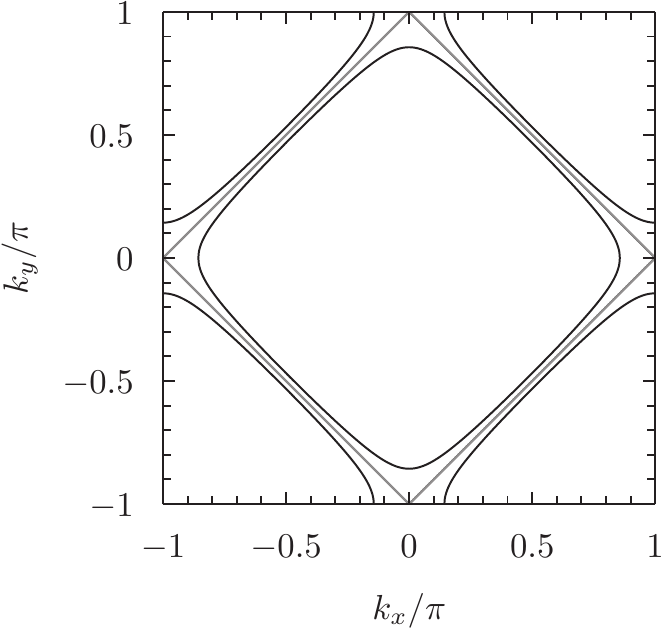}\hspace{0.09\linewidth}\includegraphics[width=0.45\linewidth]{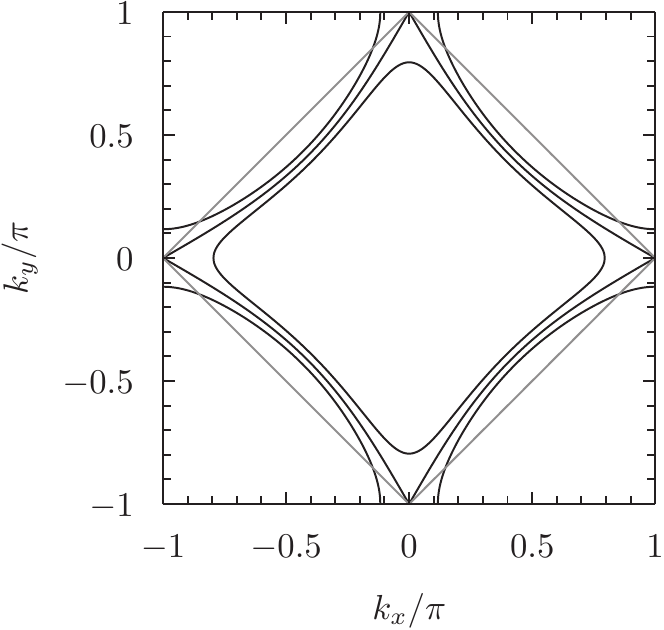}
	\caption{Fermi surfaces slightly below, at and slightly above van Hove filling for $t' = 0$ (left) and $t' = -0.25$ (right). The grey line shows the umklapp surface or magnetic Brillouin zone boundary.}
	\label{fig:Intro:FermiSurfaces}
\end{figure}
For weakly correlated fermions, competing instabilities occur only for special Fermi surface geometries. For curved and regular Fermi surfaces, \ie\ Fermi surfaces without flat pieces or saddle points where the Fermi velocity vanishes, the only weak-coupling instability of the Fermi liquid is towards superfluidity (or superconductivity)~\cite{Feldman1993}. This changes in the presence of flat Fermi surface pieces, which give rise to nesting and enhance the singularities of fermionic propagator loops in perturbation theory. This is also the case if the Fermi surface contains saddle points, which entail so-called van Hove singularities in the density of states. These play an important role not only in weakly correlated systems, but also in (strongly correlated) cuprate superconductors~\cite{Markiewicz1997}. Examples for interesting Fermi surface geometries are shown in figure~\ref{fig:Intro:FermiSurfaces}. 
The left panel shows Fermi surfaces that arise in a system with only nearest neighbour hopping for fillings slightly below, at and slightly above half-filling. In the half-filled case, the Fermi surface is perfectly nested. Below or above half-filling, the Fermi surface is curved but has nevertheless pieces where the curvature is small. The right panel shows Fermi surfaces for an intermediate value of $t' = -0.25$ for fermionic densities below, at and above van Hove filling. At van Hove filling, the saddle points of the dispersion~\eqref{eq:Intro:Dispersion} at $\boldsymbol k = (0,\pi)$ and symmetry related points are part of the Fermi surface. In the non-interacting system, this is the case for $\mu = 4 t'$. For negative (positive) values of $t'$, another class of interesting Fermi points, the so-called hot spots, appears above (below) van Hove filling. These are located at the crossing points of the Fermi surface and the umklapp surface, which is formed by 
the lines connecting $(0,\pi)$ with $(\pi,0)$ and similarly for symmetry related points. An example for a Fermi surface with hot spots is shown in the right panel of figure~\ref{fig:Intro:FermiSurfaces}. Hot spots allow for scattering processes between low energy states in which the momenta of the particles are conserved only up to a reciprocal lattice vector. These drive and couple different scattering channels, which gives rise to interesting low energy physics~\cite{Honerkamp2001a,Metlitski2010b}.

Perturbative renormalization group methods have a long history in statistical mechanics and in the study of one-dimensional fermionic systems (for reviews see~\cite{Wilson1974,Berges2002} and~\cite{Solyom1979}, respectively). They were first applied to fermionic systems in higher dimensions by mathematicians in order to obtain rigorous statements~\cite{Feldman1990,Benfatto1990,Benfatto1990a,Feldman1991} and were popularised among physicists by several authors~\cite{Bourbonnais1991,Shankar1991,Polchinski1993,Shankar1994}. The perturbative renormalization group or scaling theory originally proposed by Kadanoff~\cite{Kadanoff1966} and Wilson~\cite{Wilson1971,Wilson1971a} was reformulated later in terms of exact flow equations for generating functionals. The latter RG schemes were termed exact or functional RG because they are based on exact flow equations for generating functionals and yield flow equations for coupling functions. 
Starting from the partition function, Polchinski~\cite{Polchinski1984} derived an exact hierarchy of flow equations for bosonic amputated connected Green functions and used it to prove renormalizability of certain scalar field theories. 
A similar scheme for fermionic fields was derived by Brydges and Wright~\cite{Brydges1988}. Based on the generating functional for Wick-ordered effective interactions, an alternative hierarchy was formulated by Wieczerkowski~\cite{Wieczerkowski1988} and Salmhofer~\cite{Salmhofer1998} for bosonic and fermionic fields, respectively. 
These schemes turned out to be particularly useful for proving rigorous results. Hierarchies of flow equations for one-particle irreducible (1PI) vertex functions for bosons and fermions were derived by Wetterich~\cite{Wetterich1993} and Salmhofer and Honerkamp~\cite{Salmhofer2001}, respectively. The 1PI scheme turned out to be convenient for practical calculations because it i) contains non-trivial renormalization contributions for the determination of Fermi liquid instabilities already on one-loop level in an equation that is local in the scale parameter, ii) contains only one-particle irreducible renormalization contributions and iii) allows for a convenient treatment of self-energy insertions. 
Functional renormalization group methods proved very successful in the classification of weak-coupling instabilities of the Fermi liquid for example in the two-dimensional Hubbard model (see~\cite{Zanchi1996,Zanchi1998,Zanchi2000,Halboth2000a,Halboth2000b,Honerkamp2001a,Honerkamp2001b,Honerkamp2001c,Honerkamp2002,
Honerkamp2003,Honerkamp2004,Katanin2009,Husemann2009,Husemann2012,Giering2012} for an incomplete list and~\cite{Metzner2012} for a recent review). These studies unambiguously established the formation of $d$-wave superfluidity in the weak coupling regime, but were not able to access the symmetry broken phase. A drawback of the presently available truncations of the hierarchy of renormalization group equations for fermions is that they are restricted to weak microscopic interactions~\cite{Salmhofer2001}.

Spontaneous symmetry breaking in fermionic systems can be studied within the functional RG using several approaches that vary with respect to the fields in the effective action. The fermionic two-particle interaction can be decoupled and bosonic auxiliary fields describing collective degrees of freedom be introduced via Hubbard-Stratonovich transformations~\cite{Popov1987}. One way to proceed is to integrate out the fermions completely and to study an effective action for the order parameter fields. 
Alternatively, keeping the fermionic degrees of freedom, the RG flow can be computed for a mixed fermionic and bosonic action. Following this approach Baier~\etal~\cite{Baier2004} studied antiferromagnetism in the half-filled repulsive Hubbard model.
They demonstrated that the low-energy collective behaviour, which is described by the non-linear sigma model, can be recovered from truncated flow equations. 
Baier~\etal\ pointed out that the correct fermionic one-loop flow is captured only if the regenerated two-fermion interactions are not neglected, as is usually done in the simplest truncations, but treated for example by dynamical bosonization as suggested by Gies and Wetterich~\cite{Gies2002}. 
Floerchinger and Wetterich~\cite{Floerchinger2009} proposed an exact flow equation that allows to perform Hubbard-Stratonovich transformations continuously during the flow. Functional RG flows of mixed fermionic and bosonic actions were also used to study superfluidity in the two-dimensional attractive Hubbard model~\cite{Strack2008} or in three-dimensional continuum systems of attractively interacting fermions~\cite{Floerchinger2008b,Bartosch2009,Diehl2010}. However, the decoupling of the fermionic two-particle interaction by a Hubbard-Stratonovich transformation is ambiguous if several possible channels exist, as in the case of competing instabilities. 
The ambiguity can be resolved by keeping the microscopic fermionic interaction explicitly in the action and dynamically bosonizing only the fluctuation contributions~\cite{Friederich2010}, which requires the introduction of several bosonic fields. This approach was used to investigate antiferromagnetism and $d$-wave superfluidity~\cite{Friederich2011} in the two-dimensional repulsive Hubbard model at finite temperatures. 
Note that the choice of auxiliary fields or the truncation of the effective action is not straightforward in situations where several interaction channels compete. 

Instead of introducing bosonic degrees of freedom, spontaneous symmetry breaking can also be studied within a purely fermionic formalism, where all interaction channels are kept on equal footing. This also allows to gain insight about feedback effects of critical fluctuations to other interaction channels. 
Salmhofer~\etal~\cite{Salmhofer2004} showed that reduced models exhibiting spontaneous symmetry breaking can be solved exactly within a modified one-loop truncation of the fermionic RG hierarchy proposed by Katanin~\cite{Katanin2004a} that takes into account certain renormalization contributions from the two-loop level as additional self-energy insertions. Gersch~\etal~\cite{Gersch2008} applied the modified one-loop truncation to the attractive Hubbard model and demonstrated that it is possible to continue fermionic RG flows beyond the critical scale also in non-reduced models without employing another method for the low energy modes. This avoids issues like a dependence of the results on the scale where the method is changed that appear in hybrid approaches, for example when combining functional RG and mean-field theory (see for example~\cite{Reiss2007}). The study by Gersch~\etal~\cite{Gersch2008} yielded reasonable results for the superfluid gap even with a rather simple approximation for the fermionic 
two-particle vertex. The external pairing field could be chosen at least two orders of magnitude smaller than the final value of the superfluid gap without encountering unphysical divergences at finite scales. 
Despite these successes, the work by Gersch~\etal~\cite{Gersch2008} leaves room for improvements. First, the momentum resolution of the vertex was rather low, which is an issue due to the large values of the phase mode of the superfluid gap that occur at and below the critical scale. Second, the question about the compatibility of global Ward identities and the modified one-loop truncation remained an open problem. Third, the study did not clarify to what extent the renormalization by collective mode fluctuations is taken into account in the Katanin scheme. 
Fourth, the approximations employed by Gersch~\etal\ did not allow to clarify whether singularities in non-Cooper channels exist in the vertex in the limit of a vanishing external pairing field. These methodological issues are addressed in this thesis. Besides extending and improving the work by Gersch~\etal\ for the attractive Hubbard model, the fermionic RG is also applied to the repulsive Hubbard model in order to study its $d$-wave superfluid ground state.

\section{Thesis outline}
This thesis is organized as follows: In the first part, consisting of chapters~\ref{chap:FunctionalRG} to~\ref{chap:WICharge}, methodological developments are discussed, which are at the heart of this thesis. In the second part that includes chapters~\ref{chap:RPFM} to~\ref{chap:RepulsiveHubbard}, applications of the developed framework to several model systems are presented. In chapter~\ref{chap:Conclusion}, the thesis is summarised.

Ground state properties of correlated superfluids are studied within this thesis using the functional renormalization group method. In order to make the thesis self-contained, this method is introduced in chapter~\ref{chap:FunctionalRG} and the flow equations for fermionic 1PI vertex functions are rederived, including terms in third order in the effective interaction. This derivation is similar to those given by Salmhofer and Honerkamp~\cite{Salmhofer2001} or by Metzner~\etal~\cite{Metzner2012}.

In chapter~\ref{chap:VertexParametrization}, the structure of the Nambu two-particle vertex in a singlet superfluid is clarified. The number of independent components of the vertex is reduced to a minimum by exploiting symmetries, in particular spin rotation invariance, while discrete symmetries like inversion or time reversal symmetry yield constraints on their momentum and frequency dependence. This simplifies the identification of singular dependences of the vertex on external momenta and frequencies, allowing for the definition of interaction channels. Parts of this chapter were published in~\cite{Eberlein2010}.

The decomposition of the vertex in interaction channels forms the basis for the formulation of channel-decomposed renormalization group equations in chapter~\ref{chap:ChannelDecomposition}. These equations extend ideas by Husemann and Salmhofer~\cite{Husemann2009} to the case of symmetry breaking in the Cooper channel. They allow to isolate the singular dependences of the flow equations on external momenta and frequencies in one variable per equation, thus providing a good starting point for the formulation of approximations for the effective interactions in the channels and their efficient computation. In chapter~\ref{chap:ChannelDecomposition}, channel-decomposed RG equations on one-loop level, which are based on the modified one-loop truncation by Katanin~\cite{Katanin2004a}, and equations on two-loop level, which take all renormalization contributions to the two-particle vertex in third order in the effective interaction into account, are presented. 
Besides providing a good starting point for the formulation of approximations, the channel-decomposed flow equations yield insight into the singularity structure of the Nambu two-particle vertex in the limit where the external pairing field vanishes. In particular, the feedback of phase fluctuations on non-Cooper channels can be studied because all interaction channels are present in the purely fermionic formulation. The channel-decomposition scheme on one-loop level was published in~\cite{Eberlein2010}.

In chapter~\ref{chap:WICharge}, the question about the compatibility of truncated flow equations and global conservation laws is addressed. It is found that the Katanin scheme is compatible with the Ward identity for global charge conservation only up to terms of third order in the effective interaction. This is improved by considering all renormalization contributions to the two-particle vertex in third order in the effective interaction, \ie\ on two-loop level. It is demonstrated in chapter~\ref{chap:AttractiveHubbard} that the Ward identity can be enforced in the numerical solution of flow equations by fixing the relation between singular quantities through the Ward identity. This completes the methodological part of this thesis.

In chapter~\ref{chap:RPFM}, the exact solution of a reduced pairing and forward scattering model is presented. It yields insight into the vertex in a singlet superfluid in the limit where the external pairing field vanishes. Capturing the exact solution of this model within the channel-decomposition scheme is important for the description of the singularities associated with the critical scale for superfluidity in non-reduced models. Subsequently, small momentum transfers are allowed for and the resummation of all chains of Nambu particle-hole diagrams yields insight into the singular momentum and frequency dependence of various components of the vertex. Parts of this chapter were published in~\cite{Eberlein2010}.

In chapters~\ref{chap:AttractiveHubbard} and~\ref{chap:RepulsiveHubbard}, the application of the channel-decomposition scheme to the attractive and the repulsive Hubbard model is described, respectively. The attractive Hubbard model can be seen as a good testing ground for new approximation schemes because it has an $s$-wave superfluid ground state in a large portion of parameter space that is present already on mean-field level, while fluctuations renormalise its properties like the size of the order parameter. In chapter~\ref{chap:AttractiveHubbard}, approximations for the momentum and frequency dependence of the vertex that are applied within the channel-decomposition scheme are discussed. Subsequently, numerical results for the fermionic two-particle vertex and self-energy are presented. Their momentum and frequency dependence as well as the impact of fluctuations on their flow are of particular interest.

In chapter~\ref{chap:RepulsiveHubbard}, the computation of ground state properties of the repulsive Hubbard model is described in case $d$-wave superfluidity is the leading instability. This constitutes an extension of former instability analyses that allows to compute not only critical scales but also the properties of the $d$-wave superfluid ground state. The approximations employed in this chapter are less sophisticated than those of chapter~\ref{chap:AttractiveHubbard}. Nevertheless, it is demonstrated that the channel-decomposition scheme is able to cope with competition of instabilities and symmetry breaking into phases that are not captured by mean-field theory.

In chapter~\ref{chap:Conclusion}, the thesis is summarised and conclusions are drawn. Furthermore, some interesting directions for future research that are based on the methodological developments of this thesis are outlined.

%% file: Thesis_fRG.tex
\chapter{Functional renormalization group}
\label{chap:FunctionalRG}
In this chapter, the derivation of functional renormalization group equations for one-particle irreducible (1PI) vertex functions for fermionic fields is outlined. Flow equations in the 1PI formalism were first derived for bosonic fields by Wetterich~\cite{Wetterich1993} and for fermionic fields by Salmhofer and Honerkamp~\cite{Salmhofer2001}. The derivation in this chapter is very similar to the one presented by Metzner~\etal~\cite{Metzner2012} and uses the sign convention of the book by Negele and Orland~\cite{Negele1998}. More extensive derivations and discussions can be found in the reviews by Berges \etal~\cite{Berges2002} and Metzner~\etal~\cite{Metzner2012}.

The derivation starts from the generating functional for connected Green functions
\begin{equation}
	\mathcal G[\bar\eta,\eta] = \ln \int D\mu_Q[\bar\chi,\chi] \e{-\mathcal V[\bar\chi,\chi]-(\bar\chi,\eta)-(\bar\eta,\chi)}
\end{equation}
where $\eta$, $\bar\eta$ are Grassmann source fields and $\chi$, $\bar\chi$ Grassmann fields representing the physical fermionic degrees of freedom. $\mathcal V[\bar\chi,\chi]$ is the interaction part of the microscopic action and
\begin{equation}
	D\mu_Q[\bar\chi,\chi] = \frac{1}{\det Q} D\bar\chi D\chi \e{(\bar\chi,Q\chi)}
\end{equation}
the measure with $(-Q)$ being the kernel of the bilinear part of the action, \ie\ the inverse bare one-particle Green function $(-Q) = (G_0)^{-1}$. The `scalar product' notation $(\bar\chi,\eta) = \sum_\alpha \bar\chi_\alpha \eta_\alpha$ implies the summation over a multi-index including for example Matsubara frequencies, momenta and spin or Nambu indices. Connected Green functions are obtained after functional differentiation of $\mathcal G[\bar\eta,\eta]$ with respect to the source fields.

For the derivation of renormalization group differential equations, an infrared cutoff that suppresses low-energy modes in the functional integral is introduced in the bilinear part of the action by replacing $Q\rightarrow Q^\Lambda$. Consequently, the generating functional becomes scale-dependent, $\mathcal G\rightarrow \mathcal G^\Lambda$, and reads
\begin{equation}
	\Glam[\bar\eta,\eta] = \ln \int D\mu_{Q^\Lambda}[\bar\chi,\chi] \e{-\mathcal V[\bar\chi,\chi]-(\bar\chi,\eta)-(\bar\eta,\chi)}
\end{equation}
with the scale dependent measure
\begin{equation}
	\label{eq:fRG:measure}
	D\mu_{Q^\Lambda}[\bar\chi,\chi] = \frac{1}{\det Q^\Lambda} D\bar\chi D\chi \e{(\bar\chi,Q^\Lambda\chi)}.
\end{equation}
The regularization has to fulfil the requirement that the generating functional vanishes at an initial scale $\Lambda_0$, where all fermionic modes in the functional integral are suppressed, and that the full generating functional is recovered for $\Lambda \rightarrow 0$, where all fermionic modes are taken into account. It is usually chosen as an infrared regularization that suppresses low-energy modes at high scales. Apart from that the regularization scheme can be chosen with considerable freedom, including momentum and\,/\,or frequency cutoffs, interaction cutoffs~\cite{Honerkamp2004} and temperature cutoffs~\cite{Honerkamp2001b}. Even the external pairing field can be treated as a regulator (see chapter~\ref{chap:AttractiveHubbard}).

The functional renormalization group differential equation follows after differentiation of $\mathcal G^\Lambda$ with respect to the scale $\Lambda$
\begin{align}
	\partial_\Lambda \Glam &= \e{-\Glam} \partial_\Lambda \e{\Glam}= \nonumber\\
	&= -\partial_\Lambda(\ln\det Q^\Lambda) + \e{-\Glam}\frac{1}{\det Q^\Lambda}\int D[\bar\chi,\chi] \e{(\bar\chi,Q^\Lambda\chi)-\mathcal V[\bar\chi,\chi]-(\bar\chi,\eta)-(\bar\eta,\chi)}(\bar\chi,\dot Q^\Lambda\chi)=\nonumber\\
	&=\tr(\dot Q^\Lambda G^\Lambda_0) - \e{-\mathcal G^\Lambda}\Bigl(\dfunc{\eta},\dot Q^\Lambda\dfunc{\bar\eta}\Bigr)\e{\mathcal G^\Lambda},
\end{align}
where the first term results from differentiation of the determinant and the second from differentiation of the exponential factor in the measure~\eqref{eq:fRG:measure}. In the last equality, the relation
\begin{equation}
	\partial_\Lambda\ln\det Q^\Lambda=\partial_\Lambda\tr\ln Q^\Lambda = -\tr (G^\Lambda_0 \dot Q^\Lambda)
\end{equation}
was exploited and the $\chi$-fields were expressed through derivatives with respect to $\eta$-fields. Evaluating the remaining functional derivatives yields the functional renormalization group differential equation
\begin{equation}
\label{eq:fRG:funcRGGreen}
	\partial_\Lambda \mathcal G^\Lambda = \tr(\dot Q^\Lambda G^\Lambda_0) - \Bigl(\dfunc[\mathcal G^\Lambda]{\eta},\dot Q^\Lambda \dfunc[\mathcal G^\Lambda]{\bar\eta}\Bigr) + \tr\Bigl(\dot Q^\Lambda \secdfunc{\Glam}{\bar\eta}{\eta}\Bigr)
\end{equation}
where $\tr\bigl(\dot Q^\Lambda \secdfunc{\Glam}{\bar\eta}{\eta}\bigr) = \sum_{\alpha,\beta} \dot Q^\Lambda_{\alpha\beta} \secdfunc{\Glam}{\bar\eta_\beta}{\eta_\alpha}$. This functional renormalization group equation describes how the scale dependent generating functional $\mathcal G^\Lambda$ changes when the cutoff is successively lowered and more modes are taken into account. Renormalization group equations for connected Green functions follow from this functional renormalization group equation by expansion of both sides in the source fields and comparison of coefficients.

Starting from~\eqref{eq:fRG:funcRGGreen}, renormalization group equations for one-particle irreducible vertex functions can easily be derived. Their generating functional is the so-called effective action $\Gamma[\bar\phi,\phi]$, the Legendre transform of the generating functional for connected Green functions $\mathcal G[\bar\eta,\eta]$,
\begin{equation}
	\Gamma[\bar\phi,\phi] = -\mathcal G[\bar\eta,\eta]-(\bar\eta,\phi)-(\bar\phi,\eta),
\end{equation}
with fermionic Grassmann fields $\phi = \phi[\bar\eta,\eta]$ and $\bar\phi = \bar\phi[\bar\eta,\eta]$. The Legendre transform is an involution so that $\mathcal G[\bar\eta,\eta]$ is also the Legendre transform of $\Gamma[\bar\phi,\phi]$. The generating functionals and the fields are connected by
\begin{align}
	\bar\phi &= \frac{\delta\mathcal G}{\delta\eta}		&		\phi &= -\frac{\delta\mathcal G}{\delta \bar\eta} &
	\bar\eta &= \frac{\delta\Gamma}{\delta\phi}		&		\eta &= -\frac{\delta\Gamma}{\delta\bar\phi}.
	\label{eq:fRG:FieldConnection}
\end{align}
These relations follow from the condition that $\Gamma[\bar\phi,\phi]$ (or $\mathcal G[\bar\eta,\eta]$) is stationary under variation of $\eta$ and $\bar\eta$ (or $\bar\phi$ and $\phi$) while $\bar\phi$ and $\phi$ (or $\eta$ and $\bar\eta$) are kept fixed.

The introduction of an infrared cutoff in $\mathcal G$ also makes the effective action and the relations between $\eta$, $\bar\eta$ and $\phi$, $\bar\phi$ scale dependent while the fields $\phi$ and $\bar\phi$ are fixed. Consequently, the scale dependent generating functional for one-particle irreducible vertex functions reads
\begin{equation}
	\label{eq:fRG:Legendre}
	\Gamma^\Lambda[\bar\phi,\phi] = -\mathcal G^\Lambda[\bar\eta^\Lambda,\eta^\Lambda]-(\bar\eta^\Lambda,\phi)-(\bar\phi,\eta^\Lambda)
\end{equation}
where
\begin{align}
	\eta^\Lambda &= \eta^\Lambda[\bar\phi,\phi]	&	\bar\eta^\Lambda = \bar\eta^\Lambda[\bar\phi,\phi]
\end{align}
and
\begin{align}
	\bar\phi &= \frac{\delta\mathcal G^\Lambda}{\delta\eta^\Lambda}		&		\phi &= -\frac{\delta\mathcal G^\Lambda}{\delta \bar\eta^\Lambda} &
	\bar\eta^\Lambda &= \frac{\delta\Gamma^\Lambda}{\delta\phi}		&		\eta^\Lambda &= -\frac{\delta\Gamma^\Lambda}{\delta\bar\phi}.
\end{align}
At the initial scale $\Lambda_0$, $\Gamma^{\Lambda_0}[\bar\phi,\phi]$ equals the regularized bare action of the system. For $\Lambda = 0$, the full effective action $\Gamma^{\Lambda = 0}[\bar\phi,\phi] = \Gamma[\bar\phi,\phi]$ is obtained. After differentiation of equation~\eqref{eq:fRG:Legendre} with respect to $\Lambda$, most terms cancel, yielding
\begin{equation}
	\partial_\Lambda \Gamlam = -\partial_\Lambda \Glam = -\tr\bigl(\dot Q^\Lambda G^\Lambda_0\bigr) + \Bigl(\dfunc[\mathcal G^\Lambda]{\eta^\Lambda},\dot Q^\Lambda \dfunc[\mathcal G^\Lambda]{\bar\eta^\Lambda}\Bigr) - \tr\Bigl(\dot Q^\Lambda \secdfunc{\Glam}{\bar\eta^\Lambda}{\eta^\Lambda}\Bigr).
\end{equation}
The $\Glam$-functional on the right-hand side has to be expressed through the $\Gamlam$-functional. For the second term this is straightforward using the above relations. The last term is rewritten by exploiting the exact reciprocity relation (see for example~\cite{Negele1998})
\begin{equation}
	\label{eq:fRG:ReciprocityRelation}
	\begin{pmatrix}
		\secdfunc{\Glam}{\bar\eta}{\eta} & -\secdfunc{\Glam}{\bar\eta}{\bar\eta} \\
		-\secdfunc{\Glam}{\eta}{\eta}	&	\secdfunc{\Glam}{\eta}{\bar\eta}
	\end{pmatrix}
=\begin{pmatrix}
		\secdfunc{\Gamlam}{\bar\phi}{\phi} & \secdfunc{\Gamlam}{\bar\phi}{\bar\phi} \\
		\secdfunc{\Gamlam}{\phi}{\phi}	&	\secdfunc{\Gamlam}{\phi}{\bar\phi}
	\end{pmatrix}^{-1}
\end{equation}
(note that the entries of this matrix carry two fermionic multi-indices from the functional derivatives besides the ``field index''). This yields
\begin{equation}
	\label{eq:fRG:funcRGvertex}
\partial_\Lambda \Gamlam = -\tr(\dot Q^\Lambda G^\Lambda_0) - (\bar\phi,\dot Q^\Lambda \phi) -\tr(\dot Q^\Lambda [(\boldsymbol\delta^2 \boldsymbol\Gamma^\Lambda)^{-1}]_{11})
\end{equation}
where
\begin{equation}
	[(\boldsymbol\delta^2 \boldsymbol\Gamma^\Lambda)^{-1}]_{11} = \begin{bmatrix}\begin{pmatrix}
		\secdfunc{\Gamlam}{\bar\phi}{\phi} & \secdfunc{\Gamlam}{\bar\phi}{\bar\phi} \\
		\secdfunc{\Gamlam}{\phi}{\phi}	&	\secdfunc{\Gamlam}{\phi}{\bar\phi}
	\end{pmatrix}^{-1}\end{bmatrix}_{11} = \secdfunc{\Glam}{\bar\eta^\Lambda}{\eta^\Lambda}.
\end{equation}
In this chapter, bold symbols denote matrices in the ``particle-hole space'' spanned by the fields $\phi$, $\bar\phi$ or derivatives thereof. Expanding the effective action in the source fields, its second functional derivative can be written as
\begin{equation}
	\label{eq:fRG:second_derivative_matrix}
	\boldsymbol\delta^2 \boldsymbol\Gamma^\Lambda = \begin{pmatrix}
		\secdfunc{\Gamlam}{\bar\phi}{\phi} & \secdfunc{\Gamlam}{\bar\phi}{\bar\phi} \\
		\secdfunc{\Gamlam}{\phi}{\phi}	&	\secdfunc{\Gamlam}{\phi}{\bar\phi}
	\end{pmatrix} = -(\textbf G^\Lambda)^{-1} + \boldsymbol {\tilde\Gamma}^\Lambda = -(\textbf G^\Lambda)^{-1} (\boldsymbol 1 - \textbf G^\Lambda \boldsymbol {\tilde\Gamma}^\Lambda)
\end{equation}
where $\boldsymbol {\tilde\Gamma}^\Lambda$ is at least quadratic in the fields and $\textbf G^\Lambda$ is the (field independent) fermionic propagator. In the following, the shorthand notation
\begin{equation}
	\secdfunc{\Gamlam}{\bar\phi}{\phi} = \Gamlam_{\bar\phi\phi}
\end{equation}
is partially used for functional derivatives of the effective action or for matrix elements of $\boldsymbol {\tilde \Gamma}^\Lambda$. Because of the field independent part in $\boldsymbol\delta^2 \boldsymbol\Gamma^\Lambda$, its inverse can be computed from a geometric series,
\begin{equation}
	(\boldsymbol\delta^2 \boldsymbol\Gamma^\Lambda)^{-1} = -(\boldsymbol 1 - \textbf G^\Lambda \boldsymbol {\tilde\Gamma}^\Lambda)^{-1} \textbf G^\Lambda = -\sum_{n=0}^\infty \bigl(\textbf G^\Lambda \boldsymbol {\tilde\Gamma}^\Lambda\bigr)^n \textbf G^\Lambda.
\end{equation}
Inserting this expression into equation~\eqref{eq:fRG:funcRGvertex} yields the functional renormalization group equation for one-particle irreducible vertex functions
\begin{equation}
\label{eq:fRG:RGvertexExpansion}
\partial_\Lambda \Gamlam = -\tr(\dot Q^\Lambda G^\Lambda_0) - (\bar\phi,\dot Q^\Lambda \phi) +\tr\Bigl(\dot Q^\Lambda \Bigl[\sum_{n=0}^\infty \bigl(\mathbf G^\Lambda \boldsymbol {\tilde\Gamma}^\Lambda\bigr)^n \textbf G^\Lambda\Bigr]_{11}\Bigr).
\end{equation}
From this equation it can be inferred that the renormalization contributions to one-particle irreducible vertex functions are given by one-loop diagrams which are obtained by forming trees of vertices and closing them with the so-called single-scale propagator
\begin{equation}
	S^\Lambda = G^\Lambda \dot Q^\Lambda G^\Lambda = \partial_\Lambda \G{}|_{\Sigma^\Lambda=\text{const.}},
	\label{eq:fRG:SLdef}
\end{equation}
where $G^\Lambda$ is the regularized full one-particle Green function (see below). The renormalization group flow described by equation~\eqref{eq:fRG:funcRGvertex} has to be integrated from an initial scale $\Lambda_0$ where no modes are taken into account in the generating functional, to the scale $\Lambda = 0$ where the full generating functional is recovered. 

For bosonic fields, an equation very similar to~\eqref{eq:fRG:RGvertexExpansion} can be used as a starting point for non-perturbative RG calculations without truncation of the bosonic potential~\cite{Berges2002}. For fermionic Grassmann fields, the above equation is only meaningful after expansion in the source fields and comparison of coefficients as an equation for vertex functions. In the following, this expansion is restricted to terms with up to six Grassmann fields $\bar\phi$, $\phi$ in the effective action. Furthermore, only normal vertices (with an equal number of Grassmann creators and annihilators) are considered, making the derived equations appropriate for the normal state (in spinor representation) or a fermionic superfluid that is invariant under spin rotations around at least one axis (in Nambu representation).
Thus, the effective action is expanded as
\begin{equation}
	\label{eq:fRG:effaction_ansatz}
	\begin{split}
	\Gamlam[\bar\phi,\phi] &= \Gamlamcoeff{0}{} + \sum_{\alpha,\beta} \Gamlamcoeff{2}{\alpha\beta} \bar\phi_\alpha \phi_\beta + \frac{1}{(2!)^2} \sum_{\alpha,\beta,\gamma,\delta} \Gamlamcoeff{4}{\alpha\beta\gamma\delta} \bar\phi_\alpha \bar\phi_\beta \phi_\gamma \phi_\delta\\
		&\ +\frac{1}{(3!)^2} \sum_{\alpha,\beta,\gamma,\delta,\mu,\nu} \Gamlamcoeff{6}{\alpha\beta\gamma\delta\mu\nu} \bar\phi_\alpha \bar\phi_\beta \bar\phi_\gamma \phi_\delta \phi_\mu \phi_\nu+\ldots,
		\end{split}
\end{equation}
where the Greek (multi-) indices collect quantum numbers like for example frequencies, momenta and spin or Nambu indices. The coefficient of the quadratic term $\bar\phi\phi$ is the inverse of the full one-particle propagator
\begin{equation}
	\Gamlamcoeff{2}{\alpha\beta} = (G^\Lambda)^{-1}_{\alpha\beta}.
\end{equation}
The coefficients of $\bar\phi^n \phi^n$ are the antisymmetrized one-particle irreducible $n$-particle vertex functions. For this ansatz, the matrix propagator that appears in the flow equations reads
\begin{equation}
	\textbf G^\Lambda = \begin{pmatrix} G^\Lambda	&	0	\\	0	&	-(G^\Lambda)^T \end{pmatrix}
\end{equation}
where $(G^\Lambda)^T_{\alpha\beta} = G^\Lambda_{\beta\alpha}$.

\begin{figure}
	\centering
	\includegraphics[scale=0.75]{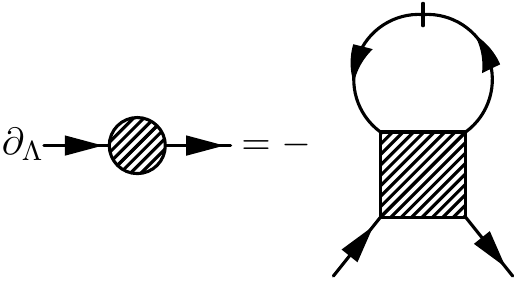}
	\caption{Diagrammatic representation of the renormalization group equation for the self-energy. The slashed line represents the single-scale propagator $S^\Lambda$.}
	\label{fig:fRG:RGDE-Selfenergy}
\end{figure}

\begin{figure}
	\centering
	a)\subfigure{\includegraphics[scale=0.9]{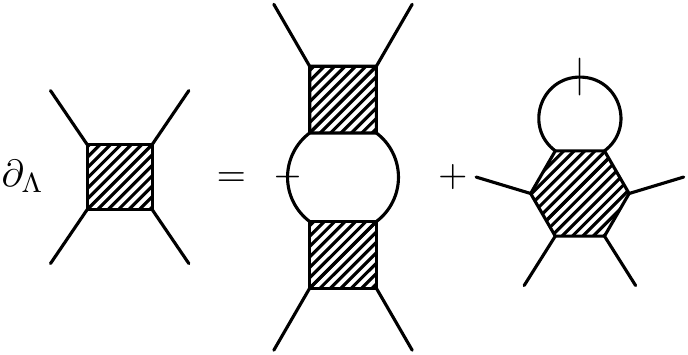}\label{fig:fRG:RGDE-2P-simp}}
	b)\subfigure{\includegraphics[scale=0.9]{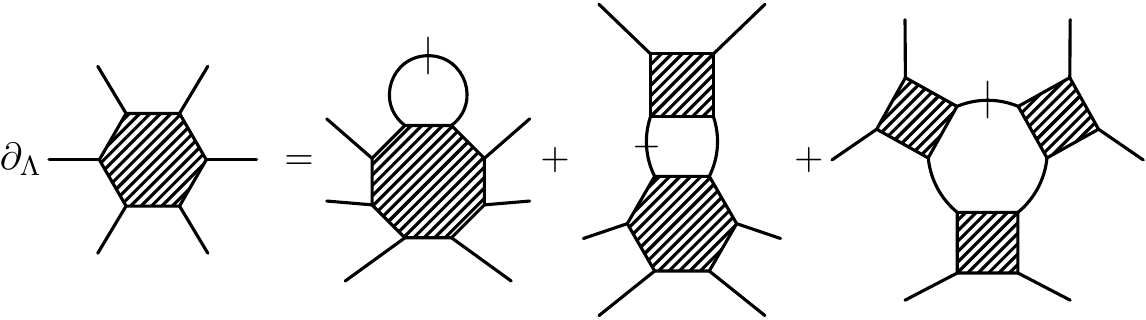}\label{fig:fRG:RGDE-3P-simp}}
	\caption{Simplified diagrammatic representation of the one-loop renormalization group equations for the two- (a) and three-particle (b) vertex illustrating the topological structure of diagrams.}
	\label{fig:fRG:RGDE-2P-3P-simp}
\end{figure}

Analytical expressions for the flow equations are obtained after inserting the ansatz for the fermionic effective action~\eqref{eq:fRG:effaction_ansatz} in equation~\eqref{eq:fRG:RGvertexExpansion} and differentiation with respect to source fields. The equations for the interaction correction to the density of the grand canonical potential, the self-energy and the two-particle vertex read
\begin{gather}
	\partial_\Lambda\Gamlamcoeff{0}{} = \tr(\dot Q^\Lambda (G^\Lambda - G^\Lambda_0)) \label{eq:fRG:Gamma0}\\
	\partial_\Lambda\Gamlamcoeff{2}{\alpha\beta} + \partial_\Lambda Q^\Lambda_{\alpha\beta} = \partial_\Lambda\Sigma^\Lambda_{\alpha\beta} = -\sum_{\gamma\delta} S^\Lambda_{\delta\gamma} \Gamlamcoeff{4}{\alpha\gamma\delta\beta}\label{eq:fRG:RGSelfenergy}\\
	\begin{split}
	\partial_\Lambda\Vertex{\alpha\beta\gamma\delta} = \sum_{a,b,c,d} & \bigl[(\SL{ab} \G{cd} + \SL{cd} \G{ab}) (\Vertex{\alpha b c \delta} \Vertex{d \beta \gamma a} - \Vertex{\beta b c \delta} \Vertex{d \alpha \gamma a}) \\
	&-\tfrac{1}{2} (\SL{ab} \G{dc} + \SL{dc} \G{ab}) \Vertex{\alpha\beta d a} \Vertex{b c \gamma \delta}\bigr] - \sum_{a,b} \SL{ba} \Gamlamcoeff{6}{\alpha\beta a b \gamma \delta}.
	\label{eq:fRG:RGvertex}
	\end{split}
\end{gather}
The three terms in the square bracket on the right hand side of equation~\eqref{eq:fRG:RGvertex} are termed direct particle-hole diagram ($\Pi^\text{PH,d}$), crossed particle-hole diagram ($\Pi^\text{PH,cr}$) and particle-particle diagram ($\Pi^\text{PP}$). The RG equation for the three-particle vertex is quite lengthy and thus not given explicitly. An approximation that includes terms up to the third order in the effective interaction is discussed below. The flow equation for the self-energy is shown diagrammatically in figure~\ref{fig:fRG:RGDE-Selfenergy}. The topological structure of diagrams that renormalise higher-order vertex functions is shown exemplarily for the two- and three-particle vertex in figures~\ref{fig:fRG:RGDE-2P-simp} and~\ref{fig:fRG:RGDE-3P-simp}.

\begin{figure}
	\centering
	\includegraphics[scale=0.9]{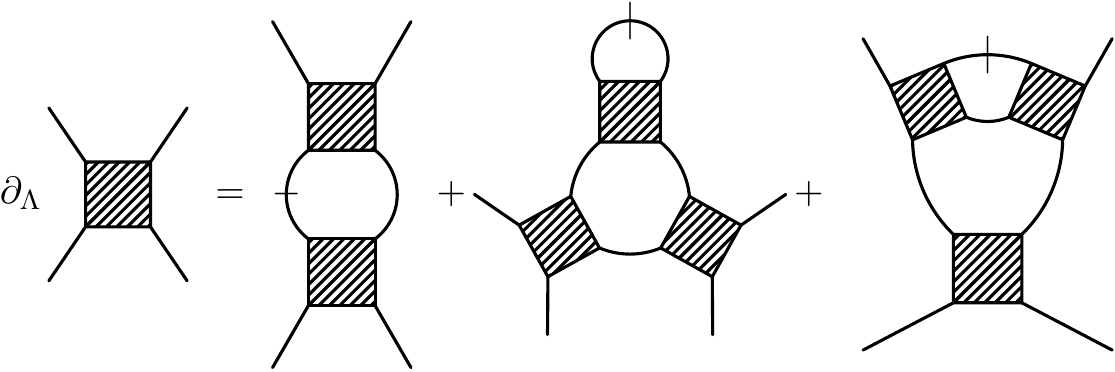}
	\caption{Simplified diagrammatic representation of the two-loop equation for the fermionic two-particle vertex. The second and the third diagram on the right hand side can be classified as two-loop contributions with non-overlapping and overlapping loops, respectively.}
	\label{fig:fRG:RGDE-2P-simp-OG3}
\end{figure}

\begin{figure}
	\centering
	\includegraphics[scale=0.9]{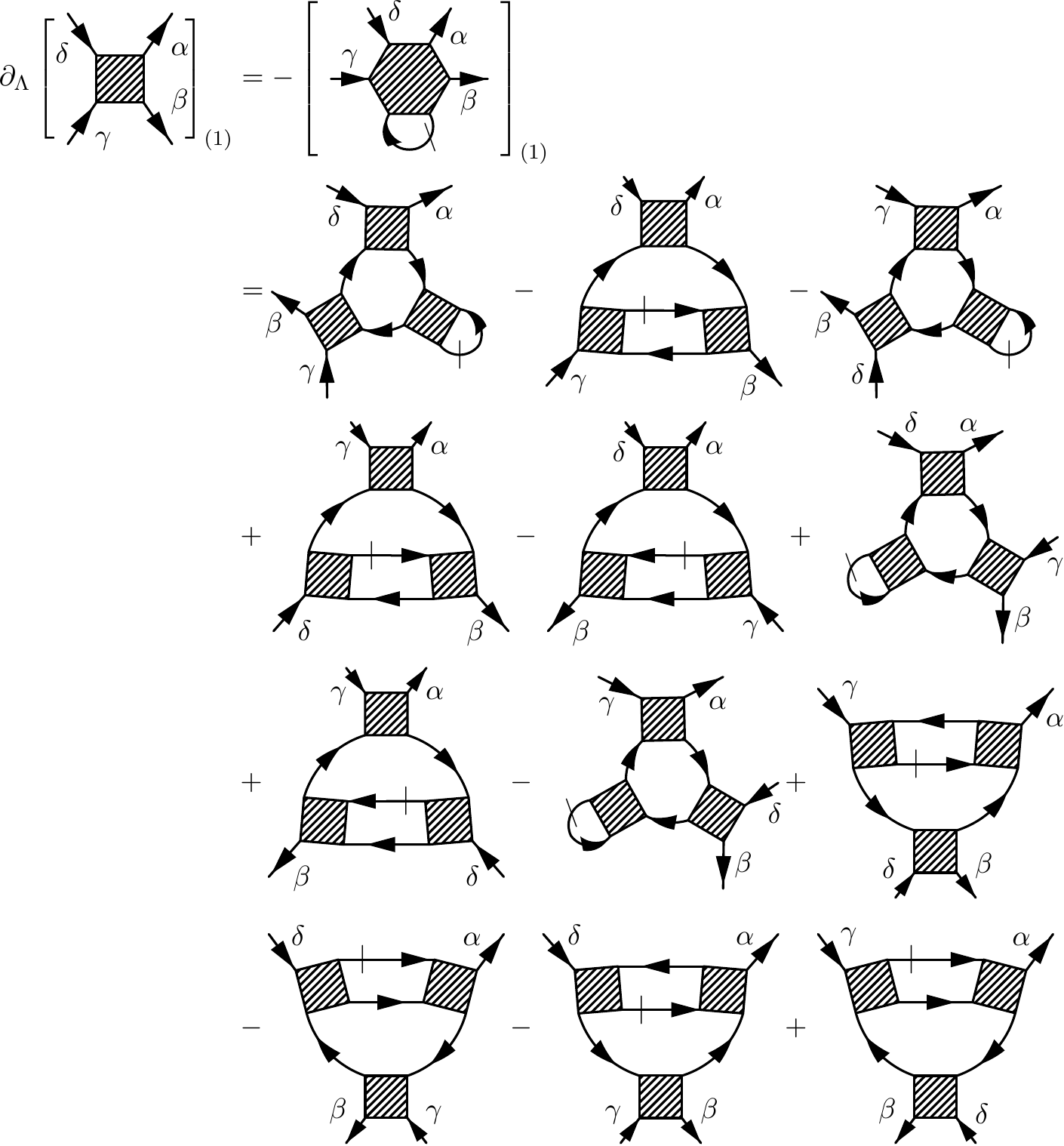}
	\caption{Diagrammatic representation of the two-loop renormalization contributions to the two-particle vertex $\Vertex{}$ as obtained from equation~\eqref{eq:fRG:Gamma6-3P-1}.}
	\label{fig:fRG:dGamma4_dL_TwoLoop1}
\end{figure}

\begin{figure}
	\centering
	\includegraphics[scale=0.9]{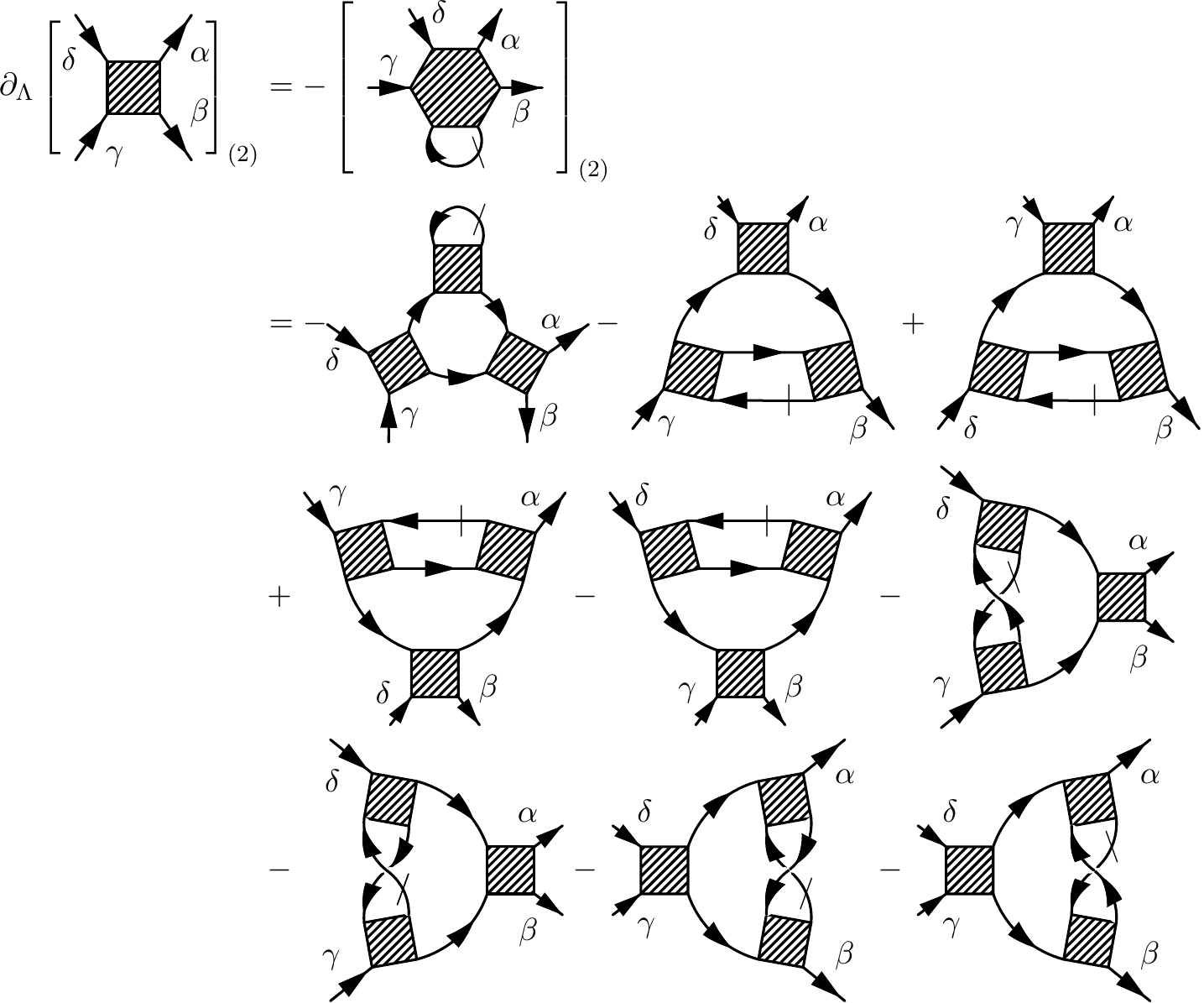}
	\caption{Diagrammatic representation of the two-loop renormalization contributions to the two-particle vertex $\Vertex{}$ as obtained from equation~\eqref{eq:fRG:Gamma6-3P-2}. Some contributions have been simplified by exploiting symmetries of the vertex under the exchange of particles (making, however, exchange symmetries of the right hand side less transparent).}
	\label{fig:fRG:dGamma4_dL_TwoLoop2}
\end{figure}

In the following, the derivation of a two-loop renormalization group equation for the fermionic two-particle vertex is presented. The result is very similar to the third-order $\dot G$-scheme by Salmhofer~\etal~\cite{Salmhofer2001}, but might be more suitable for the extension of the channel-decomposition scheme by Husemann and Salmhofer~\cite{Husemann2009} or of chapter~\ref{chap:ChannelDecomposition} to the two-loop level. For obtaining the feedback of the three-particle vertex on the flow of the two-particle vertex in third order in the effective interaction, it is sufficient to consider only renormalization contributions to the three-particle vertex that originate from the third diagram on the right hand side of the diagrammatic equation in figure~\ref{fig:fRG:RGDE-3P-simp}. The reason is that the other two diagrams are at least of fourth order in the effective interaction because the three- and four-particle vertex are at least $O((\Vertex{})^3)$ and $O((\Vertex{})^4)$, respectively. Within this 
approximation, the RG equation for the three-particle vertex is obtained after the evaluation of the functional derivatives in
\begin{equation}
	\begin{split}
	\partial_\Lambda \Gamlamcoeff{6}{\alpha\beta\mu\nu\gamma\delta} = \frac{\delta^6}{\delta\phi_\delta \delta\phi_\gamma \delta\phi_\nu \delta\bar\phi_\mu \delta\bar\phi_\beta \delta\bar\phi_\alpha}& \bigl(\tfrac{1}{3} \partial_{\Lambda,S} \tr(\G{} \Gamlamtilde_{\bar\phi \phi} \G{} \Gamlamtilde_{\bar\phi \phi} \G{} \Gamlamtilde_{\bar\phi\phi})\\
		&-\partial_{\Lambda,S} \tr(\G{} \Gamlamtilde_{\bar\phi\phi} \G{} \Gamlamtilde_{\bar\phi \bar\phi} (\G{})^T \Gamlamtilde_{\phi\phi}\bigl)|_{\phi=\bar\phi=0}
		\label{eq:fRG:dGamma6_dL_O3}
		\end{split}
\end{equation}
where $\partial_{\Lambda,S}$ is a shorthand for the operation $\partial_{\Lambda,S} \G{} = \SL{} = \partial_\Lambda \G{}|_{\Sigma^\Lambda=\text{const.}}$ and $\Gamlamtilde_{\bar\phi\phi}$ etc.~represent the matrix elements of $\boldsymbol {\tilde \Gamma}^\Lambda$ as defined in~\eqref{eq:fRG:second_derivative_matrix} containing the second functional derivatives of the effective action with respect to the fields. Up to terms of fourth order in $\Vertex{}$, it is possible to replace $\partial_{\Lambda,S} \G{}$ by $\partial_\Lambda \G{}$ (because the self-energy insertions are themselves at least of order $\Vertex{}$) and furthermore to let the scale-derivative also act on the vertices (because $\partial_\Lambda \Vertex{}$ is at least of order ${(\Vertex{})}^2$). In this manner, the right hand side of equation~\eqref{eq:fRG:dGamma6_dL_O3} can be written as a total derivative with respect to $\Lambda$ and hence integrated, yielding
\begin{equation}
	\begin{split}
	\Gamlamcoeff{6}{\alpha\beta\mu\nu\gamma\delta} = \frac{\delta^6}{\delta\phi_\delta \delta\phi_\gamma \delta\phi_\nu \delta\bar\phi_\mu \delta\bar\phi_\beta \delta\bar\phi_\alpha}& \bigl(\tfrac{1}{3} \tr(\G{} \Gamlamtilde_{\bar\phi \phi} \G{} \Gamlamtilde_{\bar\phi \phi} \G{} \Gamlamtilde_{\bar\phi\phi})\\
		&-\tr(\G{} \Gamlamtilde_{\bar\phi\phi} \G{} \Gamlamtilde_{\bar\phi \bar\phi} (\G{})^T \Gamlamtilde_{\phi\phi}\bigl)|_{\phi=\bar\phi=0}
		\end{split}
		\label{eq:fRG:Gamma6_O3}
\end{equation}
for a system with only two-particle interactions on the microscopic level (note that $G^{\Lambda_0}$ vanishes). The evaluation of the functional derivatives in the first line yields
\begin{equation}
	\begin{split}
		(\Gamlamcoeff{6}{\alpha\beta\mu\nu\gamma\delta})_{(1)} &= \tfrac{1}{3}\frac{\delta^6}{\delta\phi_\delta \delta\phi_\gamma \delta\phi_\nu \delta\bar\phi_\mu \delta\bar\phi_\beta \delta\bar\phi_\alpha} \tr(\G{} \Gamlamtilde_{\bar\phi \phi} \G{} \Gamlamtilde_{\bar\phi \phi} \G{} \Gamlamtilde_{\bar\phi\phi})|_{\phi=\bar\phi=0}\\
	= \sum_{a,b,c,d,e,f}\hspace{-1ex}& \G{ab} \G{cd} \G{ef} \bigl(\Vertex{\alpha b e \delta} \Vertex{\beta d a \nu} \Vertex{\mu f c \gamma} - \Vertex{\alpha b e \delta} \Vertex{\mu d a \nu} \Vertex{\beta f c \gamma} + \Vertex{\alpha b e \gamma} \Vertex{\mu d a \nu} \Vertex{\beta f c \delta}\\[-2ex]
	&\hspace{4em}-\Vertex{\alpha be \gamma} \Vertex{\beta d a \nu} \Vertex{\mu f c \delta} + \Vertex{\alpha b e \delta} \Vertex{\mu d a \gamma} \Vertex{\beta f c \nu} -\Vertex{\alpha b e \delta} \Vertex{\beta d a \gamma} \Vertex{\mu f c \nu}\\
	&\hspace{4em} - \Vertex{\alpha b e \gamma} \Vertex{\mu d a \delta} \Vertex{\beta f c \nu} + \Vertex{\alpha b e \gamma} \Vertex{\beta d a \delta} \Vertex{\mu f c \nu} - \Vertex{\alpha b e \nu} \Vertex{\mu d a \gamma} \Vertex{\beta f c \delta}\\
	&\hspace{4em} + \Vertex{\alpha b e \nu} \Vertex{\beta d a \gamma} \Vertex{\mu f c \delta} + \Vertex{\alpha b e \nu} \Vertex{\mu d a \delta} \Vertex{\beta f c \gamma} - \Vertex{\alpha b e \nu} \Vertex{\beta d a \delta} \Vertex{\mu f c \gamma}\bigr)
	\label{eq:fRG:Gamma6-3P-1}
	\end{split}
\end{equation}
and the second line is equal to
\begin{equation}
\begin{split}
	\hspace{-1ex}(\Gamlamcoeff{6}{\alpha\beta\mu\nu\gamma\delta})_{(2)} &= -\frac{\delta^6}{\delta\phi_\delta \delta\phi_\gamma \delta\phi_\nu \delta\bar\phi_\mu \delta\bar\phi_\beta \delta\bar\phi_\alpha}\tr(\G{} \Gamlamtilde_{\bar\phi\phi} \G{} \Gamlamtilde_{\bar\phi \bar\phi} (\G{})^T \Gamlamtilde_{\phi\phi}\bigl)|_{\phi=\bar\phi=0}\\
	= \sum_{a,b,c,d,e,f}\hspace{-1ex}& \G{ab} \G{cd} \G{ef} \bigl(\Vertex{\alpha \beta c a} \Vertex{f d \gamma \delta} \Vertex{\mu b e \nu} + \Vertex{\alpha\beta c a} \Vertex{f d \nu \gamma} \Vertex{\mu b e \delta} + \Vertex{\alpha \beta c a} \Vertex{f d \delta \nu} \Vertex{\mu b e \gamma}\\[-2ex]
	&\hspace{4em}+ \Vertex{\beta\mu c a} \Vertex{f d \gamma \delta} \Vertex{\alpha b e \nu} + \Vertex{\beta \mu c a} \Vertex{f d \nu \gamma} \Vertex{\alpha b e \delta} + \Vertex{\beta \mu c a} \Vertex{f d \delta \nu} \Vertex{\alpha b e \gamma}\\
	&\hspace{4em} + \Vertex{\mu \alpha c a} \Vertex{f d \gamma \delta} \Vertex{\beta b e \nu} + \Vertex{\mu \alpha c a} \Vertex{f d \nu \gamma} \Vertex{\beta b e \delta} + \Vertex{\mu \alpha c a} \Vertex{f d \delta \nu} \Vertex{\beta b e \gamma}\bigr).
	\label{eq:fRG:Gamma6-3P-2}
\end{split}
\end{equation}
These contributions are formally split for a more convenient diagrammatic representation, where they differ in the direction of internal propagators. While all internal propagators in the former have the same mathematical sense, they run in different directions in the latter. The insertion of this approximation for the three-particle vertex into the flow equation~\eqref{eq:fRG:RGvertex} for the two-particle vertex yields a two-loop equation for $\Vertex{}$ in $\mathcal O((\Vertex{})^3)$ that is schematised diagrammatically in figure~\ref{fig:fRG:RGDE-2P-simp-OG3}. Including external multi-indices and arrows on fermionic propagators, the two-loop contributions are depicted diagrammatically in figures~\ref{fig:fRG:dGamma4_dL_TwoLoop1} and~\ref{fig:fRG:dGamma4_dL_TwoLoop2}. Note that the (non-simplified) diagrammatic equations contain all signs and combinatorial factors that arise for the given configuration of propagators and effective interactions.


The two-loop contributions can be classified into diagrams with non-overlapping or overlapping loops (in the sense of Feldman \etal~\cite{Feldman1996}), whose topological structure is schematised in the second and the third diagram on the right hand side of the equation in figure~\ref{fig:fRG:RGDE-2P-simp-OG3}, respectively. Diagrams with non-overlapping loops arise if the single-scale propagator in the two-loop contributions in equation~\eqref{eq:fRG:RGvertex} begins and ends at the same vertex, yielding insertions of scale-differentiated self-energies. These can effectively be included in the one-loop equation by replacing single-scale propagators by scale-differentiated full propagators~\cite{Katanin2004a,Salmhofer2004} using the relation
\begin{equation}
	\partial_\Lambda G^\Lambda = S^\Lambda - G^\Lambda \partial_\Lambda\Sigma^\Lambda G^\Lambda
	\label{eq:fRG:S_Gdot_connection}
\end{equation}
that follows from~\eqref{eq:fRG:SLdef}. This yields
\begin{equation}
	\begin{split}
	\partial_\Lambda\Vertex{\alpha\beta\gamma\delta} = \sum_{a,b,c,d}& \bigl[\partial_\Lambda(\G{ab} \G{cd}) (\Vertex{\alpha b c \delta} \Vertex{d \beta \gamma a} - \Vertex{\beta b c \delta} \Vertex{d \alpha \gamma a}) \\
			& - \tfrac{1}{2} \partial_\Lambda(\G{ab} \G{dc}) \Vertex{\alpha\beta d a} \Vertex{b c \gamma \delta}\bigr] - \sum_{a,b} \bigl(\SL{ba} \Gamlamcoeff{6}{\alpha\beta a b \gamma \delta}\bigr)_\text{overlapping}
	\end{split}
	\label{eq:fRG:RGvertexTwoLoop}
\end{equation}
where the last term contains contributions in $\mathcal O((\Vertex{})^3)$ with overlapping loops, in which the single-scale propagator begins and ends at different vertices. The diagrams with overlapping loops are usually neglected due to phase space arguments, which are valid above the critical scale~\cite{Salmhofer2001}. Note that up to terms of $\mathcal O((\Vertex{})^4)$, it would be possible to replace $\SL{}$ by $\partial_\Lambda \G{}$ in the last term of~\eqref{eq:fRG:RGvertexTwoLoop}.

\begin{figure}
	\centering
	\includegraphics[scale=0.8]{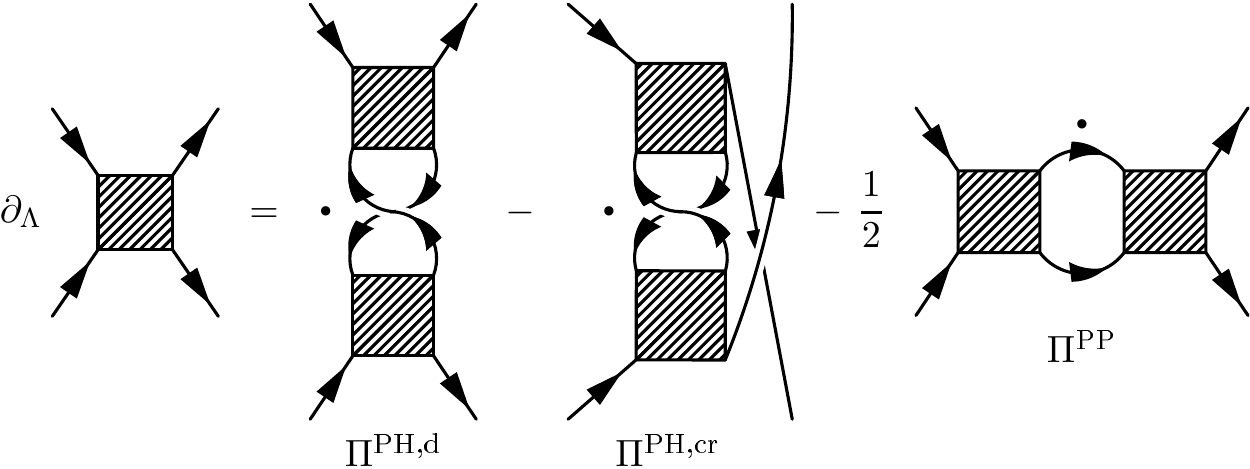}
	\caption{Diagrammatic representation of the renormalization group equation for the two-particle vertex in the modified one-loop truncation by Katanin~\cite{Katanin2004a}. The dots denote scale derivatives acting on the products of fermionic propagators.}
	\label{fig:fRG:RGDE-Vertex-OneLoop}
\end{figure}

Neglecting diagrams with overlapping loops, one arrives at the modified one-loop equation for the two-particle vertex as proposed by Katanin~\cite{Katanin2004a},
\begin{equation}
	\begin{split}
	\partial_\Lambda\Vertex{\alpha\beta\gamma\delta} &= \Pi^\text{PH,d}_{\alpha\beta\gamma\delta} - \Pi^\text{PH,cr}_{\alpha\beta\gamma\delta} - \tfrac{1}{2} \Pi^\text{PP}_{\alpha\beta\gamma\delta} \\
		&\begin{split}
			= \sum_{a,b,c,d}& \bigl[\partial_\Lambda(\G{ab} \G{cd}) (\Vertex{\alpha b c \delta} \Vertex{d \beta \gamma a} - \Vertex{\beta b c \delta} \Vertex{d \alpha \gamma a}) \\
			& - \tfrac{1}{2} \partial_\Lambda(\G{ab} \G{dc}) \Vertex{\alpha\beta d a} \Vertex{b c \gamma \delta}\bigr].
		\end{split}
	\end{split}
	\label{eq:fRG:RGvertexKatanin}
\end{equation}
This equation is shown diagrammatically in figure~\ref{fig:fRG:RGDE-Vertex-OneLoop} including all distinct diagrams and arrows on the lines. In comparison to the $\mathcal O((\Vertex{})^2)$-contributions in equation~\eqref{eq:fRG:RGvertex}, the single-scale propagators are replaced by scale-differentiated full propagators through the inclusion of self-energy feedback from the third order in $\Vertex{}$. The truncation consisting of this equation and the flow equation for the self-energy allows to solve mean-field models exactly~\cite{Salmhofer2004} and to continue RG flows to the symmetry-broken phase also in models with non-reduced interactions like the attractive Hubbard model~\cite{Gersch2008}. It serves as the basis for the one-loop channel-decomposition scheme for a singlet superfluid that is presented in section~\ref{sec:CD:OneLoop}.

Up to terms of fourth order in the two-particle vertex, equation~\eqref{eq:fRG:RGvertexTwoLoop} is equivalent to the so-called third-order-$\dot G^\Lambda$ scheme by Salmhofer~\etal~\cite{Salmhofer2004}, in which the diagram with overlapping loops involving a single-scale propagator is replaced by its total scale derivative with full propagators on the lines. The above equation seems to provide a better starting point for the extension of the channel-decomposition scheme to the two-loop level, which is discussed in section~\ref{sec:CD:TwoLoop}.

%% file: Thesis_VertexParametrization.tex
\chapter{Nambu vertex in a singlet superfluid}
\label{chap:VertexParametrization}

In this chapter, the structure of the Nambu two-particle vertex in a singlet superfluid is studied. By exploiting symmetries, the number of its independent components describing normal and anomalous effective interactions is reduced to a minimum. This simplifies the identification of the singular dependences of the vertex on external momenta and frequencies and allows for a decomposition of the vertex in interaction channels. The latter provides the starting point for the derivation of channel-decomposed renormalization group equations for the vertex, which are presented in chapter~\ref{chap:ChannelDecomposition}.

\section{General structure and symmetries}
\label{sec:VP:GeneralAnsatz}
In this section, the structure of the two-particle vertex in a singlet superfluid is studied and simplified by exploiting symmetries. In this work, the following symmetries are assumed to hold:
\begin{multicols}{2}
\begin{itemize}
	\item spin rotation invariance
	\item inversion symmetry
	\item time reversal symmetry
	\item translation invariance.
\end{itemize}
\end{multicols}
\noindent
In addition, the effective action is assumed to be Osterwalder-Schrader positive, reflecting the hermiticity of the Hamiltonian in the functional integral formalism~\cite{Osterwalder1973,Wetterich2007}. The action of the symmetry operations associated with time reversal and positivity is summarized in appendix~\ref{sec:appendix:Symmetries}.

The vertex part of the effective action of a singlet superfluid can be written as
\begin{equation}
	V[\bar\psi,\psi] = V_{(2+2)}[\bar\psi,\psi] + V_{(3+1)}[\bar\psi,\psi] + V_{(4+0)}[\bar\psi,\psi]
\end{equation}
where $\psi$,$\bar\psi$ represent fermionic Grassmann fields in spinor representation. The first term on the right hand side describes the normal effective interaction and the other terms anomalous effective interactions. The latter appear due to the breaking of the gauge symmetry for global charge conservation in a superfluid, similarly to the anomalous self-energy. They describe processes in which the number of particles is not conserved because of scattering to and out of the fermionic condensate (similar to the interacting Bose gas, see for example~\cite{Negele1998}). In spinor representation, they are described by operators with an unequal number of creation and annihilation operators. Salmhofer~\etal~\cite{Salmhofer2004} showed that linear combinations of the normal effective interaction in the Cooper channel and the anomalous (4+0)-effective interaction, which creates or annihilates four particles, describe the amplitude and phase mode of the superfluid gap. 
In the presence of other interaction channels besides the Cooper channel, anomalous (3+1)-effective interactions appear that create three particles and annihilate one or vice versa, mixing the particle-particle and the particle-hole channels~\cite{Belkhir1994}. These were also included in the fermionic RG study by Gersch~\etal~\cite{Gersch2008}, but their physical significance was not clarified completely and symmetries were not fully exploited.

Taking advantage of symmetries allows to formulate an ansatz that sheds some light on the physical meaning of the anomalous effective interactions and leads to great simplifications of their structure\,--\,in particular for a system with spin rotation invariance. The implementation of this symmetry is more transparent in spinor representation, which is why the ansatz is first formulated in this representation and then rewritten in Nambu representation. The latter representation greatly simplifies the diagrammatics in a superfluid, because only vertices with an equal number of Nambu creation and annihilation operators appear. The operators in spinor representation ($c^{(\dagger)}$) and Nambu representation ($a^{(\dagger)}$) are related through the mapping\\
\parbox{\linewidth-8mm}{\begin{align*}
	a^\dagger_{\boldsymbol k,+} &= \ccup{\boldsymbol k}	&	a_{\boldsymbol k,+} &= \caup{\boldsymbol k}\\
	a^\dagger_{\boldsymbol k,-} &= \cadown{-\boldsymbol k}	&	a_{\boldsymbol k,-} &= \ccdown{\boldsymbol k}
\end{align*}}\parbox{8mm}{\begin{eqnarray}\label{eq:SpinorNambuOperators}\end{eqnarray}}
while the Grassmann fields in spinor representation ($\psi$ and $\bar\psi$) and Nambu representation ($\phi$ and $\bar\phi$) are related by\\
\parbox{\linewidth-8mm}{\begin{align*}
	\bar \phi_{k+} &= \bar \psi_{k\uparrow}	&	\phi_{k+} &= \psi_{k\uparrow}\\
	\bar \phi_{k-} &= \psi_{-k\downarrow}	&	\phi_{k-} &= \bar\psi_{-k\downarrow}
\end{align*}}\parbox{8mm}{\begin{eqnarray}\label{eq:SpinorNambuGrassmann}\end{eqnarray}}\\
where $k = (k_0, \boldsymbol k)$ collects Matsubara frequencies and momenta. For the implementation of spin rotation invariance, it is most straightforward first to construct general spin rotation invariant interaction operators and to subsequently obtain an ansatz for the vertex part of the fermionic effective action by replacing fermionic operators by Grassmann fields and by endowing the coefficients with the most general frequency dependence. Before presenting the construction of ansätze for the anomalous effective interaction, the outlined procedure is demonstrated for the normal effective interaction. Its spin structure is well known and can for example be found in~\cite{Husemann2009, Halboth2000a}.

\subsection{Normal (2+2)-effective interaction}
\subsubsection{Spinor representation}
In this subsection, an ansatz for the normal part of the effective interaction in a singlet superfluid $V_{(2+2)}[\bar\psi,\psi]$ is constructed. It describes scattering processes in which the number of particles is conserved,
\begin{equation}
	V_{(2+2)}[\bar\psi,\psi] \propto \bar\psi\bar\psi\psi\psi.
\end{equation}
The construction starts with a general (2+2)-interaction operator that commutes with the $z$-component of spin,
\begin{equation}
	\begin{split}
	O^V=&a^V_{1234} \ccup{1} \ccup{2} \caup{3} \caup{4} + b^V_{1234} \ccdown{1} \ccdown{2} \cadown{3} \cadown{4} + d^V_{1234} \ccup{1} \ccdown{2} \caup{3} \cadown{4} + e^V_{1234} \ccdown{1} \ccup{2} \cadown{3} \caup{4}\\
	&\ + f^V_{1234} \ccup{1} \ccdown{2} \cadown{3} \caup{4} + g^V_{1234} \ccdown{1} \ccup{2} \caup{3} \cadown{4}
	\end{split}
\end{equation}
where $1$, $2$, $3$ and $4$ are dummy indices representing an arbitrary set of quantum numbers that is summed over. The coefficients $a^V$, $b^V$, $d^V$, $e^V$, $f^V$ as well as $g^V$ are arbitrary functions of the fermionic quantum numbers without any assumed symmetries. They carry the superscript $V$, because the normal (2+2)-vertex is termed $V$ below. Spin rotation invariance imposes relations between the coefficient functions, which are obtained from computing the commutator
\begin{equation}
	[S^x, O^V] \stackrel{!}{=} 0
\end{equation}
and comparison of coefficients for different operators. Solving the resulting linear system of equations yields\footnote{All dummy indices appear in the same order and are therefore suppressed, $a^V \equiv a^V_{1234}$.}
\begin{align*}
	f^V &= g^V	&	a^V &= d^V + f^V\\
	e^V &= d^V	& b^V &= d^V + f^V
\end{align*}
and reduces the number of independent coefficients to two. Thus, $O^V$ can be written as
\begin{equation}
\begin{split}
	O^V=& (d^V_{1234}+f^V_{1234}) (\ccup{1} \ccup{2} \caup{3} \caup{4} + \ccdown{1} \ccdown{2} \cadown{3} \cadown{4}) + d^V_{1234} (\ccup{1} \ccdown{2} \caup{3} \cadown{4} + \ccdown{1} \ccup{2} \cadown{3} \caup{4})\\
	&\ + f^V_{1234} (\ccup{1} \ccdown{2} \cadown{3} \caup{4} + \ccdown{1} \ccup{2} \caup{3} \cadown{4}).
	\end{split}
\end{equation}
Setting $d^V = -\alpha^V + \beta^V$ and $f = \alpha^V + \beta^V$ yields an operator whose spin structure resembles the decomposition of a two-particle interaction operator into a spin singlet part and a spin triplet part (with yet unsymmetrized coefficients):
\begin{equation}
	\begin{split}
		O^V =& \alpha^V_{1234} (\ccup{1} \ccdown{2} \cadown{3} \caup{4} + \ccdown{1} \ccup{2} \caup{3} \cadown{4} - \ccup{1} \ccdown{2} \caup{3} \cadown{4} - \ccdown{1} \ccup{2} \cadown{3} \caup{4})\\
	+ &\beta^V_{1234} \bigl[\ccup{1} \ccdown{2} \caup{3} \cadown{4} + \ccdown{1} \ccup{2} \cadown{3} \caup{4} + \ccup{1} \ccdown{2} \cadown{3} \caup{4} + \ccdown{1} \ccup{2} \caup{3} \cadown{4}\\
	&\ \ + 2(\ccup{1} \ccup{2} \caup{3} \caup{4} + \ccdown{1} \ccdown{2} \cadown{3} \cadown{4})\bigr].
	\end{split}
\end{equation}
Replacing operators through Grassmann variables
\begin{align}
	\cc{1\sigma_1} &\rightarrow	\psic{k_1 \sigma_1} &	\ca{1\sigma_1} &\rightarrow	\psia{k_1 \sigma_1},
\end{align}
adding quantum numbers including Matsubara frequencies with $k = (k_0, \boldsymbol k)$
\begin{align}
\alpha^V_{1234} &\rightarrow V^S_{k_1 k_2 k_3 k_4}	& \beta^V_{1234} &\rightarrow V^T_{k_1 k_2 k_3 k_4}
\end{align}
and symmetrizing the coefficients through the replacement
\begin{align}
	\psic{k_1 \sigma_1}\psic{k_2 \sigma_2} \psia{k_3 \sigma_3}\psia{k_4 \sigma_4} \rightarrow \frac{1}{4} (\psic{k_1 \sigma_1}\psic{k_2 \sigma_2} - \psic{k_2 \sigma_2}\psic{k_1 \sigma_1}) (\psia{k_3 \sigma_3}\psia{k_4 \sigma_4} - \psia{k_4 \sigma_4}\psia{k_3 \sigma_3})
\end{align}
in order to incorporate the indistinguishability of the fermions, the final result can be written as\footnote{Note that temperature and volume prefactors are absorbed in the summation symbols. In the ansatz for the effective action, momentum and frequency summation symbols should be read as $\sum_{k_1, k_2, k_3, k_4} (\ldots) \equiv (\beta L)^{-1} \sum_{k_1, k_2, k_3, k_4} (\ldots)$ where $\beta = T^{-1}$ is the inverse temperature and $L$ the volume of the system. In the flow equations, one factor $(\beta L)^{-1}$ appears per momentum and frequency integration.}
\begin{equation}
	V_{(2+2)}[\bar\psi,\psi] = \frac{1}{2} \sum_{k_i,\sigma_i} (S_{\sigma_1 \sigma_2 \sigma_3 \sigma_4} V^S_{k_1 k_2 k_3 k_4} + T_{\sigma_1 \sigma_2 \sigma_3 \sigma_4} V^T_{k_1 k_2 k_3 k_4}) \psic{k_1 \sigma_1}\psic{k_2 \sigma_2} \psia{k_3 \sigma_3} \psia{k_4 \sigma_4}.
	\label{eq:2p2SingletTriplet}
\end{equation}
Equation~\eqref{eq:2p2SingletTriplet} is the well-known singlet-triplet decomposition of the normal effective interaction with $S_{\sigma_1 \sigma_2 \sigma_3 \sigma_4} = \frac{1}{2}(\delta_{\sigma_1 \sigma_4} \delta_{\sigma_2 \sigma_3} - \delta_{\sigma_1 \sigma_3} \delta_{\sigma_2 \sigma_4})$ and $T_{\sigma_1 \sigma_2 \sigma_3 \sigma_4} = \frac{1}{2}(\delta_{\sigma_1 \sigma_4} \delta_{\sigma_2 \sigma_3} + \delta_{\sigma_1 \sigma_3} \delta_{\sigma_2 \sigma_4})$ being the projection operators on the spin singlet and triplet components of the interaction. After symmetrization, the singlet and the triplet vertex are symmetric and antisymmetric, respectively, under the exchange of the incoming or the outgoing particles,
\begin{gather}
	V^S_{k_1 k_2 k_3 k_4} = V^S_{k_2 k_1 k_3 k_4} = V^S_{k_1 k_2 k_4 k_3} = V^S_{k_2 k_1 k_4 k_3}\\
	V^T_{k_1 k_2 k_3 k_4} = -V^T_{k_2 k_1 k_3 k_4} = -V^T_{k_1 k_2 k_4 k_3} = V^T_{k_2 k_1 k_4 k_3}.
\end{gather}
However, for the definition of interaction channels, it is more convenient to use vertex functions that are symmetric only under the simultaneous exchange of both incoming and outgoing particles,
\begin{equation}
	V_{k_1 k_2 k_3 k_4} \stackrel{\operatorname{def}}{=} V^S_{k_1 k_2 k_3 k_4} + V^T_{k_1 k_2 k_3 k_4} = V_{k_2 k_1 k_4 k_3}.
\end{equation}
In terms of vertex functions with this exchange symmetry, the normal effective interaction reads
\begin{equation}
V_{(2+2)}[\bar\psi,\psi] = \frac{1}{4} \sum_{k_i,\sigma_i} (\delta_{\sigma_1 \sigma_4} \delta_{\sigma_2 \sigma_3} V_{k_1 k_2 k_3 k_4} - \delta_{\sigma_1 \sigma_3} \delta_{\sigma_2 \sigma_4} V_{k_1 k_2 k_4 k_3}) \psic{k_1 \sigma_1}\psic{k_2 \sigma_2} \psia{k_3 \sigma_3} \psia{k_4 \sigma_4}
\label{eq:VP:2p2spinor}
\end{equation}
and equals the ansatz used by Husemann and Salmhofer~\cite{Husemann2009}.

\hyphenation{con-straints}
The symmetries mentioned at the beginning of this section entail the following constraints on the momentum and frequency dependence of $V_{k_1 k_2 k_3 k_4}$:\\[0.5ex]
\begin{minipage}{\linewidth-1.1\mathnumlength}
\begin{center}
\begin{tabular}{ll}
	$V_{k_1 k_2 k_3 k_4} \propto \delta_{k_1 + k_2, k_3 + k_4}$	&	Translation invariance\\[0.2ex]
	$V_{k_1 k_2 k_3 k_4} = V_{k_2 k_1 k_4 k_3}$ & Exchange of particles\\[0.2ex]
	$V_{k_1 k_2 k_3 k_4} = V^\ast_{\bar k_4 \bar k_3 \bar k_2 \bar k_1}$ & Positivity\\[0.2ex]
	$V_{k_1 k_2 k_3 k_4} = V_{Rk_4 Rk_3 Rk_2 Rk_1}$ & Time reversal\\[0.2ex]
	$V_{k_1 k_2 k_3 k_4} = V_{Rk_1 Rk_2 Rk_3 Rk_4}$ & Space inversion
\end{tabular}
\end{center}
\end{minipage}
\begin{minipage}{\mathnumlength}
	\begin{eqnarray}
	\label{eq:2p2symmetries4mom}
	\end{eqnarray}
\end{minipage}\\[1ex]
where $\bar k = (-k_0, \boldsymbol k)$, $Rk = (k_0, -\boldsymbol k)$ and $^\ast$ denoting complex conjugation.

\subsubsection{Nambu representation}
The rewriting of the effective interaction~\eqref{eq:VP:2p2spinor} in Nambu representation is straightforward, albeit lengthy. Applying the relations~\eqref{eq:SpinorNambuGrassmann} to~\eqref{eq:VP:2p2spinor} yields
\begin{equation}
\begin{split}
	V_{(2+2)}[\bar\phi, \phi] = &\frac{1}{4} \sum_{k_i, s_i} \bigl[(V_{k_1 k_2 k_3 k_4} - V_{k_1 k_2 k_4 k_3}) \delta_{s_1+}\delta_{s_2+}\delta_{s_3+}\delta_{s_4+}\\
	&\ + (V_{-k_4, -k_3,-k_2,-k_1} - V_{-k_3, -k_4,-k_2,-k_1})  \delta_{s_1-}\delta_{s_2-}\delta_{s_3-}\delta_{s_4-}\\
	&\ + V_{k_1,-k_4,-k_2,k_3} \delta_{s_1+}\delta_{s_2-} \delta_{s_3 +} \delta_{s_4 -} + V_{k_2,-k_3,-k_1,k_4} \delta_{s_1-}\delta_{s_2+} \delta_{s_3 -} \delta_{s_4 +}\\
	&\ - V_{k_1,-k_3,-k_2 k_4} \delta_{s_1+}\delta_{s_2-}\delta_{s_3-}\delta_{s_4+} - V_{k_2,-k_4,-k_1,k_3}\delta_{s_1-}\delta_{s_2+}\delta_{s_3+}\delta_{s_4-}\bigr] \times\\
	&\ \ \times\phic{k_1 s_1}\phic{k_2 s_2}\phia{k_3 s_3}\phia{k_4 s_4}.
\end{split}
\label{eq:VP:2p2Nambu}
\end{equation}

\subsection{Anomalous (3+1)-effective interaction}
\subsubsection{Spinor representation}
The construction of an ansatz for the anomalous (3+1)-effective interaction that creates three particles and annihilates one or vice versa,
\begin{equation}
	V_{(3+1)}[\bar\psi,\psi] \propto \bar\psi \bar\psi \bar\psi \psi \text{\ and\ } \propto \bar\psi \psi \psi \psi,
\end{equation}
works similarly to the normal (2+2)-effective interaction. It is sufficient to consider interaction operators that create three particles and annihilate one, because the part of the operator or action that creates one particle and annihilates three can be fixed by demanding hermiticity of the interaction operator or Osterwalder-Schrader positivity of the action. For such (3+1)-operators, spin rotation invariance around the $z$-axis holds if the commutator
\begin{equation}
	[S^z, \cc{\sigma_1} \cc{\sigma_2} \cc{\sigma_3} \ca{\sigma_4}] = (\sigma_1 + \sigma_2 + \sigma_3 - \sigma_4) \cc{\sigma_1} \cc{\sigma_2} \cc{\sigma_3} \ca{\sigma_4}
\end{equation}
vanishes. This is the case if the spin of the annihilation operator equals that of one creation operator and if the spin of the other two creation operators are opposite. Constructing the linear combination of all such (3+1)-operators yields
\begin{equation}
	\begin{split}
		O^\Omega =& a^\Omega_{1234} \ccup{1} \ccup{2} \ccdown{3} \caup{4}+ b^\Omega_{1234} \ccup{1} \ccdown{2} \ccup{3} \caup{4} + d^\Omega_{1234} \ccdown{1} \ccup{2} \ccup{3} \caup{4} \\
	+& e^\Omega_{1234} \ccdown{1} \ccdown{2} \ccup{3} \cadown{4} + f^\Omega_{1234} \ccdown{1} \ccup{2} \ccdown{3} \cadown{4} + g^\Omega_{1234} \ccup{1} \ccdown{2} \ccup{3} \cadown{4}
	\end{split}
\end{equation}
with coefficients that depend on an arbitrary set of quantum numbers. They carry the superscript $\Omega$, because the anomalous (3+1)-vertex is termed $\Omega$ below. Spin rotation invariance around all axes requires
\begin{equation}
[S^x, O^\Omega] \stackrel{!}{=} 0
\end{equation}
to hold in addition. Comparing coefficients of different operators in the resulting expression leads to a system of four linearly independent equations\footnote{All dummy indices appear in the same order and are therefore suppressed, $a^\Omega \equiv a^\Omega_{1234}$.}
\begin{align*}
	f^\Omega &= -b^\Omega	&	g^\Omega &= b^\Omega + a^\Omega\\
	e^\Omega &= -a^\Omega	&	d^\Omega &= -b^\Omega - a^\Omega
\end{align*}
that reduces the number of independent coefficient functions from six to two. Therefore, the operator $O^\Omega$ simplifies to
\begin{equation}
	\begin{split}
		O^\Omega &= a^\Omega_{1234} (\ccup{1} \ccup{2} \ccdown{3} \caup{4} + \ccup{1} \ccdown{2} \ccdown{3} \cadown{4} - \ccdown{1} \ccup{2} \ccup{3} \caup{4} - \ccdown{1} \ccdown{2} \ccup{3} \cadown{4}) \\
	&+ b^\Omega_{1234} (\ccup{1}\ccdown{2}\ccup{3} \caup{4} + \ccup{1} \ccdown{2} \ccdown{3} \cadown{4} - \ccdown{1} \ccup{2} \ccup{3} \caup{4} - \ccdown{1} \ccup{2} \ccdown{3} \cadown{4})
	\end{split}
\end{equation}
with yet unsymmetrized coefficient functions. 

In order to shed some light on the physical meaning of the anomalous (3+1)-effective interactions, it is convenient to redefine the coefficient functions analogously to the singlet-triplet decomposition of the normal interaction. This rewriting is not mandatory, but does not reduce the generality of the ansatz and helps in defining interaction channels (see section~\ref{sec:VP:ChannelDecomposition}). Introducing
\begin{align*}
	a^\Omega_{1234} &= \eta^\Omega_{1234} + \iota^\Omega_{1234}		&		b^\Omega_{1234} &= \eta^\Omega_{1234} - \iota^\Omega_{1234},
\end{align*}
the operators are regrouped into
\begin{equation}
	\begin{split}
		O^\Omega &= \iota^\Omega_{1234} \sum_\sigma \cc{1\sigma} (\ccup{2} \ccdown{3} - \ccdown{2} \ccup{3}) \ca{4\sigma} \\
			&+ \eta^\Omega_{1234} \Bigl[ \sum_\sigma \epsilon_\sigma \cc{1\sigma}(\ccup{2} \ccdown{3} + \ccdown{2} \ccup{3}) \ca{4\sigma} + 2 (\ccup{1}\ccdown{2}\ccdown{3} \cadown{4} - \ccdown{1} \ccup{2} \ccup{3} \caup{4})\Bigr]
	\end{split}
\end{equation}
where $\epsilon_\uparrow=1$, $\epsilon_\downarrow=-1$. After symmetrization under those exchanges of particles that leave the spin wave-function invariant (\ie\ exchange of particles $2$ and $3$), the spin structure of the first term suggests that it describes the interaction between Cooper pairs and charge density fluctuations. The role of the second term is not so obvious because of the triplet structure of the involved operators. It is expected to describe interaction processes between particle-particle and particle-hole pairs involving a spin flip (for example the scattering of two fermions out of the condensate into a spin triplet final state). Introducing fermionic Grassmann fields and quantum numbers as above leads to the ansatz for the anomalous (3+1)-effective interaction
\begin{equation}
	\begin{split}
		V_{(3+1)}[\psic{},\psia{}] &= \frac{1}{2} \sum_{k_i} \bigl[\Omega^S_{k_1 k_2 k_3 k_4} \sum_\sigma  \psic{k_1\sigma} (\psicup{k_2}\psicdown{k_3} - \psicdown{k_2} \psicup{k_3}) \psia{k_4\sigma} \\
	&+\Omega^T_{k_1 k_2 k_3 k_4} \bigl(\sum_\sigma \epsilon_\sigma \psic{k_1\sigma} (\psicup{k_2} \psicdown{k_3} + \psicdown{k_2} \psicup{k_3}) \psia{k_4\sigma} \\
	&\ \ + 2(\psicup{k_1}\psicdown{k_2}\psicdown{k_3}\psiadown{k_4} - \psicdown{k_1}\psicup{k_2}\psicup{k_3} \psiaup{k_4})\bigr) + \conj \bigr]
	\label{eq:3p1spinor}
	\end{split}
\end{equation}
where the terms involving one creation and three annihilation operators follow from the invariance under Osterwalder-Schrader positivity (abbreviated by ``$\conj$''). The superscripts $S$ and $T$ stand for singlet and triplet following the spin structure of the particle-particle pair operators.

Differently from the normal effective interaction, it is preferable to defer the total antisymmetrization of the coefficient functions in the anomalous (3+1)-effective interaction to the Nambu representation. The reason is that the apparent symmetry of the spin part of the above ansatz under the exchange of particles is less transparent after total antisymmetrization. More importantly, some exchange symmetries of the antisymmetrized coefficient functions would disappear again in Nambu representation because of the mapping between creation and annihilation operators\footnote{The Nambu vertices are given by linear combinations of spinor vertices that transform differently under the exchange of particles. These linear combinations are less symmetric than the individual terms.}. Therefore, it seems convenient to work with $\Omega^S$ and $\Omega^T$ instead of the totally antisymmetrized coefficients and to exploit only those particle-exchange symmetries that leave the spin wave 
functions invariant, so that
\begin{align}
	\Omega^S_{k_1 k_2 k_3 k_4} &= \Omega^S_{k_1 k_3 k_2 k_4}		&		\Omega^T_{k_1 k_2 k_3 k_4} &= -\Omega^T_{k_1 k_3 k_2 k_4}.
\end{align}
These exchange symmetries also leave the total momentum of the particle-particle pair invariant. In this form, the anomalous (3+1)-effective interaction provides a good starting point for the definition of interaction channels (see section~\ref{sec:VP:ChannelDecomposition}).

The discrete symmetries mentioned at the beginning of this section imply the following relations for $\Omega^S_{k_1 k_2 k_3 k_4}$ and $\Omega^T_{k_1 k_2 k_3 k_4}$:\\[0.5ex]
\begin{minipage}{\linewidth-1.1\mathnumlength}
\begin{center}
\begin{tabular}{ll}
	$\Omega^i_{k_1 k_2 k_3 k_4} \propto \delta_{k_1+k_2+k_3,k_4}$	&	Translation invariance\\[0.2ex]
	$\Omega^S_{k_1 k_2 k_3 k_4} = \Omega^S_{k_1 k_3 k_2 k_4}$ & Exchange of particles\\[0.2ex]
	$\Omega^T_{k_1 k_2 k_3 k_4} = -\Omega^T_{k_1 k_3 k_2 k_4}$ & \\[0.2ex]
	$\Omega^i_{k_1 k_2 k_3 k_4} = {\Omega^i}^\ast_{-k_1, -k_2, -k_3, -k_4}$ & Time reversal and positivity\\[0.2ex]
	$\Omega^i_{k_1 k_2 k_3 k_4} = \Omega^i_{Rk_1 Rk_2 Rk_3 Rk_4}$ & Space inversion
\end{tabular}
\end{center}
\end{minipage}
\begin{minipage}{\mathnumlength}
	\begin{eqnarray}
	\label{eq:3p1symmetries4mom}
	\end{eqnarray}
\end{minipage}\\[1ex]
with $i\in\{S,T\}$, $\bar k = (-k_0, \boldsymbol k)$, $Rk = (k_0, -\boldsymbol k)$ and $^\ast$ denoting complex conjugation.

\subsubsection{Nambu representation}
Demanding spin rotation invariance yields substantial simplifications for the anomalous components of the Nambu vertex in a singlet superfluid. Gersch~\etal~\cite{Gersch2008} only exploited spin rotation invariance around one axis and had to deal with four functions describing the anomalous (3+1)-effective interactions, out of which two were independent because of particle-exchange symmetries and time reversal symmetry. Invariance under spin rotations around all axes reduces the number of independent functions describing the anomalous (3+1)-effective interaction in Nambu representation to only one and the discrete symmetries further constrain its momentum and frequency dependence. 

The anomalous (3+1)-effective interaction in Nambu representation is obtained from equation~\eqref{eq:3p1spinor} by use of the relations~\eqref{eq:SpinorNambuGrassmann} and antisymmetrization through
\begin{align}
	\phic{k_1 s_1}\phic{k_2 s_2} \phia{k_3 s_3}\phia{k_4 s_4} \rightarrow \frac{1}{4} (\phic{k_1 s_1}\phic{k_2 s_2} - \phic{k_2 s_2}\phic{k_1 s_1}) (\phia{k_3 s_3}\phia{k_4 s_4} - \phia{k_4 s_4}\phia{k_3 s_3}),
\end{align}
yielding
\begin{equation}
	\begin{split}
		V_{(3+1)}[\phic{},\phia{}] = &\frac{1}{4} \sum_{k_i,s_i} \bigl[\Omega_{k_1 k_2 k_3 k_4} \delta_{s_1 +} \delta_{s_2 +} \delta_{s_3 -} \delta_{s_4 +} - \Omega_{k_1 k_2 k_4 k_3} \delta_{s_1 +} \delta_{s_2 +} \delta_{s_3 +} \delta_{s_4 -} \\
	&\ + \Omega_{-k_4, -k_3, -k_1, -k_2} \delta_{s_1 +} \delta_{s_2 -} \delta_{s_3 -} \delta_{s_4 -} - \Omega_{-k_4, -k_3, -k_2, -k_1} \delta_{s_1 -} \delta_{s_2 +} \delta_{s_3 -} \delta_{s_4 -} \\
	&\ + \Omega_{-k_4, -k_3, -k_2, -k_1}^\ast \delta_{s_1 +} \delta_{s_2 -} \delta_{s_3 +} \delta_{s_4 +} - \Omega_{-k_4, -k_3, -k_1, -k_2}^\ast \delta_{s_1 -} \delta_{s_2 +} \delta_{s_3 +} \delta_{s_4 +} \\
	&\ + \Omega_{k_1 k_2 k_4 k_3}^\ast \delta_{s_1 -} \delta_{s_2 -} \delta_{s_3 -} \delta_{s_4 +} - \Omega_{k_1 k_2 k_3 k_4}^\ast \delta_{s_1 -} \delta_{s_2 -} \delta_{s_3 +} \delta_{s_4 -}\bigr]\times\\
	&\ \ \times\phic{k_1 s_1} \phic{k_2 s_2} \phia{k_3 s_3} \phia{k_4 s_4}
	\end{split}\raisetag{1em}
\end{equation}
where $\Omega_{k_1 k_2 k_3 k_4}$ is related to $\Omega^S_{k_1 k_2 k_3 k_4}$ and $\Omega^T_{k_1 k_2 k_3 k_4}$ by
\begin{equation}
	\begin{split}
	\Omega_{k_1 k_2 k_3 k_4} &= \frac{1}{2}(\Omega^S_{k_1 k_2, -k_3, k_4} + \Omega^S_{k_1, -k_3, k_2 k_4} - \Omega^S_{k_2 k_1, -k_3, k_4} - \Omega^S_{k_2, -k_3, k_1 k_4})\\
	&\ \ +\frac{1}{2}(\Omega^T_{k_1 k_2,-k_3, k_4} - \Omega^T_{k_2 k_1,-k_3, k_4} + \Omega^T_{k_2,-k_3, k_1 k_4} - \Omega^T_{k_1, -k_3, k_2 k_4}\\
	&\ \  + 2 \Omega^T_{-k_3, k_2 k_1 k_4} - 2 \Omega^T_{-k_3, k_1 k_2 k_4})=\\
&= \Omega^S_{k_1 k_2, -k_3, k_4} - \Omega^S_{k_2 k_1, -k_3, k_4} +\Omega^T_{k_1 k_2,-k_3, k_4} - \Omega^T_{k_2 k_1,-k_3, k_4} + 2 \Omega^T_{-k_3, k_2 k_1 k_4}.
	\label{eq:VP:Omega3p1Vertex}\raisetag{4em}
\end{split}
\end{equation}
The first equality holds without any assumption on the symmetries under the exchange of indices of $\Omega^i_{k_1 k_2 k_3 k_4}$ while the second equality holds for the partial symmetrization described above. $\Omega_{k_1 k_2 k_3 k_4}$ is antisymmetric with respect to the exchange of the first two indices,
\begin{equation}
	\Omega_{k_1 k_2 k_3 k_4} = -\Omega_{k_2 k_1 k_3 k_4},
\end{equation}
and time reversal invariance implies
\begin{equation}
	\Omega_{k_1 k_2 k_3 k_4} = \Omega^\ast_{-k_1, -k_2, -k_3, -k_4}.
\end{equation}
Without time reversal invariance, there is still only one independent function $\Omega_{k_1 k_2 k_3 k_4}$, but the simple relation to its complex conjugate ceases to hold.

\subsection{Anomalous (4+0)-effective interaction}
\subsubsection{Spinor representation}
For the construction of a spin-rotation invariant ansatz for the anomalous (4+0)-effective interaction, it is sufficient to consider the part of the effective interaction that creates four particles and annihilates none. As for the anomalous (3+1)-effective interaction, the other part involving four annihilation operators can be fixed subsequently by demanding Osterwalder-Schrader positivity of the action or hermiticity of the operator. The starting point is a linear combination of all interaction operators that create four particles and conserve the spin projection for example in the $z$ direction. This requirement is fulfilled if two spin up and two spin down particles are created, leading to
\begin{equation}
	\begin{split}
		O^W &= a^W_{1234} \ccup{1}\ccdown{2}\ccup{3}\ccdown{4} + b^W_{1234}\ccdown{1}\ccup{2}\ccdown{3}\ccup{4} + d^W_{1234} \ccup{1} \ccdown{2} \ccdown{3} \ccup{4} \\
	&+ e^W_{1234} \ccdown{1} \ccup{2} \ccup{3} \ccdown{4} + f^W_{1234} \ccup{1} \ccup{2} \ccdown{3} \ccdown{4} + g^W_{1234} \ccdown{1} \ccdown{2} \ccup{3} \ccup{4}.
	\end{split}
\end{equation}
In this ansatz, $a^W$, $b^W$, $d^W$, $e^W$, $f^W$ and $g^W$ are arbitrary coefficient functions without any assumed symmetries. They carry the superscript $W$, because the anomalous (4+0)-vertex is termed $W$ below. This ansatz obviously commutes with $S^z$. The total spin is conserved by this interaction operator, if it also commutes with $S^x$,
\begin{equation}
	[S^x, O^W] \stackrel{!}{=}0.
\end{equation}
This commutator yields relations among the coefficient functions. Demanding that prefactors of distinct operators vanish leads to a linear system of eight equations, out of which four are independent\footnote{All dummy indices appear in the same order and are therefore suppressed, $a^W \equiv a^W_{1234}$.}:
\begin{align*}
	a^W + e^W + g^W &= 0			&			b^W + d^W + f^W &= 0\\
	a^W + e^W + f^W &= 0			&			b^W + d^W + g^W &= 0\\
	a^W + d^W + f^W &= 0			&			b^W + e^W + g^W &= 0\\
	a^W + d^W + g^W &= 0			&			b^W + e^W + f^W &= 0.
\end{align*}
This system is for example solved by
\begin{align*}
	b^W &= a^W		&		g^W &= f^W &	e^W &= d^W		&		a^W + d^W + f^W &= 0
\end{align*}
where two coefficients out of $a^W$, $d^W$ and $f^W$ are independent. Thus, the above operator simplifies to
\begin{equation}
	\begin{split}
		O^W &= a^W_{1234} (\ccup{1} \ccdown{2}\ccup{3}\ccdown{4} + \ccdown{1}\ccup{2}\ccdown{3}\ccup{4}) + d^W_{1234} (\ccup{1} \ccdown{2} \ccdown{3} \ccup{4} + \ccdown{1}\ccup{2}\ccup{3}\ccdown{4})\\
	&+ f^W_{1234} (\ccup{1} \ccup{2} \ccdown{3} \ccdown{4} + \ccdown{1}\ccdown{2}\ccup{3}\ccup{4}).
	\end{split}
\end{equation}
Regrouping terms by defining
\begin{align}
	d^W &= \alpha^W + \beta^W		&		a^W &= \beta^W - \alpha^W		&		f^W &= -2\beta^W
	\label{eq:4p0regrouping}
\end{align}
yields
\begin{equation}
\begin{split}
	O^W &= \alpha^W_{1234} (\ccup{1}\ccdown{2}\ccdown{3}\ccup{4} + \ccdown{1} \ccup{2}\ccup{3} \ccdown{4} - \ccup{1}\ccdown{2}\ccup{3}\ccdown{4} - \ccdown{1}\ccup{2}\ccdown{3}\ccup{4}) \\
	&+ \beta^W_{1234}\bigl[\ccup{1}\ccdown{2}\ccdown{3}\ccup{4} + \ccdown{1}\ccup{2}\ccup{3}\ccdown{4} + \ccup{1}\ccdown{2}\ccup{3}\ccdown{4} + \ccdown{1}\ccup{2}\ccdown{3}\ccup{4} \\
	&\ - 2 (\ccup{1}\ccup{2}\ccdown{3}\ccdown{4} + \ccdown{1}\ccup{2}\ccup{3}\ccup{4})\bigr],
\end{split}
\end{equation}
where the spin-wave function of the operator part in the first line is odd under the exchange $1\leftrightarrow 2$ or $3\leftrightarrow 4$ and resembles singlet Cooper pairs whereas the spin-wave function of the operator part in the second line is even under these transformations and resembles triplet Cooper pairs. As for the anomalous (3+1)-effective interaction, the regrouping through~\eqref{eq:4p0regrouping} is not mandatory, but simplifies the definition of interaction channels and the channel-decomposition of the renormalization group equations\footnote{Different choices, for example with $a^W$ and $d^W$ being independent and $f^W=-a^W - d^W$ fixed, allow to regroup the operators into singlet Cooper pairs. Introducing interaction channels in the obvious way with the total momentum of the Cooper pairs as singular dependence, one runs into the problem that no term with transfer momentum $k_1+k_2$ appears in Nambu representation. 
However, such a term is required to achieve an assignment of diagrams to interaction channels according to the singular momentum dependence of fermionic loop integrals (see chapter~\ref{chap:ChannelDecomposition}). Note that this problem arises only for the anomalous (4+0)-triplet interactions that are defined below}.

Adding quantum numbers and replacing operators by Grassmann fields yields the ansatz for the anomalous (4+0)-contribution in the vertex part of the effective action:
\begin{equation}
	\begin{split}
		V_{(4+0)}[\psic{},\psia{}] &= -\frac{1}{8} \sum_{k_i,\sigma_i} \bigl[W^S_{k_1 k_2 k_3 k_4} (\delta_{\sigma_1 \sigma_4} \delta_{\sigma_2 \sigma_3} - \delta_{\sigma_1\sigma_3} \delta_{\sigma_2\sigma_4}) \psic{k_1\sigma_1} \psic{k_2\sigma_2} \psic{k_3\sigma_3} \psic{k_4\sigma_4} \\
	&+ W^T_{k_1 k_2 k_3 k_4} \bigl[(\psicup{k_1}\psicdown{k_2}+\psicdown{k_1}\psicup{k_2})(\psicdown{k_3}\psicup{k_4}+\psicup{k_3}\psicdown{k_4}) \\
	&\ - 2 (\psicup{k_1}\psicup{k_2}\psicdown{k_3}\psicdown{k_4} + \psicdown{k_1}\psicdown{k_2} \psicup{k_3} \psicup{k_4})\bigr] + \conj \bigr].
	\end{split}
	\label{eq:VP:4p0spinor}
\end{equation}
The prefactor of this ansatz is chosen in such a way that the linear combinations of normal (2+2)- and anomalous (4+0)-effective interactions that describe the amplitude and phase mode of the superfluid gap are the same as in~\cite{Salmhofer2004, Gersch2008}. The ansatz is invariant under those particle exchanges that leave the spin structure unaltered if the coefficients fulfil
\begin{equation}
	\begin{split}
		W^S_{k_1 k_2 k_3 k_4} = W^S_{k_2 k_1 k_3 k_4} = W^S_{k_1 k_2 k_4 k_3} = W^S_{k_2 k_1 k_4 k_3} = W^S_{k_4 k_3 k_2 k_1}\\
		W^T_{k_1 k_2 k_3 k_4} = -W^T_{k_2 k_1 k_3 k_4} = -W^T_{k_1 k_2 k_4 k_3} = W^T_{k_2 k_1 k_4 k_3} = W^T_{k_4 k_3 k_2 k_1}
	\end{split}
	\label{eq:symW4p0spinor}
\end{equation}
where the last identity follows after reversing the order of the four fields. The total antisymmetrization is deferred until Nambu fields are introduced because some possible exchange symmetries of the (4+0)-vertices in spinor representation are lost again in the Nambu representation due to the mapping between creation and annihilation operators. Furthermore, the above expression is most convenient for the identification of singular momentum dependences and the definition of interaction channels.

The discrete symmetries impose further constraints on the momentum and frequency dependence of $W^S_{k_1 k_2 k_3 k_4}$ and $W^T_{k_1 k_2 k_3 k_4}$:\\[0.5ex]
\begin{minipage}{\linewidth-1.1\mathnumlength}
\begin{center}
\begin{tabular}{ll}
	$W^i_{k_1 k_2 k_3 k_4} \propto \delta_{k_1+k_2+k_3+k_4,0}$	&	Translation invariance\\[0.2ex]
	$W^i_{k_1 k_2 k_3 k_4} = W^{i\ \ast}_{-k_1, -k_2, -k_3, -k_4}$ & Time reversal and positivity\\[0.2ex]
	$W^i_{k_1 k_2 k_3 k_4} = W^i_{Rk_1 Rk_2 Rk_3 Rk_4}$ & Space inversion
\end{tabular}
\end{center}
\end{minipage}
\begin{minipage}{\mathnumlength}
	\begin{eqnarray}
	\label{eq:4p0symmetries4mom}
	\end{eqnarray}
\end{minipage}\\[1ex]
where $i\in\{S,T\}$, $\bar k = (-k_0, \boldsymbol k)$, $Rk = (k_0, -\boldsymbol k)$ and $^\ast$ denoting complex conjugation.

\subsubsection{Nambu representation}
In Nambu representation, the anomalous (4+0)-effective interaction reads after applying the relations~\eqref{eq:SpinorNambuGrassmann} to equation~\eqref{eq:VP:4p0spinor} and subsequent antisymmetrization as for the anomalous (3+1)-effective interactions
\begin{equation}
	V_{(4+0)}[\bar\phi,\phi] = \frac{1}{4} \sum_{k_i} (W_{k_1 k_2 k_3 k_4} \phicp{k_1}\phicp{k_2} \phiam{k_3}\phiam{k_4} + W^\ast_{\bar k_4 \bar k_3 \bar k_2 \bar k_1} \phicm{k_1} \phicm{k_2} \phiap{k_3} \phiap{k_4})
\end{equation}
where
\begin{equation}
	\begin{split}
		W_{k_1 k_2 k_3 k_4} &= W^S_{k_1,-k_4,-k_3,k_2} - W^S_{k_1,-k_3,-k_4,k_2}\\
			&\ + W^T_{k_1,-k_4,-k_3,k_2} - W^T_{k_1,-k_3,-k_4,k_2} + 2 W^T_{k_1, k_2, -k_3, -k_4}.
	\end{split}
\end{equation}
The behaviour of $W^S_{k_1 k_2 k_3 k_4}$ and $W^T_{k_1 k_2 k_3 k_4}$ under the discrete symmetry operations and the exchange of particles gives rise to additional symmetries of the Nambu anomalous (4+0)-effective interaction,
\begin{equation}
	W_{k_1 k_2 k_3 k_4} = W^\ast_{-k_1,-k_2,-k_3,-k_4} = W_{-k_4, -k_3, -k_2, -k_1} = W^\ast_{\bar k_1, \bar k_2, \bar k_3, \bar k_4},
\end{equation}
where the first equality holds due to time reversal symmetry, the second due to exchange symmetries and the third due to the combination of time reversal and inversion symmetry.

\section{Channel-decomposition of vertex}
\label{sec:VP:ChannelDecomposition}
The effective interactions that were introduced in the last section depend on three equivalent and independent fermionic momenta and frequencies. The parametrization or approximate description of these dependences for example within the so-called $N$-patch approximation is a cumbersome task (see for example~\cite{Zanchi1998, Zanchi2000} for studies of the symmetric phase of the Hubbard model or~\cite{Gersch2008} for a study of symmetry breaking in the attractive Hubbard model). A more efficient description is possible after the identification of the singular dependences of the vertex on combinations of external momenta and frequencies, which are called transfer momenta\footnote{For the sake of brevity, in this section the term momentum usually refers to momentum \emph{and} frequency.}. This identification allows to write the vertex as a sum of several terms called interaction channels, each depending possibly singularly on a specific transfer momentum. 
The effective interactions in the channels are then parametrized in terms of a singular ``bosonic'' transfer momentum and two more regular ``fermionic'' relative momenta. 
This idea was applied by Karrasch~\etal\ to the frequency dependent vertex in the single-impurity Anderson model~\cite{Karrasch2008} and by Husemann and Salmhofer to the symmetric state of the repulsive Hubbard model~\cite{Husemann2009}. It is extended for the description of the two-particle vertex in a singlet superfluid in this work. Note that this idea of parametrization differs from a non-relativistic analogue of Mandelstam variables, which yield a description of the vertex in terms of three equivalent bosonic momenta and frequencies~\cite{Mandelstam1958}.

After the definition of interaction channels for the normal and anomalous effective interactions and the identification of their singular dependences on external momenta, the effective interactions in the channels are parametrized as boson-mediated interactions and expanded in bosonic exchange propagators and fermion-boson vertices\footnote{Note that the notion ``bosonic propagator'' is used somewhat sloppily in this work but follows the terminology introduced by Husemann~\etal~\cite{Husemann2009,Husemann2012}. The effective interactions in the channels do not necessarily satisfy the positivity requirements for `true' bosonic propagators~\cite{Husemann2012}. However, this does not impose problems within a purely fermionic formalism, but is the reason why the ``bosonic propagators'' are also called ``exchange propagators'' as in~\cite{Husemann2012}.}. 
In chapter~\ref{chap:ChannelDecomposition}, this channel-decomposition scheme serves as the basis for an efficient treatment of the singularities of the diagrams in the renormalization group equations for the two-particle vertex. 
In chapter~\ref{chap:RPFM}, the reduced pairing and forward-scattering model is solved within the channel-decomposition scheme in order to motivate the identification of singular dependences of the vertex at and below the critical scale for superfluidity.

\subsection{Definition of interaction channels}
\subsubsection{Normal (2+2)-effective interaction}
The normal effective interaction is described by an ansatz very similar to the one by Husemann and Salmhofer~\cite{Husemann2009}. Interaction channels are introduced as two-particle interactions with different spin symmetries, yielding
\begin{equation}
\begin{split}
\label{eq:VP:2p2SpinorAnsatz}
	V^{(2+2)}[\bar\psi,\psi] &= U \sum_{k_i} \delta_{k_1+k_2,k_3+k_4} \psicup{k_1}\psicdown{k_2}\psiadown{k_3}\psiaup{k_4} + \\
 &\ + \frac{1}{2}\sum_{k_i,\sigma,\sigma'} C_{k_1 k_2 k_3 k_4} \psic{k_1\sigma}\psic{k_2\sigma'}\psia{k_3\sigma'}\psia{k_4\sigma} + \\
	&\ + \frac{1}{2} \sum_{k_i,\sigma,\sigma'} P_{k_1 k_2 k_3 k_4} \psic{k_1\sigma}\psic{k_2\sigma'}\psia{k_3\sigma'}\psia{k_4\sigma}\\
	&\ + \frac{1}{2}\sum_{k_i,\sigma_i} M_{k_1 k_2 k_3 k_4} \vec\tau_{\sigma_1 \sigma_4}\cdot \vec\tau_{\sigma_2 \sigma_3} \psic{k_1\sigma_1} \psic{k_2\sigma_2} \psia{k_3\sigma_3} \psia{k_4\sigma_4}
\end{split}
\end{equation}
where the first term describes the local interaction of the Hubbard model and the other terms charge-charge, pairing and spin-spin interactions. The Hubbard interaction is not assigned to a specific channel in order to avoid ambiguities~\cite{Husemann2009}. 
Although the formulas below are stated for a local microscopic interaction as appropriate for the Hubbard model, they can easily be adapted to more general microscopic interactions by replacements like $P_{k_1 k_2 k_3 k_4} \rightarrow P^{(0)}_{k_1 k_2 k_3 k_4} + P_{k_1 k_2 k_3 k_4}$ for the Cooper channel as an example, where $P^{(0)}_{k_1 k_2 k_3 k_4}$ explicitly describes the (momentum dependent) microscopic interaction while $P_{k_1 k_2 k_3 k_4}$ absorbs fluctuation corrections, and similarly for the other channels. 
The functions $C$, $P$ and $M$ are symmetric under the simultaneous exchange of indices $1\leftrightarrow 2$ and $3\leftrightarrow 4$ and transform under the discrete symmetries as described in~\eqref{eq:2p2symmetries4mom} for $V_{k_1 k_2 k_3 k_4}$.

The singular dependences of the normal part of the vertex on external momenta and frequencies are identified as those of a Fermi liquid. In the particle-hole channel, the vertex may depend singularly on small momentum and frequency transfers due to forward-scattering (see for example~\cite{Abrikosov1964,Metzner1998}) or on large total momentum and small frequency transfers due to umklapp and backward scattering or approximate nesting of the Fermi surface, yielding enhanced $2\boldsymbol k_F$-scattering for special Fermi momenta $\boldsymbol k_F$ (see for example~\cite{Solyom1979,Altshuler1995}). The singular momentum and frequency dependence of the charge-charge and spin-spin interaction is therefore identified as
\begin{gather}
	C_{k_1 k_2 k_3 k_4} = C_{\frac{k_1+k_4}{2},\frac{k_2+k_3}{2}}(k_3-k_2)\delta_{k_1+k_2,k_3+k_4}\\
	M_{k_1 k_2 k_3 k_4} = M_{\frac{k_1+k_4}{2},\frac{k_2+k_3}{2}}(k_3-k_2)\delta_{k_1+k_2,k_3+k_4}
\end{gather}
where the transfer momentum $k_3-k_2$ describes the possibly singular ``bosonic'' momentum and frequency dependence of the channel  while $\tfrac{k_1+k_4}{2}$ and $\tfrac{k_2+k_3}{2}$ describe the more regular dependences on the ``fermionic'' relative momenta of the particle-hole pairs\footnote{At this stage it is worth illustrating why a decomposition of the effective particle-hole interaction into a singlet and a triplet part is inconvenient for the definition of channels: In this case, the transfer momentum is not invariant under the exchange of incoming \emph{or} outgoing particles,
\begin{align*}
	C_{k_1 k_2 k_3 k_4} &= C_{\frac{k_1+k_4}{2},\frac{k_2+k_3}{2}}(k_3-k_2)	& C_{k_2 k_1 k_3 k_4} &= C_{\frac{k_2+k_4}{2},\frac{k_1+k_3}{2}}(k_3-k_1)
\end{align*}
and similarly for the magnetic channel. This is different for the Cooper channel.}. The Cooper channel depends singularly on the total momentum of the incoming particle-particle pair due to repeated scattering in the vicinity of the Fermi surface (see for example~\cite{Schrieffer1964,Abrikosov1964}). The total momentum is therefore identified as the ``bosonic'' transfer momentum for the pairing channel,
\begin{gather}
	P_{k_1 k_2 k_3 k_4} = P_{\frac{k_1-k_2}{2},\frac{k_4-k_3}{2}}(k_1 + k_2)\delta_{k_1+k_2,k_3+k_4}.
	\label{eq:VSCchannel}
\end{gather}
Singularities usually arise if $k_1 + k_2$ vanishes, but the pairing interaction may also be enhanced at larger momenta for special Fermi surface geometries due to umklapp or $2 \boldsymbol k_F$-scattering. The dependences on $\tfrac{k_1-k_2}{2}$ and $\tfrac{k_4-k_3}{2}$ are expected to be more regular because they describe the relative motion of the scattered particle-particle pairs.

Inserting this ansatz in equation~\eqref{eq:VP:2p2spinor} yields
\newlength{\deltalength}
\settowidth{\deltalength}{$\delta_{k_1+k_2,k_3+k_4}$}
\begin{equation}
\begin{split}
	V_{k_1 k_2 k_3 k_4} &= \delta_{k_1+k_2,k_3+k_4} (U + P_{k_1 k_2 k_3 k_4} + C_{k_1 k_2 k_3 k_4} - 2 M_{k_1 k_2 k_4 k_3} - M_{k_1 k_2 k_3 k_4})\\
=& \delta_{k_1+k_2,k_3+k_4} \bigl[U + P_{\frac{k1-k2}{2},\frac{k_4-k_3}{2}}(k_1+k_2) + C_{\frac{k_1+k_4}{2},\frac{k_2+k_3}{2}}(k_3-k_2)\\
	&- 2 M_{\frac{k_1+k_3}{2},\frac{k_2+k_4}{2}}(k_4-k_2) - M_{\frac{k_1+k_4}{2},\frac{k_2+k_3}{2}}(k_3-k_2)\bigr].
\end{split}
\label{eq:VP:V2p2Spinor}
\end{equation}
Applying the symmetry relations~\eqref{eq:2p2symmetries4mom} to the effective interaction in every channel yields
\begin{gather}
	P_{kk'}(q) = P_{Rk,Rk'}(Rq) = P_{k',k}(q) = P^\ast_{-k',-k}(-q) = P_{-k,-k'}(q)\\ 
	C_{kk'}(q) = C_{Rk,Rk'}(Rq) = C_{kk'}(-q) = C^\ast_{-k,-k'}(q) = C_{k',k}(-q)\\ 
	M_{kk'}(q) = M_{Rk,Rk'}(Rq) = M_{kk'}(-q) = M^\ast_{-k,-k'}(q) = M_{k',k}(-q) 
\end{gather}
where the first equality follows from inversion symmetry, the second from inversion and time reversal symmetry, the third from inversion symmetry and Osterwalder-Schrader positivity and the last from the invariance under the exchange of the incoming and the outgoing particles.

In contrast to the particle-hole channel, it is favourable to decompose the effective interaction in the Cooper channel in a singlet and a triplet part. The reason is that the total momentum of the Cooper pair in~\eqref{eq:VSCchannel} is not only invariant under the exchange of the incoming \emph{and} the outgoing particles, but also under the separate exchange of the incoming \emph{or} the outgoing particles. The effective interaction in the Cooper channel is therefore written as
\begin{equation}
	P_{k_1 k_2 k_3 k_4} = P^S_{k_1 k_2 k_3 k_4} + P^T_{k_1 k_2 k_3 k_4}
\end{equation}
where
\begin{gather}
	P^S_{k_1 k_2 k_3 k_4} = P^S_{k_2 k_1 k_3 k_4} = P^S_{k_1 k_2 k_4 k_3} = P^S_{k_2 k_1 k_4 k_3}\\
	P^T_{k_1 k_2 k_3 k_4} = -P^T_{k_2 k_1 k_3 k_4} = -P^T_{k_1 k_2 k_4 k_3} = P^T_{k_2 k_1 k_4 k_3}.
\end{gather}
Bosonic transfer momenta are introduced as above,
\begin{gather}
	P^S_{k_1 k_2 k_3 k_4} = \delta_{k_1+k_2,k_3+k_4} P^S_{\frac{k_1-k_2}{2},\frac{k_4-k_3}{2}}(k_1 + k_2)\\
	P^T_{k_1 k_2 k_3 k_4} = \delta_{k_1+k_2,k_3+k_4} P^T_{\frac{k_1-k_2}{2},\frac{k_4-k_3}{2}}(k_1 + k_2)\\
	P_{kk'}(q) = P^S_{kk'}(q) + P^T_{kk'}(q),
\end{gather}
but $P^{S/T}$ posses further exchange symmetries,
\begin{gather}
	P^S_{k k'}(q) = P^S_{-k, k'}(q) = P^S_{k, -k'}(q) = P^S_{-k, -k'}(q)\\
	P^T_{k k'}(q) = -P^T_{-k, k'}(q) = -P^T_{k, -k'}(q) = P^T_{-k, -k'}(q).
\end{gather}
The singlet and triplet pairing interactions $P^{S/T}$ transform under the discrete symmetries as described above for $P$.

\subsubsection{Anomalous (3+1)-effective interaction}
Because of the presence of the singlet-pair operator in the first line of~\eqref{eq:3p1spinor}, it is expected that the (3+1) singlet vertex $\Omega^S_{k_1 k_2 k_3 k_4}$ depends singularly on the total momentum of the particle-particle pair, which equals the momentum transfer to the other particle. For the (3+1)-triplet vertex $\Omega^T_{k_1 k_2 k_3 k_4}$ the same singular dependence is assumed and it is shown in chapter~\ref{chap:ChannelDecomposition} that this identification allows for a unique assignment of the diagrams in the flow equations to the interaction channels. The bosonic and fermionic momentum dependences of the anomalous (3+1)-effective interaction are therefore defined as
\begin{gather}
	\Omega^S_{k_1 k_2 k_3 k_4} = \Omega^S_{\frac{k_1+k_4}{2},\frac{k_2-k_3}{2}}(k_2+k_3) \delta_{k_1 + k_2 + k_3,k_4}\\
\Omega^T_{k_1 k_2 k_3 k_4} = \Omega^T_{\frac{k_1+k_4}{2},\frac{k_2-k_3}{2}}(k_2+k_3) \delta_{k_1 + k_2 + k_3,k_4}.
\end{gather}
It is convenient to split the anomalous (3+1)-effective interaction into real ($X$) and imaginary ($\tilde X$) parts,
\begin{gather}
\Omega^S_{k k'}(q) = X^S_{k k'}(q) + i \tilde X^S_{k k'}(q)\\
\Omega^T_{k k'}(q) = X^T_{k k'}(q) + i \tilde X^T_{k k'}(q).
\end{gather}
Due to the exchange symmetries of the pair operators, the singlet interactions are even in the second fermionic momentum $k'$ while the triplet interactions are odd,
\begin{align}
	\Omega^S_{k, -k'}(q) &= \Omega^S_{k k'}(q)		&		\Omega^T_{k, -k'}(q) &= -\Omega^T_{k k'}(q),
\end{align}
and similarly for the real and imaginary parts. No such symmetry holds for the first fermionic momentum $k$. The discrete symmetries yield the relations
\begin{equation}
	\Omega^i_{kk'}(q) = \Omega^i_{Rk,Rk'}(Rq) = \Omega^{i\,\ast}_{-k,-k'}(-q) = \Omega^{i\,\ast}_{\bar k,\bar {k'}}(\bar q)
\end{equation}
for $i\in \{S,T\}$, where the first equality follows from inversion symmetry, the second from time reversal symmetry and the third from the combination of both discrete symmetries. Using these definitions, the Nambu vertex in equation~\eqref{eq:VP:Omega3p1Vertex} can be rewritten as
\begin{align}
	\Omega_{k_1 k_2 k_3 k_4} &= \Omega^S_{\frac{k_1+k_4}{2},\frac{k_2+k_3}{2}}(k_2-k_3) - \Omega^S_{\frac{k_2+k_4}{2},\frac{k_1+k_3}{2}}(k_1-k_3) \label{eq:VP:Omega3p1Nambu}\\
	&+ \Omega^T_{\frac{k_1+k_4}{2},\frac{k_2+k_3}{2}}(k_2-k_3) - \Omega^T_{\frac{k_2+k_4}{2},\frac{k_1+k_3}{2}}(k_1-k_3) - 2 \Omega^T_{\frac{k_4-k_3}{2},\frac{k_1-k_2}{2}}(k_1+k_2).\nonumber
\end{align}

\subsubsection{Anomalous (4+0)-effective interaction}
The presence of the pair operators in~\eqref{eq:VP:4p0spinor} suggests that $W^S_{k_1 k_2 k_3 k_4}$ and $W^T_{k_1 k_2 k_3 k_4}$ depend singularly on the total momentum and frequency of the particle-particle pairs. The interaction channels for the anomalous (4+0)-effective interaction with their singular dependence on external momenta and frequencies are therefore defined as
\begin{gather}
	W^S_{k_1 k_2 k_3 k_4} = W^S_{\frac{k_1-k_2}{2},\frac{k_4-k_3}{2}}(k_1+k_2) \delta_{k_1+k_2+k_3+k_4,0}\\
	W^T_{k_1 k_2 k_3 k_4} = W^T_{\frac{k_1-k_2}{2},\frac{k_4-k_3}{2}}(k_1+k_2) \delta_{k_1+k_2+k_3+k_4,0}
\end{gather}
with the total momentum of the pairs as bosonic and the relative momenta of the pairs as fermionic dependences. The exchange symmetries in equation~\eqref{eq:symW4p0spinor} imply
\begin{gather}
	W^S_{k k'}(q) = W^S_{-k,k'}(q) = W^S_{k, -k'}(q) = W^S_{-k,-k'}(q) = W^S_{k',k}(-q)\\
	W^T_{k k'}(q) = -W^T_{-k,k'}(q) = -W^T_{k, -k'}(q) = W^T_{-k,-k'}(q) = W^T_{k',k}(-q),
\end{gather}
while invariance under time reversal and space inversion yields
\begin{equation}
	W^i_{k k'}(q) = W^{i\,\ast}_{-k, -k'}(-q) = W^i_{Rk,Rk'}(Rq)
\end{equation}
for $i\in \{S,T\}$, respectively. The combination of time reversal symmetry and exchange symmetries yields
\begin{equation}
	W^i_{k k'}(q) = W^{i\,\ast}_{k k'}(-q) = W^{i\,\ast}_{k' k}(q).
\end{equation}
In terms of $W^i$, the anomalous (4+0)-effective interaction in Nambu representation reads
\begin{align}
		W_{k_1 k_2 k_3 k_4} &= W^S_{\frac{k_1+k_4}{2},\frac{k_2+k_3}{2}}(k_3-k_2) - W^S_{\frac{k_2+k_4}{2},\frac{k_1+k_3}{2}}(k_3-k_1)\label{eq:VP:W4p0Nambu} \\
		&+W^T_{\frac{k_1+k_4}{2},\frac{k_2+k_3}{2}}(k_3-k_2) - W^T_{\frac{k_2+k_4}{2},\frac{k_1+k_3}{2}}(k_3-k_1) + 2 W^T_{\frac{k_1-k_2}{2},\frac{k_3-k_4}{2}}(k_1+k_2).\nonumber
\end{align}

\subsubsection{Effective interactions for the amplitude and phase mode of the superfluid gap}
At and below the critical scale for superfluidity, the normal effective interaction in the Cooper channel and the anomalous (4+0)-effective interaction are singular. Specific linear combinations of these interactions describe the amplitude and phase mode of the superfluid gap~\cite{Salmhofer2004, Gersch2008}. Besides being less singular below the critical scale, these linear combinations are physically more transparent. The phase mode describes the Goldstone degree of freedom of the superfluid state and is only regularized by the external pairing field, whereas the amplitude mode is regularized below the critical scale by the generated superfluid gap. 
The replacement of normal and anomalous effective interactions in the Cooper channel by effective interactions describing the amplitude and phase mode is simplified by the isolation of singular momentum and frequency dependences in bosonic transfer momenta. No approximation is required for this step, which allows to simplify the Nambu structure of the vertex considerably at the same time.

In the singlet channel, the effective interactions describing the amplitude and phase mode of the superfluid gap $A^S_{kk'}(q)$ and $\Phi^S_{kk'}(q)$ are defined as
\begin{gather}
\begin{split}
	A^S_{kk'}(q) &= \re P^S_{kk'}(q) + \re W^S_{kk'}(q) =\\
		&= \re P^S_{k+\frac{q}{2},\frac{q}{2}-k,\frac{q}{2}-k',k'+\frac{q}{2}} + W^S_{k+\frac{q}{2},\frac{q}{2}-k,\frac{q}{2}-k',k'+\frac{q}{2}}
\end{split}\\
\begin{split}
	\Phi^S_{kk'}(q) &= \re P^S_{kk'}(q) - \re W^S_{kk'}(q)\\
	&= \re P^S_{k+\frac{q}{2},\frac{q}{2}-k,\frac{q}{2}-k',k'+\frac{q}{2}} - W^S_{k+\frac{q}{2},\frac{q}{2}-k,\frac{q}{2}-k',k'+\frac{q}{2}}.
\end{split}
\end{gather}
Here and in the following, it is assumed that the gap is real valued, which is no limitation in a system with time reversal symmetry and invariance under spatial inversions. For the triplet channel, similar combinations are introduced (with $S\rightarrow T$) in order to simplify the Nambu structure as for the singlet channel, but the terms amplitude and phase mode are not meaningful there (because the superfluid gap is assumed to have singlet symmetry). The imaginary parts are just renamed for convenience in
\begin{align}
	\nu^{i}_{kk'}(q) &= \im P^i_{kk'}(q)		&		\tilde\nu^i_{kk'}(q) &= \im W^i_{kk'}(q)
\end{align}
with $i \in \{S,T\}$. The insertion of these definitions into the above ansatz for the normal and anomalous effective interactions in the Cooper channel and the reorganization of Kronecker symbols using Pauli matrices yield
\begin{equation}
	\begin{split}
		V&_{(2+2)}[\bar\phi,\phi] + V_{(4+0)}[\bar\phi,\phi] = \\
&=\frac{1}{4}\sum_{k_i,s_i}\delta_{k_1+k_2,k_3+k_4} \Bigl\{\frac{1}{2}\sum_{i\in{S,T}}\Bigl[A^i_{\frac{k_1+k_4}{2},\frac{k_2+k_3}{2}}(k_3-k_2)\mtau{1}{s_1 s_4} \mtau{1}{s_2 s_3} + \\
	&\ + \Phi^i_{\frac{k_1+k_4}{2},\frac{k_2+k_3}{2}}(k_3-k_2) \mtau{2}{s_1 s_4} \mtau{2}{s_2 s_3} + \nu^i_{\frac{k_1+k_4}{2},\frac{k_2+k_3}{2}}(k_3-k_2) (\mtau{1}{s_1 s_4} \mtau{2}{s_2 s_3} - \mtau{2}{s_1 s_4}\mtau{1}{s_2 s_3}) -\\
	&\ - \tilde\nu^i_{\frac{k_1+k_4}{2},\frac{k_2+k_3}{2}}(k_3-k_2) (\mtau{1}{s_1 s_4} \mtau{2}{s_2 s_3} + \mtau{2}{s_1 s_4} \mtau{1}{s_2 s_3}) - (3\leftrightarrow 4)\Bigr] +\\
	&\ + A^T_{\frac{k_1-k_2}{2},\frac{k_4-k_3}{2}}(k_1+k_2) \mtau{3}{s_1 s_2} \mtau{3}{s_3 s_4} + i \nu^T_{\frac{k_1-k_2}{2},\frac{k_4-k_3}{2}}(k_1+k_2) (\mtau{0}{s_1 s_2} \mtau{3}{s_3 s_4} + \mtau{3}{s_1 s_2}\mtau{0}{s_3 s_4}) + \\
	&\ + \Phi^T_{\frac{k_1-k_2}{2},\frac{k_4-k_3}{2}}(k_1+k_2) \mtau{0}{s_1 s_2} \mtau{0}{s_3 s_4} + i \tilde\nu^T_{\frac{k_1-k_2}{2},\frac{k_4-k_3}{2}}(k_1+k_2) (\mtau{0}{s_1 s_2} \mtau{3}{s_3 s_4} - \mtau{3}{s_1 s_2}\mtau{0}{s_3 s_4}) \Bigl\}\times\\
	&\ \ \times \phic{k_1 s_1} \phic{k_2 s_2} \phia{k_3 s_3} \phia{k_4 s_4}
	\end{split}\raisetag{1em}
\end{equation}
where the Pauli matrices $\mtau{i}{}$ and the unit matrix $\mtau{0}{}$ describe the dependence on Nambu indices and ``$(3\leftrightarrow 4)$'' is a shorthand for the terms in the square brackets with indices $3$ and $4$ exchanged\footnote{In this expression, the invariance under simultaneous exchange of particles $1 \leftrightarrow 2$ and $3 \leftrightarrow 4$ of the term involving $\tilde \nu(k_3 - k_2)$ is not obvious. This is due to simplifications that arise from time reversal symmetry and invariance under space inversion. The expression can be cast in a more symmetric form by noting that
\begin{displaymath}
	\tilde \nu^{i}_{k k'}(q) = \tilde \nu^{i}_{k' k}(-q)
\end{displaymath}
holds in a system that is invariant under time reversal and space inversion.}.

\subsubsection{Effective interactions in the Nambu particle-hole and particle-particle channels}
\label{subsec:VP:EffIntChannels}
\begin{figure}
	\centering
	\includegraphics[scale=1]{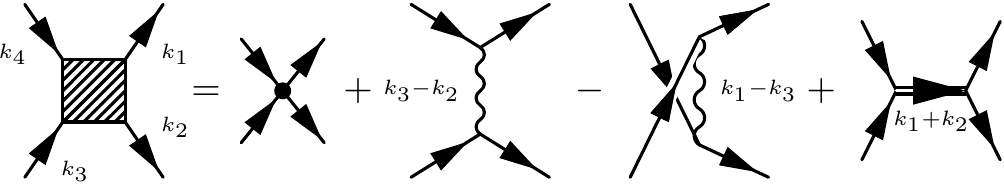}
	\caption{Diagrammatic representation of the decomposition of the Nambu two-particle vertex in bare interaction, particle-hole channels and particle-particle channel.}
	\label{fig:VP:VertexDecomposition}
\end{figure}
For later convenience, it is useful to write the Nambu two-particle vertex as a sum of effective interactions in the Nambu particle-hole and particle-particle channels,
\begin{align}
	\Vertex{s_1 s_2 s_3 s_4}(k_1 k_2 k_3 k_4) &= U_{s_1 s_2 s_3 s_4} + V^\text{PH}_{s_1 s_2 s_3 s_4}\bigl(\tfrac{k_1+k_4}{2}, \tfrac{k_2+k_3}{2}; k_3-k_2\bigr) \label{eq:VP:VertexDecomposition} \\
	& - V^\text{PH}_{s_2 s_1 s_3 s_4}\bigl(\tfrac{k_2+k_4}{2}, \tfrac{k_1+k_3}{2}; k_3-k_1\bigr) + V^\text{PP}_{s_1 s_2 s_3 s_4}\bigl(\tfrac{k_1-k_2}{2}, \tfrac{k_4-k_3}{2}; k_1+k_2\bigr),\nonumber
\end{align}
as shown diagrammatically in figure~\ref{fig:VP:VertexDecomposition}. The terms on the right hand side represent the (antisymmetrized) local microscopic interaction $U_{s_1 s_2 s_3 s_4}$ and the effective interactions in the Nambu particle-hole channel $V^\text{PH}_{s_1 s_2 s_3 s_4}(\tfrac{k_1+k_4}{2}, \tfrac{k_2+k_3}{2}; k_3-k_2)$, which appears twice in order to incorporate exchange symmetries, as well as in the Nambu particle-particle channel $V^\text{PP}_{s_1 s_2 s_3 s_4}(\tfrac{k_1-k_2}{2}, \tfrac{k_4-k_3}{2}; k_1+k_2)$. Collecting terms with the corresponding transfer momenta, one obtains
\begin{gather}
	U_{s_1 s_2 s_3 s_4} = U (\delta_{s_{1-4},+-+-} + \delta_{s_{1-4},-+-+} - \delta_{s_{1-4},+--+} - \delta_{s_{1-4},-++-})\\
	\begin{split}
	V^\text{PH}_{s_1 s_2 s_3 s_4}(k, k'; q) &= \bigl(C_{k k'}(q) + M_{k k'}(q)\bigr) \delta_{s_{1-4},++++} + \bigl(C^\ast_{k k'}(q) + M^\ast_{k k'}(q)\bigr) \delta_{s_{1-4},----}\\
	&+ \bigl(M_{k, -k'}(-q) - C_{k, -k'}(-q)\bigr) \delta_{s_{1-4},+--+}\\
	&+ \bigl(M^\ast_{k, -k'}(-q) - C^\ast_{k, -k'}(-q)\bigr) \delta_{s_{1-4},-++-}\\
	&+\sum_{i\in S,T}\bigl[\tfrac{1}{2} A^i_{k k'}(q) \mtau{1}{s_1 s_4} \mtau{1}{s_2 s_3} + \tfrac{1}{2} \Phi^i_{k k'}(q) \mtau{2}{s_1 s_4} \mtau{2}{s_2 s_3}\\
	&\quad + \tfrac{1}{2} \nu^i_{k k'}(q) (\mtau{1}{s_1 s_4} \mtau{2}{s_2 s_3} - \mtau{2}{s_1 s_4} \mtau{1}{s_2 s_3})\\
	&\quad - \tfrac{1}{2} \tilde \nu^i_{k k'}(q) (\mtau{1}{s_1 s_4} \mtau{2}{s_2 s_3} + \mtau{2}{s_1 s_4} \mtau{1}{s_2 s_3})\\
	&\quad + \Omega^i_{k k'}(-q) \delta_{s_{1-4},++-+} + \Omega^i_{k' k}(q) \delta_{s_{1-4},+++-} + \Omega^i_{k k'}(q) \delta_{s_{1-4},+-++}\\
	&\quad - \Omega^{i\,\ast}_{k' k}(-q) \delta_{s_{1-4},+---} - \Omega^{i,\ast}_{k' k}(q) \delta_{s_{1-4},---+} - \Omega^{i\,\ast}_{k k'}(q) \delta_{s_{1-4},-+--}\\
	&\quad - \Omega^{i\,\ast}_{k k'}(-q) \delta_{s_{1-4},--+-} + \Omega^i_{k' k}(-q) \delta_{s_{1-4},-+++}\bigr]
	\label{eq:VP:NambuPH}\raisetag{15pt}
	\end{split}
\end{gather}
\begin{equation}
	\begin{split}
		V^\text{PP}_{s_1 s_2 s_3 s_4}(k, k'; q) &= 2 M_{k k'}(q) \delta_{s_{1-4},+--+} + 2 M^\ast_{k k'}(q) \delta_{s_{1-4},-++-}\\
	&-2 M_{k,-k'}(q) \delta_{s_{1-4},+-+-} - 2 M^\ast_{k,-k'}(q) \delta_{s_{1-4},-+-+}\\
	&+ A^T_{k k'}(q) \mtau{3}{s_1 s_2} \mtau{3}{s_3 s_4} + \Phi^T_{k k'}(q) \mtau{0}{s_1 s_2} \mtau{0}{s_3 s_4}\\
	&+ i \nu^T_{k k'}(q) (\mtau{0}{s_1 s_2} \mtau{3}{s_3 s_4} + \mtau{3}{s_1 s_2} \mtau{0}{s_3 s_4})\\
	&+ i \tilde \nu^T_{k k'}(q) (\mtau{0}{s_1 s_2} \mtau{3}{s_3 s_4} - \mtau{3}{s_1 s_2} \mtau{0}{s_3 s_4})\\
	&- 2\Omega^T_{k' k}(q) \delta_{s_{1-4},++-+} + 2\Omega^T_{-k',k}(q) \delta_{s_{1-4},+++-} - 2\Omega^{T\,\ast}_{-k,k'}(q) \delta_{s_{1-4},+---}\\
	&+ 2\Omega^{T\,\ast}_{kk'}(q) \delta_{s_{1-4},-+--} - 2\Omega^T_{k k'}(q) \delta_{s_{1-4},+-++} + 2\Omega^T_{-k,k'}(q) \delta_{s_{1-4},-+++}\\
	&+ 2\Omega^{T\,\ast}_{k' k}(q) \delta_{s_{1-4},--+-} - 2\Omega_{-k',k}^{T\,\ast}(q) \delta_{s_{1-4},---+}
	\label{eq:VP:NambuPP}\raisetag{15pt}
	\end{split}
\end{equation}
where $\delta_{s_{1-4},abcd} = \delta_{s_1 a} \delta_{s_2 b} \delta_{s_3 c} \delta_{s_4 d}$. Note that the effective interaction in the Nambu particle-hole channel describes (spinor) particle-hole as well as (spinor) particle-particle processes and similarly for the Nambu particle-particle channel. In the above expressions for $V^\text{PH}_{s_1 s_2 s_3 s_4}$ and $V^\text{PP}_{s_1 s_2 s_3 s_4}$, all Kronecker symbols may be replaced by Pauli matrices as for the effective interactions in the Cooper channel. However, this does not simplify the Nambu structure of the expressions, wherefore the results are stated in appendix~\ref{appendix:VP} only for completeness.

\subsection{Boson propagators and fermion-boson vertices}
\label{subsec:VP:BosonProp_gFB}
In order to achieve an efficient description of the singular dependences of the vertex on momenta and frequencies, the effective interactions in the channels are described as boson-mediated interactions, similar to the work by Husemann and Salmhofer~\cite{Husemann2009} or Husemann~\etal~\cite{Husemann2012}. The idea is to expand the effective interaction in the channels in bosonic exchange propagators and fermion-boson vertices, where the former capture the singular dependence on the transfer momentum and frequency while the latter describe the more regular dependence on the fermionic relative momenta and frequencies. As an example, the amplitude mode of the superfluid gap is written as
\begin{equation}
	A_{k k'}(q) = \sum_{\alpha,\beta} A_{\alpha\beta}(q) h^A_\alpha(q,k) h^A_\beta(q, k')
\label{eq:VP:Aexpansion}
\end{equation}
with bosonic propagators $A_{\alpha\beta}(q)$ and fermion-boson vertices $h^A_\alpha(q,k)$. The (multi-) indices $\alpha$ and $\beta$ label a conveniently chosen orthonormal set of basis functions describing the dependence of the fermion-boson vertices on $q$ and $k$. In principle, different basis functions can be chosen for each channel. 

The bosonic propagators have to be approximated in a way that captures the singular momentum and frequency dependence of the channels. Their singularities can be inferred from resummations of perturbation theory, from the singularity structure of the diagrams in the RG equations or from RG calculations for example within the $N$-patch approximation. The approximations for the bosonic propagators are motivated in chapter~\ref{chap:RPFM} and described in detail in chapters~\ref{chap:AttractiveHubbard} and~\ref{chap:RepulsiveHubbard} for the Hubbard model.

The approximations for the fermion-boson vertices are discussed in the following, because they influence the symmetry properties of the bosonic propagators. The coupling between fermions and bosons is described similarly to the work by Husemann~\etal~\cite{Husemann2012} for the repulsive Hubbard model as
\begin{equation}
	h_\alpha(q,k) = f_\alpha(\boldsymbol k) g_\alpha(\boldsymbol q, k_0)
\end{equation}
for every channel. $f_\alpha(\boldsymbol k)$ are basis functions for the description of the dependence on the fermionic relative momentum. In this work, they are either chosen as lattice form factors fulfilling
\begin{equation}
	\intzwei{k} f_\alpha(\boldsymbol k) f_\beta(\boldsymbol k) = \delta_{\alpha\beta}
\end{equation}
or as Fermi surface harmonics fulfilling
\begin{equation}
	\int_{\boldsymbol k_F} \frac{ds}{L_F} f_\alpha(\boldsymbol k) f_\beta(\boldsymbol k) = \inteins{\theta} m(\theta) f_\alpha(\boldsymbol k_F(\theta)) f_\beta(\boldsymbol k_F(\theta)) = \delta_{\alpha\beta}
\end{equation}
where $m(\theta)$ is the measure for the integration along the Fermi surface (see section~\ref{sec:CD:ProjBosProp}). In a system with inversion symmetry, the basis functions $f_\alpha(\boldsymbol k)$ can be chosen as to have a fixed parity with respect to the inversion of momenta,
\begin{gather}
	f_\alpha(-\boldsymbol k) = \zeta_\alpha f_\alpha(\boldsymbol k)
\end{gather}
where $\zeta_\alpha = \pm 1$. In expansions like~\eqref{eq:VP:Aexpansion}, it is assumed that both basis functions for the momentum dependence have the same parity\footnote{For the Cooper channel this is not an approximation but follows from symmetries (see below).}, \ie\ $\zeta_\alpha = \zeta_\beta$. $g_\alpha(\boldsymbol q, k_0)$ describes the dependence of the effective interaction on the fermionic relative frequency and is assumed to be even in $\boldsymbol q$ and $k_0$ in the following. This excludes possible particle-hole or particle-particle fluctuations with an odd dependence on the fermionic frequency. Such interactions may arise in the Cooper channel~\cite{Schrieffer1994} but are expected to be subdominant. Similarly to the above example, the bosonic propagators and fermion-boson vertices for the Cooper channel are defined as
\begin{gather}
	A^{S/T}_{k k'}(q) = \sum_{\alpha,\beta} A^{S/T}_{\alpha\beta}(q) h^A_\alpha(q,k) h^A_\beta(q, k')\\
	\Phi^{S/T}_{k k'}(q) = \sum_{\alpha,\beta} \Phi^{S/T}_{\alpha\beta}(q) h^\Phi_\alpha(q,k) h^\Phi_\beta(q, k')\\
	\nu^{S/T}_{k k'}(q) = \sum_{\alpha,\beta} \nu^{S/T}_{\alpha\beta}(q) h^\nu_\alpha(q,k) h^\nu_\beta(q, k')\\
	\tilde\nu^{S/T}_{k k'}(q) = \sum_{\alpha,\beta} \tilde\nu^{S/T}_{\alpha\beta}(q) h^{\tilde\nu}_\alpha(q,k) h^{\tilde\nu}_\beta(q, k')
\end{gather}
where the sums on the right hand side run over the corresponding subset of basis functions with even or odd parity for the singlet and triplet channel, respectively. The same idea of decomposition is also applied to the effective interactions in the spinor particle-hole channel and to the anomalous (3+1)-effective interaction:
\begin{gather}
	C_{k k'}(q) = \sum_{\alpha,\beta} C_{\alpha\beta}(q) h^C_\alpha(q,k) h^C_\beta(q, k')\\
	M_{k k'}(q) = \sum_{\alpha,\beta} M_{\alpha\beta}(q) h^M_\alpha(q,k) h^M_\beta(q, k')\\
	X^{S/T}_{k k'}(q) = \sum_{\alpha,\beta} X^{S/T}_{\alpha\beta}(q) h^{X_\text{PH}}_\alpha(q, k) h^{X_A}_\beta(q, k')\\
	\tilde X^{S/T}_{k k'}(q) = \sum_{\alpha,\beta} \tilde X^{S/T}_{\alpha\beta}(q) h^{\tilde X_\text{PH}}_\alpha(q, k) h^{\tilde X_\Phi}_\beta(q, k')
\end{gather}
For the anomalous (3+1)-effective interaction, the superscript $S/T$ refers to the parity of $h^{X_A}$ or $h^{\tilde X_\Phi}$. 

Exploiting the orthogonality of the basis functions for the dependence on the fermionic momenta, symmetry properties of the bosonic propagators can be deduced from those of the effective interactions in the channels. As an example, inserting symmetry related expressions for $A_{kk'}(q)$ in
\begin{equation}
	\intzwei{k}\intzwei{k'} f_\alpha(\boldsymbol k) f_\beta(\boldsymbol k') A_{kk'}(q) = A_{\alpha\beta}(q) g^A_\alpha(k_0) g^A_\beta(k_0'),
\end{equation}
which holds for the above ansatz for $h^A_\alpha(q,k)$, yields constraints on $A_{\alpha\beta}(q)$. Symmetry properties of the other bosonic propagators follow similarly and this procedure can also be applied when using Fermi surface harmonics as basis functions for the dependences on the fermionic momenta. It should be noted that $P_{kk'}(q)$ and $W_{kk'}(q)$ are individually invariant under more symmetry operations than the linear combinations $A_{k k'}(q)$ and $\Phi_{k k'}(q)$. Time reversal invariance and the positivity requirement lead for example to two relations for $P_{kk'}(q)$, while only their combination yields a symmetry constraint on $W_{kk'}(q)$. Exploiting the symmetry relations for the effective interactions in the particle-particle channel leads to
\begin{align}
	A^{S/T}_{\alpha\beta}(q) &= A^{S/T}_{\alpha\beta}(Rq)		&		\Phi^{S/T}_{\alpha\beta}(q) &= \Phi^{S/T}_{\alpha\beta}(Rq)		&		& \text{Inversion}\\
	\nu^{S/T}_{\alpha\beta}(q) &= \nu^{S/T}_{\alpha\beta}(Rq)		&		\tilde\nu^{S/T}_{\alpha\beta}(q) &= \tilde\nu^{S/T}_{\alpha\beta}(Rq)		&		&\nonumber\\
	A^{S/T}_{\alpha\beta}(q) &= A^{S/T}_{\alpha\beta}(\bar q)		&		\Phi^{S/T}_{\alpha\beta}(q) &= \Phi^{S/T}_{\alpha\beta}(\bar q)		&		& \text{Time Rev.\ and Pos.}\\
	\nu^{S/T}_{\alpha\beta}(q) &= -\nu^{S/T}_{\alpha\beta}(\bar q)		&		\tilde\nu^{S/T}_{\alpha\beta}(q) &= -\tilde\nu^{S/T}_{\alpha\beta}(\bar q)		&		&\nonumber\\
	&		&		\tilde \nu^{S/T}_{\alpha\beta}(q) &= \tilde \nu^{S/T}_{\beta\alpha}(-q)	&		&\text{Exchange.}
\end{align}
Besides the last relation, the invariance under the exchange of particles implies that $A^{S/T}_{\alpha\beta}(q)$, $\Phi^{S/T}_{\alpha\beta}(q)$, $\nu^{S/T}_{\alpha\beta}(q)$ and $\tilde \nu^{S/T}_{\alpha\beta}(q)$ are block-diagonal with respect to the parity of the form factors, \ie\ the matrices are non-zero only if $\zeta_\alpha \zeta_\beta = 1$. The combination of the three symmetry relations for $\tilde\nu$ yields
\begin{equation}
	\tilde \nu^{S/T}_{\alpha\beta}(q) = - \tilde\nu^{S/T}_{\beta\alpha}(q),
\end{equation}
which implies that the diagonal components $\tilde\nu^{S/T}_{\alpha\alpha}$ vanish. For the particle-hole channel, the procedure described above yields
\begin{align}
	M_{\alpha\beta}(q) &= M_{\beta\alpha}(-q)			&		C_{\alpha\beta}(q) &= C_{\beta\alpha}(-q)		&		&\text{Exchange of particles}\\
	M_{\alpha\beta}(q) &= M_{\alpha\beta}(-q)			&		C_{\alpha\beta}(q) &= C_{\alpha\beta}(-q)		&		&\text{Time reversal}\\
	M_{\alpha\beta}(q) &= M^\ast_{\alpha\beta}(Rq)			&		C_{\alpha\beta}(q) &= C^\ast_{\alpha\beta}(Rq)		&		&\text{Positivity}\\
	M_{\alpha\beta}(q) &= M_{\alpha\beta}(Rq)			&		C_{\alpha\beta}(q) &= C_{\alpha\beta}(Rq)		&		&\text{Inversion}.
\end{align}
The combination of these relations implies that $M_{\alpha\beta}(q)$ and $C_{\alpha\beta}(q)$ are real, symmetric with respect to $\alpha$, $\beta$ and even functions in $q_0$ and $\boldsymbol q$. The bosonic propagators for the anomalous (3+1)-effective interaction fulfil
\begin{align}
	X^{S/T}_{\alpha\beta}(q) &= \zeta_\alpha \zeta_\beta X^{S/T}_{\alpha\beta}(R q)		&		\tilde X^{S/T}_{\alpha\beta}(q) &= \zeta_\alpha \zeta_\beta \tilde X^{S/T}_{\alpha\beta}(R q)		&		& \text{Inversion}\\
	X^{S/T}_{\alpha\beta}(q) &= X^{S/T}_{\alpha\beta}(\bar q)		&		\tilde X^{S/T}_{\alpha\beta}(q) &= -\tilde X^{S/T}_{\alpha\beta}(\bar q)		&		& \text{Time Rev.\ and Pos.}
\end{align}

Within the above approximation for the fermion-boson vertices and the bosonic propagators, the Nambu structure of the effective interactions in the Nambu particle-hole and particle-particle channel can be cast to a relatively compact form with the help of Pauli matrices. The simplifications occur because of the assumed $\boldsymbol q$-independence of the momentum dependent part of the fermion-boson vertices, $f_\alpha(\boldsymbol q, \boldsymbol k) \approx f_\alpha(\boldsymbol k)$, that allows to collect terms like $M_{k,k'}(q)$ and $M_{k,-k'}(q)$ in equations~\eqref{eq:VP:NambuPH} and~\eqref{eq:VP:NambuPP}. Inserting the expansions in boson propagators and fermion-boson vertices, the effective interactions in the Nambu particle-hole and particle-particle channels read
\begin{equation}
	\begin{split}
	V^\text{PH}_{s_1 s_2 s_3 s_4}&(k, k'; q) = \sum_{\alpha,\beta} \Bigl[ C_{\alpha\beta}(q) h^C_\alpha(q,k) h^C_\beta(q,k') \bigl(\tfrac{1 + \zeta_\alpha}{2} \mtau{3}{s_1 s_4} \mtau{3}{s_2 s_3} + \tfrac{1-\zeta_\alpha}{2} \mtau{0}{s_1 s_4} \mtau{0}{s_2 s_3}\bigr)\\
	&+ M_{\alpha\beta}(q) h^M_\alpha(q,k) h^M_\beta(q,k') \bigl(\tfrac{1 - \zeta_\alpha}{2} \mtau{3}{s_1 s_4} \mtau{3}{s_2 s_3} + \tfrac{1+\zeta_\alpha}{2} \mtau{0}{s_1 s_4} \mtau{0}{s_2 s_3}\bigr)\\
	&+ \tfrac{1}{2} A_{\alpha\beta}(q) h^A_\alpha(q, k) h^A_\beta(q, k') \mtau{1}{s_1 s_4} \mtau{1}{s_2 s_3} + \tfrac{1}{2} \Phi_{\alpha\beta}(q) h^\Phi_\alpha(q, k) h^\Phi_\beta(q, k') \mtau{2}{s_1 s_4} \mtau{2}{s_2 s_3}\\
	&+ \tfrac{1}{2} \nu_{\alpha\beta}(q) h^\nu_\alpha(q, k) h^\nu_\beta(q, k') (\mtau{1}{s_1 s_4} \mtau{2}{s_2 s_3} - \mtau{2}{s_1 s_4} \mtau{1}{s_2 s_3})\\
	&- \tfrac{1}{2} \tilde\nu_{\alpha\beta}(q) h^{\tilde\nu}_\alpha(q,k) h^{\tilde\nu}_\beta(q,k') (\mtau{1}{s_1 s_4} \mtau{2}{s_2 s_3} + \mtau{2}{s_1 s_4} \mtau{1}{s_2 s_3})\\
	&+ X_{\alpha\beta}(q) h^{X_\text{PH}}_\alpha(q, k) h^{X_A}_\beta(q, k') \mtau{3}{s_1 s_4} \mtau{1}{s_2 s_3} + X_{\beta\alpha}(q) h^{X_A}_\alpha(q, k) h^{X_\text{PH}}_\beta(q, k') \mtau{1}{s_1 s_4} \mtau{3}{s_2 s_3} \\
	&+ \tilde X_{\alpha\beta}(q) h^{\tilde X_\text{PH}}_\alpha(q, k) h^{\tilde X_\Phi}_\beta(q, k') \mtau{3}{s_1 s_4} \mtau{2}{s_2 s_3} - \tilde X_{\beta\alpha}(q) h^{\tilde X_\Phi}_\alpha(q, k) h^{\tilde X_\text{PH}}_\beta(q, k') \mtau{2}{s_1 s_4} \mtau{3}{s_2 s_3}\Bigr]
	\end{split}
	\label{eq:VP:NambuPH:gEB}
\end{equation}
and
\begin{equation}
	\begin{split}
	V^\text{PP}_{s_1 s_2 s_3 s_4}&(k, k'; q) = \\
	& =\sum_{\alpha,\beta} \Bigl[M_{\alpha\beta}(q) h^M_\alpha(q,k) h^M_\beta(q,k') \bigl((1 - \zeta_\alpha) \mtau{1}{s_1 s_2} \mtau{1}{s_3 s_4} + (1+\zeta_\alpha) \mtau{2}{s_1 s_2} \mtau{2}{s_3 s_4}\bigr)\\
	&+ A^T_{\alpha\beta}(q) h^{A,T}_\alpha(q,k) h^{A,T}_\beta(q,k') \mtau{3}{s_1 s_2} \mtau{3}{s_3 s_4} + \Phi^T_{\alpha\beta}(q) h^{\Phi,T}_\alpha(q,k) h^{\Phi,T}_\beta(q,k') \mtau{0}{s_1 s_2} \mtau{0}{s_3 s_4}\\
	&+ i \nu^T_{\alpha\beta}(q) h^{\nu,T}_\alpha(q,k) h^{\nu,T}_\beta(q,k') (\mtau{3}{s_1 s_2} \mtau{0}{s_3 s_4} + \mtau{0}{s_1 s_2} \mtau{3}{s_3 s_4})\\
	&+ i \tilde\nu^T_{\alpha\beta}(q) h^{\tilde\nu,T}_\alpha(q, k) h^{\tilde\nu,T}_\beta(q, k') (\mtau{0}{s_1 s_2} \mtau{3}{s_3 s_4} - \mtau{3}{s_1 s_2} \mtau{0}{s_3 s_4})\\
	&-2 X^T_{\alpha\beta}(q) h^{X_\text{PH},T}_\alpha(q, k) h^{X_A,T}_\beta(q, k') \mtau{1}{s_1 s_2} \mtau{3}{s_3 s_4}\\
	& - 2 X^T_{\beta\alpha}(q) h^{X_A,T}_\alpha(q, k) h^{X_\text{PH},T}_\beta(q, k') \mtau{3}{s_1 s_2} \mtau{1}{s_3 s_4}\\
	& - 2 \tilde X^T_{\alpha\beta}(q) h^{\tilde X_\text{PH},T}_\alpha(q, k) h^{\tilde X_\Phi,T}_\beta(q, k') \mtau{1}{s_1 s_2} \mtau{0}{s_3 s_4}\\
	& - 2 \tilde X^T_{\beta\alpha}(q) h^{\tilde X_\Phi,T}_\alpha(q, k) h^{\tilde X_\text{PH},T}_\beta(q, k') \mtau{0}{s_1 s_2} \mtau{1}{s_3 s_4}\Bigr],
	\end{split}
	\label{eq:VP:NambuPP:gEB}
\end{equation}
respectively. The superscript $T$ stands for triplet and indicates that only terms with fermion-boson vertices of odd parity are kept in the sums. The tensor products of Pauli matrices together with the unit matrix form an orthogonal basis in the space of 4x4-matrices and this property can be exploited for an efficient evaluation of the Nambu index sums that appear when deriving flow equations for the effective interactions in the channels. It should be noted that the expression~\eqref{eq:VP:NambuPH:gEB} and~\eqref{eq:VP:NambuPP:gEB} simplify considerably within the approximations that are applied in chapters~\ref{chap:AttractiveHubbard} and~\ref{chap:RepulsiveHubbard}.

%% file: Thesis_ChannelDecomposition.tex
\chapter{Channel-decomposed renormalization group equations}
\label{chap:ChannelDecomposition}

In chapter~\ref{chap:VertexParametrization}, a decomposition of the Nambu two-particle vertex in interaction channels that isolate the singular dependences of the vertex on external momenta and frequencies was presented. In this chapter, a reorganisation of the diagrams in the renormalization group equation for the vertex is described that serves as the basis for the formulation of approximations for the effective interactions in the channels and for their efficient computation. It is guided by the leading singular dependence of the diagrams on external momenta and the exact solution of a reduced pairing and forward scattering model (see chapter~\ref{chap:RPFM}). 
Capturing the latter within the channel-decomposition scheme is important for a correct treatment of the singularities in the Cooper channel at the critical scale for superfluidity. 

The idea of deriving channel-decomposed RG equations for the vertex by assigning diagrams to interaction channels according to their singular dependence on external momenta and frequencies was applied before to the single impurity Anderson model by Karrasch~\etal~\cite{Karrasch2008} and to the repulsive Hubbard model in the symmetric state by Husemann and Salmhofer~\cite{Husemann2009}. It is extended to the case of symmetry breaking in the Cooper channel in this chapter. 

In the derivation of channel-decomposed RG equations, the same symmetries are assumed to hold as in chapters~\ref{chap:FunctionalRG} and~\ref{chap:VertexParametrization}, making the equations suitable for the description of a spin rotation invariant singlet superfluid in Nambu representation or for a symmetric (non-superfluid) system in spinor representation. The equations reduce to the channel-decomposition scheme by Husemann and Salmhofer~\cite{Husemann2009} in the absence of symmetry breaking. 

This chapter is organized as follows. In section~\ref{sec:CD:OneLoop}, the channel-decomposition scheme on one-loop level is presented. The channel-decomposed RG equations are derived in subsection~\ref{subsec:CD:RGEOneLoop}. Estimates for the impact of phase fluctuations in the infrared on one-loop level are discussed in subsection~\ref{subsec:CD:InfraredSingOneLoop}, yielding insight into the singular behaviour of the vertex in the limit of a vanishing external pairing field. In section~\ref{sec:CD:ProjBosProp}, flow equations for bosonic propagators and fermion-boson vertices are derived from those for the effective interactions. The channel-decomposition scheme is extended to the two-loop level in section~\ref{sec:CD:TwoLoop}. The channel-decomposed flow equations are derived in subsection~\ref{subsec:CD:RGETwoLoop}. Estimates for the impact of phase fluctuations on the infrared flow on two-loop level are discussed in subsection~\ref{subsec:CD:InfraredSingTwoLoop}. 

\section{One-loop level}
\label{sec:CD:OneLoop}
In this section, the channel-decomposition scheme by Husemann and Salmhofer~\cite{Husemann2009} is extended for the description of singlet superfluidity. The channel-decomposed flow equations on one-loop level are obtained by assigning diagrams to interaction channels according to their singular dependences on external momenta and frequencies. The resulting equations capture the exact solution of the reduced pairing and forward scattering model, which is discussed in chapter~\ref{chap:RPFM}. In chapters~\ref{chap:AttractiveHubbard} and~\ref{chap:RepulsiveHubbard}, the obtained equations are applied to the attractive and the repulsive Hubbard model, respectively.

\subsection{Channel-decomposed renormalization group equations}
\label{subsec:CD:RGEOneLoop}
In chapter~\ref{chap:VertexParametrization}, an ansatz for the Nambu two-particle vertex is formulated in terms of effective interactions that depend singularly on a bosonic transfer momentum. It can be written as 
\begin{align}
	\Vertex{s_1 s_2 s_3 s_4}(k_1, k_2, &k_3, k_4) = U_{s_1 s_2 s_3 s_4} + V^{\text{PH},\Lambda}_{s_1 s_2 s_3 s_4}\bigl(\tfrac{k_1+k_4}{2}, \tfrac{k_2+k_3}{2}; k_3-k_2\bigr) \label{eq:CD:VertexDecomposition} \\
	& - V^{\text{PH},\Lambda}_{s_2 s_1 s_3 s_4}\bigl(\tfrac{k_2+k_4}{2}, \tfrac{k_1+k_3}{2}; k_3-k_1\bigr) + V^{\text{PP},\Lambda}_{s_1 s_2 s_3 s_4}\bigl(\tfrac{k_1-k_2}{2}, \tfrac{k_4-k_3}{2}; k_1+k_2\bigr)\nonumber
\end{align}
(see also equation~\eqref{eq:VP:VertexDecomposition}) where the terms on the right hand side describe the antisymmetrized local microscopic interaction $U_{s_1 s_2 s_3 s_4}$ and the effective interactions in the Nambu particle-hole channel $V^{\text{PH},\Lambda}_{s_1 s_2 s_3 s_4}(k, k'; q)$ as well as in the Nambu particle-particle channel $V^{\text{PP},\Lambda}_{s_1 s_2 s_3 s_4}(k, k'; p)$. They collect the effective interactions with the corresponding transfer momenta from the ansatz in chapter~\ref{chap:VertexParametrization}. This decomposition is illustrated diagrammatically in figure~\vref{fig:VP:VertexDecomposition}. Expressions for $V^{\text{PH},\Lambda}_{s_1 s_2 s_3 s_4}(k, k'; q)$ and $V^{\text{PP},\Lambda}_{s_1 s_2 s_3 s_4}(k, k'; p)$ for a singlet superfluid with full spin rotation invariance can be found in equations~\eqref{eq:VP:NambuPH} and~\eqref{eq:VP:NambuPP}.

\begin{figure}
	\centering
	\includegraphics[scale=1]{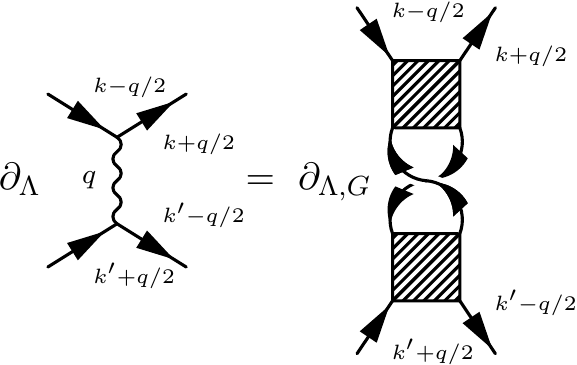}
	\caption{Diagrammatic representation of the flow equation for the effective interaction in the particle-hole channel on one-loop level. $\partial_{\Lambda,G}$ denotes differentiation of the propagator loop with respect to $\Lambda$.}
	\label{fig:CD:RGDE_PH}
	\vspace*{2em}
	\includegraphics[scale=1]{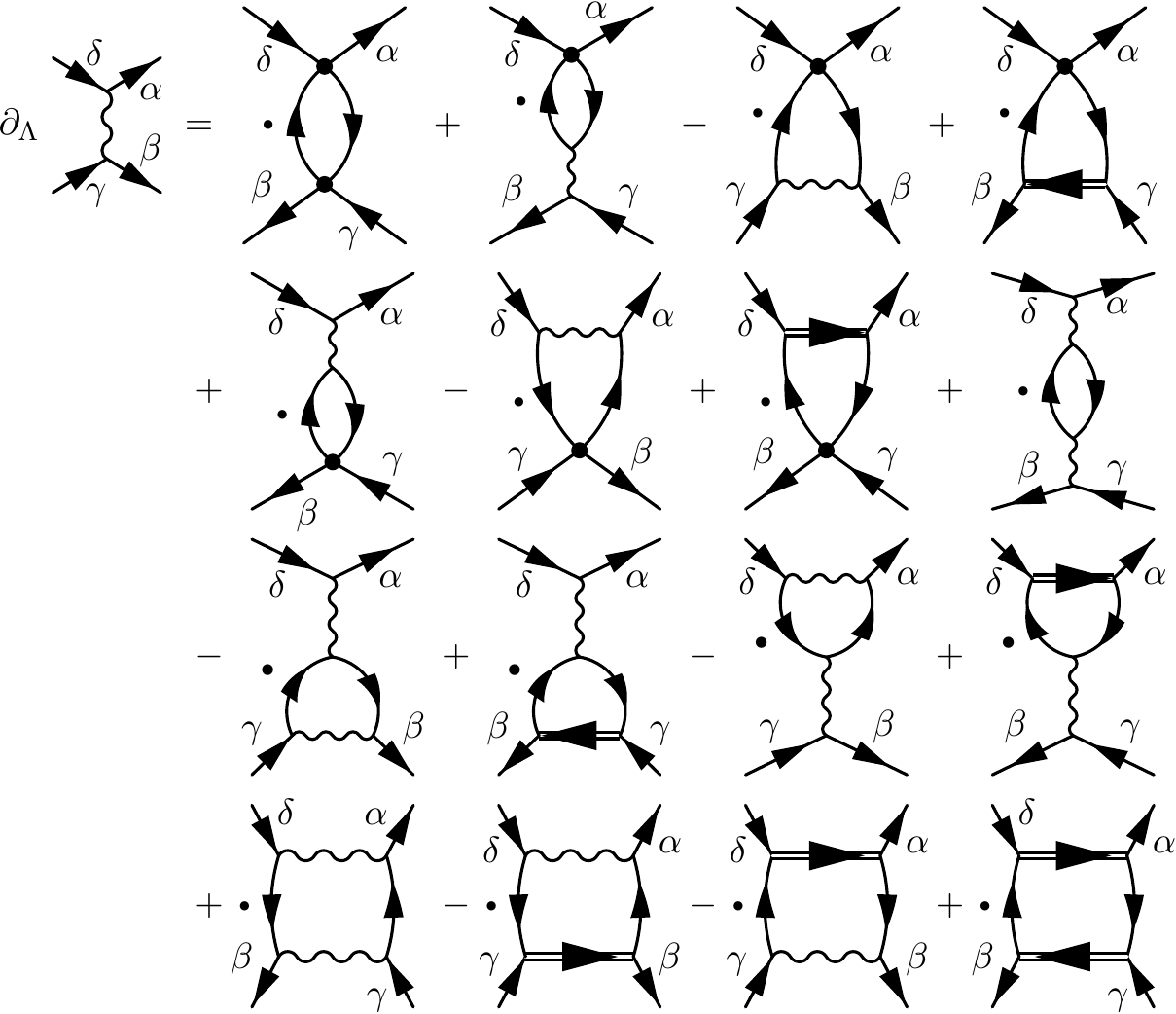}
	\caption{Detailed diagrammatic representation of the flow equation for the effective interaction in the particle-hole channel on one-loop level. The (small) dots denote $\Lambda$-derivatives acting on both internal fermionic lines.}
	\label{fig:CD:RGDE_PH_det}
\end{figure}

\begin{figure}
	\centering
	\includegraphics[scale=1]{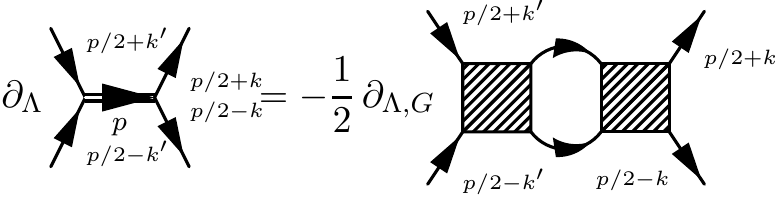}
	\caption{Diagrammatic representation of the flow equation for the effective interaction in the particle-particle channel on one-loop level. $\partial_{\Lambda,G}$ denotes differentiation of the propagator loop with respect to $\Lambda$.}
	\label{fig:CD:RGDE_PP}
	\vspace*{2em}
	\includegraphics[scale=1]{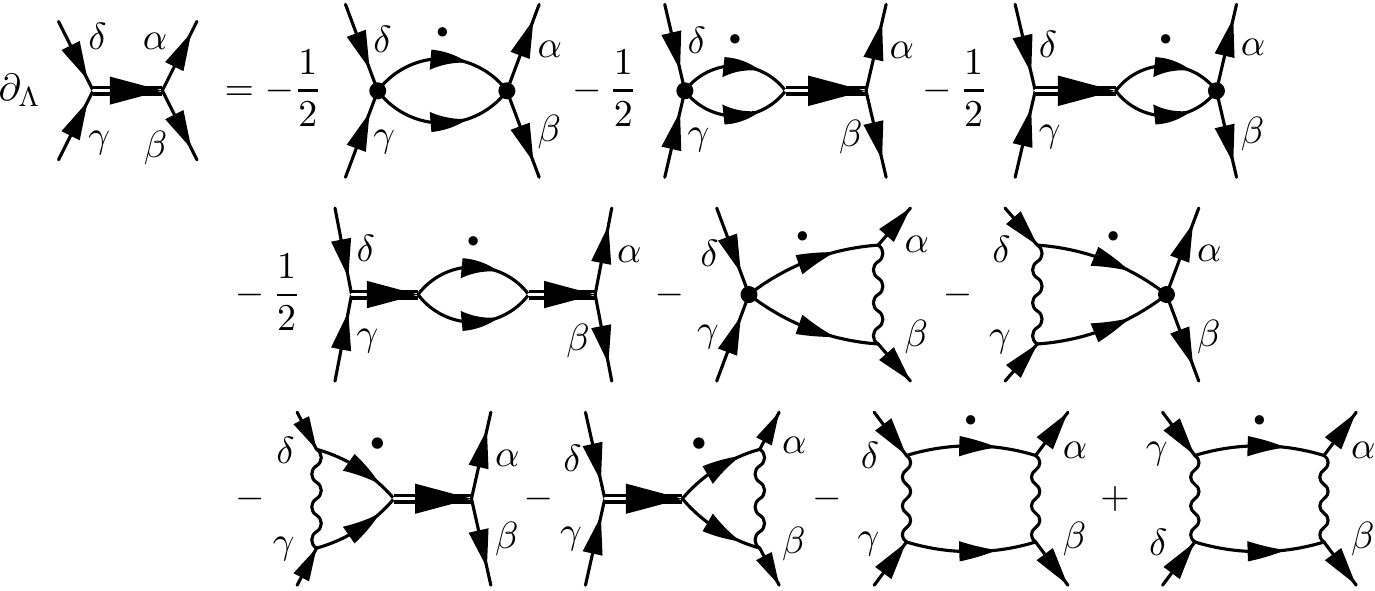}
	\caption{Detailed diagrammatic representation of the flow equation for the effective interaction in the particle-particle channel on one-loop level. The (small) dots denote $\Lambda$-derivatives acting on both internal fermionic lines.}
	\label{fig:CD:RGDE_PP_det}
\end{figure}

In order to assign the diagrams in the RG equation to interaction channels according to transfer momenta, the multi-indices in the one-loop RG equation for the two-particle vertex~\eqref{eq:fRG:RGvertexKatanin} are specialised to momenta, Matsubara frequencies and Nambu indices, yielding
\begin{gather}
	\begin{split}
	\partial_\Lambda\Vertex{s_1 s_2 s_3 s_4}(k_1, k_2, k_3, k_4) &= \Pi^\text{PH,d}_{s_1 s_2 s_3 s_4}(k_1, k_2, k_3, k_4) - \Pi^\text{PH,cr}_{s_1 s_2 s_3 s_4}(k_1, k_2, k_3, k_4)\\
	&\ - \tfrac{1}{2} \Pi^\text{PP}_{s_1 s_2 s_3 s_4}(k_1, k_2, k_3, k_4)
	\label{eq:CD:RGvertexKatanin}
	\end{split}
\intertext{where}
	\begin{split}
	\Pi^\text{PH,d}_{s_1 s_2 s_3 s_4}(k_1, k_2, k_3, k_4) &= \sum_{p,s_i'} \partial_\Lambda\bigl(G^\Lambda_{s_1' s_2'}(p) G^\Lambda_{s_3' s_4'}(q)\bigr)\\
	&\times\Vertex{s_1 s_2' s_3's_4}(k_1, p, q, k_4) \Vertex{s_4' s_2 s_3 s_1'}(q, k_2, k_3, p)\bigr|_{q = p + k_3 - k_2}
	\end{split}\\
	\begin{split}
	\Pi^\text{PH,cr}_{s_1 s_2 s_3 s_4}(k_1, k_2, k_3, k_4) &= \sum_{p,s_i'} \partial_\Lambda\bigl(G^\Lambda_{s_1' s_2'}(p) G^\Lambda_{s_3' s_4'}(q)\bigr)\\
	&\times \Vertex{s_2 s_2' s_3's_4}(k_2, p, q, k_4) \Vertex{s_4' s_1 s_3 s_1'}(q, k_1, k_3, p)\bigr|_{q = p + k_3 - k_1}
	\end{split}\\
	\begin{split}
		\Pi^\text{PP}_{s_1 s_2 s_3 s_4}(k_1, k_2, k_3, k_4) &= \sum_{p,s_i'} \partial_\Lambda\bigl(G^\Lambda_{s_1' s_2'}(p) G^\Lambda_{s_4' s_3'}(q)\bigr)\\
	&\times \Vertex{s_1 s_2 s_4's_1'}(k_1, k_2, q, p) \Vertex{s_2' s_3' s_3 s_4}(p, q, k_3, k_4)\bigr|_{q = k_1 + k_2 - p}
	\end{split}
\end{gather}
for a system with translation invariance (note that $k_4 = k_1 + k_2 - k_3$ due to momentum conservation is implicit). The fermionic loop integrands in the different contributions depend singularly on the following combinations of external momenta:
\begin{align*}
	\Pi^\text{PH,d}_{s_1 s_2 s_3 s_4}(k_1, k_2, k_3, k_4)\quad &: \quad k_3-k_2\\
	\Pi^\text{PH,cr}_{s_1 s_2 s_3 s_4}(k_1, k_2, k_3, k_4)\quad &: \quad k_3-k_1\\
	\Pi^\text{PP}_{s_1 s_2 s_3 s_4}(k_1, k_2, k_3, k_4)\quad &: \quad k_1+k_2
\end{align*}
where $k_i = (k_{0,i}, \boldsymbol k_i)$. If these transfer momenta vanish or become equal to special large momenta, the Fermi surface singularities of the fermionic propagators coincide and give rise to large or singular contributions to the flow. Above the critical scale, the two-particle vertex is non-singular and a unique assignment of diagrams to interaction channels according to the transfer momenta in the fermionic loops is possible, yielding the channel decomposition scheme by Husemann and Salmhofer~\cite{Husemann2009}. 
This assignment for the symmetric state allows to describe the generation of an attractive interaction in the $d$-wave pairing channel in the repulsive Hubbard model~\cite{Husemann2009} through the Kohn-Luttinger mechanism~\cite{Kohn1965}. At and below the critical scale, the one-loop flow equation as proposed by Katanin cannot be justified by the improved power counting arguments of Salmhofer and Honerkamp~\cite{Salmhofer2001}. 
The reason is that in this regime, the growth of the effective interactions with decreasing scale is not any more outbalanced by the reduction of the available phase space due to the fermionic single-scale propagators in diagrams with overlapping loops. An exact solution of the flow equations is nevertheless possible for reduced models~\cite{Salmhofer2004}, where higher-order contributions that are neglected in the truncation are suppressed by the restrictions on the momentum dependence of the effective interactions. 
In non-reduced models, the one-loop flow and the channel-decomposition can be justified at least in the weak-coupling regime by limitations on the phase space for scattering by singular effective interactions in the Cooper channel that arise from their momentum and frequency dependence or by the presence of an external pairing field that regularises the singularities in the Cooper channel. 
In chapter~\ref{chap:RPFM}, it is shown that the exact solution of a reduced pairing and forward scattering model is captured within the channel-decomposition scheme if the renormalization contributions in the flow equations for the vertex are assigned to the channels according to the transfer momenta in the fermionic loops. This motivates the extension of this decomposition scheme to the case of symmetry breaking below the critical scale also for non-reduced models. Comparing the singular dependences of the fermionic loops on external momenta in~\eqref{eq:CD:RGvertexKatanin} with the dependences of the vertex on transfer momenta in the flow equation
\begin{equation}
	\begin{split}
	\partial_\Lambda\Vertex{s_1 s_2 s_3 s_4}(k_1, k_2, k_3, k_4) &= \partial_\Lambda V^{\text{PH},\Lambda}_{s_1 s_2 s_3 s_4}\bigl(\tfrac{k_1+k_4}{2}, \tfrac{k_2+k_3}{2}; k_3-k_2\bigr) \\
	& - \partial_\Lambda V^{\text{PH},\Lambda}_{s_2 s_1 s_3 s_4}\bigl(\tfrac{k_2+k_4}{2}, \tfrac{k_1+k_3}{2}; k_3-k_1\bigr)\\
	& + \partial_\Lambda V^{\text{PP},\Lambda}_{s_1 s_2 s_3 s_4}\bigl(\tfrac{k_1-k_2}{2}, \tfrac{k_4-k_3}{2}; k_1+k_2\bigr)
	\end{split}
\end{equation}
and assigning diagrams to interaction channels according to the transfer momenta, one obtain
\begin{gather}
	\partial_\Lambda V^{\text{PH},\Lambda}_{s_1 s_2 s_3 s_4}\bigl(\tfrac{k_1+k_4}{2}, \tfrac{k_2+k_3}{2}; k_3-k_2\bigr) = \Pi^\text{PH,d}_{s_1 s_2 s_3 s_4}(k_1, k_2, k_3, k_4) \label{eq:CD:RGDE_PH}\\
	\partial_\Lambda V^{\text{PP},\Lambda}_{s_1 s_2 s_3 s_4}\bigl(\tfrac{k_1-k_2}{2}, \tfrac{k_4-k_3}{2}; k_1+k_2\bigr) = - \tfrac{1}{2} \Pi^\text{PP}_{s_1 s_2 s_3 s_4}(k_1, k_2, k_3, k_4). \label{eq:CD:RGDE_PP}
\end{gather}
The crossed particle-hole diagram yields equation~\eqref{eq:CD:RGDE_PH} with the indices of particles $1$ and $2$ exchanged. The flow equation~\eqref{eq:CD:RGDE_PH} is depicted diagrammatically in figure~\ref{fig:CD:RGDE_PH} in a compact and in figure~\ref{fig:CD:RGDE_PH_det} in a more detailed way. Equation~\eqref{eq:CD:RGDE_PP} is depicted in figures~\ref{fig:CD:RGDE_PP} and~\ref{fig:CD:RGDE_PP_det} in the same way.

The assignment of diagrams to interaction channels does not involve approximations by itself. However, the necessity of approximations for the momentum and frequency dependence of the vertex makes a careful assignment necessary in order to capture the leading singular behaviour at and below the critical scale for superfluidity in models with non-reduced interactions. 
Having in mind approximations like the expansion of the effective interactions in exchange propagators and fermion-boson vertices as in subsection~\ref{subsec:VP:BosonProp_gFB} together with a projection of the flow of effective interactions on exchange propagators using a small number of angular momentum channels, it is discussed in the remainder of this section that the leading singularities are indeed captured within the above decomposition scheme at least at weak coupling and for not too small external pairing fields. Furthermore, it is outlined how renormalization group equations for the effective interactions in the channels can be obtained from these equations.

\begin{figure}
	\begin{center}
		a) \subfigure{\includegraphics[scale=1]{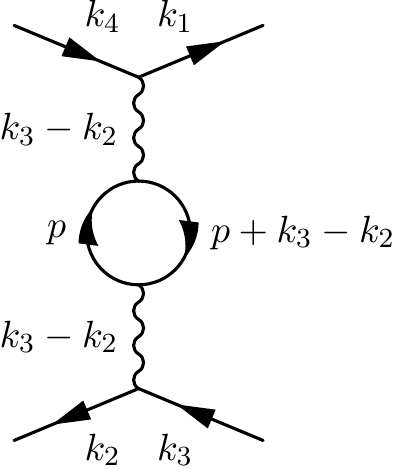}\label{fig:CD:PropRen}}\hspace{12mm}
		b) \subfigure{\includegraphics[scale=1]{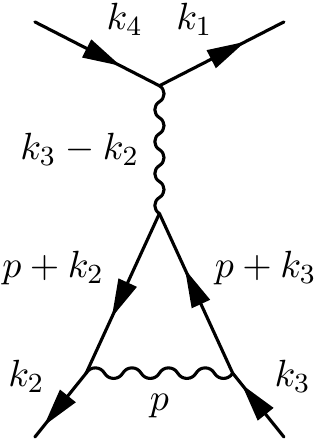}\label{fig:CD:VertCorr}}
	\end{center}
	\caption{Examples for propagator renormalization (a) and vertex correction (b) diagrams. The loop momentum and frequency $p$ is integrated over.}
\end{figure}
Inserting the ansatz~\eqref{eq:CD:VertexDecomposition} into the right hand side of equation~\eqref{eq:CD:RGvertexKatanin}, one obtains 48 diagrams (each of the three diagrams in~\eqref{eq:CD:RGvertexKatanin} involving the two-particle vertex gives rise to 16 diagrams involving the effective interactions in the channels) that can be subdivided in three classes: propagator renormalization, vertex correction and box diagrams. Each class has a different singularity structure as a function of the external momenta and is therefore discussed separately in the following. Figure~\ref{fig:CD:PropRen} shows an example of a propagator renormalization diagram. Their assignment to interaction channels is straightforward, because as shown exemplarily in the figure, the transfer momenta of the fermionic loop and of the effective interactions coincide. This is similar for the vertex-correction diagrams, of which an example is shown in figure~\ref{fig:CD:VertCorr}. 
In these diagrams, the singularities of the fermionic loop and of one effective interaction coincide, while it is not possible to bring the transfer momentum $k_3-k_2$ to the argument of the other effective interaction. These diagrams can therefore be assigned according to the singular dependence of the fermionic loop. 

\begin{figure}
	\centering
	\includegraphics[scale=1]{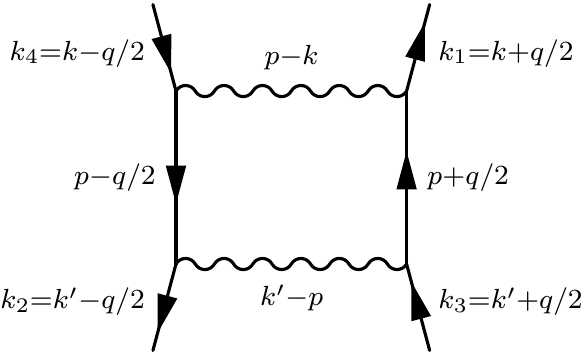}\hspace{4em}\includegraphics[scale=1]{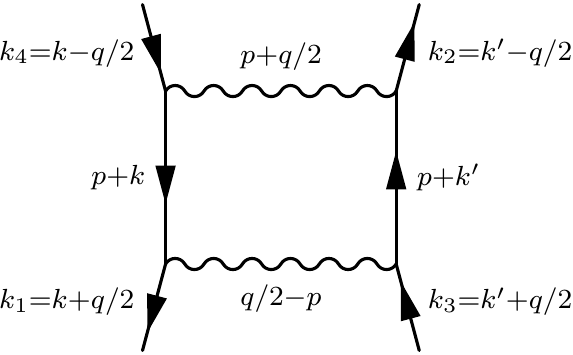}
	\caption{Example for the momentum and frequency dependence of propagators and effective interactions in box diagrams. The diagram on the right is obtained from the one on the left by exchanging momenta on outgoing lines and relabelling of internal momenta. Note that the direction of arrows on the lines is chosen exemplarily as in the Nambu particle-hole diagram. The described scattering process (for example spinor particle-particle or particle-hole) is determined by the choice of external Nambu indices. The loop momentum and frequency $p$ is integrated over.}
	\label{fig:CD:Box}
\end{figure}

The box diagrams are somewhat more complicated because depending on the choice of external momenta, the transfer momentum can be transported through the diagram either by the fermions or by the effective interactions, as shown in figure~\ref{fig:CD:Box}. Their assignment to interaction channels according to the singular dependence on external momenta is therefore not unique. This ambiguity is however lifted, if the flow is treated in a two-step process. In the first step, the fermionic regulator is removed in the presence of an external pairing field that regularises the singularities of the effective interactions in the Cooper channel. 
In the weak-coupling regime, fluctuation contributions involving effective interactions in the Cooper channel are further suppressed by the narrowness of the support around their point-singularities as a function of the transfer momentum. For this step it is plausible to assign the box diagrams in such a way that the transfer momentum is transported through the diagram by the fermionic propagators. 
As discussed in chapter~\ref{chap:AttractiveHubbard} and appendix~\ref{sec:Appendix:UToZero}, this assignment also allows for a correct treatment of the detrimental effects of (spinor) particle-hole fluctuations on the superfluid gap in the weak-coupling limit. Furthermore, in case the RG contributions to the effective interactions are projected on bosonic propagators, this assignment may lead to a higher degree of consistency and an improvement of the fulfilment of the Ward identity for the global $U(1)$ charge symmetry. The reason is that the approximations due to the projection of fluctuation contributions in the flow equations for the self-energy or for the vertex are comparable for this assignment in case the self-energy is expanded with respect to the same basis functions as the fermion-boson vertices\footnote{This assertion is based on numerical experiments for the attractive Hubbard model within the approximations of chapter~\ref{chap:AttractiveHubbard}. 
For this model, different assignments of the terms that involve anomalous fermionic propagators have been tested,  which are expected to give rise to the dominant contributions from the box diagrams below the critical scale. In comparison to the assignment scheme discussed in the main text, the assignment of the box diagram contributions that involve anomalous propagators according to the bosonic singularity yielded very similar results for most quantities. However, the deviation of the gap as obtained from the RG equation from the value fulfilling the cutoff Ward identity at the same scale came out larger than for the channel-decomposition scheme discussed in the main text. The difference between the results is attributed in particular to the projection of the flow of the effective interactions on bosonic propagators with a limited number of angular momentum channels.}. 
In the second step, the external pairing field is treated as an additional regulator that is removed in a flow. This allows for a controlled treatment of the singularities that arise from the effective interactions in the Cooper channel. However, it is shown in subsection~\ref{subsec:CD:InfraredSingOneLoop} that the box diagrams do not give rise to infrared singularities if the external pairing field is sent to zero in a flow after integrating out the fermions, independently from the choice of the external momenta. Thus, the above assignment according to the transfer momenta in the fermionic loops captures the singular behaviour of the one-loop flow.

In the following, two ways of obtaining renormalization group equations for the various effective interactions in a singlet superfluid with full spin rotation invariance are discussed. Eventually, both lead to the same equations within the same framework of approximations for the vertex. The first way demonstrates more transparently that all diagrams in the flow equation for the two-particle vertex can be assigned uniquely to effective interactions. The second option allows for a more efficient evaluation of Nambu index sums, especially in case the effective interactions are decomposed in bosonic propagators and fermion-boson vertices as presented in subsection~\ref{subsec:VP:BosonProp_gFB}.

As discussed in section~\ref{sec:VP:GeneralAnsatz}, the Nambu two-particle vertex in a spin rotation-invariant system is described by three independent coefficient functions for the normal, anomalous (3+1)- and anomalous (4+0)-effective interactions. Flow equations for these functions are obtained after specifying the external Nambu indices in~\eqref{eq:CD:RGvertexKatanin}. The choice $s_1 = +$, $s_2 = -$, $s_3 = +$ and $s_4 = -$ yields a flow equation for the normal effective interaction,
\begin{gather}
\begin{split}
	\partial_\Lambda V^\Lambda_{k_1,-k_4,-k_2 k_3} &= \partial_\Lambda\Vertex{+-+-}(k_1, k_2, k_3, k_4) \\
	&= \Pi^\text{PH,d}_{+-+-}(k_1, k_2, k_3, k_4) - \Pi^\text{PH,cr}_{+-+-}(k_1, k_2, k_3, k_4) - \tfrac{1}{2} \Pi^\text{PP}_{+-+-}(k_1, k_2, k_3, k_4)\\
	&= \partial_\Lambda P^\Lambda_{\frac{k_1+k_4}{2},\frac{k_2+k_3}{2}}(k_3-k_2) - 2 \partial_\Lambda M^\Lambda_{\frac{k_1-k_2}{2},\frac{k_3-k_4}{2}}(k_1+k_2)\\
	&\ + \partial_\Lambda C^\Lambda_{\frac{k_1+k_3}{2},-\frac{k_2+k_4}{2}}(k_1-k_3) - \partial_\Lambda M^\Lambda_{\frac{k_1+k_3}{2},-\frac{k_2+k_4}{2}}(k_1-k_3),
\end{split}
\end{gather}
where $V^\Lambda_{k_1 k_2 k_3 k_4}$ from~\eqref{eq:VP:V2p2Spinor} is used. Equating terms with the same transfer momenta yields
\begin{gather}
	\partial_\Lambda P^\Lambda_{\frac{k_1+k_4}{2},\frac{k_2+k_3}{2}}(k_3 - k_2) = \Pi^\text{PH,d}_{+-+-}(k_1, k_2, k_3, k_4)\\
	\partial_\Lambda M^\Lambda_{\frac{k_1-k_2}{2},\frac{k_3-k_4}{2}}(k_1+k_2) = \tfrac{1}{4} \Pi^\text{PP}_{+-+-}(k_1, k_2, k_3, k_4)\\
	\partial_\Lambda C^\Lambda_{\frac{k_1+k_3}{2},-\frac{k_2+k_4}{2}}(k_1-k_3) - \partial_\Lambda M^\Lambda_{\frac{k_1+k_3}{2},-\frac{k_2+k_4}{2}}(k_1-k_3) = - \Pi^\text{PH,cr}_{+-+-}(k_1, k_2, k_3, k_4).
\end{gather}
After renaming of external momenta and reexpressing of $M^\Lambda$ in the last line, the equations read
\begin{gather}
	\partial_\Lambda P^\Lambda_{k k'}(q) = \Pi^\text{PH,d}_{+-+-}(k+\tfrac{q}{2}, k'-\tfrac{q}{2}, k'+\tfrac{q}{2}, k-\tfrac{q}{2})\label{eq:CD:PhiSC_Flow}\\
	\partial_\Lambda M^\Lambda_{k k'}(q) = \tfrac{1}{4} \Pi^\text{PP}_{+-+-}(k+\tfrac{q}{2}, \tfrac{q}{2}-k, k'+\tfrac{q}{2}, \tfrac{q}{2}-k')\\
	\begin{split}
	\partial_\Lambda C^\Lambda_{k k'}(q) =& \tfrac{1}{4} \Pi^\text{PP}_{+-+-}(k+\tfrac{q}{2}, \tfrac{q}{2}-k, k'+\tfrac{q}{2}, \tfrac{q}{2}-k') \\
	& - \Pi^\text{PH,cr}_{+-+-}(k+\tfrac{q}{2}, -\tfrac{q}{2}-k', k-\tfrac{q}{2}, \tfrac{q}{2}-k').
	\end{split}
\end{gather}
The choice $s_1 = +$, $s_2 = +$, $s_3 = -$ and $s_4 = +$ leads to a flow equation for the anomalous (3+1)-effective interaction $\Omega^\Lambda_{k_1 k_2 k_3 k_4}$ in~\eqref{eq:VP:Omega3p1Nambu},
\begin{gather}
	\begin{split}
	\partial_\Lambda\Omega^\Lambda_{k_1 k_2 k_3 k_4} &= \partial_\Lambda \Vertex{++-+}(k_1, k_2, k_3, k_4) \\
	&= \Pi^\text{PH,d}_{++-+}(k_1, k_2, k_3, k_4) - \Pi^\text{PH,cr}_{++-+}(k_1, k_2, k_3, k_4) - \tfrac{1}{2} \Pi^\text{PP}_{++-+}(k_1, k_2, k_3, k_4)\\
	&= \partial_\Lambda\Omega^{S,\Lambda}_{\frac{k_1+k_4}{2},\frac{k_2+k_3}{2}}(k_2-k_3) - \partial_\Lambda\Omega^{S,\Lambda}_{\frac{k_2+k_4}{2},\frac{k_1+k_3}{2}}(k_1-k_3)\\
	&+ \partial_\Lambda\Omega^{T,\Lambda}_{\frac{k_1+k_4}{2},\frac{k_2+k_3}{2}}(k_2-k_3) - \partial_\Lambda\Omega^{T,\Lambda}_{\frac{k_2+k_4}{2},\frac{k_1+k_3}{2}}(k_1-k_3) \\
	&- 2 \partial_\Lambda\Omega^{T,\Lambda}_{\frac{k_4-k_3}{2},\frac{k_1-k_2}{2}}(k_1+k_2).
	\end{split}
\end{gather}
Equations for the singlet and triplet parts follow after comparison of transfer momenta, yielding
\begin{gather}
	\partial_\Lambda\Omega^{S,\Lambda}_{\frac{k_1+k_4}{2},\frac{k_2+k_3}{2}} (k_2-k_3) + \partial_\Lambda \Omega^{T,\Lambda}_{\frac{k_1+k_4}{2},\frac{k_2+k_3}{2}}(k_2 - k_3) = \Pi^\text{PH,d}_{++-+}(k_1, k_2, k_3, k_4)\\
	\partial_\Lambda \Omega^{T,\Lambda}_{\frac{k_4-k_3}{2},\frac{k_1-k_2}{2}} (k_1+k_2) = \tfrac{1}{4} \Pi^\text{PP}_{++-+}(k_1, k_2, k_3, k_4)
\end{gather}
for the Nambu direct particle-hole and particle-particle diagrams. The Nambu crossed particle-hole diagram and the effective interactions with transfer momentum $k_1-k_3$ lead to an equation that is equivalent to the first of the two. After renaming of external momenta, the anomalous (3+1)-effective interaction can be computed from
\begin{gather}
	\partial_\Lambda\Omega^{T,\Lambda}_{k k'}(q) = \tfrac{1}{4} \Pi^\text{PP}_{++-+}(k'+\tfrac{q}{2}, \tfrac{q}{2} - k', \tfrac{q}{2} - k, k + \tfrac{q}{2})\\
	\begin{split}
	\partial_\Lambda\Omega^{S,\Lambda}_{k k'}(q) =& \Pi^\text{PH,d}_{++-+}(k-\tfrac{q}{2}, k'+\tfrac{q}{2}, k'-\tfrac{q}{2}, k + \tfrac{q}{2}) \\
	&- \tfrac{1}{4} \Pi^\text{PP}_{++-+}(k'+\tfrac{q}{2}, \tfrac{q}{2}-k', \tfrac{q}{2}-k, k + \tfrac{q}{2}).
	\end{split}
\end{gather}
Analogously, the scale dependence of the anomalous (4+0)-effective interaction is obtained from the scale derivative of~\eqref{eq:VP:W4p0Nambu},
\begin{equation}
	\begin{split}
	\partial_\Lambda W^\Lambda_{k_1 k_2 k_3 k_4} &= \partial_\Lambda\Vertex{++--}(k_1, k_2, k_3, k_4)\\
		&= \Pi^\text{PH,d}_{++--}(k_1, k_2, k_3, k_4) - \Pi^\text{PH,cr}_{++--}(k_1, k_2, k_3, k_4) - \tfrac{1}{2} \Pi^\text{PP}_{++--}(k_1, k_2, k_3, k_4)\\
		&= \partial_\Lambda W^{S,\Lambda}_{\frac{k_1+k_4}{2},\frac{k_2+k_3}{2}}(k_3-k_2) - \partial_\Lambda W^{S,\Lambda}_{\frac{k_2+k_4}{2},\frac{k_1+k_3}{2}}(k_3-k_1)\\
		&\ +\partial_\Lambda W^{T,\Lambda}_{\frac{k_1+k_4}{2},\frac{k_2+k_3}{2}}(k_3-k_2) - \partial_\Lambda W^{T,\Lambda}_{\frac{k_2+k_4}{2},\frac{k_1+k_3}{2}}(k_3-k_1) \\
		&\ + 2 \partial_\Lambda W^{T,\Lambda}_{\frac{k_1-k_2}{2},\frac{k_3-k_4}{2}}(k_1+k_2).
	\end{split}
\end{equation}
The comparison of transfer momenta of diagrams and effective interactions yields
\begin{gather}
	\partial_\Lambda W^{S,\Lambda}_{\frac{k_1+k_4}{2},\frac{k_2+k_3}{2}}(k_3-k_2) + \partial_\Lambda W^{T,\Lambda}_{\frac{k_1+k_4}{2},\frac{k_2+k_3}{2}}(k_3-k_2) = \Pi^\text{PH,d}_{++--}(k_1, k_2, k_3, k_4)\\
	\partial_\Lambda W^{T,\Lambda}_{\frac{k_1-k_2}{2},\frac{k_3-k_4}{2}}(k_1+k_2) = -\tfrac{1}{4}\Pi^\text{PP}_{++--}(k_1, k_2, k_3, k_4)
\end{gather}
for the Nambu direct particle-hole and particle-particle diagrams, while the Nambu crossed particle-hole diagram yields an equation that is equivalent to the first. After the renaming of external momenta, the equations read
\begin{gather}
	\begin{split}
		\partial_\Lambda W^{S,\Lambda}_{k k'}(q) =& \Pi^\text{PH,d}_{++--}(k+\tfrac{q}{2}, k'-\tfrac{q}{2}, k'+\tfrac{q}{2}, k-\tfrac{q}{2}) \\
		& - \tfrac{1}{4} \Pi^\text{PP}_{++--}(k+\tfrac{q}{2},\tfrac{q}{2}-k, \tfrac{q}{2}-k', k'+\tfrac{q}{2})
	\end{split}\\
	\partial_\Lambda W^{T,\Lambda}_{k k'}(q) = \tfrac{1}{4} \Pi^\text{PP}_{++--}(k+\tfrac{q}{2}, \tfrac{q}{2}-k, \tfrac{q}{2}-k', k'+\tfrac{q}{2})\label{eq:CD:Upsilon_T_Flow}
\end{gather}
where the exchange symmetries of $W^{T,\Lambda}_{k k'}(q)$ were exploited in the last line in order to eliminate a minus sign. Equations for $A^\Lambda_{k k'}(q)$, $\Phi^\Lambda_{k k'}(q)$, $\nu^\Lambda_{k k'}(q)$ and $\tilde\nu^\Lambda_{k k'}(q)$ follow from suitable linear combinations of these equations. It should be noted that this set of equations is not unique: Because of spin rotation invariance, different sets of external Nambu indices could be chosen in order to derive flow equations for the same quantities. In terms of Nambu particle-hole and particle-particle diagrams, these differ from the above set of equations. However, after the insertion of the effective interactions the detailed equations are the same due to symmetries.

Alternatively, the flow equations can be obtained by projection on Pauli matrices. This is particularly useful when the effective interactions are expanded in exchange propagators and fermion-boson vertices as described in subsection~\ref{subsec:VP:BosonProp_gFB}. A glimpse on equations~\eqref{eq:VP:NambuPH} and~\eqref{eq:VP:NambuPP} shows that due to symmetries (in particular spin rotation invariance), it is sufficient to consider the flow equation for the effective interaction in the Nambu particle-hole channel. In $V^{\text{PH},\Lambda}_{s_1 s_2 s_3 s_4}(k,k'; q)$ every interaction channel is uniquely attached to a specific product of Pauli matrices, so that it is possible to obtain the renormalization contributions for that channel by projection. The Pauli matrices together with the unit matrix form a basis in the space of $2\times 2$ matrices that is orthogonal with respect to the scalar product
\begin{equation}
	<A,B> =\frac{1}{2} \tr(A^\dagger B) = \frac{1}{2} \sum_{s_1, s_2} (A^\dagger)_{s_2 s_1} B_{s_1 s_2}
\end{equation}
where $A$ and $B$ are $2\times2$ matrices, that is,
\begin{equation}
	<\mtau{i}{},\mtau{j}{}> =\frac{1}{2} \tr((\mtau{i}{})^\dagger \mtau{j}{}) = \delta_{ij}.
\end{equation}
This projection scheme can be extended to tensor products of the unit matrix and Pauli matrices, which form a basis in the space of $4\times4$ matrices. In this basis, the effective interactions can be rewritten by defining
\begin{equation}
	V^{\text{PH},\Lambda}_{ij}(k, k'; q) = \frac{1}{4} \sum_{s_i} \mtau{i}{s_4 s_1} \mtau{j}{s_3 s_2} V^{\text{PH},\Lambda}_{s_1 s_2 s_3 s_4}(k, k'; q)
\end{equation}
for the Nambu particle-hole and similarly for the Nambu particle-particle channel. As an example, the effective interaction that describes amplitude fluctuations of the superfluid gap is obtained by projection with $\mtau{1}{}$,
\begin{equation}
	V^{\text{PH},\Lambda}_{11}(k, k'; q) = \frac{1}{4} \sum_{s_i} \mtau{1}{s_4 s_1} \mtau{1}{s_3 s_2} V^{\text{PH},\Lambda}_{s_1 s_2 s_3 s_4}(k, k'; q) = \frac{1}{2} \sum_{\alpha, \beta}A^\Lambda_{\alpha\beta}(q) h^{A,\Lambda}_\alpha(q,k) h^{A,\Lambda}_\beta(q,k').
\end{equation}
Flow equations for $V^{\text{PH},\Lambda}_{ij}(k, k'; q)$ follow from
\begin{equation}
\begin{split}
	\partial_\Lambda V^{\text{PH},\Lambda}_{ij}(k, k'; q) &= \frac{1}{4} \sum_{s_i} \mtau{i}{s_4 s_1} \mtau{j}{s_3 s_2} \partial_\Lambda V^{\text{PH},\Lambda}_{s_1 s_2 s_3 s_4}(k, k'; q)\\
		& = \frac{1}{4} \sum_{s_i} \mtau{i}{s_4 s_1} \mtau{j}{s_3 s_2} \Pi^\text{PH,d}_{s_1 s_2 s_3 s_4}(k+\tfrac{q}{2}, k'-\tfrac{q}{2}, k'+\tfrac{q}{2}, k-\tfrac{q}{2})
\end{split}
\end{equation}
where
\begin{equation}
	\begin{split}
	\partial_\Lambda V^{\text{PH},\Lambda}_{s_1 s_2 s_3 s_4}&(k, k'; q) = \Pi^\text{PH,d}_{s_1 s_2 s_3 s_4}(k+\tfrac{q}{2}, k'-\tfrac{q}{2}, k'+\tfrac{q}{2}, k-\tfrac{q}{2})\\
	&= \sum_{p, s_i'} \partial_\Lambda(G^\Lambda_{s_1' s_2'}(p-\tfrac{q}{2}) G^\Lambda_{s_3' s_4'}(p+\tfrac{q}{2})) \Vertex{s_1 s_2' s_3' s_4}(k+\tfrac{q}{2}, p-\tfrac{q}{2}, p+\tfrac{q}{2}, k-\tfrac{q}{2}) \\
	&\hspace{2cm} \times \Vertex{s_4' s_2 s_3 s_1'}(p+\tfrac{q}{2}, k'-\tfrac{q}{2}, k'+\tfrac{q}{2}, p-\tfrac{q}{2}).
	\end{split}
\end{equation}
In order to ``perform'' the sums over Nambu indices, it is convenient to define
\begin{gather}
	C^\Lambda_{ij}(q; k, k') = \frac{1}{2} \sum_{s_i} \mtau{i}{s_4 s_1} \mtau{j}{s_3 s_2} \Vertex{s_1 s_2 s_3 s_4}(k+\tfrac{q}{2}, k'-\tfrac{q}{2}, k'+\tfrac{q}{2}, k-\tfrac{q}{2})\\
	L^\Lambda_{ij}(p, q) = \frac{1}{2} \sum_{s_i} \mtau{i}{s_1 s_4} \mtau{j}{s_2 s_3} G^\Lambda_{s_3 s_1}(p-\tfrac{q}{2}) G^\Lambda_{s_4 s_2}(p+\tfrac{q}{2})\label{eq:CD:Lij}
\end{gather}
where $C^\Lambda_{ij}$ inherit the exchange symmetry
\begin{equation}
	C^\Lambda_{ij}(q; k, k') = C^\Lambda_{ji}(-q; k', k)
\end{equation}
from the vertex and where $L^\Lambda_{ij}$ are defined in such a way that the flow equation can be written as a matrix product. This yields
\begin{equation}
	\label{eq:CD:RGOneLoop}
	\partial_\Lambda V^{\text{PH},\Lambda}_{ij}(k, k'; q) = \frac{1}{2} \sum_{p, m, n} C^\Lambda_{im}(q; k, p) \partial_\Lambda L^\Lambda_{mn}(p, q) C^\Lambda_{nj}(q; p, k').
\end{equation}
The evaluation of the sums in this expression is more convenient than the summation over Nambu indices, especially in a numerical implementation. A similar projection scheme can be applied to the effective interaction in the Nambu particle-particle channel. Expressions for $C^\Lambda_{ij}$ and $L^\Lambda_{ij}$ that result within the approximations of chapters~\ref{chap:AttractiveHubbard} and~\ref{chap:RepulsiveHubbard} for the attractive and repulsive Hubbard model can be found in subsection~\ref{subsec:Appendix:OneLoopContributions} in the appendix.

The flow equations for exchange propagators and fermion-boson vertices follow from
\begin{gather}
	\sum_{\alpha,\beta} \partial_\Lambda \Bigl[A^\Lambda_{\alpha\beta}(q) \hFB{A}{\alpha}{q}{k} \hFB{A}{\beta}{q}{k'}\Bigr] = 2 \partial_\Lambda \VPH{11}{k, k'; q} \label{eq:CD:AdecompFlow}\\
	\sum_{\alpha,\beta} \partial_\Lambda \Bigl[\Phi^\Lambda_{\alpha\beta}(q) \hFB{\Phi}{\alpha}{q}{k} \hFB{\Phi}{\beta}{q}{k'}\Bigr] = 2 \partial_\Lambda \VPH{22}{k, k'; q}\\
	\sum_{\alpha,\beta} \partial_\Lambda \Bigl[\nu^\Lambda_{\alpha\beta}(q) \hFB{\nu}{\alpha}{q}{k} \hFB{\nu}{\beta}{q}{k'}\Bigr] = \partial_\Lambda \VPH{12}{k, k'; q} - \partial_\Lambda \VPH{21}{k, k'; q}\\
	\sum_{\alpha,\beta} \partial_\Lambda \Bigl[\tilde \nu^\Lambda_{\alpha\beta}(q) \hFB{\tilde \nu}{\alpha}{q}{k} \hFB{\tilde \nu}{\beta}{q}{k'}\Bigr] = -\partial_\Lambda \VPH{12}{k, k'; q} - \partial_\Lambda \VPH{21}{k, k'; q}\\
	\sum_{\alpha,\beta} \partial_\Lambda \Bigl[X^\Lambda_{\alpha\beta}(q) \hFB{X_\text{PH}}{\alpha}{q}{k} \hFB{X_A}{\beta}{q}{k'}\Bigr] = \partial_\Lambda \VPH{31}{k, k'; q}\\
	\sum_{\alpha,\beta} \partial_\Lambda \Bigl[\tilde X^\Lambda_{\alpha\beta}(q) \hFB{\tilde X_\text{PH}}{\alpha}{q}{k} \hFB{\tilde X_\Phi}{\beta}{q}{k'}\Bigr] = \partial_\Lambda \VPH{32}{k, k'; q}\\
	\begin{split}
		\sum_{\alpha,\beta} \partial_\Lambda \Bigl[C^\Lambda_{\alpha\beta}(q) \hFB{C}{\alpha}{q}{k} \hFB{C}{\beta}{q}{k'} \tfrac{1+\zeta_\alpha}{2}\Bigr]& \\
		+\sum_{\alpha,\beta} \partial_\Lambda \Bigl[M^\Lambda_{\alpha\beta}(q) \hFB{M}{\alpha}{q}{k} \hFB{M}{\beta}{q}{k'} \tfrac{1-\zeta_\alpha}{2}\Bigr]& = \partial_\Lambda \VPH{33}{k, k'; q}
	\end{split}\\
	\begin{split}
		\sum_{\alpha,\beta} \partial_\Lambda \Bigl[C^\Lambda_{\alpha\beta}(q) \hFB{C}{\alpha}{q}{k} \hFB{C}{\beta}{q}{k'} \tfrac{1-\zeta_\alpha}{2}\Bigr] &\\
		+\sum_{\alpha,\beta} \partial_\Lambda \Bigl[M^\Lambda_{\alpha\beta}(q) \hFB{M}{\alpha}{q}{k} \hFB{M}{\beta}{q}{k'} \tfrac{1+\zeta_\alpha}{2}\Bigr] &= \partial_\Lambda \VPH{00}{k, k'; q}
	\end{split}
\end{gather}
by projection as described in the next section (note that $\zeta_\alpha = \zeta_\beta$ is assumed in the derivations in chapter~\ref{chap:VertexParametrization}). For a numerical implementation, it is convenient to rewrite~\eqref{eq:CD:RGOneLoop} as
\begin{equation}
	\label{eq:CD:RGOneLoop2}
	\partial_\Lambda V^{\text{PH},\Lambda}_{ij}(k, k'; q) = \frac{1}{2} \sum_{p, m, n} C^\Lambda_{im}(q; k, p) \partial_\Lambda L^\Lambda_{mn}(p, q) C^\Lambda_{jn}(-q; k', p).
\end{equation}
by exploiting the exchange symmetry of $C^\Lambda_{ij}$. The external fermionic momenta $k$ and $k'$ then appear in the first ``fermionic'' argument of $C^\Lambda$, which simplifies the projection on the bosonic propagators.

\subsection{Absence of non-Cooper infrared singularities on one-loop level in a fermionic \texorpdfstring{$s$}{s}-wave superfluid}
\label{subsec:CD:InfraredSingOneLoop}
In this subsection, the impact of phase fluctuations on the one-loop RG flow below the critical scale is discussed. For an attractively interacting Fermi system with an $s$-wave superfluid ground state, it is shown that the box diagrams are non-singular if the flow is treated in a two-step process with the elimination of the fermionic regulator in the first step and the removal of the external pairing field in the second step. 
Thus, no singular feedback to other channels is possible and the singularities of the one-loop flow are those found by resumming all chains of Nambu particle-hole bubble diagrams (see chapter~\ref{chap:RPFM} and in particular section~\ref{sec:RPFM:RPA}). This justifies the channel-decomposition on one-loop level with an assignment of diagrams according to the transfer momenta in the fermionic loops as discussed in subsection~\ref{subsec:CD:RGEOneLoop}.
In order to obtain this result, relatively simple estimates and ansätze are used for the fermionic propagators and effective interactions. These can however be justified by Ward identities (see for example~\cite{Schrieffer1964,Castellani1997a,Pistolesi2004}) or numerical results for the attractive Hubbard model as a prototypical model (see chapter~\ref{chap:AttractiveHubbard}). 
Finite prefactors are usually omitted for simplicity and convenience as they do not influence the singular behaviour. The assumptions and estimates seem appropriate to judge the singular behaviour of the values of effective interactions (for example at the extrema of the exchange propagators). However, they are not accurate enough for conclusions on the behaviour for example of bosonic wave-function renormalization constants parametrizing the dependence of the vertex on transfer momenta and frequencies, which are often influenced by subtle cancellations.

The most singular renormalization contributions to non-Cooper channels on one-loop level arise from box diagrams involving the phase mode that have the form
\begin{equation}
	\intdrei{p} \partial_\Lambda\bigl(\G{s_1 s_2}(p-\tfrac{q}{2}) \G{s_3 s_4}(p+\tfrac{q}{2})\bigr) \Phi^\Lambda(p+k) \Phi^\Lambda(p+k')
\end{equation}
where $k$ and $k'$ are the external fermionic momenta and $\Phi^\Lambda$ is the propagator for the phase mode in the $s$-wave channel. The vertex-correction diagrams yield less singular contributions to the non-Cooper channels, as can be shown with estimates similar to those presented below. The above contributions appear in the flow equations for all channels and the fermionic loop integrands decide on their impact similar to BCS coherence factors. They  have the largest impact on the flow for vanishing transfer momentum $q = 0$, equal external fermionic momenta $k = k'$, vanishing external fermionic frequency and a fermionic momentum on the Fermi surface $k = (0, \boldsymbol k_F)$. In order to gauge the contribution of the above diagram in case the external pairing field is removed during the fermionic flow, a particular regularization scheme is assumed for concreteness, so that estimates for the scale-derivative of the fermionic self-energy can be obtained. 
In this section, an additive frequency regulator is used as in chapters~\ref{chap:AttractiveHubbard} and~\ref{chap:RepulsiveHubbard}, which replaces the linear frequency term of the inverse bare fermionic propagator roughly by the scale $\Lambda$ for small frequencies,
\begin{equation}
	R^\Lambda(k_0) = i \sgn(k_0) \sqrt{\Lambda^2 + k_0^2} - i k_0 \stackrel{\text{def}}{=} i \tilde R^\Lambda(k_0) - i k_0.
\end{equation}
However, the choice of the fermionic regularization is not expected to influence the singular infrared behaviour of the diagrams. For small transfer momenta $q$, the propagator for the phase mode is expected to behave like
\begin{equation}
	\Phi^\Lambda(q) \sim - \frac{1}{\Delta^\Lambda_{(0)} + Z^\Lambda_\Phi q_0^2 + A^\Lambda_\Phi {\boldsymbol q}^2}
\end{equation}
where $\Delta^\Lambda_{(0)}$ is the external pairing field. This dependence on the external pairing field and the transfer momentum can be obtained from a resummation of chains of Nambu particle-hole bubbles (see section~\ref{sec:RPFM:RPA}) and can be justified non-perturbatively by Ward identities (see chapter~\ref{chap:WICharge} or the work by Castellani~\etal~\cite{Castellani1997a} as well as Pistolesi~\etal~\cite{Pistolesi2004}). In particular, it follows that $Z^\Lambda_\Phi$ and $A^\Lambda_\Phi$ are finite.

For the above choice of momenta, the contribution of the box diagram reduces to
\begin{equation}
	\beta^\Lambda \sim \intdrei{p} \partial_\Lambda\bigl(\G{s_1 s_2}(p+k)\G{s_3 s_4}(p+k)\bigr) (\Phi^\Lambda(p))^2,
\end{equation}
independently from whether the diagram is assigned according to the fermionic or the bosonic singularity. In order to obtain the infrared-scaling behaviour of $\dot G^\Lambda$, it is necessary to estimate $\dot\Sigma^\Lambda$. Looking at the flow equations for the self-energy, which can be found in appendix~\ref{sec:Appendix:RGDESelfenergy}, one finds that the most singular fluctuation contribution involves an integral over the single-scale propagator and the propagator for the phase mode of the gap,
\begin{equation}
	\intdrei{p} \SL{}(k+p) \Phi^\Lambda(p).
\end{equation}
The absolute value of the single-scale propagator is bounded from above due to the presence of the gap and decays fast enough at high frequencies in order to assure convergence of the integral in the ultraviolet. In a two-dimensional system and at zero temperature, the integral exists even for $\Delta^\Lambda_{(0)} \rightarrow 0$ because the singularity of the Goldstone propagator is integrable. Under the assumption that the effective interactions for charge forward-scattering $C^\Lambda(0)$ and amplitude fluctuations of the superfluid gap $A^\Lambda(0)$ remain finite in the infrared (an assumption to be justified below), the low-energy behaviour of $\dot\Sigma^\Lambda$ is dominated by the behaviour of the numerator of the single-scale propagator. For small frequencies, one finds
\begin{equation}
	\SL{}(p) \sim \max\Bigl((\tilde R^\Lambda(p_0) - \im \Sigma^\Lambda(p)) \partial_\Lambda \tilde R^\Lambda(p_0), \Delta^\Lambda(p) \partial_\Lambda \Delta^\Lambda_{(0)}\Bigr).
\end{equation}
Assuming that the low-frequency behaviour of the imaginary part of the normal self-energy can be approximated by $\im \Sigma^\Lambda(p) \approx p_0 - (Z^\Lambda_f)^{-1} p_0$, where $Z^\Lambda_f$ is the fermionic quasiparticle weight, this result can sloppily be approximated up to finite proportionality constants by
\begin{equation}
	\SL{}(p) \sim \max(\Lambda, \partial_\Lambda\Delta^\Lambda_{(0)}).
\end{equation}
Note that if necessary, the high frequency decay of the single-scale propagator has to be considered in order to make integrals finite in the ultraviolet. This yields the estimate $\dot \Sigma^\Lambda \sim \max(\Lambda, \partial_\Lambda\Delta^\Lambda_{(0)})$. 

Within the above assumptions and approximations, a rough estimate for the contribution of the box diagram reads
\begin{equation}
	\beta^\Lambda \sim \max(\Lambda, \partial_\Lambda\Delta^\Lambda_{(0)}) \intdrei{p} (\Phi^\Lambda(p))^2.
\end{equation}
Ignoring the periodicity of the lattice and extending the momentum integrations to infinity, which seems to be reasonable because the biggest contributions arise from a small vicinity of $p = 0$, the integration can easily be performed, yielding
\begin{equation}
	\beta^\Lambda \sim \frac{\max(\Lambda, \partial_\Lambda\Delta^\Lambda_{(0)})}{A^\Lambda_\phi\sqrt{Z^\Lambda_\phi\Delta^\Lambda_{(0)}}}.
\end{equation}
The behaviour of this result for $\Lambda \rightarrow 0$ and $\Delta^\Lambda_{(0)} \rightarrow 0$ is discussed for three different cases in the following.
\begin{enumerate}
	\item Fermionic flow in the presence of an external pairing field ($\Lambda$ is sent to zero while $\partial_\Lambda \Delta^\Lambda_{(0)} = 0$ and $\Delta^\Lambda_{(0)} = \text{const.}$): In this case, the contribution of the box diagram scales to zero with the fermionic scale,
	\begin{equation}
		\beta^\Lambda \sim \frac{\Lambda}{\sqrt{\Delta_{(0)}}} \rightarrow 0
	\end{equation}
	for $\Lambda \rightarrow 0$. Note that phase fluctuations may give rise to large renormalization contributions although the diagram scales to zero, if the external pairing field is chosen very small.
	\item $\Delta_{(0)}$-flow after $\Lambda$-flow ($\Lambda = 0$ and $\partial_\Lambda\Delta^\Lambda_{(0)} \neq 0$): In this case, the contribution of the box diagram reads
	\begin{equation}
		\beta^\Lambda \sim \frac{\partial_\Lambda \Delta^\Lambda_{(0)}}{\sqrt{\Delta^\Lambda_{(0)}}}.
	\end{equation}
	Writing $\partial_\Lambda \alpha^\Lambda = \beta^\Lambda$ and exploiting the chain-rule, one can readily show that the contribution of the box diagram to the dummy variable $\alpha$ is finite in the pairing field flow:
	\begin{gather*}
		\partial_\Lambda \alpha^\Lambda = \frac{\partial\alpha^\Lambda}{\partial\Delta^\Lambda_{(0)}} \frac{\partial\Delta^\Lambda_{(0)}}{\partial\Lambda} = \beta^\Lambda \sim \frac{\partial_\Lambda \Delta^\Lambda_{(0)}}{\sqrt{\Delta^\Lambda_{(0)}}}
	\end{gather*}
	or equivalently
	\begin{gather}
		\frac{\partial\alpha(\Delta_{(0)})}{\partial\Delta_{(0)}} = (\Delta_{(0)})^{-1/2}.
	\end{gather}
	The singularity of the right hand side does not result in singular contributions to the vertex because the pairing field flow is integrated over a finite interval for $\Delta_{(0)}$. Thus, no singularities arise in the one-loop flow from the box diagrams. The same result can be obtained after making an ansatz for the scale dependence of the external pairing field. 
	\item $\Delta_{(0)}$-flow during $\Lambda$-flow ($\Lambda > 0$ and $\partial_\Lambda\Delta^\Lambda_{(0)} \neq 0$): If the external pairing field is removed in the fermionic flow, the strength of the singularity of the right hand side depends on the scale dependence of the external pairing field. Choosing 
	\begin{equation}
		\Delta^\Lambda_{(0)} \sim \Lambda^2,
	\end{equation}
	one obtains
	\begin{equation}
		\beta^\Lambda \sim \mathcal O(1),
	\end{equation}
	which is integrated over a finite $\Lambda$-interval. If the external pairing field is removed slower, for example as
	\begin{equation}
		\Delta^\Lambda_{(0)} \sim \Lambda,
	\end{equation}
	the box diagram scales in the infrared as
	\begin{equation}
		\beta^\Lambda \sim \Lambda^{-1/2},
	\end{equation}
	which is an integrable singularity in the $\Lambda$-flow.
\end{enumerate}
In summary, the above discussion reveals that the box diagrams may be singular when the fermionic regulator or the external pairing field is removed, but that they do not result in singular contributions to the effective interactions in the sense of infrared divergences of couplings. However, the degree of singularity of the box diagrams depends on how the fermionic regulator or the external pairing field is removed. The most convenient scheme seems to be a two-step procedure in which the fermionic regulator is removed in the first flow and the external pairing field in the second one. It is noteworthy that the above estimate was computed for a particular momentum on the Fermi surface. The projection of the diagrams on the flow of bosonic propagators would further decrease the degree of singularity in case the expansion of the effective interaction is restricted to a small number of basis functions for the dependence on the fermionic momenta.

\hyphenation{sub-sec-tion}
\section{Projection on bosonic propagators}
\label{sec:CD:ProjBosProp}
In the last section, channel-decomposed renormalization group equations for the two-particle vertex on one-loop level were derived. These are used to compute the renormalization group flow of the effective interactions in the channels, which can be parametrized as boson-mediated interactions. In this section, the extraction of the flow of the bosonic propagators and the fermion-boson vertices from the RG equation for the effective interactions is discussed.
The presentation is schematic because the procedure applies to the Nambu particle-hole as well as Nambu particle-particle channel and is applicable to the formulation of the flow equations in equations~\eqref{eq:CD:PhiSC_Flow} to~\eqref{eq:CD:Upsilon_T_Flow} or~\eqref{eq:CD:RGOneLoop} to~\eqref{eq:CD:RGOneLoop2}.

In the following, the effective interactions are approximated as described in subsection~\ref{subsec:VP:BosonProp_gFB} using the decomposition
\begin{equation}
	\begin{split}
	V^\Lambda(k,k'; q) &= \sum_{\alpha,\beta} V^\Lambda_{\alpha\beta}(q) h^\Lambda_\alpha(q, k) h^\Lambda_\beta(q, k')\\
	&= \sum_{\alpha,\beta} V^\Lambda_{\alpha\beta}(q) g^\Lambda_\alpha(\boldsymbol q, k_0) f_\alpha(\boldsymbol k) g^\Lambda_\beta(\boldsymbol q, k_0') f_\beta(\boldsymbol k')
	\end{split}
\end{equation}
where $V^\Lambda$ represents an exchange propagator in the particle-hole or particle-particle channel. The form factors for the dependence on the fermionic momenta are assumed to be scale-independent. The flow equation reads schematically
\begin{equation}
	\partial_\Lambda\Bigl(\sum_{\alpha,\beta} V^\Lambda_{\alpha\beta}(q) g^\Lambda_\alpha(\boldsymbol q, k_0) f_\alpha(\boldsymbol k) g^\Lambda_\beta(\boldsymbol q, k_0') f_\beta(\boldsymbol k')\Bigr) = \partial_\Lambda V^\Lambda(k,k'; q)
\end{equation}
where the right hand side represents equations~\eqref{eq:CD:PhiSC_Flow} to~\eqref{eq:CD:Upsilon_T_Flow} or~\eqref{eq:CD:RGOneLoop} to~\eqref{eq:CD:RGOneLoop2}. In the first step, the orthogonality of the momentum-dependent form factors is exploited in order to get rid of the dependence on $\boldsymbol k$ and $\boldsymbol k'$ on both sides of the equations. The projection is first discussed for $f_\alpha(\boldsymbol k)$ being lattice form factors and second for $f_\alpha(\boldsymbol k)$ being Fermi surface harmonics.

Husemann and Salmhofer~\cite{Husemann2009} proposed to extract the renormalization contributions to the bosonic propagators by expansion of the dependence of the effective interactions on the fermionic momenta with respect to lattice form factors and projection with the help of their orthogonality property
\begin{equation}
	\intzwei{k} f_\alpha(\textbf k) f_\beta(\textbf k) = \delta_{\alpha\beta}
\end{equation}
where the integral is over the first Brillouin zone. Exploiting this relation yields
\begin{equation}
	\partial_\Lambda\bigl(V^\Lambda_{\alpha\beta}(q) g^\Lambda_\alpha(\boldsymbol q, k_0) g^\Lambda_\beta(\boldsymbol q, k_0')\bigr) = \intzwei{k} \intzwei{k'} f_\alpha(\boldsymbol k) f_\beta(\boldsymbol k') \partial_\Lambda V^\Lambda(k, k'; q).
\label{eq:CD:BZav}
\end{equation}
This projection scheme is referred to in the following as ``Brillouin zone averaging'' scheme. Taking into account the $s$- and $d$-wave channels, the channel-decomposition combined with this projection scheme allows to capture the leading instabilities of the repulsive Hubbard model at van Hove filling~\cite{Husemann2009}. In principle, the approximation for the effective interaction can systematically be improved through the inclusion of additional lattice form factors. However, the possible number of form factors is relatively small in practice due to the increasing numerical complexity. At and below the critical scale for superfluidity, strongly peaked functions have to be projected in the fluctuation contributions to the flow. Restricting to the $s$-wave channel, only the Brillouin zone averaged propagators effectively contribute and the resulting ``local approximation'' might lead to an underestimation of fluctuation contributions in the RG flow.

Alternatively, it is possible to extract the renormalization contributions to the fermion-boson vertices and the bosonic propagators by averaging the external fermionic momenta over the Fermi surface. This improves the approximation for the low energy effective interactions in the fluctuation contributions in comparison to the above projection scheme when using a small number of basis functions for the fermionic momentum dependences. The reason is that averaging over Fermi momenta yields an expansion for the momentum dependence of the effective interactions tangential to the Fermi surface, while the (in the sense of power counting) irrelevant momentum dependence perpendicular to the Fermi surface is neglected. The basis functions $f_\alpha(\boldsymbol k)$ have to fulfil the orthogonality condition
\begin{equation}
	\int_{\boldsymbol k_F}\negthickspace \frac{ds}{L_F} f_\alpha(\boldsymbol k) f_\beta(\boldsymbol k) = \inteins{\theta} m(\theta) f_\alpha(\boldsymbol k_F(\theta)) f_\beta(\boldsymbol k_F(\theta)) = \delta_{\alpha\beta}
\end{equation}
where the integration runs along the Fermi surface. $L_F$ denotes the length of the Fermi surface, $ds$ a line element along the Fermi surface and $m(\theta)$ the measure for this line integration expressed as an integration over the angle $\theta$. The measure reads
\begin{equation}
	m(\theta) = \frac{1}{\inteins{\theta} | \partial \boldsymbol k(\theta)|} |\partial \boldsymbol k(\theta)|
\end{equation}
where
\begin{equation}
	|\partial \boldsymbol k(\theta)| = \sqrt{(\partial_\theta k_F(\theta))^2 + (k_F(\theta))^2}
\end{equation}
for $\boldsymbol k_F(\theta)$ being a parametrization of the Fermi surface in the first Brillouin zone and $k_F(\theta) = |\boldsymbol k_F(\theta)|$ being the Fermi momentum in the direction $\theta$. The scheme outlined below is equivalent to conventional Fermi surface averaging for vanishing transfer momenta $\boldsymbol q = 0$. For finite transfer momenta $\boldsymbol q \neq 0$, generically not all four external momenta $\boldsymbol k\pm \tfrac{\boldsymbol q}{2}$ and $\boldsymbol k'\pm \tfrac{\boldsymbol q}{2}$ lie on the Fermi surface (this is possible only in special cases due to momentum conservation). In this case, the Fermi surface averaging can be performed approximately by averaging over all external momenta separately.

In order to motivate this projection scheme, consider the contribution of the box diagram in figure~\vref{fig:CD:Box} for vanishing momentum transfer $q = 0$. Suppressing fermion-boson vertices and form factors on the right-hand side of the equation for brevity (the full result is given below), it reads in a schematic way
\begin{equation}
	\label{eq:CD:Box}
	[\partial_\Lambda V^\Lambda(k, k'; q = 0)]_\text{Box} = \intdrei{p} \partial_\Lambda\bigl(G^\Lambda(p) G^\Lambda(p)\bigr) V^\Lambda(p-k) V^\Lambda(k'-p)
\end{equation}
where $V^\Lambda$ represents the exchange propagator for some interaction channel\footnote{$V^\Lambda$ is assumed to be even in momentum and frequency for the sake of simplicity in the presentation. The generalization for exchange propagators that are odd in particular in the transfer frequency is straightforward.}. Suppose the effective interaction $V^\Lambda(q)$ is strongly peaked at $q = 0$ and the scale $\Lambda$ is relatively small. Then, the scale-differentiated fermionic propagators effectively restrict the loop momentum $p$ to small frequencies $|p_0| \lesssim \Lambda$ and to momenta $\boldsymbol p$ in the vicinity of the Fermi surface. 
Consequently, the dominant contributions to the flow will come from small fermionic frequencies $k_0$, $k_0'$ and fermionic momenta $\boldsymbol k$, $\boldsymbol k'$ close to the Fermi surface. It seems therefore reasonable to project the renormalization contributions on the flow of the exchange propagator and the fermion-boson vertices by averaging the external momenta over the Fermi surface,
\begin{equation}
	\begin{split}
	\label{eq:CD:Box:q0}
	[\partial_\Lambda (V^\Lambda_{\alpha\beta}&(0) g^\Lambda_\alpha(0, k_0) g^\Lambda_\beta(0, k_0'))]_\text{Box} = \intdrei{p} \inteins{\theta} m(\theta) f_\alpha(\boldsymbol k_F(\theta)) \inteins{\theta'} m(\theta') f_\beta(\boldsymbol k_F(\theta')) \times \\
	& \times \partial_\Lambda\bigl(G^\Lambda(p) G^\Lambda(p)\bigr) V^\Lambda(p_0 - k_0, \boldsymbol p - \boldsymbol k_F(\theta))V^\Lambda(p_0 - k_0', \boldsymbol p - \boldsymbol k_F(\theta')).
	\end{split}
\end{equation}

For finite transfer momenta $\boldsymbol q$, the contribution of the box diagram reads schematically
\begin{equation}
	\label{eq:CD:Boxq}
	[\partial_\Lambda V^\Lambda(k, k'; q)]_\text{Box} = \intdrei{p} \partial_\Lambda\bigl(G^\Lambda(p-\tfrac{q}{2}) G^\Lambda(p+\tfrac{q}{2})\bigr) V^\Lambda(p-k) V^\Lambda(k'-p).
\end{equation}
The effective interaction $V^\Lambda(p-k)$ connects states $k-q/2$ with $p-q/2$ and $k+q/2$ with $p+q/2$ (see figure~\vref{fig:CD:Box}). Thus, it can effectively be written as
\begin{equation}
	V^\Lambda(p_0 - k_0, \boldsymbol p - \boldsymbol k) = \frac{1}{2} \bigl[V^\Lambda(p_0 - k_0, (\boldsymbol p - \boldsymbol q/2) - (\boldsymbol k - \boldsymbol q/2)) + V^\Lambda(p_0 - k_0, (\boldsymbol p + \boldsymbol q/2) - (\boldsymbol k + \boldsymbol q/2))\bigr]
\end{equation}
(and similarly for $V^\Lambda(k'-p)$). At small scales, the scale-differentiated fermionic propagators restrict $p-\tfrac{q}{2}$ and $p+\tfrac{q}{2}$ to small frequencies and to a small vicinity of the Fermi surface. Assuming that $V^\Lambda(q)$ is strongly peaked around $q = 0$, the diagram yields the largest contributions if $\boldsymbol k\pm \boldsymbol q/2$ also lie close to the Fermi surface. Considering $\boldsymbol k\pm \boldsymbol q/2$ as Fermi momenta and averaging over the terms with $\boldsymbol k_F = \boldsymbol k + \boldsymbol q/2$ and $\boldsymbol k_F = \boldsymbol k - \boldsymbol q/2$ separately, the projected contribution of the box diagram can be approximated as
\begin{equation}
	\begin{split}
	\label{eq:CD:Box:q}
	[\partial_\Lambda (V^\Lambda_{\alpha\beta}(q) g^\Lambda_\alpha(\boldsymbol q, k_0) &g^\Lambda_\beta(\boldsymbol q, k_0'))]_\text{Box} = \intdrei{p}  \partial_\Lambda\bigl(G^\Lambda(p-\tfrac{q}{2}) G^\Lambda(p+\tfrac{q}{2})\bigr)\times \\
	& \times \frac{1}{2}\bigl[V^\Lambda_{\alpha,\text{FS}}(p_0 - k_0, \boldsymbol p - \boldsymbol q/2) + V^\Lambda_{\alpha,\text{FS}}(p_0 - k_0, \boldsymbol p + \boldsymbol q/2)\bigr] \times\\
	& \times \frac{1}{2}\bigl[V^\Lambda_{\beta,\text{FS}}(p_0 - k_0', \boldsymbol p - \boldsymbol q / 2) + V^\Lambda_{\beta,\text{FS}}(p_0 - k_0', \boldsymbol p + \boldsymbol q / 2)\bigr]
	\end{split}
\end{equation}
where
\begin{equation}
	V^\Lambda_{\alpha,\text{FS}}(p_0, \boldsymbol p) = \inteins{\theta} m(\theta) f_\alpha(\boldsymbol k_F(\theta)) V^\Lambda(p_0, \boldsymbol p - \boldsymbol k_F(\theta)).
\end{equation}
Note that the approximation of separately averaging over all external momenta is only necessary in the fluctuation contributions contained in the vertex correction and box diagrams. Including all contributions, in case the fermionic momenta are averaged over the Fermi surface, the renormalization contributions to the exchange propagators and fermion-boson vertices thus follow from
\begin{equation}
\begin{split}
	\partial_\Lambda\bigl[V^\Lambda_{\alpha\beta}(q) g^\Lambda_\alpha(\boldsymbol q, k_0) g^\Lambda_\beta(\boldsymbol q, k_0')\bigr] &= \int_{\boldsymbol k_F}\negthickspace \frac{ds}{L_F} \int_{\boldsymbol k'_F}\negthickspace \frac{ds'}{L_F} f_\alpha(\boldsymbol k_F) f_\beta(\boldsymbol k'_F) \partial_\Lambda V^\Lambda(k, k'; q)\\
	&= \inteins{\theta} \inteins{\theta'} m(\theta) f_\alpha(\boldsymbol k_F(\theta)) m(\theta') f_\beta(\boldsymbol k_F(\theta')) \partial_\Lambda V^\Lambda(k, k'; q),
\label{eq:CD:FSav}
\end{split}
\end{equation}
where the right hand side is to be understood as described above.

Both projection schemes yield flow equations in which both sides depend on the fermionic frequencies $k_0$, $k_0'$ as well as on the bosonic transfer momentum $q$,
\begin{equation}
\begin{split}
	\partial_\Lambda\bigl[V^\Lambda_{\alpha\beta}(q) &g^\Lambda_\alpha(\boldsymbol q, k_0) g^\Lambda_\alpha(\boldsymbol q, k_0')\bigr] = \int_{\boldsymbol k,\alpha} \int_{\boldsymbol k',\beta} \partial_\Lambda V^\Lambda(k, k'; q)\\
	& = \partial_\Lambda V^\Lambda_{\alpha\beta}(q) g^\Lambda_\alpha(\boldsymbol q, k_0) g^\Lambda_\alpha(\boldsymbol q, k_0') + V^\Lambda_{\alpha\beta}(q) \partial_\Lambda g^\Lambda_\alpha(\boldsymbol q, k_0) g^\Lambda_\alpha(\boldsymbol q, k_0')\\
	&\quad + V^\Lambda_{\alpha\beta}(q) g^\Lambda_\alpha(\boldsymbol q, k_0) \partial_\Lambda g^\Lambda_\alpha(\boldsymbol q, k_0').
\end{split}
\end{equation}
The right hand side of the first line is a shorthand for the projected flow equation for the effective interaction in equations~\eqref{eq:CD:BZav} or~\eqref{eq:CD:FSav}. Flow equations for the bosonic propagator and the fermion-boson vertices follow after specification of the fermionic frequencies, very similar to the procedure in~\cite{Husemann2012}. Using the normalization condition $g^\Lambda_\alpha(\boldsymbol q, k_0 = 0) = 1$, the flow equation for the bosonic propagator follows after setting $k_0 = k_0' = 0$,
\begin{equation}
	\partial_\Lambda V^\Lambda_{\alpha\beta}(q) = \int_{\boldsymbol k,\alpha} \int_{\boldsymbol k',\beta} \partial_\Lambda V^\Lambda(k, k'; q)|_{k_0 = k_0' = 0}
\end{equation}
due to $\partial_\Lambda g^\Lambda_\alpha(\boldsymbol q, k_0 = 0) = 0$. In order to obtain flow equations for the fermion-boson vertices, $k_0$ is chosen non-zero and $k_0'$ set to zero, yielding\footnote{In case the flow is started at an initial scale $\Lambda_0$ where $V^\Lambda_{\alpha\beta}(q) = 0$, this flow equation for the fermion-boson vertex is formally ill-defined because a zero appears in the denominator. Husemann~\etal~\cite{Husemann2012} discussed two ways how to resolve this problem in practice. It does not appear in case the high-energy modes are treated perturbatively as in chapters~\ref{chap:AttractiveHubbard} and~\ref{chap:RepulsiveHubbard}.}
\begin{equation}
	\partial_\Lambda g^\Lambda_\alpha(\boldsymbol q, k_0) = \frac{1}{V^\Lambda_{\alpha\beta}(q)} \Bigl[\int_{\boldsymbol k,\alpha} \int_{\boldsymbol k',\beta} \partial_\Lambda V^\Lambda(k, k'; q)|_{k_0' = 0} - g^\Lambda_\alpha(\boldsymbol q, k_0) \partial_\Lambda V^\Lambda_{\alpha\beta}(q)\Bigr]_{q_0 = 0}.
\end{equation}
The choice $k_0 = 0$, $k_0' \neq 0$ yields a similar equation.

Consider as an example the renormalization contributions to the amplitude mode in case Fermi surface harmonics are used as basis function. After projection, equation~\eqref{eq:CD:AdecompFlow} reads
\begin{equation}
	\partial_\Lambda\bigl[A^\Lambda_{\alpha\beta}(q) g^{A,\Lambda}_\alpha(\boldsymbol q, k_0) g^{A,\Lambda}_\beta(\boldsymbol q, k_0')\bigr] = 2 \int_{\boldsymbol k_F}\negthickspace \frac{ds}{L_F} f_\alpha(\boldsymbol k_F) \int_{\boldsymbol k'_F}\negthickspace \frac{ds'}{L_F} f_\beta(\boldsymbol k_F') \partial_\Lambda \VPH{11}{k, k'; q}
\end{equation}
and inserting~\eqref{eq:CD:RGOneLoop2} on the right hand side yields
\begin{equation}
	= \sum_{p, m, n} \int_{\boldsymbol k_F}\negthickspace \frac{ds}{L_F} f_\alpha(\boldsymbol k_F) C^\Lambda_{im}(q; k, p) \partial_\Lambda L^\Lambda_{mn}(p, q) \int_{\boldsymbol k'_F}\negthickspace \frac{ds'}{L_F} f_\beta(\boldsymbol k_F') C^\Lambda_{jn}(-q; k', p).
\end{equation}
Defining
\begin{equation}
	C^\Lambda_{im,\alpha}(q; k_0, p) = \int_{\boldsymbol k_F}\negthickspace \frac{ds}{L_F} f_\alpha(\boldsymbol k_F) C^\Lambda_{im}(q; k, p)|_{k = (k_0, \boldsymbol k_F)},
\end{equation}
the renormalization contributions to the exchange propagator and the fermion-boson vertices can be written as
\begin{equation}
	\partial_\Lambda\bigl[A^\Lambda_{\alpha\beta}(q) g^{A,\Lambda}_\alpha(\boldsymbol q, k_0) g^{A,\Lambda}_\beta(\boldsymbol q, k_0')\bigr] = \sum_{p,m,n} C^\Lambda_{im,\alpha}(q; k_0, p) \partial_\Lambda L^\Lambda_{mn}(p, q) C^\Lambda_{jn,\beta}(-q; k_0', p).
\end{equation}
The renormalization contributions to $A^\Lambda_{\alpha\beta}(q)$ or $g^{A,\Lambda}_\alpha(\boldsymbol q, k_0)$ follow from this equation after choosing the external fermionic frequencies as explained above. The projected $C^\Lambda$-functions are easily obtained from the expressions stated in section~\ref{subsec:Appendix:OneLoopContributions} in the appendix.

\section{Two-loop level}
\label{sec:CD:TwoLoop}
In this section, the channel-decomposition scheme is extended to the two-loop level. This is interesting from a methodological point of view because the comparison of the obtained equations with those of a mixed fermion-boson RG (for example in~\cite{Strack2008}) provides some insight about the fluctuation contributions that are included in the fermionic one- and two-loop renormalization group schemes. The two-loop channel-decomposition scheme is applied to the attractive Hubbard model in subsection~\ref{subsec:AH:MomFreqTwoLoop}, where some numerical results are presented.

A Fermi system with attractive interactions behaves very similarly to an interacting Bose gas below the critical scale for superfluidity~\cite{Strack2008}. The reason is that the fermions are gapped and the remaining low-energy degrees of freedom are collective fluctuations of the order parameter. The bosonic propagators in the ansatz in chapter~\ref{chap:VertexParametrization} should therefore behave as in an interacting Bose gas (see for example~\cite{Nepomnyashchii1975,Nepomnyashchii1978,Castellani1997a, Pistolesi2004}). As discussed in subsection~\ref{subsec:CD:InfraredSingOneLoop}, this low-energy physics is not fully captured by the fermionic one-loop truncation because the singularities of the diagrams are not strong enough. 
The one-loop box diagrams contain ``bosonic'' loops formed out of two propagators for the phase mode, which could give rise to singular behaviour. However, their contributions are multiplied with a scale factor or an external pairing field stemming from the fermionic single-scale propagator, so that no singularities are caused when integrating the flow.
In order to recover the correct infrared behaviour, the fermionic equivalent of a bosonic single-scale propagator is required in the diagrams. In this section, it is demonstrated that the relevant contributions can be found in a two-loop truncation of the fermionic RG, \ie\ on the three-particle level.

\subsection{Channel-decomposed renormalization group equations}
\label{subsec:CD:RGETwoLoop}
\begin{figure}
	\centering
	\includegraphics{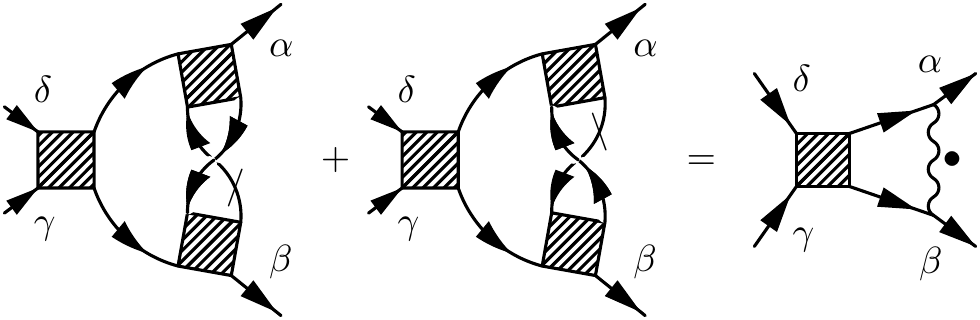}
	\caption{Illustration of the reorganisation of two-loop diagrams with overlapping loops in a one-loop diagram with a scale-differentiated effective interaction that represents the one-loop result without $\dot\Sigma^\Lambda$-insertions.}
	\label{fig:CD:TwoLoopDiagramCollection}
\end{figure}
\begin{figure}
	\centering
	\includegraphics[width=0.9\linewidth]{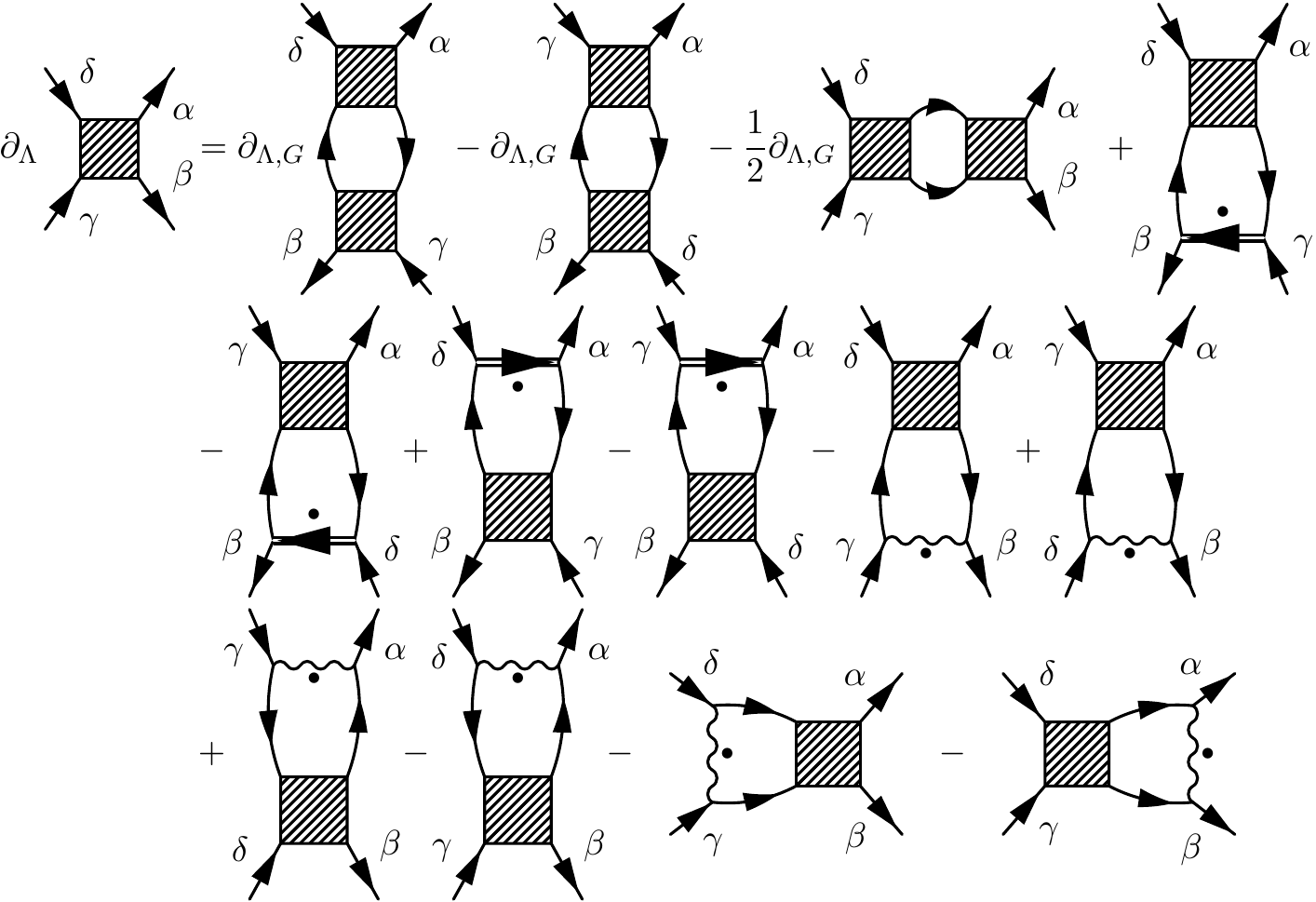}
	\caption{Diagrammatic representation of the two-loop flow equation for the two-particle vertex after reorganising diagrams as shown in figure~\ref{fig:CD:TwoLoopDiagramCollection}. $\partial_{\Lambda,G}$ denotes $\Lambda$-derivatives acting on the fermionic propagator loops. Effective interactions with dots represent their one-loop scale-derivative without $\dot \Sigma^\Lambda$-insertions.}
	\label{fig:CD:TwoLoopVertices}
\end{figure}

The starting point for the extension of the channel-decomposition scheme to the two-loop level is the two-loop RG equation for the fermionic two-particle vertex~\eqref{eq:fRG:RGvertexTwoLoop}. The aim is to reorganise the diagrams in third order in the effective interaction in a way that highlights the physical significance of different terms and possibly allows for an efficient numerical implementation. In the two-loop contributions with overlapping loops shown in the third diagram on the right hand side of the schematic equation in figure~\vref{fig:fRG:RGDE-2P-simp-OG3}, the insertion of two vertices that are connected by a single-scale and a full propagator is reminiscent of the one-loop RG contribution to the two-particle vertex (without $\dot\Sigma^\Lambda$-insertions). 
Thus, the question arises whether it is possible to replace all such insertions by one-loop scale-derivatives of effective interactions, as shown exemplarily in figure~\ref{fig:CD:TwoLoopDiagramCollection}. It turns out that such a reorganization is indeed possible for all contributions because for every choice of external labels and directions of arrows on the lines, there exist two two-loop diagrams that differ only in the position of the fermionic single-scale propagator in the inner loop that connects the same vertices. These insertions form exactly the one-loop contributions to the scale-derivative of the two-particle vertex, see equations~\eqref{eq:CD:RGDE_PH} and~\eqref{eq:CD:RGDE_PP} or figures~\ref{fig:CD:RGDE_PH} and~\ref{fig:CD:RGDE_PP}. The result of this reorganization is shown in figure~\ref{fig:CD:TwoLoopVertices}, where $\partial_{\Lambda,G}$ acts only on the fermionic propagators and where dots on ``bosonic'' lines represent the one-loop result for the scale-differentiated effective 
interaction in the unmodified 1PI scheme.

\begin{figure}
	\centering
	\includegraphics[width=0.9\linewidth]{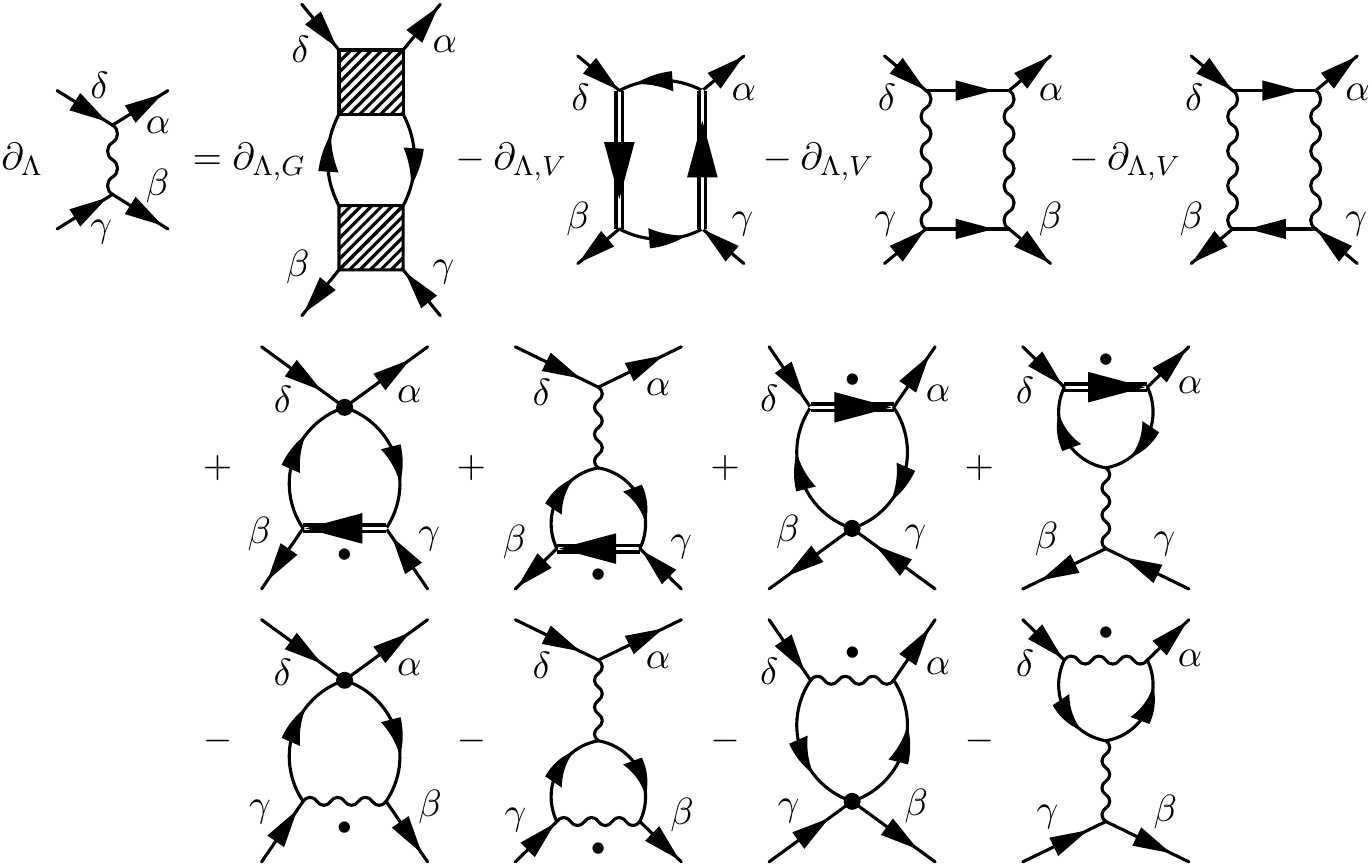}
	\caption{Diagrammatic representation of the two-loop flow equation for the effective interaction in the (Nambu) particle-hole channel. The first line contains the one-loop contributions and the two-loop two-boson exchange contributions. The second and the third line describe two-loop vertex corrections.}
	\label{fig:CD:TwoLoopVPH}
\end{figure}

\begin{figure}
	\centering
	\includegraphics[width=0.9\linewidth]{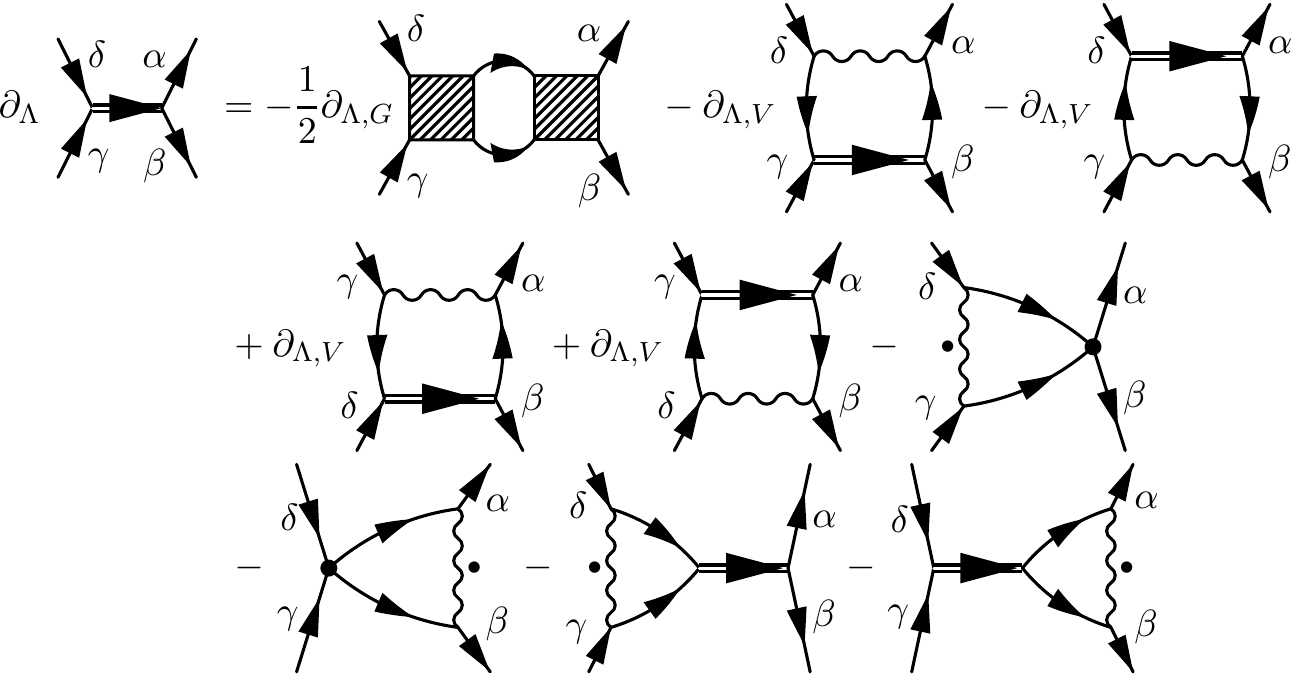}
	\caption{Diagrammatic representation of the two-loop flow equation for the effective interaction in the (Nambu) particle-particle channel. The first line and the first two diagrams in the second line contain the one-loop contributions and the two-loop two-boson exchange diagrams. The other diagrams contain two-loop vertex corrections.}
	\label{fig:CD:TwoLoopVPP}
\end{figure}

In order to highlight the singularity structure of the diagrams and to assign them to interaction channels, it is necessary to insert the decomposition of the two-particle vertex in interaction channels as shown in figure~\vref{fig:VP:VertexDecomposition} into the two-loop equation. The structure of the diagrams on two-loop level is very similar to that on one-loop level, but with the scale-derivative acting on the effective interactions instead of on the fermionic propagators. Furthermore, no two-loop propagator renormalization diagrams exist, which is a simple consequence of the topological structure of the two-loop diagrams. With a similar argument as on one-loop level, the two-loop vertex-correction diagrams are assigned according to the transfer momentum in the fermionic loop that coincides with the transfer momentum of one effective interaction. 
The two-loop box diagrams have the same multiple singularities as those on one-loop level, see figure~\ref{fig:CD:Box}. Differently to the one-loop case, it is argued that on two-loop level the box diagrams should be assigned according to their bosonic singularity. The reason is that the two-loop diagrams become only important close to the critical scale~\cite{Salmhofer2001} and below, where a strong momentum dependence has already developed. Furthermore, Goldstone mode fluctuations give rise to infrared singularities in the two-loop box diagrams when the external pairing field is removed (see below), while the fermionic propagator in an $s$-wave superfluid is bounded.  
Using simple estimates for the contributions of Goldstone mode fluctuations and analogies to the partially bosonized RG, it can be shown that the singularities that give rise to the infrared behaviour of an interacting Bose gas are contained in the two-loop box diagrams (see subsection~\ref{subsec:CD:InfraredSingTwoLoop}). Consequently, they should be regarded as effective two-boson exchange diagrams. This yields the channel-decomposed RG equations for the effective interactions in the particle-hole and particle-particle channel on two-loop level as shown diagrammatically in figures~\ref{fig:CD:TwoLoopVPH} and~\ref{fig:CD:TwoLoopVPP}, respectively. In these diagrams, $\partial_{\Lambda,V}$ or the dot on ``bosonic'' lines act only on the effective interactions, representing the one-loop result (without $\dot\Sigma^\Lambda$-insertions) for their scale-derivatives.

Analytical expressions for the two-loop flow equations are only given for the particle-hole channel, which is sufficient for a system with full spin rotation invariance. The scale-derivatives of effective interactions that appear in the two-loop contributions are given by
\begin{gather}
	\partial_{\Lambda,V} V^{\text{PH},\Lambda}_{s_1 s_2 s_3 s_4}\bigl(\tfrac{k_1+k_4}{2}, \tfrac{k_2+k_3}{2}; k_3-k_2\bigr) = \Pi^\text{PH,d}_{s_1 s_2 s_3 s_4}(k_1, k_2, k_3, k_4)|_{\partial_\Lambda \rightarrow \partial_{\Lambda,S}} \label{eq:CD:RGDE_PH_dLdV}\\
	\partial_{\Lambda,V} V^{\text{PP},\Lambda}_{s_1 s_2 s_3 s_4}\bigl(\tfrac{k_1-k_2}{2}, \tfrac{k_4-k_3}{2}; k_1+k_2\bigr) = - \tfrac{1}{2} \Pi^\text{PP}_{s_1 s_2 s_3 s_4}(k_1, k_2, k_3, k_4)|_{\partial_\Lambda \rightarrow \partial_{\Lambda,S}}, \label{eq:CD:RGDE_PP_dLdV}
\end{gather}
where $\partial_\Lambda \rightarrow \partial_{\Lambda,S}$ indicates that the scale-derivative acts only on the fermionic regulator. The two-loop flow equation for the effective interaction in the particle-hole channel then reads
\begin{align}
	\partial_\Lambda &V^{\text{PH},\Lambda}_{s_1 s_2 s_3 s_4}(k, k'; q) = \Pi^\text{PH,d}_{s_1 s_2 s_3 s_4}(k + \tfrac{q}{2}, k' - \tfrac{q}{2}, k' + \tfrac{q}{2}, k - \tfrac{q}{2}) \label{eq:CD:RGDE_PH_TwoLoop}\\
	& + \sum_{p, s'_i} \bigl(U_{s_1 s'_2 s'_3 s_4} + V^{\text{PH},\Lambda}_{s_1 s'_2 s'_3 s_4}(k, p; q)\bigr) \G{s_1' s_2'}(p-\tfrac{q}{2}) \G{s_3' s_4'}(p + \tfrac{q}{2}) \times \nonumber \\
	&\qquad \qquad \times \partial_{\Lambda,V}\Vertex{s_4' s_2 s_3 s_1'}(p + \tfrac{q}{2}, k' - \tfrac{q}{2}, k' + \tfrac{q}{2}, p - \tfrac{q}{2}) \nonumber \\
	& + \sum_{p, s_i'} \partial_{\Lambda,V} \Vertex{s_1 s_2' s_3' s_4}(k + \tfrac{q}{2}, p - \tfrac{q}{2}, p + \tfrac{q}{2}, k - \tfrac{q}{2}) \G{s_1' s_2'}(p-\tfrac{q}{2}) \G{s_3' s_4'}(p + \tfrac{q}{2}) \times \nonumber \\
	&\qquad \qquad \times \bigl(U_{s_4' s_2 s_3 s_1'} + V^{\text{PH},\Lambda}_{s_4' s_2 s_3 s_1'}(p, k'; q)\bigr) \nonumber \\
	& - \sum_{p, s_i'} \G{s_1' s_2'}(p-k) \G{s_3' s_4'}(p-k') \partial_{\Lambda,V}\Bigl( V^{\text{PP},\Lambda}_{s_1 s_2' s_3' s_3}(k - \tfrac{p}{2} + \tfrac{q}{4}, k' - \tfrac{p}{2} + \tfrac{q}{4}; p + \tfrac{q}{2}) \times \nonumber \\
	&\qquad \qquad \times V^{\text{PP},\Lambda}_{s_2 s_4' s_1' s_4}(k' - \tfrac{p}{2} - \tfrac{q}{4}, k - \tfrac{p}{2} - \tfrac{q}{4}; p - \tfrac{q}{2}) \Bigr) \nonumber \\
	& - \sum_{p, s_i'} \G{s_1' s_2'}(p+k) \G{s_3' s_4'}(k'-p) \partial_{\Lambda,V} \Bigl( V^{\text{PH},\Lambda}_{s_1 s_2 s_3' s_1'}(k + \tfrac{p}{2} + \tfrac{q}{4}, k' - \tfrac{p}{2} - \tfrac{q}{4}; \tfrac{q}{2} - p) \times \nonumber \\
	&\qquad \qquad \times V^{\text{PH},\Lambda}_{s_2' s_4' s_3 s_4}(k + \tfrac{p}{2} - \tfrac{q}{4}, k' - \tfrac{p}{2} + \tfrac{q}{4}; p + \tfrac{q}{2})\Bigr) \nonumber \\
	& - \sum_{p, s_i'} \G{s_1' s_2'}(p+k) \G{s_3' s_4'}(p+k') \partial_{\Lambda,V} \Bigr( V^{\text{PH},\Lambda}_{s_1 s_4' s_3 s_1'}(k + \tfrac{p}{2} + \tfrac{q}{4}, k' + \tfrac{p}{2} + \tfrac{q}{4}; \tfrac{q}{2} - p) \times \nonumber \\
	&\qquad \qquad \times V^{\text{PH},\Lambda}_{s_2' s_2 s_3' s_4}(k + \tfrac{p}{2} - \tfrac{q}{4}, k' + \tfrac{p}{2} - \tfrac{q}{4}; p + \tfrac{q}{2}) \Bigr) \nonumber
\end{align}
where
\begin{equation}
	\begin{split}
	\partial_{\Lambda,V} \Vertex{s_1 s_2 s_3 s_4}(k_1, k_2, k_3, k_4) &= - \partial_{\Lambda,V} V^{\text{PH},\Lambda}_{s_2 s_1 s_3 s_4}\bigl(\tfrac{k_2 + k_4}{2}, \tfrac{k_1 + k_3}{2}; k_3 - k_1\bigr)\\
	&\ + \partial_{\Lambda,V} V^{\text{PP},\Lambda}_{s_1 s_2 s_3 s_4}\bigl(\tfrac{k_1 - k_2}{2}, \tfrac{k_4 - k_3}{2}; k_1 + k_2\bigr).
	\end{split}
\end{equation}
The first line of~\eqref{eq:CD:RGDE_PH_TwoLoop} contains the one-loop contributions, the second and third lines the two-loop vertex corrections and the last three lines the two-loop two-boson exchange contributions. 

In a numerical implementation, the evaluation of the flow equations on two-loop level proceeds in two steps. In the first, the one-loop contributions involving single-scale propagators and $\dot \Sigma^\Lambda$-insertions are evaluated separately. In the second step, the two-loop contributions involving $\partial_{\Lambda,V} \Vertex{}$ are computed using the one-loop result. 
In order to obtain two-loop flow equations for exchange propagators and fermion-boson vertices, equation~\eqref{eq:CD:RGDE_PH_TwoLoop} is further evaluated as described in sections~\ref{subsec:CD:RGEOneLoop} and~\ref{sec:CD:ProjBosProp}. 
Analytical expressions for the two-loop vertex correction diagrams can easily be read off from those on one-loop level, because they differ only in the objects the scale-derivatives act on. For the two-loop two-boson exchange contributions, analytical expressions are given in section~\ref{subsec:Appendix:TwoLoopContributions} in the appendix. 

Analytical estimates for the singular infrared behaviour of a fermionic $s$-wave superfluid on two-loop level, which were recently completed and are discussed in the next subsection, indicate that the one-loop box diagrams might not be assigned optimally in the two-loop channel-decomposition scheme. In comparison to the one-loop level, these diagrams become more singular in the two-loop flow due to self-energy insertions (see subsection~\ref{subsec:CD:InfraredSingTwoLoop}) and may lead to logarithmic singularities in certain interaction channels in the limit where the external pairing field vanishes due to ``bosonic'' loops involving two propagators for the phase mode and anomalous fermionic propagators. In an improved version of the two-loop channel-decomposition scheme, the one-loop box diagrams with anomalous fermionic propagators should therefore be assigned as two-boson exchange contributions according to their bosonic singularity. 
The one-loop box diagrams with normal fermionic propagators should nevertheless be assigned according to the fermionic singularity. This is important above the critical scale in order to capture the Kohn-Luttinger effect~\cite{Kohn1965} within the channel-decomposition scheme~\cite{Husemann2009}. Below the critical scale and in particular in pairing field flows, the one-loop box diagrams with normal propagators are expected to be less important and remain non-singular in the limit where the external pairing field vanishes.

Note that this change in the assignment of diagrams is not expected to have a significant impact on two-loop flows with fixed or flowing external pairing fields as long as the latter is not too small. In the weak-coupling regime and on one-loop level, it has been checked for the attractive Hubbard model that the results of flows with fixed external pairing fields are indeed similar irrespective of whether the box diagrams with anomalous propagators are assigned according to their bosonic or fermionic singularity. Due to the weak singularity of the one-loop box diagrams, the issue about their assignment seems to become relevant only in the limit where the external pairing field vanishes. In the discussion of this limit in the next subsection, the one-loop box diagrams are therefore considered to be assigned according to their bosonic singularity. They then always appear together with more singular two-loop contributions and do not give rise to singularities beyond those discussed in the next subsection.

\subsection{Infrared singularities due to phase fluctuations in a fermionic \texorpdfstring{$s$}{s}-wave superfluid}
\label{subsec:CD:InfraredSingTwoLoop}
In order to investigate the feedback of phase fluctuations on other interaction channels, the estimates from subsection~\ref{subsec:CD:InfraredSingOneLoop} are extended to the two-loop level in the following. It is demonstrated that the singular infrared behaviour of the amplitude mode that is expected in a fermionic $s$-wave superfluid~\cite{Strack2008} is captured in the fermionic RG on two-loop level. Besides it is found that no singular feedback to non-Cooper channels exists.

In order to obtain these results, a pairing field flow is considered in which the fermionic regulator vanishes and the external pairing field is treated as the flow parameter. The bosonic exchange propagators for the $s$-wave Cooper channel and the imaginary part of the anomalous (3+1)-effective interaction are parametrized by the following ansätze at small transfer momenta and frequencies:
\begin{gather}
	\Phi^\Lambda(q) \sim -\frac{1}{\Delta^\Lambda_{(0)} + Z^\Lambda_\Phi q_0^2 + A^\Lambda_\Phi \boldsymbol q^2} \label{eq:CD:InfraredSingTwoLoop:PhiAnsatz}\\
	\nu^\Lambda(q) \sim -\frac{q_0}{\Delta^\Lambda_{(0)} + Z^\Lambda_\nu q_0^2 + A^\Lambda_\nu \boldsymbol q^2}\\
	\tilde X^\Lambda(q) \sim \frac{q_0}{\Delta^\Lambda_{(0)} + Z^\Lambda_{\tilde X} q_0^2 + A^\Lambda_{\tilde X} \boldsymbol q^2}\\
	A^\Lambda(q) \sim -\frac{1}{\sqrt{\Delta^\Lambda_{(0)} + Z^\Lambda_A q_0^2 + A^\Lambda_A \boldsymbol q^2}}.
\end{gather}
The first three ansätze are motivated by a resummation of all chains of Nambu particle-hole diagrams (see chapter~\ref{chap:RPFM}) and can be justified non-perturbatively using Ward identities~\cite{Castellani1997a,Pistolesi2004}. The ansatz for $A^\Lambda(q)$ is inspired by the results by Pistolesi~\etal~\cite{Pistolesi2004} as well as Strack~\etal~\cite{Strack2008}. It reproduces the singular infrared scaling of the amplitude mode that was described by Strack~\etal~\cite{Strack2008} in terms of divergent wave function renormalization factors. The above-mentioned works indicate that the coefficients $A^\Lambda_i$ and $Z^\Lambda_i$ in the ansätze remain finite. Furthermore, it is assumed that all other exchange propagators remain finite in the limit where the external pairing field vanishes, which has to be justified below. The above ansätze are inserted into~\eqref{eq:CD:RGDE_PH_TwoLoop} together with appropriate ansätze for the low-energy behaviour of the fermionic propagators. When evaluating the 
renormalization contributions for vanishing transfer momentum $q$ and for external fermionic momenta on an almost circular Fermi surface for simplicity, the components of the fermionic propagator behave as
\begin{gather}
	G^\Lambda_{+-}(k+p) = F^\Lambda(k+p) \approx \frac{1}{\Delta^\Lambda} \label{eq:CD:Ansatz_F}\\
	G^\Lambda_{++}(k+p) = G^\Lambda(k+p) \approx \frac{i p_0 + v_F \boldsymbol p \cdot \boldsymbol e_{\boldsymbol k_F}}{(\Delta^\Lambda)^2} \label{eq:CD:Ansatz_G}
\end{gather}
for small $p$ and $k = (0,\boldsymbol k_F)$ together with appropriate ultraviolet cutoffs, where $v_F$ is the Fermi velocity and $e_{\boldsymbol k_F}$ a unit vector pointing in the direction of $\boldsymbol k_F$. Note that the shape of the Fermi surface is not expected to influence the conclusions of this section. When averaging external fermionic momenta over the Fermi surface, the term $ v_F \boldsymbol p \cdot \boldsymbol e_{\boldsymbol k_F}$ has to be replaced by a term $\sim \boldsymbol p^2$ for an almost circular Fermi surface. 

The most interesting two-loop contributions arise from the two-loop two-boson exchange diagrams. The reason is that the two-loop vertex-correction diagrams are non-singular for non-Cooper channels and are less singular than the other renormalization contributions in the Cooper channel. This can be seen by estimating the integral
\begin{equation}
	\intdrei{q} \partial_{\Lambda,V} \Phi^\Lambda(q)
\end{equation}
which appears in the most singular contributions from vertex corrections on two-loop level. An estimate for $\partial_{\Lambda,V} \Phi^\Lambda(q)$ can be obtained from~\eqref{eq:CD:InfraredSingTwoLoop:PhiAnsatz} by letting $\partial_{\Lambda,V}$ act only on the external pairing field. Extending the momentum and frequency integration in the resulting integral to infinity, it yields
\begin{equation}
\partial_\Lambda \Delta^\Lambda_{(0)} \intdrei{q} \frac{1}{(\Delta^\Lambda_{(0)} + Z^\Lambda_\Phi q_0^2 + A^\Lambda_\Phi \boldsymbol q^2)^2} \sim \frac{\partial_\Lambda \Delta^\Lambda_{(0)}}{\sqrt{\Delta^\Lambda_{(0)}}}.
\end{equation}
Inserting for example the ansatz $\Delta^\Lambda_{(0)} \sim \Lambda^2$ for the scale dependence of $\Delta^\Lambda_{(0)}$, the integral turns out to be non-singular. For non-Cooper channels, it is multiplied by non-singular effective interactions in the flow equations. 

The evaluation of the two-loop two-boson (TB) exchange contributions is only demonstrated for the amplitude mode, density forward scattering and the real part of the anomalous (3+1)-effective interaction, because these lead to the most interesting results. The estimation of the contributions to the other channels can be done analogously. The full renormalization contributions of the two-loop two-boson exchange diagrams can be found in subsection~\ref{subsec:Appendix:TwoLoopContributions} in the appendix\footnote{The expressions given in the appendix are valid in case the external fermionic momenta are averaged over the Brillouin zone or the Fermi surface. In comparison to the evaluation for external momenta on the Fermi surface, some contributions involving magnetic propagators cancel due to averaging. However, these cancellations do not influence the most singular terms involving propagators in the Cooper channel.}, while only the most singular terms are shown in the text in order to keep the 
discussion clear. 
The renormalization contributions to the amplitude mode can be read off from~\eqref{eq:Appendix:V11TL} in the appendix, where the most singular contribution yields
\begin{equation}
	\partial_\Lambda A^\Lambda(0)|_\text{TB} \sim (\Delta^\Lambda)^{-2} \intdrei{p} \partial_{\Lambda,V}\Bigl((A^\Lambda(p))^2 + (\Phi^\Lambda(p))^2\Bigr) + \ldots
	\label{eq:CD:TwoLoop:TBE11}
\end{equation}
The prefactor $(\Delta^\Lambda)^{-2}$ arises from anomalous fermionic propagators. Differentiating the above ansätze for the exchange propagators in the Cooper channel with respect to the scale dependence of $\Delta^\Lambda_{(0)}$ and inserting the resulting expressions into this equation, it is found that the term $\sim \partial_{\Lambda,V} (\Phi^\Lambda(p))^2$ yields the largest contribution, leading to
\begin{equation}
	\partial_\Lambda A^\Lambda(0)|_\text{TB} \sim \frac{\partial_\Lambda \Delta^\Lambda_{(0)}}{(\Delta^\Lambda_{(0)})^{3/2}}.
\end{equation}
All other contributions receive extra momentum or frequency factors from the fermionic or bosonic propagators that weaken the singularities. Using the chain rule, this equation can be rewritten as
\begin{equation}
	\partial_{\Delta_{(0)}} A^{\Delta_{(0)}}(0)|_\text{TB} \sim (\Delta_{(0)})^{-3/2}.
\end{equation}
Because all other contributions to $\partial_\Lambda A^\Lambda(0)$ are less singular, it yields
\begin{equation}
	A^{\Delta_{(0)}}(0) \sim (\Delta_{(0)})^{-1/2}
\end{equation}
for the singular behaviour of the amplitude mode in the limit where the external pairing field vanishes. Using the ansatz $\Delta^\Lambda_{(0)} = \Lambda^2$ in order to mimic the infrared scaling of the phase mode in the work by Strack~\etal~\cite{Strack2008}, this result is equivalent to
\begin{equation}
	A^\Lambda(0) \sim \Lambda^{-1},
\end{equation}
as found by Strack~\etal

A similar estimate shows that the density forward scattering channel does not receive singular feedback from phase fluctuations. The most singular contribution in~\eqref{eq:Appendix:V33TL} in the appendix reads
\begin{equation}
	\partial_\Lambda C^\Lambda(0)|_\text{TB} \sim  \intdrei{p} \re G^\Lambda(k + p) \re G^\Lambda(k' + p)\partial_{\Lambda,V}\Bigl((A^\Lambda(p))^2 + (\Phi^\Lambda(p))^2\Bigr) + \ldots
	\label{eq:CD:TwoLoop:TBE33}
\end{equation}
Evaluating the right hand side for $k = k' = (0, \boldsymbol k_F)$, it is found that the integral is non-singular because the fermionic propagators contribute an extra factor $\sim \boldsymbol p^2$ in the integrand (or a higher power if the external fermionic momenta are averaged over the Fermi surface) in comparison to~\eqref{eq:CD:TwoLoop:TBE11}. This contribution is thus finite, but larger than the terms represented by the ellipsis, so that
\begin{equation}
	\partial_\Lambda C^\Lambda(0)|_\text{TB} \sim \mathcal O(1).
\end{equation}

The corresponding contribution~\eqref{eq:Appendix:V13TL} to the real part of the anomalous (3+1)-effective interaction is potentially more singular, because in the dominant term one normal fermionic propagator is replaced by an anomalous one in comparison to~\eqref{eq:CD:TwoLoop:TBE33}. It reads
\begin{equation}
	\partial_\Lambda X^\Lambda(0)|_\text{TB} \sim  \intdrei{p} F^\Lambda(k + p) \re G^\Lambda(k' + p)\partial_{\Lambda,V}\Bigl((\Phi^\Lambda(p))^2 - (A^\Lambda(p))^2\Bigr) + \ldots
	\label{eq:CD:TwoLoop:TBE13}
\end{equation}
Evaluating this expression for $k = k' = (0, \boldsymbol k_F)$, a simple scaling analysis suggests that it leads to a logarithmic singularity, $X^{\Delta_{(0)}}(0) \sim \log \Delta_{(0)}$. However, a closer inspection reveals that the integral vanishes due to $\re G^\Lambda(k' + p)|_{k = (0, \boldsymbol k_F)} \sim \boldsymbol p \cdot \boldsymbol e_{\boldsymbol k_F}$ being an odd function in $\boldsymbol p$ in a small vicinity of the Fermi surface. In case the external fermionic momenta are averaged over the Fermi surface, the averaged $\re G^\Lambda(k' + p)|_{k = (0, \boldsymbol k_F)}$ becomes quadratic in $|\boldsymbol p|$, which suffices to suppress the logarithmic singularity.

In subsection~\ref{subsec:CD:InfraredSingOneLoop}, it was assumed that the amplitude mode $A^\Lambda(0)$ is non-singular, leading to $\partial_\Lambda \Delta^\Lambda \sim \partial_\Lambda \Delta^\Lambda_{(0)}$. This changes on two-loop level, so that a comment about the one-loop propagator renormalization contributions with self-energy insertions is in order. Repeating the estimates for the self-energy from subsection~\ref{subsec:CD:InfraredSingOneLoop} with $A^\Lambda(0) \sim (\Delta^\Lambda_{(0)})^{-1/2}$, one obtains
\begin{equation}
	\partial_\Lambda \Delta^\Lambda \sim A^\Lambda(0) \partial_\Lambda \Delta^\Lambda_{(0)} \sim (\Delta^\Lambda_{(0)})^{-1/2} \partial_\Lambda \Delta^\Lambda_{(0)} + \ldots
	\label{eq:CD:TwoLoop_dDelta_dL}
\end{equation}
Using the chain rule, this equation can readily be integrated, yielding
\begin{equation}
	\Delta^{\Delta_{(0)}} \sim \Delta_{(0)}^{1/2} + \operatorname{const.}, 
\end{equation}
where the integration constant is the anomalous self-energy in case of spontaneous symmetry breaking. For extrapolating pairing field flows to $\Delta_{(0)} \rightarrow 0$, it is convenient to insert for example the ansatz $\Delta^\Lambda_{(0)} \sim \Lambda^2$ for the scale dependence of the external pairing field. The right hand side of~\eqref{eq:CD:TwoLoop_dDelta_dL} is then of the order of one and $\Delta^\Lambda$ becomes linear in $\Lambda$. The change in the infrared scaling of $\partial_\Lambda \Delta^\Lambda$ modifies the propagator renormalization contributions, for example
\begin{equation}
	\partial_\Lambda A^\Lambda(0)|_\text{Prop. Renorm.} = (A^\Lambda(0))^2 \partial_{\Lambda,\Sigma} L^\Lambda_{11}(0) + \ldots \sim (A^\Lambda(0))^2 \partial_\Lambda \Delta^\Lambda \sim (A^\Lambda(0))^3 \partial_\Lambda \Delta^\Lambda_{(0)}
\end{equation}
for the amplitude mode where $\partial_{\Lambda,\Sigma}$ acts only on the fermionic self-energy. This contribution is as singular as the one that arises from phase fluctuations. For the real part of the anomalous (3+1)-effective interaction,
\begin{equation}
	\partial_\Lambda X^\Lambda(0)|_\text{Prop. Renorm.} = \bigl(A^\Lambda(0) C^\Lambda(0) + (X^\Lambda(0))^2\bigr) \partial_{\Lambda,\Sigma} L^\Lambda_{13}(0) + \ldots,
\end{equation}
the right hand side could lead to a logarithmic singularity. However, it has a vanishing prefactor as for the two-loop contributions. The reason is that $\partial_{\Lambda,\Sigma} L^\Lambda_{13}(0)$ vanishes in the infrared due to the approximate particle-hole symmetry of the ansatz for the normal propagator~\eqref{eq:CD:Ansatz_G}.

It should be noted that the propagator renormalization contributions to the phase mode involving self-energy insertions,
\begin{equation}
	\begin{split}
	\partial_\Lambda \Phi^\Lambda(0)|_\text{Prop. Renorm.} &= (\Phi^\Lambda(0))^2 \partial_{\Lambda,\Sigma} L^\Lambda_{22}(0) + \ldots \sim (\Phi^\Lambda(0))^2 \partial_\Lambda \Delta^\Lambda\\
		& \sim (\Delta^\Lambda_{(0)})^{-5/2} \partial_\Lambda \Delta^\Lambda_{(0)},
	\end{split}
	\label{eq:CD:PropRenormSigma:Phi}
\end{equation}
have a \emph{stronger} singularity than the corresponding one-loop contributions,
\begin{equation}
	\partial_\Lambda \Phi^\Lambda(0)|_\text{One-loop} \sim (\Delta^\Lambda_{(0)})^{-2} \partial_\Lambda\Delta^\Lambda_{(0)}.
\end{equation}
This could spoil the scaling behaviour $\Phi^\Lambda(0) \sim (\Delta^\Lambda_{(0)})^{-1}$ in the two-loop RG flow, which would violate the Ward identity for global charge conservation. However, in the numerical evaluation of the two-loop flow for the attractive Hubbard model (which is discussed in subsection~\ref{subsec:AH:MomFreqTwoLoop}), the scaling $\Phi^\Lambda(0) \sim (\Delta^\Lambda_{(0)})^{-1}$ was observed down to the smallest accessible external pairing fields. This may be a consequence of cancellations arising from Ward identities.

As discussed at the end of the last subsection, the one-loop box diagrams also become more singular on two-loop level due to $\partial_\Lambda \Delta^\Lambda \sim \mathcal (\Delta^\Lambda_{(0)})^{-1/2} \partial_\Lambda \Delta^\Lambda_{(0)}$ instead of $\partial_\Lambda \Delta^\Lambda \sim \partial_\Lambda \Delta^\Lambda_{(0)}$, which alters the estimate for these diagrams from subsection~\ref{subsec:CD:InfraredSingOneLoop} to
\begin{equation}
	\beta^\Lambda|_\text{Two-Loop flow} \sim (\Delta^\Lambda_{(0)})^{-1} \partial_\Lambda \Delta^\Lambda_{(0)}.
\end{equation}
The one-loop box diagrams would therefore lead to logarithmic singularities in various channels if they are assigned according to the transfer momenta in the fermionic loops. However, due to this ``bosonic'' singularity, the one-loop box diagrams with anomalous fermionic propagators should be assigned to interaction channels according to their bosonic singularity (\ie\ as one-loop two-boson exchange diagrams) as suggested at the end of subsection~\ref{subsec:CD:RGETwoLoop}. In this case, these diagrams would lead to logarithmically singular contributions to the amplitude mode, which are, however, less singular than those from phase fluctuations on two-loop level, and would not give rise to singularities in other channels.

\section{Conclusion: Theoretical framework}
In this chapter, channel-decomposed renormalization group equations for the Nambu two-particle vertex in a singlet superfluid were presented, in which the singular dependence on momenta and frequencies is isolated in only one variable per interaction channel. They thus form a good starting point for the formulation of approximations for the effective interactions in the channels and their efficient computation. 

The channel-decomposed equations on one-loop level extend work for the symmetric phase by Husemann and Salmhofer~\cite{Husemann2009} to the superfluid phase. They are based on the modified one-loop truncation by Katanin~\cite{Katanin2004a} and allow to solve mean-field models exactly~\cite{Eberlein2010}. The two-loop channel-decomposition scheme allows to efficiently take into account the renormalization contributions to the two-particle vertex in third order in the effective interaction with overlapping loops that are neglected in the modified one-loop truncation. This is accomplished by reexpressing these contributions as effective one-loop diagrams involving the scale-derivative of the effective interactions as computed on one-loop level.

The assignment of diagrams to interaction channels in the flow equations is facilitated by organizing the description of spontaneous symmetry breaking within the fermionic RG as a two step process: In the first step, fermionic modes are integrated out in the presence of an external pairing field that regularizes the singularities at and below the critical scale. In the second step, the external pairing field is regarded as a regulator that is removed in a flow. This allows to control the size of effective interactions during the flow and to study the effects of long-range phase fluctuations in a systematic way.

Using simple estimates for the renormalization contributions on one- and two-loop level, the singularity structure of the fermionic two-particle vertex in an $s$-wave superfluid was clarified. In the limit where the external pairing field vanishes, the singular behaviour of the vertex on one-loop level is the same as in a resummation of all chains of Nambu particle-hole bubble diagrams. The infrared divergence of the amplitude mode that is expected in a fermionic superfluid at zero temperature due to phase fluctuations~\cite{Strack2008} is captured on two-loop level. No singular feedback of pairing fluctuations to non-Cooper channels is found.

%% file: Thesis_WICharge.tex
\chapter{Truncated flow equations and global conservation laws}
\label{chap:WICharge}
The underlying symmetries of a theory lead to constraints on the Green functions~\cite{Ward1950}. Baym and Kadanoff~\cite{Baym1961} and Baym~\cite{Baym1962} emphasised the importance of conservation laws for the computation of physical observables, in particular of transport properties, and identified necessary conditions on Green functions for being compatible with the symmetries of the system. In addition, they suggested a construction scheme for so-called conserving approximations, and Baym~\cite{Baym1962} realized their connection to the concept of $\Phi$-derivability. These works were devoted to local conservation laws and the associated Ward identities, which are usually incompatible with truncated flow equations, as Enss discussed in some detail~\cite{Enss2005}. 
Katanin~\cite{Katanin2004a} showed within the 1PI formalism for fermions that local Ward identities are violated by terms of some order in the effective interaction that depends on the specific truncation. In a truncation where the $n$-particle vertex is set to zero, local Ward identities are violated by terms of $n$-th order in the effective interaction, \ie\ on the truncation level. The fulfilment of local Ward identities in the one-loop truncation can be improved by taking into account the self-energy feedback from the third order in the effective interaction~\cite{Katanin2004a}. Within this so-called Katanin scheme, local Ward identities are violated by terms of third order in the effective interaction with overlapping loops only. Besides improving the fulfilment of Ward identities, this scheme solves mean-field models exactly~\cite{Salmhofer2004, Gersch2005} (see also chapter~\ref{chap:RPFM}).

Going beyond the Katanin scheme by including higher orders in the truncation is a cumbersome task for fermionic systems. There are vital alternatives if the motivation for this is only to improve the Ward identity fulfilment and not to take higher-order fluctuation effects into account. Schütz~\etal~\cite{Schuetz2005} obtained the exact solution of the one-dimensional Tomonaga-Luttinger model by introducing collective fields and closing of the hierarchy of renormalization group equations through Dyson-Schwinger equations as well as the Ward identities associated with the separate conservation of charge at each Fermi point. Blaizot~\etal~\cite{Blaizot2011} derived a closed set of flow equations for $\phi^4$-theory in three-dimensions within a $\Phi$-derivable approximation and pointed out that its solution might be easier than that of the self-consistency equations associated with the same Luttinger-Ward functional. 
The two-particle irreducible (2PI) RG formalism~\cite{Wetterich2007} may also be of interest because it is directly connected to the Luttinger-Ward functional of $\Phi$-derivable approximations. However, truncations of the 2PI hierarchy of flow equations are not necessarily conserving~\cite{Dupuis2005}.

The situation for Ward identities that arise from global conservation laws is not so clear. These impose way less stringent constraints on the theory and may be fulfilled even if local Ward identities are violated. Their compatibility with truncated flow equations is of interest, because it decides on whether truncated flow equations may generate massless Goldstone modes in case of spontaneous symmetry breaking. From this point of view, the work by Gersch~\etal~\cite{Gersch2008} is encouraging because it demonstrated that the Katanin scheme yields reasonable results for the flow of the superfluid gap and its phase mode also in the attractive Hubbard model. 
Furthermore, it is noteworthy that the external pairing field could be chosen around two orders of magnitude smaller than the final value of the superfluid gap in that study without running into unphysical divergences. 
However, this numerical study did not conclusively answer the question about the fulfilment of global Ward identities because besides possible violations due to the truncation of the hierarchy of flow equations, further violations may arise from approximations on the momentum and frequency dependence of the self-energy and the two-particle vertex. 

In this chapter, the compatibility of truncated flow equations with the Ward identities associated with the global $U(1)$ charge symmetry is discussed. First, functional Ward identities for connected Green functions and one-particle irreducible vertex functions are derived. The latter were presented before by Salmhofer~\etal~\cite{Salmhofer2004}. Subsequently, Ward identities relating the one-particle Green function with connected two-particle Green functions or the anomalous self-energy with the two-particle vertex are derived. A scheme is presented that allows to obtain global Ward identities for higher order vertex functions from the corresponding flow equations, which is applied for the derivation of truncated Ward identities for the anomalous components of the two-particle vertex. 
The compatibility of truncated flow equations and Ward identities is subsequently discussed under the assumption that the two-particle vertex fulfils the Ward identity that is obtained by truncating the hierarchy of Ward identities on the same level of approximation as that of the flow equations. It follows that the flow of the superfluid gap in the Katanin scheme is incompatible with the Ward identity. The reason is the ``incomplete'' consideration of contributions in third order in the effective interaction. Under the above assumption for the vertex, compatibility of the flow equation and the Ward identity for the gap would be achieved if all contributions in third order in the effective interaction are taken into account in the flow equation and the Ward identity for the vertex.

The notation in this chapter is the same as in chapter~\ref{chap:FunctionalRG} and Nambu representation (see chapter~\ref{chap:VertexParametrization}) is used if not stated differently. The scale dependence of objects is suppressed for notational convenience but can easily be added by replacing all objects by their scale-dependent counter-parts, yielding so-called cutoff Ward identities~\cite{Enss2005}.

\section{Ward identity for connected Green functions}
The derivation of a functional Ward identity for connected Green functions starts from their generating functional
\begin{equation}
	\mathcal G[\eta,\bar\eta, \Delta_{(0)}, \bar\Delta_{(0)}] = \ln \int D\mu_Q(\bar\chi,\chi) \e{-\mathcal V[\bar\chi,\chi]-(\bar\chi,\eta)-(\bar\eta,\chi) - (\bar\chi,\Delta_{(0)} \chi) - (\bar\chi,\bar\Delta_{(0)} \chi)}
\end{equation}
where $\chi$, $\bar\chi$ are fermionic Nambu Grassmann fields representing the microscopic degrees of freedom and the external pairing field
\begin{align}
	\Delta_{(0),\alpha\beta} &= \Delta_{(0)}(k_\alpha) \delta_{k_\alpha,k_\beta} \delta_{s_\alpha,+} \delta_{s_\beta,-}		&  \bar\Delta_{(0),\alpha\beta} &= \bar\Delta_{(0)}(k_\alpha) \delta_{k_\alpha,k_\beta} \delta_{s_\alpha,-} \delta_{s_\beta,+}
\end{align}
acts as an additional source term. Being bilinear in the fields, it could be absorbed in the measure but is kept explicit in the generating functional for convenience. In the effective action, the external pairing field is absorbed in the anomalous self-energy, specifying its initial conditions in a flow. Greek symbols represent multi-indices that collect for example momenta, frequencies and Nambu indices, $\alpha = (k_\alpha,s_\alpha)$.

The invariance of the system under global $U(1)$ gauge transformations requires the action and the measure in the above functional integral to be invariant under the transformation
\begin{equation}
	\begin{split}
	\bar\chi \rightarrow \bar\chi' &= \bar\chi \e{i \mtau{3}{} \theta}\\
	\chi \rightarrow \chi' &= \e{-i \mtau{3}{} \theta} \chi
	\end{split}
\end{equation}
with $\theta$ being the gauge phase and $\mtau{3}{}$ being a Pauli matrix that acts on the Nambu index of fields. The latter is connected to the operator for the fermionic density in terms of Nambu fields,
\begin{gather}
	n_i =   \phicp{i}\phiap{i} - \phicm{i}\phiam{i} = \sum_{s,s'} \mtau{3}{s s'} \phic{i s} \phia{i s'}\\
	n = \sum_k (\phicp{k}\phiap{k} - \phicm{k}\phiam{k}) = \sum_{k,s,s'} \mtau{3}{s s'} \phic{k s} \phia{k s'}
\end{gather}
in real and momentum space, respectively. The generating functional is invariant under the above transformation, if the source fields are transformed according to
\begin{align}
	\begin{aligned}
		\eta &\rightarrow \eta' = \e{-i \mtau{3}{} \theta} \eta\\
		\Delta_{(0)} &\rightarrow \Delta_{(0)}' = \e{-2 i \theta} \Delta_{(0)}
	\end{aligned}
	&&
	\begin{aligned}
		\bar\eta &\rightarrow \bar\eta' = \bar\eta \e{i \mtau{3}{} \theta}\\
		\bar\Delta_{(0)} &\rightarrow \bar\Delta_{(0)}' = \e{2 i \theta} \bar\Delta_{(0)}.
	\end{aligned}
\end{align}

The invariance of $\mathcal G$ under the gauge transformation can be stated as the condition
\begin{equation}
	\frac{d}{d\theta} \mathcal G[\eta', \bar\eta', \Delta_{(0)}', \bar\Delta_{(0)}'] = 0
\end{equation}
or equivalently
\begin{equation}
	\frac{d}{d\theta} \mathcal G\bigl[\e{-i\mtau{3}{} \theta} \eta, \bar\eta \e{i\mtau{3}{} \theta}, \e{-2i \theta} \Delta_{(0)}, \e{2i \theta} \bar\Delta_{(0)}\bigr]_{\theta = 0} = 0.
\end{equation}
Accomplishing the derivatives using the chain rule and taking care of multi-indices in the ``scalar product notation'' yields the functional Ward identity for connected Green functions
\begin{equation}
	\tfrac{1}{2}\Bigl(\eta, \mtau{3}{}\frac{\delta \mathcal G}{\delta \eta}\Bigr) -\tfrac{1}{2}\Bigl(\bar\eta, \mtau{3}{}\frac{\delta \mathcal G}{\delta \bar\eta}\Bigr) + \tr \Bigl(\Delta_{(0)}^T, \frac{\delta \mathcal G}{\delta \Delta_{(0)}}\Bigr) - \tr \Bigl(\bar\Delta_{(0)}^T, \frac{\delta \mathcal G}{\delta \bar\Delta_{(0)}}\Bigr) = 0
	\label{eq:WardIdentity:FuncConnCorr}
\end{equation}
where $\tr (\Delta_{(0)}^T, \frac{\delta \mathcal G}{\delta \Delta_{(0)}}) = \sum_{\alpha,\beta} \Delta_{(0),\alpha\beta} \frac{\delta \mathcal G}{\delta \Delta_{(0),\alpha\beta}}$ and similarly for the last term. Applying functional derivatives with respect to $\eta$ or $\bar\eta$ at vanishing fermionic sources yields Ward identities for connected Green functions.

The Ward identity that connects the anomalous propagator with connected two-particle Green functions follows after functional differentiation with respect to $\bar\eta_{k,-}$ and $\eta_{k,+}$, yielding
\begin{equation}
	-\langle \psia{k,-} \psic{k,+}\rangle = \sum_{k'} \bigl[\bar\Delta_{(0)}(k') \langle \psia{k,-} \psic{k,+}; \psic{k',-} \psia{k',+}\rangle - \Delta_{(0)}(k') \langle \psia{k,-} \psic{k,+}; \psic{k',+}\psia{k',-} \rangle \bigr]
\end{equation}
where $\langle A; B\rangle = \langle AB \rangle - \langle A \rangle \langle B \rangle$. This Ward identity expresses Goldstone's theorem and yields information on singularities of two-particle Green functions: Spontaneous symmetry breaking is obtained if the external pairing field is sent to zero $\Delta_{(0)} \rightarrow 0$ while the anomalous fermionic propagator remains finite. The Ward identity then requires some components of the two-particle Green function to diverge like $1 / \Delta_{(0)}$.

\section{Ward identity for one-particle irreducible vertex functions}
\subsection{Functional Ward identity}
In order to have exact relations connecting those quantities that are computed in the functional renormalization group for one-particle irreducible vertex functions, it is convenient to derive a functional Ward identity for these quantities from equation~\eqref{eq:WardIdentity:FuncConnCorr}. The necessary steps are similar to those that led in chapter~\ref{chap:FunctionalRG} from the flow equation for the generating functional for connected Green functions to the one for the effective action: The source fields and the corresponding derivatives of the generating functional for connected Green functions have to be rewritten in terms of their conjugate fields and corresponding derivatives of the effective action. Derivatives with respect to the external pairing field are most conveniently expressed through derivatives with respect to the fermionic sources,
\begin{equation}
\begin{split}
	\frac{\delta\mathcal G}{\delta \Delta_{(0),\alpha\beta}} &= \Bigl(\frac{\delta\mathcal G}{\delta \eta_\alpha} \frac{\delta\mathcal G}{\delta \bar\eta_\beta} + \frac{\delta^2\mathcal G}{\delta \eta_\alpha \delta\bar\eta_\beta}\Bigr) \delta_{k_\alpha,k_\beta} \delta_{s_\alpha,+} \delta_{s_\beta,-}\\
	\frac{\delta\mathcal G}{\delta \bar \Delta_{(0),\alpha\beta}} &= \Bigl(\frac{\delta\mathcal G}{\delta \eta_\alpha} \frac{\delta\mathcal G}{\delta \bar\eta_\beta} + \frac{\delta^2\mathcal G}{\delta \eta_\alpha \delta\bar\eta_\beta}\Bigr) \delta_{k_\alpha,k_\beta} \delta_{s_\alpha,-} \delta_{s_\beta,+},
\end{split}
\end{equation}
and then treated accordingly. Defining the shorthand
\begin{equation}
	\delta\Delta_{(0),\alpha\beta} = \Delta_{(0),\alpha\beta} - \bar\Delta_{(0),\alpha\beta},
\end{equation}
equation~\eqref{eq:WardIdentity:FuncConnCorr} can be rewritten as
\begin{equation}
	\tfrac{1}{2}\Bigl(\eta, \mtau{3}{}\frac{\delta\mathcal G}{\delta\eta}\Bigr) - \tfrac{1}{2}\Bigl(\bar\eta, \mtau{3}{} \frac{\delta\mathcal G}{\delta\bar\eta}\Bigr) + \Bigl(\frac{\delta\mathcal G}{\delta\eta},\delta\Delta_{(0)} \frac{\delta\mathcal G}{\delta\bar\eta}\Bigr) + \tr\Bigl(\delta\Delta_{(0)}^T \frac{\delta^2\mathcal G}{\delta\eta\delta\bar\eta}\Bigr) = 0.
\end{equation}
Expressing $\mathcal G$ through $\Gamma$ by using relations~\eqref{eq:fRG:Legendre} and~\eqref{eq:fRG:ReciprocityRelation} from chapter~\ref{chap:FunctionalRG}, the functional Ward identity for the effective action reads
\begin{equation}
	\tfrac{1}{2}\Bigl(\bar\phi, \mtau{3}{} \frac{\delta\Gamma}{\delta\bar\phi}\Bigr) + \tfrac{1}{2}\Bigl(\frac{\delta\Gamma}{\delta\phi},\mtau{3}{} \phi\Bigr) =  (\bar\phi,\delta\Delta_{(0)} \phi) +  \tr\Bigl(\delta\Delta_{(0)} \bigl[(\boldsymbol\delta^2\boldsymbol\Gamma)^{-1}\bigr]_{11}\Bigr).
	\label{eq:WardIdentity:Vertizes}
\end{equation}
The application of functional derivatives with respect to $\bar\phi$ and $\phi$ yields Ward identities for one-particle irreducible vertex functions. Cutoff Ward identities that are valid in the presence of a regulator at intermediate scales $\Lambda$ follow after endowing the effective action $\Gamma$ and possibly the external pairing field $\Delta_{(0)}$ with a scale dependence (similar to chapter~\ref{chap:FunctionalRG} for the former and chapter~\ref{chap:AttractiveHubbard} for the latter).

\subsection{Ward identity for superfluid gap}
\label{sec:WICharge:WIGap}
Of particular interest is the Ward identity for the anomalous self-energy that obtains after differentiation of~\eqref{eq:WardIdentity:Vertizes} with respect to $\bar\phi_\alpha$ and $\phi_\beta$ at vanishing source fields. It relates the anomalous self-energy to the two-particle vertex and reads
\begin{equation}
	\label{eq:WICharge:WIGap}
	\tfrac{1}{2}(\mtau{3}{\alpha\alpha} - \mtau{3}{\beta\beta}) \Gamma^{(2)}_{\alpha\beta} = \Delta_{(0),\alpha\beta} - \bar\Delta_{(0),\alpha\beta} + \sum_{\gamma,\delta,\mu,\nu} (\Delta_{(0),\gamma\delta} - \bar\Delta_{(0),\gamma\delta}) G_{\delta\nu} G_{\mu\gamma} \Gamma^{(4)}_{\alpha\nu\mu\beta}.
\end{equation}
As for the anomalous Green function, sustaining a finite anomalous self-energy in the limit of a vanishing external pairing field requires some components of the vertex to diverge inversely proportional to the external pairing field. This expresses Goldstone's theorem and implies the existence of long-range (or ``massless'') excitations due to the spontaneous breaking of the global $U(1)$ charge symmetry.

Specialising the multi-indices to $\alpha = (k, +)$ as well as $\beta = (k,-)$ and using $\Gamma^{(2)}_{(k,+),(k,-)} =\Delta(k)$, one obtains after explicitly writing out the Nambu summations
\begin{align}
		\Delta(k)& - \Delta_{(0)}(k) = \\
& \sum_{p} \Bigl[\Delta_{(0)}(p) \Bigl(G_{-+}(p) G_{++}(p) \Gamma^{(4)}_{+++-}(k,p,p,k) + (G_{-+}(p))^2 \Gamma^{(4)}_{++--}(k,p,p,k) \nonumber \\
	&\hspace{2em}+ G_{--}(p) G_{++}(p) \Gamma^{(4)}_{+-+-}(k,p,p,k) + G_{--}(p) G_{-+}(p) \Gamma^{(4)}_{+---}(k,p,p,k)\Bigr) \nonumber \\
	&-\bar\Delta_{(0)}(p) \Bigl(G_{++}(p) G_{+-}(p) \Gamma^{(4)}_{+++-}(k,p,p,k) + (G_{+-}(p))^2 \Gamma^{(4)}_{+-+-}(k,p,p,k) \nonumber \\
	&\hspace{2em} + G_{++}(p) G_{--}(p) \Gamma^{(4)}_{++--}(k,p,p,k) + G_{+-}(p) G_{--}(p) \Gamma^{(4)}_{+---}(k,p,p,k)\Bigr)\Bigr]. \nonumber 
\end{align}
This expression simplifies considerably if the phase of the superfluid gap equals that of the external pairing field, $\Delta_{(0)}(k) / |\Delta_{(0)}(k)| = \Delta(k) / |\Delta(k)|$, because then relations similar to $\Delta_{(0)}(k) G_{-+}(k) = \bar\Delta_{(0)}(k) G_{+-}(k)$ hold and all terms involving anomalous (3+1)-effective interactions cancel, yielding
\begin{align}
		\Delta(k)& - \Delta_{(0)}(k) = \\
&\sum_{p} \Bigl[\Delta_{(0)}(p) \Bigl((G_{-+}(p))^2 \Gamma^{(4)}_{++--}(k,p,p,k) + G_{++}(p) G_{--}(p) \Gamma^{(4)}_{+-+-}(k,p,p,k)\Bigr) \nonumber \\
	&-\bar\Delta_0(p) \Bigl((G_{+-}(p))^2 \Gamma^{(4)}_{+-+-}(k,p,p,k) + G_{++}(p) G_{--}(p) \Gamma^{(4)}_{++--}(k,p,p,k)\Bigr)\Bigr]. \nonumber 
\end{align}
If the external pairing field is chosen real, the Ward identity simplifies further and reads
\begin{equation}
	\Delta(k) - \Delta_{(0)}(k) =  \sum_{p} \Delta_{(0)}(p) L^\Lambda_{22}(p, 0) \bigl(\Gamma^{(4)}_{+-+-}(k,p,p,k) - \Gamma^{(4)}_{++--}(k,p,p,k)\bigr)
	\label{eq:WICharge:WIGap:GapReal}
\end{equation}
where $L^\Lambda_{22}(p, 0) = G_{++}(p) G_{--}(p) - (G_{+-}(p))^2$. Inserting the expressions for the vertex from chapter~\ref{chap:VertexParametrization} on the right hand side and keeping only terms that do not vanish in case the external pairing field is formally sent to zero, the Ward identity for the ground state reduces to
\begin{equation}
	\Delta(k) - \Delta_0(k) =  \sum_{p} \Delta_{(0)}(p) L^\Lambda_{22}(p, 0) \Phi_{kp}(0) + \ldots
\end{equation}
where the ellipsis represents terms that vanish for $\Delta_{(0)}\rightarrow 0$ (note that the fluctuation contributions are integrable in two dimensions and at zero temperature). This relation indicates that the momentum and frequency dependence of the superfluid gap arises mostly from the dependence of the phase mode of the superfluid gap on the relative momenta and frequencies or from the fermion-boson vertices after an appropriate expansion if the external pairing field is small.

\subsection{Ward identity for higher vertex functions}
\label{subsec:WIHigherVertex}
The functional Ward identity~\eqref{eq:WardIdentity:Vertizes} has a strong formal resemblance to the functional flow equation~\eqref{eq:fRG:funcRGvertex},
\begin{displaymath}
\partial_\Lambda \Gamlam = -\tr(\dot Q^\Lambda G^\Lambda_0) - (\bar\phi,\dot Q^\Lambda \phi) -\tr(\dot Q^\Lambda [(\boldsymbol\delta^2 \boldsymbol\Gamma^\Lambda)^{-1}]_{11}),
\end{displaymath}
which can be exploited for the derivation of Ward identities for higher order vertex functions. The left hand side of the Ward identity~\eqref{eq:WardIdentity:Vertizes} follows after straightforward evaluation of the functional derivatives for the effective action. The right hand side is obtained by replacing the matrix for the inverse bare propagator $\dot Q^\Lambda$ in the flow equation by the matrix $-\delta\Delta_0$. This implies that diagrammatic representations of Ward identities have the same topological structure as the flow equations and in particular that there is an infinite hierarchy of Ward identities.

In order to be useful, the infinite hierarchy of Ward identities has to be truncated as that of the flow equations. Approximating the three-particle vertex as in chapter~\ref{chap:FunctionalRG}, the Ward identity for the two-particle vertex can be closed and read of from the two-loop flow equations with the help of the above replacement rules. This truncated Ward identity becomes exact for reduced models in the thermodynamic limit due to the same arguments as for the flow equations (see~\cite{Salmhofer2004} or chapter~\ref{chap:RPFM}). In the presence of fluctuations, the truncation of the hierarchy of Ward identities can be justified similarly to the truncation of the hierarchy of flow equations. 

\section[Incompatibility of truncated flow equations and global Ward identities]{Incompatibility of truncated flow equations and global Ward identities in the modified one-loop truncation}
\label{sec:WICharge:Incomp}
In this section, the question about the fulfilment of the Ward identity for global charge conservation in the Katanin scheme is addressed. The idea is to derive a flow equation for the superfluid gap from the scale derivative of the corresponding cutoff Ward identity. The resulting flow equation should equal the one in the 1PI hierarchy if the flow of the anomalous self-energy is compatible with the Ward identity. Scale derivatives of the fermionic propagator and the vertex are determined from the flow equations in the Katanin scheme. This generates higher-order contributions in the two-particle vertex that have to be eliminated by exploiting the Ward identities for the gap and the vertex. The use of the Ward identity for the gap has to be justified \emph{a posteriori} by showing that the flow is compatible with the Ward identity. The Ward identity for the two-particle vertex involves the three-particle vertex and thus has to be truncated. 
In this section, it is truncated on the same level of approximation as the flow equation and is obtained from the latter as described in subsection~\ref{subsec:WIHigherVertex}. It is assumed that the flow of the vertex in the Katanin scheme is compatible with the truncated Ward identity. This is a subtle issue that is discussed in more detail below. 
Under these assumptions, it is possible to show that the contributions in the scale-differentiated Ward identity in first order in the vertex equal those in the flow equation for the anomalous self-energy in the 1PI scheme while all contributions in second order in the vertex vanish. However, there remain some third order contributions with overlapping loops, which lead to an incompatibility of truncated flow equations and Ward identities. This incompatibility arises from the incomplete consideration of contributions in third order in the two-particle vertex.

The starting point for the derivation of this result is the Ward identity for the superfluid gap~\eqref{eq:WICharge:WIGap} written in the form
\begin{equation}
	\delta\Delta^\Lambda_{\alpha\beta} = \delta\Delta^\Lambda_{(0),\alpha\beta} + \sum_{\gamma, \delta, \mu, \nu} \Vertex{\alpha\gamma\delta\beta} G^\Lambda_{\delta\mu} \delta\Delta^\Lambda_{(0),\mu\nu} G^\Lambda_{\nu\gamma}
\end{equation}
where $\delta\Delta$ has the same Nambu matrix structure as $\delta\Delta_{(0)}$, but the external pairing field is replaced by the superfluid gap. Differentiation with respect to the scale $\Lambda$ yields
\begin{equation}
	\begin{split}
	\partial_\Lambda \delta\Delta^\Lambda_{\alpha\beta} = \partial_\Lambda\delta\Delta^\Lambda_{(0),\alpha\beta} + \sum_{\gamma,\delta,\mu,\nu} &\Bigl(\partial_\Lambda\Vertex{\alpha\gamma\delta\beta} \G{\delta\mu} \delta\Delta^\Lambda_{(0),\mu\nu} \G{\nu\gamma} + \Vertex{\alpha\gamma\delta\beta} \delta\Delta^\Lambda_{(0),\mu\nu} \partial_\Lambda(\G{\delta\mu} \G{\nu\gamma})\\
	&\ + \Vertex{\alpha\gamma\delta\beta} \partial_\Lambda\delta\Delta^\Lambda_{(0),\mu\nu} \G{\delta\mu} \G{\nu\gamma} \Bigr),
	\end{split}
\end{equation}
where $\partial_\Lambda\Vertex{}$ is given by the modified one-loop equation~\eqref{eq:fRG:RGvertexKatanin},
\begin{equation}
	\begin{split}
		\partial_\Lambda \delta\Delta^\Lambda_{\alpha\beta} = \partial_\Lambda\delta\Delta^\Lambda_{(0),\alpha\beta}& + \sum_{\gamma,\delta,\mu,\nu} \Bigl( \bigl(\Pi^\text{PH,d}_{\alpha\gamma\delta\beta} - \Pi^\text{PH,cr}_{\alpha\gamma\delta\beta} - \tfrac{1}{2} \Pi^\text{PP}_{\alpha\gamma\delta\beta}\bigr) \G{\delta\mu} \delta\Delta^\Lambda_{(0),\mu\nu} \G{\nu\gamma} \\
			&+ \Vertex{\alpha\gamma\delta\beta} \delta\Delta^\Lambda_{(0),\mu\nu} \partial_\Lambda(\G{\delta\mu} \G{\nu\gamma}) + \Vertex{\alpha\gamma\delta\beta} \partial_\Lambda\delta\Delta^\Lambda_{(0),\mu\nu} \G{\delta\mu} \G{\nu\gamma}\Bigr).
	\end{split}
\end{equation}
The contribution from the Nambu direct particle-hole diagram and the first term in the second line,
\begin{equation}
	\begin{split}
		\sum_{\gamma,\delta,\mu,\nu} \Bigl(\Pi^\text{PH,d}_{\alpha\gamma\delta\beta} \G{\delta\mu} \delta\Delta^\Lambda_{(0),\mu\nu} \G{\nu\gamma} + \Vertex{\alpha\gamma\delta\beta} \delta\Delta^\Lambda_{(0),\mu\nu} \partial_\Lambda(\G{\delta\mu} \G{\nu\gamma})\Bigr)=\\
		\sum_{\gamma,\delta,\mu,\nu} \Bigl(\sum_{a,b,c,d} \partial_\Lambda(\G{ab} \G{cd}) \Vertex{\alpha b c \beta} \Vertex{d \gamma \delta a} \G{\delta\mu} \delta\Delta^\Lambda_{(0),\mu\nu} \G{\nu\gamma} + \Vertex{\alpha\gamma\delta\beta} \delta\Delta^\Lambda_{(0),\mu\nu} \partial_\Lambda(\G{\delta\mu} \G{\nu\gamma})\Bigr)
	\end{split}
\end{equation}
can be simplified after some renaming of summation indices by exploiting the Ward identity for the superfluid gap,
\begin{equation}
	= \sum_{\gamma,\delta,\mu,\nu} \Vertex{\alpha\gamma\delta\beta} \partial_\Lambda(\G{\delta\mu} \G{\nu\gamma}) \underbrace{\Bigl(\delta\Delta^\Lambda_{(0),\mu\nu} + \sum_{a, b, c, d} \Vertex{\mu d a \nu} \G{ab} \delta\Delta^\Lambda_{(0),bc} \G{cd} \Bigr)}_{\delta\Delta^\Lambda_{\mu\nu}}.
\end{equation}
Thus, the scale-differentiated Ward identity reduces to
\begin{equation}
	\begin{split}
	\partial_\Lambda\delta\Delta^\Lambda_{\alpha\beta} = \partial_\Lambda\delta\Delta^\Lambda_{(0),\alpha\beta} + \sum_{\gamma,\delta,\mu,\nu} &\Bigl(\Vertex{\alpha\gamma\delta\beta} \partial_\Lambda(\G{\delta\mu} \G{\nu\gamma}) \delta\Delta^\Lambda_{\mu\nu} + \Vertex{\alpha\gamma\delta\beta} \partial_\Lambda\delta\Delta^\Lambda_{(0),\mu\nu} \G{\delta\mu} \G{\nu\gamma}\\
	&\ - \bigl(\Pi^\text{PH,cr}_{\alpha\gamma\delta\beta} + \tfrac{1}{2} \Pi^\text{PP}_{\alpha\gamma\delta\beta} \bigr) \G{\delta\mu} \delta\Delta^\Lambda_{(0),\mu\nu} \G{\nu\gamma}\Bigr)
	\end{split}
\end{equation}
and after the replacement of scale-differentiated propagators by single-scale propagators and self-energy insertions (using~\eqref{eq:fRG:S_Gdot_connection}), it reads
\begin{equation}
	\begin{split}
	=\partial_\Lambda\delta\Delta^\Lambda_{(0),\alpha\beta} &+ \sum_{\gamma,\delta,\mu,\nu} \Bigl( \Vertex{\alpha\gamma\delta\beta} \delta\Delta^\Lambda_{\mu\nu} \partial_{\Lambda,S} (\G{\delta\mu} \G{\nu\gamma}) + \Vertex{\alpha\gamma\delta\beta} \partial_\Lambda\delta\Delta^\Lambda_{(0),\mu\nu} \G{\delta\mu} \G{\nu\gamma} + \\
	&\hspace{3em} +\sum_{a,b,c,d} \Vertex{\alpha\gamma\delta\beta} \Vertex{a c d b} \SL{dc} \bigl(\G{\delta a} \G{b\mu} \delta\Delta^\Lambda_{\mu\nu} \G{\nu\gamma} + \G{\delta\mu} \delta\Delta^\Lambda_{\mu\nu} \G{\nu a} \G{b \gamma}\bigr)\\
		&\hspace{3em} - \bigl(\Pi^\text{PH,cr}_{\alpha\gamma\delta\beta} + \tfrac{1}{2} \Pi^\text{PP}_{\alpha\gamma\delta\beta} \bigr) \G{\delta\mu} \delta\Delta_{(0),\mu\nu} \G{\nu\gamma}\Bigr),
	\end{split}
	\label{eq:WICharge:dWI_dLambda_preWIGamma}
\end{equation}
where $\partial_{\Lambda,S} G^\Lambda = S^\Lambda$. The contributions in the second line of this equation can be rewritten by exploiting the (truncated) Ward identity for the two-particle vertex, which is obtained from the flow equation for the two-particle vertex including two-loop contributions with non-overlapping loops\footnote{Or equivalently by replacing the matrix $\dot G^\Lambda$ in the flow equation in the Katanin scheme by the matrix $-G^\Lambda \delta\Delta^\Lambda G^\Lambda$ (which exploits the Ward identity for the gap).} as described in subsection~\ref{subsec:WIHigherVertex}. This yields after some renaming of summation indices
\begin{equation}
	\begin{split}
		\partial_\Lambda\delta\Delta^\Lambda_{\alpha\beta} &= \partial_\Lambda\delta\Delta^\Lambda_{(0),\alpha\beta} -\sum_{\gamma,\delta,\mu,\nu} \Vertex{\alpha\gamma\delta\beta} \Bigl(\tfrac{1}{2} (\mtau{3}{\alpha\alpha} + \mtau{3}{\gamma\gamma} - \mtau{3}{\delta\delta} - \mtau{3}{\beta\beta}) \SL{\delta\gamma} \\
	&\hspace{10em}- \delta\Delta^\Lambda_{\mu\nu} \partial_{\Lambda,S}(\G{\delta\mu} \G{\nu\gamma}) - \partial_\Lambda\delta\Delta^\Lambda_{(0),\mu\nu} \G{\delta\mu} \G{\nu\gamma}\Bigr)\\
		&\quad + \sum_{a,b,c,d,\delta,\gamma} \bigl(\Vertex{\alpha d \delta a} \Vertex{b \gamma c \beta} + \Vertex{\alpha\gamma c a} \Vertex{b d \delta \beta} + \Vertex{\alpha b c \delta} \Vertex{\gamma d a \beta}\bigr)\times\\
		&\hspace{4em} \times \G{cd} \bigl(\SL{ab} (\G{} \delta\Delta^\Lambda \G{})_{\delta\gamma} - \partial_\Lambda\G{ab} (\G{} \delta\Delta_{(0)} \G{})_{\delta\gamma}\bigr).
	\end{split}
\end{equation}
Using relations like $\Delta_{+-}^\Lambda(p) \bigl(G^\Lambda_{+-}(p) G^\Lambda_{-+}(p) - G^\Lambda_{++}(p) G^\Lambda_{--}(p)\bigr) = G^\Lambda_{+-}(p)$, it is easy to show that the first line equals the flow equation for the anomalous self-energy with appropriate signs as in $\delta\Delta^\Lambda$,
\begin{equation}
	\begin{split}
		\partial_\Lambda\delta\Delta^\Lambda_{\alpha\beta} &= \tfrac{1}{2} (\mtau{3}{\alpha\alpha} - \mtau{3}{\beta\beta}) \bigl(\partial_\Lambda\delta\Delta^\Lambda_{(0),\alpha\beta} - \sum_{\gamma,\delta,\mu,\nu} \Vertex{\alpha\gamma\delta\beta} \SL{\delta\gamma}\bigr)\\
		&\quad + \sum_{a,b,c,d,\delta,\gamma} \bigl(\Vertex{\alpha d \delta a} \Vertex{b \gamma c \beta} + \Vertex{\alpha\gamma c a} \Vertex{b d \delta \beta} + \Vertex{\alpha b c \delta} \Vertex{\gamma d a \beta}\bigr)\times\\
		&\hspace{4em} \times \G{cd} \bigl(\SL{ab} (\G{} \delta\Delta^\Lambda \G{})_{\delta\gamma} - \partial_\Lambda\G{ab} (\G{} \delta\Delta^\Lambda_{(0)} \G{})_{\delta\gamma}\bigr).
	\end{split}
	\label{eq:WICharge:dWI_dLambda_final}
\end{equation}
The second line contains contributions in third order in the effective interaction, because
\begin{equation}
	\begin{split}
	\SL{ab} (\G{} \delta\Delta^\Lambda \G{})_{\delta\gamma} - &\partial_\Lambda\G{ab} (\G{} \delta\Delta^\Lambda_{(0)} \G{})_{\delta\gamma} = \\
		& = \SL{ab} \bigl(G^\Lambda(\delta\Delta^\Lambda - \delta\Delta^\Lambda_{(0)}) G^\Lambda\bigr)_{\delta\gamma} - (\dot G^\Lambda - S^\Lambda)_{ab} (G^\Lambda \delta\Delta^\Lambda_{(0)} G^\Lambda)_{\delta\gamma}
	\end{split}
\end{equation}
is at least of first order in the effective interaction due to
\begin{align}
	\begin{split}
	(\dot G^\Lambda - S^\Lambda)_{ab} &= \sum_{c,d,e,f} \G{ae} \Vertex{ecdf} \SL{dc} \G{fb}\\
	(\delta\Delta^\Lambda - \delta\Delta^\Lambda_{(0)})_{ab} &= \sum_{c,d} \Vertex{acdb} (G^\Lambda \delta\Delta^\Lambda_{(0)} G^\Lambda)_{dc}.
	\end{split}
\end{align}
The third order contributions have different topological structures and are therefore not expected to cancel. It thus follows that
\begin{equation}
	\partial_\Lambda \delta\Delta^\Lambda_{\alpha\beta} = \tfrac{1}{2} (\mtau{3}{\alpha\alpha} - \mtau{3}{\beta\beta}) \partial_\Lambda \Delta^\Lambda_{\alpha\beta} + \mathcal O\bigl((\Vertex{})^3\bigr),
\end{equation}
if the hierarchies of flow equations and Ward identities are truncated by setting contributions to the two-particle vertex from the third order in the effective interaction with overlapping loops to zero and if the vertex fulfils the truncated Ward identity on the same level of approximation.

The latter assumption is discussed in the following. The reexamination of the above derivation reveals a surprising result: If the hierarchies of flow equations and Ward identities were closed by setting the three-particle vertex to zero, the Ward identity for global charge conservation connecting the gap and the vertex would be fulfilled according to the above calculation and for the underlying assumptions. However, within this truncation it is not possible to meaningfully continue the flow to the symmetry broken state even in the reduced BCS model~\cite{Salmhofer2004}, because the vertex diverges at a finite scale for small external pairing fields. The reason for this discrepancy is that, contrary to the above assumption, the truncated Ward identity for the vertex in the standard one-loop truncation is violated even by singular terms and the above conclusion therefore not valid. 
The above result for the Katanin scheme is therefore to be understood in the sense that even if the vertex fulfilled the truncated Ward identity, the flow of the self-energy would not be compatible with the Ward identity. Additional remainder terms appear in~\eqref{eq:WICharge:dWI_dLambda_final} if the vertex does not obey the truncated Ward identity. The question arises whether the remainder terms appearing in equation~\eqref{eq:WICharge:dWI_dLambda_final} or those from the truncated Ward identity for the vertex give larger contributions, or whether these contributions are singular in the limit of a vanishing external pairing field.

The above result for the unmodified one-loop scheme poses the question whether the flow of the anomalous self-energy and the Ward identity would be compatible under the above assumptions in case \emph{all} third-order contributions (\ie\ also those with overlapping loops) were taken into account in the flow equations and the Ward identity. The answer to this question is indeed affirmative, as a straightforward but lengthy calculation shows. It would be interesting to know whether the complete consideration of the third-order contributions to the two-particle vertex is also favourable for the fulfilment of the Ward identity for the vertex. This is conceivable because the two-loop approximation amounts to the use of the same approximation for the three-particle vertex in the flow equation and the Ward identity, which may lead to a higher degree of consistency. 
If the vertex fulfilled the truncated Ward identity within the two-loop scheme, Goldstone's theorem would be respected and a massless Goldstone mode generated in the limit of a vanishing external pairing field. Otherwise, the terms violating the Ward identity for the vertex due to the truncation would give rise to remainder terms of fourth order in the effective interaction in the scale-differentiated Ward identity for the gap (similar to the case of local Ward identities, see~\cite{Katanin2004a}). In summary, besides allowing to describe the singular infrared physics of a fermionic superfluid at zero temperature, the two-loop truncation may possibly improve the fulfilment of global Ward identities.

%% file: Thesis_RedPairFWModel.tex
\chapter{Reduced pairing and forward scattering model}
\label{chap:RPFM}
In this chapter, a model with reduced interactions in the pairing and forward scattering channels is studied. The modified one-loop truncation described in chapter~\ref{chap:FunctionalRG} yields the exact flow for that model. Analysing the flow yields insight into the role of various anomalous interactions and their singularities at and below the critical scale for superfluidity. Besides it is shown that the exact solution is contained within the channel-decomposition scheme described in chapter~\ref{chap:ChannelDecomposition}. After discussing the exact solution of the reduced model, small momentum and frequency transfers are allowed for in the interactions and the vertex is computed by resumming all chains of Nambu particle-hole diagrams. This serves as the starting point for the formulation of approximations for the singular momentum and frequency dependences of the vertex.

The reduced model is specified by the action
\begin{equation}
	\label{eq:RPFM:Action}
	S[\psic{},\psia{}] = S_0[\psic{},\psia{}] + V_{p}[\psic{},\psia{}] + V_{c}[\psic{},\psia{}] + V_{m}[\psic{},\psia{}]
\end{equation}
that consists of a quadratic term $S_0[\psic{},\psia{}]$ and several interaction terms $V_{p/c/s}[\psic{},\psia{}]$.
\begin{equation}
	\label{eq:RPFM:KinEnergy}
	S_0[\psic{},\psia{}] = \sum_{k,\sigma} \psic{k\sigma}(-i k_0 + \xi(\boldsymbol k))\psia{k\sigma} + \sum_k \Delta_{(0)}(k) (\psicup{k} \psicdown{-k} + \psiadown{-k} \psiaup{k})
\end{equation}
contains the kinetic energy $\xi(\boldsymbol k) = \epsilon(\boldsymbol k) - \mu$ measured from the chemical potential $\mu$ and an external pairing field $\Delta_{(0)}(k)$. The latter breaks the global $U(1)$ charge symmetry explicitly and spontaneous symmetry breaking is obtained in the limit $\Delta_{(0)}\rightarrow 0$. The interaction terms describe reduced interactions in the pairing, charge forward scattering and magnetic forward scattering channel, respectively:
\begin{gather}
	V_{p}[\bar\psi,\psi] = \frac{1}{2} \sum_{k, k', \sigma,\sigma'} V(k, k') \psic{k\sigma}\psic{-k\sigma'}\psia{-k'\sigma'}\psia{k'\sigma}\label{eq:RPFM:V0SC}\\
	V_{c}[\bar\psi,\psi] = \frac{1}{2}\sum_{k_i,\sigma,\sigma'} F_c(k,k') \psic{k\sigma}\psic{k'\sigma'}\psia{k'\sigma'}\psia{k\sigma} \label{eq:RPFM:F0C}\\
	V_{m}[\bar\psi,\psi] = \frac{1}{2}\sum_{k,k',\sigma_i} F_m(k, k') \vec\tau_{\sigma_1\sigma_4}\cdot\vec\tau_{\sigma_2\sigma_3} \psic{k\sigma_1}\psic{k'\sigma_2}\psia{k'\sigma_3}\psia{k\sigma_4}.\label{eq:RPFM:F0S}
\end{gather}
They are of the form~\eqref{eq:VP:2p2SpinorAnsatz} but restricted to interaction processes with zero momentum transfer,
\begin{align}
	P(k, k'; q) &= V(k, k') \delta_{q,0}	&	C(k, k'; q) &= F_c(k, k') \delta_{q,0}	& M(k, k'; q) &= F_m(k, k') \delta_{q,0}
\end{align}
and correspond to a spin-rotation invariant normal vertex of the form~\eqref{eq:VP:2p2spinor} with
\begin{equation}
\begin{split}
	V_{k_1 k_2 k_3 k_4} &= V(k_1,k_4) \delta_{k_1,k_2} \delta_{k_3,-k_4} + F_c(k_1, k_2) \delta_{k_1,k_4} \delta_{k_2,k_3} \\
	&- F_m(k_1,k_2) (2\delta_{k_1,k_3} \delta_{k_2,k_4} + \delta_{k_1,k_4} \delta_{k_2,k_3})
\end{split}
\end{equation}
on the microscopic level.

Due to the restricted momentum dependence of the interaction terms, the model~\eqref{eq:RPFM:Action} is exactly solvable in the thermodynamic limit, similarly to the reduced BCS model~\cite{Muehlschlegel1962} in which only $V_{p}$ contributes and whose RG flow was discussed extensively within the 1PI scheme by Salmhofer~\etal~\cite{Salmhofer2004}. Gersch~\cite{Gersch2007} discussed the model~\eqref{eq:RPFM:Action} without magnetic forward scattering, $F_m = 0$, within a resummation of all contributing Feynman diagrams.

\section{Exact integral equation and Ward identity}
Due to the restricted momentum dependence of the interactions in the reduced pairing and forward scattering model, a straightforward generalization of the arguments given by Salmhofer \etal~\cite{Salmhofer2004} shows that all contributions to $\Gamma^{(2)\,\Lambda}$ and $\Gamma^{(4)\,\Lambda}$ that are discarded in the truncation consisting of equations~\eqref{eq:fRG:RGSelfenergy} and~\eqref{eq:fRG:RGvertexKatanin} vanish in the thermodynamic limit. The restricted momentum dependence of the interaction terms leads to the following constraints on the external momenta of the various contributions to the one-loop flow of the Nambu two-particle vertex in~\eqref{eq:CD:RGvertexKatanin} in the thermodynamic limit:
\begin{gather}
	\Pi^\text{PH,d}_{s_1 s_2 s_3 s_4}(k_1, k_2, k_3, k_4) \propto \delta_{k_1,k_4} \delta_{k_2,k_3}\\
	\Pi^\text{PH,cr}_{s_1 s_2 s_3 s_4}(k_1, k_2, k_3, k_4) \propto \delta_{k_1,k_3} \delta_{k_2,k_4}\\
	\Pi^\text{PP}_{s_1 s_2 s_3 s_4}(k_1, k_2, k_3, k_4) \propto \delta_{k_1,-k_2} \delta_{k_3,-k_4}.
\end{gather}
It is sufficient to consider the flow equation for the Nambu vertex with $k_1 = k_4$ and $k_2 = k_3$, since the non-vanishing matrix elements for other choices of momenta follow from symmetries (in particular spin rotation invariance). Choosing $k_1 = k_4 \stackrel{\text{def}}{=} k$ and $k_2 = k_3 \stackrel{\text{def}}{=} k'$, only the direct Nambu particle-hole term contributes and the flow equation simplifies to
\begin{equation}
	\begin{split}
	\partial_\Lambda \Vertex{s_1 s_2 s_3 s_4}&(k, k', k', k) = \Pi^\text{PH,d}_{s_1 s_2 s_3 s_4}(k, k', k', k)\\
	&=\sum_{p, s_i'} \partial_\Lambda(G^\Lambda_{s_1' s_2'}(p) G^\Lambda_{s_3' s_4'}(p)) \Vertex{s_1 s_2' s_3' s_4}(k, p, p, k) \Vertex{s_4' s_2 s_3 s_1'}(p, k', k', p).
	\label{eq:RPFM:RG}
	\end{split}
\end{equation}
This differential equation is equivalent to the Bethe-Salpeter-like integral equation
\begin{equation}
	\begin{split}
		\Vertex{s_1 s_2 s_3 s_4}&(k, k', k', k) = \Gamma^{(4) \Lambda_0}_{s_1 s_2 s_3 s_4}(k, k', k', k) \\
	&+ \sum_{p, s_i'} \Gamma^{(4) \Lambda_0}_{s_1 s_2' s_3' s_4}(k, p, p, k) G^\Lambda_{s_1' s_2'}(p) G^\Lambda_{s_3' s_4'}(p) \Vertex{s_4' s_2 s_3 s_1'}(p, k', k', p)
		\label{eq:RPFM:VertexBetheSalpeter}
	\end{split}
\end{equation}
that sums up all chains of Nambu particle-hole bubble diagrams (note that $G^{\Lambda_0}_{s_1 s_2}(p) = 0$). Inserting this implicit solution for $\Vertex{}$ into the flow equation for the self-energy~\eqref{eq:fRG:RGSelfenergy} and exploiting the relation between the single-scale propagator and the full propagator differentiated with respect to the scale~\eqref{eq:fRG:S_Gdot_connection}, the flow of the self-energy can be integrated, yielding
\begin{equation}
	\Sigma^\Lambda_{s_1 s_2}(k) - \Sigma^{\Lambda_0}_{s_1 s_2}(k) = -\smashoperator{\sum_{k',s_3,s_4}} G^\Lambda_{s_4 s_3}(k') \VertexNull{s_1 s_3 s_4 s_2}(k, k', k', k).
	\label{eq:RPFM:SelfenergyMeanField}
\end{equation}
This is just the well-known mean-field equation for the self-energy, which is exact in the reduced model.

The exact solution determined by equations~\eqref{eq:RPFM:VertexBetheSalpeter} and~\eqref{eq:RPFM:SelfenergyMeanField} fulfils the Ward identity for global charge conservation~\cite{Salmhofer2004} (see subsection~\ref{sec:WICharge:WIGap} for a brief derivation),
\begin{equation}
	\Delta^\Lambda(k) - \Delta_{(0)}(k) = \sum_{k',s,s'} \Delta_{(0)}(k') \bigl(G^\Lambda_{s+}(k') G^\Lambda_{-s'}(k') - G^\Lambda_{s-}(k') G^\Lambda_{+s'}(k')\bigr) \Vertex{+s's-}(k, k', k', k),
\end{equation}
which connects the anomalous self-energy $\Delta^\Lambda(k)$ with the two-particle vertex. This Ward identity implies that some component of the Nambu vertex diverges $\sim \Delta_{(0)}^{-1}$ in case of spontaneous symmetry breaking in order to yield a finite superfluid gap in the limit $\Delta_{(0)} \rightarrow 0$.

In the reduced model, the vertex with external momenta $k_1 = k_4 \stackrel{\text{def}}{=} k$ and $k_2 = k_3 \stackrel{\text{def}}{=} k'$ equals the microscopic interaction plus the effective interaction in the Nambu particle-hole channel,
\begin{equation}
	\Vertex{s_1 s_2 s_3 s_4}(k, k', k', k) = \Gamma^{(4)\,\Lambda_0}_{s_1 s_2 s_3 s_4}(k, k', k', k) + V^{\text{PH},\Lambda}_{s_1 s_2 s_3 s_4}(k, k'; 0).
\end{equation}
Comparing the flow equation~\eqref{eq:RPFM:RG} with the one for the effective interaction in the Nambu particle-hole channel within the channel-decomposition scheme~\eqref{eq:CD:RGDE_PH} for the same choice of external momenta, one finds that both coincide. This implies that the exact solution of the reduced pairing and forward scattering model is indeed captured within the channel-decomposition scheme.

\hyphenation{equa-tion}
\section{Explicit solution}
An explicit solution of the Bethe-Salpeter equation~\eqref{eq:RPFM:VertexBetheSalpeter} for the Nambu two-particle vertex can be obtained for a separable momentum dependence of the interaction terms. In this special case the singularities of the Nambu vertex become particularly transparent.

The effective interaction in the Nambu particle-hole channel $V^{\text{PH},\Lambda}_{s_1 s_2 s_3 s_4}(k, k')$ is of the form~\eqref{eq:VP:NambuPH} supplemented by a delta function that restricts the momentum transfer $q$ to zero. In the following, the bare coupling functions $V$, $F_c$ and $F_m$ as well as the external pairing field $\Delta_{(0)}$ are assumed to be real and frequency independent. Then the Nambu vertex $\Vertex{}$ and the self-energy $\Sigma^\Lambda$ are also real and frequency independent, such that the complex conjugation operations in~\eqref{eq:VP:NambuPH} can be omitted. In this and the following section, the microscopic interaction is included in the effective interaction in the Nambu particle-hole channel for convenience because only one channel is present,
\begin{equation}
	V^{\text{PH},\Lambda_0}_{s_1 s_2 s_3 s_4}(k, k'; 0) = \Gamma^{(4)\,\Lambda_0}_{s_1 s_2 s_3 s_4}(k, k', k', k).
\end{equation}
Dependences on vanishing transfer momenta are usually suppressed in this section, for example $L_{ij}(p) \equiv L_{ij}(p, q = 0)$ or $A^\Lambda(k,k') \equiv A^\Lambda(k, k'; q = 0)$. Using the 4x4-matrix representation of the Nambu vertex from chapter~\ref{chap:ChannelDecomposition}, the integral equation~\eqref{eq:RPFM:VertexBetheSalpeter} can be written in matrix form as
\begin{equation}
	\hat V^{\text{PH},\Lambda}(k,k') = \hat V^{\text{PH},\Lambda_0}(k,k') + \sum_{p} \hat V^{\text{PH},\Lambda_0}(k,p) (2 \hat L^\Lambda(p)) \hat V^{\text{PH},\Lambda}(p, k')
\end{equation}
where $(\hat V^{\text{PH},\Lambda}(k,k'))_{ij} = V^{\text{PH},\Lambda}_{ij}(k,k')$ and similarly for the other matrices. In matrix representation, $\hat V^{\text{PH},\Lambda}(k,k')$ reads
\begin{equation}
	\hat V^{\text{PH},\Lambda}(k,k') =%
	\begin{pmatrix} \tfrac{1}{2} F^\Lambda_+(k,k')	&		0		&		0		&		0\\
	0		&		\tfrac{1}{2} A^\Lambda(k,k')		&		0		&		X^\Lambda(k',k)\\
	0		&		0		&		\tfrac{1}{2} \Phi^\Lambda(k,k')		&		0\\
	0		&		X^\Lambda(k,k')		&		0		&		\tfrac{1}{2} F^\Lambda_-(k,k')
	                                   	\end{pmatrix}
\end{equation}
where
\begin{gather*}
	F^\Lambda_+(k,k') = C^\Lambda(k,k') + M^\Lambda(k,k') + M^\Lambda(k,-k') - C^\Lambda(k,-k')\\
	F^\Lambda_-(k,k') = C^\Lambda(k,k') + M^\Lambda(k,k') - M^\Lambda(k,-k') + C^\Lambda(k,-k').
\end{gather*}
For the reduced interactions~\eqref{eq:RPFM:V0SC} to~\eqref{eq:RPFM:F0S}, the bare interaction in the Nambu particle-hole channel reads
\begin{equation}
	\label{eq:RPFM:V0_MF}
	\hat V^{\text{PH},\Lambda_0}(k,k') =%
	\begin{pmatrix} \tfrac{1}{2} F^{\Lambda_0}_+(k,k')	&		0		&		0		&		0\\
	0		&		\tfrac{1}{2} V(k,k')		&		0		&		0\\
	0		&		0		&		\tfrac{1}{2} V(k,k')		&		0\\
	0		&		0		&		0		&		\tfrac{1}{2} F^{\Lambda_0}_-(k,k')
	                                   	\end{pmatrix}.
\end{equation}
The momentum dependence of the bare interactions is assumed to factorize in the following,
\begin{equation}
	\begin{split}
		V(k,k') &= g^{(0)} f_{p}(\boldsymbol k) f_{p}(\boldsymbol k')\\
		F_c(k,k') &= g^{(0)}_c f_{c}(\boldsymbol k) f_{c}(\boldsymbol k')\\
		F_m(k,k') &= g^{(0)}_m f_{m}(\boldsymbol k) f_{m}(\boldsymbol k'),
	\end{split}
\end{equation}
where $f_{p}(\boldsymbol k)$, $f_c(\boldsymbol k)$ and $f_m(\boldsymbol k)$ are arbitrary reflection invariant form factors. The following result can easily be generalized to frequency dependent bare interactions with real $f_i$ that fulfil $f_i(k) = f_i(\bar k) = f_i(Rk)$ for all channels by replacing $f_i(\boldsymbol k)$ with $f_i(k)$ in the formulas. The momentum dependence of the flowing interactions then also factorizes,
\begin{align}
	A^\Lambda(k,k') &= g^\Lambda_a f_p(\boldsymbol k) f_p(\boldsymbol k')\\
	\Phi^\Lambda(k,k') &= g^\Lambda_\phi f_p(\boldsymbol k) f_p(\boldsymbol k')\\
	C^\Lambda(k,k') &= g^\Lambda_c f_c(\boldsymbol k) f_c(\boldsymbol k')\\
	M^\Lambda(k,k') &= g^\Lambda_m f_m(\boldsymbol k) f_m(\boldsymbol k')\\
	X^\Lambda(k,k') &= g^\Lambda_x f_c(\boldsymbol k) f_p(\boldsymbol k')\\
	F^\Lambda_+(k,k') &= 2 g^\Lambda_m f_m(\boldsymbol k) f_m(\boldsymbol k')\\
	F^\Lambda_-(k,k') &= 2 g^\Lambda_c f_c(\boldsymbol k) f_c(\boldsymbol k').
\end{align}

The momentum dependences in the integral equation~\eqref{eq:RPFM:VertexBetheSalpeter} can be eliminated after introducing the matrix
\begin{equation}
\label{eq:RPFM:fMatrix}
\hat f(\boldsymbol k)= 
	\begin{pmatrix}
		f_m(\boldsymbol k)		&		0		&		0		&		0\\
		0		&		f_p(\boldsymbol k)		&		0		&		0\\
		0		&		0		&		f_p(\boldsymbol k)		&		0\\
		0		&		0		&		0		&		f_c(\boldsymbol k)
	\end{pmatrix},
\end{equation}
which allows to rewrite the integral equation as
\begin{align}
	\hat V^{\text{PH},\Lambda}(k,k') &\stackrel{\text{def}}{=} \hat f(\boldsymbol k) \hat V^{\text{PH},\Lambda} \hat f(\boldsymbol k') \label{eq:RPFM:IntegralEquation}\\
	&=  \hat f(\boldsymbol k) \hat V^{\text{PH},\Lambda_0} \hat f(\boldsymbol k') + \hat f(\boldsymbol k) \hat V^{\text{PH},\Lambda_0} \sum_{p} \hat f(\boldsymbol p) (2 \hat L^\Lambda(p)) \hat f(\boldsymbol p) \hat V^{\text{PH},\Lambda} \hat f(\boldsymbol k')\nonumber
\end{align}
or equivalently as
\begin{equation}
	\label{eq:RPFM:BetheSalpeter:Matrix}
	\hat V^{\text{PH},\Lambda} = \hat V^{\text{PH},\Lambda_0} + \hat V^{\text{PH},\Lambda_0} (2 \hat L^\Lambda_f)  \hat V^{\text{PH},\Lambda},
\end{equation}
where $\hat L_f$ collects the fermionic loop integrands including the $\boldsymbol p$-dependent form factors. It is defined as
\begin{equation}
	\hat L^\Lambda_f = \sum_p \hat f(\boldsymbol p) \hat L^\Lambda(p) \hat f(\boldsymbol p) = \begin{pmatrix}
		l^\Lambda_m		&		0		&		0		&		0\\
		0		&		l^\Lambda_a		&		0		&		l^\Lambda_x\\
		0		&		0		&		l^\Lambda_\phi		&		0\\
		0		&		l^\Lambda_x		&		0		&		l^\Lambda_c
		\end{pmatrix}
		\label{eq:RPFM:MFBubbles}
\end{equation}
with matrix elements
\begin{equation}
\begin{split}
	l^\Lambda_m &= \sum_p f^2_m(\boldsymbol p) L^\Lambda_{00}(p) = \sum_p f^2_m(\boldsymbol p) \bigl[G^\Lambda_{++}(p)^2 + G^\Lambda_{+-}(p)^2\bigr]\\
	l^\Lambda_a &= \sum_p f^2_p(\boldsymbol p) L^\Lambda_{11}(p) = \sum_p f^2_p(\boldsymbol p) \bigl[G^\Lambda_{++}(p) G^\Lambda_{--}(p) + G^\Lambda_{+-}(p)^2\bigr]\\
	l^\Lambda_\phi &= \sum_p f^2_p(\boldsymbol p) L^\Lambda_{22}(p) =\sum_p f^2_p(\boldsymbol p) \bigl[G^\Lambda_{++}(p) G^\Lambda_{--}(p) - G^\Lambda_{+-}(p)^2\bigr]\\
	l^\Lambda_x &= \sum_p f_c(\boldsymbol p) f_p(\boldsymbol p) L^\Lambda_{13}(p) = 2 \sum_p f_c(\boldsymbol p) f_p(\boldsymbol p) G^\Lambda_{++}(p) G^\Lambda_{+-}(p)\\
	l^\Lambda_c &= \sum_p f^2_c(\boldsymbol p) L^\Lambda_{33}(p) = \sum_p f^2_c(\boldsymbol p) \bigl[G^\Lambda_{++}(p)^2 - G^\Lambda_{+-}(p)^2\bigr].
\end{split}
\label{eq:RPFM:MFBubblesEntries}
\end{equation}
The explicit solution for the flowing couplings is obtained by solving the matrix equation~\eqref{eq:RPFM:BetheSalpeter:Matrix} for $\hat V^{\text{PH},\Lambda}$, yielding
\begin{align}
	g^\Lambda_a &= \frac{(g^{(0)}_c)^{-1} - 2 l^\Lambda_c}{2 d^\Lambda},\label{eq:RPFM:gA}\\
	g^\Lambda_\phi &= \frac{1}{(g^{(0)})^{-1} - l^\Lambda_\phi},\\
	g^\Lambda_x &= \frac{l^\Lambda_x}{2 d^\Lambda},\\
	g^\Lambda_c &= \frac{(g^{(0)})^{-1} - l^\Lambda_a}{2 d^\Lambda},\\
	g^\Lambda_m &= \frac{1}{(g^{(0)}_m)^{-1} - 2 l^\Lambda_m},
\end{align}
where
\begin{equation}
	d^\Lambda =  \bigl[(g^{(0)})^{-1}-l^\Lambda_a\bigr] \bigl[(2 g^{(0)}_c)^{-1} - l^\Lambda_c\bigr] - (l^\Lambda_x)^2.
\end{equation}
Note that $g^\Lambda_m$ and $g^\Lambda_\phi$ are coupled to other interactions only indirectly via the propagators while $g^\Lambda_a$, $g^\Lambda_c$ and $g^\Lambda_x$ are coupled also directly.

Inserting the bare interaction with separable momentum dependence in the mean-field equation, one finds that the interaction-driven part of the self-energy adopts the momentum dependence of the form factors,
\begin{align}
	\Delta^\Lambda(k) - \Delta_{(0)}(k) &= -g^{(0)} f_p(\boldsymbol k) \sum_{p} f_p(\boldsymbol p) G^\Lambda_{+-}(p) \label{eq:RPFM:GapEquation}\\
	\Sigma^\Lambda(k) - \Sigma_{(0)}(k) &= - 2 g^{(0)}_c f_c(\boldsymbol k) \sum_p f_c(\boldsymbol p) G^\Lambda_{++}(p)
\end{align}
where $\Sigma^\Lambda_{+-}(k) = \Delta^\Lambda(k) - \Delta_{(0)}(k)$ and $\Sigma^\Lambda_{++}(k) = \Sigma^\Lambda(k)$. Assuming also $\Delta_{(0)}(k) = \delta_{(0)} f_p(\boldsymbol k)$ and $\Sigma_{(0)}(k) = \sigma_{(0)} f_{c}(\boldsymbol k)$, one obtains
\begin{align}
	\Delta^\Lambda(k) &= \delta^\Lambda f_p(\boldsymbol k)\\
	\Sigma^\Lambda(k) &= \sigma^\Lambda f_{c}(\boldsymbol k).
\end{align}

In the following, the behaviour of various components of the Nambu vertex as a function of the flow parameter $\Lambda$ is discussed, in particular the singularities near and below the critical scale $\Lambda_c$ for spontaneous breaking of the global $U(1)$ symmetry, at which the vertex diverges for $\Delta_{(0)} \rightarrow 0$. It is assumed that superfluidity is the only instability of the system and no instabilities occur in the forward scattering channel, \ie\ $g^\Lambda_c$ and $g^\Lambda_m$ remain finite for all $\Lambda$. For $\Lambda > \Lambda_c$ and $\Delta_{(0)} = 0$ the anomalous propagator $G^\Lambda_{+-}(k) = F^\Lambda(k)$ vanishes such that $l^\Lambda_a = l^\Lambda_\phi$, $l^\Lambda_x = 0$ and $l^\Lambda_c = l^\Lambda_m$. The anomalous components of the Nambu vertex also vanish and
\begin{equation}
	g^\Lambda_a = g^\Lambda_\phi = \frac{1}{(g^{(0)})^{-1} - l^\Lambda_a}.
\end{equation}
The critical scale is defined as the scale where
\begin{equation}
	l^{\Lambda_c}_a = - \sum_p f^2_p(\boldsymbol p) G^{\Lambda_c}(p) G^{\Lambda_c}(-p) = (g^{(0)})^{-1},
\end{equation}
where $G^\Lambda(p) = G^\Lambda_{++}(p)$, such that $g^{\Lambda_c}_a$ diverges.

For $\Lambda < \Lambda_c$ and $\Delta_{(0)}\rightarrow 0$ or for any $\Lambda$ and $\Delta_{(0)} \neq 0$, anomalous components appear in the propagator and the vertex. Using the relation
\begin{equation}
	F^\Lambda(k) = \Delta^\Lambda(k) [G^\Lambda(k) G^\Lambda(-k) + F^\Lambda(k)^2],
\end{equation}
the gap equation~\eqref{eq:RPFM:GapEquation} can be written in the form
\begin{equation}
	\delta^\Lambda - \delta_{(0)} = g^{(0)} \delta^\Lambda l^\Lambda_\phi.
\end{equation}
Inserting this into the solution for $\Phi^\Lambda(k, k')$ yields
\begin{equation}
	\label{eq:RPFM:PhiDelta0}
	\Phi^\Lambda(k, k') = \frac{\delta^\Lambda}{\delta_{(0)}} V(k,k').
\end{equation}
Note that the same relation holds also in the reduced BCS model~\cite{Salmhofer2004} and that it is neither affected by the forward scattering nor the anomalous (3+1)-effective interaction.

In contrast to the phase component $\Phi^\Lambda$ of the vertex, the amplitude component $A^\Lambda$ is regularized by the gap below the critical scale. Slightly below $\Lambda_c$, it behaves as
\begin{equation}
	A^\Lambda(k,k') \propto \frac{1}{\frac{\delta_{(0)}}{\delta^\Lambda} + \mathcal O ((\delta^\Lambda)^2)}.
\end{equation}
For $\Delta_{(0)}\rightarrow 0$, this component is therefore of order $(\delta^\Lambda)^{-2}$ slightly below $\Lambda_c$. The anomalous (3+1)-interaction behaves as
\begin{equation}
	X^\Lambda(k, k') \propto \frac{\delta^\Lambda}{\frac{\delta_{(0)}}{\delta^\Lambda} + \mathcal O ((\delta^\Lambda)^2)}
\end{equation}
and is therefore of order $(\delta^\Lambda)^{-1}$ slightly below $\Lambda_c$ for $\Delta_{(0)}\rightarrow 0$.

The above results are illustrated by plotting the flow of the self-energy and the components of the Nambu vertex for a specific choice of model parameters and at zero temperature. The kinetic energy $\xi(\boldsymbol k)$ enters only via the density of states, which is chosen as a constant $D(\xi) = 1$ with $|\xi| \leq 1$. All form factors in the interaction terms are chosen as unity, $f_p(\boldsymbol k) = f_c(\boldsymbol k) = f_m(\boldsymbol k) = 1$ and the bare couplings as $g^{(0)} = -0.5$, $g^{(0)}_c = 0.2$ and $g^{(0)}_m = 0$. The qualitative features of the flow do not depend on the size of the couplings. The magnetic coupling is chosen zero because the magnetic channel is decoupled from the rest. The flow is computed for a sharp cutoff acting on the bare kinetic energy, such that $|\xi| > \Lambda$ in the fermionic propagators. Since all bare couplings are momentum independent, also the flowing self-energy and the various components of the flowing Nambu vertex are momentum independent.

\begin{figure}
	\centering
	\includegraphics[width=0.66\linewidth]{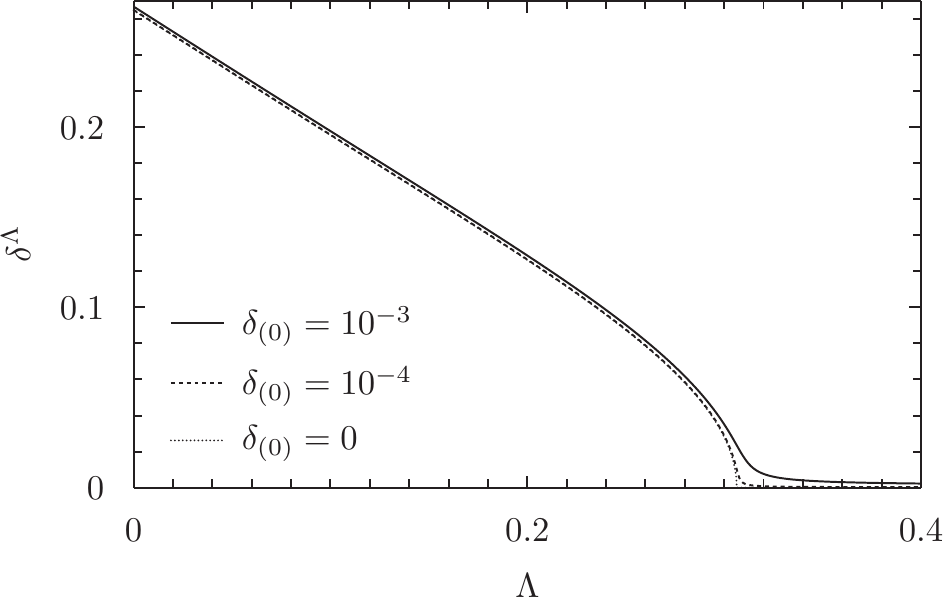}
	\caption{Flow of the superfluid gap $\delta^\Lambda$ for various values of the external pairing field $\delta_{(0)}$.}
	\label{fig:RPFM:Delta}
\end{figure}
\begin{figure}
	\centering
	\includegraphics[width=0.8\linewidth]{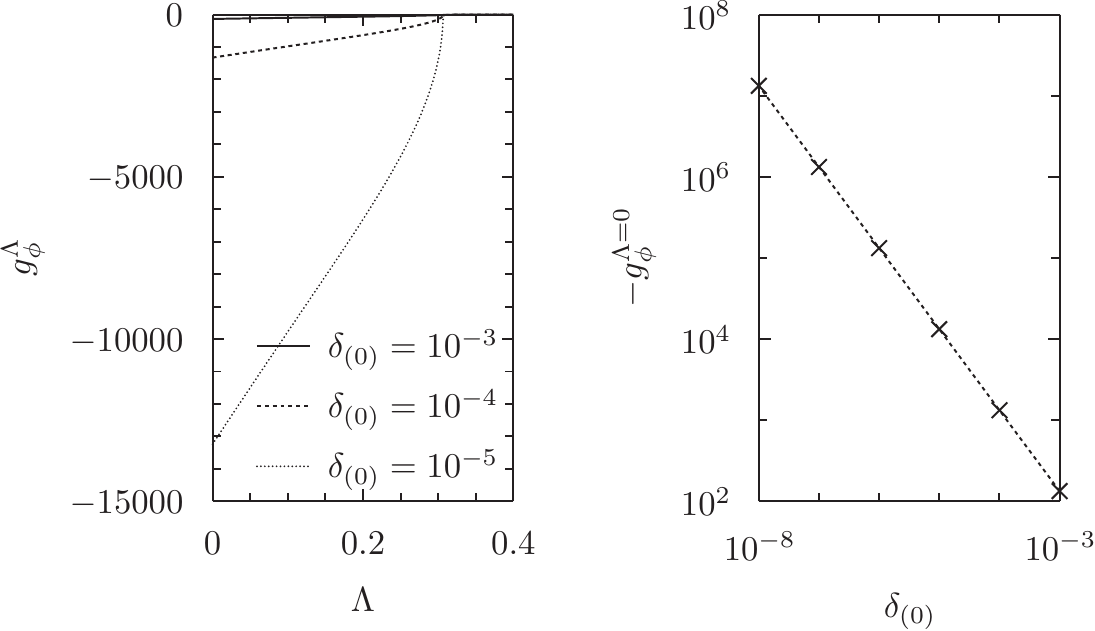}
	\caption{Left: Flow of the phase component of the Nambu vertex $\Phi^\Lambda = g^\Lambda_\phi$ for various values of $\delta_{(0)}$. Right: $g^{\Lambda = 0}_\phi$ as a function of $\delta_{(0)}$. Note that $\Phi^\Lambda$ is momentum independent due to the choice $f_p(\boldsymbol k) = 1$.}
	\label{fig:RPFM:Phi}
\end{figure}
\begin{figure}
	\centering
	\includegraphics[width=0.8\linewidth]{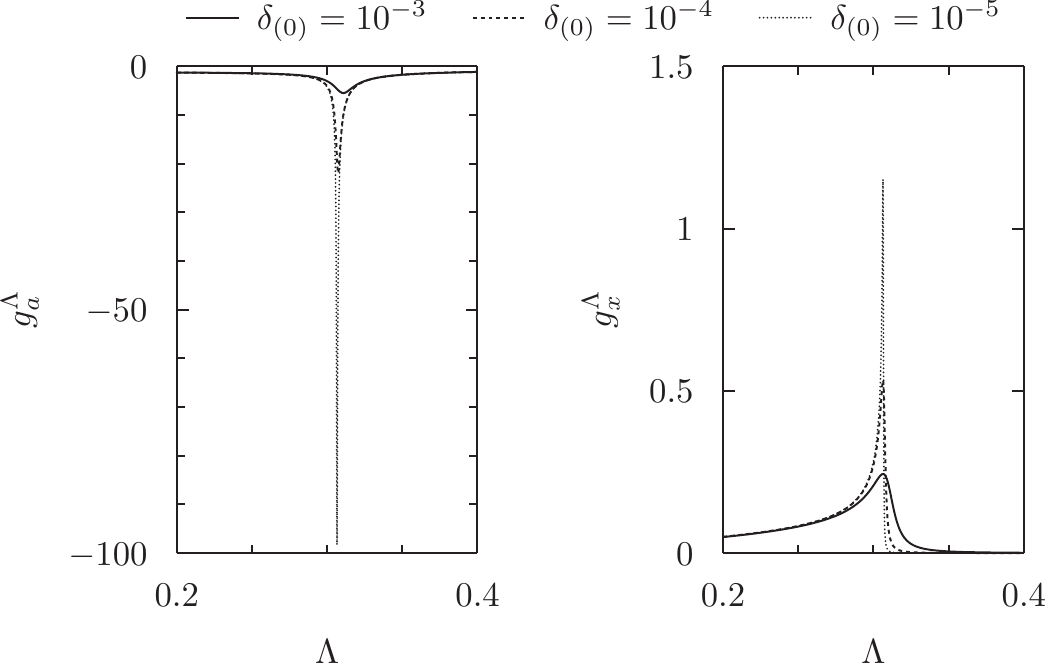}
	\caption{Flow of the amplitude component of the Nambu vertex $A^\Lambda = g^\Lambda_a$ (left) and of the anomalous (3+1)-effective interaction $X^\Lambda = g^\Lambda_x$ (right) for various values of $\delta_{(0)}$. Note that $A^\Lambda$ and $X^\Lambda$ are momentum independent due to the choice $f_p(\boldsymbol k) = f_c(\boldsymbol k) = 1$.}
	\label{fig:RPFM:AX}
\end{figure}
\begin{figure}
	\centering
	\includegraphics[width=0.66\linewidth]{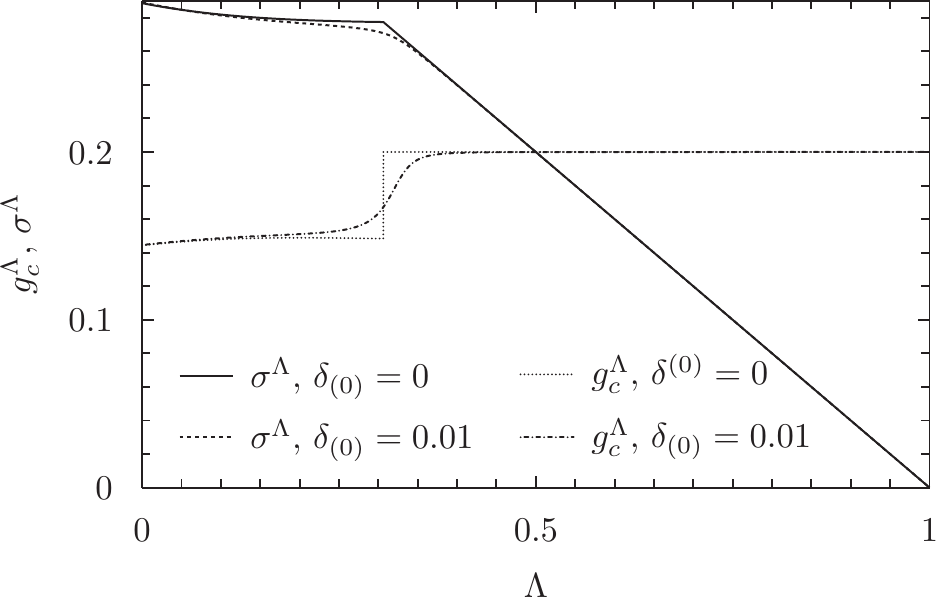}
	\caption{Flow of the normal self-energy $\sigma^\Lambda$ and the effective interaction in the charge-forward scattering channel $C^\Lambda = g^\Lambda_c$ for various values of $\delta_{(0)}$. Note that $C^\Lambda$ is momentum independent due to the choice $f_c(\boldsymbol k) = 1$.}
	\label{fig:RPFM:SigmaPhiC}
\end{figure}

Figure~\ref{fig:RPFM:Delta} shows the flow of the gap $\delta^\Lambda$ for various values of the external pairing field $\delta_{(0)}$. Note the sharp onset of $\delta^\Lambda$ at the critical scale $\Lambda_c$ for $\delta_{(0)} = 0$. This singularity is obviously smeared out for $\delta_{(0)} > 0$. The flow of the phase component $\Phi^\Lambda$ of the Nambu vertex is shown in the left panel of figure~\ref{fig:RPFM:Phi} and its value at the end of the flow as a function of $\delta_{(0)}$ in the right panel. Note that $\Phi^\Lambda$ has the same shape as $\delta^\Lambda$ and that $\Phi^{\Lambda=0}$ diverges as $(\delta_{(0)})^{-1}$ for small $\delta_{(0)}$, as described by equation~\eqref{eq:RPFM:PhiDelta0}. Figure~\ref{fig:RPFM:AX} shows the flow of $A^\Lambda$ and $X^\Lambda$. Both diverge at the critical scale $\Lambda_c$  for $\delta_{(0)} \rightarrow 0$, but decrease again for $\Lambda < \Lambda_c$ due to the growing gap. 
Note that $X^\Lambda$ is much smaller than $A^\Lambda$ on all scales. Figure~\ref{fig:RPFM:SigmaPhiC} shows the flow of the normal component of the self-energy and of the effective interaction for charge forward scattering. For the chosen multiplicative momentum cutoff, $C^\Lambda$ does not flow above the critical scale.

\section{Resummation approach for small transfer momenta}
\label{sec:RPFM:RPA}
This section briefly discusses a generalization of the model~\eqref{eq:RPFM:Action} allowing for small momentum and frequency transfers in the interaction. The vertex is obtained by resumming chains of Nambu particle-hole diagrams using a generalization of the integral equation~\eqref{eq:RPFM:VertexBetheSalpeter}. This is referred to as random-phase approximation (RPA) in the following and is equivalent to a solution of the channel-decomposed flow equations for the vertex where only the propagator renormalization contribution is kept. Differently from the reduced model, the Bethe-Salpeter like integral equation does not yield the exact solution for the generalization. However, it yields some insight into the properties of collective excitations. 
The focus of this section is not on their dynamical properties which are well-known\,--\,see for example the textbook by Popov~\cite{Popov1987} or more recent discussions by de Melo~\etal~\cite{Melo1993}, Belkhir and Randeria~\cite{Belkhir1994} and Gersch~\etal~\cite{Gersch2008}\,--\, but on singularities of the vertex for vanishing momentum and frequency transfer in the limit where the external pairing field goes to zero.

The generalized model is defined by an action similar to~\eqref{eq:RPFM:Action}, consisting of the bare action~\eqref{eq:RPFM:KinEnergy} and interaction terms that generalize to
\begin{gather}
	V_p[\bar\psi,\psi] = \frac{1}{2} \sum_{k, k', q, \sigma,\sigma'} V(k, k'; q) \psic{k+q/2,\sigma}\psic{q/2-k,\sigma'}\psia{q/2-k',\sigma'}\psia{q/2+k',\sigma}\label{eq:RPFM:V0SC_RPA}\\
	V_{c}[\bar\psi,\psi] = \frac{1}{2}\sum_{k, k',q,\sigma,\sigma'} F_c(k, k'; q) \psic{k+q/2,\sigma}\psic{k'-q/2,\sigma'}\psia{k'+q/2,\sigma'}\psia{k-q/2,\sigma} \label{eq:RPFM:F0C_RPA}\\
	V_{m}[\bar\psi,\psi] = \frac{1}{2}\sum_{k,k',q,\sigma_i} F_m(k, k'; q) \vec\tau_{\sigma_1\sigma_4}\cdot\vec\tau_{\sigma_2\sigma_3} \psic{k+q/2,\sigma_1}\psic{k'-q/2,\sigma_2}\psia{k'+q/2,\sigma_3}\psia{k-q/2,\sigma_4},\label{eq:RPFM:F0S_RPA}
\end{gather}
where $V(k, k'; q)$ and $F_i(k, k'; q)$ are peaked around $q = 0$. The latter implies that the RPA for the Nambu particle-hole channel yields a reasonable approximation for the momentum and frequency dependence of the vertex, which becomes exact in the limit where the momentum and frequency transfer is restricted to zero (yielding the reduced model). Using the matrix formulation for the Nambu vertex from chapter~\ref{chap:ChannelDecomposition}, the integral equation for the effective interaction in the Nambu particle-hole channel reads
\begin{equation}
	\hat V^{\text{PH},\Lambda}(k, k'; q) = \hat V^{\text{PH}}_{(0)}(k, k'; q) + \sum_p \hat V^{\text{PH}}_{(0)}(k, p; q) (2 \hat L(p, q)) \hat V^{\text{PH},\Lambda}(p, k'; q)
\end{equation}
where hats denote matrices and $\hat V^{\text{PH}}_{(0)}(k, k'; q)$ is a straightforward generalization of~\eqref{eq:RPFM:V0_MF}. Similarly to the reduced model, $\hat V^{\text{PH}}_{(0)}$ is absorbed into $\hat V^{\text{PH},\Lambda}$ in this section because only one channel is present. The singularities of the effective interaction become particularly transparent for a separable dependence on the fermionic momenta $k$ and $k'$,
\begin{equation}
	\begin{split}
	V(k, k'; q) &= V(q) f_p(k) f_p(k')\\
	F_c(k, k'; q) &= F_c(q) f_c(k) f_c(k')\\
	F_m(k, k'; q) &= F_m(q) f_m(k) f_m(k'),
	\end{split}
\end{equation}
where $f_{i}(k) = f_i(\bar k) = \zeta_i f_i(Rk)$ with $\zeta_i = \pm 1$ being the parity. For the above bare interactions, the flowing interactions also factorize
\begin{align}
	C^\Lambda(k, k'; q) &= C^\Lambda(q) f_c(k) f_c(k')\\
	M^\Lambda(k, k'; q) &= M^\Lambda(q) f_m(k) f_m(k')\\
	A^\Lambda(k, k'; q) &= A^\Lambda(q) f_p(k) f_p(k')\\
	\Phi^\Lambda(k, k'; q) &= \Phi^\Lambda(q) f_p(k) f_p(k')\\
	\nu^\Lambda(k, k'; q) &= \nu^\Lambda(q) f_p(k) f_p(k')\\
	X^\Lambda(k, k'; q) &= X^\Lambda(q) f_{\text{PH}}(k) f_p(k')\\
	\tilde X^\Lambda(k, k'; q) &= \tilde X^\Lambda(q) f_{\text{PH}}(k) f_p(k'),
\end{align}
with scale-independent form factors and $f_{\text{PH}}(k) = \tfrac{1+\zeta}{2} f_c(k) + \tfrac{1-\zeta}{2} f_m(k)$. The basis functions for the dependence on the fermionic momenta have been chosen in a way that allows to eliminate them from the integral equation by a matrix multiplication (see below). Due to symmetries, $\nu^\Lambda(q)$ and $\tilde X^\Lambda(q)$ are odd functions of the ``bosonic'' frequency $q_0$, while all other propagators are even.

The effective interaction in the Nambu particle-hole channel with a separable dependence on the fermionic momenta can be written in matrix form similarly to the reduced model,
\begin{equation}
	\hat V^{\text{PH},\Lambda}(k, k'; q) = \hat f(k) \hat V^{\text{PH},\Lambda}(q) \hat f(k')
\end{equation}
where $\hat f(k)$ is the matrix in~\eqref{eq:RPFM:fMatrix} with $f_c$ and $f_m$ replaced by the parity dependent $f^{\text{PH}}$. The matrix for $\hat V^{\text{PH},\Lambda}(q)$ reads
\begin{equation}
	\hat V^{\text{PH},\Lambda}(q) =%
	\begin{pmatrix} \tfrac{1}{2} F^{\Lambda}_+	&		0		&		0		&		0\\
	0		&		\tfrac{1}{2} A^\Lambda		&		\tfrac{1}{2} \nu^\Lambda		&		X^\Lambda\\
	0		&		-\tfrac{1}{2} \nu^\Lambda		&		\tfrac{1}{2} \Phi^\Lambda		&		-\tilde X^\Lambda\\
	0		&		X^\Lambda		&		\tilde X^\Lambda		&		\tfrac{1}{2} F^{\Lambda}_-
	                                   	\end{pmatrix}(q)
\end{equation}
where
\begin{align}
	F^\Lambda_+(q) &= (1-\zeta) C^\Lambda(q) + (1+\zeta) M^\Lambda(q)\\
	F^\Lambda_-(q) &= (1+\zeta) C^\Lambda(q) + (1-\zeta) M^\Lambda(q).
\end{align}
Note that the block diagonal structure of $\hat V^{\text{PH},\Lambda}(q)$ is a consequence of the expansion of the effective interactions in the channels with respect to $q$-independent basis functions $f_i$ for the fermionic momenta (see subsection~\ref{subsec:VP:BosonProp_gFB}). In the more general case, for example the upper right or lower left corner of the matrix would contain combinations of the imaginary parts of the effective interaction in the charge and magnetic channels. Note further that the effective interactions in the magnetic and charge channel change their position in the matrix depending on the parity of the basis functions. This implies that anomalous (3+1)-effective interactions couple the singlet pairing and the singlet charge channel for $\zeta = 1$ while they couple the triplet pairing and the triplet magnetic channel for $\zeta = -1$ on the RPA level.

The dependence on the fermionic momenta can be eliminated from the integral equation for the effective interaction similarly to the reduced model, resulting in
\begin{equation}
	\hat V^{\text{PH},\Lambda}(q) = \hat V^{\text{PH}}_{(0)}(q) + \hat V^{\text{PH}}_{(0)}(q) (2 \hat L_f(q)) \hat V^{\text{PH},\Lambda}(q)
\end{equation}
where 
\begin{equation}
	\hat L^\Lambda_f(q) = \sum_p \hat f(p) \hat L^\Lambda(p, q) \hat f(p).
\end{equation}
The entries of this matrix are straightforward generalizations of~\eqref{eq:RPFM:MFBubblesEntries}, where $\hat L^\Lambda(p)$ in~\eqref{eq:RPFM:MFBubbles} has to be replaced by $\hat L^\Lambda(p, q)$. In comparison to~\eqref{eq:RPFM:MFBubblesEntries}, two additional loop integrals appear for the imaginary parts,
\begin{equation}
	\begin{split}
		l^\Lambda_\nu(q) &= \sum_p f_p^2(p) L^\Lambda_{12}(p, q) = L_{f,12}(q) = -L_{f,21}(q)\\
		l^\Lambda_{\tilde x}(q) &= \sum_p f_p(p) f_\text{PH}(p) L^\Lambda_{23}(p, q) = L_{f,23}(q) = - L_{f,32}(q).
	\end{split}
\end{equation}
The above integral equation can then easily be solved for $\hat V^{\text{PH},\Lambda}(q)$, yielding
\begin{equation}
	\label{eq:RPFM:RPAVertexMatrix}
	\hat V^{\text{PH},\Lambda}(q) = \bigl((\hat V^{\text{PH}}_{(0)}(q))^{-1} - 2 \hat L_f(q)\bigr)^{-1}
\end{equation}
for $q$ such that $\hat V^{\text{PH}}_{(0)}(q) \neq 0$.

The following discussion is restricted to the even parity channel in which superfluidity occurs and it is assumed that $\min_q V(q)$ is located at $q = 0$. Furthermore, it is assumed that no other instability is present in the particle-hole channel. The block diagonal structure of $\hat V^{\text{PH},\Lambda}$ and $\hat L^\Lambda_{f}$ allows to immediately read of the effective interaction in the magnetic channel,
\begin{equation}
	M^\Lambda(q) = \bigl((F_m(q))^{-1} - 2 l^\Lambda_m(q)\bigr)^{-1}.
\end{equation}
The other non-zero entries are obtained after inverting the matrix in
\begin{equation}
	\begin{pmatrix}
	 	A^\Lambda		&		\nu^\Lambda		&		2 X^\Lambda\\
	-\nu^\Lambda		&		\Phi^\Lambda		&		-2 \tilde X^\Lambda\\
	2 X^\Lambda		&		2 \tilde X^\Lambda		&		2 C^\Lambda
	 \end{pmatrix}(q) = 
	-\begin{pmatrix}
-V^{-1} + l_a^\Lambda		&		l_\nu^\Lambda		&		l_x^\Lambda\\
-l_\nu^\Lambda		&		 -V^{-1} + l_\phi^\Lambda		&		l_{\tilde x}^\Lambda\\
l_x^\Lambda		&		-l_{\tilde x}^\Lambda		&		(-2 F_c)^{-1} + l_c^\Lambda
	                                 \end{pmatrix}^{-1}(q).
\label{eq:RPFM:RPAMatrix}
\end{equation}
For $q = 0$, the structure of this equation becomes the same as in the reduced model because $l^\Lambda_\nu$ and $l^\Lambda_{\tilde X}$ are odd functions in $q_0$. The explicit solution for the components reads
\begin{align}
	A^\Lambda &= -\frac{1}{d^\Lambda_\text{RPA}} \Bigl(\bigl(-V^{-1} + l_\phi^\Lambda\bigr) \bigl((-2 F_c)^{-1} + l_c^\Lambda\bigr) + \bigl(l_{\tilde x}^\Lambda\bigr)^2\Bigr)\\
	\Phi^\Lambda &= -\frac{1}{d^\Lambda_\text{RPA}} \Bigl(\bigl(-V^{-1} + l_a^\Lambda\bigr) \bigl((-2 F_c)^{-1} + l_c^\Lambda\bigr) - \bigl(l_x^\Lambda\bigr)^2\Bigr) \label{eq:RPFM:phiRPA}\\
	\nu^\Lambda &= \frac{1}{d^\Lambda_\text{RPA}} \Bigl( l_\nu^\Lambda \bigl((-2 F_c)^{-1} + l_c^\Lambda\bigr) + l_x^\Lambda l_{\tilde x}^\Lambda\Bigr)\\
	C^\Lambda &= -\frac{1}{2d^\Lambda_\text{RPA}} \Bigl(\bigl(-V^{-1} + l_a^\Lambda\bigr) \bigl(-V^{-1} + l_\phi^\Lambda\bigr) + \bigl(l_\nu^\Lambda\bigr)^2\Bigr)\\
	X^\Lambda &= \frac{1}{2d^\Lambda_\text{RPA}} \Bigl(\bigl(-V^{-1} + l_\phi^\Lambda\bigr) l^\Lambda_x - l_\nu^\Lambda l_{\tilde x}^\Lambda\Bigr)\\
	\tilde X^\Lambda &= -\frac{1}{2d^\Lambda_\text{RPA}} \Bigl(l_\nu^\Lambda l_x^\Lambda + \bigl(-V^{-1} + l_a^\Lambda\bigr) l_{\tilde x}^\Lambda \Bigr)
\end{align}
where the dependences on $q$ are suppressed on both sides of the equations for brevity and
\begin{equation}
\begin{split}
	d^\Lambda_\text{RPA} &= \Bigl[\bigl(-V^{-1} + l_a^\Lambda\bigr) \bigl((-2 F_c)^{-1} + l_c^\Lambda\bigr) - \bigl(l_x^\Lambda\bigr)^2\Bigr] \bigl(-V^{-1} + l_\phi^\Lambda\bigr)\\
	& + 2 l_\nu^\Lambda l^\Lambda_x l^\Lambda_{\tilde x} + \bigl(-V^{-1} + l_a^\Lambda\bigr) \bigl(l_{\tilde x}^\Lambda\bigr)^2 + \bigl((-2 F_c)^{-1} + l_c^\Lambda\bigr) \bigl(l_\nu^\Lambda\bigr)^2.
\end{split}
\end{equation}
For $q = 0$, the result for $\Phi^\Lambda$ reduces to
\begin{equation}
	\Phi^\Lambda(0) = \bigl(V^{-1}(0) - l_\phi^\Lambda(0)\bigr)^{-1},
\label{eq:RPFM:PhiRPA}
\end{equation}
implying that the anomalous (3+1)-effective interactions do not influence $\Phi^\Lambda(q)$ for $q = 0$ but for $q \neq 0$. They thus renormalize the momentum and frequency dependence of the phase mode of the superfluid gap.

In RPA, the anomalous self-energy follows from the same mean-field gap equation as in the reduced model and its momentum and frequency dependence equals that of $f_p$. Assuming $\Delta_{(0)}(k) = \Delta_{(0)} f_p(k)$, the gap equation reads
\begin{equation}
	\Delta^\Lambda - \Delta_{(0)} = - V(0) \intdrei{p} f_p(p) G^\Lambda_{+-}(p) = V(0) \Delta^\Lambda l_\phi^\Lambda(0).
\end{equation}
Inserting this result into~\eqref{eq:RPFM:PhiRPA}, one finds that the phase mode at $q = 0$ diverges as
\begin{equation}
	\Phi^\Lambda(0) = \frac{\Delta^\Lambda}{\Delta_{(0)}} V(0)
\end{equation}
for $\Delta_{(0)} \rightarrow 0$ and $\Lambda < \Lambda_c$, which is the same relation as in the mean-field model.

This is not the only singular dependence of the effective interaction on $\Delta_{(0)}$. Additional singular dependences appear in the imaginary parts of the normal effective interaction in the pairing channel $\nu^\Lambda$ and the anomalous (3+1)-effective interaction $\tilde X^\Lambda$. Their singular behaviour becomes most transparent after an expansion of the numerators and the denominator in the above expressions with respect to small momentum and frequency transfers at $\Lambda = 0$, where no regulator induced dependences appear (quantities without denoted scale dependence are for $\Lambda = 0$ in the following). $d_\text{RPA}(q)$ is expanded to $\mathcal O(q_0^2)$ and $\mathcal O(\boldsymbol q^2)$, while the numerators are expanded only up to $\mathcal O(q_0)$ and $\mathcal O(\boldsymbol q^0)$. This yields the approximation
\begin{equation}
	d_\text{RPA}(q) \approx d_0 + d_1 q_0^2 + d_2 \boldsymbol q^2
\end{equation}
for the denominator where
\begin{gather}
	d_0 = -V^{-1} \frac{\Delta_{(0)}}{\Delta} \Bigl[\bigl(-V^{-1} + l_a\bigr) \bigl((-2 F_c)^{-1} + l_c\bigr) - \bigl(l_x\bigr)^2\Bigr](0)\\
	d_1 = \frac{1}{2} \partial_{q_0}^2 d_\text{RPA}(q)|_{q = 0}\\
	d_2 = \frac{1}{2 d} \Delta_{\boldsymbol q} d_\text{RPA}(q)|_{q = 0}
\end{gather}
with $\Delta_{\boldsymbol q} = \sum_{i=1}^d \partial_{q_i}^2$ being the Laplace operator in $d$ dimensions. Note that $d_1$ and $d_2$ remain finite in the limit $\Delta_{(0)} \rightarrow 0$ for an $s$- and $d$-wave superfluid. The numerators in $\nu(q)$ and $\tilde X(q)$ can be approximated as
\begin{gather}
	 \Bigl(l_\nu^\Lambda \bigl((-2 F_c)^{-1} + l_c^\Lambda\bigr) + l_x^\Lambda l_{\tilde x}^\Lambda\Bigr)(q) \approx \beta_\nu q_0\\
	\Bigl(l_\nu^\Lambda l_x^\Lambda + \bigl(-V^{-1} + l_a^\Lambda\bigr) l_{\tilde x}^\Lambda\Bigr)(q) \approx \beta_{\tilde X} q_0
\end{gather}
for small $q_0$ where
\begin{gather}
	\beta_\nu = \partial_{q_0} \Bigl(l_\nu^\Lambda \bigl((-2 F_c)^{-1} + l_c^\Lambda\bigr) + l_x^\Lambda l_{\tilde x}^\Lambda\Bigr)(q)|_{q = 0}\\
	\beta_{\tilde X} = \partial_{q_0} \Bigl(l_\nu^\Lambda l_x^\Lambda + \bigl(-V^{-1} + l_a^\Lambda\bigr) l_{\tilde x}^\Lambda\Bigr)(q) |_{q = 0}
\end{gather}
are finite in the limit of a vanishing external pairing field. Consequently, it is possible to approximate the imaginary parts of the effective interactions for small transfer momenta and frequencies $q$ by
\begin{gather}
	\nu(q) \approx \frac{\beta_\nu q_0}{d_0 + d_1 q_0^2 + d_2 |\boldsymbol q|^2} \stackrel{\Delta_{(0)}\rightarrow 0}{\longrightarrow} \frac{\beta_\nu q_0}{d_1 q_0^2 + d_2 |\boldsymbol q|^2} \label{eq:RPFM:nu}\\
	\tilde X(q) \approx -\frac{\beta_{\tilde X} q_0}{d_0 + d_1 q_0^2 + d_2 |\boldsymbol q|^2} \stackrel{\Delta_{(0)}\rightarrow 0}{\longrightarrow} -\frac{\beta_{\tilde X} q_0}{d_1 q_0^2 + d_2 |\boldsymbol q|^2} \label{eq:RPFM:tildeX}.
\end{gather}
The phase mode behaves for small $q$ like
\begin{equation}
	\Phi^\Lambda(q) \approx - \frac{\alpha_\phi}{d_0 + d_1 q_0^2 + d_2 |\boldsymbol q|^2} \stackrel{\Delta_{(0)}\rightarrow 0}{\longrightarrow} -\frac{\alpha_\phi}{d_1 q_0^2 + d_2 |\boldsymbol q|^2},
\end{equation}
where $\alpha_\phi$ is the numerator of~\eqref{eq:RPFM:phiRPA} evaluated at $q = 0$. For small $q$ this approximation is equivalent to an expansion of the matrix elements of the right-hand side of~\eqref{eq:RPFM:RPAMatrix} up to second order in the transfer momentum and frequency before the computation of the inverse.

After sending $\Delta_{(0)}$ to zero, the limits $q_0\rightarrow 0$ and $|\boldsymbol q| \rightarrow 0$ do not commute any more for $\nu(q)$ and $\tilde X(q)$, which diverge $\sim q_0^{-1}$ when the frequency is sent to zero after the momentum. This behaviour of the $\nu$-propagator is well-known for superfluids (see for example~\cite{Popov1987,Castellani1997a,Pistolesi2004}). The singular behaviour of the imaginary part of the anomalous (3+1)-effective interaction in a superfluid with short-range interactions in the charge channel seems not to be known in the published literature.

%% file: Thesis_AttractiveHubbard.tex
\chapter{Attractive Hubbard model}
\label{chap:AttractiveHubbard}

In this chapter, the channel-decomposition scheme of chapter~\ref{chap:ChannelDecomposition} is applied to study the ground state of the two-dimensional attractive Hubbard model with Hamiltonian~\eqref{eq:Intro:HubbardHamiltonian} and $U < 0$. Its microscopic action is given by
\begin{equation}
	S[\bar\psi,\psi] = \sum_{k,\sigma} \bar\psi_{k\sigma}(-ik_0 + \xi(\boldsymbol k))\psi_{k\sigma} + U \sum_{k,k',q} \psicup{k+q} \psicdown{k'-q} \psiadown{k'} \psiaup{k}
\end{equation}
where the negative $U$ describes a local attraction between fermions that occupy the same lattice site and $\xi(\boldsymbol k)$ is the fermionic dispersion in~\eqref{eq:Intro:Dispersion} (Note that $t\equiv 1$ is used as the unit of energy.). This model is of relevance for the description of narrow-band systems with local non-retarded attractive interactions (see~\cite{Micnas1990} for a review) or for experiments with gases of ultracold fermionic atoms (see~\cite{Bloch2008,Esslinger2010} for reviews). As a model for the study of correlated fermions, it is of interest also because its ground state shows a crossover from the physics of weakly bound Cooper pairs in the weak-coupling regime to that of strongly bound bosonic molecules in the strong-coupling regime~\cite{Eagles1969,Leggett1980,Nozieres1985}. 
This BCS-BEC crossover attracted a lot of interest in the past decades, which qualifies the model as a good testing ground for new approximation schemes. Note however that this crossover was mostly studied in three-dimensional continuum systems, which could be obtained from the attractive Hubbard model in the low-density limit. 
At finite temperatures, the attractive Hubbard model can be employed for the study of pseudogap phenomena or of Kosterlitz-Thouless physics (see for example~\cite{Randeria1992,Allen1999,Kyung2001,Moreo1991}).

This chapter is organized as follows. The approximations that are applied to the fermionic self-energy and two-particle vertex within the channel-decomposition scheme are described in sections~\ref{sec:AH:ApproxSelfEnergy} and~\ref{sec:AH:ApproxVertex}, respectively. Numerical results for flows of these objects are discussed for different fermionic densities and interactions in section~\ref{sec:AH:NumResults} together with representative examples for their dependence on momenta and frequencies. This chapter is summarised and some conclusions are drawn in section~\ref{sec:AH:Conclusion}

\section{Approximation for fermionic propagator}
\label{sec:AH:ApproxSelfEnergy}
The fermionic propagator is parametrized as
\begin{gather}
	(\hat G^\Lambda)^{-1}(k) = (\hat G^\Lambda_0)^{-1}(k) + \hat\Sigma^\Lambda(k)
\end{gather}
where 
\begin{gather}
	(\hat G^\Lambda_0)^{-1}(k) = \begin{pmatrix} -(i k_0+\R{}(k)) + \xi(\boldsymbol k) + \delta\xi^\Lambda(\boldsymbol k)	&	\Delta_{(0)}(k) \\ \Delta_{(0)}(k)	&	-(i k_0 + \R{}(k)) - \xi(\boldsymbol k) - \delta\xi^\Lambda(\boldsymbol k) \end{pmatrix}
\end{gather}
is the regularized bare propagator in Nambu representation and
\begin{gather}
	\hat \Sigma^\Lambda(k) =  \begin{pmatrix} \Sigma^\Lambda_{++}(k) & \Sigma^\Lambda_{+-}(k) \\ \Sigma^\Lambda_{-+}(k) & \Sigma^\Lambda_{--}(k)\end{pmatrix} = \begin{pmatrix} \Sigma(k)	&	\Delta^\Lambda(k) - \Delta_{(0)}(k)	\\	\Delta^\Lambda(k) - \Delta_{(0)}(k)	&	-\Sigma(-k) \end{pmatrix}
\end{gather}
is the fermionic self-energy. Its anomalous component $\Delta^\Lambda(k)$ is chosen real and the second equality follows from symmetries. $R^\Lambda(k)$ is a regulator for the singularities of the fermionic propagator and is chosen as an additive frequency regulator,
\begin{equation}
	\label{eq:AH:Regulator}
	R^\Lambda(k) = i \sgn(k_0) \sqrt{k_0^2+\Lambda^2} - i k_0 \stackrel{\text{def}}{=} i \tR{}(k) - i k_0.
\end{equation}
It effectively replaces small frequencies $k_0$ with $|k_0| \ll \Lambda$ by $\sgn(k_0)\Lambda$ and is non-continuous at $k_0 = 0$. This choice is preferred over a (step-function like) Litim regulator~\cite{Litim2000}, because a sharp cutoff cannot be exploited for reducing the number of integrations in the renormalization contributions in the Katanin scheme. The reason is that the diagrams with scale-differentiated self-energy insertions contribute also for higher frequencies independently from the choice of the cutoff. Furthermore, the Litim cutoff would not result in frequency-independent integrands even for small $k_0$ with $|k_0| \ll \Lambda$ in case the self-energy depends non-linearly on frequency.

The external pairing field $\Delta_{(0)}(k)$ breaks the global $U(1)$ charge symmetry explicitly and is chosen independent of momentum and frequency in this chapter, $\Delta_{(0)}(k) = \Delta_{(0)}$. It is constant (independent of $\Lambda$) during the integration over the fermionic modes for $\Lambda > 0$ and can be removed subsequently in an additional flow where the fermionic regulator $R^\Lambda(k)$ vanishes and where $\Delta_{(0)}$ serves as the flow parameter. In such a pairing field flow, the external pairing field acts as an additive mass regulator for the phase mode, because the Ward identity for the global $U(1)$ symmetry dictates
\begin{equation}
	(-\Phi^\Lambda(0))^{-1} = m_\Phi^\Lambda \sim \Delta_{(0)}.
\end{equation}

$\delta\xi^\Lambda(\boldsymbol k)$ is a counterterm that accounts for deformations of the Fermi surface. It is defined by
\begin{equation} 
	\delta\xi^\Lambda(\boldsymbol k) = -\Sigma^\Lambda(k_0 = 0,\boldsymbol k).
\end{equation}
A general description of the counterterm method can be found in the book by Nozières~\cite{Nozieres1964}, more recent applications in perturbation theory and within the fRG in~\cite{Feldman1996,Neumayr2003} and~\cite{Gersch2008}, respectively. Note that $\Delta_{(0)}(k)$ and $\delta\xi^\Lambda(\boldsymbol k)$ are part of the bare action and propagator so that their scale-derivatives contribute to the single-scale propagator.

The full propagator reads
\begin{equation}
	\hat G^\Lambda(k) = \begin{pmatrix} G^\Lambda_{++}(k) & G^\Lambda_{+-}(k) \\ G^\Lambda_{-+}(k) & G^\Lambda_{--}(k)\end{pmatrix} = \begin{pmatrix} G^\Lambda(k) & F^\Lambda(k) \\ F^\Lambda(k) & -G^\Lambda(-k)\end{pmatrix},
\end{equation}
where the last equality holds due to symmetries. The normal and anomalous component are given by
\begin{gather}
	G^\Lambda(k) = \frac{i (\tilde R^\Lambda(k) - \im \Sigma^\Lambda(k)) + \xi(\boldsymbol k) + \delta\xi(\boldsymbol k) + \re\Sigma^\Lambda(k)}{D^\Lambda(k)}\\
	F^\Lambda(k) = \frac{\Delta^\Lambda(k)}{D^\Lambda(k)}
\end{gather}
where
\begin{gather}
	D^\Lambda(k) = \bigl(\tilde R^\Lambda(k) - \im \Sigma^\Lambda(k)\bigr)^2 + \bigl(\xi(\boldsymbol k) + \delta\xi^\Lambda(\boldsymbol k) + \re \Sigma^\Lambda(k)\bigr)^2 + \bigl(\Delta^\Lambda(k)\bigr)^2.
\end{gather}
For the attractive Hubbard model, the momentum dependence of the normal self-energy $\Sigma^\Lambda(k) = \hat \Sigma^\Lambda_{++}(k)$, the anomalous self-energy $\Delta^\Lambda(k) = \hat \Sigma^\Lambda_{+-}(k)$ and the counterterm are neglected,
\begin{gather}
	\Sigma^\Lambda(k) = \Sigma^\Lambda(k_0)\\
	\Delta^\Lambda(k) = \Delta^\Lambda(k_0)\\
	\delta\xi^\Lambda(\boldsymbol k) = \delta\xi^\Lambda. 
\end{gather}
The frequency dependence of the self-energy is discretised on a geometric grid of around 30 frequencies between $k_0 = 0$ and $k_0 \approx 350$. The grid points are computed from
\begin{gather}
	q_{0,i} = \delta + \alpha q_{0,i-1} = \delta \frac{\alpha^i - 1}{\alpha - 1}
\label{eq:AH:GeometricGrid}
\end{gather}
for $i \geq 1$ and $q_{0,0} = 0$ with typical parameters being $\alpha = 1.5$ and $\delta = 10^{-3}$. 
Neglecting the momentum dependence is most appropriate away from half- or van Hove-filling, when the Fermi surface is a slightly deformed circle. In this approximation, the counterterms fixes the Fermi level, while deformations of the Fermi surface are neglected. Note that the normal self-energy cannot simply be neglected because this may entail the violation of the Ward identity for global charge conservation. This can be seen in the exactly solvable reduced pairing and forward scattering model. 
Giering and Salmhofer~\cite{Giering2012} studied the impact of the momentum dependence of the normal self-energy on the flow of the two-particle vertex in the symmetric phase of the repulsive Hubbard model at van Hove filling and found only very small changes of the stopping scales or the hopping amplitudes. 
Neumayr and Metzner~\cite{Neumayr2003} and Gersch~\etal~\cite{Gersch2008} studied the momentum dependence of the anomalous self-energy tangential to the Fermi surface within renormalized perturbation theory and fermionic one-loop RG for $U=-2$, respectively. Both found only a small variation of the anomalous self-energy along the Fermi surface, in the former work even close to half-filling. Note however that an exact theorem connects the on-site and the nearest-neighbour component of $s$-wave gap in the Hubbard model~\cite{Zhang1990}.

The renormalization contributions to the self-energy are computed as Fermi surface averages. The anomalous self-energy is for example computed from
\begin{equation}
	\partial_\Lambda\Delta^\Lambda(k_0) = \int_{\boldsymbol k_F} \negthickspace\frac{ds}{L_F} \partial_\Lambda \Delta^\Lambda(k_0, \boldsymbol k_F)
\end{equation}
and analogously for the normal self-energy. This is similar to the projection of the flow of the effective interactions on that of the exchange propagators described in section~\ref{sec:CD:ProjBosProp}. The above approximation can be regarded as an expansion of the momentum dependence of the self-energy tangential to the Fermi surface with respect to orthonormal basis functions (for example Fermi surface harmonics, see~\cite{Allen1976}) where only the first term in the expansion is kept. The flow of the counterterm is obtained from the condition that the normal self-energy vanishes for $k_0 = 0$ and for momenta on the Fermi surface for all scales,
\begin{equation}
	\partial_\Lambda\delta\xi^\Lambda + \partial_\Lambda\re\Sigma^\Lambda(k_0 = 0) \stackrel{!}{=} 0.
	\label{eq:AH:CounterTermDef}
\end{equation}
At the scale $\Lambda$, the counterterm and the self-energy are known and only their scale-derivatives have to be calculated from equation~\eqref{eq:AH:CounterTermDef}, which is a linear integral equation for $\partial_\Lambda\delta\xi^\Lambda(\boldsymbol k)$. In case the momentum dependence of the normal self-energy and of the counterterm are neglected, the linear integral equation can be solved analytically.

For the regulator~\eqref{eq:AH:Regulator}, the fermionic propagator is strictly zero only for $\Lambda\rightarrow\infty$. One therefore has to introduce an upper cutoff $\Lambda_0$, from which the RG flow is integrated numerically, by hand. The high-energy flow for modes with $\Lambda > \Lambda_0$ can be integrated approximately. $\Lambda_0$ is chosen of the order of several times the bandwidth (usually around $\Lambda_0 = 100-150$) so that the coupling between channels is negligible for the high-energy modes. These can then be treated in mean-field and random-phase approximation\footnote{Note that random-phase approximation refers to the resummation of all chains of Nambu particle-hole bubble diagrams, as in chapter~\ref{chap:RPFM}.} in the presence of an infrared-regulator, which provides a starting point for the low-energy flow that fulfils the Ward identity for global charge conservation within the numerical accuracy. 
At the scale $\Lambda_0$, the self-energy and the counterterm are determined self-consistently in first order in the microscopic interaction $U$ and are independent of momentum and frequency\footnote{For $\Lambda \rightarrow \infty$, the initial conditions for the self-energy and the counterterm read $\Delta^{\Lambda\rightarrow \infty} = \Delta_{(0)}$, $\Sigma^{\Lambda\rightarrow \infty} = 0$ and $\delta\xi^{\Lambda\rightarrow \infty} = 0$.},
\begin{gather}
	\Delta^{\Lambda_0}(k) = \Delta^{\Lambda_0} = -U \intdrei{p} F^{\Lambda_0}(p) + \Delta_{(0)}\\
	\Sigma^{\Lambda_0}(k) = \Sigma^{\Lambda_0} = -U \intdrei{p} \e{i p_0 \eta} G^{\Lambda_0}(p) |_{\eta\rightarrow 0^+} = -\delta\xi^{\Lambda_0}.
	\label{eq:AH:InitialConditionsSigma}
\end{gather}
For the above choice of $\Lambda_0$, the correction to the external pairing field is rather small. In contrast, the normal self-energy and the counterterm are of the order of $U$ because of the Hartree term. They read
\begin{equation}
	\Sigma^{\Lambda_0} = -\delta\xi^{\Lambda_0} = \frac{U}{2} \intzwei{p} \left(1 - \frac{\xi(\boldsymbol p)}{\sqrt{\Lambda_0^2 + \xi(\boldsymbol p)^2 + \Delta^{\Lambda_0}(p)^2}}\right),
\end{equation}
which reduces to $\Sigma^{\Lambda_0} \approx U/2$ if $\Lambda_0$ is sufficiently large but finite. This contribution does not influence the flow of self-energies and vertices at all, because in their flow equations, either $\partial_\Lambda \delta\xi^\Lambda(\boldsymbol p)$ or $\delta\xi^\Lambda(\boldsymbol p) + \re\Sigma^\Lambda(p)$ appear. However, it would influence physical observables that are computed from the flow of the interaction correction to the grand canonical potential.

In order to compare fRG results for the attractive Hubbard model with results from other methods, an approximation for the fermionic density of the interacting system is needed. The reason is that the flow of the anomalous and the normal self-energy or equivalently of the counterterm lead to a renormalization of the relation between the chemical potential $\mu$ and the fermionic density $n$. In the following, the fermionic density is computed from the full propagator through the relation
\begin{equation}
	n = -2 \intdrei{p} \e{i p_0 0^+} G^{\Lambda=0}(p).
\end{equation}
This is possible because, firstly, shifts of the Fermi surface are avoided with the counterterm, secondly the neglected momentum dependence of the self-energy is expected to be small away from half- or van Hove-filling and thirdly because the frequency dependence of the self-energy is taken into account. Note that within this approximation, the density of the system is only known after evaluating the flow. In order to avoid numerically expensive repetitions of flows, a chemical potential is accepted if the density of the system at the end of the flow agrees with the requested density to within $3\%$.

\section{Approximation for two-particle vertex}
\label{sec:AH:ApproxVertex}
In this section, the approximations for the vertex in the attractive Hubbard model that are applied within the framework of chapter~\ref{chap:VertexParametrization} are described. In order to parametrize the vertex as a boson mediated interaction, it is decomposed in bosonic propagators and fermion-boson vertices. For $U < 0$, the expansion is restricted to the $s$-wave channel with $f_s(\boldsymbol k) = 1$. This approximation seems to be justified for the attractive Hubbard model because away from half-filling the ground state is always an $s$-wave superfluid and the pairing interaction is dominant from the outset. At weak-coupling, bosonic propagators in higher angular momentum channels and off-diagonal components should be negligible because they are generated only in $\mathcal O(U^3)$ or higher.

The $s$-wave approximation substantially simplifies the flow equations because all triplet terms can be omitted and the effective interaction in the Nambu particle-particle channel reduces to the magnetic one (see equation~\eqref{eq:VP:NambuPP:gEB}). Within this approximation, the fermion-boson vertices depend only on frequency. The parametrization of the different channels is described in more detail in the following, first for the effective interactions in the Cooper channel as well as for the imaginary part of the anomalous (3+1)-effective interaction and second for the effective interactions in the spinor particle-hole channel as well as the real part of the anomalous (3+1)-effective interaction. 

At the scale $\Lambda_0$, the fluctuation corrections to the microscopic interaction are computed as described in section~\ref{sec:RPFM:RPA} by summing up all chains of Nambu particle-hole diagrams, yielding
\begin{equation}
\begin{split}
	\hat V^{\text{PH},\Lambda_0}(q) = [\hat 1 - 2 \hat U \hat L^{\Lambda_0}(q)]^{-1} \hat U - \hat U
\end{split}
\end{equation}
where $(\hat V^{\text{PH},\Lambda}(q))_{ij} = V^{\text{PH},\Lambda}_{ij}(q)$ and $(\hat U)_{ij} = \frac{1}{2} U \delta_{ij} \operatorname{diag}(-1, 1, 1, 1)$. Within the approximations described above, the effective interaction in the Nambu particle-particle channel is fixed by symmetry. The fermion-boson vertices at the scale $\Lambda_0$ are independent of frequency and set to one.

\subsubsection{Particle-particle channel and imaginary part of (3+1)-effective interaction}
The description of the momentum and frequency dependence of the effective interactions in the spinor particle-particle channel is discussed together with that of the imaginary part of the anomalous (3+1)-effective interactions, because their singularity structure is very similar. Within the approximations described above, the effective interactions reduce to
\begin{gather}
	A^\Lambda_{kk'}(q) = A^\Lambda(q) g^{A,\Lambda}(k_0) g^{A,\Lambda}(k_0')\\
	\Phi^\Lambda_{kk'}(q) = \Phi^\Lambda(q) g^{\Phi,\Lambda}(k_0) g^{\Phi,\Lambda}(k_0')\\
	\nu^\Lambda_{kk'}(q) = \nu^\Lambda(q) g^{\nu,\Lambda}(k_0) g^{\nu,\Lambda}(k_0')\\
	\tilde X^\Lambda_{k k'}(q) = \tilde X^\Lambda(q) g^{\tilde X_\text{PH},\Lambda}(k_0) g^{\tilde X_\Phi,\Lambda}(k_0')\\
	\tilde \nu^\Lambda_{kk'}(q) = 0
\end{gather}
where the dependence of the fermion-boson vertices on $\boldsymbol q$ is neglected. As additional simplifications, $g^{\nu,\Lambda}$ is approximated by $g^{\Phi,\Lambda}$, which is justified because $\nu^\Lambda(q)$ has a small impact on the results in the coupling range considered and because $g^{A,\Lambda}$ and $g^{\Phi,\Lambda}$ differ only slightly on all scales in the weak-coupling regime. Besides, $g^{\tilde X_\text{PH},\Lambda}$ and $g^{\tilde X_\Phi,\Lambda}$ are approximated by $g^{C,\Lambda}$ and $g^{\Phi,\Lambda}$, respectively. The flow of $g^{A,\Lambda}$ and $g^{\Phi,\Lambda}$ is evaluated at $q = 0$, the evaluation of $g^{C,\Lambda}$ is described below. The imaginary part of the anomalous (4+0)-effective interaction $\tilde\nu^\Lambda$ vanishes because of the restriction to exchange propagators that are diagonal with respect to the form factor index due to symmetries.

The exchange propagators $A^\Lambda(q)$, $\Phi^\Lambda(q)$, $\nu^\Lambda(q)$ and $\tilde X^\Lambda(q)$ are singular for small momenta and frequencies $q$ at and below the critical scale for superfluidity, as can be seen from the resummation of diagrams in the Nambu particle-hole channel in chapter~\ref{chap:RPFM}. It is therefore important to accurately describe the neighbourhood of $q = 0$. This could be done by expanding the inverse bosonic propagators around their extrema, as it is usually done for bosonized theories. As an example, the propagator for the amplitude mode could be described by the ansatz
\begin{gather}
	-(A^\Lambda(q))^{-1} = m^\Lambda_A + A^\Lambda_A \boldsymbol q^2 + Z^\Lambda_A q_0^2+\ldots,
\end{gather}
which can be justified by power counting arguments after partial bosonization or resummations of perturbation theory as in chapter~\ref{chap:RPFM} in the presence of a gap. In order to improve the description of processes with intermediate transfer momenta and frequencies and to allow for more general momentum and frequency dependences, the low order polynomials are replaced by general functions, which are discretised on a grid of frequencies and momenta. The amplitude and phase mode of the superfluid gap are described by
\begin{gather}
	-(A^\Lambda(q))^{-1} = m^\Lambda_A(q_0) + F^\Lambda_A(\boldsymbol q)\\
	-(\Phi^\Lambda(q))^{-1} = m^\Lambda_\Phi(q_0) + F^\Lambda_\Phi(\boldsymbol q),
\end{gather}
where the frequency dependent ``masses'' $m^\Lambda_{A/\Phi}(q_0)$ and ``kinetic energies'' $F^\Lambda_{A/\Phi}(\boldsymbol q)$ are defined as
\begin{gather}
	m^\Lambda_A(q_0) = -\bigl(A^\Lambda(q_0,\boldsymbol q = 0)\bigr)^{-1}\\
	F^\Lambda_A(\boldsymbol q) = -\bigl(A^\Lambda(q_0 = 0, \boldsymbol q)\bigr)^{-1} + \bigl(A^\Lambda(q_0 = 0, \boldsymbol q = 0)\bigr)^{-1}
\end{gather}
and similarly for the phase mode\footnote{Note that $(A^\Lambda(q_0 = 0, \boldsymbol q = 0))^{-1}$ could also be added to $m^\Lambda_A$ instead of $F^\Lambda_A$ in the ansatz.}. The imaginary parts $\nu^\Lambda(q)$ and $\tilde X^\Lambda(q)$ are odd functions in $q_0$, motivating the ansatz
\begin{gather}
	-q_0 (\nu^\Lambda(q))^{-1} = m^\Lambda_\nu(q_0) + F^\Lambda_\nu(\boldsymbol q)\\
	q_0 (\tilde X^\Lambda(q))^{-1} = m^\Lambda_{\tilde X}(q_0) + F^\Lambda_{\tilde X}(\boldsymbol q)
\end{gather}
where
\begin{gather}
	m^\Lambda_\nu(q_0) = -\frac{q_0}{\nu^\Lambda(q_0, \boldsymbol q = 0)}\\
	m^\Lambda_{\tilde X}(q_0) = \frac{q_0}{\tilde X^\Lambda(q_0, \boldsymbol q = 0)}\\
	F^\Lambda_\nu(\boldsymbol q) = -\frac{q_{0,1}}{\nu^\Lambda(q_{0,1}, \boldsymbol q)} + \frac{q_{0,1}}{\nu^\Lambda(q_{0,1}, \boldsymbol q = 0)}\\
	F^\Lambda_{\tilde X}(\boldsymbol q) = \frac{q_{0,1}}{\tilde X^\Lambda(q_{0,1}, \boldsymbol q)} - \frac{q_{0,1}}{\tilde X^\Lambda(q_{0,1}, \boldsymbol q = 0)}
\end{gather}
and $q_{0,1}$ being the lowest non-zero frequency in the grid. Within this approximation, the flow equations for the exchange propagators have to be solved for $q_0 = 0$ or $q_0 = q_{0,1}$ and finite $\boldsymbol q$ or for finite $q_0$ and $\boldsymbol q = 0$. The momentum dependence of the imaginary parts is computed at the lowest non-zero frequency $q_{0,1}$ in the grid because of the antisymmetry of these propagators\footnote{In order to avoid problems arising from the limited numerical accuracy of integration routines occurring at very small frequencies and for finite momenta, it is favourable to exploit the approximately linear scaling of these propagators as a function of small frequencies
\begin{equation}
	\nu^\Lambda(q_0, \boldsymbol q) \approx s \nu^\Lambda(q_0/s, \boldsymbol q)
\end{equation}
(and similarly for $\tilde X^\Lambda(q_0, \boldsymbol q)$) by evaluating the renormalization contributions to the momentum dependence at a finite frequency well below the maximum of $|\nu^\Lambda(q_0, \boldsymbol 0)|$ or $|\tilde X^\Lambda(q_0, \boldsymbol 0)|$ (which roughly occurs for $q_0 \sim O(\Lambda)$ or $q_0 \sim O(\Delta_{(0)}^{1/2})$) and subsequent rescaling.}.  The above ansatz then provides an ``interpolation scheme'' for obtaining the momentum and frequency dependence of the propagators away from the sets of momenta and frequencies where they are computed. It yields a good description of the vertex as obtained from a resummation of all chains of Nambu particle-hole bubbles.

In comparison to a three-dimensional discretisation, this approximation allows to reduce the numerical effort for the solution of the flow equations considerably while the singular momentum and frequency dependence is still captured. Besides, it is more flexible in the description of generated momentum and frequency dependences than a simple parametrization and allows for example for terms that are linear in the frequency $\propto |q_0|$. Such terms are induced for the above choice of the regulator in all exchange propagators at intermediate and high scales.

\begin{figure}
	\centering
	\includegraphics[scale=0.9]{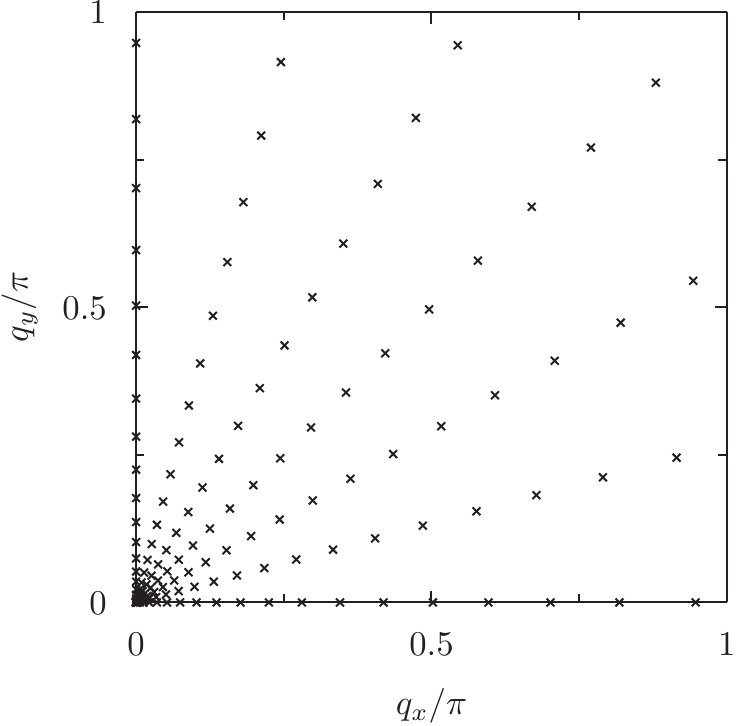}
	\caption{Example for the distribution of momentum grid points in the first quadrant of the Brillouin zone for the particle-particle channel.}
	\label{fig:AH:PPgrid}
\end{figure}

In order to accurately describe the singularities in the Cooper channel, the grid points should be denser near $q_0 = 0$ and $\boldsymbol q = 0$ and can be sparser for large frequencies and momenta. The frequency dependence of the exchange propagators is therefore computed on a geometric grid of 30 to 40 frequencies between $q_0 = 0$ and $q_0 \approx 350$. The grid points are computed as in equation~\eqref{eq:AH:GeometricGrid} with typical parameters being $\alpha = 1.45$ and $\delta = 5\cdot 10^{-4}$. The fermion-boson vertices are evaluated for the same frequencies. For frequencies above the maximal frequency in the grid, the propagators are smoothly suppressed to zero. The maximal frequency in the grid is chosen high enough in order to assure insensitivity of the results to its choice or to the high-frequency cutoff procedure. The momentum dependence is computed on a grid in polar coordinates with the centre at $\boldsymbol q = 0$. The angular dependence is computed for seven angles in the 
first quadrant of the Brillouin zone and 25 to 30 points are used in the radial direction with $q_{r,n} \sim n^3$ so that many points are placed in the vicinity of $\boldsymbol q = 0$. An example for a typical distribution of grid points is shown in figure~\ref{fig:AH:PPgrid}. 
This grid yields a satisfactory description of the momentum dependence even for momenta in the vicinity of $\boldsymbol q = (\pi,\pi)$ where the exchange propagators in the Cooper channel vary slowly away from half- or van Hove-filling. Note that due to the symmetries of the system, it is sufficient to compute the propagators only for one octant of the Brillouin zone. The propagators for non-grid frequencies and momenta are determined by cubic and bicubic spline interpolation, respectively. The above approximation together with this discretisation scheme yields an efficient approximation for the singular dependences of the exchange propagators. Roughly 150 couplings per propagator suffice to yield a good momentum and frequency resolution even if the exchange propagators become very large.

For the above ansatz, it is assumed that the functions $m^\Lambda_i$ and $F^\Lambda_i$ are positive. However, for certain parameters (usually close to half-filling or for stronger interactions), $\nu^\Lambda(q_{0,1}, \boldsymbol q)$ may change sign as a function of momentum in the vicinity of $\boldsymbol q = (\pi, 0)$ and symmetry related points. This is not problematic from a physical point of view, but leads to problems within the above approximation (because it may entail artificial zeros in the denominators). In the spirit of the above ansatz, such complications can be circumvented as long as the extremum of $|\nu^\Lambda(q_{0,1}, \boldsymbol q)|$ is at $q = (q_{0,1}, \boldsymbol 0)$ by adding the ``mass'' and the ``kinetic energy'' with an appropriate sign in such a way that the result equals the computed values for $q = (q_{0,1}, \boldsymbol q)$ or $q = (q_0, \boldsymbol 0)$. 
It should be noted that $|\nu^\Lambda(q_0, \boldsymbol q)|$ is very small in the regions where the sign changes occur, so that no significant impact on the results is expected.

\subsubsection{Particle-hole channel and real part of (3+1)-effective interaction}
In this section, the description of the momentum and frequency dependence of the effective interactions in the (spinor) particle-hole channel and of the real part of the anomalous (3+1)-effective interaction is discussed. These describe forward scattering for small and (incommensurate) density-wave fluctuations for large transfer momenta. Similarly to the Cooper channel, the aim is to capture the singular momentum and frequency dependence of the bosonic propagators near important points in momentum and frequency space. This requires a good description of a neighbourhood of $q = 0$ for forward scattering and of incommensurate peaks, which are located on the Brillouin zone boundary for the fermionic densities considered in this chapter, for density wave fluctuations. Within the approximations described above, the effective interactions reduce to
\begin{gather}
	C^\Lambda_{kk'}(q) = C^\Lambda(q) g^{C,\Lambda}(k_0) g^{C,\Lambda}(k_0')\\
	M^\Lambda_{kk'}(q) = M^\Lambda(q) g^{M,\Lambda}(k_0) g^{M,\Lambda}(k_0')\\
	X^\Lambda_{kk'}(q) = X^\Lambda(q) g^{X_\text{PH},\Lambda}(k_0) g^{X_A,\Lambda}(k_0')
\end{gather}
where the momentum dependence of the fermion-boson vertices is neglected. As an additional simplification, $g^{X_\text{PH},\Lambda}$ and $g^{X_A,\Lambda}$ are approximated by $g^{C,\Lambda}$ and $g^{A,\Lambda}$, respectively.

In order to allow for different approximations for forward scattering and density wave fluctuations, processes with small and large momentum transfers are described separately. They are assembled for example in the density channel as
\begin{equation}
	C^\Lambda(q) = C^\Lambda_{\boldsymbol 0}(q) \Theta(q_\text{max} - |\boldsymbol q|) + C^\Lambda_{\boldsymbol \pi}(q) \Theta(|\boldsymbol q| - q_\text{max})
\end{equation}
and similarly for the other propagators. $q_\text{max}$ is typically chosen of the order of $\pi / 3$ and the ($2\pi$-periodized) theta functions are replaced by smooth partitions of unity in the numerical implementation. The approximations for the (in-) commensurate density wave fluctuations are very similar to those for the Cooper channel and motivated by an expansion of the momentum and frequency dependence of the inverse propagators in low order polynomials at the (in-) commensurate peaks,
\begin{gather}
	-(M^\Lambda_{\boldsymbol \pi})^{-1}(q) = m^\Lambda_M(q_0) + F^\Lambda_M(\boldsymbol q)\\
	-(C^\Lambda_{\boldsymbol \pi})^{-1}(q) = m^\Lambda_C(q_0) + F^\Lambda_C(\boldsymbol q)\\
	(X^\Lambda_{\boldsymbol \pi})^{-1}(q) = m^\Lambda_X(q_0) + F^\Lambda_X(\boldsymbol q)
\end{gather}
where
\begin{gather}
	m^\Lambda_M(q_0) = -\frac{1}{M^\Lambda_{\boldsymbol \pi}(q_0, \boldsymbol q = \boldsymbol Q^\Lambda_M)}\\
	m^\Lambda_C(q_0) = -\frac{1}{C^\Lambda_{\boldsymbol \pi}(q_0, \boldsymbol q = \boldsymbol Q^\Lambda_C)}\\
	m^\Lambda_X(q_0) = \frac{1}{X^\Lambda_{\boldsymbol \pi}(q_0, \boldsymbol q = \boldsymbol Q^\Lambda_X)}
\end{gather}
\begin{gather}
	F^\Lambda_M(\boldsymbol q) = - \frac{1}{M^\Lambda_{\boldsymbol \pi}(q_0 = 0, \boldsymbol q)} + \frac{1}{M^\Lambda_{\boldsymbol \pi}(q_0 = 0, \boldsymbol q = \boldsymbol Q^\Lambda_M)}\\
	F^\Lambda_C(\boldsymbol q) = - \frac{1}{C^\Lambda_{\boldsymbol \pi}(q_0 = 0, \boldsymbol q)} + \frac{1}{C^\Lambda_{\boldsymbol \pi}(q_0 = 0, \boldsymbol q = \boldsymbol Q^\Lambda_C)}\\
	F^\Lambda_X(\boldsymbol q) = \frac{1}{X^\Lambda_{\boldsymbol \pi}(q_0 = 0, \boldsymbol q)} - \frac{1}{X^\Lambda_{\boldsymbol \pi}(q_0 = 0, \boldsymbol q = \boldsymbol Q^\Lambda_X)}.
\end{gather}
Note that the positions of the incommensurate peaks of magnetic, charge and anomalous channel, $\boldsymbol Q^\Lambda_M$, $\boldsymbol Q^\Lambda_C$ and $\boldsymbol Q^\Lambda_X$, respectively, differ in general. For intermediate fermionic densities, they move quite far away from $\boldsymbol q = \boldsymbol \pi$ during the flow. The momentum and frequency dependence at the incommensurate peaks may become singular due to $2 k_F$ scattering~\cite{Altshuler1995}. However, it is smeared in the presence of a superfluid gap and is pronounced only for small microscopic interactions $U$ where the gap is small. Away from half-filling, the momentum dependence of the propagators for incommensurate density wave fluctuations can thus be discretised on a simple (not necessarily equidistant) Cartesian grid. 
A good description of the momentum dependence is obtained for 15-25 points per momentum coordinate in one quadrant of the Brillouin zone. Due to symmetries, the propagators have to be computed only for one octant of the Brillouin zone. For the frequency dependence, the same grid is used as for the bosonic propagators in the Cooper channel. 
Very close to half-filling, the use of a polar grid with centre $\boldsymbol q = \boldsymbol \pi$ may become more economic in order to obtain a better description of interaction processes with momentum transfer $\boldsymbol \pi$. Note that for some parameters, the propagator for anomalous (3+1)-fluctuations $X^\Lambda(q)$ may also be plagued by sign changes in its momentum dependence near $(\pi,0)$ and symmetry related points. These can be captured within the spirit of the above approximation by adding the frequency dependent mass with an appropriate sign as described above for the imaginary part of the effective interaction in the Cooper channel. In regions where the sign changes occur, $X^\Lambda(0, \boldsymbol q)$ is very small and should thus not have a significant impact on the results.

The momentum and frequency dependence of the inverse propagators for forward scattering cannot be separated as for the other interaction channels due to the finite-scale manifestations of Landau damping. Furthermore, a parametrization as a boson mediated interaction as above is complicated by the fact that the effective interactions in the magnetic and the anomalous (3+1)-channel change sign for small $\boldsymbol q$ and finite $q_0$. In order to capture these effects, the momentum and frequency dependence is discretised on the same footing. However, a three-dimensional discretisation seems not to be feasible due to the required numerical effort. Away from van Hove filling, a two-dimensional discretisation is possible because the propagators have an approximate radial symmetry in their momentum dependence around $\boldsymbol q = \boldsymbol 0$, so that
\begin{gather}
	M^\Lambda_{\boldsymbol 0}(q_0, \boldsymbol q) \approx M^\Lambda(q_0,|\boldsymbol q|)\\
	C^\Lambda_{\boldsymbol 0}(q_0, \boldsymbol q) \approx C^\Lambda(q_0,|\boldsymbol q|)\\
	X^\Lambda_{\boldsymbol 0}(q_0, \boldsymbol q) \approx X^\Lambda(q_0,|\boldsymbol q|)
\end{gather}
holds for small $|\boldsymbol q|$. These approximations capture the singular momentum and frequency dependence of the channels and provide a good approximation at least for the vertex that is obtained from a resummation of all chains of Nambu particle-hole bubble diagrams. In the numerical implementation, the frequency dependence is discretised on the same grid as for the other channels and the momentum dependence is discretised with around ten momentum points between $\boldsymbol q = \boldsymbol 0$ and $\boldsymbol q = (q_\text{max}, 0)$. Note that such a radial approximation breaks down close to van Hove filling, where incommensurate peaks appear near $\boldsymbol q = 0$ (see for example~\cite{Holder2012,Husemann2012b}). At quarter-filling, it has been checked that the results for the forward scattering propagators do not depend sensitively on the direction in which the evaluation points for the momentum dependence are chosen, thus justifying the radial approximation.

The frequency dependence of the fermion-boson vertices is evaluated for the same grid of frequencies as for the exchange propagators. Its renormalization is determined at the position of the incommensurate peaks $\boldsymbol q = \boldsymbol Q^\Lambda_M$ and $\boldsymbol Q^\Lambda_C$, which are the extrema of the corresponding exchange propagators. The dependence of the fermion-boson vertices on the transfer momentum $\boldsymbol q$ is neglected, because the difference between the fermion-boson vertices for forward scattering and for incommensurate density-wave fluctuations was found to be small above the critical scale in the parameter range considered below. This is in agreement with findings by Husemann~\etal~\cite{Husemann2012} for the symmetric phase of the repulsive Hubbard model. Below the critical scale, the fermion-boson vertices for the particle-hole channel as evaluated above are only weakly renormalized.

\section{Numerical results}
\label{sec:AH:NumResults}
In this section, numerical results for properties of the superfluid ground state of the two-dimensional attractive Hubbard model, which were obtained within the channel-decomposition scheme for the flow equations and the vertex, are presented. After describing the numerical setup in subsection~\ref{subsec:AH:NumSetup}, results for the vertex and the self-energy on one-loop level are discussed in subsection~\ref{subsec:AH:MomFreqOneLoop}. This is followed by a discussion of the violation of the Ward identity for global charge conservation in the numerical calculations and its impact on the results in subsection~\ref{subsec:AH:WI}. 
Furthermore, the implementation of a coordinate projection scheme that allows to enforce the Ward identity in the numerical solution of the flow equations is described (see for example~\cite{Ascher1994}) and flows at fixed external pairing fields are compared to pairing field flows. Subsequently, the impact of fluctuations on two-loop level on the flow of exchange propagators and self-energies is examined in subsection~\ref{subsec:AH:MomFreqTwoLoop}. This sets the stage for the discussion of the impact of fluctuations on the superfluid gap. 

Most results in this section were obtained by using the coordinate projection scheme in order to enforce the Ward identity. If not stated differently, lines in plots are guides to the eyes. For quantities at the end of the fermionic flow, the dependence on $\Lambda$ is usually suppressed in the notation for convenience, for example $A^{\Lambda = 0}(q) \equiv A(q)$. Typical parameters are microscopic interactions in the range $|U|= 1-3$, next-nearest neighbour hoppings $t' = -0.1$, $0$, $0.1$, fermionic densities $n = 0.5$ or $0.78$ and external pairing fields ranging from $\Delta_{(0)} = \Delta_\text{MF}/1000$ to $\Delta_\text{MF}/50$, where $\Delta_\text{MF}$ is the mean-field gap for the same parameters and at the same density. In pairing field flows on one-loop level that include the frequency dependence of the vertex, the external pairing field could at least be reduced to $\Delta_\text{MF}/5000$ without encountering divergences. 
Some results are presented for a nearly half-filled system with $n = 0.95$. This parameter space is restricted with regards to larger couplings or fillings closer to half-filling by the applicability of the parametrization of the exchange propagators as presented in section~\ref{sec:AH:ApproxVertex}, which is discussed in some detail in section~\ref{sec:AH:Conclusion}.

The projection of the flow equations for the effective interactions on those for the bosonic propagators is performed within the ``Fermi surface averaging'' scheme (see section~\ref{sec:CD:ProjBosProp}) if not stated differently. This scheme was compared to projection by ``Brillouin zone averaging'' for $|U| = 2 - 3$ at fixed external pairing field on one-loop level and the differences in the results were found to be relatively small. 
Above the critical scale, the choice of the projection scheme leaves the flow of the dominant couplings (the extrema of the exchange propagators) almost unaffected, while noticeable differences arise away from the extrema (\ie\ for subdominant couplings). Below the critical scale, the choice of the projection scheme has also some impact on the flow of the exchange propagators in the Cooper channel at the extrema, but only a minor impact on the flow of the superfluid gap. These differences arise likely from the better treatment of the strong momentum dependence of the exchange propagators in the pairing channel in the ``Fermi surface averaging'' scheme. The reason is that this scheme provides a better approximation for the effective interactions between states in the vicinity of the Fermi surface in case a small number of form factors is used. 
This is why the renormalization contributions to the exchange propagators are obtained by averaging external momenta over the Fermi surface in the following, in particular in pairing field or two-loop flows.

\hyphenation{im-prove-ments}
\subsection{Numerical setup}
\label{subsec:AH:NumSetup}
By discretising the momentum and frequency dependence of the self-energy and the vertex, the functional renormalization group equations are transformed into a system of non-linear ordinary differential equations (ODE). For the discretisation as described above, roughly 2000 coupled equations have to be solved with three-dimensional integrals on the right hand sides. The numerical solution of this ODE system constitutes a considerable effort. Strongly depending on the parameters, it requires between a day and a week on 16 to 24 CPU cores for one-loop calculations or even longer for two-loop calculations. 

When evaluating the right hand side of the flow equations, the projection integrals over the external fermionic momenta can be carried out and tabulated before integrating over the loop momentum and frequency, thus resulting in three-dimensional integrals. The use of the ``Fermi surface averaging scheme'' for the projection prevents from ``pre-computing'' the integrations over the loop momentum in order to reduce the loop integrals to one-dimensional ones over the loop frequency, as it was done by Husemann and Salmhofer~\cite{Husemann2009}, Husemann~\etal~\cite{Husemann2012} and Giering and Salmhofer~\cite{Giering2012}. Besides, the ``momentum integrated loop integrals'' would have to be tabulated in a four-dimensional space of frequencies for all loops with a high resolution, which seems quite expensive with respect to memory. The three-dimensional loop integrals are therefore carried out with an adaptive integration algorithm for integrands of moderate dimension~\cite{Genz1980, Berntsen1991}. 
An implementation of this algorithm for C/C++ is available in the public domain~\cite{Cubature2012}. 
This algorithm is much faster than three iterated one-dimensional integrations, however, at the prize that the achievable accuracies are somewhat lower. They are nevertheless sufficient for the computation of the right hand side of the ODE system as long as the accuracy goals of the ODE solver are not too strict. The relative accuracy goal for the integration routine was chosen as $\epsilon_\text{rel} = 10^{-3}$ together with a tiny absolute accuracy goal, which should provide some security. It has been checked that different algorithms yield the same result within the accuracy goals for typical integrands, scales and microscopic parameters.

The system of ordinary differential equations is solved using routines from the GNU Scientific Library~\cite{Gough2009}. The accuracy of the numerical results was benchmarked by solving flows with a third-order Runge-Kutta routine and a fifth-order Runge-Kutta routine and the error tolerances were adjusted in such a way that the differences between the results were small (corresponding to absolute and relative error tolerances of about $10^{-4}$ for the fifth-order Runge-Kutta routine or of about $5\cdot 10^{-3}$ for the third-order Runge-Kutta routine). The third-order Runge-Kutta routine was used for most of the calculations because it performed better regarding the runtime and the scale-resolution.

This numerical framework has been tested for the exactly solvable reduced pairing and forward scattering model by comparing the obtained numerical solution of the flow with the exact solution. For external pairing fields that were two orders of magnitude smaller than the final gap, only tiny differences were found between the two solutions, which arose from the finite accuracy in numerical routines. In particular, the violation of the Ward identity due to numerical errors that accumulated during the entire flow was of the order of the accuracy goals of the numerical routines. 
For smaller external pairing fields (for example $\sim \Delta_\text{MF} / 1000$), the Ward identity was violated noticeably below the critical scale due to numerical errors. These arose due to the very pronounced singularities at the critical scale in the reduced model for such small external pairing fields and could be cured with the coordinate projection scheme discussed below. Note that from the numerical point of view, the situation improves for flows \emph{with} fluctuations, because the singularities at the critical scale are significantly broadened in that case and numerical errors do not seem to influence the results even for the smallest (fixed) external pairing fields considered without coordinate projection ($\sim \Delta_\text{MF} / 1000$).

\subsection{Vertex and self-energy on one-loop level}
\label{subsec:AH:MomFreqOneLoop}
In this section, numerical results for the vertex and the self-energy on one-loop level are presented. Their flow as well as their dependence on momenta and frequencies are discussed, firstly for the vertex and secondly for the self-energy. The results were obtained by numerically solving the flow equations from section~\ref{subsec:CD:RGEOneLoop}. Note that negative exchange propagators give rise to attractive effective interactions within the employed sign convention.

\subsubsection{Vertex on one-loop level}
\label{subsubsec:AH:VertexOneLoop}
\begin{figure}
	\centering
	\includegraphics{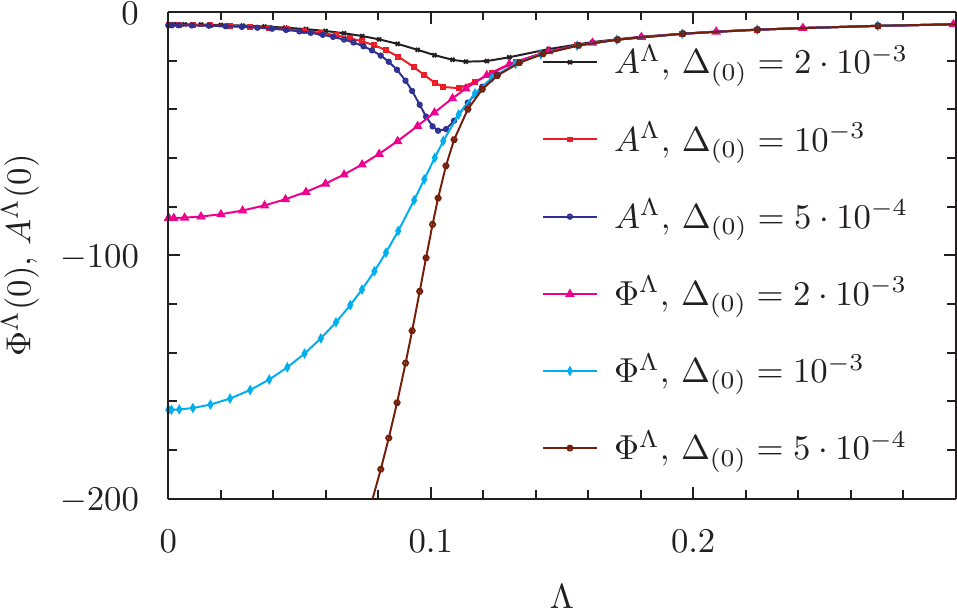}
	\caption{Scale dependence of the amplitude mode $A^\Lambda(0)$ and the phase mode $\Phi^\Lambda(0)$ of the superfluid gap for different external pairing fields, $U = -2$, $t' = -0.1$ and $n = 1/2$.}
	\label{fig:AH:APhiFlow}
\end{figure}
The flow of the singular effective interactions in the Cooper channel on one-loop level is qualitatively similar to that of the corresponding couplings in the reduced pairing and forward scattering model in chapter~\ref{chap:RPFM} or the reduced BCS model~\cite{Salmhofer2004}. In comparison to mean-field flows, the critical scale and the magnitude of the propagators are diminished by fluctuations and the singularities at the critical scale are broadened. This can be seen in figure~\ref{fig:AH:APhiFlow} that shows the flow of the amplitude and phase mode of the superfluid gap for different external pairing fields for $U=-2$ and $t'=-0.1$ at quarter-filling. The largest external pairing field in the figure, $\Delta_{(0)} = 2\cdot 10^{-3}$, corresponds to $\Delta_\text{MF} / 100$ (or roughly $\Delta^{\Lambda = 0} / 50$). 
For $U=-2$, stable numerical calculations without artificial singularities could be performed with external pairing fields that are at least three orders 
of magnitude smaller than the mean-field gap. The exchange propagator for the phase mode at the end of the flow $\Phi^{\Lambda=0}(0)$ grows inversely proportional to the external pairing field. In contrast, the exchange propagator for the amplitude mode $A^\Lambda(0)$ depends only weakly on $\Delta_{(0)}$ at the end of the flow on one-loop level. The influence of the external pairing field on the flow of $A^\Lambda(0)$ is strongest near the critical scale, where the flow of $A^\Lambda(0)$ follows $\Phi^\Lambda(0)$ the longer the smaller $\Delta_{(0)}$. The scale dependence of $A^\Lambda(0)$ and $\Phi^\Lambda(0)$ is qualitatively similar for all fillings and couplings considered.

\begin{figure}
	\centering
	\includegraphics[width=0.4\linewidth]{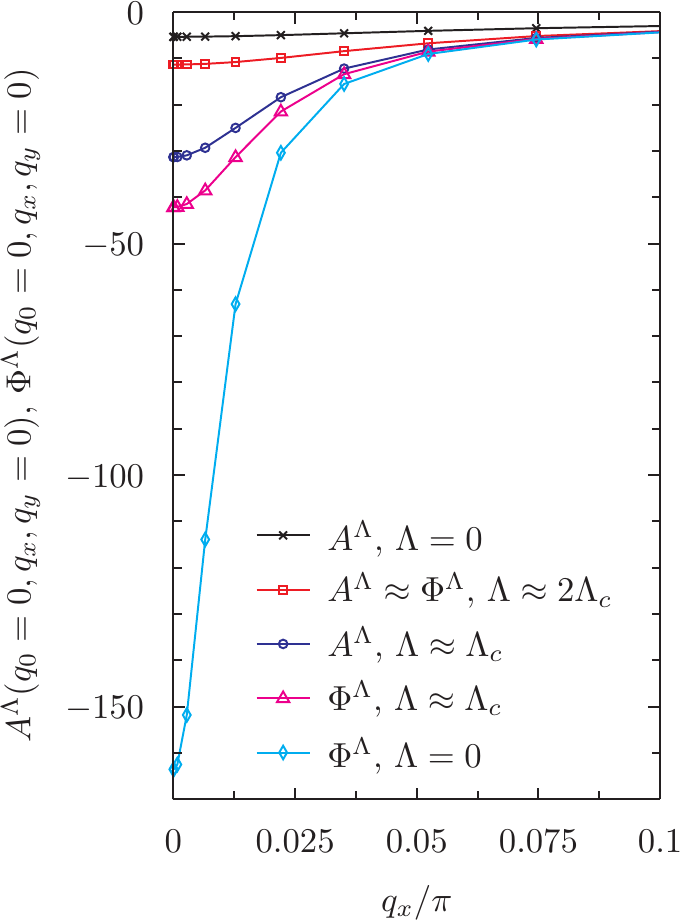}\hspace{0.02\linewidth}\includegraphics[width=0.4\linewidth]{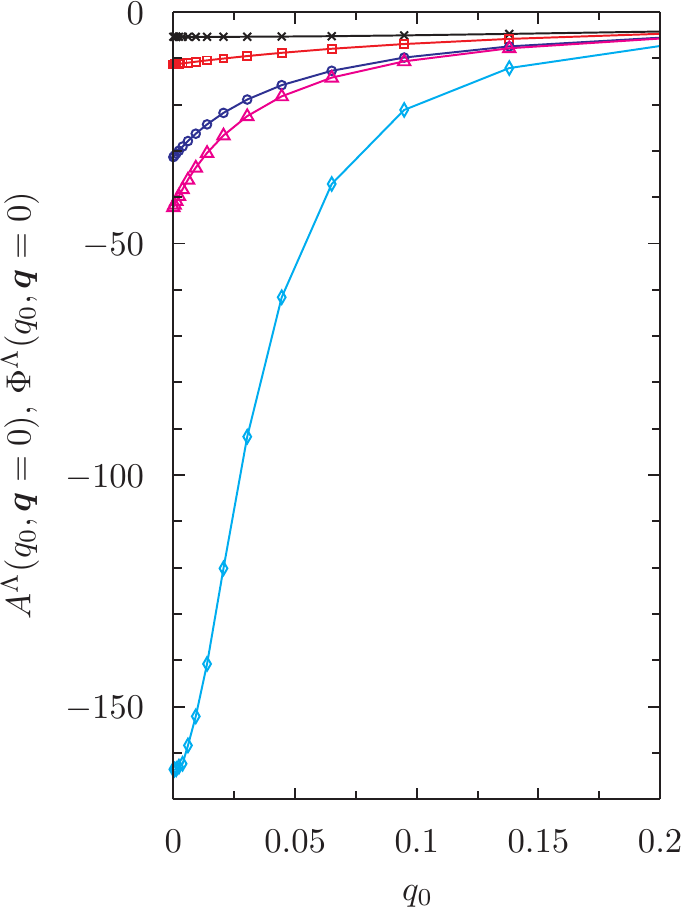}
	\caption{Momentum (left) and frequency (right) dependence of the exchange propagators for the amplitude $A^\Lambda$ and phase $\Phi^\Lambda$ mode of the superfluid gap for $U = -2$, $n = 0.5$, $t' = -0.1$ and $\Delta_{(0)} \approx \Delta_\text{MF}/200$}
	\label{fig:AH:phi_A_mom_freq}
\end{figure}
A good overall picture of the exchange propagators $A^\Lambda(q)$ and $\Phi^\Lambda(q)$ is obtained by looking at their momentum and frequency dependence for several scales $\Lambda$, as shown in figure~\ref{fig:AH:phi_A_mom_freq}. The figure shows their dependence on small transfer momenta and frequencies for $U = -2$ and $t' = -0.1$ at quarter-filling above the critical scale ($\Lambda \approx 2\Lambda_c$), at the critical scale ($\Lambda \approx \Lambda_c$) and at the end of the flow. For small momenta, the propagators are almost radially symmetric around $\boldsymbol q = \boldsymbol 0$ and to a good approximation quadratic in $|\boldsymbol q|$. 
The frequency dependence of $(A^\Lambda(q_0, \boldsymbol 0))^{-1}$ and $(\Phi^\Lambda(q_0, \boldsymbol 0))^{-1}$ at small frequencies is well described by a quadratic ansatz of the form $a^\Lambda + b^\Lambda |q_0| + c^\Lambda q_0^2$. The linear term $\sim |q_0|$ is dominant above the critical scale, as can be seen in the right panel of figure~\ref{fig:AH:phi_A_mom_freq}, and is at least in parts induced by the chosen regulator. Below the critical scale, the quadratic term becomes dominant and the coefficient of the linear term small. Depending on the parameters, at small scales the coefficient of the linear term in the quadratic fit for $(A^\Lambda(q_0, \boldsymbol 0))^{-1}$ may assume (small) negative values due to the renormalization of the amplitude mode by phase fluctuations or numerical artefacts. 
The frequency dependence of $\Phi(q_0,\boldsymbol q = \boldsymbol 0)$ at small $q_0$ is, up to small deviations at the lowest frequencies which are considered as numerical artefacts, consistent with a Lorentzian like decay. The dependence of $\Phi(q)$ on small frequencies and momenta is therefore consistent with a linear dispersion of the phase mode in the limit $\Delta_{(0)}\rightarrow 0$, as expected in a fermionic superfluid~\cite{Popov1987,Belkhir1994}. The decay of the exchange propagators at high frequencies is also quadratic, but the coefficient differs from that for small frequencies. 
Such a behaviour was also found by Husemann~\etal~\cite{Husemann2012} for the exchange propagators in the repulsive Hubbard model. $\Phi^\Lambda(q)$ reaches large values below the critical scale but decays very fast away from $q = 0$. This fast decay restricts the phase space for pairing fluctuations in the weak-coupling regime. 
Their impact is overestimated in particular for $\Phi(q)$ if the frequency dependence of the vertex is neglected.

\begin{figure}
	\centering
	\includegraphics{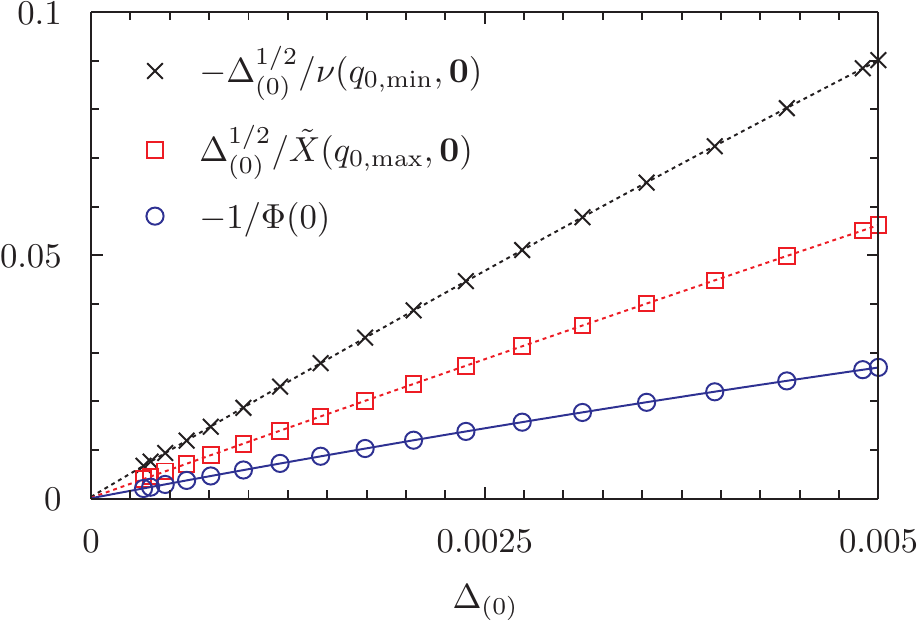}
	\caption{Singular dependence of the phase mode $\Phi$ and the imaginary part of the effective interactions in the Cooper channel $\nu$ as well as in the anomalous (3+1)-channel $\tilde X$ on the external pairing field $\Delta_{(0)}$ for $U = -2$, $t' = -0.1$ and $n = 0.5$ as obtained from a pairing field flow. The lines are quadratic fits that miss the point of origin by an amount of the order of $10^{-4}$ along the ordinate.}
	\label{fig:AH:VPHdelta0Scaling}
\end{figure}
The above-mentioned singular dependence of $\Phi$ on the external pairing field is shown in figure~\ref{fig:AH:VPHdelta0Scaling} together with the singular dependence of the imaginary parts of the normal effective interaction in the Cooper channel $\nu$ and of the anomalous (3+1)-effective interaction $\tilde X$ as obtained from a pairing field flow for $U = -2$ and $t' = -0.1$ at quarter-filling. The inverse propagators for the phase mode is proportional to the external pairing field and the imaginary parts are rescaled in such a way that a linear dependence on the external pairing field appears. The dependence of the imaginary parts on small frequencies is well captured by
\begin{equation}
	\nu(q_0,\boldsymbol 0) \sim -\frac{q_0}{\Delta_{(0)} + \alpha q_0^2}
\end{equation}
(see~\eqref{eq:RPFM:nu}) and similarly for $\tilde X$. As a function of momentum, $\nu$ and $\tilde X$ decay very rapidly away from $\boldsymbol q = \boldsymbol 0$. Determining the position of the extrema of the exchange propagators as a function of frequency and evaluating them at the extremum yields the scaling relation
\begin{equation}
	\nu(q_{0,\text{min}}, \boldsymbol 0) \sim (\Delta_{(0)})^{-1/2}\quad \text{for} \quad q_{0,\text{min}} = (\Delta_{(0)} / \alpha)^{1/2}
\end{equation}
and similarly for $\tilde X$ where $q_{0,\text{min}}$ is the position of the extremum. This scaling relation holds as long as the coefficient $\alpha$ varies weakly with $\Delta_{(0)}$. The other exchange propagators, which are discussed in the following, approach finite values for $\Delta_{(0)}\rightarrow 0$. The singular behaviour of the vertex on one-loop level in the limit of a vanishing external pairing field is therefore the same as in a resummation of all chains of Nambu particle-hole bubble diagrams (see section~\ref{sec:RPFM:RPA}).

\begin{figure}
	\centering
	\includegraphics{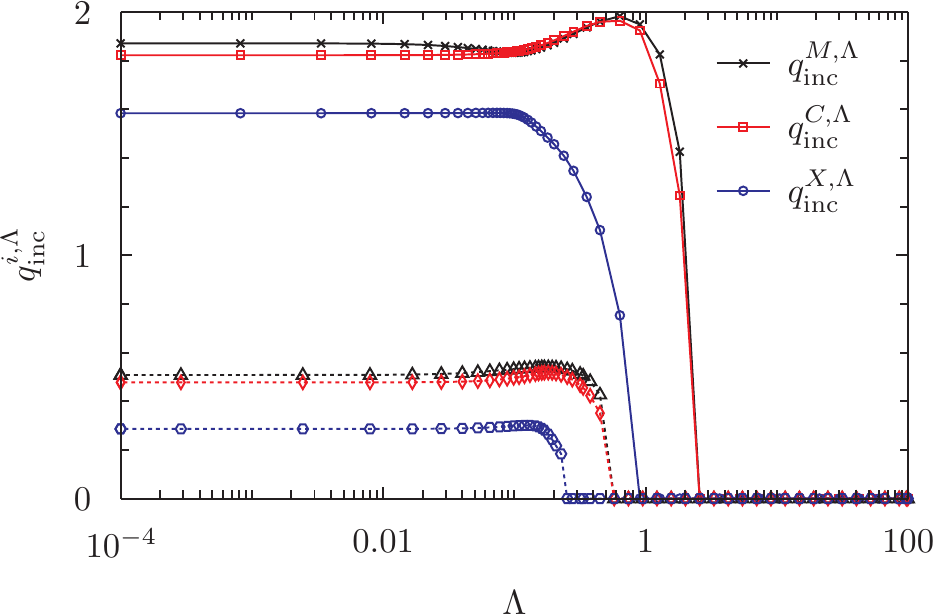}
	\caption{Scale dependence of the distance of the incommensurate peaks from $(\pi,\pi)$ along the Brillouin zone boundary in the magnetic, charge and anomalous (3+1)-density wave channels. The upper curves are for $U = -2$, $t' = -0.1$ and $n = 0.5$, the lower curves for $U = -2$, $t'=0$ and $n=0.78$.}
	\label{fig:AH:Qin_pos_flow}
\end{figure}

\begin{figure}
	\centering
	\includegraphics{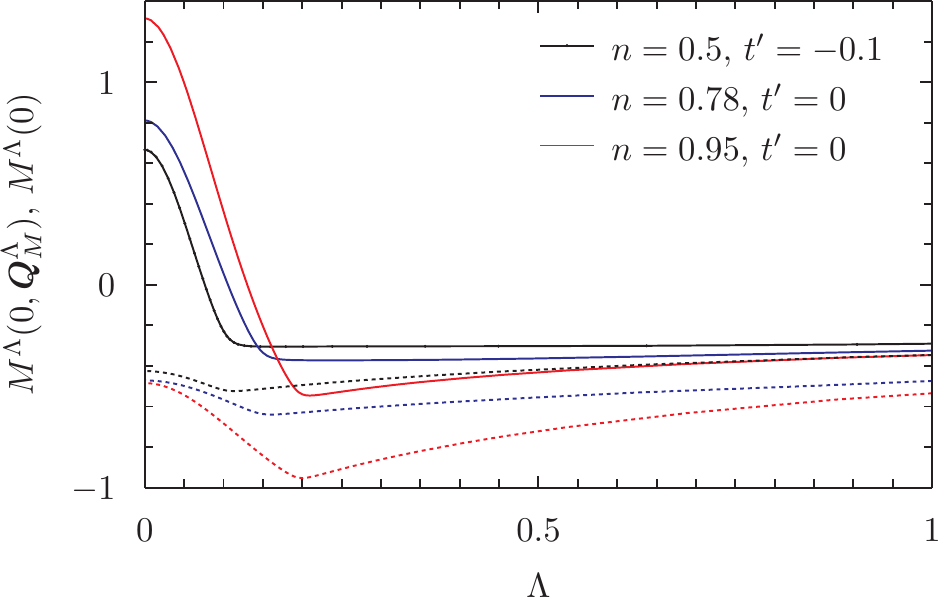}
	\caption{Scale dependence of the effective interaction for magnetic forward scattering $M^\Lambda(0)$ (full curves) and (in-) commensurate spin density wave fluctuations $M^\Lambda(0,\boldsymbol Q^\Lambda_M)$ (dashed curves) for different fermionic densities $n$ and $U = -2$.}
	\label{fig:AH:PhMFlow}
\end{figure}

\begin{figure}
	\centering
	\includegraphics[width=0.43\linewidth]{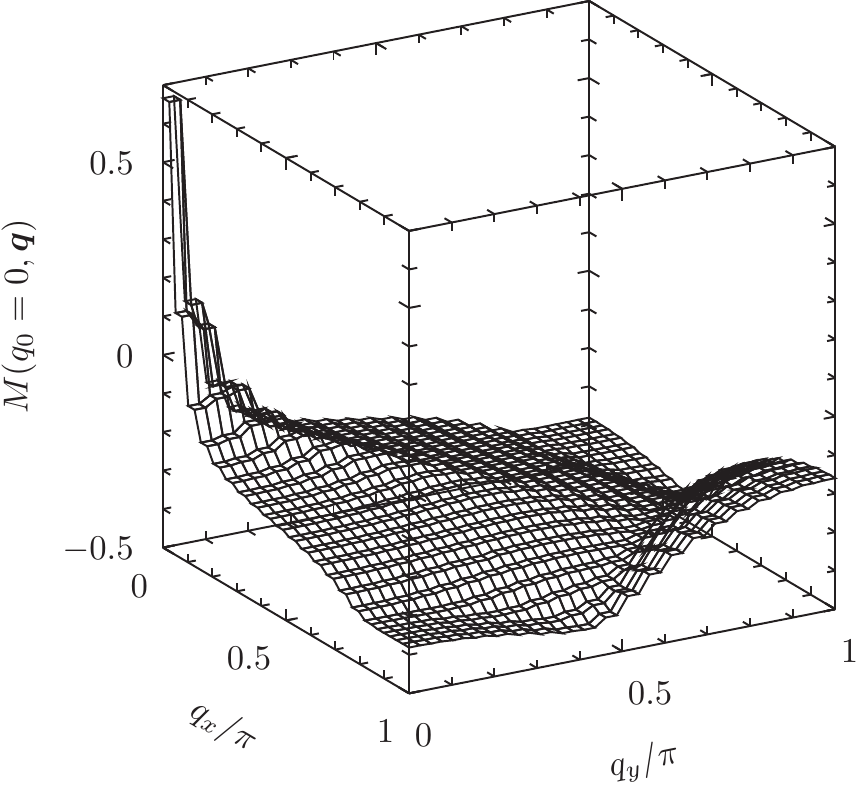}\hspace{0.02\linewidth}\includegraphics[width=0.43\linewidth]{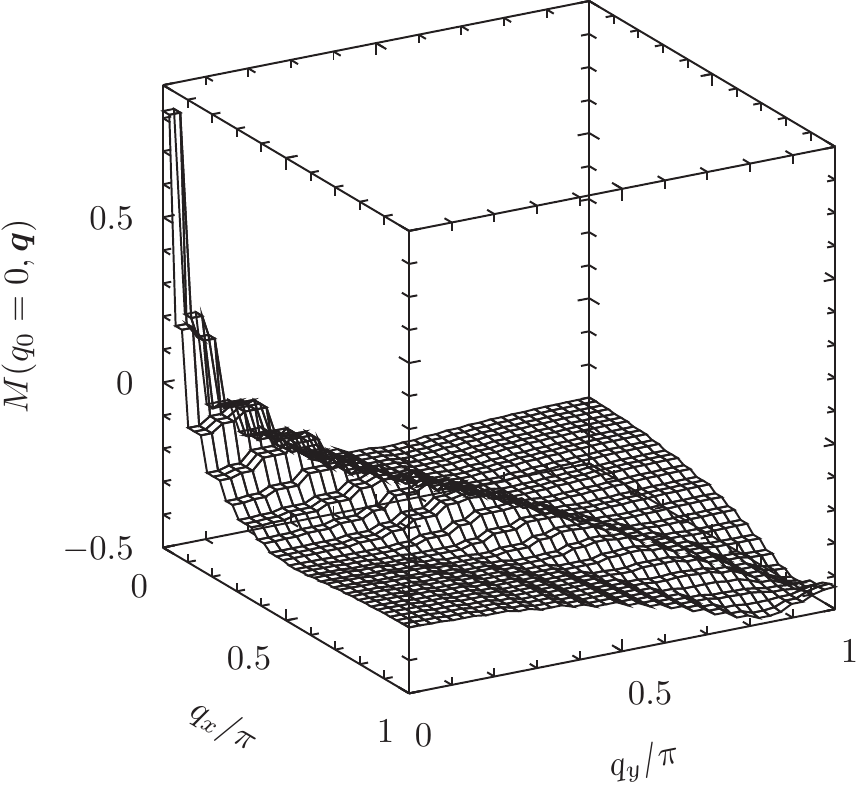}
	\caption{Momentum dependence of the exchange propagator in the magnetic channel $M(q_0 = 0, \boldsymbol q)$ for $U = -2$ at the end of the flow for $n = 0.5$, $t'=-0.1$ (left) and $n = 0.78$, $t' = 0$ (right).}
	\label{fig:AH:Mqxy}
\end{figure}

\begin{figure}
	\centering
	\includegraphics{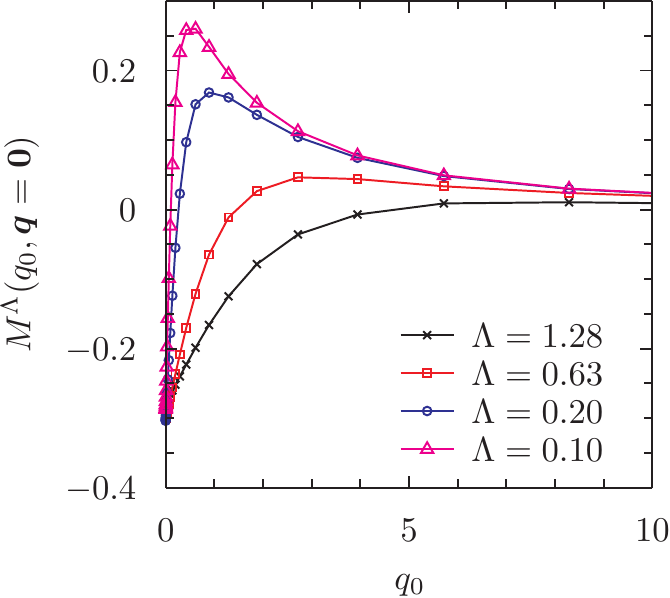} \includegraphics{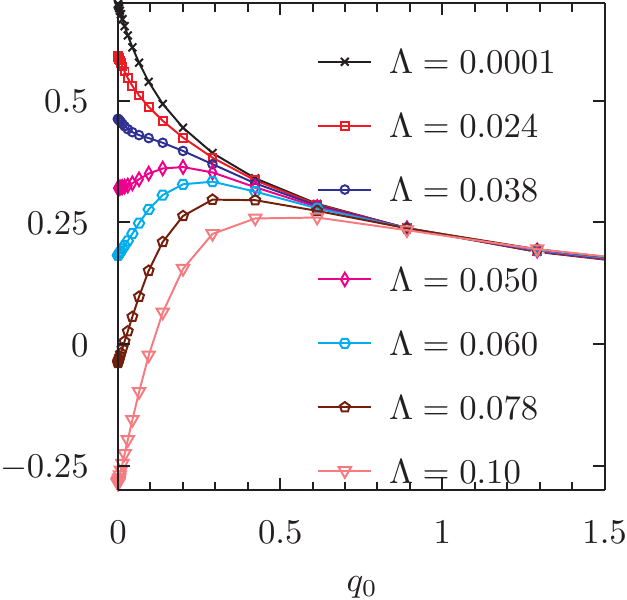}
	\caption{Frequency dependence of the effective interaction for magnetic forward scattering for different scales above (left) and below (right) the critical scale for $U=-2$, $t' = -0.1$ and $n=1/2$.}
	\label{fig:AH:PhM_FW_q0}
\end{figure}

\begin{figure}
	\centering
	\includegraphics{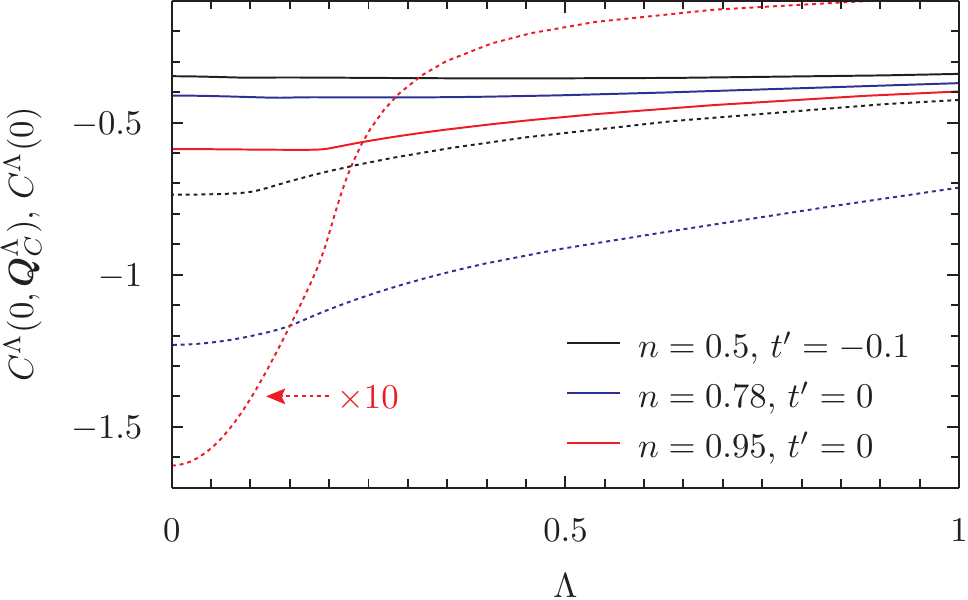}
	\caption{Scale dependence of the effective interaction for charge forward scattering $C^\Lambda(0)$ (full curves) and (incommensurate) charge density wave fluctuations $C^\Lambda(0,\boldsymbol Q^\Lambda_C)$ (dashed curves) for different fermionic densities $n$ and $U = -2$. For $n = 0.95$, $C^\Lambda(0,\boldsymbol Q^\Lambda_C)$ is divided by a factor of 10 for convenience.}
	\label{fig:AH:PhCFlow}
\end{figure}

\begin{figure}
	\centering
	\includegraphics[width=0.43\linewidth]{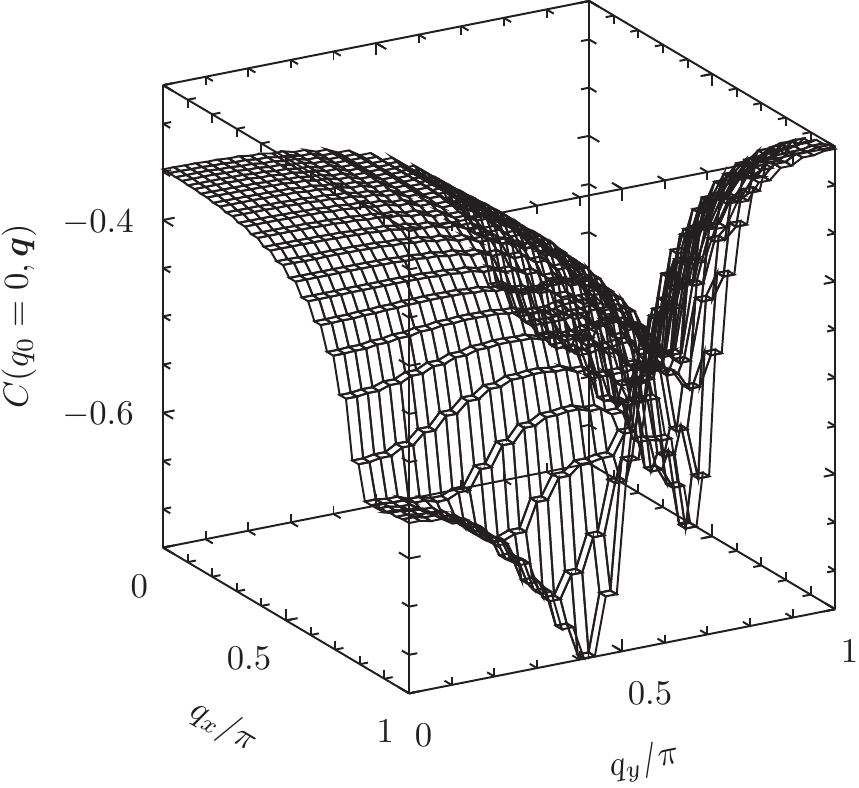}\hspace{0.02\linewidth}\includegraphics[width=0.43\linewidth]{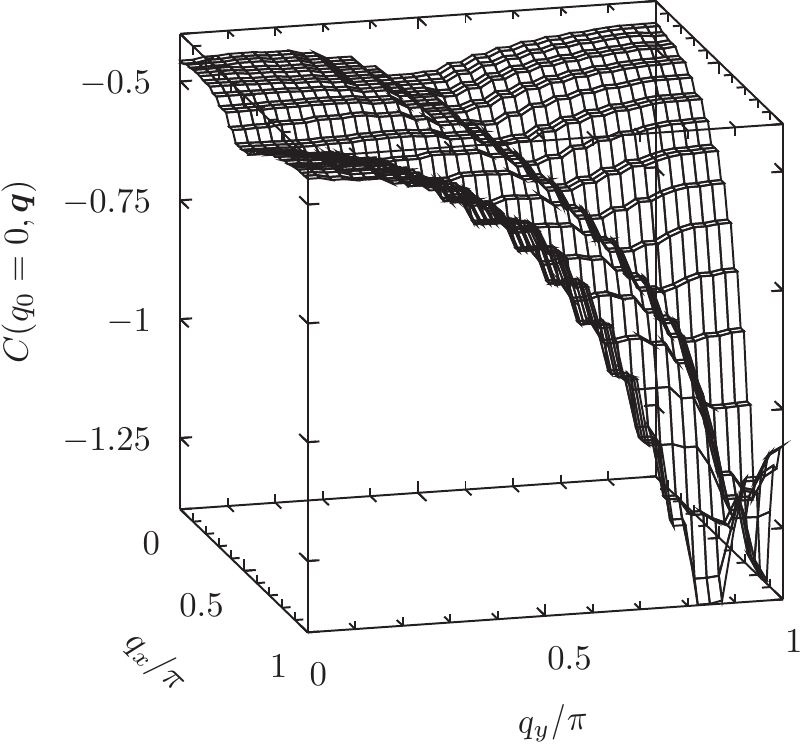}
	\caption{Momentum dependence of the exchange propagator in the charge channel $C(q_0 = 0, \boldsymbol q)$ for $U = -2$ at the end of the flow for $n = 0.5$, $t'=-0.1$ (left) and $n = 0.78$, $t' = 0$ (right).}
	\label{fig:AH:Cqxy}
\end{figure}

\begin{figure}
	\centering
	\includegraphics{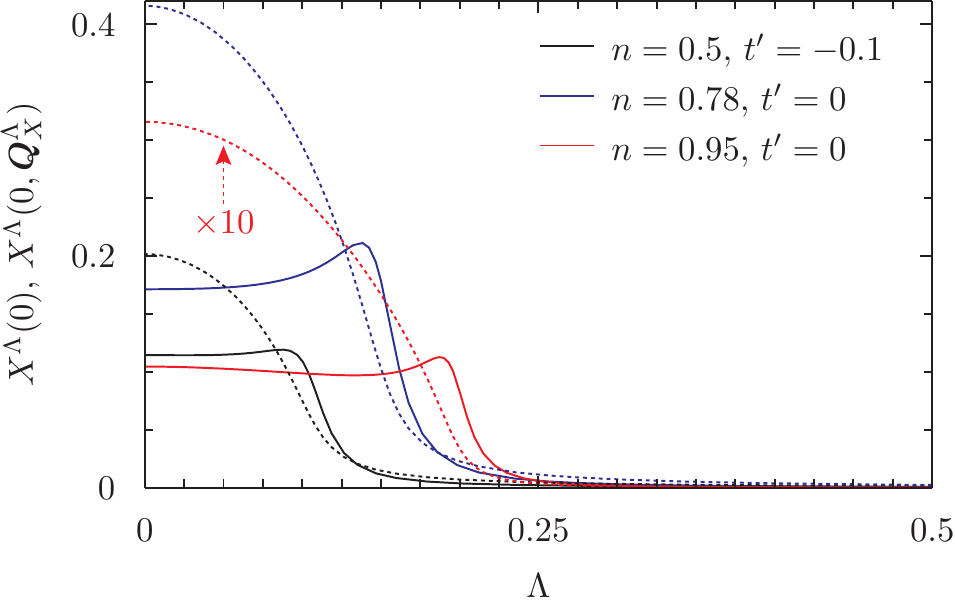}
	\caption{Scale dependence of the effective interaction for anomalous (3+1)-forward scattering $X^\Lambda(0)$ (full curves) and (incommensurate) anomalous (3+1)-density wave fluctuations $X^\Lambda(0,\boldsymbol Q^\Lambda_X)$ (dashed curves) for different fermionic densities $n$ and $U = -2$. For $n = 0.95$, $X^\Lambda(0,\boldsymbol Q^\Lambda_X)$ is divided by a factor of 10 for convenience. The external pairing fields $\Delta_{(0)}$ are $\tfrac{\Delta^\text{MF}}{200}$, $\tfrac{\Delta^\text{MF}}{150}$ and $\tfrac{\Delta^\text{MF}}{250}$ for $n = 0.5$, $0.78$ and $0.95$, respectively.}
	\label{fig:AH:XFlow}
\end{figure}

\begin{figure}
	\centering
	\includegraphics[width=0.43\linewidth]{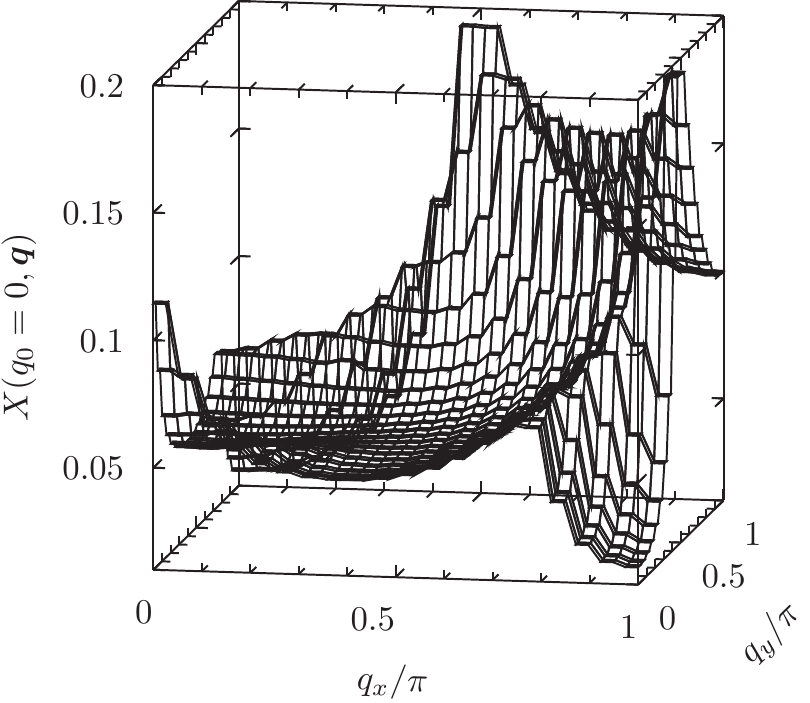}\hspace{0.02\linewidth}\includegraphics[width=0.43\linewidth]{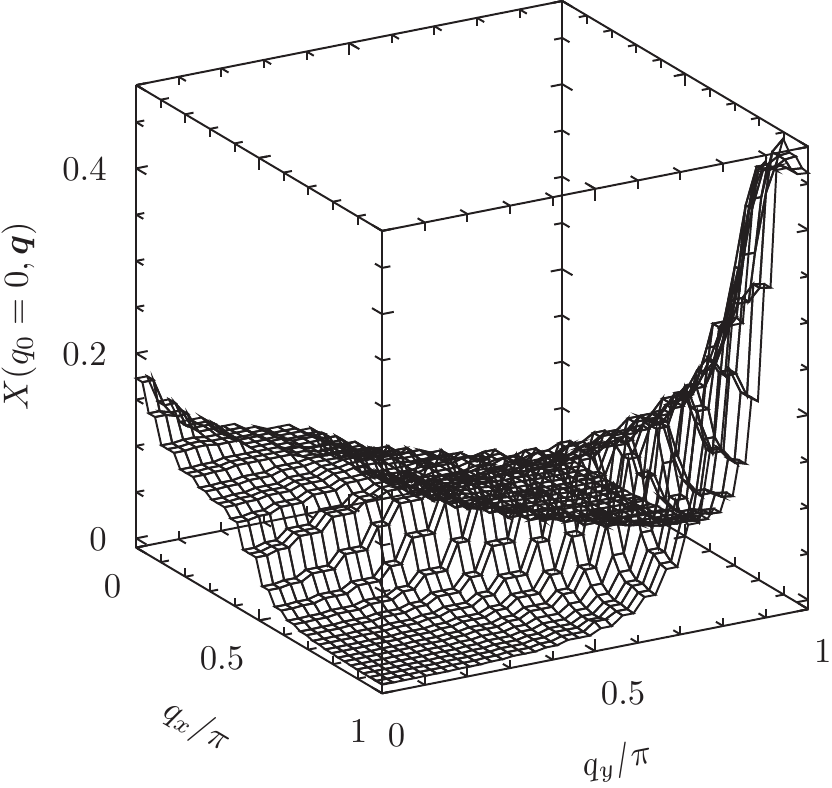}
	\caption{Momentum dependence of the anomalous (3+1)-exchange propagator $X(q_0 = 0, \boldsymbol q)$ for $U = -2$ at the end of the flow for $n = 0.5$, $t'=-0.1$ (left) and $n = 0.78$, $t' = 0$ (right).}
	\label{fig:AH:Xqxy}
\end{figure}

The scale dependence of the exchange propagators in the magnetic and density channels and for the real part of the anomalous (3+1)-effective interaction depends qualitatively stronger on $U$ or on the filling than for the Cooper channel. The flow of the couplings for forward scattering and (incommensurate) density wave fluctuations are shown in figures~\ref{fig:AH:PhMFlow},~\ref{fig:AH:PhCFlow} and~\ref{fig:AH:XFlow} for the magnetic, density and anomalous (3+1)-channel, respectively. Figures~\ref{fig:AH:Mqxy},~\ref{fig:AH:PhM_FW_q0},~\ref{fig:AH:Cqxy} and~\ref{fig:AH:Xqxy} show their momentum and frequency dependences during or at the end of the flow. 
In the density range considered in this chapter and at low scales, the exchange propagators for density wave fluctuations develop incommensurate peaks which are located on the Brillouin zone boundary. 
Their distance $q^{i,\Lambda}_\text{inc}$ from $\boldsymbol q = \boldsymbol \pi$ is shown in figure~\ref{fig:AH:Qin_pos_flow} as a function of $\Lambda$ for $n = 0.5$ and $n = 0.78$. It was determined from the extremum of the interpolated momentum dependence of the propagators and is accurate only to within the distance between grid points (roughly $\pi / N_\text{PH}$ where $N_\text{PH} \sim 25$ is the number of grid points in one direction). Note that this distance may differ for the three channels, in particular below the critical scale. 

Figure~\ref{fig:AH:PhMFlow} shows the scale dependence of the effective interaction for spin forward scattering and incommensurate spin density wave fluctuations for different fillings at $U = -2$. Above the critical scale, the fluctuation corrections to the microscopic interaction in the magnetic channel are negative (\ie\ attractive) but remain smaller than $|U|$ in absolute value. The reason is that the microscopic interaction $U$ is repulsive in the magnetic channel. Despite being weak, magnetic fluctuations play an important role in reducing the size of the effective interactions in the Cooper channel or the superfluid gap. Below the critical scale, the magnetic exchange propagator gets suppressed for small frequencies and even becomes repulsive for small $\boldsymbol q$. This sign change is driven by pairing fluctuations, because $M(q = 0)$ vanishes in RPA due to the presence of the gap. 
As can be seen in figure~\ref{fig:AH:Mqxy}, which shows the magnetic exchange propagator at the end of the flow, the peak at $q = 0$ is narrow and not very high for $U = -2$ so that it does not have a significant impact on the flow of other channels. This changes for larger couplings, where the peak gets broader and increases in height. The effective interaction for spin density wave fluctuations remains small even close to half-filling. Its incommensurate peaks move towards $\boldsymbol q = \boldsymbol \pi$ with increasing filling as expected.

While $M^\Lambda(0)$ changes sign below the critical scale, $M^\Lambda(q_0,\boldsymbol 0)$ changes sign already on higher scales for finite $q_0$. This can be seen in the left panel of figure~\ref{fig:AH:PhM_FW_q0}. Such an effect was also found by Husemann~\etal~\cite{Husemann2012} and by Giering and Salmhofer~\cite{Giering2012} in the frequency dependence of the exchange propagator for charge forward scattering in the symmetric phase of the repulsive Hubbard model within similar approximations. At van Hove filling and with the normal self-energy neglected, it may result in a scattering instability where the exchange propagator diverges at a finite transfer frequency $q_0$, thus describing a singular interaction between (imaginary) time modulated densities~\cite{Husemann2012}. 
When including the frequency dependence of the imaginary part of the normal self-energy, the scattering instability is suppressed, but the exchange propagator for the density channel may nevertheless reach large values at finite transfer frequencies for $\boldsymbol q = \boldsymbol 0$~\cite{Giering2012}. 
The sign change at finite frequencies in the magnetic exchange propagator in the attractive Hubbard model seems to be a related phenomenon. The reason is that the negative $U$ in the attractive model is repulsive in the magnetic channel and this is also the case for the positive $U$ in the density channel in the repulsive model. 
It is therefore interesting to follow the evolution of the frequency dependence of the magnetic exchange propagator for $\boldsymbol q = \boldsymbol 0$ below the critical scale in the attractive model, which was not possible in the study by Husemann~\etal\ or by Giering and Salmhofer for the propagator in the density channel of the repulsive model. 
The scale dependence of $M^\Lambda(q_0, \boldsymbol 0)$ for $\Lambda < \Lambda_c$ is shown in the right panel of figure~\ref{fig:AH:PhM_FW_q0}, where $M^\Lambda(q_0, \boldsymbol 0)$ remains relatively small in value because the system is away from van Hove filling. Below the critical scale, the opening of the superfluid gap and pairing fluctuations push the low-frequency part of the exchange propagator to positive values. At the end of the flow, $M(q_0, \boldsymbol 0)$ is repulsive for all frequencies $q_0$ and its extremum is located at $q_0 = 0$.

The exchange propagator $C^\Lambda(q)$ gets enhanced during the flow due to the resummation of chains of particle-hole bubbles in comparison to second order perturbation theory, because of the negative $U$ being attractive in the density channel. This enhancement is however small away from half-filling, as can be seen by comparing figure~\ref{fig:AH:PhCFlow} that shows the flow of the couplings for density forward scattering and charge density wave fluctuations to figure~\ref{fig:AH:PhMFlow} for the magnetic channel or in figure~\ref{fig:AH:Cqxy} that shows the momentum dependence of the exchange propagator in the density channel at the end of the flow for $U = -2$ and $n = 0.5$ or $n = 0.78$. For these couplings and fermionic densities, the exchange propagator remains smaller than the microscopic interaction in absolute value. 
This changes very close to half-filling, where $U + 2 C^\Lambda(0,\boldsymbol \pi)$ becomes comparable to $U + P^\Lambda(0, \boldsymbol 0)$ above the critical scale, but never exceeds it. At half-filling and for $t' = 0$, both interaction channels become degenerate~\cite{Micnas1990}. 

The exchange propagator for the real part of the anomalous (3+1)-channel $X^\Lambda(q)$ remains smaller than that in the density channel on all scales for all couplings, densities and external pairing fields considered. In comparison to a resummation of all chains of Nambu particle-hole bubble diagrams, it is suppressed and its singularity at the critical scale considerably broadened by fluctuations, in particular at $q = 0$. 
Nevertheless it has to be taken into account in order to avoid artefacts like non-monotonic flows of $\Phi^\Lambda(0)$ even for small interactions $U$. While the (incommensurate) peaks in $X(0, \boldsymbol q)$ grow when approaching half-filling, $X(0)$ first increases and then starts to decrease with increasing filling. A similar behaviour is obtained in random-phase approximation (as described in chapter~\ref{chap:RPFM}), where in particular $X(0)$ vanishes in the presence of particle-hole symmetry.
The real part of the anomalous (3+1)-effective interaction $X(q)$ depends only weakly on the external pairing field. This is similar to the findings for the reduced pairing and forward scattering model in chapter~\ref{chap:RPFM}, but contradicts results by Gersch~\cite{Gersch2007} for the attractive Hubbard model, who finds a weak divergence for $\Delta_{(0)} \rightarrow 0$.

\begin{figure}
	\centering
	\includegraphics[scale=.75]{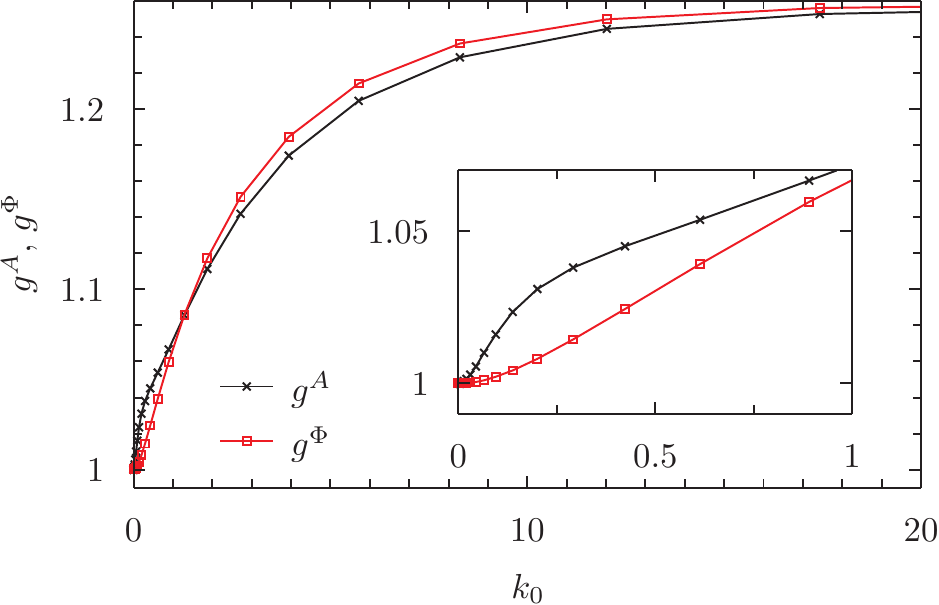}
	\includegraphics[scale=.75]{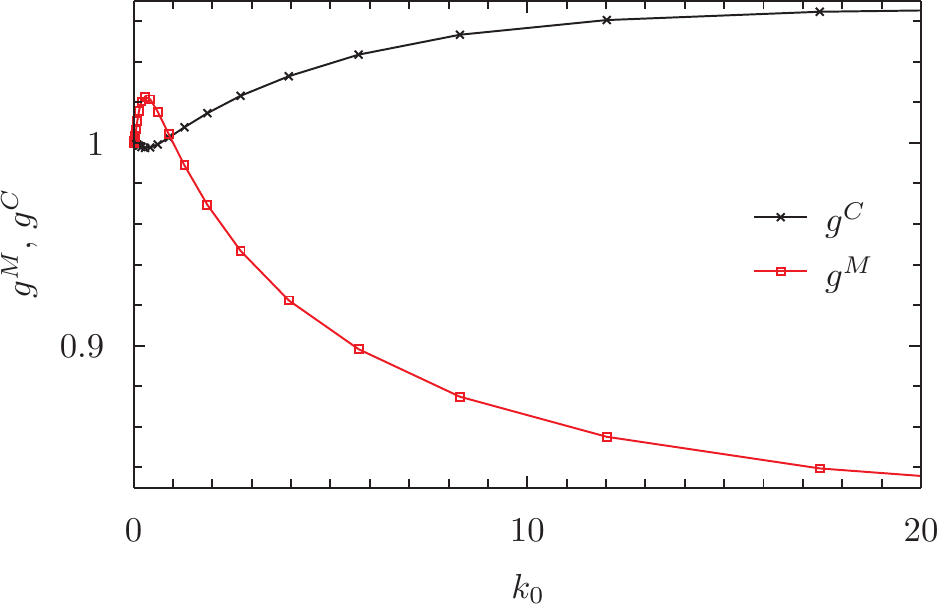}
	\caption{Frequency dependence of the fermion-boson vertices in the Cooper channel (left) and in the particle-hole channel (right) for $U = -2$, $t' = -0.1$ and $n = 1/2$.}
	\label{fig:AH:gFB}
\end{figure}
The exchange propagators are coupled to the fermions via fermion-boson vertices whose frequency dependence is renormalized during the flow. It is shown for $\Lambda = 0$, $U=-2$, $t'=-0.1$ and quarter-filling in figure~\ref{fig:AH:gFB}. The fermion-boson vertices are even functions of the fermionic frequency $k_0$ within the employed approximations and normalized to one at $k_0 = 0$. Their frequency dependence is similar to that found by Husemann~\etal~\cite{Husemann2012} for the repulsive Hubbard model, albeit with repulsive and attractive channels exchanged due to the different sign of the microscopic interaction $U$. The fermion-boson vertices for the Cooper channel $g^A$ and $g^\Phi$ show a similar frequency dependence with a monotonic increase at low and intermediate frequencies and a saturation at around the value for the highest frequency in the plot towards high frequencies. $g^A$ increases somewhat faster for small frequencies, as shown in the inset. 
$g^A$ and $g^\Phi$ differ stronger with increasing $U$. 
The fermion-boson vertex for the density channel $g^C$ shows a small dip at low frequencies and increases towards high frequencies, where it saturates at roughly the value for the highest frequency in the plot. The fermion-boson vertex for the magnetic channel $g^M$ shows a small peak at low frequencies and decays towards high frequencies to a value that roughly equals that at the highest frequency in the plot. The small peak appears only below the critical scale and may be related to the suppression of the incommensurate peak of the magnetic exchange propagator at small transfer frequencies. 
Except for this peak in $g^M$, the fermion-boson vertices show a qualitatively similar frequency dependence on all scales. The renormalization of the fermion-boson vertices in particular in the Cooper channel has a strong impact on the frequency dependence of $\Delta(k_0)$ (see below), but not on its value at $k_0 = 0$. It has also some impact on the flow of $\Phi^\Lambda$ and thereby on the Ward identity fulfilment (see subsection~\ref{subsec:AH:WI}). The other self-energies and exchange propagators are only weakly influenced by the frequency dependence of the fermion-boson vertices. This is similar to the findings by Husemann~\etal~\cite{Husemann2012} for the repulsive Hubbard model in the symmetric state.

\subsubsection{Self-energy on one-loop level}
\label{subsec:AH:SelfenergyOneLoop}
\begin{figure}
	\centering
	\includegraphics{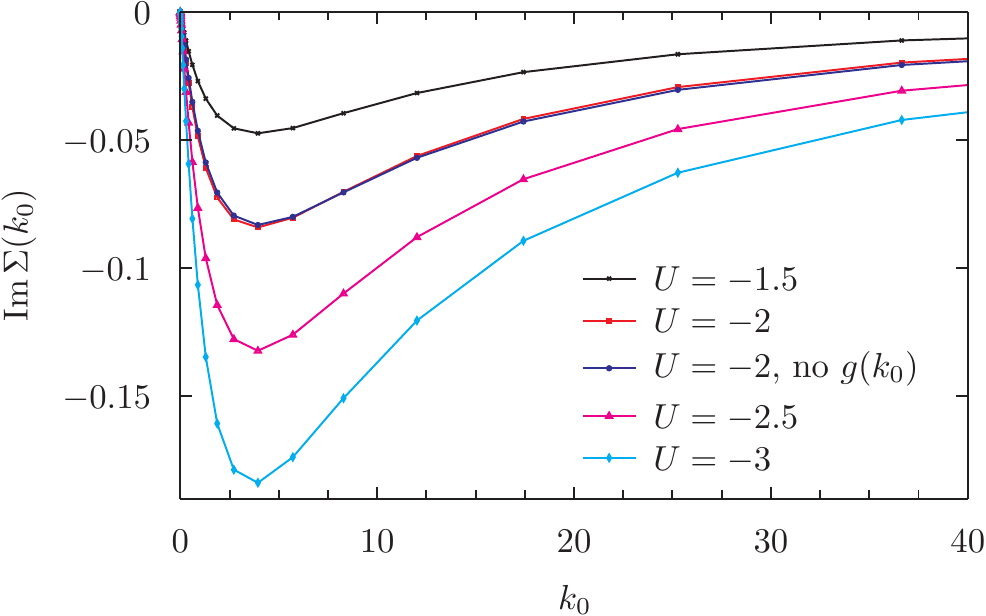}
	\caption{Frequency dependence of the imaginary part of the normal self-energy $\im \Sigma(k_0)$ for different interactions $U$ for $n = 1/2$ and $t' = -0.1$. The result labelled ``no $g(k_0)$'' was obtained without frequency dependent renormalization of the fermion-boson vertices.}
	\label{fig:AH:ImSigma_k0}
\end{figure}
\begin{figure}
	\centering
	\includegraphics{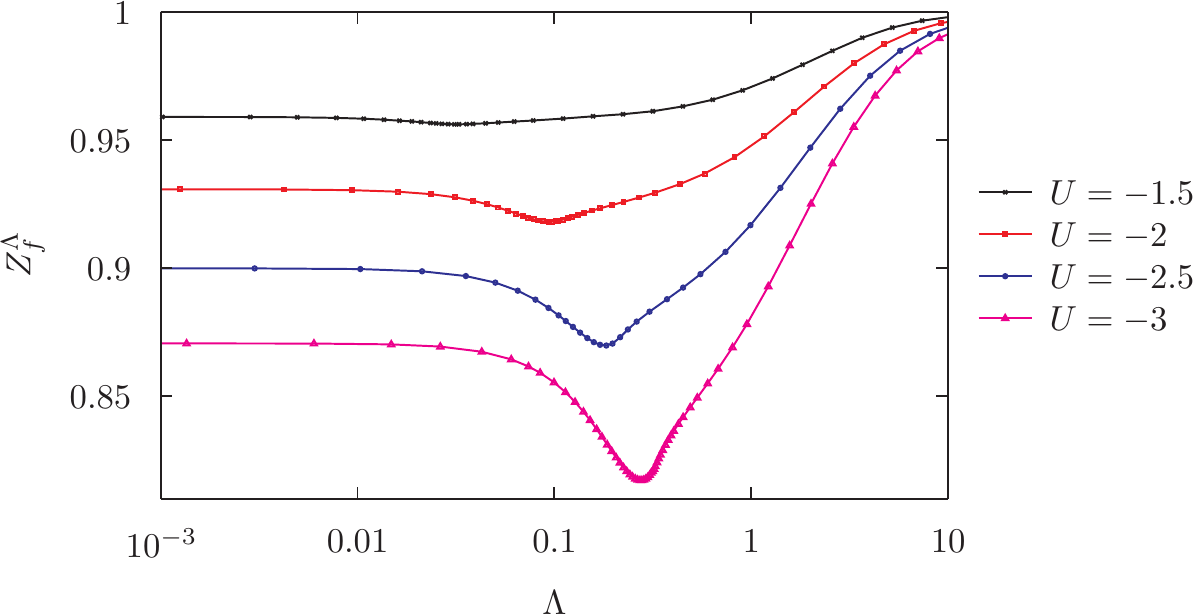}
	\caption{Scale dependence of the fermionic quasi-particle weight $Z^\Lambda_f$ for different interactions $U$ for $n = 1/2$ and $t' = -0.1$.}
	\label{fig:AH:Zf_flow}
\end{figure}
In the parameter range that is considered in this chapter, the ground state of the attractive Hubbard model is a fermionic superfluid formed out of weakly renormalized fermionic quasiparticles. The renormalization of the quasiparticles occurs already at high and intermediate scales and is therefore discussed before the anomalous self-energy, which is generated at lower scales. Figure~\ref{fig:AH:ImSigma_k0} and~\ref{fig:AH:Zf_flow} show the imaginary part of the normal self-energy $\im \Sigma^\Lambda(k_0)$ (which is an odd function in $k_0$) for different interactions $U$ at $\Lambda = 0$ and the flow of the fermionic quasiparticle weight at the Fermi energy, respectively. The latter is determined from $\im \Sigma^\Lambda(k_0)$ through
\begin{equation}
	Z^\Lambda_f = \bigl(1 - \partial_{k_0} \im \Sigma^\Lambda(k_0)\bigr)^{-1} \bigr|_{k_0 = 0},
\end{equation}
where the derivative is approximated by finite differences of the flowing self-energy at the lowest frequencies. At higher frequencies, the curves in figure~\ref{fig:AH:ImSigma_k0} almost coincide after rescaling with $U^{-2}$. At the lowest frequencies, the dependence of $\im \Sigma(k_0)$ on $U$ is not quadratic, as can be seen in figure~\ref{fig:AH:Zf_flow} for the quasiparticle weight. The results for $U = -2$ in figure~\ref{fig:AH:ImSigma_k0} show that the frequency-dependent renormalization of the fermion-boson vertices has almost no impact on $\im \Sigma^\Lambda(k_0)$. 
The minima of $Z^\Lambda_f$ in figure~\ref{fig:AH:Zf_flow} occur somewhat below the critical scale and are expected to be a consequence of using an additive frequency regulator. They arise from a small reduction of $\im \Sigma^\Lambda(k_0)$ below the critical scale that may be a consequence of partially taking into account the renormalization of the self-energy by low energy modes below the gap already on higher scales. 
Despite the imaginary part of the normal self-energy being small and the quasiparticle weight at the Fermi energy being only slightly suppressed, the renormalization of the quasiparticles has a significant impact on the size of the superfluid gap even for weak interactions (see subsection~\ref{subsec:AH:GapRed}).

The imaginary part of the normal self-energy or the quasi-particle weight become anisotropic for non-circular Fermi surfaces. In order to justify the neglect of the momentum dependence of $\im \Sigma^\Lambda$, the renormalization of $Z^\Lambda_f$ has been evaluated for different momenta on the Fermi surface using the momentum-independent self-energy in the fermionic propagators. This yields information on the anisotropy that would arise from the momentum dependence of the Fermi velocity or the low-energy effective interaction. The result is that the anisotropy stays small at quarter-filling in the range of couplings considered. Even for $U=-2.5$, $t'=0$ and $n = 0.95$ (corresponding to $\mu = -0.11$), the anisotropy between the quasiparticle weights at the Fermi points $(k_{F,x}, 0)$ and $(k_{F,xy},k_{F,xy})$ remains smaller than $6\%$, thus justifying the neglect of the momentum dependence of $\im \Sigma^\Lambda$. 
For larger couplings and closer to half- or van Hove-filling, the anisotropy grows and it may become more appropriate to evaluate the self-energy at the Fermi point with the smallest Fermi velocity.

The real part of the normal self-energy (not shown) has only a minor influence on the results. The relevant combination $\delta\xi + \re \Sigma(k_0)$, which appears in the fermionic propagator, vanishes for $k_0 = 0$ by definition, grows quadratically for small $k_0$ and saturates at a positive value towards high frequencies that is of the order of the maximum of $|\im \Sigma(k_0)|$. After rescaling with $U^{-2}$, the frequency dependence of $\delta\xi + \re \Sigma(k_0)$ for different values of $U$ coincide to a good approximation for all frequencies. Note that $\delta\xi + \re \Sigma(k_0)$ is irrelevant in the sense of power counting due to its quadratic dependence on $k_0$ for small frequencies.

\begin{figure}
	\centering
	\includegraphics{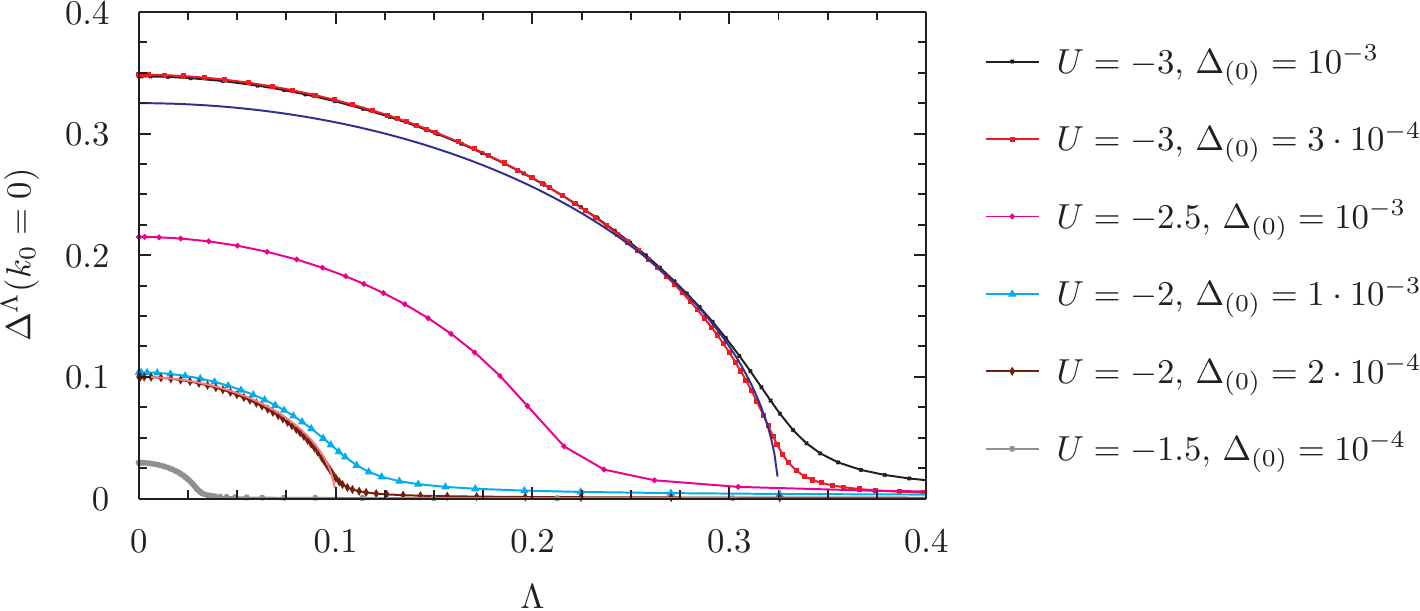}
	\caption{Scale dependence of the superfluid gap $\Delta^\Lambda(k_0 = 0)$ for different interactions $U$ for $n = 1/2$ and $t' = -0.1$. The flows with the smaller external pairing fields for $U = -2$ and $-3$ are compared to the mean-field like scale dependence $\Delta^\Lambda(0) = \sqrt{\Lambda_c^2 - \Lambda^2}$ (lines without points).}
	\label{fig:AH:DeltaFlowU}
\end{figure}
At lower scales, the large effective interaction in the Cooper channel promotes the gap from values of the order of the small external pairing field to sizeable values. This is shown in figure~\ref{fig:AH:DeltaFlowU} for different interactions $U$. For the chosen cutoff and in the weak-coupling regime, the gap at the end of the flow $\Delta(k_0 = 0)$ roughly equals the critical scale $\Lambda_c$ as determined from the maximum of $|A^\Lambda(0)|$ as a function of $\Lambda$,
\begin{equation} 
	\Delta^{\Lambda = 0}(k_0 = 0) \approx \Lambda_c.
	\label{eq:AH:Gap:CriticalScale}
\end{equation}
Furthermore, for $\Lambda < \Lambda_c$ and small external pairing fields, the scale dependence of the gap roughly follows
\begin{equation}
	\Delta^\Lambda(k_0 = 0) \approx \sqrt{(\Delta^{\Lambda=0}(k_0 = 0))^2 - \Lambda^2} \approx \sqrt{\Lambda^2_c - \Lambda^2}.
	\label{eq:AH:Gap:ScaleDependenceMF}
\end{equation}
On the mean-field level in the weak-coupling regime (where the change of the normal self-energy below the critical scale is negligible), these relations hold for the chosen regulator and for any fermionic dispersion or density of states as equalities. They can be read off immediately from the mean-field equation for $\Delta_{(0)} \rightarrow 0$,
\begin{equation}
	\Delta^\Lambda = -U \intdrei{p} F^\Lambda(p) = -U \intdrei{p} \frac{\Delta^\Lambda}{p_0^2 + \Lambda^2 + \xi(\boldsymbol p)^2 + {\Delta^\Lambda}^2},
\label{eq:AH:Gap:Lambda}
\end{equation}
after dividing both sides through $\Delta^\Lambda$. The only $\Lambda$-dependence of the right hand side then arises from the sum $\Lambda^2 + {\Delta^\Lambda}^2$ in the denominator, which thus has to be constant below the critical scale. Noting that $\Delta^{\Lambda_c} = 0$, the constant can be fixed for $\Lambda = 0$ or $\Lambda = \Lambda_c$, yielding $\Lambda^2 + {\Delta^\Lambda}^2 = (\Delta^{\Lambda = 0})^2 = \Lambda^2_c$ and therefore
\begin{equation}
	\Delta^\Lambda = \pm \sqrt{\Lambda_c^2 - \Lambda^2}
\end{equation}
for $\Lambda < \Lambda_c$. 
For the additive frequency regulator and on mean-field level, this result implies the singular behaviour $\Delta^{\Lambda \lesssim \Lambda_c} \sim \sqrt{\Lambda_c - \Lambda}$ for $\Lambda \lesssim \Lambda_c$, which is the same as for a sharp momentum~\cite{Salmhofer2004} or a sharp frequency~\cite{Strack2008} cutoff. Note however that the behaviour for $\Lambda \rightarrow 0$ differs for the additive or multiplicative cutoffs. The numerical results for the smallest external pairing fields indicate that the singular behaviour of $\Delta^\Lambda$ for $\Lambda \lesssim \Lambda_c$ on one-loop level is the same as in mean-field approximation at least for $|U| \lesssim 2$. 
In the weak-coupling regime and for small $\Delta_{(0)}$, the scale dependence of the gap is excellently described by~\eqref{eq:AH:Gap:ScaleDependenceMF}, and~\eqref{eq:AH:Gap:CriticalScale} holds to a very good approximation. 
The validity of the mean-field relation between the gap and the critical scale on one-loop level was also noted by Gersch~\etal~\cite{Gersch2008}. However, for other cutoffs than the one employed in this chapter, the equality~\eqref{eq:AH:Gap:CriticalScale} holds only up to a factor of one (see for example~\cite{Salmhofer2004}). For larger couplings, the scale dependence of the gap deviates increasingly from the mean-field dependence and the gap seems to increase slower just below the critical scale but faster for $\Lambda \rightarrow 0$. This could be caused by phase fluctuations that tend to suppress the amplitude mode below the critical scale on one-loop level, which in turn slows down the growth of $\Delta^\Lambda$, and to enhance the fermionic gap in the flow equation for the anomalous self-energy. The latter was also found by Strack~\cite{Strack2009}, who noted that Goldstone fluctuations enhance the fermionic gap. 
The good agreement in the weak-coupling regime may be a consequence of an approximate ``renormalized mean-field structure'' in the flow equations. In this regime, pairing fluctuations are expected to have only a small impact on the size of the gap due to phase-space restrictions that arise from the fast decay of the corresponding propagators in momentum and frequency space. 
Furthermore, the Fermi surface average of the effective interactions in the particle-hole channel flows only relatively slowly below $\Lambda_c$. This indicates that the dominant fluctuation contributions in the flow equation for $\Delta^\Lambda$ and $A^\Lambda(0)$ could roughly be absorbed into a weakly scale-dependent effective $U$. The flow then becomes similar to that of a renormalized mean-field model. This picture becomes more accurate with decreasing $U$.

\begin{figure}
	\centering
	\includegraphics{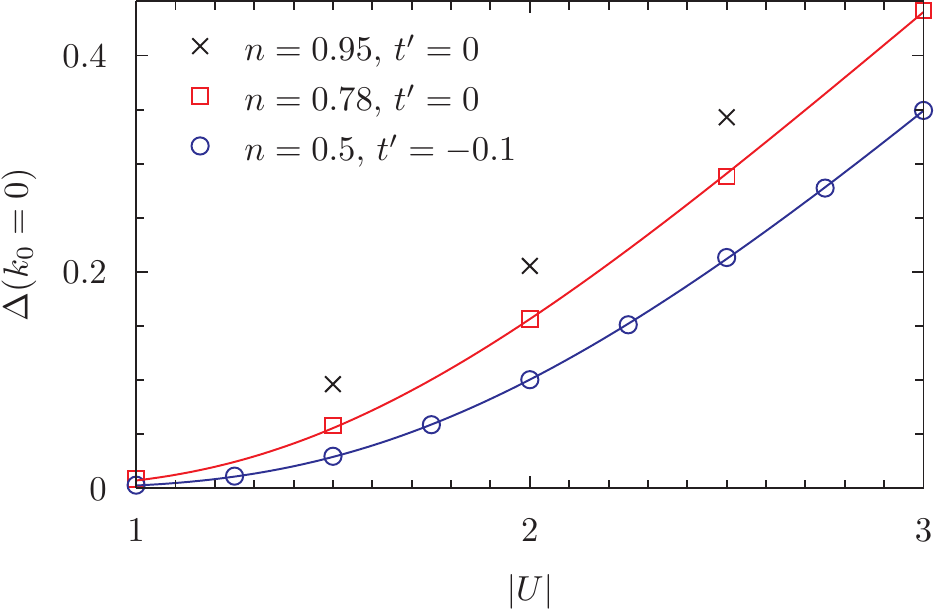}
	\caption{Dependence of the superfluid gap $\Delta(k_0 = 0)$ on the microscopic interaction $U$ for $\Lambda = 0$ and $\Delta_{(0)} \rightarrow 0$ (from quadratic extrapolations of pairing field flows). The lines for the two lower densities are fits to $\Delta = a \exp(-b / |U|)$.}
	\label{fig:AH:Delta_U}
\end{figure}
Despite the increasing deviations of the scale dependence of $\Delta^\Lambda(0)$ from the mean-field result with increasing $|U|$, the gap at the end of the flow follows the weak-coupling $U$-dependence 
\begin{equation}
	\Delta = a \exp(-b / |U|)
	\label{eq:AH:DeltaSmallU}
\end{equation}
to a good approximation in the coupling range considered. This can be seen in figure~\ref{fig:AH:Delta_U} that shows $\Delta^{\Lambda = 0}(0)$ for different interactions $U$ together with fits to~\eqref{eq:AH:DeltaSmallU}. Note that the prefactor $a$ is renormalized in comparison to the mean-field result. This is discussed in section~\ref{subsec:AH:GapRed}, or in appendix~\ref{sec:Appendix:UToZero} for the limit $|U|\rightarrow 0$. The good agreement implies that the attractive Hubbard model is well inside the BCS regime in the coupling range considered, in agreement with other methods~\cite{Bauer2009,Tamura2012}.

\begin{figure}
	\centering
	\includegraphics{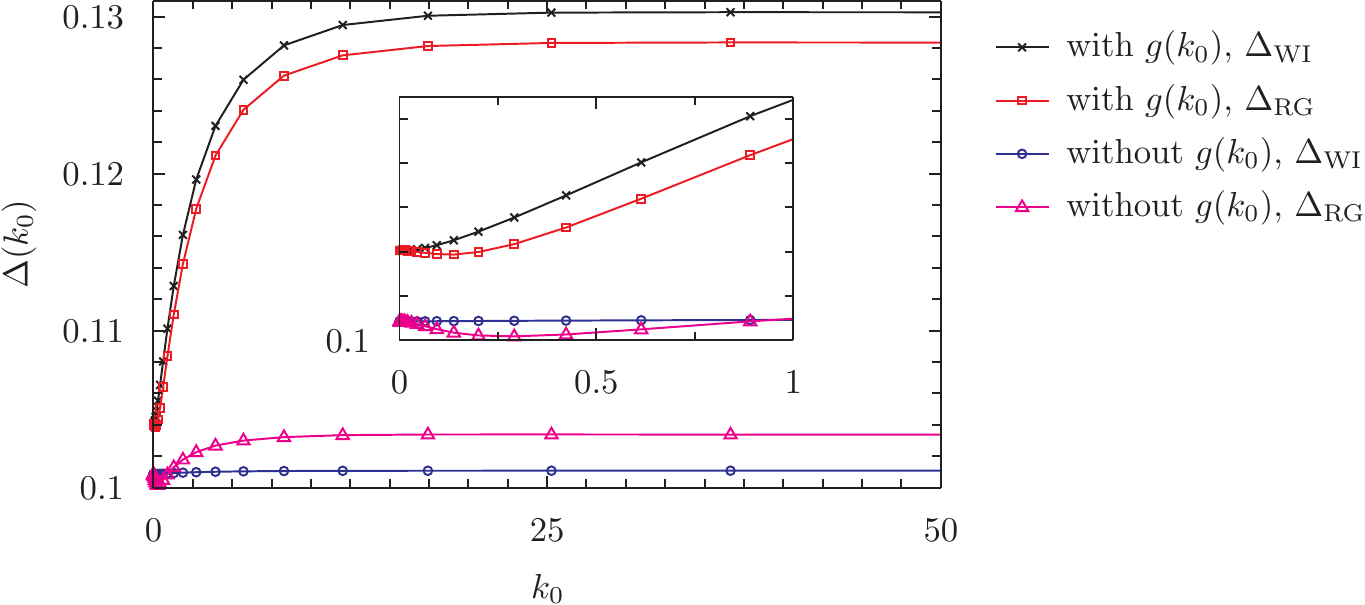}
	\caption{Comparison between the frequency dependence of the anomalous self-energy $\Delta(k_0)$ as computed with fermion-boson vertices (``with $g(k_0)$'') and without (``without $g(k_0)$'') for $U = -2$, $t' = -0.1$ and $n = 1/2$. ``$\Delta_\text{RG}$'' labels the frequency dependence as obtained from the flow equations, while ``$\Delta_\text{WI}$'' labels results that were obtained by inserting the vertex and the normal self-energy in the Ward identity.}
	\label{fig:AH:Delta_k0_gFB}
\end{figure}
In contrast to the frequency dependence of the normal self-energy, the frequency dependence of the anomalous self-energy is strongly affected by the renormalization of the fermion-boson vertices. This can be seen in figure~\ref{fig:AH:Delta_k0_gFB} that shows $\Delta(k_0)$ in case the renormalization of the fermion-boson vertices is taken into account (``with $g(k_0)$'') or neglected (``without $g(k_0)$''). Results labelled ``$\Delta_\text{RG}$'' were obtained from the renormalization group equation while those labelled ``$\Delta_\text{WI}$'' were computed from the Ward identity at the end of the flow (using the normal self-energy and the vertex from the RG). 
Obviously, a reasonable frequency dependence of $\Delta(k_0)$ is only obtained in case the frequency dependence of the fermion-boson vertices is taken into account. This can be understood from the Ward identity for global charge conservation in the limit of a small (momentum and frequency independent) external pairing field,
\begin{equation}
	\Delta^\Lambda(k) = \Delta_{(0)} \Phi^\Lambda(0) h^{\Phi,\Lambda}(0,k) \intdrei{p} h^{\Phi,\Lambda}(0,p) L^\Lambda_{22}(p,0) + \ldots,
\end{equation}
where the ellipsis represents terms that vanish for $\Delta_{(0)}\rightarrow 0$ (see equation~\eqref{eq:AH:WardIdentityFull} for the Ward identity including all contributions that appear within the approximations employed in this chapter). Within the approximation that effective interactions are decomposed into bosonic exchange propagators and fermion-boson vertices, the dependence of the right hand side on $k$ in the limit $\Delta_{(0)} \rightarrow 0$ stems solely from the fermion-boson vertex. This finding can also be understood from the structure of the flow equation for the anomalous self-energy: The strong increase of the gap at the critical scale is caused by the (regularized) singularity of $A^\Lambda(0)$. 
Neglecting the fermion-boson vertices, the contribution in the flow equation involving $A^\Lambda(0)$ would give rise to a momentum and frequency independent gap and a dependence on $k$ would only arise from the fluctuation contributions. These are, however, non-singular and much smaller than the leading term proportional to $A^\Lambda(0)$ due to phase space restrictions as well as the small size of the effective interactions in the particle-hole channel. It has been checked at quarter-filling that only an indeed tiny momentum dependence of the superfluid gap along the Fermi surface would be generated within the employed approximations. The latter was also found in the $N$-patch study by Gersch~\etal~\cite{Gersch2008}.

\subsection{Ward identity for global charge conservation}
\label{subsec:AH:WI}
In section~\ref{sec:WICharge:Incomp} evidence is provided that the Ward identity for global charge conservation is violated in the Katanin truncation. Further Ward identity violating contributions may arise from the approximations for the momentum and frequency dependence of self-energies and vertex functions that are applied in order to make the numerical solution of the flow equations tractable. Within the approximations of section~\ref{sec:AH:ApproxSelfEnergy} and~\ref{sec:AH:ApproxVertex}, the Ward identity that relates the anomalous self-energy, the external pairing field and the vertex reads
\begin{equation}
	\begin{split}
		\Delta^\Lambda(k) &= \Delta_{(0)} + \Delta_{(0)} \intdrei{p} L^\Lambda_{22}(p,0)\Bigl[U + \Phi^\Lambda(0) g^{\Phi,\Lambda}(k_0) g^{\Phi,\Lambda}(p_0)\\
	&\quad + \tfrac{1}{2} \bigl(A^\Lambda(p-k) g^{A,\Lambda}\bigl(\tfrac{k_0+p_0}{2}\bigr)^2 - \Phi^\Lambda(p-k) g^{\Phi,\Lambda}\bigl(\tfrac{k_0+p_0}{2}\bigr)^2\bigr)\\
	&\quad + C^\Lambda(p-k) g^{C,\Lambda}\bigl(\tfrac{k_0+p_0}{2}\bigr)^2 - 3 M^\Lambda(p-k) g^{M,\Lambda}\bigl(\tfrac{k_0+p_0}{2}\bigr)^2 \Bigr]
	\label{eq:AH:WardIdentityFull}
	\end{split}
\end{equation}
This Ward identity is violated if the scaling relation $\Phi(q = 0) \propto \Delta_{(0)}^{-1}$ is not fulfilled for small $\Delta_{(0)}$ or equivalently if the gap $\Delta$ deviates from the gap $\Delta_\text{WI}$ that would fulfil the Ward identity for a given vertex and normal self-energy. Besides the question about the severity of the violation of the Ward identity, it is also of interest where it arose from, that is whether the truncation or the approximations are the main source of the violation of the Ward identity in practical calculations. 
The first part of this section discusses the impact of different approximations on the violation of the Ward identity. The result is that the violation of the Ward identity becomes more severe with increasing interaction $U$ and that it is also influenced by the approximations for the non-singular dependences of the vertex. All results in the first part of this section were obtained within the Katanin scheme without further modifications. In the second part of this section, the application of a so-called coordinate projection scheme for enforcing invariants or conservation laws in the numerical solution of ordinary differential equations is discussed. This scheme allows to avoid artefacts of the violation of the Ward identity like non-monotonic flows of $\Phi^\Lambda(0)$ also for larger couplings $U$.

\begin{figure}
	\centering
	\includegraphics{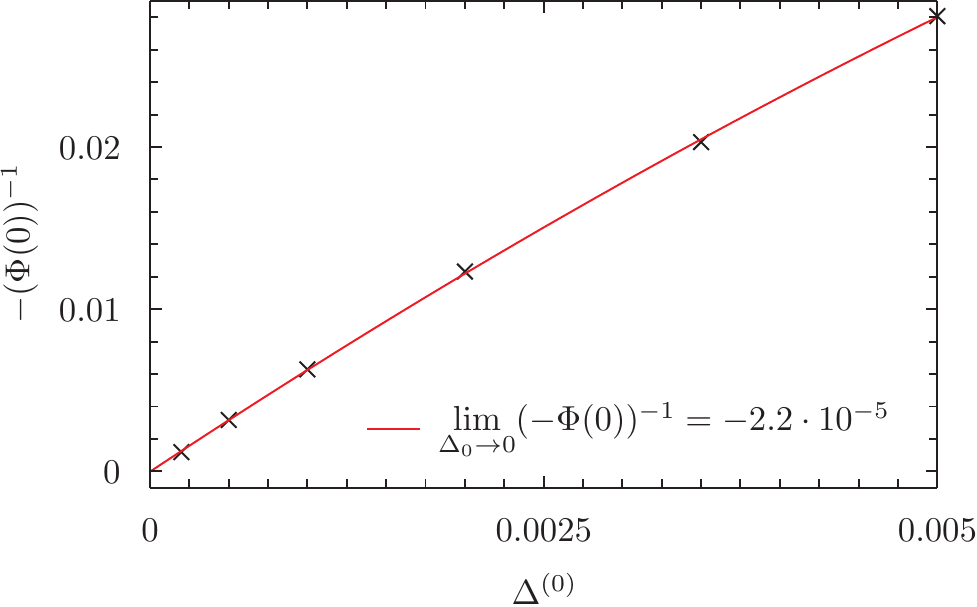}
	\caption{Dependence of the phase mode of the superfluid gap $\Phi(q = 0)$ on the external pairing field $\Delta_{(0)}$ for $U = -2$, $t' = -0.1$ and $n = 1/2$ as determined from fermionic flows with fixed $\Delta_{(0)}$ and without enforcing the Ward identity. The latter implies $\Phi(0) \propto \Delta_{(0)}^{-1}$ for small $\Delta_{(0)}$. The line is a quadratic fit.}
	\label{fig:AH:phi_inv_scaling}
\end{figure}
Before discussing the impact of different approximations on the Ward identity fulfilment, it is interesting to know how severe the deviations are or how small the external pairing field could be chosen without running into unphysical divergences. Solving the flow equations for $U = -2$, $t' = -0.1$ and $n = 1/2$ within the approximations of sections~\ref{sec:AH:ApproxSelfEnergy} and~\ref{sec:AH:ApproxVertex} for different external pairing fields and extrapolating the ``Goldstone mass'' $m_\Phi = (-\Phi(0))^{-1}$ to $\Delta_{(0)} = 0$ as shown in figure~\ref{fig:AH:phi_inv_scaling}, it is found that $m_\Phi$ would vanish and $\Phi(0)$ diverge already for a finite external pairing field. According to the quadratic fit in figure~\ref{fig:AH:phi_inv_scaling}, the divergence of $\Phi(0)$ would occur at an external pairing field of the order $10^{-5}$. 
This is roughly one order of magnitude smaller than the smallest external pairing fields of order $\Delta_\text{MF} / 1000 \approx 2\cdot 10^{-4}$ that were used in numerical calculations for the above parameters. For this latter external pairing field, the exchange propagator for the phase mode at the end of the flow $\Phi(0)$ is roughly six percent larger than required for the fulfilment of the Ward identity. For larger external pairing fields, $\Phi(0)$ typically comes out too small. Note that the situation rapidly worsens for larger interactions.

\begin{figure}
	\centering
	\includegraphics[width=0.6\linewidth]{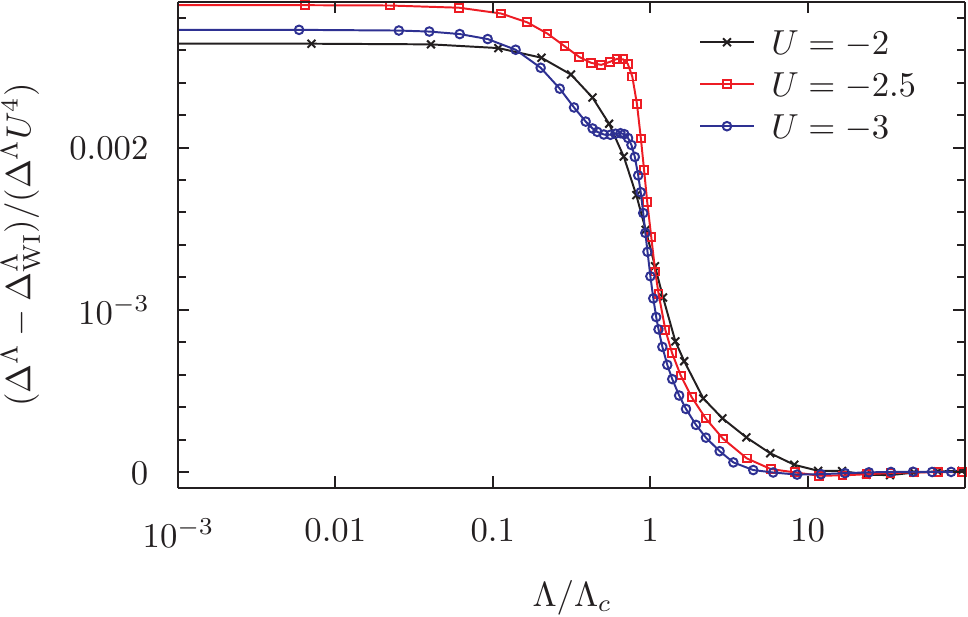}
	\caption{Dependence of the relative deviation of the superfluid gap $\Delta^\Lambda(k_0 = 0)$ from the Ward identity value $\Delta^\Lambda_\text{RG}(k_0 = 0)$ on the RG scale $\Lambda$ and the interaction strength $U$. The external pairing field is $\Delta_{(0)} \approx \Delta_\text{MF} / 200$ in all calculations.}
	\label{fig:AH:WIvioU}
\end{figure}
The discussion in section~\ref{sec:WICharge:Incomp} implies that the violation of the Ward identity within the modified one-loop truncation depends strongly on the interaction $U$ and that the above-mentioned deviations grow rapidly with $U$. In subsection~\ref{subsec:AH:MomFreqOneLoop}, it was argued that the RG flow becomes more mean-field like with decreasing microscopic interaction because of the shrinking phase space for pairing fluctuations and the absence of singularities in the particle-hole channel. A simple approximation for the flow equations allows to recover the perturbative solution for the gap involving the particle-particle irreducible vertex in second order in $U$, which is expected to be asymptotically exact in the limit $U\rightarrow 0$ (see appendix~\ref{sec:Appendix:UToZero}). It is thus expected that the Ward identity fulfilment improves for smaller interactions. 
For finite $U$, the relative deviation of the gap $\Delta^\Lambda$ from the Ward identity value $\Delta^\Lambda_\text{WI}$ roughly grows with $U^4$. This can be seen in figure~\ref{fig:AH:WIvioU} that shows the scale dependence of the relative deviation rescaled with $U^4$ as a function of $\Lambda / \Lambda_c$ at quarter-filling. The curves for $|U| < 2$ deviate from those shown in the figure after rescaling due to numerical errors caused by the smallness of critical scales and gaps. For $n = 0.78$, the agreement between the rescaled curves is worse, nevertheless rescaling with $U^4$ yields a better agreement than rescaling with for example $U^3$. 
The relative deviation of the gap from the Ward identity value thus depends strongly on the microscopic interaction $U$, but the scaling exponent should not be taken too serious: It may differ from the $U^3$-scaling that is expected in lowest order from the discussion in section~\ref{sec:WICharge:Incomp} and could be influenced by the truncation of the hierarchy of flow equations for the effective interactions or the approximations for the non-singular dependences of the vertex or the self-energy.

\begin{figure}
	\centering
	\includegraphics{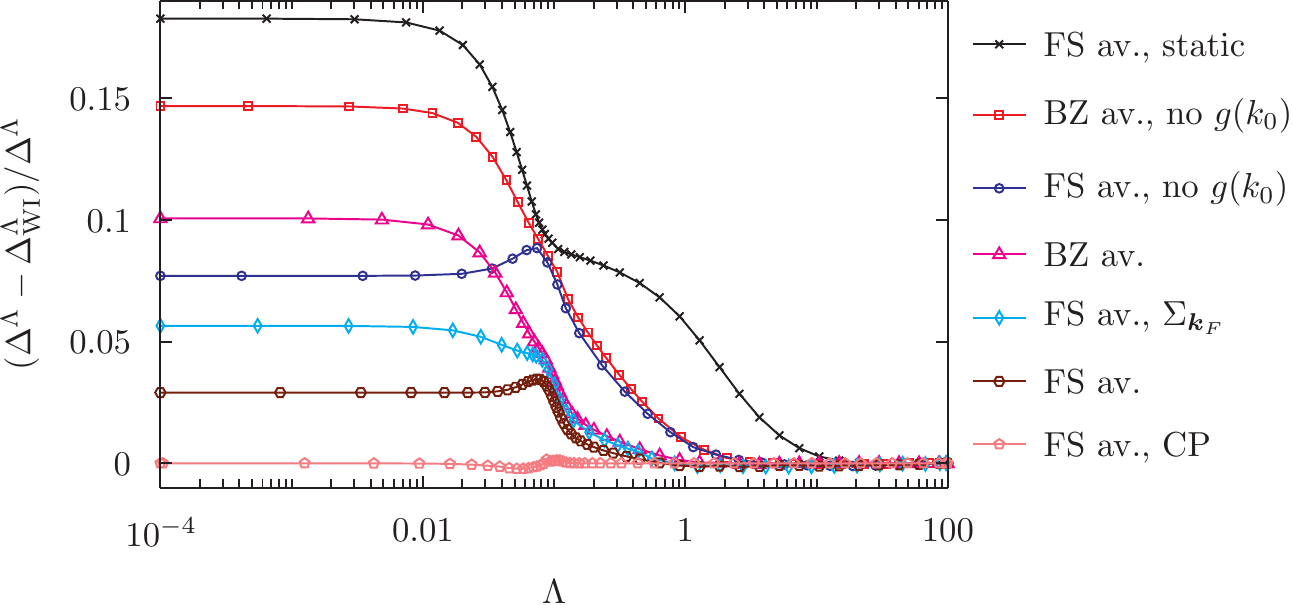}
	\caption{Dependence of the relative deviation of the superfluid gap $\Delta^\Lambda(k_0 = 0)$ from the Ward identity value $\Delta^\Lambda_\text{RG}(k_0 = 0)$ on the scale and different approximations (see text) for $U = -2$, $t' = -0.1$ and $n = 1/2$.}
	\label{fig:AH:WIvio_approx}
\end{figure}
This is also true for the prefactor, as can be seen in figure~\ref{fig:AH:WIvio_approx}. This figure compares the relative deviation of the gap $\Delta^\Lambda$ from the value $\Delta^\Lambda_\text{WI}$ that would fulfil the cutoff Ward identity~\eqref{eq:AH:WardIdentityFull} (using the normal self-energy and the vertex from the RG) for different approximations. The results are for $U = -2$, $t' = -0.1$, $\Delta_{(0)} = \Delta_\text{MF} / 200$ and $n = 1/2$, but similar trends were also observed for other parameters. The labels ``FS av.'' and ``BZ av.'' imply that the RG contributions to the exchange propagators were obtained by projection through averaging external fermionic momenta over the Fermi surface or over the Brillouin zone (as described in subsection~\ref{subsec:VP:BosonProp_gFB}), respectively. 
Note that the RG contributions to the self-energy were computed for the Fermi momentum with the smallest Fermi velocity $(k_{\text{F},x},0)$ for the results labelled ``BZ av.'', while they were computed as Fermi surface average in the calculations labelled ``FS av.'' (except the result labelled $\Sigma_{\boldsymbol k_F}$, where the self-energy was also evaluated for $(k_{\text{F},x},0)$). The label ``static'' refers to calculations with frequency-independent vertex and self-energy. Calculations without frequency-dependent renormalization of the fermion boson vertices are labelled by ``no $g(k_0)$''. The curve labelled ``CP'' uses coordinate projection in order to enforce the Ward identity (as described below) and shows the deviation that is generated during one step of the ODE solver before projection or that accumulates during several steps of the ODE solver. 
The figure should \emph{not} be misunderstood in the sense that the gap as obtained from the RG equations depends sensitively on the employed approximation. The improved calculations using the coordinate projection scheme show that it is the exchange propagator for the phase mode that is susceptible to violations of the Ward identity. For the chosen external pairing field, $\Phi^\Lambda(0)$ typically comes out too small and the relative deviation as shown in the figure~\ref{fig:AH:WIvio_approx} is therefore positive. 
The relative deviation grows monotonically above the critical scale $\Lambda_c \approx 0.1$ in all approximations. Below the critical scale, it typically saturates or continues to grow. The slight decreases seen in the figure are usually a consequence of the fact that the gap grows faster than the violation of the Ward identity just below the critical scale.

The ``static'' approximation does the worst job regarding the Ward identity fulfilment. Taking into account the frequency dependence of the exchange propagators yields improvements on all scales. This is expected in particular for lower scales where the neglect of the frequency dependence of the phase mode is an arguable approximation with regards to the large values it achieves. It is interesting to compare the result where the projection on the flow of exchange propagators is done via averaging external fermionic momenta over the entire Brillouin zone or the Fermi surface. Above $\Lambda_c$, the violation of the Ward identity evolves similarly in both approximations and this is also true for the self-energy and the vertex. 
However, below $\Lambda_c$, the violation of the Ward identity as measured in the figure saturates when averaging external momenta over the Fermi surface, while it continues to grow when averaging external momenta over the whole Brillouin zone. For $U = -2$, the critical scale is relatively small, so that only a narrow vicinity of the Fermi surface effectively contributes to the low energy flow. 
Averaging external momenta over the Fermi surface should therefore yield a better approximation for the effective interaction between the relevant low-energy states, in particular because $\Phi^\Lambda(q)$ depends strongly on the transfer momentum. The evaluation of the RG contributions to the self-energy for a momentum on the Fermi surface should lead to an even better consideration of fluctuation contributions. At least for the parameters in the figure, it does not improve the Ward identity fulfilment\,--\,presumable because different approximations are used for the effective interactions in the flow equations for the self-energy and the vertex. This assertion is substantiated by the finding that the Ward identity fulfilment is improved if the flow of the self-energy is also computed in a local approximation (not shown in the figure). 
The frequency-dependent renormalization of the fermion-boson vertices leads to improvements in particular at high scales. At these scales, contributions from higher frequencies, where the fermion-boson vertices differ from one, should have a significant impact. At low scales, the dominant contributions to the flow arise from small frequencies, where the fermion-boson vertices typically show a relatively weak quadratic dependence on the fermionic frequency and thus differ only mildly from one (see figure~\ref{fig:AH:gFB}).

The violation of the Ward identity as described above suggests that it is not possible to obtain a massless Goldstone boson or spontaneous symmetry breaking within the above approximations in the modified one-loop truncation as proposed by Katanin. The reason is that the scaling relation $\Phi(0) \sim 1 / \Delta_{(0)}$ is violated so that no reasonable limit $\Delta_{(0)} \rightarrow 0$ exists. This is a major obstacle for going to larger interactions $U$, where phase fluctuations become more important and the effects of the violation of the Ward identity more severe. Because improvements of the parametrization did not reduce the violation of the Ward identity to reasonably small values for larger $U$ and because the discussion in section~\ref{sec:WICharge:Incomp} implies that the Ward identity is already violated by the truncation, it would be desirable to enforce the Ward identity in the numerical solution of the flow equations. 
From the point of view of mathematics, the Ward identity is an invariant that together with the flow equations forms a system of differential algebraic equations (DAE). Such systems can be solved numerically with special algorithms for which numerical implementations are available\footnote{An implementation of a DAE-solver in C/C++ is for example contained in the SUNDIALS suite~\cite{Hindmarsh2005}.}. However, the numerical solution of the flow equations using such a DAE solver turned out to be infeasible due to the relatively large number of flowing couplings within the employed approximation scheme\footnote{The reason is that DAE solvers require the computation of the Jacobian of the DAE system, which involves a much higher number of integral evaluations than the evaluation of the right hand side of the flow equations.}. 
A numerically cheaper alternative is provided by coordinate projection or stabilization schemes~\cite{Shampine1986,Gear1989,Ascher1994,Shampine1999}. 
Most results in this work are obtained by using the coordinate projection scheme suggested by Ascher~\etal~\cite{Ascher1994} that combines the numerical solution of the ODE system with a projection of the flowing quantities on the manifolds that are spanned by the invariants (the Ward identities) in a way that the projected solution stays as close to the original solution of the ODE system as possible and that deviations from the invariant manifold are damped exponentially. Schematically, the idea is as follows. Suppose one is looking for the solution of a (system) of ordinary differential equations
\begin{equation}
	\partial_t y = f(t, y)
\end{equation}
under the constraint that
\begin{equation}
	h(t, y(t)) = 0
\end{equation}
holds for all $t$. The numerical solution for $y(t)$ will typically violate\footnote{This is usually true for non-linear constraints even if the ODE system is compatible with the invariant~\cite{Shampine1986}, although the resulting deviations may be very small.} the invariant $h$. It can be enforced by modifying the ODE system to
\begin{equation}
	\partial_t y = f(t, y) - H^T (H H^T)^{-1} h(t, y(t))
\end{equation}
where $H = \partial_y \circ h = \partial_{y_i} h_j$. The second term does not alter the solution of the ODE system if the invariant is fulfilled. Otherwise, it projects the solution towards the manifold that is spanned by the invariant and ensures that the violation of the invariant is of the order of the numerical error if the ODE system and the invariant are compatible~\cite{Shampine1999}. Note that the solution of the projected ODE system may be seen as the solution of the DAE system with a rather crude approximation for the Jacobian and only one Newton iteration for solving the DAE system per discrete step of the ODE solver. In this work, the coordinate projection is applied for the Goldstone mass $m^\Lambda_\Phi = -1/\Phi^\Lambda(0)$, the amplitude mass $m^\Lambda_A = -1/A^\Lambda(0)$ and the gap at zero frequency $\Delta^\Lambda(0)$. The concrete implementation of the coordinate projection scheme is discussed in appendix~\ref{sec:Appendix:CPWI}. 
If desired, the projection could be improved by taking into account more variables. 
However, this choice is expected to capture the largest terms in the gradient of the Ward identity with respect to the flowing couplings. Even the derivative of the Ward identity with respect to the amplitude mass is quite small.

\begin{figure}
	\centering
	\begin{minipage}{0.485\linewidth}
		\includegraphics[width=\linewidth]{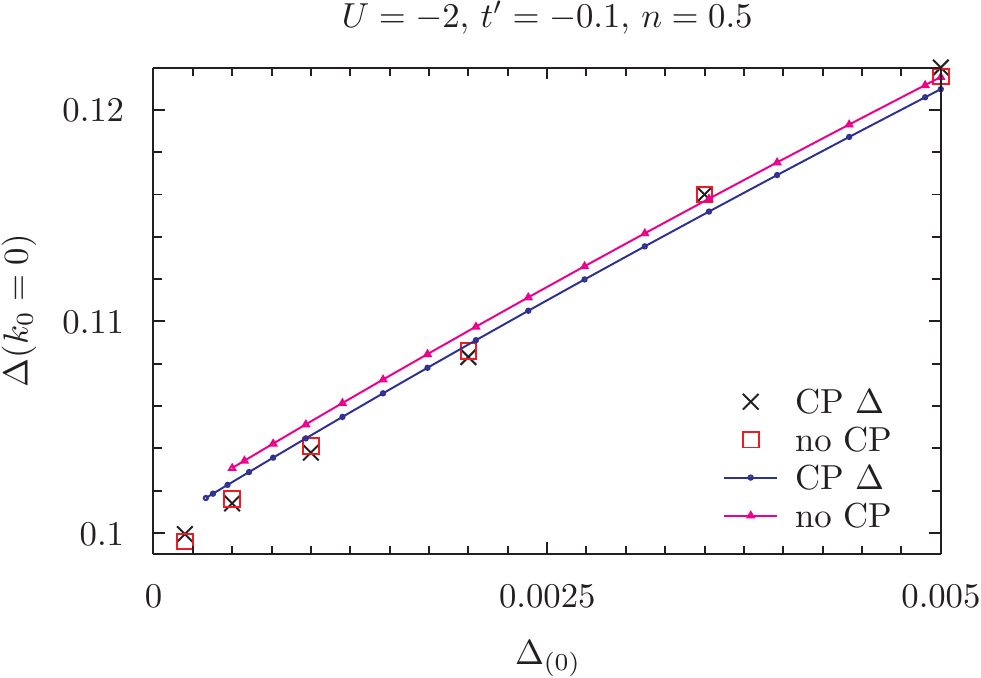}\vspace{2mm}
		\includegraphics[width=\linewidth]{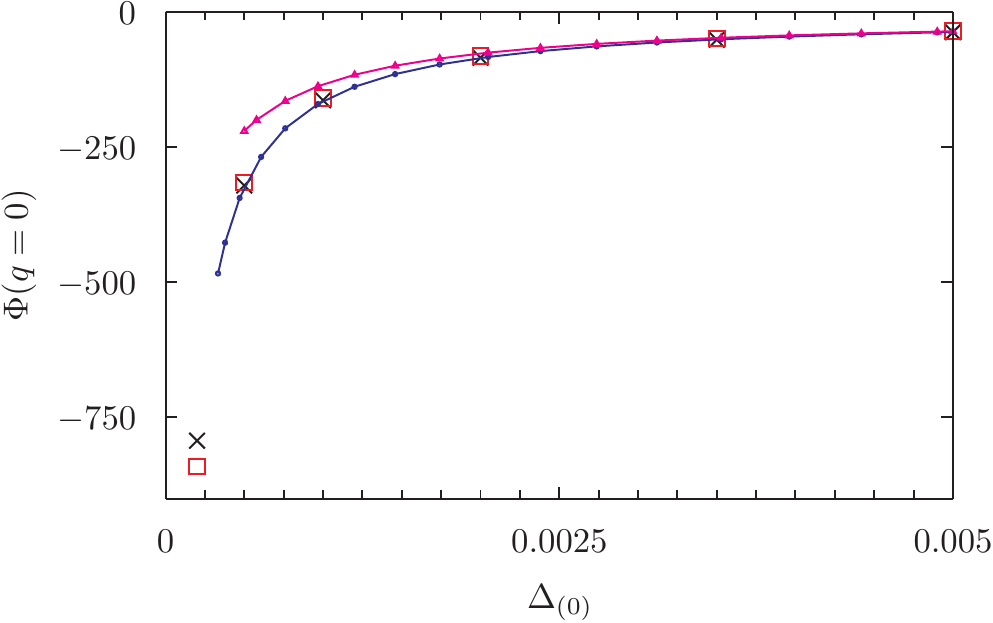}\vspace{2mm}		
		\includegraphics[width=\linewidth]{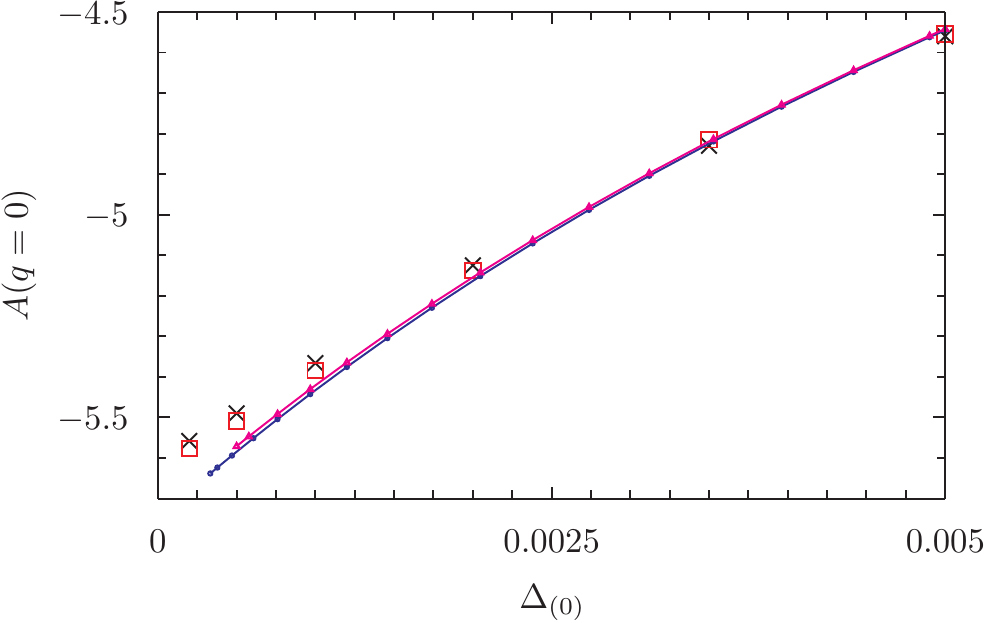}
	\end{minipage}
	\hspace{0.01\linewidth}
	\begin{minipage}{0.485\linewidth}
		\includegraphics[width=\linewidth]{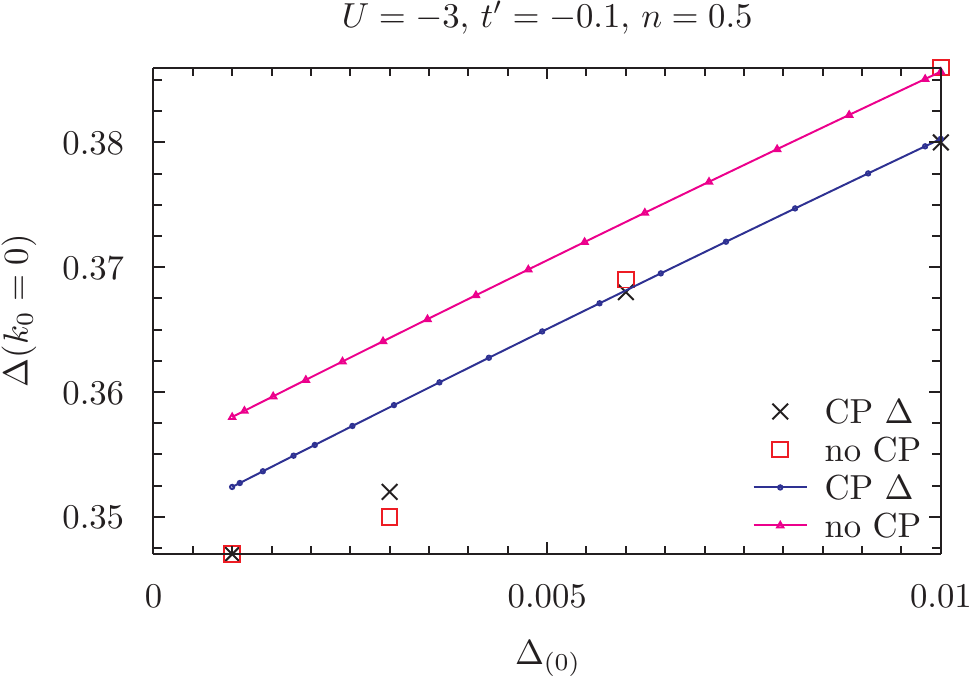}\vspace{2mm}
		\includegraphics[width=\linewidth]{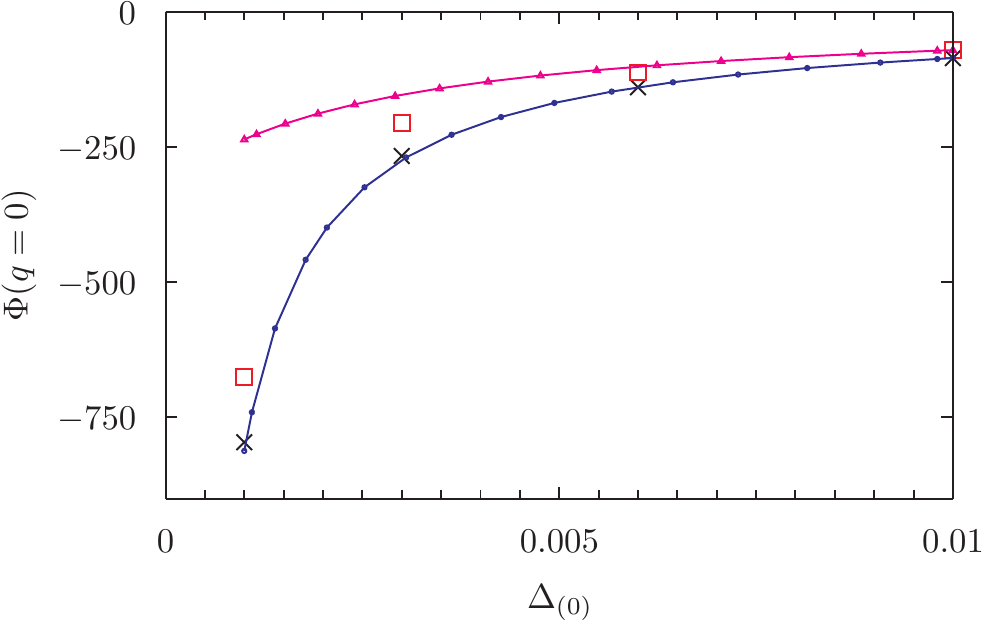}\vspace{2mm}		
		\includegraphics[width=\linewidth]{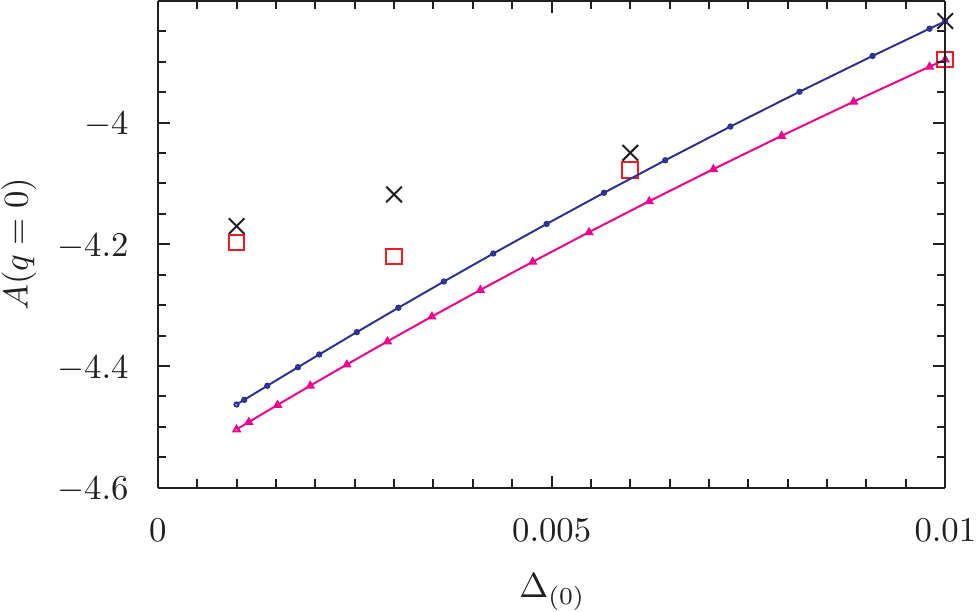}
	\end{minipage}
	\caption{Comparison between the anomalous self-energy $\Delta$ (top) and the propagators for the phase mode $\Phi$ (centre) as well as the amplitude mode $A$ (bottom) as obtained at the end of fermionic flows with fixed external pairing field $\Delta_{(0)}$ (symbols) and from pairing field flows (lines) evaluated with (``CP $\Delta$'') or without (``no CP'') coordinate projection for $n = 1/2$ and $t' = -0.1$. The left column is for $U = -2$ and the right column for $U = -3$. The colours are the same in all plots.}
	\label{fig:AH:Delta0FlowComparison}
\end{figure}
The effects of enforcing the Ward identity can be seen in figure~\ref{fig:AH:Delta0FlowComparison}. It compares results of fermionic flows with fixed external pairing fields and pairing field flows that were evaluated with (labelled ``CP $\Delta$'') or without (labelled ``no CP'') coordinate projection for enforcing the Ward identity for the gap. Results are shown for the gap at zero frequency $\Delta(k_0 = 0)$, the propagator for the phase mode $\Phi(q = 0)$ and the propagator for the amplitude mode $A(q = 0)$ at zero transfer momentum and frequency for a quarter-filled system. The left column is for $U = -2$, the right column for $U = -3$. For $U = -2$, the results with or without projection show good agreement for fixed or flowing external pairing fields. 
This is a simple consequence of the fact that the violation of the Ward identity (measured for example as relative deviation of the gap) is small for this interaction strength.  The inverse propagator for the phase mode in the pairing field flow would, however, extrapolate to a finite value instead of zero in the limit $\Delta_{(0)}\rightarrow 0$. For the larger interaction, the quantitative agreement between flows with or without coordinate projection slightly worsens. In summary, enforcing the Ward identity for the gap improved its fulfilment while the flow was not changed qualitatively (also for the quantities that are not shown).

\subsection{Vertex and self-energy on two-loop level}
\label{subsec:AH:MomFreqTwoLoop}
In subsection~\ref{subsec:AH:MomFreqOneLoop}, the vertex and the self-energy as obtained from one-loop RG flows into the symmetry broken phase were discussed. In chapter~\ref{chap:ChannelDecomposition}, it is argued that the one-loop approximation is justifiable below the critical scale as long as the phase space for pairing fluctuations (of the amplitude or phase mode) is limited by the fast decay of the corresponding exchange propagators away from $q = 0$, which is the case at least in the weak-coupling regime. The two-loop channel-decomposition scheme presented in subsection~\ref{subsec:CD:RGETwoLoop} allows to efficiently take into account fluctuation contributions on two-loop level, so that the assertion on the validity of the one-loop approximation can be checked.
Besides, as motivated in subsection~\ref{subsec:CD:InfraredSingTwoLoop}, phase fluctuations on two-loop level drive the infrared divergence of the amplitude mode that is expected for a fermionic superfluid. In this subsection some numerical results of two-loop flows for the vertex and the self-energy for fixed and flowing external pairing fields are presented and compared with results of the one-loop approximation. The result is that for $|U| = 2-3$, the flows do not change qualitatively in the presence of a not too small external pairing field (for example $\sim \Delta_\text{MF} / 300$). Quantitatively, the critical scale and the gap are further reduced by fluctuations on two-loop level, which is discussed in more detail in the next section for various parameters.

\begin{figure}
	\centering
	\includegraphics{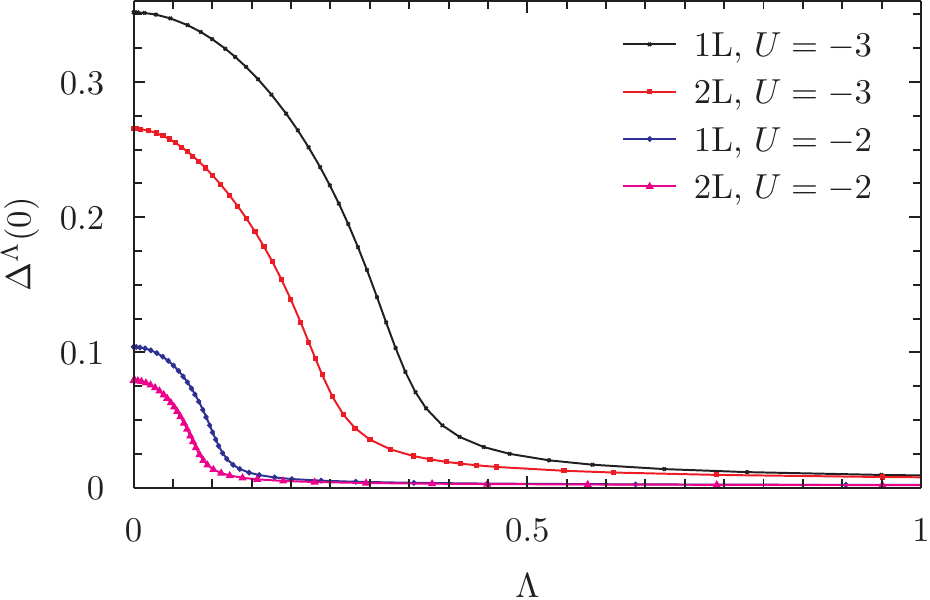}
	\caption{Comparison between one- (1L) and two-loop (2L) flows of the superfluid gap $\Delta^\Lambda(k_0 = 0)$ in a quarter-filled system with $t' = -0.1$ for $U = -2$ and $U = -3$.}
	\label{fig:AH:Delta_OneTwoLoop}
\end{figure}
The two-loop flows were numerically evaluated within the same numerical framework and using the same approximations as on one-loop level, except that the frequency-dependence of the fermion-boson vertices was neglected. The parametrization of the momentum and frequency dependence of the vertex as discussed in section~\ref{sec:AH:ApproxVertex} also provides an ansatz for the one-loop scale derivative of the vertex, which appears in the two-loop flow equations, simply by differentiation of the parametrization with respect to $\Lambda$. It should be noted that more pronounced deviations of the momentum and frequency dependence of the exchange propagators from that of simple bosonic propagators appear on two-loop level. 
The stronger renormalization of the effective interactions leads to sign changes in the frequency dependence of for example the amplitude as well as the phase mode at zero transfer momentum and of the magnetic exchange propagator at the incommensurate peak. These occur however at frequencies that are much higher than the relevant scales (typically $q_0 \gtrsim 20$ at small scales) and 
are not expected to have a significant impact on the results. They are treated within the spirit of the approximations described in section~\ref{sec:AH:ApproxVertex} by adding the momentum dependent part to the frequency dependent mass with an appropriate sign that avoids artificial singularities (Note that the exchange propagators as well as their scale-derivatives coincide with the computed values despite this modification.).

\begin{figure}
	\centering
	\includegraphics{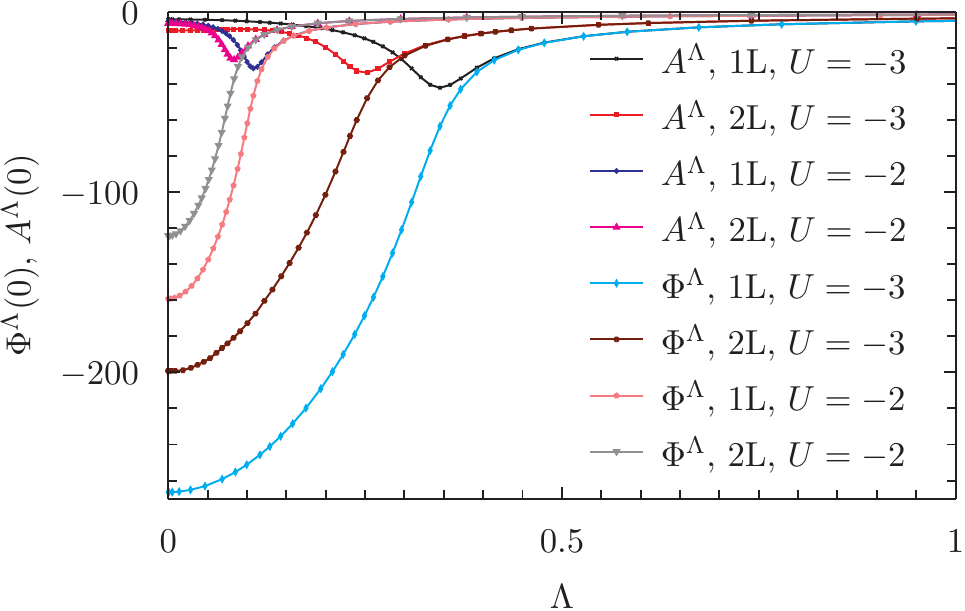}
	\caption{Comparison between one- (1L) and two-loop (2L) flows of the amplitude $A^\Lambda(q = 0)$ and phase $\Phi^\Lambda(q = 0)$ mode of the superfluid gap in a quarter-filled system with $t' = -0.1$ for $U = -2$ and $U = -3$.}
	\label{fig:AH:APhi_OneTwoLoop}
\end{figure}

The two-loop flows of the gap or of the amplitude and phase mode at a fixed external pairing field resemble one-loop flows with a larger external pairing field and a smaller microscopic interaction qualitatively. This can be seen in figures~\ref{fig:AH:Delta_OneTwoLoop} and~\ref{fig:AH:APhi_OneTwoLoop}, respectively. The same conclusion holds for the flow of the extrema of $\nu^\Lambda(q_0,\boldsymbol 0)$ and $\tilde X^\Lambda(q_0, \boldsymbol 0)$. The fermionic quasi-particle weight also shows a very similar scale dependence on one- and two-loop level. Notable changes occur in the spinor particle-hole channel for small momentum and frequency transfers. The effective interaction for charge forward-scattering is strongly suppressed at the critical scale or may even change sign. 
The frequency dependence of $C^\Lambda(q_0, \boldsymbol 0)$ below the critical scale resembles the frequency dependence of the effective interaction for magnetic forward scattering at intermediate scales (see figure~\ref{fig:AH:PhM_FW_q0}). The effective interaction for magnetic forward scattering shows a sign change at $q = 0$ during the flow as on one-loop level, but the height of the positive peak below the critical scale is strongly reduced. The real part of the anomalous (3+1)-effective interaction for small $q$ is also strongly suppressed in comparison to the one-loop level.

\begin{figure}
	\centering
	\includegraphics{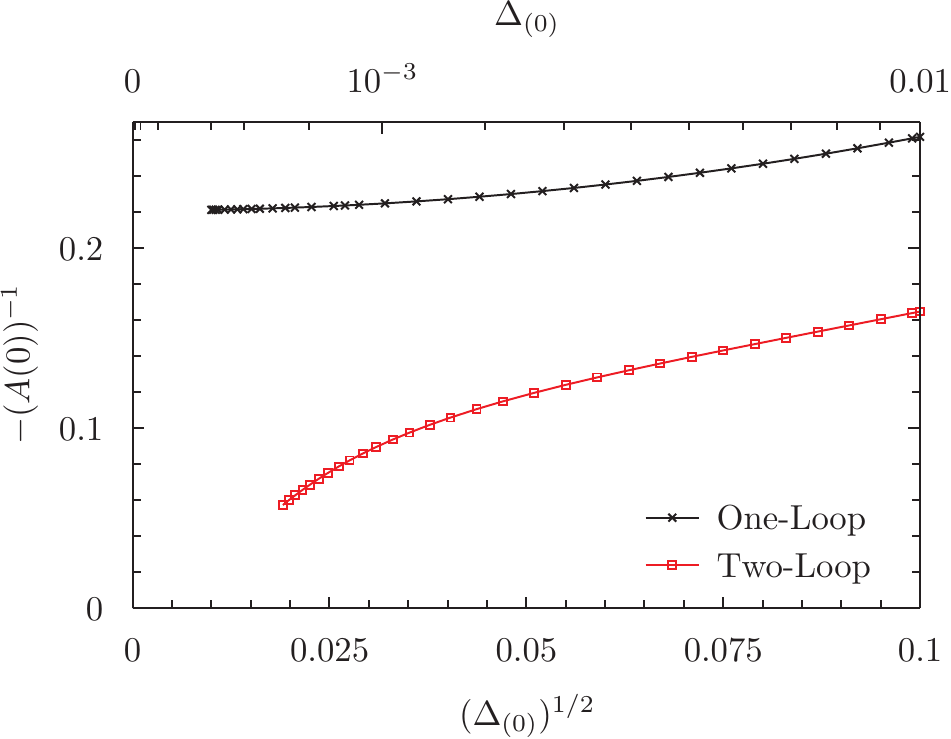}
	\caption{Dependence of the inverse of the propagator for the amplitude mode $A(0)$ on the external pairing field $\Delta_{(0)}$ on one- and two-loop level for a quarter-filled system with $U = -3$ and $t' = -0.1$.}
	\label{fig:AH:A_OneTwoLoop}
\end{figure}
The most prominent change in comparison to the one-loop approximation is expected in the two-loop flow of the propagator for the amplitude mode at $q = 0$. The discussion in subsections~\ref{subsec:CD:InfraredSingOneLoop} and~\ref{subsec:CD:InfraredSingTwoLoop} indicates that $A(0)$ saturates at a finite value for $\Delta_{(0)} \rightarrow 0$ on one-loop level while it diverges as 
\begin{displaymath}
A(0) \propto (\Delta_{(0)})^{-1/2}
\end{displaymath}
on two-loop level. Figure~\ref{fig:AH:A_OneTwoLoop} compares pairing field flows of $-(A(0))^{-1}$ on one- and two-loop level for $U = -3$, $n = 0.5$ and $t' = -0.1$, in which the external pairing field was reduced from $\Delta_{(0)} = 0.01 \approx \Delta_\text{MF} / 60$ by roughly a factor of $30$ during the flow. The numerical results show that the amplitude mode is strongly enhanced on two-loop level. Furthermore, -$(A(0))^{-1}$ on two-loop level bends downwards with decreasing external pairing field while it seems to saturate on one-loop level. However, from the numerical results it is not possible to judge whether the expected linear scaling in $(\Delta_{(0)})^{-1/2}$ for $\Delta_{(0)} \rightarrow 0$ can be reproduced\footnote{At the end of the two-loop flow, the correlation length of the phase mode $\xi_\Phi$ can be estimated by fitting the ansatz
\begin{displaymath}
	\Phi(q_0 = 0, \boldsymbol q) = \frac{\Phi(0)}{1 + \xi_\Phi^2 \boldsymbol q^2}
\end{displaymath}
to numerical data at small momenta. For the flow shown in figure~\ref{fig:AH:A_OneTwoLoop}, this yields $\xi_\Phi \approx 40$ lattice constants, which might be too small for resolving the infrared singularities that are expected in a fermionic superfluid~\cite{Strack2008}.}. 
Besides, the exchange propagators $C^\Lambda$ and $X^\Lambda$ near $q = 0$ also get slightly enhanced in absolute value with decreasing $\Delta_{(0)}$, but it is not possible to infer whether they remain finite or become singular in the limit $\Delta_{(0)}\rightarrow 0$. 

This would require flows for smaller external pairing fields, which are, however, not accessible within the present framework of approximations due to purely technical issues: The aforementioned deviations of the frequency dependence of the phase mode from Lorentzian like decay at small frequencies, which are present already at the end of the fermionic flow and considered as numerical artefacts (see~\ref{subsubsec:AH:VertexOneLoop}), grow during the two-loop pairing field flow. For small external pairing fields, this leads to a plateau in $\Phi(q_0,\boldsymbol 0)$ at small $q_0$, so that the description of the infrared behaviour becomes unreliable. It is expected that these issues would be absent if the pairing field flow were started with a quadratic fit to $\Phi^{\Lambda = 0}(q_0, \boldsymbol 0)$ as its frequency dependence instead of the computed one.

\subsection{Reduction of superfluid gap by fluctuations}
\label{subsec:AH:GapRed}

In this section, the impact of fluctuations on the superfluid gap is discussed and results of renormalization group calculations within the one- and two-loop channel-decomposition scheme are compared with those of other methods and mean-field theory. The latter is known to overestimate the size of order-parameters because the detrimental effect of fluctuations is not taken into account. The mean-field gap $\Delta_\text{MF}$ is obtained by solving
\begin{align}
	\Delta &= \Delta(\mu) = -U \intdrei{p} F(p)		&		n = n(\mu,\Delta) = -2 \intdrei{p} \e{i p_0 0^+} G(p)
\end{align}
self-consistently at the same density as in the interacting system. For Fermi systems with attractive contact interactions, it is well-known that fluctuations reduce the critical temperature even in the weak-coupling limit~\cite{Gorkov1961}. The ground state gap is affected in the same way, so that the ratio between the ground state gap and the critical temperature does not change in comparison to mean-field theory~\cite{Martin-Rodero1992}. The reduction of the order parameter at finite interactions was addressed in many studies. 
For the attractive Hubbard model it was for example studied with perturbative methods~\cite{Georges1991,Martin-Rodero1992,Neumayr2003}, several variants of functional RG~\cite{Reiss2007,Gersch2008,Strack2008}, QMC~\cite{Singer1996} or DMFT~\cite{Garg2005,Bauer2009}. The reduction of the critical temperature by fluctuations was addressed in the two-dimensional attractive Hubbard model for example using QMC~\cite{Paiva2004}, the $T$-matrix approximation~\cite{Keller1999} or DMFT~\cite{Toschi2005} and in three-dimensional continuum models for example using perturbative methods~\cite{Gorkov1961} or the functional RG~\cite{Floerchinger2008b}.

\begin{table}
	\centering
	\begin{tabular}{c|ccc}
			&	$n = 0.95$, $t' = 0$	&	$n = 0.8$, $t' = 0$	&	$n = 0.5$, $t' = -0.1$\\ \hline
		$\Delta / \Delta_\text{MF}$ 	&	0.367	&	0.284	&	0.30
	\end{tabular}
	\caption{Ratio between the superfluid gap as obtained with the particle-particle irreducible vertex in second order in $U$ and the mean-field result in the limit $U \rightarrow 0$ for different parameters. These results were obtained as described in appendix~\ref{sec:Appendix:UToZero}.}
	\label{tab:AH:DeltaRatioU0}
\end{table}
\begin{table}
	\centering
	\begin{tabular}{c|cccc}
		$m$	&	0	&	0.025	&	0.1	& 0.25\\ \hline
		$m_s / m_{s,\text{cl}}$	&	0.6	&	0.65	&	0.73	&	0.77
	\end{tabular}
	\caption{Staggered magnetization $m_s$ divided by the classical result $m_{s,\text{cl}}$ as a function of the uniform magnetization $m$ in the Heisenberg model in a magnetic field. The data is taken from figure 9 in~\cite{Luescher2009}, where it was obtained from quantum Monte Carlo simulations and linear spin-wave theory (with very good agreement between both methods).}
	\label{tab:AH:DeltaRatioUInf}
\end{table}
For small microscopic interactions, the reduction of the gap by particle-hole fluctuations can be estimated within second order perturbation theory for the particle-particle irreducible vertex~\cite{Georges1991,Martin-Rodero1992,Neumayr2003}. This approximation is asymptotically exact in the limit $U \rightarrow 0$ and can be reproduced (non-self-consistently) from the one-loop RG using a rather simple approximation (see appendix~\ref{sec:Appendix:UToZero}). Some results for the ratio $\Delta / \Delta_\text{MF}$ in the limit of a vanishing interaction $U$ are given in table~\ref{tab:AH:DeltaRatioU0}. 
In the opposite strong-coupling limit for $U \rightarrow -\infty$, the attractive Hubbard model with a fermionic density $n$ and $t' = 0$ can be mapped to the Heisenberg model in a magnetic field with a uniform magnetization $m = \tfrac{1}{2}(1 - n)$~\cite{Micnas1990}. 
The gap ratio $\Delta / \Delta_\text{MF}$ of the attractive Hubbard model in that limit translates to the ratio between the staggered magnetization $m_s$ and the corresponding classical result $m_{s,\text{cl}}$ of the Heisenberg model. Some results for $m_s / m_{s,\text{cl}}$ from Lüscher and Läuchli~\cite{Luescher2009} are presented in table~\ref{tab:AH:DeltaRatioUInf}. The reduction of the staggered magnetization in the Heisenberg model without magnetic field by fluctuations has been studied by QMC simulations~\cite{Trivedi1989} and in $1/S$-expansion~\cite{Singh1989, Igarashi1992, Hamer1992}, yielding $m_s / m_{s,\text{cl}} = 0.6$.

\begin{figure}
	\centering
	\includegraphics{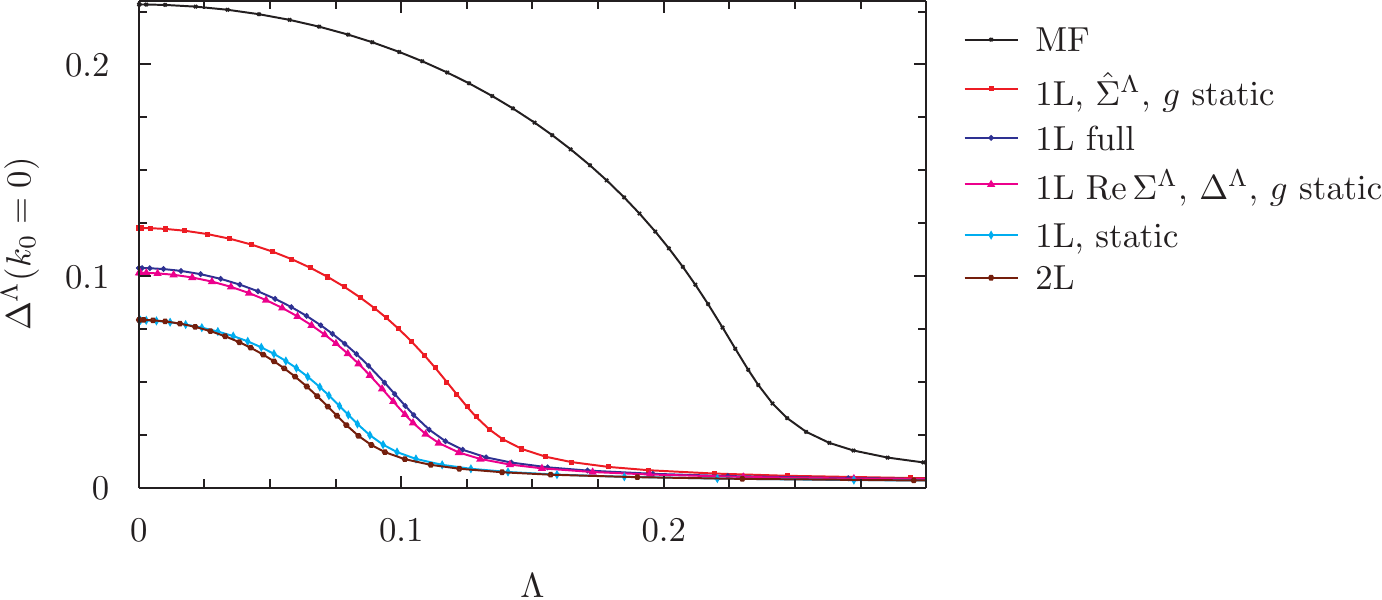}
	\caption{Influence of different approximations (see text) on the flow of the anomalous self-energy $\Delta^\Lambda(k_0 = 0)$ for $U = -2$, $n = 1/2$ and $t' = -0.1$.}
	\label{fig:AH:DeltaFlowApprox}
\end{figure}
Before presenting results for the ratio $\Delta / \Delta_\text{MF}$ as a function of the interaction $|U|$ for different fillings, the impact of several approximations for the self-energy and the vertex on the flow of $\Delta^\Lambda(0)$ is discussed. Figure~\ref{fig:AH:DeltaFlowApprox} shows the scale dependence of the gap for $U = -2$ and $t' = -0.1$ at quarter-filling for an external pairing field $\Delta_{(0)} = 10^{-3} \approx \Delta_\text{MF} / 200$. The result labelled ``MF'' was obtained by neglecting fluctuations and yields the mean-field approximation for $\Lambda = 0$. All other results include fluctuation contributions and used different approximations for the frequency dependence of the self-energy and the vertex. 
The almost indistinguishable third and fourth curves from the top were obtained within the approximations described in sections~\ref{sec:AH:ApproxSelfEnergy} and~\ref{sec:AH:ApproxVertex} (labelled ``1L, full'') and by neglecting the frequency dependence of the real part of the normal self-energy $\re \Sigma^\Lambda$, of the gap $\Delta^\Lambda$ and of the fermion-boson vertices $g^\Lambda$ (but taking the frequency dependence of the exchange propagators and $\im \Sigma^\Lambda$ into account). The result labelled ``1L, $\hat \Sigma$, $g$ static'' was obtained by also neglecting the frequency dependence of $\im \Sigma^\Lambda$. The result labelled ``1L, static'' was obtained for frequency-independent self-energy and exchange propagators. Almost indistinguishable from the static one-loop result is the flow including fluctuations on two-loop level (labelled ``2L''). In comparison to the mean-field flow, the critical scale as well the gap are reduced when taking fluctuations into account. The overall shape of 
the curves is, however, very similar in all cases. 
The frequency-dependence of $\re \Sigma^\Lambda$, $\Delta^\Lambda$ and of the fermion-boson vertices has almost no impact on $\Delta(0)$ for this coupling (note that their frequency dependence is irrelevant in the sense of power counting). This is different for the imaginary part of the normal self-energy, whose neglect increases the critical scale and the gap by roughly 20\,\%, despite the fact that $\im \Sigma^\Lambda(k_0)$ is small. Fluctuations on two-loop level reduce the gap and the critical scale significantly in comparison to the full one-loop flow with frequency dependences considered. 
This is also true if the frequency dependence of the vertex and the self-energy are neglected completely. Remarkably, the critical scales and gaps for the static one-loop and the two-loop flows are very similar. These tendencies were also found for other parameters. Note however that the stronger suppression of $\Lambda_c$ and $\Delta^\Lambda(0)$ in the static approximation in comparison to the full one-loop flow may be model- or regularization-dependent. Husemann~\etal~\cite{Husemann2012} and Giering and Salmhofer~\cite{Giering2012} found a different order of the critical scales in the repulsive Hubbard model at van Hove filling. Using a similar channel-decomposition scheme for the vertex and a so-called $\Omega$-regularization, they found that the critical scales are lower when taking the frequency dependence of the vertex and of the imaginary part of the normal self-energy into account.

\begin{figure}
	\centering
	a)\subfigure{\includegraphics[scale=0.95]{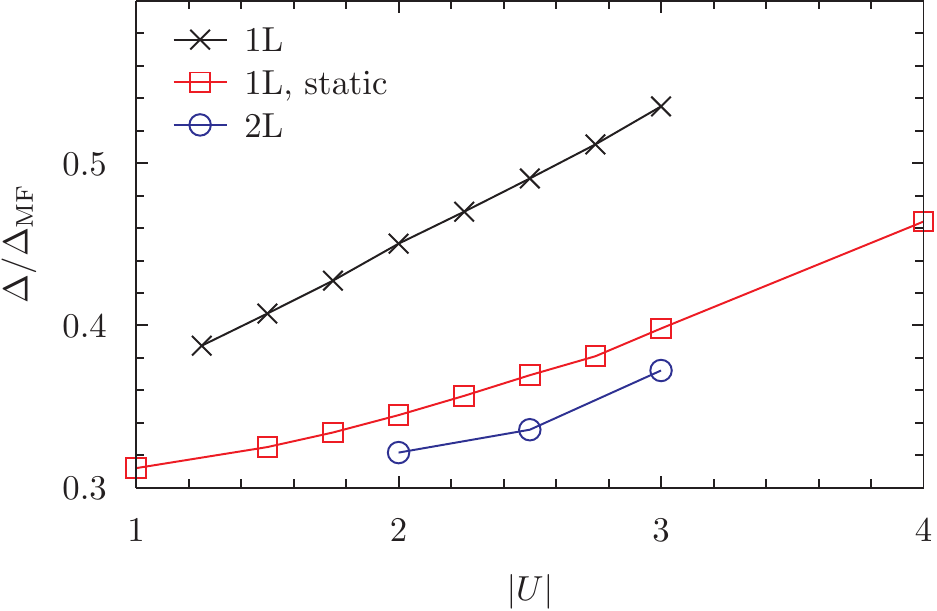}\label{fig:AH:Gapreduction_n_subfig5}}
	b)\subfigure{\includegraphics[scale=0.95]{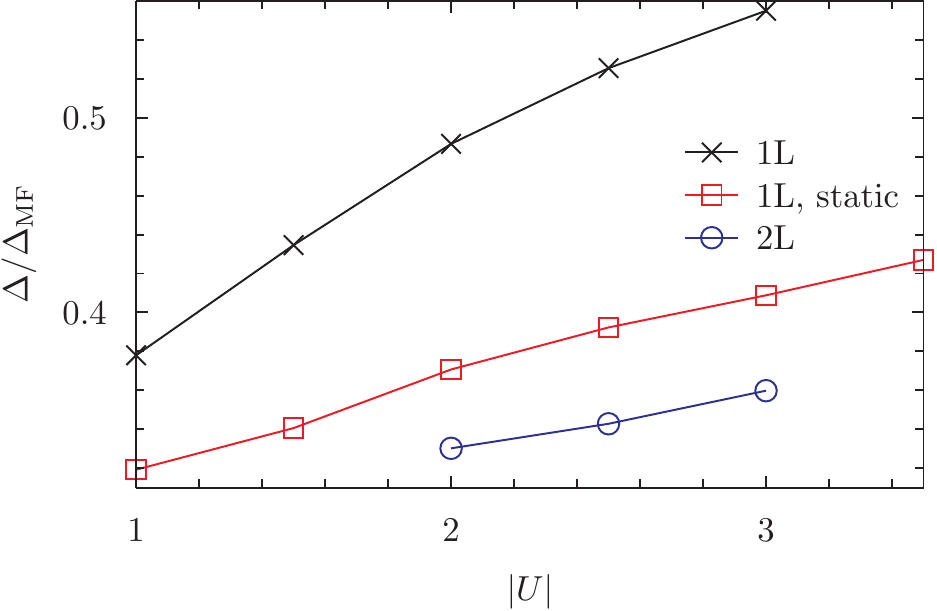}\label{fig:AH:Gapreduction_n_subfig8}}
	c)\subfigure{\includegraphics[scale=0.95]{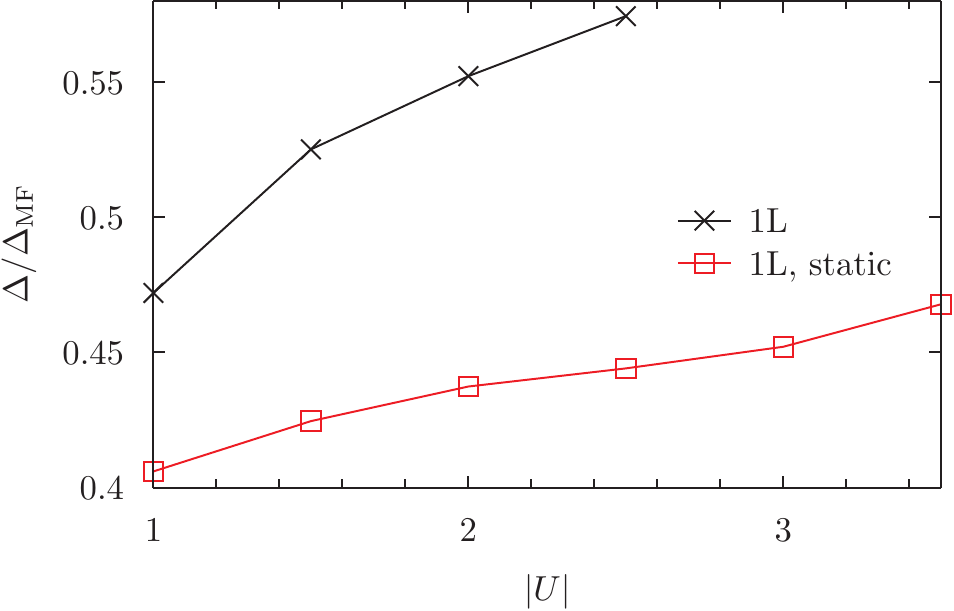}\label{fig:AH:Gapreduction_n_subfig1}}
	\caption{Reduction of the superfluid gap $\Delta(k_0 = 0)$ by fluctuations on one- (1L) and two-loop (2L) level in comparison to the mean-field result $\Delta_\text{MF}$ as a function of the interaction $|U|$ for (a) $n = 1/2$, $t' = -0.1$, (b) $n = 0.78$, $t' = 0$ and (c) $n = 0.95$, $t' = 0$. The frequency dependence of the vertex and the self-energy was included for results labelled ``1L'' and ``2L'', while it was neglected for results labelled ``1L, static''.}
	\label{fig:AH:Gapreduction_n}
\end{figure}
In the following, the reduction of the gap by fluctuations is compared for different fillings as a function of the microscopic interaction $U$ for the static one-loop approximation, the one-loop approximation with frequency dependences and the two-loop approximation. Figure~\ref{fig:AH:Gapreduction_n} shows the ratio $\Delta / \Delta_\text{MF}$ for the quarter-filled system with $t' = -0.1$, at an intermediate fermionic density of $n = 0.78$ with $t' = 0$ and for the almost half-filled system with $n = 0.95$ and $t' = 0$. The gaps were obtained from pairing field flows and subsequent quadratic extrapolation to $\Delta_{(0)} \rightarrow 0$.

For all parameters, the reduction of the superfluid gap by fluctuations is strongest in the weak-coupling regime and becomes weaker with increasing $|U|$. 
This trend is more pronounced on one-loop level with frequency-dependent vertex and self-energy than in the static one-loop and the two-loop approximations. The increase of $\Delta / \Delta_\text{MF}$ with $|U|$ is expected from the behaviour in the limits $U \rightarrow 0$ and $U \rightarrow -\infty$, as can be seen from tables~\ref{tab:AH:DeltaRatioU0} and~\ref{tab:AH:DeltaRatioUInf}, respectively. The data in the figures suggests that the gap ratio may extrapolate in the limit $U \rightarrow 0$ to a similar value for all approximations, which is close to the asymptotic weak-coupling result. For larger couplings, all results for $\Delta / \Delta_\text{MF}$ are smaller than the expected values for the limit $U \rightarrow -\infty$. However, the increasing spread between the one- and the two-loop results hints at a rapidly growing importance of fluctuation contributions beyond the Katanin truncation.

The obtained results agree well with values reported in the literature. For $U = -2$, good agreement of $\Delta / \Delta_\text{MF}$ is found with the results obtained by Gersch~\cite{Gersch2007} and Gersch~\etal~\cite{Gersch2008} within the fermionic one-loop RG using the Katanin scheme or by Reiss~\cite{Reiss2006} and Reiss~\etal~\cite{Reiss2007} using a combination of functional RG and mean-field approximation. Gersch~\etal~\cite{Gersch2008} obtained an almost constant gap ratio at quarter-filling in the coupling range considered. Reiss~\cite{Reiss2006} found the same increase of $\Delta / \Delta_\text{MF}$ as in figures~\ref{fig:AH:Gapreduction_n_subfig5} to~\subref{fig:AH:Gapreduction_n_subfig1}, albeit at somewhat different parameters. 
The results by Gersch seem not to approach the asymptotic weak-coupling result, which might however be a consequence of the use of the so-called $N$-patch approximation for the vertex~\cite{Zanchi1998} with only twelve patches along the Fermi surface per momentum argument. Interestingly, the static approximation within the channel-decomposition scheme leads to a stronger reduction of the gap in comparison to the results by Gersch, although the low-energy description of the vertex is expected to be the same in both cases. This discrepancy might be related to the low momentum resolution for the singular dependences of the vertex in the study by Gersch or to the different choices of the cutoff (the gaps in the study by Gersch~\etal\ were mostly obtained from an interaction flow using the counterterm method). 
Garg~\etal~\cite{Garg2005} studied the attractive Hubbard model at quarter-filling within the DMFT and obtained $\Delta / \Delta_\text{MF} = 0.33 - 0.66$ for $|U| = 1.5 - 3$, which is similar to the results obtained in this work, albeit with a stronger increase of the ratio with $|U|$. Bauer~\etal~\cite{Bauer2009} used the same method and obtained $\Delta / \Delta_\text{MF} = 0.386$ for $U = -1.4$ at quarter-filling, which is very close to the one-loop result with included frequency dependence of the vertex and the self-energy. Neumayr and Metzner~\cite{Neumayr2003} used renormalized perturbation theory in second order and obtained $\Delta / \Delta_\text{MF} \approx 1/3$ for $U = -2$, $n = 0.9$ and $t' = -0.15$ (van Hove filling). This ratio is similar to the weak-coupling expectation for $U \rightarrow 0$, but smaller than the one- and two-loop RG results obtained at $n = 0.78$ or $n = 0.95$ for the same interaction. 
The reduction of the gap by fluctuations for intermediate couplings could be compared to the results of the Quantum Monte Carlo study by Singer~\etal~\cite{Singer1996}. For $n = 0.2$ and $U = -4$, they found a reduction of the gap to around one half of the mean-field value. A similar reduction is obtained for $n = 0.5$ from the static one-loop approximation or by extrapolating the two-loop result to $U = -4$. The extrapolated one-loop result with included frequency dependence is somewhat larger. 
Note that Singer~\etal\ used in their study a relatively small lattice (with $12\times 12$ sites), on which phase fluctuations with long wave length are suppressed. Strack~\etal~\cite{Strack2008} found $\Delta / \Delta_\text{MF} = 0.25$ at quarter-filling for $|U| = 2-4$ in a one-loop RG study for a mixed fermion-boson action that was obtained via a Hubbard-Stratonovich transformation from the attractive Hubbard model.

\section{Conclusion: Attractive Hubbard model}
\label{sec:AH:Conclusion}
In this chapter, the ground state of the attractive Hubbard model in the weak-coupling regime was investigated using the parametrization of the Nambu two-particle vertex from chapter~\ref{chap:VertexParametrization} and the channel-decomposed renormalization group equations from chapter~\ref{chap:ChannelDecomposition}. The Nambu two-particle vertex was described as a boson mediated interaction, where the momentum and frequency dependence of the exchange propagators was approximated using an efficient 2+1-dimensional discretisation. The fermionic self-energy was approximated as frequency dependent but independent of momentum. The description of the singular dependences of the vertex on external momenta and frequencies within the channel-decomposition scheme extends and improves work by Gersch~\etal~\cite{Gersch2008}, who neglected frequency dependences and described momentum dependences within the $N$-patch approximation using a small number of patches. 

Following the discussions in chapter~\ref{chap:WICharge}, the fulfilment of the Ward identity for global charge conservation in numerical solutions of the flow equations was investigated. The Ward identity was fulfilled to a good approximation at weak coupling, but the violation grew rapidly with the microscopic interaction and depended also weakly on the approximations for the non-singular dependences of the vertex. The Ward identity could however be enforced on the flow by fixing the relation between the gap and the mass of the phase mode through the Ward identity instead of the flow equations. This allowed to numerically integrate one- or two-loop flows with external pairing fields that were three or two orders of magnitude smaller than the final gap, respectively, without encountering unphysical divergences. 
Alternatively, a somewhat larger external pairing field could be used in the fermionic flow, which could subsequently be reduced by a factor of at least ten in a pairing field flow. Note that this works even on two-loop level where the violation of the Ward identity is even larger when it is not enforced in the flow.

Treating the external pairing field as a regulator allowed to study the singular behaviour of the Nambu two-particle vertex in the limit where the external pairing field vanishes. On one-loop level, the vertex behaved as in a resummation of all chains of Nambu particle-hole bubble diagrams. On two-loop level, the amplitude mode received strong renormalizations from phase fluctuations. However, the accessible external pairing fields were too large to numerically reproduce the asymptotic infrared behaviour of a fermionic superfluid or an interacting Bose gas that was shown to be captured within the two-loop truncation in subsection~\ref{subsec:CD:InfraredSingTwoLoop}.

The reduction of the superfluid gap by fluctuations was studied for different fillings and microscopic interactions. Results were obtained on one- and two-loop level including frequency dependences of the self-energy and exchange propagators as well as on one-loop level within a static approximation. The different approximations yielded similar results in the weak-coupling regime, but the spread became larger with increasing interaction strength. This indicates that two-loop fluctuations become at least quantitatively important when going to larger interactions. Besides, in the coupling range considered the gap on one-loop level followed a BCS-like dependence on the interaction strength with a renormalized prefactor. These results are in good agreement with values reported in the literature.

An interesting question concerns the coupling range of applicability of the modified one-loop scheme by Katanin in the symmetry broken phase. It captures exact solutions of reduced models exhibiting spontaneous symmetry breaking and this allows to justify the continuation of RG flows beyond the critical scale in the weak-coupling regime although certain effective interactions become large. The one-loop truncation with frequency dependences neglected seems to break down when the microscopic interaction surmounts one half of the bandwidth. In this case, a divergence of the effective interaction for magnetic forward scattering occurred at finite scales below the critical scale for superfluidity. This divergence can be traced back to a mutual reinforcement of the growth of the effective interaction for magnetic forward scattering, the superfluid gap and the phase mode. 
It is expected to arise from the overestimation of phase fluctuations due to the neglect of the frequency dependence of exchange propagators. When including the frequency dependence, the flow equations could be numerically integrated for $|U| \leq 3$. For larger couplings, the employed discretisation scheme broke down below the critical scale due to sign changes in the exchange propagators or due to maxima being renormalized to local minima and vice versa. This is however a merely technical problem that arises from the specific discretisation scheme for the exchange propagators and could be overcome with larger computational power. Due to these issues, no statement about the range of applicability of the modified one-loop scheme by Katanin is possible in case the frequency dependence of exchange propagators is taken into account. 

%% file: Thesis_RepulsiveHubbard.tex
\chapter{Repulsive Hubbard model}
\label{chap:RepulsiveHubbard}

In this chapter, the two-dimensional repulsive Hubbard model is studied at zero temperature using the one-loop channel-decomposition scheme of chapter~\ref{chap:ChannelDecomposition}. Its microscopic action is the same as in chapter~\ref{chap:AttractiveHubbard} except that the local interaction is repulsive, $U > 0$. The repulsive version of the Hubbard model is more interesting than the attractive one in the sense that it exhibits competition of instabilities and superfluidity cannot be obtained in a simple mean-field approximation on the microscopic level. The repulsive Hubbard model has been extensively studied as a model for the electronic structure of copper-oxide planes in cuprate materials (for reviews on its knows properties, see for example~\cite{Baeriswyl1995,Bulut2002}). 
Very different methods indicate the existence of $d$-wave pairing correlations or superfluidity\footnote{Note that it is very common to speak about superconductivity in the repulsive Hubbard model. In this work, the notion superfluidity is preferred because no long-range Coulomb interaction is present so that the phase mode in the symmetry-broken state is expected to be a Goldstone mode with a linear dispersion\,--\,as briefly discussed in chapter~\ref{chap:Introduction}.} near half- or van Hove filling. Examples are the FLEX approximation~\cite{Bickers1989,Monthoux1994}, instability analyses within the functional RG~\cite{Zanchi1998,Zanchi2000,Halboth2000b,Honerkamp2001b,Husemann2009,Katanin2009,Katanin2010,Husemann2012,Friederich2010}, a combination of functional RG and mean-field theory that allowed to study the ground state phase diagram~\cite{Reiss2007} or a functional RG study using dynamical bosonization in order to obtain the phase diagram at finite temperatures~\cite{Friederich2011}. 
These studies were devoted to the repulsive Hubbard model in the weak-coupling regime. For intermediate interactions, numerical evidence for $d$-wave superfluidity or pairing was obtained from cluster extensions of dynamical mean-field theory~\cite{Maier2000,Maier2005,Kancharla2008,Sentef2011,Gull2012-arXiv} or variational QMC~\cite{Yamaji1998,Baeriswyl2009}. The results of (non-variational) QMC studies for $t' = 0$ are controversial: Earlier works found an increase in the $d$-wave pairing susceptibility when reducing the temperature~\cite{White1989,Loh1990}, while a more recent work finds no sign of $d$-wave superfluidity~\cite{Aimi2007}. However, in the presence of a finite next-nearest neighbour hopping $t'$, a Kosterlitz-Thouless transition was detected at low temperatures~\cite{Yanagisawa2010}.

Husemann and Salmhofer~\cite{Husemann2009} studied the repulsive Hubbard model in the symmetric phase using a similar channel-decomposition scheme and approximations for the vertex that are comparable to those employed in this chapter. They confirmed that the Kohn-Luttinger mechanism for pairing from electronic interactions~\cite{Kohn1965} is captured within the channel-decomposition scheme and found that antiferromagnetic spin fluctuations generate an attractive interaction in the $d$-wave pairing channel. The aim of this chapter is to extend the study by Husemann and Salmhofer to the symmetry broken phase in case $d$-wave superfluidity is the leading instability. 
In this case, the $d$-wave superfluid gap in the ground state is obtained as a function of the interaction $U$, the next-nearest neighbour hopping $t'$ and the fermionic density $n$ or the Fermi level $\epsilon_F$. If $d$-wave superfluidity is not the leading instability, the flow has to be stopped at some scale $\Lambda_c$ where the effective interactions become large. The approximations for the self-energy and the vertex are similar to those of chapter~\ref{chap:AttractiveHubbard}, but less sophisticated because dependences on frequency are neglected. 

This chapter is organized as follows. The approximations for the self-energy and the exchange propagators that are applied within the channel-decomposition scheme are described in more detail in sections~\ref{sec:RH:ApproxSelfEnergy} and~\ref{sec:RH:ApproxVertex}. Results for the dependence of the critical scale and the pairing gap on various control parameters are presented in section~\ref{sec:RH:Results} together with representative flows of the self-energy and the exchange propagators as well as examples for their momentum dependence at the end of the flow.

\section{Approximation for fermionic propagator}
\label{sec:RH:ApproxSelfEnergy}
The description of the fermionic propagator is similar to the one used in section~\ref{sec:AH:ApproxSelfEnergy}. The fermionic kinetic energy is given by the tight-binding dispersion~\eqref{eq:Intro:Dispersion} and the additive frequency regulator~\eqref{eq:AH:Regulator} is used in this chapter, too. The fermionic self-energy is independent of frequency due to the neglect of the frequency dependence of the vertex (see section~\ref{sec:RH:ApproxVertex}). Differently from chapter~\ref{chap:AttractiveHubbard}, the momentum dependence of the anomalous self-energy has to be taken into account for the repulsive Hubbard model because the gap has nodes. In this chapter, the momentum dependence of the anomalous self-energy is described by the ansatz
\begin{equation}
	\Delta^\Lambda(k) = \Delta^\Lambda f_d(\boldsymbol k)
\end{equation}
where $f_d(\boldsymbol k) = \cos k_x - \cos k_y$ is a form factor with $d_{x^2-y^2}$-symmetry. The external pairing field is chosen as $\Delta_{(0)}(k) = \Delta_{(0)} f_d(\boldsymbol k)$. Note that for this ansatz, the maximal gap in the propagator occurs at $\boldsymbol k = (0, \pi)$ and symmetry-related points and its absolute value is given by $2\Delta^\Lambda$. The flow of $\Delta^\Lambda$ is determined from the $d$-wave projection of the flow equation for the anomalous self-energy at zero fermionic frequency,
\begin{equation}
	\partial_\Lambda \Delta^\Lambda = \intzwei{k} f_d(\boldsymbol k) \partial_\Lambda \Delta^\Lambda(k_0 = 0, \boldsymbol k).
\label{eq:RHM:Delta_Flow}
\end{equation}
In this chapter, the flow of the self-energy is determined from Brillouin zone averages and not from Fermi surface averages. The motivations for choosing this projection scheme are that it is simpler from a computational point of view and that the differences between the schemes turned out to be small for the self-energy and the leading couplings in the attractive Hubbard model. 
The same projection scheme is used for the extraction of the flow of the exchange propagators from that of the effective interactions\footnote{Besides, the use of the same projection scheme for the flow of the self-energy and the effective interactions yielded a higher degree of consistency from the point of view of Ward identity fulfilment in the attractive Hubbard model (see subsection~\ref{subsec:AH:WI}).} (see section~\ref{sec:RH:ApproxVertex}), too. The above ansatz can be seen as an expansion of the anomalous self-energy with respect to lattice form factors of $d_{x^2-y^2}$-symmetry in which only the lowest order term is kept. 

The imaginary part of the normal self-energy is zero due to the neglect of the frequency dependence of the vertex, $\im \Sigma^\Lambda(k) = 0$. Its real part is approximated as a momentum and frequency independent constant $\re \Sigma^\Lambda(k) \approx \re \Sigma^\Lambda$ that describes a shift of the Fermi surface. Its flow is determined from the $s$-wave projection of the flow of the normal self-energy at vanishing frequency,
\begin{equation}
	\partial_\Lambda \re \Sigma^\Lambda = \intzwei{k} \partial_\Lambda \re\Sigma^\Lambda(k_0 = 0, \boldsymbol k).
\end{equation}
Note that $\re \Sigma^\Lambda$ cannot simply be neglected because this may entail a violation of the Ward identity for global charge conservation even in reduced models. The shift of the Fermi surface resulting from $\re \Sigma^\Lambda$ could in principle be treated within the counter term method as in chapter~\ref{chap:AttractiveHubbard}. However, in case a momentum and frequency independent counter term was considered, the linear equation for its determination did not have a unique solution at van Hove filling for all scales in the range of couplings considered below\footnote{This problem might however be an artefact of neglecting the momentum dependence of the self-energy and of the counter-term.}. 
The above approximation for the normal self-energy can be seen as arising from an expansion of its momentum dependence with respect to lattice form factors, where only the first term is kept. Within the approximations for the vertex of section~\ref{sec:RH:ApproxVertex}, the higher order terms, which renormalize the hopping amplitudes, are expected to be small. They were computed and indeed found to be small within the functional RG using the $N$-patch scheme by Honerkamp and Salmhofer~\cite{Honerkamp2001a} or using the channel-decomposition scheme for the symmetric state by Giering and Salmhofer~\cite{Giering2012} for similar model parameters as in this chapter.

The numerical integration of the flow equations was started at a scale $\Lambda_0$ that is of the order of several times the bandwidth ($\Lambda_0 \approx 100 - 150$). The initial condition for the normal self-energy is determined self-consistently in first order in $U$ as in chapter~\ref{chap:AttractiveHubbard}. The anomalous self-energy at this scale is approximated by the external pairing field, $\Delta^{\Lambda_0} = \Delta_{(0)}$.

In section~\ref{sec:RH:Results}, the numerical results are compared with those of other methods mostly at the same fermionic density. The latter is obtained from the full fermionic propagator as explained in chapter~\ref{chap:AttractiveHubbard}. However, due to the neglect of the frequency dependence of the self-energy in this chapter, the density is expected to be less accurate than that obtained there. Within the approximations for the fermionic propagator described above, a Fermi level of the interacting system can be defined at the scale $\Lambda$ as the chemical potential shifted by the real part of the normal self-energy,
\begin{equation}
	\epsilon_F^\Lambda = \mu - \re \Sigma^\Lambda.
\end{equation}
The (interacting) system is at van Hove filling if $\epsilon_F = \epsilon_F^{vH} = 4t'$.

\section{Approximation for two-particle vertex}
\label{sec:RH:ApproxVertex}
In this chapter, the vertex in the repulsive Hubbard model is parametrized as a boson-mediated interaction within the channel-decomposition scheme of chapter~\ref{chap:VertexParametrization}. The frequency dependence of the exchange propagators as well as of the fermion-boson vertices is neglected. The dependence of the effective interactions on the fermionic momenta is described by scale-independent lattice form factors. For the amplitude mode as an example, the expansion of the effective interaction reduces to
\begin{equation}
	A^\Lambda_{k k'}(q) = \sum_{\alpha,\beta} A^\Lambda_{\alpha\beta}(\boldsymbol q) f_\alpha (\boldsymbol k) f_\beta (\boldsymbol k),
\label{eq:RHM:expansion}
\end{equation}
and similarly for the other channels. In the sum on the right hand side, only the $s$- and $d_{x^2-y^2}$-wave channels with form factors
\begin{align}
	f_s(\boldsymbol k) &= 1		&		f_d(\boldsymbol k) &= \cos k_x - \cos k_y
\end{align}
are kept. These approximations were also applied by Husemann and Salmhofer in their study of the symmetric state of the repulsive Hubbard model~\cite{Husemann2009}. The restriction to the $s$- and $d$-wave channels is motivated by the results for the effective interactions in all previous fRG studies of the Hubbard model for the model parameters that are of interest in this chapter (see for example the review by Metzner~\etal~\cite{Metzner2012}). In principle, the right hand side of~\eqref{eq:RHM:expansion} contains diagonal and off-diagonal terms with respect to the form factor index. However, Husemann and Salmhofer found that the off-diagonal terms in the expansion of the normal effective interaction remain very small even at van Hove filling~\cite{Husemann2009,Husemann2009Dis}. The off-diagonal components of the propagators for the normal and the anomalous (4+0)-effective interactions are therefore neglected. The off-diagonal components of the anomalous (3+1)-effective interactions should however be kept. 
They are necessary for exactly solving a reduced model with pairing and forward-scattering interactions in the $s$- and $d$-wave channels with an instability towards $d$-wave pairing within the functional RG (see appendix~\ref{sec:Appendix:SDRPFM}). In such a model, the off-diagonal exchange propagators in the anomalous (3+1)-channel lead to a coupling between the $s$-wave forward scattering channel and the $d$-wave pairing channel and vice versa, while the diagonal anomalous (3+1)-exchange propagators vanish\footnote{This can be understood from equation~\eqref{eq:RPFM:MFBubblesEntries} in chapter~\ref{chap:RPFM}: In a $d$-wave superfluid with unbroken fourfold rotation symmetry of the lattice, the anomalous (3+1)-bubble $l^\Lambda_x$ is non-zero only if one of the two form factors $f_c$ or $f_p$ is of $s$-wave and the other of $d$-wave symmetry due to the presence of the $d$-wave gap in the anomalous propagator.}. 
The diagonal exchange propagators in the anomalous (3+1)-channel are therefore expected to be small also in the repulsive Hubbard model and are thus neglected. The frequency dependence of the vertex and the self-energy is neglected because Husemann~\etal~\cite{Husemann2012} and Giering and Salmhofer~\cite{Giering2012} found that it does not change the leading instabilities at van Hove filling for model parameters similar to those studied in this chapter. In the static approximation, the imaginary parts of the effective interaction $\nu^\Lambda_{ss}$, $\nu^\Lambda_{dd}$, $\tilde X^\Lambda_{sd}$ and $\tilde X^\Lambda_{ds}$ vanish. In summary, the following exchange propagators 
are taken into account:
\begin{align}
	A^\Lambda_{ii}&(\boldsymbol q)		&		\Phi^\Lambda_{ii}&(\boldsymbol q)		&		C^\Lambda_{ii}&(\boldsymbol q)		&		M^\Lambda_{ii}&(\boldsymbol q)		&		X^\Lambda_{sd}&(\boldsymbol q)		&		X^\Lambda_{ds}&(\boldsymbol q)
	\label{eq:RH:Propagators}
\end{align}
where $i\in \{s,d\}$. In the following, one of the form factor indices of diagonal exchange propagators is dropped for convenience, for example $A^\Lambda_{dd}(q) \equiv A^\Lambda_d(q)$. In the normal state above the critical scale, the $d$-wave forward scattering interactions $M^\Lambda_d$ and $C^\Lambda_d$ are subleading and remain small for the model parameters that are of interest in this chapter~\cite{Husemann2012b}. However, they were kept in order to study possible feedback from $d$-wave pairing fluctuations below the critical scale. 

The momentum dependence of the exchange propagators was discretised as described in section~\ref{sec:AH:ApproxVertex}. A grid in polar coordinates with centre $\boldsymbol q = \boldsymbol 0$ was used for the propagators in the spinor particle-particle channel and for $X^\Lambda_{sd}$ in order to efficiently describe their sharp peaks at this point. The propagators in the spinor particle-hole channel and $X^\Lambda_{ds}$ were discretised on a non-equidistant Cartesian grid, whose grid points were adapted for a good description of the incommensurate peaks of $M^\Lambda_s(\boldsymbol q)$. The renormalization contributions to the exchange propagators were obtained from the flow equations for the effective interactions at vanishing fermionic frequencies by orthogonal projection using lattice form factors for the dependence on fermionic momenta.

The effective interaction in the $s$-wave channel at the scale $\Lambda_0$, where the numerical solution of the flow equations was started, was determined as described in chapter~\ref{chap:AttractiveHubbard} by resumming all chains of Nambu particle-hole diagrams in the presence of a regulator. The bosonic propagators in the $d$-wave channel at this scale were set to zero, as they are much smaller than those in the $s$-wave channel.

\section{Numerical results on one-loop level}
\label{sec:RH:Results}
In this section, numerical results for the Hubbard model are presented that were obtained within the channel-decomposition scheme on one-loop level. The dependence of the critical scales for (incommensurate) antiferromagnetism or $d$-wave superfluidity and of the $d$-wave superfluid gap on the microscopic interaction $U$, the Fermi level $\epsilon_F$, the next-nearest neighbour hopping $t'$ or the fermionic density $n$ are discussed together with representative flows. For the chosen values of $t'$ and $\mu$, only instabilities towards $d$-wave superfluidity and antiferromagnetism occurred. The obtained results are compared with the literature at the end of this section. Note that lines in plots are guides to the eye if not stated differently. The dependence on $\Lambda$ is suppressed in the notation for quantities at the end of the flow for convenience, for example $A^{\Lambda = 0}_d(\boldsymbol q) \equiv A_d(\boldsymbol q)$.

The flow equations for the self-energy and the exchange propagators were numerically integrated using the same computational framework as described in subsection~\ref{subsec:AH:NumSetup} for the attractive Hubbard model. In order to enforce the Ward identity for global charge conservation, which reads
\begin{equation}
	\begin{split}
		\Delta^\Lambda(\boldsymbol k) &= \Delta_0(\boldsymbol k) + \sum_{\alpha \in \{s, d\}} \intdrei{p} \Delta_0(\boldsymbol p) L^\Lambda_{22}(p, 0)\Bigl[U \delta_{\alpha s} + \Phi^\Lambda_{\alpha}(\boldsymbol 0) f_\alpha(\boldsymbol k) f_\alpha(\boldsymbol p)\\
	&\ + \Bigl(\tfrac{1}{2} \bigl(A^\Lambda_\alpha(\boldsymbol p - \boldsymbol k) - \Phi^\Lambda_\alpha(\boldsymbol p - \boldsymbol k) \bigr) + C^\Lambda_\alpha(\boldsymbol p - \boldsymbol k) - 3 M^\Lambda_\alpha(\boldsymbol p - \boldsymbol k)\Bigr) f_\alpha(\tfrac{\boldsymbol k + \boldsymbol p}{2})^2 \Bigr]
	\label{eq:RH:WardIdentityFull}\raisetag{3.5em}
	\end{split}
\end{equation}
within the approximations of sections~\ref{sec:RH:ApproxSelfEnergy} and~\ref{sec:RH:ApproxVertex}, in the flow, the coordinate projection scheme discussed in subsection~\ref{subsec:AH:WI} was also employed in the calculations presented in this chapter. 
The projection was done for $\Delta^\Lambda$, $\re \Sigma^\Lambda$, $\Phi^\Lambda_d(\boldsymbol 0)$ as well as $A^\Lambda_d(\boldsymbol 0)$ and was implemented as described in appendix~\ref{sec:Appendix:CPWI}. 

For the parameters discussed in the following, the flows were either integrated until $\Lambda = 0$ was reached in case the leading instability was towards $d$-wave superfluidity, or stopped at some critical or stopping scale $\Lambda_c$ where the magnetic exchange propagator fulfilled
\begin{equation}
	\max_{\boldsymbol q} |M^\Lambda_s(\boldsymbol q)| \gtrsim 100.
\label{eq:RH:StoppingCriterion}
\end{equation}
In the absence of an external symmetry breaking field, a magnetic instability would occur if $M^\Lambda_s(\boldsymbol q)$ diverges. The critical scales obtained with the above stopping criterion yield a good estimate for this divergence scale. Note that for the external pairing fields employed in this chapter, which were chosen roughly two orders of magnitude smaller than the final gap, the $d$-wave pairing interaction did not reach the stopping criterion~\eqref{eq:RH:StoppingCriterion} at the critical scale for superfluidity. In the absence of an external pairing field, this would however be the case at a scale that is only slightly smaller than the $\Lambda_c$ as obtained from the maximum of the flowing $|A^\Lambda_d(\boldsymbol 0)|$, because the pairing interaction grows very rapidly for $\Lambda \gtrsim \Lambda_c$. 
Besides, it has been checked that the correlation lengths (as obtained from fitting Lorentzians to the momentum dependence around the peaks) were significantly larger in the $d$-wave pairing channel than in the magnetic channel if an instability towards $d$-wave superfluidity occurred. 

\hyphenation{fermi-onic}
In case antiferromagnetism is the dominant instability, the Fermi level and the fermionic density at the critical scale are taken as estimates for their value at the end of the flow. This allows to plot the dependence of the critical scales for antiferromagnetism and for pairing on the Fermi level or the fermionic density together, yielding an impression of the evolution of typical energy scales for both instabilities. In the weak-coupling regime and for the chosen regulator, the critical scale for antiferromagnetism provides an estimate for the size of the antiferromagnetic gap. 
For $t' = 0$ and $n = 1$, they become equal on the mean-field level\footnote{For $t' = 0$ and $n = 1$, the Fermi surface is perfectly nested, $\epsilon(\boldsymbol k + \boldsymbol \pi) = -\epsilon(\boldsymbol k)$, and the chemical potential is fixed by particle-hole symmetry. The mean-field equation for the antiferromagnetic gap then becomes equal to that for the superfluid gap in the attractive Hubbard model. This is not surprising because the repulsive and attractive Hubbard model on a bipartite lattice can be mapped onto each other~\cite{Emery1976,Micnas1990}. For the attractive case, the equality between the critical scale and the superfluid gap on the mean-field level was established in chapter~\ref{chap:AttractiveHubbard} for the chosen regulator for any lattice and fermionic density in the weak-coupling regime.}. 
For finite $t'$, this estimate however becomes unreliable close to half-filling where $n(\mu)$ is a non-analytic function even on the mean-field level~\cite{Reiss2006}. 
This non-analytic behaviour cannot be reproduced if the flow is stopped at a finite scale, so that the assignment of estimates for the gap to chemical potentials or densities becomes ambiguous. 
The equality of critical scales and gaps for antiferromagnetism away from half-filling was confirmed on the mean-field level by comparing flows without fluctuations to results by Reiss~\cite{Reiss2006}. It was also established in the weak-coupling regime on one-loop level for $s$-wave superfluidity in subsection~\ref{subsec:AH:SelfenergyOneLoop} and is also valid to a good approximation for $d$-wave superfluidity (see below). In the latter case, the critical scale $\Lambda_c$ equals the maximal $d$-wave gap (that is found at $\boldsymbol k = (0,\pi)$ within the employed approximations) and not some Fermi surface average.

\begin{figure}
	\centering
	\includegraphics{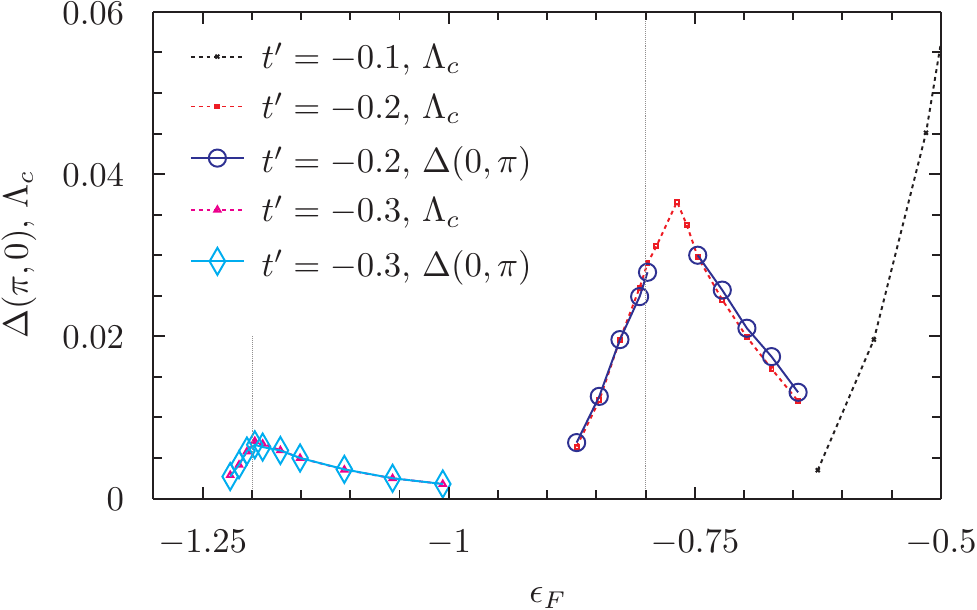}
	\caption{Critical scale $\Lambda_c$ and $d$-wave pairing gap $\Delta(\boldsymbol k = (0,\pi))$ as a function of the Fermi level $\epsilon_F$ for $U = 2.5$ and various $t'$. Regions where only $\Lambda_c$ (dashed lines) is shown have dominant (incommensurate) antiferromagnetic correlations. Regions where also $\Delta(0,\pi)$ (solid lines with symbols) is shown have a $d$-wave superfluid ground state. The vertical lines mark van Hove filling where $\epsilon_F = \epsilon_F^{vH} = 4t'$.}
	\label{fig:RH:PD2.5_tp}
\end{figure}

\begin{figure}
	\centering
	\includegraphics[width=\linewidth]{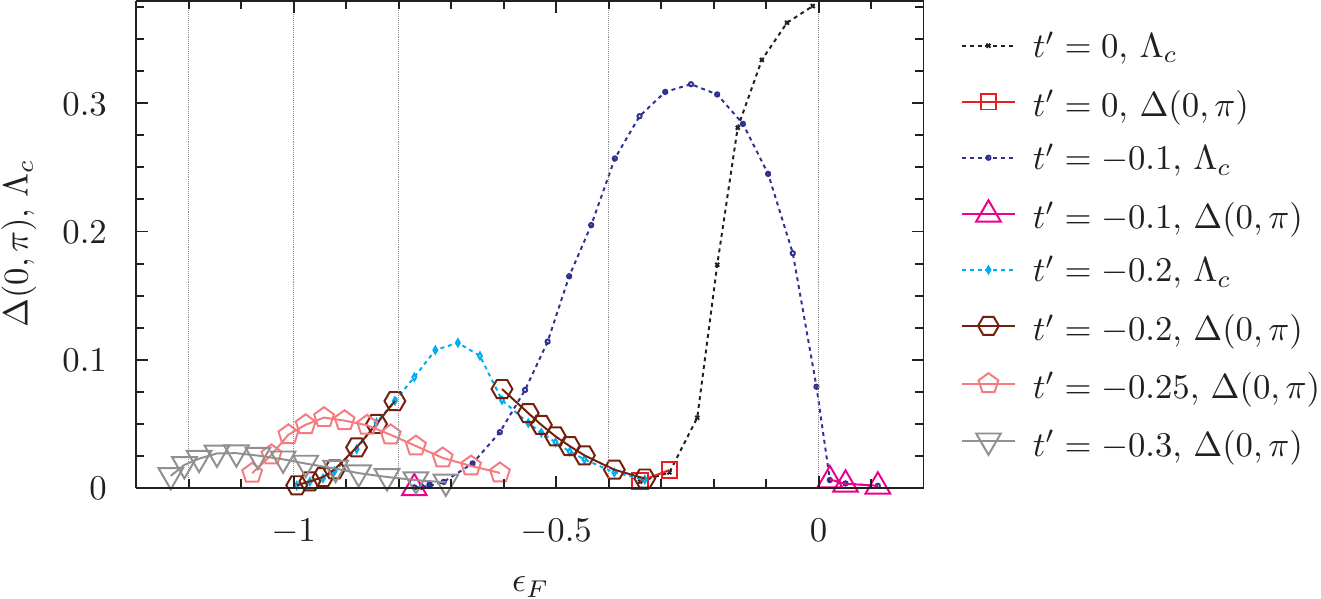}\\[2mm]
	\includegraphics[width=0.49\linewidth]{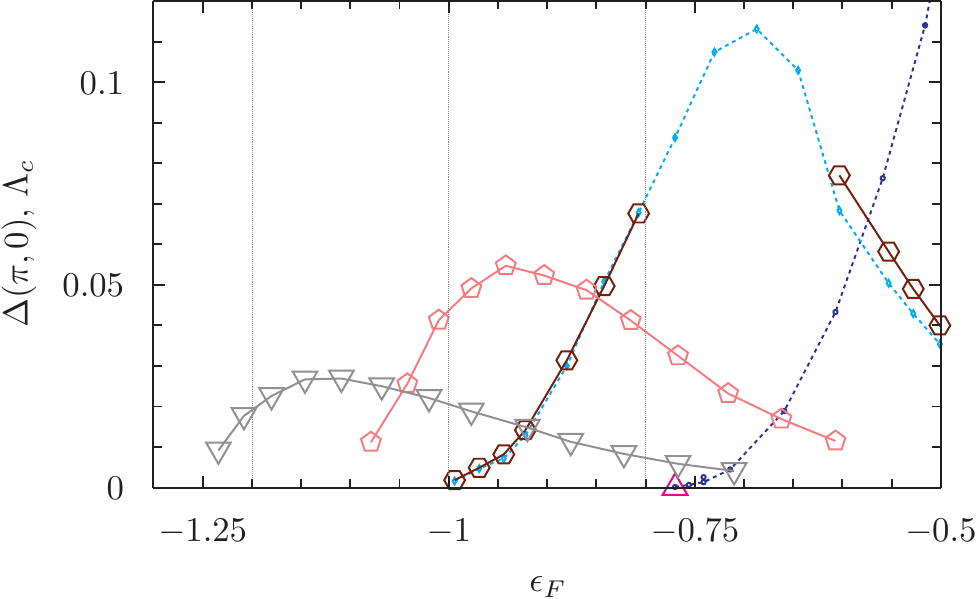}\hspace*{\fill}\includegraphics[width=0.46\linewidth]{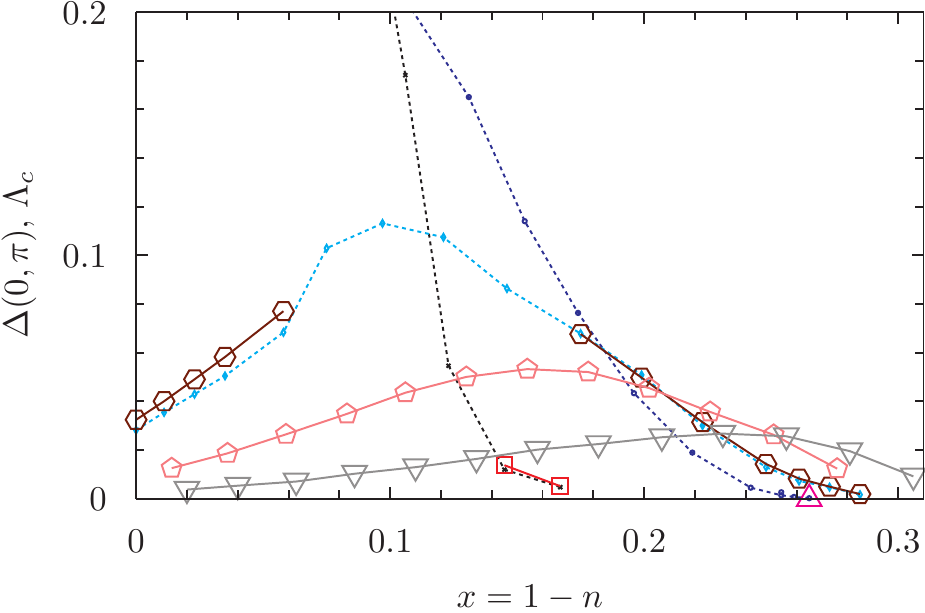}
	\caption{Critical scale $\Lambda_c$ and $d$-wave pairing gap $\Delta(\boldsymbol k = (0,\pi))$ for $U = 3$ and various $t'$ as a function of the Fermi level $\epsilon_F$ (upper panel and enlarged view on region with van Hove points in lower left panel) and of the hole doping away from half-filling $x = 1 - n$ (lower right panel). Regions where only $\Lambda_c$ (dashed lines) is shown have dominant (incommensurate) antiferromagnetic correlations. Regions where $\Delta(0,\pi)$ (solid lines with symbols) is shown have a $d$-wave superfluid ground state.}
	\label{fig:RH:PD3_tp}
\end{figure}

The approximate equality of the critical scale for $d$-wave pairing $\Lambda_c$ and of the maximal $d$-wave gap at the end of the flow $\Delta(0,\pi)$ can be seen in figures~\ref{fig:RH:PD2.5_tp} and~\ref{fig:RH:PD3_tp}. These figures show the dependence of the critical scales $\Lambda_c$ for superfluidity or antiferromagnetism and of the maximal pairing gap $\Delta(0,\pi)$ on the Fermi level $\epsilon_F$ for various next-nearest neighbour hoppings $t'$ for $U = 2.5$ and $U = 3$, respectively. In case superfluidity is the leading instability, either $\Delta(0,\pi)$ and $\Lambda_c$ are shown together or only $\Delta(0,\pi)$ is shown for the sake of clarity of the plot. In case antiferromagnetism is the leading instability, only the critical scale is shown. 
The critical scales could also be obtained from the functional RG for the symmetric phase. Those for antiferromagnetism were determined for comparison with other renormalization group schemes and for the substantiation of the discussion. Figures~\ref{fig:RH:PD2.5_tp} and~\ref{fig:RH:PD3_tp} are discussed in detail in the following together with representative flows of self-energies and exchange propagators.

\begin{figure}
	\centering
	\includegraphics[width=0.49\linewidth]{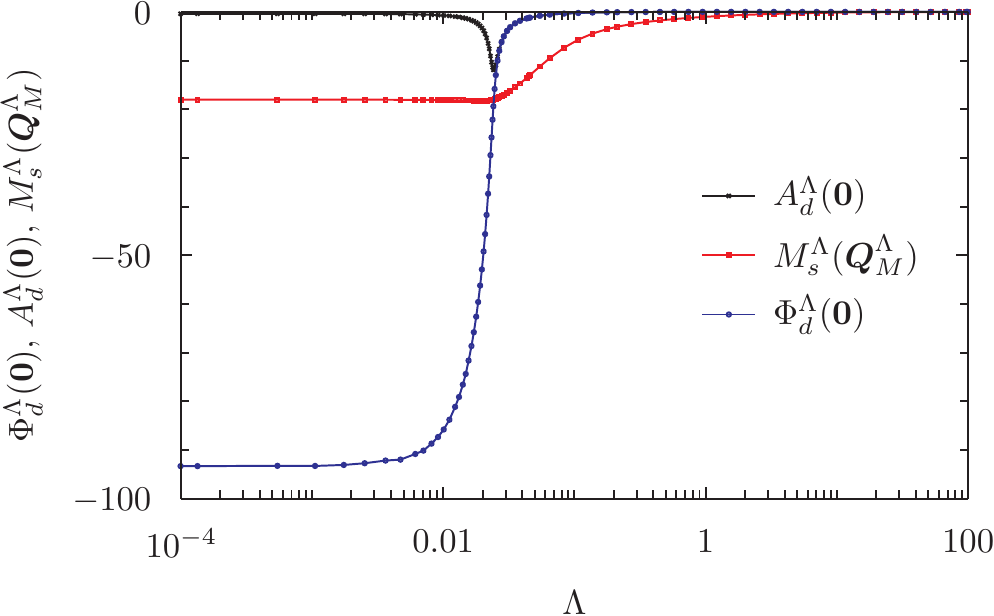}\hspace*{\fill}\includegraphics[width=0.49\linewidth]{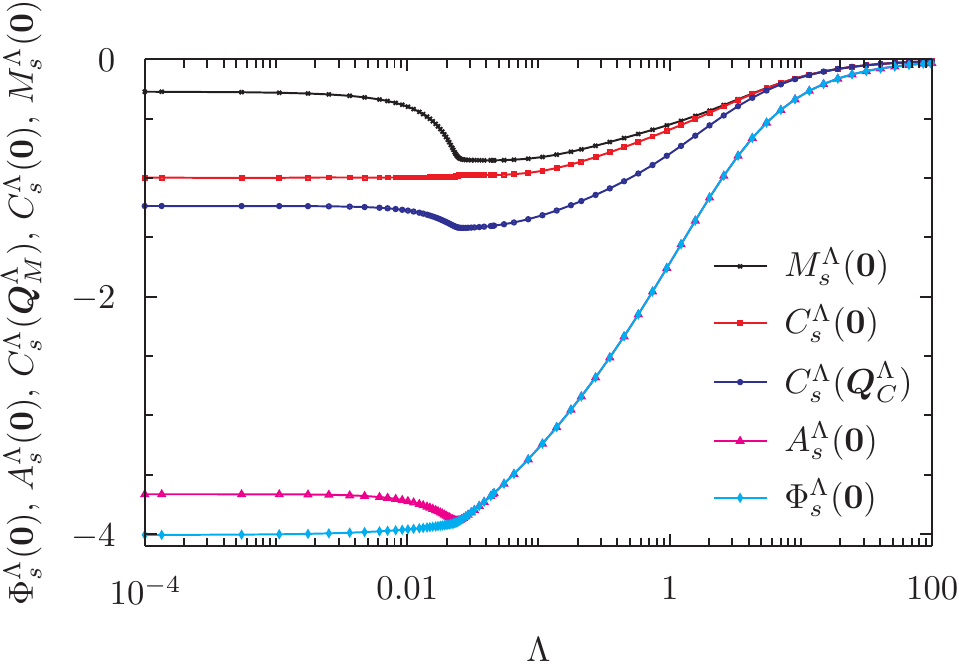}
	\caption{Representative flow to the $d$-wave superfluid state showing the scale dependence of leading interactions in the left panel and of subleading interactions in the right panel. Parameters are $U = 2.5$, $t' = -0.2$ and $\mu = 0.4$ corresponding to $\epsilon_F = -0.72$ or $n = 0.88$ and $\Delta_{(0)} = 5\cdot 10^{-5} \approx \Delta/250$.}
	\label{fig:RH:APhiM_U2c5_tp0c2}
\end{figure}
\begin{figure}
	\centering
	\includegraphics[width=0.6\linewidth]{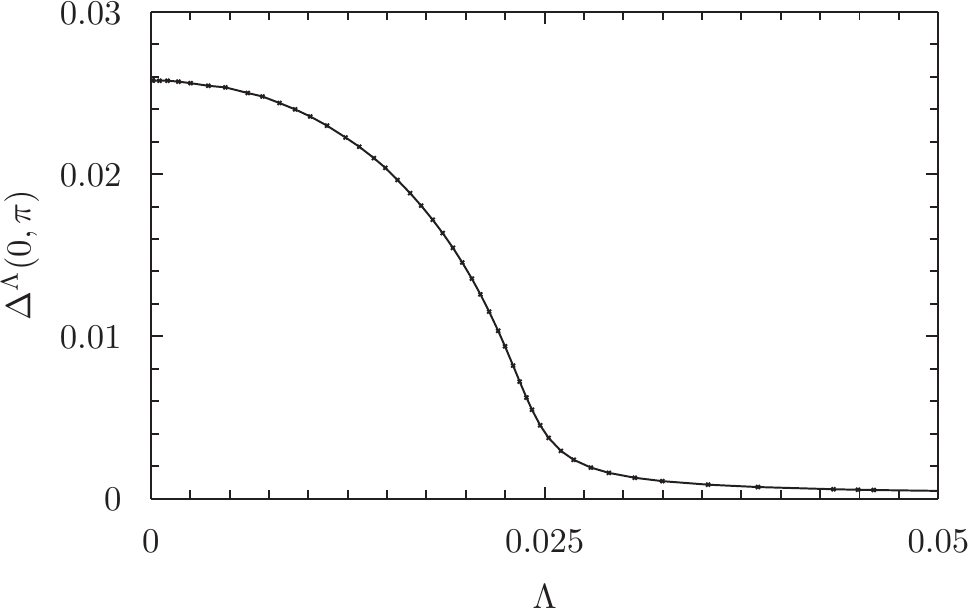}
	\caption{Flow of the $d$-wave pairing gap $\Delta(\boldsymbol k = (0,\pi))$ for the same parameters as in figure~\ref{fig:RH:APhiM_U2c5_tp0c2}.}
	\label{fig:RH:Propagators_U25_tp02}
\end{figure}
For $U = 2.5$, the dependence of critical scales and gaps on the Fermi level was determined for $t' = -0.1$, $-0.2$ and $-0.3$, and is shown in figure~\ref{fig:RH:PD2.5_tp}. For $t' = -0.1$, only instabilities towards incommensurate antiferromagnetism were found for fermionic densities below van Hove filling. It is expected that $d$-wave superfluidity appears for smaller $\epsilon_F$, but there the incommensurate peaks in the magnetic exchange propagator become very sharp and hard to describe accurately using the employed discretisation of momentum space. The behaviour for $t' = -0.2$ is more interesting. For Fermi levels slightly above van Hove filling, the system is unstable towards (commensurate) antiferromagnetism while it shows $d$-wave superfluidity for smaller and larger fermionic densities (or Fermi levels). 
Mean-field theory predicts commensurate antiferromagnetic order for $U = 2.25$, $t' = -0.2$ and $4t' < \epsilon_F < 0$~\cite{Reiss2006} and this is expected to be similar for $U = 2.5$. Above van Hove filling, antiferromagnetic order is found in mean-field theory if $U$ exceeds a critical value. On one-loop level, fluctuations suppress the effective interaction so that the antiferromagnetic instability disappears when $\epsilon_F$ is sufficiently larger than $\epsilon_F^{vH}$. The antiferromagnetic fluctuations are nevertheless strong and drive an instability towards $d$-wave pairing. 
This can be seen in figure~\ref{fig:RH:APhiM_U2c5_tp0c2}, which shows the flow of the exchange propagators in the $d$-wave pairing channel for $\boldsymbol q = \boldsymbol 0$ as well as the magnetic exchange propagator at $\boldsymbol q = \boldsymbol \pi$ in the left panel and the flow of some subleading interactions\footnote{In the parameter regime studied in this chapter, other subleading exchange propagators like those in the $d$-wave particle-hole channel remained considerably smaller than $U$ on all scales and are therefore not shown. The scale dependence of the (singular) anomalous exchange propagator $X^\Lambda_{sd}(\boldsymbol q)$ at $\boldsymbol q = \boldsymbol 0$ is qualitatively similar to that of $X^\Lambda(0)$ in the attractive Hubbard model, but it also remained small.} in the right panel for $U = 2.5$, $t' = -0.2$ and $\epsilon_F = -0.72$ (slightly above van Hove filling). 
Note that the exchange propagators in the $s$-wave pairing channel $A^\Lambda_s$ and $\Phi^\Lambda_s$ are of the order of $U$ at low scales. They weaken the (repulsive) $s$-wave component of the pairing vertex and furthermore oppose the tendency of the system to become antiferromagnetic. The flow of the superfluid gap for the same parameters as in figure~\ref{fig:RH:APhiM_U2c5_tp0c2} is shown in figure~\ref{fig:RH:Propagators_U25_tp02}. For the larger $t' = -0.3$, no instability towards antiferromagnetism was found. Due to the stronger frustration of antiferromagnetic fluctuations, a larger effective interaction would be required to drive a magnetic instability. The system therefore has a $d$-wave superfluid ground state with the maximum of $\Delta(0,\pi)$ occurring slightly above van Hove filling. This is different from the weak-coupling behaviour in the limit $U\rightarrow 0$ as determined by Raghu~\etal~\cite{Raghu2010}. 
Instead of a maximum at van Hove filling, Raghu~\etal\ find a pronounced dip in the pairing strength in the $d$-wave channel, and $p$-wave pairing becomes the dominant instability\footnote{Note however that the approach by Raghu~\etal\ is not valid right at van Hove filling, because they treat the particle-hole channel perturbatively and only the particle-particle channel within the perturbative renormalization group. At van Hove filling, \emph{all} channels have to be treated on equal footing due to competition of instabilities~\cite{Zanchi2000}. Treating all channels in an unbiased way for $U = 1$, the functional RG study by Halboth and Metzner~\cite{Halboth2000b} yielded a \emph{maximum} of $\Lambda_c$ at van Hove filling.}.

\begin{figure}
	\centering
	\subfigure{\includegraphics[scale=0.8]{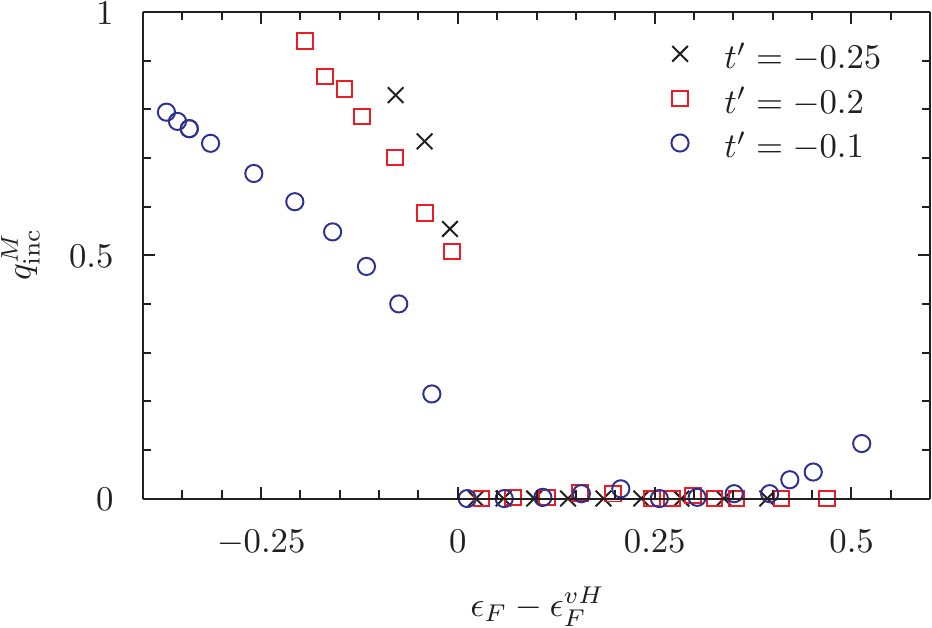}}\hspace*{0.04\linewidth}\subfigure{\includegraphics[scale=0.8]{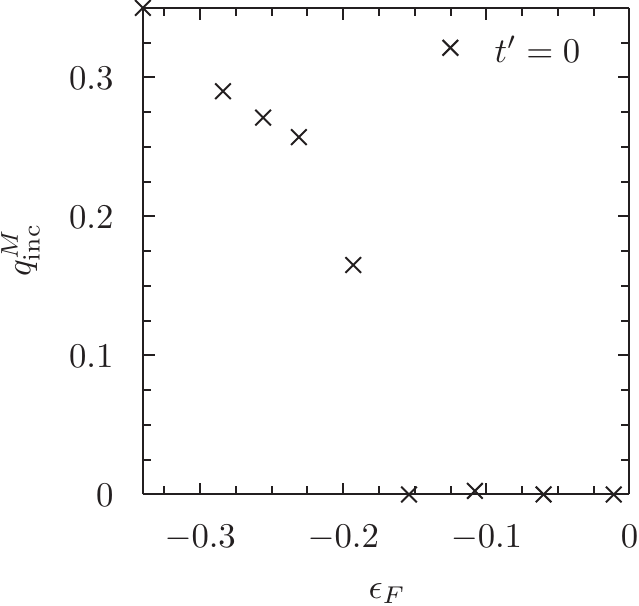}}
	\caption{Distance of the incommensurate peaks $\boldsymbol Q_M = (\pi \pm q^M_\text{inc}, \pi)$, $(\pi, \pi \pm q^M_\text{inc})$ of the magnetic exchange propagator $M^\Lambda_s(\boldsymbol q)$ from $\boldsymbol q = \boldsymbol \pi$ as a function of the Fermi level measured as $\epsilon_F - \epsilon^{vH}_F$ for $U = 3$ and various $t'$ either for $\Lambda = 0$ or $\Lambda_c$. $q^M_\text{inc}$ is only an estimate in case incommensurate antiferromagnetism is the leading instability. The right plot is symmetric around $\epsilon_F = 0$.}
	\label{fig:RH:Qinc_epsF}
\end{figure}

The dependence of $\Lambda_c$ and of $\Delta(0,\pi)$ on $\epsilon_F$ for $U = 3$ is shown for various $t'$ in the upper panel of figure~\ref{fig:RH:PD3_tp}. The lower left panel shows an enlarged view on the region around the van Hove points at $\epsilon_F^{vH} = 4t'$. The lower right panel shows the same data as the upper panel, but as a function of the hole doping away from half-filling $x = 1 - n$. The dependence of $\Lambda_c$ and of $\Delta(0,\pi)$ on $\epsilon_F$ is qualitatively similar to the case of $U = 2.5$ shown in figure~\ref{fig:RH:PD2.5_tp}. However, the extension of antiferromagnetic or $d$-wave superfluid regions changes considerably when going to the larger interaction. For $t' = 0$ and $t' = -0.1$, antiferromagnetism is the leading instability in a broad range of fermionic densities or Fermi levels around van Hove filling\footnote{Note that the results for $t' = 0$ are symmetry around $\epsilon_F = 0$ due to particle-hole symmetry.}. 
For $t' = -0.2$, antiferromagnetism is found in a small region of fillings above van Hove filling. The instabilities are towards commensurate antiferromagnetism for $\epsilon_F^{vH} < \epsilon_F < 0$ and towards incommensurate antiferromagnetism otherwise. The character of the dominant antiferromagnetic fluctuations is determined in the same way except for $t' = 0$. This can be seen in figure~\ref{fig:RH:Qinc_epsF} that shows the distance $q^M_\text{inc}$ of the incommensurate peaks in $M^\Lambda_s(\boldsymbol q)$, which are located on the Brillouin zone boundary, from $\boldsymbol q = \boldsymbol \pi$ as a function of the Fermi level measured from van Hove filling, $\epsilon_F - \epsilon_F^{vH}$. For $t' = 0$, the transition from commensurate to incommensurate antiferromagnetic correlations appears at a finite doping away from half-filling, consistent with results by Krahl~\etal~\cite{Krahl2009b}. 
For $t' = 0$, $-0.1$ and $-0.2$, the ground state is a $d$-wave superfluid when the critical scales for antiferromagnetism become small on both sides of the antiferromagnetic dome. It is interesting to note that the $d$-wave pairing gaps for $\epsilon_F > \epsilon_F^{vH}$ (above the antiferromagnetic dome) are slightly larger than those found for $\epsilon_F < \epsilon_F^{vH}$.  For $t' = -0.25$ and $-0.3$, no antiferromagnetic instability is found. 
The ground state shows $d$-wave superfluidity for all chemical potentials considered, but antiferromagnetic fluctuations are nevertheless strong (in particular for $t' = -0.25$). The maximal values of $\Lambda_c$ and of $\Delta(0,\pi)$ are found in all cases for Fermi levels considerably above van Hove filling. Besides, the superfluid domes are asymmetric around their maximum and $\Delta(0,\pi)$ drops slower when increasing the Fermi level beyond that of the maximum.

The behaviour described above can be understood from properties of the Fermi surface. Of particular interest are the saddle points of the dispersion at $\boldsymbol k = (0,\pm \pi)$, $(\pm\pi,0)$ and the so-called hot spots where the Fermi surface intersects the umklapp surface. The latter consists of the lines that connect $(0,\pi)$ with $(\pi,0)$ and similarly for symmetry related points, as shown in figure~\ref{fig:Intro:FermiSurfaces}. At the saddle points, the gradient of the fermionic dispersion vanishes, giving rise to a logarithmically singular density of states and stronger singularities of the particle-hole and particle-particle bubbles. These enhance scattering processes with transfer momenta $\boldsymbol q = \boldsymbol 0$ or $\boldsymbol \pi$ that connect the saddle points, leading to a competition of instabilities at low energy scales~\cite{Schulz1987,Lederer1987,Dzyaloshinskii1987} or even to a truncation of the Fermi surface at the hot spots~\cite{Furukawa1998}.
Hot spots exist for Fermi levels between $\epsilon_F = \epsilon_F^{vH}$ (where they coincide with the saddle points) and $\epsilon_F = 0$ (where they merge on the Brillouin zone diagonals). Their existence enlarges the low energy phase space for scattering processes with momentum transfers $\boldsymbol q = \boldsymbol \pi$ and allows for umklapp scattering between Fermi points. The relevance of the interplay between hot spots and saddle points has been pointed out by Honerkamp~\etal~\cite{Honerkamp2001a,Honerkamp2002}.

\begin{figure}
	\centering
	\includegraphics{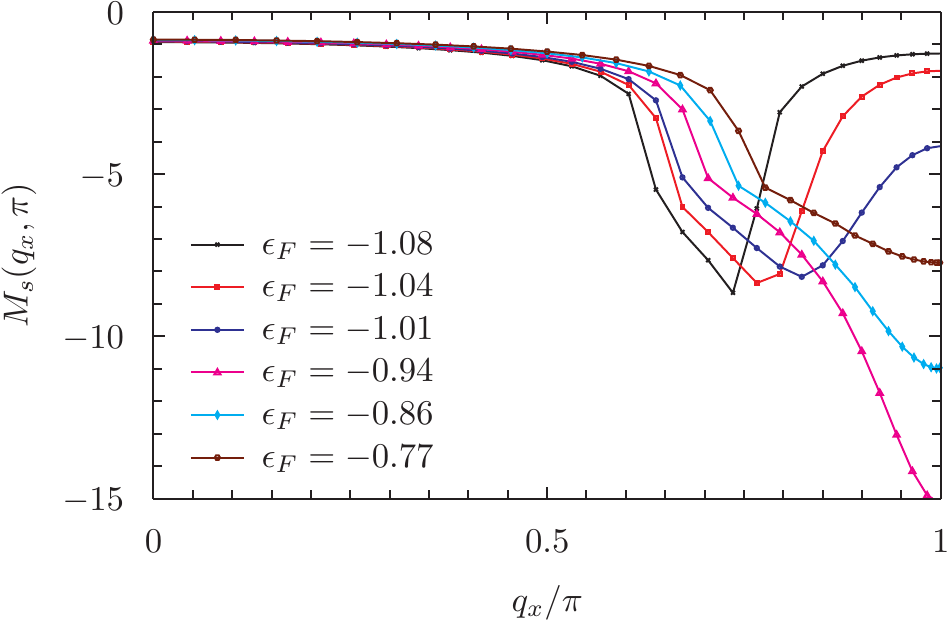}
	\caption{Cuts through the magnetic exchange propagator $M_s(\boldsymbol q)$ along the Brillouin zone boundary at the end of the flow for $U = 3$ and $t' = -0.25$.}
	\label{fig:RH:M_qx_U3_tp-0c25}
\end{figure}

For $\epsilon_F < \epsilon_F^{vH}$, the Fermi surface lies inside the umklapp surface and no hot spots exist. For $t' = 0$ and $-0.1$, the approximate nesting of the Fermi surface leads to large critical scales for (incommensurate) antiferromagnetism and $d$-wave pairing becomes the leading instability only far below van Hove filling. For the larger negative values of $t'$, magnetic fluctuations are increasingly frustrated due to the stronger curvature of the Fermi surface and the resulting inferior approximate nesting. For $t' \leq -0.2$, the effective interaction is not strong enough to drive an instability towards incommensurate antiferromagnetism. The nevertheless strong antiferromagnetic fluctuations give rise to $d$-wave pairing with sizeable gaps. The critical scales shrink quickly when going further away from van Hove filling because of the worsening coupling of the incommensurate antiferromagnetic fluctuations to the $d$-wave pairing channel. 
The latter is caused by the increasingly large distance of the incommensurate peaks from $\boldsymbol q = \boldsymbol \pi$, which is shown in figure~\ref{fig:RH:Qinc_epsF} and for $t' = -0.25$ also in figure~\ref{fig:RH:M_qx_U3_tp-0c25}.

\begin{figure}
	\centering
	\includegraphics[width=0.42\linewidth]{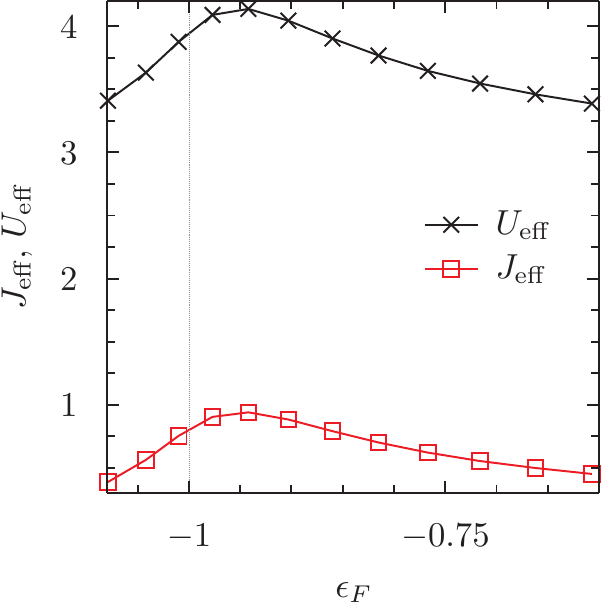}\hspace*{0.04\linewidth}\includegraphics[width=0.44\linewidth]{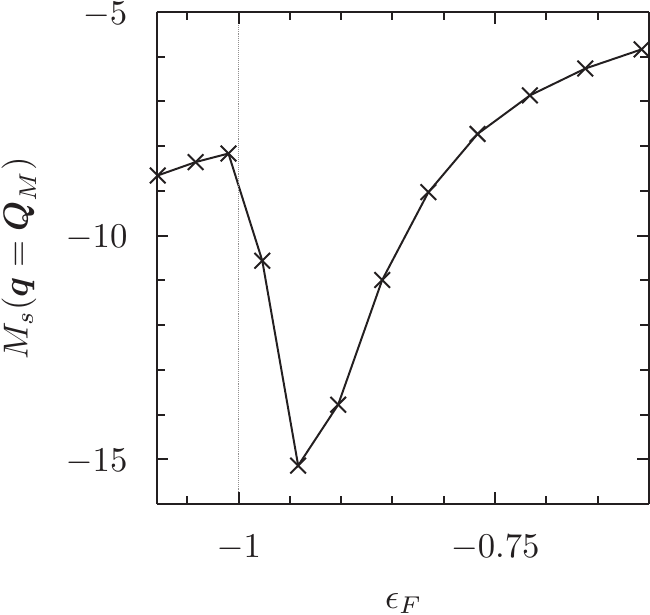}
	\caption{Renormalized local repulsion $U_\text{eff}$ and nearest-neighbour exchange integral $J_\text{eff}$ (left panel) as well as value of the magnetic exchange propagator $M_s(\boldsymbol q)$ at the (in-) commensurate peak (right panel) at the end of the flow as a function of the Fermi level $\epsilon_F$ for $U = 3$ and $t' = -0.25$. The grey dashed vertical lines mark van Hove filling.}
	\label{fig:RH:UeffJeff_MinM}
\end{figure}
For $\epsilon_F > \epsilon_F^{vH}$, the dependence of the critical scales and gaps on the Fermi level is characterized by the existence of hot spots and their interplay with the saddle point regions. As mentioned above, scattering processes between the hot spots and between the saddle points enhance effective interactions with momentum transfer $\boldsymbol q = \boldsymbol \pi$. In the presence of a sizeable interaction $U$, the saddle points can significantly contribute to the flow at intermediate scales although they lie below the Fermi level. The flow is therefore not only driven by the saddle points and hot spots, but by an entire region around them. The increased phase space outbalances the growing distance between the saddle points and the Fermi level as well as the decreasing density of states at the Fermi level, leading to a strong mutual reinforcement of different channels, in particular between antiferromagnetic and $d$-wave pairing fluctuations. 
This can be seen in figure~\ref{fig:RH:UeffJeff_MinM} that shows the renormalized local repulsion $U_\text{eff}$ and the effective nearest-neighbour exchange interaction $J_\text{eff}$ as extracted from the vertex\footnote{These quantities are defined by
\begin{equation}
	\begin{split}
	U^\Lambda_\text{eff} &= \intzwei{k} \intzwei{k'} \intzwei{q} V^{S,\Lambda}_{k+q/2, k'-q/2, k'+q/2, k-q/2} |_{k_0 = k_0' = q_0 = 0}\\
		&= \intzwei{q} \Bigl[U + \tfrac{1}{2} \bigl(A^\Lambda_s(\boldsymbol q) + \Phi^\Lambda_s(\boldsymbol q)\bigr) + C^\Lambda_s(\boldsymbol q) - 3 M^\Lambda_s(\boldsymbol q)\Bigr]
	\end{split}
\end{equation}
\begin{align}
		J^\Lambda_\text{eff} &= \tfrac{1}{12} \intzwei{k} \intzwei{k'} \intzwei{q} (\cos q_x + \cos q_y) \sum_{\sigma_i} \vec \tau_{\sigma_4 \sigma_1} \cdot \vec \tau_{\sigma_3 \sigma_2} \times \\
	&\quad\times (V^\Lambda_{k+q/2, k'-q/2, k'+q/2, k-q/2} \delta_{\sigma_1 \sigma_4} \delta_{\sigma_2 \sigma_3} - V^\Lambda_{k+q/2, k'-q/2, k-q/2, k'+q/2} \delta_{\sigma_1 \sigma_3} \delta_{\sigma_2 \sigma_4})|_{k_0 = k_0' = q_0 = 0}\nonumber\\
	&= \intzwei{q} (\cos q_x + \cos q_y) M^\Lambda_s(\boldsymbol q) - \tfrac{1}{4} \intzwei{q} \Bigl[\tfrac{1}{2} \bigl( A^\Lambda_d(\boldsymbol q) + \Phi^\Lambda_d(\boldsymbol q)\bigr) + C^\Lambda_d(\boldsymbol q) - M^\Lambda_d(\boldsymbol q)\Bigr].\nonumber
\end{align}}
at the end of the flow in the left panel and the value of $M_s(\boldsymbol q)$ at the (in-) commensurate peak in the right panel (showing in parts the same data as in figure~\ref{fig:RH:M_qx_U3_tp-0c25}). At van Hove filling, the incommensurate peaks in $M_s(\boldsymbol q)$ merge without significantly increasing the value at the peak. However, $M_s(\boldsymbol q = \boldsymbol \pi)$ is strongly enhanced above van Hove filling, and decreases quickly in absolute value after passing through the extremum. $U_\text{eff}$ and $J_\text{eff}$ are largest slightly above van Hove filling, which is a consequence of the enhancement of magnetic fluctuations in the presence of hot spots. $J_\text{eff}$ is not enhanced as strongly as $M_s(\boldsymbol \pi)$, because the growth of the latter occurs mainly via the formation of a pronounced peak at $\boldsymbol q = \boldsymbol \pi$. 
Qualitatively similar results were also obtained for other values of $t'$. The amplification of magnetic fluctuations due to the interplay between saddle points and hot spots is the reason why the maximal critical scales and gaps for antiferromagnetism and $d$-wave pairing occur \emph{above} van Hove filling. The distance between the Fermi level with the maximal critical scale or gap and van Hove filling increases with increasing $U$, as can be seen by comparing figures~\ref{fig:RH:PD2.5_tp} and~\ref{fig:RH:PD3_tp}. The reason is that the saddle point region becomes larger if the hot spots move away from the saddle points, from which the system can profit for stronger interactions. This is different from the weak-coupling regime studied for example by Halboth and Metzner~\cite{Halboth2000b}, where the maximal critical scales occur precisely at van Hove filling. 
The asymmetric shape of the superfluid dome can also be understood in the above saddle point plus hot spot scenario. This picture suggests that the magnetic exchange propagator for $\epsilon_F > \epsilon_F^{vH}$ should consist of a broad contribution around $\boldsymbol q = \boldsymbol \pi$ and a peak at this transfer momentum. This is indeed the case as can be seen in figure~\ref{fig:RH:M_qx_U3_tp-0c25} for $\epsilon_F > 4t'$ ($=-1.0$ for the parameters in the figure). With increasing $\epsilon_F > \epsilon_F^{vH}$, the peak decreases while the broad contribution is almost unchanged, giving rise to the slower decrease of the pairing gap for Fermi levels beyond that for the maximal gap for $t' = -0.3$ and $-0.25$. Note that the broad contribution in $M_s(\boldsymbol q)$ is more pronounced for $t' = -0.3$ than for $t' = -0.25$. 
For the smaller $t' = -0.1$ and $-0.2$, it is not visible and $M_s(\boldsymbol q)$ just shows a (nevertheless broad) peak at $\boldsymbol q = \boldsymbol \pi$. The above mechanism for the interplay between saddle points and hot spots becomes ineffective at the upper end of the antiferromagnetic dome for $t' = -0.1$, thus giving rise to the fast drop of the critical scale. Note that this drop takes place close to the value of $\epsilon_F$ where the hot spots disappear. The vanishing of the antiferromagnetic instability for Fermi levels above van Hove filling at $t' = -0.1$ and $-0.2$ again gives way to $d$-wave pairing, which is caused by the nevertheless strong antiferromagnetic fluctuations in this parameter region. 
Note that $d$-wave superfluidity was also detected on both sides of the antiferromagnetic instability in the weak-coupling regime by Halboth and Metzner~\cite{Halboth2000b} within the functional RG and in the intermediate correlation regime by Kancharla~\etal~\cite{Kancharla2008} within cellular DMFT\@. 
Although the last two paragraphs referred mostly to the results for $U = 3$, the interpretation applies equally well to the case of $U = 2.5$.

The question arises whether phase fluctuations play a role in the suppression of $\Lambda_c$ or $\Delta(0,\pi)$ for fillings above the one with the maximal gap for the values of $U$ considered in this chapter. Such a mechanism was suggested for the suppression of the critical temperature for superconductivity in underdoped cuprates by Emery and Kivelson~\cite{Emery1995a,Emery1995b}. For the ground state, the inspection of the flow equations shows that the effect of quantum phase fluctuations in the repulsive Hubbard model on one-loop level is similar to that in the attractive case (see subsection~\ref{subsec:AH:SelfenergyOneLoop}). It is found that quantum phase fluctuations suppress $A^\Lambda_d$ and therewith the growth of $\Delta^\Lambda$ just below the critical scale while they tend to enhance the growth of $\Delta^\Lambda$ for $\Lambda\rightarrow 0$. The net effect of quantum phase fluctuations on one-loop level is therefore not obvious, but is expected to be small in the coupling range considered. However, it is reasonable that thermal phase fluctuations have a more drastic effect on the critical temperature than quantum phase fluctuations on the ground state gap or critical scale. 
Besides it is worth mentioning that Emery and Kivelson addressed the question about the reduction of the critical temperature for a doped insulator where the condensate density is proportional to the hole concentration and the Coulomb interaction poorly screened. In the coupling range considered in this chapter and in the presence of a sufficiently negative $t'$, the system is in contrast not a doped insulator and $n = 1$ thus not a distinguished filling. At half-filling and for $t' = -0.3$, $-0.25$ or $-0.2$, the system rather is a $d$-wave superfluid, as shown in the lower right panel of figure~\ref{fig:RH:PD3_tp}. 

An interesting question about the phase diagram of the two-dimensional Hubbard model is whether it shows coexistence of antiferromagnetism and $d$-wave superfluidity. Such a behaviour was found by Reiss~\etal~\cite{Reiss2007} within a combination of RG and mean-field theory and by Kancharla~\etal~\cite{Kancharla2008} within cellular dynamical mean-field theory, while Friederich~\etal~\cite{Friederich2011} did not find indications for a coexistence region within the functional RG using dynamical bosonization. The channel-decomposition scheme for symmetry breaking in the Cooper channel does not allow to fully address this issue because the antiferromagnetic phase cannot be accessed. 
However, no secondary instabilities towards antiferromagnetism were found once the system became superfluid, as can be seen in figures~\ref{fig:RH:PD2.5_tp} and~\ref{fig:RH:PD3_tp}. 
For all parameters considered, the antiferromagnetic peak in $M^\Lambda_s(\boldsymbol q)$ either stopped to grow, continued to grow only slowly or was even suppressed slightly when the superfluid gap opened. Superfluid regions with antiferromagnetism as secondary instability should therefore be smaller than the distance between the closest points with dominant instability towards antiferromagnetism or $d$-wave pairing in figures~\ref{fig:RH:PD2.5_tp} and~\ref{fig:RH:PD3_tp}. This result may however be sensitive to improvements of the approximations like the inclusion of (dynamical) self-energy corrections or the averaging scheme for the extraction of the renormalization contributions to the exchange propagators.

An important issue about superfluidity in the repulsive Hubbard model is whether a more negative next-nearest neighbour hopping $t'$ enhances the tendency towards pairing as well as the critical temperature. It is related to the problem whether a more negative $t'$ or a larger hopping range may increase the critical temperature for superconductivity in cuprate superconductors. This was suggested by Raimondi~\etal~\cite{Raimondi1996,Feiner1996} and Pavarini~\etal~\cite{Pavarini2001} after phenomenological analyses of multi-band models for various cuprate materials. The variational Quantum Monte Carlo study of the one-band Hubbard model by Yamaji~\etal~\cite{Yamaji1998} supports this phenomenological trend and indicates the existence of an optimal $t'$ for pairing\footnote{In the sense that the condensation energy is maximal. The largest gap is found for $t' \approx -0.2$.} between $-0.1$ and $-0.15$ for $U = 8$ and $n = 0.84$. 
However, numerical studies of the repulsive Hubbard model within the DCA on small clusters by Maier~\etal~\cite{Maier2000} or of the $t$-$t'$-$J$ model within the density matrix renormalization group (DMRG) for few-leg ladders by White~\etal~\cite{White1999} or Martins~\etal~\cite{Martins2001} yield the opposite trend. 
In functional RG studies at van Hove filling~\cite{Honerkamp2001b,Husemann2009}, a certain threshold value for $-t'$ is necessary in order to find instabilities towards $d$-wave superfluidity in the ground state, because antiferromagnetism is the leading instability for small $-t'$. Above this threshold value, the critical scale for superfluidity decreases when $-t'$ increases. At van Hove filling, a similar trend is also found in this work. Considering also fermionic densities away from van Hove filling, the largest gaps arise for $t' = -0.2$. For smaller and larger values of $t'$, the maximal (as a function of the density) superfluid gaps decrease.

It was mentioned above that on mean-field level, the critical scale for antiferromagnetism for $t' = 0$ and $\epsilon_F = 0$ equals the antiferromagnetic gap. Assuming that this equality holds also on one-loop level, the resulting estimate for the antiferromagnetic gap can be compared with the result of other methods. Figure~\ref{fig:RH:PD3_tp} shows that $\Lambda_c \approx 0.38$ for $U = 3$. This value can be compared with the mean-field gap $\Delta^\text{AF}_\text{MF} = 0.85$ and the critical scale $\Lambda_c^\text{RPA} = 0.87$ in the absence of fluctuation contributions. The latter two differ due to the finite right hand side of the stopping criterion~\eqref{eq:RH:StoppingCriterion}. It follows that $\Lambda_c / \Lambda_c^\text{RPA} \approx 0.44$, suggesting that the antiferromagnetic gap is reduced to roughly one half of the mean-field value due to fluctuations on one-loop level. 
This result compares well with the corresponding ratio for the attractive Hubbard model close to half-filling in chapter~\ref{chap:AttractiveHubbard} and is slightly larger than the ratio for the antiferromagnetic order parameter as obtained by Hirsch~\cite{Hirsch1985} from QMC simulations of the half-filled repulsive Hubbard model. Furthermore, the estimate for the size of the antiferromagnetic gap is in very good agreement with the result obtained by Bauer~\etal~\cite{Bauer2009} using DMFT\@. For $U = 2$,  $t' = 0$ and $n = 1$, the channel-decomposition scheme yields a critical scale for antiferromagnetism $\Lambda_c \approx 0.17$, which is in good agreement with the result by Reiss~\cite{Reiss2006} from a combination of RG and mean-field theory or with the above-mentioned studies.

The repulsive Hubbard model was studied by several authors within the fermionic RG in the symmetric phase in a parameter regime similar to the one in this work. Using the $N$-patch approximation, the model was studied by Honerkamp and Salmhofer~\cite{Honerkamp2001b} at van Hove filling and by Honerkamp~\etal~\cite{Honerkamp2001a}, Honerkamp~\cite{Honerkamp2001c} and Katanin~\cite{Katanin2010} away from van Hove filling. Using the channel-decomposition scheme, it was studied at van Hove filling by Husemann and Salmhofer~\cite{Husemann2009}, Husemann~\etal~\cite{Husemann2012} as well as Giering and Salmhofer~\cite{Giering2012}. In these studies, the leading instabilities for $U = 3$ were either found to be the same, or discrepancies appearing close to the boundaries of regions with different dominant instabilities might be attributed to the choice of different regularization schemes or stopping criteria. 
The leading instabilities found in this work are in good agreement with the result of the above-mentioned RG studies. 
However, the critical scales differ (also between the above-mentioned works) by a factor of up to three, which can at least in parts be attributed to the choice of different regularization schemes or stopping criteria\footnote{Discrepancies in the critical scales of up to a factor of two due to the choice of different regularization schemes were also noted by Honerkamp~\etal~\cite{Honerkamp2004}.}. 
The extension of the $d$-wave superfluid phase for $U = 3$ and $t' = -0.3$ as a function of the Fermi level in figure~\ref{fig:RH:PD3_tp} is similar to that found by Honerkamp and Salmhofer~\cite{Honerkamp2001a}. Friederich~\etal~\cite{Friederich2011} studied the repulsive Hubbard model at finite temperatures using dynamical bosonization. They allowed for symmetry breaking in the $d$-wave pairing as well as the antiferromagnetic channel and presented a finite temperature phase diagram for $U = 3$ and $t' = -0.1$. The Fermi level that marks the transition from a $d$-wave superfluid to an antiferromagnetic state in the work by Friederich~\etal\ at higher temperatures compares well with the one shown in figure~\ref{fig:RH:PD3_tp}. However, at the lowest temperatures they find a superfluid phase with a sizeable extension while the one found in this work is rather small. 
Furthermore the critical scales for antiferromagnetism as shown in figure~\ref{fig:RH:PD3_tp} are larger than the pseudocritical temperatures found by Friederich~\etal, while the superfluid gaps for this value of $t'$ in figure~\ref{fig:RH:PD3_tp} are much smaller than their critical temperatures for superfluidity. 

In variational QMC studies by Yamaji~\etal~\cite{Yamaji1998} or Baeriswyl~\etal~\cite{Baeriswyl2009}, $d$-wave superfluidity was found in the ground state of the Hubbard model with $t' = 0$ for $U = 4$ and $n = 0.84$ in the former and for $U = 8$ and $0.05 \lesssim x \lesssim 0.2$ in the latter. Besides, Yamaji~\etal\ found that commensurate spin-density wave order breaks down for $n < 0.84$ at $U = 8$ and $t' = -0.1$, and is replaced by $d$-wave pairing for smaller fillings. For $U = 4$, $t' = 0$ and at a small but finite temperature, the self-consistent FLEX approximation predicts a rather quick breakdown of antiferromagnetism when the system is doped away from half-filling, followed by a sizeable $d$-wave superfluid phase between $x\sim 0.06$ and $\sim 0.18$~\cite{Bickers1989,Bickers1991a} or $d$-wave superfluidity with a sizeable gap and critical temperature for $x = 0.125$~\cite{Monthoux1994,Pao1994}. 
Cluster extensions of DMFT like DCA~\cite{Maier2000,Maier2005} or cellular DMFT~\cite{Kancharla2008, Sentef2011} find $d$-wave superfluidity in a broad range of interactions, dopings and next-nearest neighbour hoppings at a finite temperature on small clusters. In particular, both methods find a sizeable critical temperature or order parameter for $d$-wave pairing even for $t' = 0$ when the system is doped sufficiently away from half-filling. Note that going to a larger value of $U$ may not help to enlarge the parameter space for pairing within the static one-loop approximation (leaving aside the question whether this is justifiable). This can be seen by comparing figure~\ref{fig:RH:PD2.5_tp} and~\ref{fig:RH:PD3_tp}, showing that the parameter space with dominant instability towards antiferromagnetism grows with increasing $U$.

\hyphenation{anti-ferro-mag-ne-tism}
\section{Conclusion: Repulsive Hubbard model}
In this chapter, the channel-decomposition scheme for the renormalization group equations and the vertex in a singlet superfluid was applied to the repulsive Hubbard model. The dependence of the critical scale and the superfluid gap on the microscopic interaction, on the fermionic density and the next-nearest neighbour hopping were computed on one-loop level. This constitutes the first continuation of RG flows within a purely fermionic formalism to a symmetry broken phase that is not captured by mean-field theory.

The channel-decomposition scheme provided a good starting point for an efficient and physically transparent description of the effective interactions. However, the employed approximations leave room for improvements. First, a static approximation was used for the exchange propagators as well as the self-energy and the momentum dependence of the latter was parametrized in a rather simple way. 
Second, the ground state in the parameter regime where antiferromagnetism is the leading instability is not accessible within the present scheme. Third, it would be preferable to extract the renormalization contributions to the exchange propagators from the flow of the effective interactions by averaging external fermionic momenta over the Fermi surface instead of over the entire Brillouin zone. Although the differences in the results for the two projection schemes were found to be small in the attractive Hubbard model in chapter~\ref{chap:AttractiveHubbard}, the choice of the projection scheme may enhance the critical scales and gaps for $d$-wave pairing in the repulsive case because the pairing interaction is generated by fluctuations that mostly involve fermionic low-energy states.

Nevertheless, a good agreement of the leading instabilities and a qualitative agreement of the critical scales with those of other fermionic one-loop RG studies was obtained and the superfluid gaps are in reasonable agreement with results of other methods like DMFT, DCA or FLEX\@. However, in comparison to FLEX or variational QMC, notable differences are present for small values of $t'$ in that the fermionic RG on one-loop level yields a slower breakdown of antiferromagnetism with doping and a weaker tendency towards $d$-wave pairing.

Interestingly, the largest gaps (and critical scales) were not found at, but slightly above van Hove filling. The reason is the interplay between scattering processes that involve the saddle points and the hot spots on the Fermi surface, which leads to enhanced antiferromagnetic fluctuations and a mutual reinforcement of the $d$-wave pairing and the magnetic channel at low scale.

%% file: Thesis_Conclusion.tex
\chapter{Conclusions}
\label{chap:Conclusion}
In this thesis, fluctuation effects in two-dimensional fermionic superfluids at zero temperature were studied. For this purpose, an ansatz for the Nambu two-particle vertex that incorporates symmetries was formulated and its singular dependences on external momenta and frequencies were identified. This allowed for the derivation of channel-decomposed renormalization group equations, which served as the basis for the formulation of approximations for the effective interactions and their efficient computation. The channel-decomposition scheme was applied to the attractive and repulsive Hubbard model. The methodological advancements and the results of applications are summarized together with interesting directions for future research in the following.

\section{Efficient description of vertex and channel-decomposition scheme}
At the heart of this thesis is an ansatz for the fermionic two-particle vertex in a singlet superfluid. By exploiting symmetries, in particular spin rotation invariance, the number of its independent component functions was reduced to only three, which describe normal and anomalous effective interactions. Discrete symmetries yielded further constraints on their momentum and frequency dependence. The singular dependence of these functions on external momenta and frequencies was identified. This allowed for the definition of interaction channels, for which functional renormalization group equations were derived within the one-particle irreducible formalism by assigning renormalization contributions to interaction channels according to their leading singular dependence on external momenta and frequencies. 
The resulting channel-decomposed renormalization group equations constitute an extension of ideas by Karrasch~\etal~\cite{Karrasch2008} and of work by Husemann and Salmhofer~\cite{Husemann2009} first for the description of symmetry breaking in the particle-particle channel and second to the two-loop level. 

The renormalization contributions on one-loop level were assigned to the interaction channels according to the transfer momenta and frequencies in the fermionic loops, because their singularities give rise to a singular dependence of the vertex at and below the critical scale. The renormalization contributions on two-loop level were assigned according to the transfer momenta in the exchange propagators, because the two-loop contributions become important only when a significant momentum dependence has developed in the vertex, \ie\ close to and below the critical scale. In the channel-decomposed flow equations the singular dependence on momenta and frequencies is isolated in one variable per channel, thus providing a good basis for the formulation of approximations for the effective interactions in the channels and their efficient computation. 
The flow equations on two-loop level were reformulated as effective one-loop equations in which the scale-derivatives of the effective interactions on one-loop level appear. This could make the two-loop RG a manageable tool for the investigation of systems with competing instabilities.

The flow to the symmetry broken phase was treated by a two-step procedure. In the first step, the fermionic modes were integrated out in the presence of an external pairing field. This allowed treating the singularities at the critical scale for symmetry breaking while the size of the effective interactions was limited. Subsequently, the external pairing field was considered as a regulator that was removed in a second flow. The external pairing field then mainly acted as a regulator for the phase mode of the superfluid gap, allowing to investigate the effects of phase fluctuations in a controlled way.

The channel-decomposition scheme yielded insight into the singular behaviour of the two-particle vertex in a fermionic superfluid in the limit where the external pairing field vanishes. Estimates of the size of renormalization contributions showed that the singular behaviour on one-loop level is the same as in a resummation of all chains of Nambu particle-hole bubble diagrams. The exact asymptotic infrared behaviour of an interacting Bose gas is therefore not captured within the fermionic one-loop RG\@. The reason is that the appearing diagrams involving Goldstone mode fluctuations are not singular enough to generate the expected divergence of the amplitude mode. This is different on two-loop level. The estimates further indicate that non-Cooper channels do not receive singular feedback from phase fluctuations even on two-loop level.

\section{Attractive Hubbard model}
\label{sec:Disc:AHM}

\subsubsection{Key results}
In chapter~\ref{chap:AttractiveHubbard}, the ground state of the attractive Hubbard model in two dimensions was studied within the channel-decomposition scheme on one- and two-loop level. This extends previous work by Gersch~\etal~\cite{Gersch2008} first by including the frequency dependence of the vertex as well as of the self-energy and second by considering fluctuations on two-loop level. It was demonstrated that the channel-decomposition scheme indeed allows for an efficient and accurate description of the singular dependences of the vertex on momenta and frequencies. From the methodological point of view, the first fermionic two-loop RG study of symmetry breaking as well as the first such study on one- and two-loop level that considered the full frequency dependence of the vertex and the self-energy were performed. 

Using the external pairing field as a regulator proved to be a convenient tool for the study of the effects of phase fluctuations. Within the employed framework of approximations, the external pairing field on one-loop level could be chosen at least three orders of magnitude smaller than the final value of the gap without running into unphysical divergences at finite scales. In pairing field flows a somewhat larger external pairing field could be reduced by almost two orders of magnitude. 
On one-loop level, it was found that the singular behaviour of the vertex in the limit of a vanishing external pairing field is the same as in a resummation of all chains of Nambu particle-hole bubble diagrams, confirming the analytical estimates. On two-loop level, the amplitude mode is strongly enhanced for small external pairing fields in comparison to the one-loop level. However, results for smaller external pairing fields would be required in order to numerically reproduce the expected infrared scaling of the amplitude mode or the absence of singularities in non-Cooper channels that were found in analytical estimates. The combination of the analytical estimates in chapter~\ref{chap:ChannelDecomposition} with the numerical study in chapter~\ref{chap:AttractiveHubbard} provided a comprehensive understanding of the interaction vertex in a fermionic $s$-wave superfluid.

The presented study yielded results for the superfluid gap and its reduction by fluctuations in comparison to the mean-field approximation that are in good agreement with values reported in the literature. The reduction of the gap is caused by particle-hole and collective mode fluctuations.

\subsubsection{Criticism}
The main points of criticism on the application of the channel-decomposition scheme to the attractive Hubbard model concern the discretisation scheme for the exchange propagators. While the 2+1-dimensional discretisation\footnote{Computation of the frequency dependence at a singular momentum and of the momentum dependence at a singular frequency as described in section~\ref{sec:AH:ApproxVertex}.} allowed for an efficient and flexible description of the momentum and frequency dependence of the exchange propagators in the weak-coupling regime, it turned out to be inadequate for larger couplings. However, the problems were of purely technical origin, arose from sign changes or from maxima being renormalized into local minima and could be cured with larger computational power. 
These issues hindered the computation of RG flows for larger microscopic interactions. Furthermore, the results of one- and two-loop pairing field flows at the largest couplings considered depended to some extent on the size of the initial external pairing field, indicating that the approximations for the vertex and the self-energy leave room for improvements.

It should be noted that the study of fluctuation effects in models with only one obviously dominant channel appears to be easier after partial bosonization (via Hubbard-Stratonovich transformations). Functional RG studies for mixed fermion-boson actions that were parametrized with a few running couplings allowed for example to capture the singular infrared behaviour of a fermionic superfluid in the two-dimensional attractive Hubbard model~\cite{Strack2008} or the renormalization of the BCS-BEC crossover by collective mode and particle-hole fluctuations in a three-dimensional continuum model for attractively interacting fermions~\cite{Floerchinger2008b}. However, the presented study made it possible to describe the singular infrared behaviour \emph{and} the renormalization of the superfluid ground state by particle-hole fluctuations in a unified framework at least in the BCS regime.

\subsubsection{Outlook}
An interesting question that was not answered in this thesis concerns the range of interactions where the fermionic one-loop RG for symmetry breaking is applicable. The one-loop scheme by Katanin allows for the exact solution of reduced models and this can be seen as an important ingredient for going to larger interactions. The reason is that the solution of the BCS gap equation together with an adjustment of the fermionic density already captures many aspects of the BCS-BEC crossover~\cite{Eagles1969,Leggett1980,Nozieres1985}. Furthermore, in three-dimensional continuum systems, important collective fluctuation effects on the BEC side are already captured by taking Gaussian fluctuations around the saddle point into account~\cite{Diener2008}. Besides, the thermodynamics of the crossover from large to small pairs is well captured within the self-consistent $T$-matrix approximation~\cite{Keller1999}. 
However, the BCS-BEC crossover in the attractive Hubbard model takes place for microscopic interactions that are comparable to the bandwidth (see for example~\cite{Scalettar1989,Randeria1992,Singer1996,Keller2001,Tamura2012}). 

Under which circumstances systems with such a large microscopic interaction can be treated within simple truncations of the functional RG hierarchy is a subject of debate~\cite{Salmhofer2007}. The issue is that the involved strong coupling problem is different from the one encountered within the fermionic RG for symmetry breaking in the weak-coupling regime, which was addressed in this thesis. In the latter, the renormalization contributions from the ``critical'' exchange propagators or effective interactions are constrained by the available phase space. Furthermore, the most singular part of the flow is similar to that of a reduced model and thus well-captured within the Katanin scheme. 
In contrast, if the microscopic interaction is of the order of the bandwidth, the effective interactions are large from the outset and all states thus strongly coupled. State of the art truncations of the fermionic RG may possibly be justifiable if a single channel is dominant, but not in the case of competition of instabilities~\cite{Salmhofer2001,Metzner2012}. 

The ideal route for the investigation of the question about the applicability of the Katanin scheme therefore seems to be the attractive Hubbard model at low densities. In this case, the momentum dependence of all exchange propagators becomes radially symmetric to a very good approximation and a discretisation of momentum and frequency dependences is possible on equal footing and numerically feasible. This allows to circumvent the technical problems encountered in this thesis and to concentrate on the question of applicability of the truncation. The results for the low-density limit can easily be compared to those of studies of two-dimensional continuum models (see for example~\cite{Bertaina2011}). The question about the range of applicability is certainly also of interest for the two-loop channel-decomposition scheme and can be addressed within a similar framework.

Another problem that merits further examination is the infrared behaviour within the two-loop scheme. In this thesis, it was not possible to numerically confirm that the amplitude mode diverges in the limit of a vanishing external pairing field as expected in a fermionic superfluid or whether non-Cooper channels indeed do not receive singular feedback from phase fluctuations. The fermionic two-loop scheme may be advantageous for studying the singularity structure of the two-particle vertex in a superfluid in comparison to approaches that use partial bosonization because all interaction channels are present during the entire flow. However, tackling this problem requires improvements of the approximations for the vertex, possibly with simplified parametrizations for the singular exchange propagators, so that smaller external pairing fields can be reached in the pairing field flows.

The extension of the work in this thesis to finite temperatures is also of interest. At higher temperatures, a pseudogap phase without symmetry breaking but with gaps in the one-particle and magnetic excitation spectra is expected~\cite{Keller2001,Kyung2001}. A pseudogap can be obtained from the functional RG in the symmetric phase if the full frequency dependence of the fermionic self-energy is taken into account at finite temperatures~\cite{Katanin2004b,Rohe2005b}. At lower temperatures, a Kosterlitz-Thouless phase with quasi long-range superfluid order is expected~\cite{Kosterlitz1973,Kosterlitz1974,Nagaosa1999}. This phase is peculiar because it shows algebraic order parameter correlations due to massless Goldstone excitations without spontaneous symmetry breaking~\cite{Nagaosa1999} and the superfluid density shows a universal jump at the transition point~\cite{Nelson1977}. 

Starting from bosonic effective actions, the Kosterlitz-Thouless transition was studied within the field-theoretic RG~\cite{Amit1980} or within the functional RG~\cite{Graeter1995,Gersdorff2001}. An RG study for a fermionic model was presented by Krahl and Wetterich~\cite{Krahl2007} using the functional RG for a mixed fermion-boson action. Krahl and Wetterich reproduced all characteristics of the Kosterlitz-Thouless transition except the jump in the superfluid density by using a rather simple truncation of the effective action. In the functional RG studies~\cite{Graeter1995,Gersdorff2001,Krahl2007}, a finite renormalized order parameter was generated at intermediate scales, while the unrenormalized order parameter flowed to zero in the infrared. 

This gives a first idea about how to cope with the technical challenges associated with the Kosterlitz-Thouless phase, namely the jump in the superfluid density at the transition point and the formation of a massless Goldstone mode without spontaneous symmetry breaking, within the fermionic RG\@. They may be accomplished by generating an anomalous self-energy at finite scales, which subsequently has to be suppressed to zero by fluctuations. In the fermionic RG, a finite external pairing field would be required in order to regularize the Goldstone fluctuations and has to be sent to zero in a pairing field flow. These considerations indicate the need to take into account the full momentum and frequency dependence of the normal self-energy in order to be able to describe gapped fermionic degrees of freedoms in the absence of an anomalous expectation value.

\section{Repulsive Hubbard model}
\label{sec:Disc:RHM}

\subsubsection{Key results}
In chapter~\ref{chap:RepulsiveHubbard}, the channel-decomposition scheme for a singlet superfluid on one-loop level was applied to the two-dimensional repulsive Hubbard model at zero temperature. In case $d$-wave superfluidity was the leading instability, the RG flows were continued beyond the critical scale into the symmetry-broken phase. This extends former instability analyses for the Hubbard model, allowing to study properties of the $d$-wave superfluid ground state. The presented study constitutes the first continuation of purely fermionic RG flows to a symmetry-broken phase that is not captured by mean-field theory. 

The $d$-wave superfluid gap as well as the critical scales for pairing and (incommensurate) antiferromagnetism were determined as a function of the microscopic interaction, the fermionic density (or equivalently the Fermi level) and the next-nearest neighbour hopping. The leading instabilities are in good agreement and the critical scales in qualitative agreement with those of other fermionic one-loop RG studies. Furthermore, the superfluid gaps are in reasonable agreement with results of other methods like DMFT, DCA and FLEX for similar parameters. However, for small values of the next-nearest neighbour hopping, in the fermionic one-loop RG the tendency of the system towards $d$-wave pairing was found to be weaker and the phase space with dominant instability towards (incommensurate) antiferromagnetism was found to be larger than for example in the FLEX approximation or in variational QMC simulations. 

The next-nearest neighbour hopping has a significant impact on the dominant instability or the size of the $d$-wave pairing gap in the range of microscopic interactions considered in this work. Below half-filling, the maximal $d$-wave superfluid gaps appear for intermediate (negative) values of the next-nearest neighbour hopping, suggesting the existence of an optimal value for pairing. Interestingly, as a function of the fermionic density at fixed interaction and (negative) next-nearest neighbour hopping, the largest critical scales and gaps appear for fillings slightly above van Hove filling. This is attributed to the amplification of antiferromagnetic fluctuations through umklapp scattering in between hot spots and the large available phase space for antiferromagnetic and pairing fluctuations due to the presence of saddle points in the dispersion close to the Fermi level.

\subsubsection{Criticism}
The presented study of $d$-wave superfluidity within the channel-decomposition scheme has some sore points. First, only an order parameter in the $d$-wave pairing channel was provided, but not in the magnetic channel. The competition or coexistence of $d$-wave superfluid and antiferromagnetic order or the occurrence of secondary instabilities could therefore only be investigated in parts. Second, the approximations for the vertex and the self-energy leave room for improvements. Most notably, an improved approximation should take the frequency dependence of the exchange propagators and of the self-energy into account. 

The renormalization contributions to the exchange propagators and the self-energy should furthermore be determined for external fermionic momenta on the Fermi surface as it was done for the attractive Hubbard model. This yields a better approximation for the low-energy effective interaction and may enhance the tendency towards pairing, as the latter mainly involves fermionic states near the Fermi surface. The impact of this change on the phase diagram could however be small, as the leading instabilities were found to be the same with similar critical scales in studies using the $N$-patch approximation~\cite{Honerkamp2001a} and the channel-decomposition scheme~\cite{Husemann2009}.

\subsubsection{Outlook}
Despite the above-mentioned weaknesses, the channel-decomposition scheme and the employed approximations provide a good starting point for more sophisticated studies of ground state properties of the repulsive Hubbard model within the fermionic RG\@. Such a study should include the frequency dependence of the vertex and the self-energy, as they provide valuable information on the dynamics of collective excitations and the renormalization of fermionic quasi-particles.

A very interesting direction for future research is the fusion of the channel-decomposition scheme for singlet superfluidity with the recently proposed such scheme for antiferromagnetism by Maier and Honerkamp~\cite{Maier2012}. This would allow for an improved study of the competition or coexistence of antiferromagnetism and $d$-wave superfluidity. The scheme proposed by Maier and Honerkamp is appropriate for the description of commensurate antiferromagnetism and has only been applied to mean-field models so far. In a large portion of phase space, the dominant magnetic instability is, however, towards incommensurate antiferromagnetism, which is difficult to describe within a purely fermionic formalism because the fermionic propagator becomes non-diagonal with respect to momentum~\cite{Krahl2009b}. 
Alternatively, in case the leading instability is not towards superfluidity the RG approach could be combined with a self-consistent resummation of perturbation theory for the low-energy modes as done by Reiss~\etal~\cite{Reiss2007}. The self-energy and the vertex as obtained from the RG then serve as the input for the self-consistency problem. In comparison to the work by Reiss~\etal, the channel-decomposition scheme allows to include the frequency dependence of the vertex and the self-energy. This makes it possible to investigate self-energy effects and the renormalization of quasi-particles in the self-consistency problem for the low-energy modes. The inclusion of the frequency dependence of the vertex would furthermore allow to study fluctuation contributions on two-loop level.

In this thesis, the two-loop channel-decomposition scheme was only applied to the attractive Hubbard model away from van Hove or half-filling. In this case, only the Cooper channel is singular and no competition of instabilities appears. The two-loop contributions then do not change the flow qualitatively above the critical scale. This may be different for stronger interactions or in the presence of van Hove singularities close to the Fermi level. The latter implies that the arguments for the justification of one-loop flows cease to be valid because the bounds for the feedback of the three-particle vertex to the two-particle vertex break down~\cite{Salmhofer2001}. 
In the presence of van Hove singularities or stronger interactions, two-loop fluctuations are therefore expected to lead at least to larger quantitative changes of critical scales and gaps, and may also shift the boundaries between regions with different leading instabilities. It would be interesting to study the impact of two-loop fluctuations in particular on the critical scales for incommensurate antiferromagnetism. Their strong suppression may change the obtained phase diagram for small next-nearest neighbour hoppings and could increase the tendency towards pairing. 
The application of the two-loop scheme to the repulsive Hubbard model is also of interest because the $d$-wave superfluid gap is affected by contributions to the particle-particle irreducible vertex in fourth order in the microscopic interaction even in the weak-coupling limit~\cite{Nomura2003,Yanase2003}. Yanase~\etal~\cite{Yanase2003} point out a subtle interplay between vertex corrections and self-energy insertions. It is therefore not obvious whether fluctuations on two-loop level enhance or suppress $d$-wave superfluidity.

%% file: Thesis_DeutscheZusammenfassung.tex
\chapter{\foreignlanguage{ngerman}{Deutsche Zusammenfassung}}
\selectlanguage{ngerman}
\hyphenation{fer-mi-on-ischen}
\hyphenation{kri-tischen}
In dieser Dissertation werden Grundzustandseigenschaften von zweidimensionalen fermionischen Supraflüssigkeiten im Rahmen der funktionalen Renormierungsgruppenmethode untersucht. Dazu wird ein Ansatz für den fermionischen Zweiteilchenvertex formuliert, welcher eine effiziente und genaue Beschreibung der wichtigen singulären Impulsabhängigkeiten erlaubt. Für den Ansatz werden Flussgleichungen hergeleitet, die als Ausgangspunkt für die Formulierung von Näherungen für den Vertex sowie zu dessen Berechnung dienen. Die methodischen Entwicklungen  werden auf das attraktive und das repulsive Hubbard-Modell angewendet, um deren Grundzustandseigenschaften zu untersuchen. Diese Modellsysteme sind relevant für die Simulation von Experimenten mit ultrakalten Atomen in optischen Gittern oder für die Beschreibung der elektronischen Struktur der Kupferoxidebenen in Cupraten.

Die Arbeit ist in zwei Hauptteile gegliedert. In Teil~\ref{part:TheoFramework}, der die Kapitel~\ref{chap:FunctionalRG} bis~\ref{chap:WICharge} umfasst, werden methodische Entwicklungen beschrieben. Diese werden in Teil~\ref{part:Applications} in den Kapiteln~\ref{chap:RPFM} bis~\ref{chap:RepulsiveHubbard} auf verschiedene Modellsysteme angewendet.

In Kapitel~\ref{chap:FunctionalRG} wird die funktionale Renormierungsgruppenmethode für fermionische Systeme vorgestellt und die Herleitung von Flussgleichungen für einteilchenirreduzible Vertexfunktionen skizziert. Renormierungsgleichungen für den fermionischen Zweiteilchenvertex werden in zwei Trunkierungen hergeleitet, im Katanin-Schema~\cite{Katanin2004a} sowie in einer Erweiterung, die alle Renormierungsbeiträge in der dritten Ordnung in der effektiven Wechselwirkung berücksichtigt. Bei der Katanin-Trunkierung handelt es sich um eine Ein-Schleifen-Trunkierung, bei der Erweiterung um Flussgleichungen auf Zwei-Schleifen-Niveau.

In Kapitel~\ref{chap:VertexParametrization} wird die Struktur des fermionischen Zweiteilchenvertex in einer fermionischen Singulett-Supraflüssigkeit untersucht. Durch Ausnutzen von Symmetrien\,--\,insbesondere Spinrotationsinvarianz\,--\,wird der Vertex vereinfacht, so dass drei Funktionen zur Beschreibung der Impuls- und Frequenzabhängigkeit der normalen und anomalen Komponenten ausreichen. Dies erleichtert die Identifikation von singulären Abhängigkeiten des Vertex von externen Impulsen und Frequenzen und ermöglicht die Zerlegung des Vertex in eine Summe von Termen, genannt Wechselwirkungskanäle, die jeweils singulär von einem Transfer\-impuls abhängen. Als Ausgangspunkt für eine effiziente Parametrisierung werden die effektiven Wechselwirkungen in den Kanälen in bosonische Austauschpropagatoren und Fermion-Boson-Vertizes zur Beschreibung der singulären bzw.\ regulären Abhängigkeiten der Vertex zerlegt.

Für die effektiven Wechselwirkungen in den Kanälen werden in Kapitel~\ref{chap:ChannelDecomposition} Renormierungsgleichungen hergeleitet. Dazu werden die singulären Abhängigkeiten der Renormierungsbeiträge zum Zweiteilchenvertex von externen fermionischen Impulsen und Frequenzen analysiert. Dies erlaubt die Zuordnung von Renormierungsbeiträgen zu Wechselwirkungskanälen gemäß den führenden singulären Abhängigkeiten. Diese sogenannten nach Wechselwirkungskanälen zerlegten Flussgleichungen bilden einen guten Ausgangspunkt für die Formulierung von Näherungen für die effektiven Wechselwirkungen in den Kanälen sowie für deren effiziente Berechnung, da die singuläre Abhängigkeit von externen Impulsen und Frequenzen in eine Variable pro Gleichung isoliert ist. 
Diese Zerlegung der Flussgleichungen nach Wechselwirkungskanälen erweitert die Arbeit von Husemann und Salmhofer~\cite{Husemann2009} für die Fortsetzung von Renormierungsgruppenflüssen jenseits der kritischen Skala für Suprafluidität auf Ein-Schleifen-Niveau, sowie auf Zwei-Schleifen-Niveau. 
Mit Hilfe von analytischen Abschätzungen für die führenden Fluktuationsbeiträge wird gezeigt, dass das singuläre Verhalten des Vertex im Grenzfall eines verschwindenden externen Paarungsfeldes auf Ein-Schleifen-Niveau das gleiche ist wie in einer Resummation aller Ketten von Nambu-Teilchen-Loch-Schleifendiagrammen. Auf Zwei-Schleifen-Niveau führen Phasenfluktuationen des suprafluiden Ordnungsparameters (bzw.\ der Gapfunktion) zu einer starken Renormierung der Amplitudenmode, wie sie im Grundzustand einer fermionischen Supraflüssigkeit~\cite{Strack2008} oder eines wechselwirkenden Bose-Gases~\cite{Castellani1997a,Pistolesi2004} erwartet wird. In den weiteren Wechselwirkungskanälen neben dem Cooper-Kanal treten keine weiteren Infrarotsingularitäten der effektiven Wechselwirkungen auf.

In Kapitel~\ref{chap:WICharge} wird die Kompatibilität von trunkierten Flussgleichungen mit der Ward-Identität für globale Ladungserhaltung untersucht. Diese ist von Interesse, weil die Erfüllung der Ward-Identität die Generierung einer masselosen Goldstone-Mode in der symmetriegebrochenen Phase garantieren würde. Das Ergebnis ist, dass die von Katanin~\cite{Katanin2004a} vorgeschlagene Trunkierung nur bis auf Beiträge in dritter Ordnung in der effektiven Wechselwirkung mit der Ward-Identität für globale Ladungserhaltung kompatibel ist. 
In Kapitel~\ref{chap:AttractiveHubbard} wird jedoch gezeigt, dass die Erfüllung der Ward-Identität in der Lösung der Flussgleichungen erzwungen werden kann, indem die Relation zwischen suprafluidem Ordnungsparameter und Masse der Phasenmode über die Ward-Identität und nicht über die Flussgleichungen festgelegt wird. Dies schließt den methodischen Teil der Dissertation ab. In den weiteren Kapiteln wird die Anwendung des Kanalzerlegungsschemas für den Vertex und die Flussgleichungen auf verschiedene Modellsysteme diskutiert.

In Kapitel~\ref{chap:RPFM} wird ein reduziertes Modell mit anziehender Paarungswechselwirkung und Vorwärtsstreuung untersucht. Dieses ist im Rahmen der Molekularfeldnäherung bzw.\ einer Resummation aller Ketten von Nambu-Teilchen-Loch-Schleifendiagrammen exakt lösbar und kann auch im Rahmen der von Katanin vorgeschlagenen Trunkierung der Flussgleichungen exakt gelöst werden. Die Lösung des Modells gewährt Einblicke in die Struktur des Vertex und dessen Singularitäten im Grenzfall eines verschwindenden externen Paarungsfeldes. Im zweiten Teil des Kapitels werden kleine endliche Impuls- und Frequenzüberträge im Vertex zugelassen. Dadurch ist das Modell nicht mehr exakt lösbar, durch die Resummation von Diagrammen können jedoch Informationen über die dynamischen Eigenschaften des Vertex sowie dessen singuläres Verhalten bei endlichen Transferimpulsen und -frequenzen gewonnen werden. Die Ergebnisse dieses Kapitels gehen in die in den Kapiteln~\ref{chap:AttractiveHubbard} und~\ref{chap:RepulsiveHubbard} 
formulierten Näherungen für den Vertex ein.

In Kapitel~\ref{chap:AttractiveHubbard} wird das attraktive Hubbard-Modell als Modell für Fermionen mit einer anziehenden lokalen Wechselwirkung untersucht. Dieses besitzt in einem großen Teil des Parameterraums einen suprafluiden Grundzustand mit $s$-Wellen-Symmetrie und kann daher als gutes Testsystem für neue Näherungsverfahren angesehen werden. Die Rechnungen in diesem Kapitel erweitern die Arbeit von Gersch~\etal~\cite{Gersch2008} in mehreren Aspekten. Die Zerlegung des Vertex nach Wechselwirkungskanälen ermöglicht eine genaue und effiziente Beschreibung der singulären Impulsabhängigkeiten des Vertex sowie von dessen Frequenzabhängigkeit. 
Dies ist wichtig, da insbesondere die mit der Phasenmode des Ordnungsparameters assoziierte effektive Wechselwirkung unterhalb der kritischen Skala sehr groß wird. 
Zudem wurde die Frequenzabhängigkeit der fermionischen Selbstenergie berücksichtigt. Neben umfangreichen Ergebnissen in Ein-Schleifen-Näherung werden auch einige Ergebnisse der Zwei-Schleifen-Näherung präsentiert. 

Das singuläre Verhalten des Vertex für kleine externe Paarungsfelder wird in einem Zwei-Schritt-Verfahren untersucht: Im ersten Schritt werden die fermionischen Moden im Beisein eines externen Paarungsfeldes ausintegriert. Im zweiten Schritt wird das externe Paarungsfeld als Regulator betrachtet und in einem weiteren Fluss entfernt. Auf Ein-Schleifen-Niveau werden die analytischen Abschätzungen aus Kapitel~\ref{chap:ChannelDecomposition} für das singuläre Verhalten des Vertex im Grenzfall eines verschwindenden externen Paarungsfeldes bestätigt. 
Auf Zwei-Schleifen-Niveau wird die Amplitudenmode bei kleinen externen Paarungsfeldern deutlich verstärkt, die numerisch handhabbaren externen Paarungsfelder sind jedoch zu groß um das analytisch gefundene singuläre Verhalten zu reproduzieren. Die analytischen Abschätzungen aus Kapitel~\ref{chap:ChannelDecomposition} und die numerischen Ergebnisse aus Kapitel~\ref{chap:AttractiveHubbard} ergeben zusammen ein umfassendes Verständnis des Zweiteilchenvertex in einer fermionischen Supraflüssigkeit mit $s$-Wellen-Symmetrie.

Neben der Impuls- und Frequenzabhängigkeit des Vertex sowie der Frequenzabhängigkeit der Selbstenergie ist auch der Einfluss von Fluktuationen auf die Größe des suprafluiden Ordnungsparameters von Interesse. Im behandelten Bereich von Wechselwirkungen ist das System eine Supraflüssigkeit im BCS-Regime, Vertex und Selbstenergie sind jedoch stark renormiert. Die Ein-Schleifen-Rechnung ergibt Ordnungsparameter, die je nach Parametern zwischen 38 und 55\,\% des Ergebnisses der Molekularfeldnäherung betragen. Auf Zwei-Schleifen-Niveau ist die Reduktion etwas stärker und die Ordnungsparameter betragen zwischen 30 und 40\,\% des Ergebnisses der Molekularfeldnäherung. Diese Ergebnisse sind in guter Übereinstimmung mit Resultaten in der Literatur.

In Kapitel~\ref{chap:RepulsiveHubbard} wird das repulsive Hubbard-Modell als Modell für Fermionen mit einer abstoßenden lokalen Wechselwirkung untersucht. Dieses zeigt konkurrierende Instabilitäten und die Paarungswechselwirkung, die bei niedrigen Skalen zu Suprafluidität führen kann, wird bei hohen und mittleren Skalen durch antiferromagnetische Fluktuationen generiert. Im betrachteten Parameterbereich treten Instabilitäten zu Antiferromagnetismus oder $d$-Wellen-Suprafluidität auf. Im letzteren Fall wird der Renormierungsgruppenfluss in Ein-Schleifen-Näherung in die symmetriegebrochene Phase fortgesetzt, was die Untersuchung des suprafluiden Grundzustands ermöglicht. Dies erweitert Instabilitätsanalysen im Rahmen der funktionalen Renormierungsgruppenmethode und stellt die erste Fortsetzung von rein fermionischen Renormierungsgruppenflüssen in eine symmetriegebrochene Phase, die nicht in Molekularfeldnäherung für das mikroskopische Modell existiert, dar. In diesem Kapitel wird die Impulsabhängigkeit des Vertex untersucht, seine Frequenzabhängigkeit sowie die der Selbstenergie werden vernachlässigt. Im Rahmen dieser Näherung wird der Ordnungsparameter für $d$-Wellen-Suprafluidität als Funktion der mikroskopischen Wechselwirkung, der Amplitude für das Hüpfen zu übernächsten Nachbarn und der fermionischen Dichte untersucht. 
Die maximalen Ordnungsparameter für $d$-Wellen-Suprafluidität treten unterhalb von Halbfüllung bei mittelgroßen negativen Werten der Amplitude für das Hüpfen zu übernächsten Nachbarn (gemessen in Einheiten der Amplitude für das Hüpfen zu nächsten Nachbarn) auf. 
Als Funktion der fermionischen Dichte und bei festgehaltener Wechselwirkung sowie (negativer) Amplitude für das Hüpfen zu übernächsten Nachbarn tritt das Maximum des suprafluiden Ordnungsparameters interessanterweise bei fermionischen Dichten oberhalb von van Hove-Füllung auf. Die Ursachen hierfür sind die Verstärkung der antiferromagnetischen Fluktuationen durch Umklappstreuung zwischen ``Hot spots'' bei niedrigen Skalen sowie der große Phasenraum für antiferromagnetische und Paarungsfluktuationen durch die Nähe der Sattelpunkte der fermionischen Dispersion zur Fermifläche. Die Ergebnisse für den Ordnungsparameter stimmen qualitativ mit den Resultaten anderer Methoden für ähnliche Parameter überein.

In Kapitel~\ref{chap:Conclusion} werden die Ergebnisse dieser Arbeit zusammengefasst und kritisch diskutiert. Zudem werden interessante Fragestellungen für mögliche Forschungsvorhaben, die auf den methodischen Entwicklungen dieser Dissertation aufbauen, erläutert.

\selectlanguage{british}

%% file: Thesis_Appendix.tex
\chapter{Symmetry properties of fermionic Green and vertex functions}
\label{sec:appendix:Symmetries}
In this appendix, the behaviour of fermionic Green and vertex functions under time reversal and the conjugation operation that is associated with Osterwalder-Schrader positivity is summarised. These symmetry operations yield useful constraints on the momentum and frequency dependence of Green and vertex functions.

\section{Osterwalder-Schrader positivity}
\label{subsec:OSpos}
The Hamiltonian in quantum mechanics is a hermitian operator so that the time evolution as described by Schrödingers's equation is unitary. In the imaginary time functional integral formalism, this is reflected by the so-called Osterwalder-Schrader positivity of the (effective) action. Positivity requires the action to be invariant under the transformation~\cite{Osterwalder1973,Wetterich2007}
\begin{align}
	\theta (\psi_{k \sigma}) &= \bar\psi_{\bar k \sigma}		&		\theta(\bar\psi_{k \sigma}) &= \psi_{\bar k \sigma}\\
	\theta (\psi_\sigma(\tau,\boldsymbol r)) &= \bar\psi_\sigma(-\tau,\boldsymbol r)		&		\theta(\bar\psi_\sigma(\tau,\boldsymbol r)) &= \psi_\sigma(-\tau,\boldsymbol r)
\end{align}
with $\theta^2 = 1$ and $\bar k = (-k_0, \boldsymbol k)$ for $k = (k_0, \boldsymbol k)$, which is accompanied by complex conjugation and reordering of Grassmann fields similar to taking the adjoint of an operator. The transformation acts on spinor and Nambu fields in the same way.

For a compact notation of the transformation laws for normal and anomalous Green and vertex functions in spinor notation, a charge index is defined by
\begin{align}
	\psic{k\sigma} &= \psia{k, + ,\sigma}		&		\psia{k \sigma} &= \psia{k, - ,\sigma},
\end{align}
which is flipped under the $\theta$-conjugation, $\theta(\psia{k, c, \sigma}) = \psia{\bar k, \bar c, \sigma}$ with $\bar c = -$ for $c = +$ and $\bar c = +$ for $c = -$. Using this charge index notation, the fermionic two-particle vertex transforms according to
\begin{equation}
	\theta[\Gamma^{(4)}(k_1 c_1 \sigma_1, k_2 c_2 \sigma_2, k_3 c_3 \sigma_3, k_4 c_4 \sigma_4)] = {\Gamma^{(4)}}^\ast(\bar k_4 \bar c_4 \sigma_4, \bar k_3 \bar c_3 \sigma_3, \bar k_2 \bar c_2 \sigma_2, \bar k_1 \bar c_1 \sigma_1)
\end{equation}
and similarly for other vertex functions. The behaviour of Green functions under the $\theta$-conjugation is very similar, for example
\begin{gather}
	\theta[G(k_1 c_1 \sigma_1, k_2 c_2 \sigma_2)] = G^\ast(\bar k_2 \bar c_2 \sigma_2, \bar k_1 \bar c_1 \sigma_1)\\
	\theta[G(k_1 c_1 \sigma_1, k_2 c_2 \sigma_2, k_3 c_3 \sigma_3, k_4 c_4 \sigma_4)] = G^\ast(\bar k_4 \bar c_4 \sigma_4, \bar k_3 \bar c_3 \sigma_3, \bar k_2 \bar c_2 \sigma_2, \bar k_1 \bar c_1 \sigma_1)
\end{gather}
for the one- and two-particle Green functions. Osterwalder-Schrader positivity requires Green functions to be invariant under $\theta$, for example
\begin{equation}
	\theta[G(k_1 c_1 \sigma_1, k_2 c_2 \sigma_2)] = G(k_1 c_1 \sigma_1, k_2 c_2 \sigma_2) = G^\ast(\bar k_2 \bar c_2 \sigma_2, \bar k_1 \bar c_1 \sigma_1)
\end{equation}
for the one-particle Green function. Similar relations also apply for the vertices and Green functions in Nambu representations where $\sigma_i$ have to be read as Nambu indices and $\sum c_i = 0$ for a system with spin rotation invariance around at least one axis. These relations can be exploited in making an ansatz for the part of the anomalous interactions with ``more creation than annihilation operators'' and completion of the other terms by requiring Osterwalder-Schrader positivity.

For the quadratic part of the fermionic action in a system with translation invariance (in spinor notation),
\begin{equation}
	S_0[\psic{},\psia{}] = \sum_{k,\sigma} (-ik_0 + \xi(\boldsymbol k) + \Sigma(k)) \psic{k\sigma}\psia{k\sigma} + \sum_k (\Delta(k) \psicup{k} \psicdown{-k} + \tilde\Delta(k) \psiadown{-k}\psiaup{k}),
\end{equation}
Osterwalder-Schrader positivity implies
\begin{align}
	\Sigma(k) &= \Sigma^\ast(\bar k)	&	\tilde\Delta(k) = \Delta^\ast(\bar k)
\end{align}
for the self-energies. Relations for vertex functions follow in a similar way. As an example, consider the normal effective interaction
\begin{equation}
	V_{(2+2)}[\psic{},\psia{}] = \tfrac{1}{4} \sum_{k_i,\sigma_i} \Vertex{\sigma_1 \sigma_2 \sigma_3 \sigma_4}(k_1 k_2 k_3 k_4) \psic{k_1\sigma_1}\psic{k_2\sigma_2} \psia{k_3\sigma_3} \psia{k_4\sigma_4}.
\end{equation}
The invariance under the above transformation yields after some renaming of indices
\begin{equation}
	\Gamma^{(4)}_{\sigma_1 \sigma_2 \sigma_3 \sigma_4}(k_1 k_2 k_3 k_4) = \Gamma^{(4)^\ast}_{\sigma_4 \sigma_3 \sigma_2 \sigma_1}(\bar k_4 \bar k_3 \bar k_2 \bar k_1)
\end{equation}
where
\begin{equation}
\Gamma^{(4)}_{\sigma_1 \sigma_2 \sigma_3 \sigma_4}(k_1 k_2 k_3 k_4) = \Gamma^{(4)}(k_1,+,\sigma_1;k_2,+,\sigma_2;k_3,-,\sigma_3;k_4,-,\sigma_4).
\end{equation}
Similar relations hold in Nambu representation.

\section{Invariance under time reversal}
The antiunitary transformation of time reversal in the framework of quantum mechanics is presented very comprehensibly in the book by Messiah~\cite{Messiah1975}. Transformation laws for fermionic operators can be found in~\cite{Sigrist1991} and for fermionic field operators in~\cite{Banyai1994}. The presentation in this section is similar to that given in~\cite{Halboth1999} and addresses Green and vertex functions for imaginary times or frequencies.

The antiunitary time reversal operator $K$ is defined through its action on the fundamental observables of position, momentum and spin~\cite{Messiah1975}, which transform according to
\begin{align}
	K\boldsymbol r K^\dagger &= \boldsymbol r	&	K\boldsymbol p K^\dagger &= -\boldsymbol p	&	K \boldsymbol s K^\dagger  = -\boldsymbol s.
\end{align}
Being antiunitary, the operator K can be written as a product of a unitary operator and the operator for complex conjugation. The above relations are fulfilled for
\begin{equation}
	K = \e{-i \pi s_y /\hbar} K_0 = \e{-i \pi s_y /\hbar} K_{0,\boldsymbol p} S_0,
\end{equation}
where $K_0$ and $K_{0,\boldsymbol p}$ are the operators for complex conjugation in real and momentum space, respectively, and $S_0$ is the operator for spatial inversion. The fermionic creation and annihilation operators transform under $K$ according to
\begin{align}
	K \ca{\boldsymbol k \sigma} K^\dagger &= \epsilon_\sigma \ca{-\boldsymbol k,\bar\sigma}		&		K \cc{\boldsymbol k \sigma} K^\dagger &= \epsilon_\sigma \cc{-\boldsymbol k,\bar\sigma}
\end{align}
with $\epsilon_{\uparrow} = 1$, $\epsilon_{\downarrow} = -1$ and $\bar\sigma=\uparrow$ for $\sigma=\downarrow$ and vice versa. The asymmetry in the transformation of operators with spin-$\uparrow$ and spin-$\downarrow$ is a peculiarity of fermions, for which the application of $K^2$ does not yield unity, but $(-1)^N$ where $N$ is the number of particles in the system~\cite{Messiah1975}. The imaginary time Heisenberg operators, defined by
\begin{align}
	\cc{\boldsymbol k\sigma}(\tau) &= \e{H \tau} \cc{\boldsymbol k\sigma}\e{-H\tau}	&	\ca{\boldsymbol k\sigma}(\tau) &= \e{H \tau} \ca{\boldsymbol k\sigma}\e{-H\tau},
\end{align}
transform according to
\begin{align}
K \ca{\boldsymbol k \sigma}(\tau) K^\dagger &= \epsilon_\sigma \ca{-\boldsymbol k,\bar\sigma}(\tau)		&		K \cc{\boldsymbol k \sigma}(\tau) K^\dagger &= \epsilon_\sigma \cc{-\boldsymbol k,\bar\sigma}(\tau).
\end{align}
Note that the imaginary time is not inverted under the action of $K$, in contrast to real time Heisenberg operators. In Nambu representation, the above transformation laws read
\begin{align}
	K a_{\boldsymbol k s} K^\dagger &= \epsilon_s a^\dagger_{\boldsymbol k \bar s}		&		K a^\dagger_{\boldsymbol k s} K^\dagger &= \epsilon_s a_{\boldsymbol k \bar s}
\end{align}
with $\epsilon_+ = 1$ and $\epsilon_- = -1$ and similarly for the imaginary time Heisenberg operators. Note that in Nambu representation, creation and annihilation operators are mapped onto each other under time reversal due to the inversion of the spin projection.

Transformation laws for spinor or Nambu Grassmann fields follow immediately from the properties of the field operators,
\begin{align}
	K\psic{k\sigma} K^\dagger &= \epsilon_\sigma \psic{-k,\bar \sigma}	& K\psia{k\sigma} K^\dagger &= \epsilon_\sigma \psia{-k,\bar \sigma}\\
	K\phic{k s} K^\dagger &= \epsilon_s \phia{k,\bar s}	& K\phia{k s} K^\dagger &= \epsilon_s \phic{k,\bar s}.
\end{align}

The behaviour of Green functions under time reversal can be derived from their definition as expectation values of time ordered operators. Assuming that the Hamiltonian $H$ commutes with $K$ and that time reversal invariance is not broken, besides the eigenvectors $|n\rangle$ of $H$ with $H|n\rangle = E_n |n\rangle$, the time reversed eigenstates $K|n\rangle$ are also eigenstates of $H$ with the same eigenvalue,
\begin{equation}
	H(K|n\rangle) = K(H|n\rangle) = E_n(K|n\rangle).
\end{equation}
In a grand-canonical ensemble, the expectation value of some operator $A$ is computed via
\begin{align}
	\langle A\rangle = \tr(\rho A) &= \sum_n \rho_n \langle n|A|n\rangle	&\text{with} &&	\rho_n &= \frac{1}{Z} \e{-\beta(E_n - \mu N_n)} \in \mathbb R
\end{align}
and $Z = \tr(\exp(-\beta (H - \mu N)))$. Evaluating the trace with respect to the basis spanned by $K|n\rangle$ or $K^\dagger|n\rangle$,
\begin{equation}
	\langle A\rangle = \tr(\rho A) = \sum_n (\langle n|K^\dagger) \rho A K^\dagger|n\rangle
\end{equation}
and exploiting $(\langle n|K^\dagger)\rho = \rho_n (\langle n|K^\dagger)$ as well as the antilinearity\footnote{For a brief overview of properties of antilinear operators, see the book by Messiah~\cite{Messiah1975}.} of $K$, one obtains
\begin{equation}
	\langle A\rangle =  \sum_n \rho_n (\langle n|K^\dagger) A K^\dagger|n\rangle = \sum_n \rho_n \langle n|(KAK^\dagger)|n\rangle^\ast = \langle K A K^\dagger\rangle^\ast.
\end{equation}
Thus, in a system with time reversal invariance, the relation
\begin{equation}
	\langle A \rangle = \langle K A K^\dagger \rangle^\ast
\end{equation}
holds. Choosing for example $A = T\{\ca{\boldsymbol k \sigma}(\tau) \cc{\boldsymbol k\sigma}\}$ where $T$ is the time ordering operator that orders along the imaginary time axis, one obtains a constraint on the momentum and frequency dependence of the normal one-particle Green function in a system with time reversal invariance. Relations for higher order Green functions follow similarly and imply relations for vertex functions.

Using the charge index notation in spinor representation, the action of the time reversal operator can be expressed as
\begin{gather}
	K[G(k_1 c_1 \sigma_1, k_2 c_2 \sigma_2)] = \epsilon_{\sigma_1} \epsilon_{\sigma_2} G(Rk_2 \bar c_2 \bar \sigma_2, Rk_1 \bar c_1 \bar \sigma_1)\\
	\begin{split}
	K[G(k_1 c_1 \sigma_1, k_2 c_2 \sigma_2, k_3 &c_3 \sigma_3, k_4 c_4 \sigma_4)] =\\
	&=\epsilon_{\sigma_1} \epsilon_{\sigma_2} \epsilon_{\sigma_3} \epsilon_{\sigma_4} G(Rk_4 \bar c_4 \bar \sigma_4, Rk_3 \bar c_3 \bar \sigma_3, Rk_2 \bar c_2 \bar \sigma_2, Rk_1 \bar c_1 \bar \sigma_1)
	\end{split}
\end{gather}
for the normal and anomalous one- and two-particle Green functions and as
\begin{gather}
	\begin{split}
	K[\Gamma^{(4)}(k_1 c_1 \sigma_1, k_2 c_2 &\sigma_2, k_3 c_3 \sigma_3, k_4 c_4 \sigma_4)] = \\
	&= \epsilon_{\sigma_1} \epsilon_{\sigma_2} \epsilon_{\sigma_3} \epsilon_{\sigma_4} \Gamma^{(4)}(Rk_4 \bar c_4 \bar \sigma_4, Rk_3 \bar c_3 \bar \sigma_3, Rk_2 \bar c_2 \bar \sigma_2, Rk_1 \bar c_1 \bar \sigma_1)
	\end{split}
\end{gather}
for normal and anomalous two-particle vertex functions. Similar relations hold for the self-energy and higher-order Green as well as vertex functions. In a time reversal invariant system, the Green functions have to be invariant under $K$, for example
\begin{equation}
	K[G(k_1 c_1 \sigma_1, k_2 c_2 \sigma_2)] = G(k_1 c_1 \sigma_1, k_2 c_2 \sigma_2),
\end{equation}
yielding constraints on the momentum and frequency dependence of Green and vertex functions. This implies $\Delta(k) \in \mathbb R$ in the above ansatz for the quadratic part of the fermionic action. In order to constrain the momentum and frequency dependence of the anomalous components of the vertex in spinor representation, it is useful to combine time reversal invariance with positivity, because this yields a relation between components with the same charge indices:
\begin{gather}
	\begin{split}
	\Gamma^{(4)}(k_1 c_1 \sigma_1, k_2 c_2 &\sigma_2, k_3 c_3 \sigma_3, k_4 c_4 \sigma_4) = \\
	&= \epsilon_{\sigma_1} \epsilon_{\sigma_2} \epsilon_{\sigma_3} \epsilon_{\sigma_4} \Gamma^{(4)\,\ast}(-k_1 c_1 \bar \sigma_1, -k_2 c_2 \bar \sigma_2, -k_3 c_3 \bar \sigma_3, -k_4 c_4 \bar \sigma_4).
	\end{split}
\end{gather}

In Nambu notation, the transformation of the vertex under time reversal reads somewhat differently because the order of the arguments does not change. The reason is that besides the reordering of operators due to the complex conjugation, their order is also changed because of the flipping of spin indices and the subsequent mapping from Nambu creation to annihilation operators and vice versa. In Nambu representation, the above relation for the two-particle vertex reads
\begin{equation}
	\Gamma^{(4)}_\text{Nambu}(k_1 s_1, k_2 s_2, k_3 s_3, k_4 s_4) = \epsilon_{s_1} \epsilon_{s_2} \epsilon_{s_3} \epsilon_{s_4} \Gamma^{(4)}_\text{Nambu}(-k_1 \bar s_1, -k_2 \bar s_2, -k_3 \bar s_3, -k_4 \bar s_4)
\end{equation}
for a system with time reversal invariance where $\bar s = -$ for $s = +$ and vice versa. Note that the charge index is omitted because all Nambu vertices that appear in this work are normal (with an equal number of Nambu creation and annihilation operators). The label ``Nambu'' has been added to avoid confusion as the rest of this section mostly deals with spinor representation.

\chapter{Parametrization of Nambu vertex}
\label{appendix:VP}
In subsection~\ref{subsec:VP:EffIntChannels}, the dependence of the Nambu two-particle vertex on the external Nambu indices is partially reformulated using Pauli matrices. This reformulation is completed for all channels in the following, but it does not simplify the structure of the effective interactions in the (spinor) particle-hole channel in the general case. Simplifications occur only after the decomposition of the effective interactions in bosonic propagators and fermion-boson vertices as in subsection~\ref{subsec:VP:BosonProp_gFB}. For the normal effective interaction in the Cooper channel and the anomalous (3+1)-effective interaction, the sum over singlet and triplet parts is suppressed for brevity in this section and they are to be understood for example as
\begin{equation}
	A_{k k'}(q) = A^S_{k k'}(q) + A^T_{k k'}(q).
\end{equation}
In case either the singlet or the triplet part are present, it is specified explicitly. The effective interaction in the Nambu particle-hole channel then reads
\begin{gather}
	V^\text{PH}_{s_1 s_2 s_3 s_4}(k, k'; q) = \sum_{i,j} V^\text{PH}_{ij}(k, k'; q) \mtau{i}{s_1 s_4} \mtau{j}{s_2 s_3}
\end{gather}
where
\begin{align}
	V^\text{PH}_{00}(k, k'; q) &= \tfrac{1}{2} \bigl(\re C_{kk'}(q) + \re M_{kk'}(q) + \re M_{k,-k'}(-q) - \re C_{k,-k'}(-q) \bigr)\\
	V^\text{PH}_{01}(k, k'; q) &= \tfrac{i}{2} \bigl( \tilde X_{k k'}(q) + \tilde X_{k k'}(-q) \bigr) = V^\text{PH}_{10}(k', k; q)\\
	V^\text{PH}_{02}(k, k'; q) &= -\tfrac{i}{2} \bigl(X_{k k'}(q) - X_{k k'}(-q) \bigr) = -V^\text{PH}_{20}(k', k; q)\\
	V^\text{PH}_{03}(k, k'; q) &= \tfrac{i}{2} \bigl(\im C_{k k'}(q) + \im M_{k k'}(q) + \im C_{k,-k'}(-q) - \im M_{k,-k'}(-q)\bigr)\\
	V^\text{PH}_{11}(k, k'; q) &= \tfrac{1}{2} A_{k k'}(q)\\
	V^\text{PH}_{12}(k, k'; q) &= \tfrac{1}{2} \bigl(\nu_{k k'}(q) - \tilde \nu_{k k'}(q)\bigr)\\
	V^\text{PH}_{13}(k, k'; q) &= \tfrac{1}{2} \bigl(X_{k' k}(q) + X_{k' k}(-q)\bigr) = V^\text{PH}_{31}(k', k; q)\\
	V^\text{PH}_{21}(k, k'; q) &= -\tfrac{1}{2} \bigl(\nu_{k k'}(q) + \tilde \nu_{k k'}(q)\bigr)\\
	V^\text{PH}_{22}(k, k'; q) &= \tfrac{1}{2} \Phi_{k k'}(q)\\
	V^\text{PH}_{23}(k, k'; q) &= -\tfrac{1}{2} \bigl(\tilde X_{k' k}(q) - \tilde X_{k' k}(-q)\bigr) = -V^\text{PH}_{32}(k', k; q)\\
	V^\text{PH}_{30}(k, k'; q) &= \tfrac{i}{2} \bigl(\im C_{k k'}(q) + \im M_{k k'}(q) - \im C_{k,-k'}(-q) + \im M_{k,-k'}(-q)\bigr)\\
	V^\text{PH}_{33}(k, k'; q) &= \tfrac{1}{2} \bigl(\re C_{k k'}(q) + \re M_{k k'}(q) - \re M_{k,-k'}(-q) + \re C_{k,-k'}(-q) \bigr).
\end{align}
For the effective interaction in the Nambu particle-particle channel one obtains
\begin{gather}
	V^\text{PP}_{s_1 s_2 s_3 s_4}(k, k'; q) = \sum_{i,j} V^\text{PP}_{ij}(k, k'; q) \mtau{i}{s_1 s_2} \mtau{j}{s_3 s_4}
\end{gather}
where
\begin{align}
	V^\text{PP}_{00}(k, k'; q) &= \Phi^T_{k k'}(q)\\
	V^\text{PP}_{01}(k, k'; q) &= -i\bigl(\tilde X^T_{k' k}(q) - \tilde X^T_{-k',k}(q)\bigr) = V^\text{PP}_{10}(k', k; q)\\
	V^\text{PP}_{02}(k, k'; q) &= i \bigl(X^T_{k' k}(q) + X^T_{-k',k}(q)\bigr) = - V^\text{PP}_{20}(k', k; q)\\
	V^\text{PP}_{03}(k, k'; q) &= i\bigl(\nu^T_{k k'}(q) + \tilde \nu^T_{k k'}(q)\bigr)\\
	V^\text{PP}_{30}(k, k'; q) &= i\bigl(\nu^T_{k k'}(q) - \tilde \nu^T_{k k'}(q)\bigr)\\
	V^\text{PP}_{11}(k, k'; q) &= \re M_{k k'}(q) - \re M_{k,-k'}(q)\\
	V^\text{PP}_{12}(k, k'; q) &= \im M_{k,-k'}(q) + \im M_{k k'}(q)\\
	V^\text{PP}_{21}(k, k'; q) &= \im M_{k,-k'}(q) - \im M_{k k'}(q)\\
	V^\text{PP}_{22}(k, k'; q) &= \re M_{k k'}(q) + \re M_{k,-k'}(q)\\
	V^\text{PP}_{13}(k, k'; q) &= X^T_{-k,k'}(q) - X^T_{k k'}(q) = V^\text{PP}_{31}(k', k; q)\\
	V^\text{PP}_{23}(k, k'; q) &= \tilde X^T_{k k'}(q) + \tilde X^T_{-k, k'}(q) = - V^\text{PP}_{32}(k',k; q)\\
	V^\text{PP}_{33}(k, k'; q) &= A^T_{k k'}(q).
\end{align}

\chapter{Flow equations for fermionic self-energy}
\label{sec:Appendix:RGDESelfenergy}
In this section, flow equations for the fermionic self-energy are presented within the framework of approximations of chapters~\ref{chap:VertexParametrization} and~\ref{chap:ChannelDecomposition}. The expressions are valid in case the expansion of the effective interactions in exchange propagators and fermion-boson vertices is restricted to fermion-boson vertices that can be written as
\begin{equation}
	h^i_\alpha(q, k) = g^i_\alpha(\boldsymbol q, k_0) f_\alpha(\boldsymbol k)
\end{equation}
where
\begin{equation}
	f_\alpha(\boldsymbol k) = f_\alpha(-\boldsymbol k),
\end{equation}
\ie\ all contributions with form factors of odd parity are neglected. The projection of the momentum dependence has to be performed as described in chapters~\ref{chap:AttractiveHubbard} or~\ref{chap:RepulsiveHubbard}.

The flow equation for the anomalous self-energy reads
\begin{align}
	\partial_\Lambda&\Sigma^\Lambda_{+-}(k) = \partial_\Lambda (\Delta^\Lambda(k) - \Delta^\Lambda_{(0)}(k)) = -\sum_{s_1,s_2}\intdrei{p} \SL{s_1 s_2}(p) \Vertex{+s_2 s_1 -}(k,p,p,k)\\
	&=-\intdrei{p} \Bigl[\SL{+-}(p) \bigl(\Vertex{+-+-}(k,p,p,k) + \Vertex{++--}(k,p,p,k)\bigr) \nonumber\\
	&\hspace*{2cm}+ \re \SL{++}(p) \bigl(\Vertex{+++-}(k,p,p,k) - \Vertex{+---}(k,p,p,k)\bigr)\\
	&\hspace*{2cm}+ i \im \SL{++}(p) \bigl(\Vertex{+++-}(k,p,p,k) + \Vertex{+---}(k,p,p,k)\bigr)\Bigr]\nonumber\\
	&= -\sum_{\alpha,\beta}\intdrei{p} \bigl[\SL{+-}(p) \bigl(U \delta_{\alpha0} \delta_{\beta0} + A^\Lambda_{\alpha\beta}(0) h^{A,\Lambda}_\alpha(0,k) h^{A,\Lambda}_\beta(0,p)\bigr)\nonumber\\
	&\hspace{2cm} + 2 \re \SL{++}(p) X^\Lambda_{\alpha\beta}(0) h^{X_C,\Lambda}_\alpha(0,p) h^{X_A,\Lambda}_\beta(0,k)\bigr]\nonumber\\
	&\quad - \sum_{\alpha,\beta}\intdrei{p} \Bigl[2\im \SL{++}(p) \tilde X^\Lambda_{\alpha\beta}(p-k) h^{\tilde X_C,\Lambda}_\alpha(p-k,\tfrac{k+p}{2}) h^{\tilde X_\Phi,\Lambda}_\beta(p-k,\tfrac{k+p}{2})\nonumber\\
	&\hspace{2cm} +\SL{+-}(p) \Bigl(\tfrac{1}{2} \Phi^\Lambda_{\alpha\beta}(p-k) h^{\Phi,\Lambda}_\alpha(p-k,\tfrac{k+p}{2}) h^{\Phi,\Lambda}_\beta(p-k,\tfrac{k+p}{2})\\
	&\hspace{4cm} - \tfrac{1}{2} A^\Lambda_{\alpha\beta}(p-k) h^{A,\Lambda}_\alpha(p-k,\tfrac{k+p}{2}) h^{A,\Lambda}_\beta(p-k,\tfrac{k+p}{2})\nonumber\\
	&\hspace{4cm} + C^\Lambda_{\alpha\beta}(p-k) h^{C,\Lambda}_\alpha(p-k,\tfrac{k+p}{2}) h^{C,\Lambda}_\beta(p-k,\tfrac{k+p}{2}) \nonumber \\
	&\hspace{4cm} - 3 M^\Lambda_{\alpha\beta}(p-k) h^{M,\Lambda}_\alpha(p-k,\tfrac{k+p}{2}) h^{M,\Lambda}_\beta(p-k,\tfrac{k+p}{2})\Bigr)\nonumber\\
	&\hspace{2cm} -2\re \SL{++}(p) X^\Lambda_{\alpha\beta}(p-k) h^{X_C,\Lambda}_\alpha(p-k,\tfrac{k+p}{2}) h^{X_A,\Lambda}_\beta(p-k,\tfrac{k+p}{2})\Bigr]\nonumber.
	\label{eq:Appendix:DeltaFull}
\end{align}
Note that the scale derivative of the external pairing field appears explicitly in this equation in a pairing field flow as indicated by $\partial_\Lambda \Delta^\Lambda_{(0)}(k)$.

The flow equation for the normal self-energy including convergence generating factors reads
\begin{equation}
\partial_\Lambda \Sigma^\Lambda_{++}(k) = \partial_\Lambda \Sigma^\Lambda(k) = -\sum_{s_1,s_2}\intdrei{p} \e{i 0^+ \mtau{3}{s_1 s_2}} S^\Lambda_{s_1 s_2}(p) \Vertex{+s_2 s_1 +}(k,p,p,k).
\end{equation}
In the numerical solution of the flow equations in chapters~\ref{chap:AttractiveHubbard} and~\ref{chap:RepulsiveHubbard}, a frequency regulator is used and the integration over the high energy modes is performed perturbatively. For the RG flow of the low energy modes, the convergence generating factors can then be dropped because the fermionic single scale propagator decays fast enough at high frequencies. It is then convenient to decompose the normal self-energy in a real and an imaginary part. Their flow is determined by
\begin{align}
	\partial_\Lambda &\im\Sigma^\Lambda_{++}(k) = \partial_\Lambda \im\Sigma^\Lambda(k) = -\sum_{s_1,s_2}\intdrei{p} \im\bigl(S^\Lambda_{s_1 s_2}(p) \Vertex{+s_2 s_1 +}(k,p,p,k)\bigr)\\
	&=-\intdrei{p} \Bigl[\re S^\Lambda_{++}(p) \im\bigl(\Vertex{++++}(k,p,p,k) - \Vertex{+--+}(k,p,p,k)\bigr) \nonumber\\
	&\hspace*{1cm}+ \im S^\Lambda_{++}(p) \re\bigl(\Vertex{++++}(k,p,p,k) + \Vertex{+--+}(k,p,p,k)\bigr)\\
	&\hspace*{1cm}+ S^\Lambda_{+-}(p) \im\bigl(\Vertex{++-+}(k,p,p,k) + \Vertex{+-++}(k,p,p,k)\bigr)\Bigr]\nonumber\\
	&= \sum_{\alpha,\beta}\intdrei{p} \Bigl[\re S^\Lambda_{++}(p) \nu^\Lambda_{\alpha\beta}(p-k) h^{\nu,\Lambda}_\alpha(p-k,\tfrac{k+p}{2}) h^{\nu,\Lambda}_\beta(p-k,\tfrac{k+p}{2}) \nonumber\\
	&\hspace{2cm}- 2 S^\Lambda_{+-}(p) \tilde X^\Lambda_{\alpha\beta}(p-k) h^{\tilde X_C,\Lambda}_\alpha(p-k,\tfrac{k+p}{2}) h^{\tilde X_\Phi,\Lambda}_\beta(p-k,\tfrac{k+p}{2}) \nonumber \\
	&\hspace{2cm}+\im S^\Lambda_{++}(p) \Bigl(\tfrac{1}{2} A^\Lambda_{\alpha\beta}(p-k) h^{A,\Lambda}_\alpha(p-k,\tfrac{p+k}{2}) h^{A,\Lambda}_\beta(p-k,\tfrac{p+k}{2})\\
	&\hspace{4cm} + \tfrac{1}{2} \Phi^\Lambda_{\alpha\beta}(p-k) h^{\Phi,\Lambda}_\alpha(p-k,\tfrac{p+k}{2}) h^{\Phi,\Lambda}_\beta(p-k,\tfrac{p+k}{2})\nonumber\\
	&\hspace{4cm} + C^\Lambda_{\alpha\beta}(p-k) h^{C,\Lambda}_\alpha(p-k,\tfrac{k+p}{2}) h^{C,\Lambda}_\beta(p-k,\tfrac{k+p}{2}) \nonumber \\
	&\hspace{4cm} + 3 M^\Lambda_{\alpha\beta}(p-k) h^{M,\Lambda}_\alpha(p-k,\tfrac{k+p}{2}) h^{M,\Lambda}_\beta(p-k,\tfrac{k+p}{2})\Bigl)\Bigr]\nonumber
\end{align}
for the imaginary part and
\begin{align}
	\partial_\Lambda& \re\Sigma^\Lambda_{++}(k) = \partial_\Lambda \re\Sigma^\Lambda(k) = -\sum_{s_1,s_2}\intdrei{p} \re\bigl(\SL{s_1 s_2}(p) \Vertex{+s_2 s_1 +}(k,p,p,k)\bigr)\\
&=-\intdrei{p} \Bigl[\re \SL{++}(p) \re\bigl(\Vertex{++++}(k,p,p,k) - \Vertex{+--+}(k,p,p,k)\bigr) \nonumber\\
	&\hspace*{1cm}- \im \SL{++}(p) \im\bigl(\Vertex{++++}(k,p,p,k) + \Vertex{+--+}(k,p,p,k)\bigr)\\
	&\hspace*{1cm}+ \SL{+-}(p) \re\bigl(\Vertex{++-+}(k,p,p,k) + \Vertex{+-++}(k,p,p,k)\bigr)\Bigr]\nonumber\\
&= -\sum_{\alpha,\beta}\intdrei{p} \bigl[\SL{++}(p) \bigl(U\delta_{\alpha0} \delta_{\beta0} + 2 C^\Lambda_{\alpha\beta}(0) h^{C,\Lambda}_\alpha(0,k) h^{C,\Lambda}_\beta(0,p)\bigr)\nonumber\\
	&\hspace{2cm} +2 \SL{+-}(p) X^\Lambda_{\alpha\beta}(0) h^{C,\Lambda}_\alpha(0,k) h^{A,\Lambda}_\beta(0,p)\bigr]\nonumber\\
	&\quad - \sum_{\alpha,\beta} \intdrei{p} \Bigl[-\im S^\Lambda_{++}(p) \nu^\Lambda_{\alpha\beta}(p-k) h^{\nu,\Lambda}_\alpha(p-k,\tfrac{k+p}{2}) h^{\nu,\Lambda}_\beta(p-k,\tfrac{k+p}{2})\\
	&\hspace{2cm} +\re \SL{++}(p) \Bigl(\tfrac{1}{2} A^\Lambda_{\alpha\beta}(p-k) h^{A,\Lambda}_\alpha(p-k,\tfrac{k+p}{2}) h^{A,\Lambda}_\beta(p-k,\tfrac{k+p}{2}) \nonumber \\
	&\hspace{4cm} + \tfrac{1}{2} \Phi^\Lambda_{\alpha\beta}(p-k) h^{\Phi,\Lambda}_\alpha(p-k,\tfrac{k+p}{2}) h^{\Phi,\Lambda}_\beta(p-k,\tfrac{k+p}{2})\nonumber\\
	&\hspace{4cm} - C^\Lambda_{\alpha\beta}(p-k) h^{C,\Lambda}_\alpha(p-k,\tfrac{k+p}{2}) h^{C,\Lambda}_\beta(p-k,\tfrac{k+p}{2}) \nonumber\\
	&\hspace{4cm} - 3 M^\Lambda_{\alpha\beta}(p-k) h^{M,\Lambda}_\alpha(p-k,\tfrac{k+p}{2}) h^{M,\Lambda}_\beta(p-k,\tfrac{k+p}{2})\Bigr)\nonumber\\
	&\hspace{2cm} -2 \SL{+-}(p) X^\Lambda_{\alpha\beta}(p-k) h^{C,\Lambda}_\alpha(p-k,\tfrac{k+p}{2}) h^{A,\Lambda}_\beta(p-k,\tfrac{k+p}{2})\Bigr]\nonumber
\end{align}
for the real part.

The Ward identity connecting the gap and the vertex is not a flow equation, nevertheless it is included in this section for completeness. Within the approximations of chapters~\ref{chap:VertexParametrization} and~\ref{chap:ChannelDecomposition}, for a real gap and under the above assumptions on the fermion-boson vertices, equation~\eqref{eq:WICharge:WIGap:GapReal} reads
\begin{align}
	\Delta^\Lambda(k) - \Delta_{(0)}(k) &= \intdrei{p} \Delta_{(0)}(p) L^\Lambda_{22}(p,0) (\Vertex{+-+-}(k,p,p,k) - \Vertex{++--}(k,p,p,k)) \\
	&= \intdrei{p} \sum_{\alpha,\beta} \Delta_{(0)}(p) L^\Lambda_{22}(p, 0) \Bigl[ U \delta_{\alpha,0}\delta_{\beta,0} + \Phi^\Lambda_{\alpha\beta}(0) h^{\Phi,\Lambda}_\alpha(0, k) h^{\Phi,\Lambda}_\beta(0,p) \nonumber \\
	& \hspace{2cm} + \tfrac{1}{2} A^\Lambda_{\alpha\beta}(p-k) h^{A,\Lambda}_\alpha(p-k, \tfrac{p+k}{2}) h^{A,\Lambda}_\beta(p-k, \tfrac{p+k}{2}) \nonumber \\
	& \hspace{2cm} - \tfrac{1}{2} \Phi^\Lambda_{\alpha\beta}(p-k) h^{\Phi,\Lambda}_\alpha(p-k, \tfrac{p+k}{2}) h^{\Phi,\Lambda}_\beta(p-k, \tfrac{p+k}{2}) \\
	& \hspace{2cm} + C^\Lambda_{\alpha\beta}(p-k) h^{C,\Lambda}_\alpha(p-k, \tfrac{p+k}{2}) h^{C,\Lambda}_\beta(p-k, \tfrac{p+k}{2})\nonumber \\
	& \hspace{2cm} - 3 M^\Lambda_{\alpha\beta}(p-k) h^{M,\Lambda}_\alpha(p-k, \tfrac{p+k}{2}) h^{M,\Lambda}_\beta(p-k, \tfrac{p+k}{2})\Bigr]\nonumber.
\end{align}

\chapter{Flow equations for bosonic exchange propagators}

\section{One-loop contributions}
\label{subsec:Appendix:OneLoopContributions}
In this section, the $C$-functions that appear in equation~\eqref{eq:CD:RGOneLoop} are stated within the framework of approximations of subsection~\ref{subsec:VP:BosonProp_gFB}. The expressions are valid under the assumptions for the fermion-boson vertices described in appendix~\ref{sec:Appendix:RGDESelfenergy}.

\subsection*{Definition of $C$-functions}
\begin{equation}
	\begin{split}
	C^\Lambda_{ij}(q; k, p) &= \frac{1}{2} \sum_{s_i} \mtau{i}{s_4 s_1} \mtau{j}{s_3 s_2} \Vertex{s_1 s_2 s_3 s_4}(k+\tfrac{q}{2},p-\tfrac{q}{2},p+\tfrac{q}{2},k-\tfrac{q}{2})\\
	&= \frac{1}{2}\sum_{s_i} \mtau{i}{s_4 s_1} \mtau{j}{s_3 s_2} \bigl[U_{s_1 s_2 s_3 s_4} + V^{PH}_{s_1 s_2 s_3 s_4}(q; k, p)\\
		&\hspace{1cm} -V^{PH}_{s_2 s_1 s_3 s_4}(p-k; \tfrac{k+p-q}{2}, \tfrac{k+p+q}{2}) + V^{PP}_{s_1 s_2 s_3 s_4}(k+p; \tfrac{k-p+q}{2}, \tfrac{k-p-q}{2})\bigr]
	\end{split}
\end{equation}

\begin{equation}
	\begin{split}
		C^\Lambda_{00}(q; k, p) &= \sum_{\alpha,\beta} \Bigl[-U \delta_{\alpha,0} \delta_{\beta,0} + 2 M^\Lambda_{\alpha\beta}(q) \hFB{M}{\alpha}{q}{k} \hFB{M}{\beta}{q}{p}\\
	& + 2 M^\Lambda_{\alpha\beta}(k+p) \hFB{M}{\alpha}{k+p}{\tfrac{k-p+q}{2}} \hFB{M}{\beta}{k+p}{\tfrac{k-p-q}{2}}\\
	& - M^\Lambda_{\alpha\beta}(p-k) \hFB{M}{\alpha}{p-k}{\tfrac{k+p-q}{2}} \hFB{M}{\beta}{p-k}{\tfrac{k+p+q}{2}}\\
	& - C^\Lambda_{\alpha\beta}(p-k) \hFB{C}{\alpha}{p-k}{\tfrac{k+p-q}{2}} \hFB{C}{\beta}{p-k}{\tfrac{k+p+q}{2}}\\
	& - \tfrac{1}{2} A^\Lambda_{\alpha\beta}(p-k) \hFB{A}{\alpha}{p-k}{\tfrac{k+p-q}{2}} \hFB{A}{\beta}{p-k}{\tfrac{k+p+q}{2}}\\
	& - \tfrac{1}{2} \Phi^\Lambda_{\alpha\beta}(p-k) \hFB{\Phi}{\alpha}{p-k}{\tfrac{k+p-q}{2}} \hFB{\Phi}{\beta}{p-k}{\tfrac{k+p+q}{2}}\Bigr]
	\end{split}
\end{equation}

\begin{equation}
	\begin{split}
		C^\Lambda_{11}(q; k, p) &= \sum_{\alpha,\beta} \Bigl[U \delta_{\alpha,0} \delta_{\beta,0} + A^\Lambda_{\alpha\beta}(q) \hFB{A}{\alpha}{q}{k} \hFB{A}{\beta}{q}{p}\\
	& - 2 M^\Lambda_{\alpha\beta}(k+p) \hFB{M}{\alpha}{k+p}{\tfrac{k-p+q}{2}} \hFB{M}{\beta}{k+p}{\tfrac{k-p-q}{2}}\\
	& - M^\Lambda_{\alpha\beta}(p-k) \hFB{M}{\alpha}{p-k}{\tfrac{k+p-q}{2}} \hFB{M}{\beta}{p-k}{\tfrac{k+p+q}{2}}\\
	& + C^\Lambda_{\alpha\beta}(p-k) \hFB{C}{\alpha}{p-k}{\tfrac{k+p-q}{2}} \hFB{C}{\beta}{p-k}{\tfrac{k+p+q}{2}}\\
	& - \tfrac{1}{2} A^\Lambda_{\alpha\beta}(p-k) \hFB{A}{\alpha}{p-k}{\tfrac{k+p-q}{2}} \hFB{A}{\beta}{p-k}{\tfrac{k+p+q}{2}}\\
	& + \tfrac{1}{2} \Phi^\Lambda_{\alpha\beta}(p-k) \hFB{\Phi}{\alpha}{p-k}{\tfrac{k+p-q}{2}} \hFB{\Phi}{\beta}{p-k}{\tfrac{k+p+q}{2}}\Bigr]
	\end{split}
\end{equation}

\begin{equation}
	\begin{split}
		C^\Lambda_{22}(q; k, p) &= \sum_{\alpha,\beta} \Bigl[U \delta_{\alpha,0} \delta_{\beta,0} + \Phi^\Lambda_{\alpha\beta}(q) \hFB{\Phi}{\alpha}{q}{k} \hFB{\Phi}{\beta}{q}{p}\\
	& - 2 M^\Lambda_{\alpha\beta}(k+p) \hFB{M}{\alpha}{k+p}{\tfrac{k-p+q}{2}} \hFB{M}{\beta}{k+p}{\tfrac{k-p-q}{2}}\\
	& - M^\Lambda_{\alpha\beta}(p-k) \hFB{M}{\alpha}{p-k}{\tfrac{k+p-q}{2}} \hFB{M}{\beta}{p-k}{\tfrac{k+p+q}{2}}\\
	& + C^\Lambda_{\alpha\beta}(p-k) \hFB{C}{\alpha}{p-k}{\tfrac{k+p-q}{2}} \hFB{C}{\beta}{p-k}{\tfrac{k+p+q}{2}}\\
	& + \tfrac{1}{2} A^\Lambda_{\alpha\beta}(p-k) \hFB{A}{\alpha}{p-k}{\tfrac{k+p-q}{2}} \hFB{A}{\beta}{p-k}{\tfrac{k+p+q}{2}}\\
	& - \tfrac{1}{2} \Phi^\Lambda_{\alpha\beta}(p-k) \hFB{\Phi}{\alpha}{p-k}{\tfrac{k+p-q}{2}} \hFB{\Phi}{\beta}{p-k}{\tfrac{k+p+q}{2}}\Bigr]
	\end{split}
\end{equation}

\begin{equation}
	\begin{split}
		C^\Lambda_{33}(q; k, p) &= \sum_{\alpha,\beta} \Bigl[U \delta_{\alpha,0} \delta_{\beta,0} + 2 C^\Lambda_{\alpha\beta}(q) \hFB{C}{\alpha}{q}{k} \hFB{C}{\beta}{q}{p}\\
	& - 2 M^\Lambda_{\alpha\beta}(k+p) \hFB{M}{\alpha}{k+p}{\tfrac{k-p+q}{2}} \hFB{M}{\beta}{k+p}{\tfrac{k-p-q}{2}}\\
	& - M^\Lambda_{\alpha\beta}(p-k) \hFB{M}{\alpha}{p-k}{\tfrac{k+p-q}{2}} \hFB{M}{\beta}{p-k}{\tfrac{k+p+q}{2}}\\
	& - C^\Lambda_{\alpha\beta}(p-k) \hFB{C}{\alpha}{p-k}{\tfrac{k+p-q}{2}} \hFB{C}{\beta}{p-k}{\tfrac{k+p+q}{2}}\\
	& + \tfrac{1}{2} A^\Lambda_{\alpha\beta}(p-k) \hFB{A}{\alpha}{p-k}{\tfrac{k+p-q}{2}} \hFB{A}{\beta}{p-k}{\tfrac{k+p+q}{2}}\\
	& + \tfrac{1}{2} \Phi^\Lambda_{\alpha\beta}(p-k) \hFB{\Phi}{\alpha}{p-k}{\tfrac{k+p-q}{2}} \hFB{\Phi}{\beta}{p-k}{\tfrac{k+p+q}{2}}\Bigr]
	\end{split}
\end{equation}

\begin{equation}
	\begin{split}
		C^\Lambda_{13}(q; k, p) &= \sum_{\alpha\beta} \Bigl[2 X^\Lambda_{\beta\alpha}(q) \hFB{X_A}{\alpha}{q}{k} \hFB{X_C}{\beta}{q}{p} \\
	& - X^\Lambda_{\alpha\beta}(p-k) \bigl(\hFB{X_C}{\alpha}{p-k}{\tfrac{k+p-q}{2}} \hFB{X_A}{\beta}{p-k}{\tfrac{k+p+q}{2}} \\
	& \hspace{1cm} + \hFB{X_C}{\alpha}{p-k}{\tfrac{k+p+q}{2}} \hFB{X_A}{\beta}{p-k}{\tfrac{k+p-q}{2}}\bigr)\Bigr]
	\end{split}
\end{equation}

\begin{equation}
	\begin{split}
		C^\Lambda_{31}(q; k, p) &= \sum_{\alpha\beta} \Bigl[2 X^\Lambda_{\alpha\beta}(q) \hFB{X_C}{\alpha}{q}{k} \hFB{X_A}{\beta}{q}{p} \\
	& - X^\Lambda_{\alpha\beta}(p-k) \bigl(\hFB{X_C}{\alpha}{p-k}{\tfrac{k+p-q}{2}} \hFB{X_A}{\beta}{p-k}{\tfrac{k+p+q}{2}} \\
	& \hspace{1cm} + \hFB{X_C}{\alpha}{p-k}{\tfrac{k+p+q}{2}} \hFB{X_A}{\beta}{p-k}{\tfrac{k+p-q}{2}}\bigr)\Bigr]
	\end{split}
\end{equation}

\begin{equation}
\begin{split}
	C^\Lambda_{12}(q; k, p) = -\sum_{\alpha,\beta} \Bigl[&\nu^\Lambda_{\alpha\beta}(q) \hFB{\nu}{\alpha}{q}{k} \hFB{\nu}{\beta}{q}{p} - \tilde \nu^\Lambda_{\alpha\beta}(q) \hFB{\tilde \nu}{\alpha}{q}{k} \hFB{\tilde \nu}{\beta}{q}{p}\\
	& + \tilde\nu^\Lambda_{\alpha\beta}(p-k) \hFB{\tilde \nu}{\alpha}{p-k}{\tfrac{k+p-q}{2}} \hFB{\tilde \nu}{\beta}{p-k}{\tfrac{k+p+q}{2}}\Bigr]
\end{split}
\end{equation}

\begin{equation}
\begin{split}
	C^\Lambda_{21}(q; k, p) = -\sum_{\alpha,\beta} \Bigl[&-\nu^\Lambda_{\alpha\beta}(q) \hFB{\nu}{\alpha}{q}{k} \hFB{\nu}{\beta}{q}{p} - \tilde \nu^\Lambda_{\alpha\beta}(q) \hFB{\tilde \nu}{\alpha}{q}{k} \hFB{\tilde \nu}{\beta}{q}{p}\\
	& + \tilde\nu^\Lambda_{\alpha\beta}(p-k) \hFB{\tilde \nu}{\alpha}{p-k}{\tfrac{k+p-q}{2}} \hFB{\tilde \nu}{\beta}{p-k}{\tfrac{k+p+q}{2}}\Bigr]
\end{split}
\end{equation}

\begin{equation}
	\begin{split}
		C^\Lambda_{23}(q; k, p) &= \sum_{\alpha,\beta} \Bigl[-2 \tilde X^\Lambda_{\beta\alpha}(q) \hFB{\tilde X_\Phi}{\alpha}{q}{k} \hFB{\tilde X_C}{\beta}{q}{p} \\
	& - \tilde X^\Lambda_{\alpha\beta}(p-k) \bigl(\hFB{\tilde X_C}{\alpha}{p-k}{\tfrac{k+p-q}{2}} \hFB{\tilde X_\Phi}{\beta}{p-k}{\tfrac{k+p+q}{2}} \\
	& \hspace{1cm} - \hFB{\tilde X_C}{\alpha}{p-k}{\tfrac{k+p+q}{2}} \hFB{\tilde X_\Phi}{\beta}{p-k}{\tfrac{k+p-q}{2}}\bigr)\Bigr]
	\end{split}
\end{equation}

\begin{equation}
	\begin{split}
		C^\Lambda_{32}(q; k, p) &= \sum_{\alpha,\beta} \Bigl[2 \tilde X^\Lambda_{\beta\alpha}(q) \hFB{\tilde X_\Phi}{\alpha}{q}{k} \hFB{\tilde X_C}{\beta}{q}{p} \\
	& - \tilde X^\Lambda_{\alpha\beta}(p-k) \bigl(\hFB{\tilde X_C}{\alpha}{p-k}{\tfrac{k+p-q}{2}} \hFB{\tilde X_\Phi}{\beta}{p-k}{\tfrac{k+p+q}{2}} \\
	& \hspace{1cm} - \hFB{\tilde X_C}{\alpha}{p-k}{\tfrac{k+p+q}{2}} \hFB{\tilde X_\Phi}{\beta}{p-k}{\tfrac{k+p-q}{2}}\bigr)\Bigr]
	\end{split}
\end{equation}

\begin{equation}
\begin{split}
	C^\Lambda_{01}(q; k, p) &= i \sum_{\alpha,\beta} \tilde X^\Lambda_{\alpha\beta}(p-k) \bigl(\hFB{\tilde X_C}{\alpha}{p-k}{\tfrac{k+p-q}{2}} \hFB{\tilde X_\Phi}{\beta}{p-k}{\tfrac{k+p+q}{2}}\\
	&\hspace{1cm} + \hFB{\tilde X_C}{\alpha}{p-k}{\tfrac{k+p+q}{2}} \hFB{\tilde X_\Phi}{\beta}{p-k}{\tfrac{k+p-q}{2}}\bigr)\\
	&= -C^\Lambda_{10}(q; k, p)
\end{split}
\end{equation}

\begin{equation}
\begin{split}
	C^\Lambda_{02}(q; k, p) &= -i \sum_{\alpha,\beta} X^\Lambda_{\alpha\beta}(p-k) \bigl(\hFB{X_C}{\alpha}{p-k}{\tfrac{k+p-q}{2}} \hFB{X_A}{\beta}{p-k}{\tfrac{k+p+q}{2}}\\
	&\hspace{1cm} - \hFB{X_C}{\alpha}{p-k}{\tfrac{k+p+q}{2}} \hFB{X_A}{\beta}{p-k}{\tfrac{k+p-q}{2}}\bigr)\\
	&= -C^\Lambda_{20}(q; k, p)
\end{split}
\end{equation}

\begin{equation}
\begin{split}
	C^\Lambda_{03}(q; k, p) &= -i \sum_{\alpha,\beta} \nu^\Lambda_{\alpha\beta}(p-k) \hFB{\nu}{\alpha}{p-k}{\tfrac{k+p-q}{2}} \hFB{\nu}{\beta}{p-k}{\tfrac{k+p+q}{2}} \\
	&= -C^\Lambda_{30}(q; k, p)
\end{split}
\end{equation}

\subsection*{Fermionic loop integrands}
The fermionic loop integrands that appear in the flow equation~\eqref{eq:CD:RGOneLoop} read
\begin{align}
	L^\Lambda_{00}(p, q) &= F^\Lambda(p_1) F^\Lambda(p_2) - \im G^\Lambda(p_1) \im G^\Lambda(p_2) + \re G^\Lambda(p_1) \re G^\Lambda(p_2)\\
	L^\Lambda_{01}(p, q) &= L^\Lambda_{10}(p, q) = i \bigl(F^\Lambda(p_1) \im G^\Lambda(p_2) + \im G^\Lambda(p_1) F^\Lambda(p_2)\bigr)\\
	L^\Lambda_{02}(p, q) &= -L^\Lambda_{20}(p, q) = i \bigl(\re G^\Lambda(p_1) F^\Lambda(p_2) - F^\Lambda(p_1) \re G^\Lambda(p_2)\bigr)\\
	L^\Lambda_{03}(p, q) &= L^\Lambda_{30}(p, q) = i \bigl(\re G^\Lambda(p_1) \im G^\Lambda(p_2) + \im G^\Lambda(p_1) \re G^\Lambda(p_2)\bigr)\\
	L^\Lambda_{11}(p, q) &= F^\Lambda(p_1) F^\Lambda(p_2) - \im G^\Lambda(p_1) \im G^\Lambda(p_2) - \re G^\Lambda(p_1) \re G^\Lambda(p_2)\\
	L^\Lambda_{22}(p, q) &= -F^\Lambda(p_1) F^\Lambda(p_2) - \im G^\Lambda(p_1) \im G^\Lambda(p_2) - \re G^\Lambda(p_1) \re G^\Lambda(p_2)\\
	L^\Lambda_{33}(p, q) &= -F^\Lambda(p_1) F^\Lambda(p_2) - \im G^\Lambda(p_1) \im G^\Lambda(p_2) + \re G^\Lambda(p_1) \re G^\Lambda(p_2)\\
	L^\Lambda_{12}(p, q) &= -L^\Lambda_{21}(p, q) = \im G^\Lambda(p_1) \re G^\Lambda(p_2) - \re G^\Lambda(p_1) \im G^\Lambda(p_2)\\
	L^\Lambda_{13}(p, q) &= L^\Lambda_{31}(p, q) = \re G^\Lambda(p_1) F^\Lambda(p_2) + F^\Lambda(p_1) \re G^\Lambda(p_2)\\
	L^\Lambda_{23}(p, q) &= -L^\Lambda_{32}(p, q) = \im G^\Lambda(p_1) F^\Lambda(p_2) - F^\Lambda(p_1) \im G^\Lambda(p_2)
\end{align}
where $p_1 = p - q/2$ and $p_2 = p + q/2$.

\section{Two-loop contributions}
\label{subsec:Appendix:TwoLoopContributions}
\newcommand{\pmq}{p_1}
\newcommand{\ppq}{p_2}
In this section, analytical expressions for the two-loop two-boson exchange (TB) contributions that appear in the last three lines of~\eqref{eq:CD:RGDE_PH_TwoLoop} are presented. They are derived for exchange propagators that are diagonal with respect to the form factor index and for frequency-independent fermion-boson vertices. The arguments of the exchange propagators are abbreviated as
\begin{align}
	p_1 &= p - \tfrac{q}{2}		&		p_2 &= p + \tfrac{q}{2}.
\end{align}
In order to write the equations as compact as possible, effective three-boson vertices are defined by
\begin{equation}
	\gamma_{aim;\nu\alpha\mu}^\Lambda(p, q) = \frac{1}{2} \intzwei{k} f_\nu(\boldsymbol k + \tfrac{\boldsymbol p}{2} - \tfrac{\boldsymbol q}{4}) f_\alpha(\boldsymbol k) f_\mu(\boldsymbol k + \tfrac{\boldsymbol p}{2} + \tfrac{\boldsymbol q}{4}) \tr(\mtau{a}{} \mtau{i}{} \mtau{m}{} \G{}(p+k))|_{k_0 = 0}
\end{equation}
where $a$, $i$ and $m$ label the Pauli matrix basis for describing the Nambu index structure and $\nu$, $\alpha$ as well as $\mu$ form factors. $k_0$ is set to zero because the flow equations are evaluated for vanishing fermionic frequencies in case the frequency dependence of the fermion-boson vertices is neglected. In this form, $\gamma^\Lambda(p, q)$ is appropriate if lattice form factors are used as basis set to describe the dependence of the effective interactions on the fermionic momenta. Expressions that are appropriate in case Fermi surface harmonics are used can be obtained easily by replacing $\boldsymbol k \rightarrow \boldsymbol k_F \pm \boldsymbol q / 2$, integration of the Fermi momenta $\boldsymbol k_F$ over the Fermi surface and averaging over the two signs of $\boldsymbol q$ (very similar to the procedure described in section~\ref{sec:CD:ProjBosProp}). After evaluating the trace over the Pauli matrices and the propagator, three effective three-boson vertices appear:
\begin{gather}
	\gamma^{\re G,\Lambda}_{\nu\alpha\mu}(p, q) = \intzwei{k} f_\nu(\boldsymbol k + \tfrac{\boldsymbol p}{2} - \tfrac{\boldsymbol q}{4}) f_\alpha(\boldsymbol k) f_\mu(\boldsymbol k + \tfrac{\boldsymbol p}{2} + \tfrac{\boldsymbol q}{4}) \re \G{}(p + k)|_{k_0 = 0}\\
	\gamma^{\im G,\Lambda}_{\nu\alpha\mu}(p, q) = \intzwei{k} f_\nu(\boldsymbol k + \tfrac{\boldsymbol p}{2} - \tfrac{\boldsymbol q}{4}) f_\alpha(\boldsymbol k) f_\mu(\boldsymbol k + \tfrac{\boldsymbol p}{2} + \tfrac{\boldsymbol q}{4}) \im \G{}(p + k)|_{k_0 = 0}\\
	\gamma^{F,\Lambda}_{\nu\alpha\mu}(p, q) = \intzwei{k} f_\nu(\boldsymbol k + \tfrac{\boldsymbol p}{2} - \tfrac{\boldsymbol q}{4}) f_\alpha(\boldsymbol k) f_\mu(\boldsymbol k + \tfrac{\boldsymbol p}{2} + \tfrac{\boldsymbol q}{4}) F^\Lambda(p + k)|_{k_0 = 0}.
\end{gather}

The renormalization contributions then read
\begin{align}
		\partial_\Lambda &V^{\text{PH},\Lambda}_{00}(q,\alpha)|_\text{TB} = -\partial_{\Lambda,V} \intdrei{p}\sum_{\mu,\nu} \Bigl[-4 \gamma^{\im G,\Lambda}_{\nu\alpha\mu}(p, q)^2 M^\Lambda_\mu(\pmq) M^\Lambda_\nu(\ppq) \nonumber \\
		&+\gamma^{F,\Lambda}_{\nu\alpha\mu}(p, q)^2 \bigl(A^\Lambda_\mu(\pmq) M^\Lambda_\nu(\ppq) + M^\Lambda_\mu(\pmq) A^\Lambda_\nu(\ppq)\bigr) \nonumber\\
		&+ 4\gamma^{F,\Lambda}_{\nu\alpha\mu}(p, q) \gamma^{\re G,\Lambda}_{\nu\alpha\mu}(p, q) \bigl(X^\Lambda_\mu(\pmq) M^\Lambda_\nu(\ppq) + M^\Lambda_\mu(\pmq) X^\Lambda_\nu(\ppq)\bigr) \nonumber\\
		&+ 2\gamma^{\re G,\Lambda}_{\nu\alpha\mu}(p, q)^2 \bigl(C^\Lambda_\mu(\pmq) M^\Lambda_\nu(\ppq) + M^\Lambda_\mu(\pmq) C^\Lambda_\nu(\ppq)\bigr)\Bigr]\label{eq:Appendix:V00TL}
\end{align}

\begin{align}
		\partial_\Lambda &V^{\text{PH},\Lambda}_{11}(q,\alpha)|_\text{TB} = -\partial_{\Lambda,V} \intdrei{p}\sum_{\mu,\nu} \Bigl[\gamma^{F,\Lambda}_{\nu\alpha\mu}(p, q)^2 \Bigl(\tfrac{1}{2}\bigl( A^\Lambda_\mu(\pmq) A^\Lambda_\nu(\ppq) + \Phi^\Lambda_\mu(\pmq) \Phi^\Lambda_\nu(\ppq)\bigr) \nonumber \\
		&\qquad\qquad\qquad +6 M^\Lambda_\mu(\pmq) M^\Lambda_\nu(\ppq) + 2 C^\Lambda_\mu(\pmq) C^\Lambda_\nu(\ppq) + \nu^\Lambda_\mu(\pmq) \nu^\Lambda_\nu(\ppq) \nonumber \\
		&\qquad\qquad\qquad - 4 X^\Lambda_\mu(\pmq) X^\Lambda_\nu(\ppq) - 4 \tilde X^\Lambda_\mu(\pmq) \tilde X^\Lambda_\nu(\ppq)\Bigr) \nonumber \\
		&+ \gamma^{\re G,\Lambda}_{\nu\alpha\mu}(p, q)^2 \Bigl(A^\Lambda_\mu(\pmq) C^\Lambda_\nu(\ppq) + C^\Lambda_\mu(\pmq) A^\Lambda_\nu(\ppq) + 4 X^\Lambda_\mu(\pmq) X^\Lambda_\nu(\ppq)\Bigr) \nonumber \\
		&+ \gamma^{\im G,\Lambda}_{\nu\alpha\mu}(p, q)^2 \Bigl(4\tilde X^\Lambda_\mu(\pmq) \tilde X^\Lambda_\nu(\ppq) - \Phi^\Lambda_\mu(\pmq) C^\Lambda_\nu(\ppq) - C^\Lambda_\mu(\pmq) \Phi^\Lambda_\nu(\ppq)\Bigr) \nonumber \\
		&- 2\gamma^{\im G,\Lambda}_{\nu\alpha\mu}(p, q) \gamma^{\re G,\Lambda}_{\nu\alpha\mu}(p, q) \Bigl(2\bigl(\tilde X^\Lambda_\mu(\pmq) X^\Lambda_\nu(\ppq) + X^\Lambda_\mu(\pmq) \tilde X^\Lambda_\nu(\ppq)\bigr) \nonumber \\
		&\qquad\qquad\qquad + C^\Lambda_\mu(\pmq) \nu^\Lambda_\nu(\ppq) + \nu^\Lambda_\mu(\pmq) C^\Lambda_\nu(\ppq)\Bigr) \nonumber \\
		&+ 2\gamma^{F,\Lambda}_{\nu\alpha\mu}(p, q) \gamma^{\re G,\Lambda}_{\nu\alpha\mu}(p, q) \Bigl(X^\Lambda_\mu(\pmq) \bigl(A^\Lambda_\nu(\ppq)- 2 C^\Lambda_\nu(\ppq)\bigr) + \nu^\Lambda_\mu(\pmq) \tilde X^\Lambda_\nu(\ppq) \nonumber \\
		&\qquad\qquad\qquad + \bigl(A^\Lambda_\mu(\pmq) - 2 C^\Lambda_\mu(\pmq)\bigr)X^\Lambda_\nu(\ppq) + \tilde X^\Lambda_\mu(\pmq)\nu^\Lambda_\nu(\ppq)\Bigr) \nonumber \\
		&+ 2\gamma^{F,\Lambda}_{\nu\alpha\mu}(p, q) \gamma^{\im G,\Lambda}_{\nu\alpha\mu}(p, q) \Bigl(\tilde X^\Lambda_\mu(\pmq) \bigl(\Phi^\Lambda_\nu(\ppq)+2 C^\Lambda_\nu(\ppq)\bigr) - \nu^\Lambda_\mu(\pmq) X^\Lambda_\nu(\ppq) \nonumber \\
		&\qquad\qquad\qquad + \bigl(\Phi^\Lambda_\mu(\pmq) + 2 C^\Lambda_\mu(\pmq)\bigr)\tilde X^\Lambda_\nu(\ppq) - X^\Lambda_\mu(\pmq)\nu^\Lambda_\nu(\ppq)\Bigr)\Bigr]\label{eq:Appendix:V11TL}
\end{align}

\begin{align}
		\partial_\Lambda &V^{\text{PH},\Lambda}_{22}(q,\alpha)|_\text{TB} = -\partial_{\Lambda,V} \intdrei{p}\sum_{\mu,\nu} \Bigl[\gamma^{F,\Lambda}_{\nu\alpha\mu}(p, q)^2 \Bigl(\nu^\Lambda_\mu(\pmq) \nu^\Lambda_\nu(\ppq) \nonumber \\
		&\qquad\qquad\qquad + \tfrac{1}{2} \bigl(A^\Lambda_\mu(\pmq) \Phi^\Lambda_\nu(\ppq) + \Phi^\Lambda_\mu(\pmq)A^\Lambda_\nu(\ppq)\bigr)\Bigr) \nonumber \\
		& + \gamma^{\re G,\Lambda}_{\nu\alpha\mu}(p, q)^2 \Bigl(C^\Lambda_\mu(\pmq) \Phi^\Lambda_\nu(\ppq) + \Phi^\Lambda_\mu(\pmq) C^\Lambda_\nu(\ppq) + 4 \tilde X^\Lambda_\mu(\pmq) \tilde X^\Lambda_\nu(\ppq)\Bigr) \nonumber \\
		& - \gamma^{\im G,\Lambda}_{\nu\alpha\mu}(p, q)^2 \Bigr(A^\Lambda_\mu(\pmq) C^\Lambda_\nu(\ppq) + C^\Lambda_\mu(\pmq) A^\Lambda_\nu(\ppq) - 4 X^\Lambda_\mu(\pmq) X^\Lambda_\nu(\ppq)\Bigr) \nonumber \\
		& - 2 \gamma^{\re G,\Lambda}_{\nu\alpha\mu}(p, q) \gamma^{\im G,\Lambda}_{\nu\alpha\mu}(p, q) \Bigl(\nu^\Lambda_\mu(\pmq) C^\Lambda_\nu(\ppq) + C^\Lambda_\mu(\pmq) \nu^\Lambda_\nu(\ppq) \nonumber \\
		& \qquad\qquad\qquad - 2\bigl(X^\Lambda_\mu(\pmq) \tilde X^\Lambda_\nu(\ppq) + \tilde X^\Lambda_\mu(\pmq) X^\Lambda_\nu(\ppq)\bigr)\Bigr) \nonumber \\
		&+ 2 \gamma^{\re G,\Lambda}_{\nu\alpha\mu}(p, q) \gamma^{F,\Lambda}_{\nu\alpha\mu}(p, q)\Bigl(\nu^\Lambda_\mu(\pmq) \tilde X^\Lambda_\nu(\ppq) + \tilde X^\Lambda_\mu(\pmq) \nu^\Lambda_\nu(\ppq) \nonumber \\
		& \qquad\qquad\qquad+ X^\Lambda_\mu(\pmq) \Phi^\Lambda_\nu(\ppq) + \Phi^\Lambda_\mu(\pmq) X^\Lambda_\nu(\ppq)\Bigr) \nonumber \\
		&- 2 \gamma^{\im G,\Lambda}_{\nu\alpha\mu}(p, q) \gamma^{F,\Lambda}_{\nu\alpha\mu}(p, q) \Bigl(X^\Lambda_\mu(\pmq) \nu^\Lambda_\nu(\ppq) + \nu^\Lambda_\mu(\pmq) X^\Lambda_\nu(\ppq) \nonumber \\
		& \qquad\qquad\qquad - A^\Lambda_\mu(\pmq) \tilde X^\Lambda_\nu(\ppq) - \tilde X^\Lambda_\mu(\pmq) A^\Lambda_\nu(\ppq)\Bigr)\Bigr]
\end{align}

\begin{align}
		\partial_\Lambda &V^{\text{PH},\Lambda}_{33}(q,\alpha)|_\text{TB} = -\partial_{\Lambda,V} \intdrei{p}\sum_{\mu,\nu} \Bigl[\gamma^{F,\Lambda}_{\nu\alpha\mu}(p, q)^2 \Bigl(4 X^\Lambda_\mu(\pmq) X^\Lambda_\nu(\ppq) \nonumber \\
		& \qquad\qquad \qquad + A^\Lambda_\mu(\pmq) C^\Lambda_\nu(\ppq) + C^\Lambda_\mu(\pmq) A^\Lambda_\nu(\ppq)\Bigr) \nonumber \\
		&+ \gamma^{\re G,\Lambda}_{\nu\alpha\mu}(p, q)^2 \Bigl(\tfrac{1}{2} \big(A^\Lambda_\mu(\pmq) A^\Lambda_\nu(\ppq) + \Phi^\Lambda_\mu(\pmq) \Phi^\Lambda_\nu(\ppq)\bigr) + 6 \Phi^{M,\Lambda}_\mu(\pmq) M^\Lambda_\mu(\ppq) \nonumber \\
		& \qquad \qquad \qquad + 2 C^\Lambda_\mu(\pmq) C^\Lambda_\mu(\ppq) - 4 X^\Lambda_\mu(\pmq) X^\Lambda_\nu(\ppq) - \nu^\Lambda_\mu(\pmq) \nu^\Lambda_\nu(\ppq) \nonumber \\
		& \qquad\qquad\qquad + 4 \tilde X^\Lambda_\mu(\pmq) \tilde X^\Lambda_\nu(\ppq)\Bigr) \nonumber \\
		&+ \gamma^{\im G,\Lambda}_{\nu\alpha\mu}(p, q)^2 \Bigl(\nu^\Lambda_\mu(\pmq) \nu^\Lambda_\nu(\ppq) - \tfrac{1}{2} \bigl(A^\Lambda_\mu(\pmq) \Phi^\Lambda_\nu(\ppq) + \Phi^\Lambda_\mu(\pmq) A^\Lambda_\nu(\ppq))\Bigr) \nonumber \\
		&+ \gamma^{\im G,\Lambda}_{\nu\alpha\mu}(p, q) \gamma^{\re G,\Lambda}_{\nu\alpha\mu}(p, q) \Bigl(4\bigl(X^\Lambda_\mu(\pmq) \tilde X^\Lambda_\nu(\ppq) + \tilde X^\Lambda_\mu(\pmq) X^\Lambda_\nu(\ppq)\bigr) \nonumber \\
		&\qquad\qquad\qquad - \nu^\Lambda_\mu(\pmq) \bigl(A^\Lambda_\nu(\ppq) + \Phi^\Lambda_\nu(\ppq)\bigr)- \bigl(A^\Lambda_\mu(\ppq) + \Phi^\Lambda_\mu(\ppq)\bigr) \nu^\Lambda_\nu(\ppq)\Bigr) \nonumber \\
		&+ 2\gamma^{\re G,\Lambda}_{\nu\alpha\mu}(p, q) \gamma^{F,\Lambda}_{\nu\alpha\mu}(p, q) \Bigl(\nu^\Lambda_\mu(\pmq) \tilde X^\Lambda_\nu(\ppq) + \tilde X^\Lambda_\mu(\pmq) \nu^\Lambda_\nu(\ppq) \nonumber \\
		&\qquad\qquad\qquad - X^\Lambda_\mu(\pmq) \bigl(A^\Lambda_\nu(\ppq) - 2 C^\Lambda_\nu(\ppq)\bigr) - \bigl(A^\Lambda_\mu(\pmq) - 2 C^\Lambda_\mu(\pmq)\bigr) X^\Lambda_\nu(\ppq)\Bigr) \nonumber \\
		& + 2 \gamma^{\im G,\Lambda}_{\nu\alpha\mu}(p, q) \gamma^{F,\Lambda}_{\nu\alpha\mu}(p, q) \Bigl(A^\Lambda_\mu(\pmq) \tilde X^\Lambda_\nu(\ppq) + \tilde X^\Lambda_\mu(\pmq) A^\Lambda_\nu(\ppq) \nonumber \\
		& \qquad\qquad\qquad + X^\Lambda_\mu(\pmq) \nu^\Lambda_\nu(\ppq) + \nu^\Lambda_\mu(\pmq) X^\Lambda_\nu(\ppq)\Bigr)\label{eq:Appendix:V33TL}
\end{align}

\begin{align}
		\partial_\Lambda &V^{\text{PH},\Lambda}_{13}(q,\alpha)|_\text{TB} = -\partial_{\Lambda,V} \intdrei{p}\sum_{\mu,\nu} \Bigl[\gamma^{F,\Lambda}_{\nu\alpha\mu}(p, q)^2 \Bigl(\nu^\Lambda_\mu(\pmq) \tilde X^\Lambda_\nu(\ppq) + \tilde X^\Lambda_\mu(\pmq) \nu^\Lambda_\nu(\ppq) \nonumber \\
		&\qquad\qquad\qquad + X^\Lambda_\mu(\pmq) \bigl(A^\Lambda_\nu(\ppq) - 2 C^\Lambda_\nu(\ppq)\bigr) + \bigl(A^\Lambda_\mu(\pmq) - 2 C^\Lambda_\mu(\pmq)\bigr) X^\Lambda_\nu(\ppq)\Bigr) \nonumber \\
		& + \gamma^{\re G,\Lambda}_{\nu\alpha\mu}(p, q)^2 \Bigl(\nu^\Lambda_\mu(\pmq) \tilde X^\Lambda_\nu(\ppq) + \tilde X^\Lambda_\mu(\pmq) \nu^\Lambda_\nu(\ppq) \nonumber \\
		&\qquad\qquad\qquad - X^\Lambda_\mu(\pmq) \bigl(A^\Lambda_\nu(\ppq) - 2 C^\Lambda_\nu(\ppq)\bigr) - \bigl(A^\Lambda_\mu(\pmq) - 2 C^\Lambda_\mu(\pmq)\bigr) X^\Lambda_\nu(\ppq)\Bigr) \nonumber \\
		& + \gamma^{\im G, \Lambda}_{\nu\alpha\mu}(p, q)^2 \Bigl(X^\Lambda_\mu(\pmq) \Phi^\Lambda_\nu(\ppq) + \Phi^\Lambda_\mu(\pmq) X^\Lambda_\nu(\ppq) - \tilde X^\Lambda_\mu(\pmq) \nu^\Lambda_\nu(\ppq) - \nu^\Lambda_\mu(\pmq) \tilde X^\Lambda_\nu(\ppq)\Bigr) \nonumber \\
		&+ \gamma^{\im G, \Lambda}_{\nu\alpha\mu}(p, q) \gamma^{\re G, \Lambda}_{\nu\alpha\mu}(p, q) \Bigl(2 \bigl(X^\Lambda_\mu(\pmq) \nu^\Lambda_\nu(\ppq) + \nu^\Lambda_\mu(\pmq) X^\Lambda_\mu(\ppq)\bigr) \nonumber \\
		&\qquad\qquad\qquad + \tilde X^\Lambda_\mu(\pmq) \bigl(A^\Lambda_\nu(\ppq) + \Phi^\Lambda_\nu(\ppq) - 2 C^\Lambda_\nu(\ppq)\bigr) \nonumber \\
		& \qquad\qquad\qquad + \bigl(A^\Lambda_\mu(\pmq) + \Phi^\Lambda_\mu(\pmq) - 2 C^\Lambda_\mu(\pmq)\bigr) \tilde X^\Lambda_\nu(\ppq)\Bigr) \nonumber \\
		& + \gamma^{\re G, \Lambda}_{\nu\alpha\mu}(p, q) \gamma^{F, \Lambda}_{\nu\alpha\mu}(p, q) \Bigr(\tfrac{1}{2} \Phi^\Lambda_\mu(\pmq) \Phi^\Lambda_\nu(\ppq) + 8 X^\Lambda_\mu(\pmq) X^\Lambda_\nu(\pmq) \nonumber \\
		&\qquad\qquad\qquad - \tfrac{1}{2} \bigl(A^\Lambda_\mu(\pmq) - 2 C^\Lambda_\mu(\pmq)\bigr) \bigl(A^\Lambda_\nu(\ppq) - 2 C^\Lambda_\nu(\ppq)\bigr) + 6 M^\Lambda_\mu(\pmq) M^\Lambda_\nu(\ppq)\Bigr) \nonumber \\
		& - \gamma^{\im G, \Lambda}_{\nu\alpha\mu}(p, q) \gamma^{F, \Lambda}_{\nu\alpha\mu}(p, q) \Bigl(4 \bigl(X^\Lambda_\mu(\pmq) \tilde X^\Lambda_\nu(\ppq) + \tilde X^\Lambda_\mu(\pmq) X^\Lambda_\nu(\ppq)\bigr) \nonumber \\
		&\qquad\qquad\qquad + \tfrac{1}{2}\nu^\Lambda_\mu(\pmq) \bigl(\Phi^\Lambda_\nu(\ppq) - A^\Lambda_\nu(\ppq) + 2 C^\Lambda_\nu(\ppq)\bigr) \nonumber \\
		& \qquad\qquad\qquad + \tfrac{1}{2}\bigl(\Phi^\Lambda_\mu(\pmq) - A^\Lambda_\mu(\pmq) + 2 C^\Lambda_\mu(\pmq)\bigr) \nu^\Lambda_\nu(\ppq)\Bigr)\label{eq:Appendix:V13TL}
\end{align}

\begin{align}
		\partial_\Lambda &V^{\text{PH},\Lambda}_{12}(q,\alpha)|_\text{TB} = -\partial_{\Lambda,V} \intdrei{p} \sum_{\mu,\nu} \Bigl[\gamma^{F,\Lambda}_{\nu\alpha\mu}(p, q)^2 \Bigl(2 \bigl(\tilde X^\Lambda_\mu(\pmq) X^\Lambda_\nu(\ppq) - X^\Lambda_\mu(\pmq) \tilde X^\Lambda_\nu(\ppq)\bigr) \nonumber \\
	&\qquad\qquad\qquad + \tfrac{1}{2}\bigl(A^\Lambda_\mu(\pmq) + \Phi^\Lambda_\mu(\pmq)\bigr) \nu^\Lambda_\nu(\ppq) - \tfrac{1}{2} \nu^\Lambda_\mu(\pmq) \bigl(A^\Lambda_\nu(\ppq) + \Phi^\Lambda_\nu(\ppq)\bigr)\Bigr) \nonumber \\
	&+\gamma^{\re G,\Lambda}_{\nu\alpha\mu}(p, q)^2 \Bigl(C^\Lambda_\mu(\pmq) \nu^\Lambda_\nu(\ppq) - \nu^\Lambda_\mu(\pmq) C^\Lambda_\nu(\ppq) \nonumber \\
	& \qquad\qquad\qquad + 2\bigl(X^\Lambda_\mu(\pmq) \tilde X^\Lambda_\nu(\ppq) - \tilde X^\Lambda_\mu(\pmq) X^\Lambda_\nu(\ppq)\bigr)\Bigr) \nonumber \\
	&+ \gamma^{\im G,\Lambda}_{\nu\alpha\mu}(p, q)^2 \Bigl(\nu^\Lambda_\mu(\pmq) C^\Lambda_\nu(\ppq) - C^\Lambda_\mu(\pmq) \nu^\Lambda_\nu(\ppq) \nonumber \\
	& \qquad\qquad\qquad + 2\bigl(X^\Lambda_\mu(\pmq) \tilde X^\Lambda_\nu(\ppq) - \tilde X^\Lambda_\mu(\pmq) X^\Lambda_\nu(\ppq)\bigr)\Bigr) \nonumber \\
	&+ \gamma^{\im G,\Lambda}_{\nu\alpha\mu}(p, q) \gamma^{\re G,\Lambda}_{\nu\alpha\mu}(p, q) \Bigl(C^\Lambda_\mu(\pmq) \bigl(A^\Lambda_\nu(\ppq) + \Phi^\Lambda_\nu(\ppq)\bigr) - \bigl(A^\Lambda_\mu(\pmq) + \Phi^\Lambda_\mu(\pmq)\bigr) C^\Lambda_\nu(\ppq)\Bigr) \nonumber \\
	&+\gamma^{F,\Lambda}_{\nu\alpha\mu}(p, q) \gamma^{\re G,\Lambda}_{\nu\alpha\mu}(p, q) \Bigl(2\bigl(X^\Lambda_\mu(\pmq) \nu^\Lambda_\nu(\ppq) - \nu^\Lambda_\mu(\pmq) X^\Lambda_\nu(\ppq)\bigr) \nonumber \\
	&\qquad\qquad\qquad + \bigl(A^\Lambda_\mu(\pmq) + \Phi^\Lambda_\mu(\pmq) - 2 C^\Lambda_\mu(\pmq)\bigr) \tilde X^\Lambda_\nu(\ppq) \nonumber \\
	&\qquad\qquad\qquad - \tilde X^\Lambda_\mu(\pmq) \bigl(A^\Lambda_\nu(\ppq) + \Phi^\Lambda_\nu(\ppq) - 2 C^\Lambda_\nu(\ppq)\bigr)\Bigr) \nonumber \\
	&+\gamma^{F,\Lambda}_{\nu\alpha\mu}(p, q) \gamma^{\im G,\Lambda}_{\nu\alpha\mu}(p, q) \Bigl(2 \bigl(\tilde X^\Lambda_\mu(\pmq) \nu^\Lambda_\nu(\ppq) - \nu^\Lambda_\mu(\pmq) \tilde X^\Lambda_\nu(\ppq)\bigr) \nonumber \\
	&\qquad\qquad\qquad - \bigl(A^\Lambda_\mu(\pmq) + \Phi^\Lambda_\mu(\pmq) + 2 C^\Lambda_\mu(\pmq)\bigr) X^\Lambda_\nu(\ppq) \nonumber \\
	&\qquad\qquad\qquad + X^\Lambda_\mu(\pmq) \bigl(A^\Lambda_\nu(\ppq) + \Phi^\Lambda_\nu(\ppq) + 2 C^\Lambda_\nu(\ppq)\bigr)\Bigr)\Bigr]
\end{align}

\begin{align}
		\partial_\Lambda &V^{\text{PH},\Lambda}_{23}(q,\alpha)|_\text{TB} = -\partial_{\Lambda,V} \intdrei{p} \sum_{\mu,\nu} \Bigl[\gamma^{F,\Lambda}_{\nu\alpha\mu}(p, q)^2 \Bigl(\tilde X^\Lambda_\mu(\pmq) A^\Lambda_\nu(\ppq) - A^\Lambda_\mu(\pmq) \tilde X^\Lambda_\nu(\ppq) \nonumber \\
	&\qquad\qquad\qquad + \nu^\Lambda_\mu(\pmq) X^\Lambda_\nu(\ppq) - X^\Lambda_\mu(\pmq) \nu^\Lambda_\nu(\ppq)\Bigr) \nonumber \\
	&+\gamma^{\re G,\Lambda}_{\nu\alpha\mu}(p, q)^2 \Bigl(X^\Lambda_\mu(\pmq) \nu^\Lambda_\nu(\ppq) - \nu^\Lambda_\mu(\pmq) X^\Lambda_\nu(\ppq) \nonumber \\
	&\qquad\qquad\qquad +\tilde X^\Lambda_\mu(\pmq) \bigl(\phi^\Lambda_\nu(\ppq) + 2 C^\Lambda_\nu(\ppq)\bigr) - \bigl(\phi^\Lambda_\nu(\pmq) + 2 C^\Lambda_\nu(\pmq)\bigr) \tilde X^\Lambda_\nu(\ppq)\Bigr) \nonumber \\
	&+\gamma^{\im G,\Lambda}_{\nu\alpha\mu}(p, q)^2 \Bigl(\nu^\Lambda_\mu(\pmq) X^\Lambda_\nu(\ppq) - X^\Lambda_\mu(\pmq) \nu^\Lambda_\nu(\ppq) + A^\Lambda_\mu(\pmq) \tilde X^\Lambda_\nu(\ppq) - \tilde X^\Lambda_\mu(\pmq) A^\Lambda_\nu(\ppq)\Bigr) \nonumber \\
	&+ \gamma^{\im G,\Lambda}_{\nu\alpha\mu}(p, q) \gamma^{\re G,\Lambda}_{\nu\alpha\mu}(p, q) \Bigl(2 \bigl(\nu^\Lambda_\mu(\pmq) \tilde X^\Lambda_\nu(\ppq) - \tilde X^\Lambda_\mu(\pmq) \nu^\Lambda_\nu(\ppq)\bigr) \nonumber \\
	& \qquad\qquad\qquad + X^\Lambda_\mu(\pmq) \bigl(A^\Lambda_\nu(\ppq) + \Phi^\Lambda_\nu(\ppq) + 2 C^\Lambda_\nu(\ppq)\bigr) \nonumber \\
	& \qquad\qquad\qquad - \bigl(A^\Lambda_\mu(\pmq) + \Phi^\Lambda_\mu(\pmq) + 2 C^\Lambda_\mu(\pmq)\bigr) X^\Lambda_\nu(\ppq)\Bigr) \nonumber \\
	& + \gamma^{\re G,\Lambda}_{\nu\alpha\mu}(p, q) \gamma^{F,\Lambda}_{\nu\alpha\mu}(p, q) \Bigl(4 \bigl(\tilde X^\Lambda_\mu(\pmq) X^\Lambda_\nu(\ppq) - X^\Lambda_\mu(\pmq) \tilde X^\Lambda_\nu(\ppq)\bigr) \nonumber \\
	& \qquad\qquad\qquad + \tfrac{1}{2} \nu^\Lambda_\mu(\pmq) \bigl(\Phi^\Lambda_\nu(\ppq) - A^\Lambda_\nu(\ppq) + 2 C^\Lambda_\nu(\ppq)\bigr) \nonumber \\
	& \qquad\qquad\qquad - \tfrac{1}{2} \bigl(\Phi^\Lambda_\mu(\pmq) - A^\Lambda_\mu(\pmq) + 2 C^\Lambda_\mu(\pmq)\bigr) \nu^\Lambda_\nu(\ppq)\Bigr) \nonumber \\
	& + \gamma^{\im G,\Lambda}_{\nu\alpha\mu}(p, q) \gamma^{F,\Lambda}_{\nu\alpha\mu}(p, q) \Bigl( \tfrac{1}{2} A^\Lambda_\mu(\pmq) \bigl(\Phi^\Lambda_\nu(\ppq) + 2 C^\Lambda_\nu(\ppq)\bigr) \nonumber \\
	& \qquad\qquad\qquad + \tfrac{1}{2} \bigl(\Phi^\Lambda_\mu(\pmq) + 2 C^\Lambda_\mu(\pmq)\bigr) A^\Lambda_\nu(\ppq)\Bigr)\Bigr].
\end{align}

\chapter{Attractive Hubbard model with \texorpdfstring{$U\rightarrow 0$}{U -> 0}}
\label{sec:Appendix:UToZero}

This appendix briefly reports on the solution of the attractive Hubbard model in the limit of a vanishing interaction, $U\rightarrow 0$. The first part presents a perturbative solution similar to the one by Mart\'in-Rodero and Flores~\cite{Martin-Rodero1992}. The second part discusses how the perturbative result can be recovered in the framework of the modified one-loop truncation within a very simple approximation.

\section{Perturbation theory}
In the limit of a vanishing interaction $U$, the effect of particle-hole fluctuations on the superfluid gap can be treated perturbatively. For that purpose, the superfluid gap is computed self-consistently from a renormalized gap equation,
\begin{gather}
	\Delta(k) = -\intdrei{p} \Gamma^\text{PP,irr}_{+-+-}(k,p,p,k) F(p).
	\label{eq:BCS_U2}
\end{gather}
It differs from the gap equation in mean-field approximation in that the bare interaction $U$ is replaced by the particle-particle irreducible vertex $\Gamma^\text{PP,irr}$. In the weak-coupling limit $U\rightarrow 0$, it suffices to perform a double expansion of the right hand side up to the second order in $U$ and the first order in $\Delta$~\cite{Martin-Rodero1992}. 
This yields
\begin{gather}
	\Gamma^\text{PP,irr}_{+-+-}(k,p,p,k) = U - U^2 L_\text{PH}(k+p)
\end{gather}
for the particle-particle irreducible vertex, where $L_\text{PH}(q)$ is the particle-hole bubble\footnote{Note that up to terms of order $\mathcal O(\Delta)$, there is no difference between the contributions from magnetic ($L_{00}$) and density ($L_{33}$) fluctuations, \ie\ $L_\text{PH} = L_{00} = L_{33}$.}. For a self-consistent solution in $\mathcal O(U^2)$ for a larger interaction, it would be necessary to take all contributions in the anomalous self-energy and furthermore the normal self-energy into account. The self-consistent result including the leading corrections in $\Delta$ can be found in~\cite{Martin-Rodero1992}. Evaluating the above equation for $k_0 = 0$ and assuming the gap to be local, one arrives at
\begin{gather}
	\Delta = \intzwei{p} \Delta(p_0 = 0, \boldsymbol p) = -\intdrei{p} F(p) \bigl(U - U^2 \intzwei{k} L_\text{PH}(p_0, \boldsymbol k + \boldsymbol p)\bigr).
\end{gather}
After performing the frequency integrations, the gap equation reads
\begin{gather}
	\Delta = -U\Delta \intzwei{p} \frac{a(E(\boldsymbol p))}{2 E(\boldsymbol p)},
	\label{eq:selfconU2}
\end{gather}
where $E(\boldsymbol p) = \sqrt{\Delta^2+\xi^2(\boldsymbol p)}$. In the weak-coupling limit and away from van Hove filling, the function $a(\omega)$ can be computed with bare propagators, yielding
\begin{gather}
	a(\omega) = 1 + U \intzwei{k} \intzwei{k'} \frac{\Theta(\xi(\boldsymbol k)) \Theta(-\xi(\boldsymbol k')) + \Theta(-\xi(\boldsymbol k)) \Theta(\xi(\boldsymbol k'))}{|\omega| + |\xi(\boldsymbol k)| + |\xi(\boldsymbol k')|} < 1.
	\label{eq:a}
\end{gather}
Note that $a(\omega)$ is non-analytic at $\omega = 0$ (for a constant density of states one obtains $a(\omega)-a(0)\sim \omega \log(\omega)$), but this does not have a noticeable influence on the result for the gap. Numerically solving equation~\eqref{eq:selfconU2} for $U=-2$, $t' = -0.1$ and $\mu=-1.5$ (corresponding to quarter-filling), one obtains $\Delta = 0.071$ when $O(U^2)$-corrections are included and $\Delta_\text{BCS} = 0.22$ without them. Particle-hole fluctuations thus reduce the gap to about one third of the mean-field value,
\begin{equation}
	\frac{\Delta}{\Delta_\text{BCS}} = 0.32.
\end{equation}
In $\mathcal O(U^2)$, this ratio depends only weakly on $U$ even for moderate couplings.

In order to obtain a simple expression for the suppression of the gap in the weak-coupling limit, the energy dependence of $a(\omega)$ in~\eqref{eq:selfconU2} or equivalently in~\eqref{eq:a} can be neglected. This results in equation~\eqref{eq:BCS_U2} with $\Gamma^\text{PP,irr}_{+-+-}(k,p,p,k)$ being replaced by $U_\text{eff} = U a(0)$ with $|U a(0)| < |U|$, yielding a reduction of the effective interaction in comparison to the mean-field approximation. Writing
\begin{equation}
	a(0) = 1 - |U| \alpha_0 \rho_0,
\end{equation}
where $\rho_0$ is the density of states at the Fermi level and
\begin{equation}
	\alpha_0 \rho_0 = \intzwei{k} \intzwei{k'} \frac{\Theta(\xi(\boldsymbol k)) \Theta(-\xi(\boldsymbol k')) + \Theta(-\xi(\boldsymbol k)) \Theta(\xi(\boldsymbol k'))}{|\xi(\boldsymbol k)| + |\xi(\boldsymbol k')|},
\end{equation}
it is possible to obtain a simple result for $\Delta / \Delta_\text{BCS}$ in the limit $U\rightarrow 0$. Solving the gap equation with the reduced interaction to logarithmic precision for a constant density of states yields
\begin{equation}
	\Delta = 2 D \exp\Bigl(-\frac{1}{|U| \rho_0 (1-|U| \alpha_0 \rho_0)}\Bigr) = 2 D \exp\Bigl(-\frac{1}{|U_\text{eff}| \rho_0 }\Bigr) \stackrel{U\rightarrow 0}{\approx} \Delta_\text{BCS} \exp(-\alpha_0),
	\label{eq:BCS_Uto0}
\end{equation}
where $D$ is half of the bandwidth. At quarter filling and for $t' = -0.1$, one obtains $\alpha_0 \rho_0 = 0.172$ and $\rho_0 = 0.145$, so that $\alpha_0 = 1.18$ and $\exp(-\alpha_0) = 0.31$. Equation~\eqref{eq:BCS_Uto0} therefore predicts a reduction of the superfluid gap by particle-hole fluctuations to roughly one third of the mean-field value in the limit $U \rightarrow 0$. The same result is obtained from the solution of equation~\eqref{eq:selfconU2} even for $U=-2$. Note that for this latter value of $U$, the left hand side of equation~\eqref{eq:BCS_Uto0} yields a larger reduction of the gap to one fifth of the mean-field value.

\section{Equivalent approximation for the RG}
In this section, it is demonstrated how the above perturbative result can be obtained from the RG\@. For that purpose, very simple approximations are applied to the flow equations for the superfluid gap and for the amplitude mode: First, the superfluid gap is approximated as local and frequency independent. Second, only particle-hole fluctuation corrections to the Cooper channel are kept and all other fluctuation effects are neglected. This yields the following set of equations:
\begin{gather}
	\partial_\Lambda \Delta^\Lambda = - \intdrei{p} S^\Lambda_{+-}(p) \Bigl[U + A^\Lambda(0) - \intzwei{k} \bigl(3 M^\Lambda(p_0, \boldsymbol k) - C^\Lambda(p_0, \boldsymbol k)\bigr)\Bigr]\\
	\partial_\Lambda A^\Lambda(0) = \intdrei{p} \partial_\Lambda L^\Lambda_{11}(p,0) \Bigl[U + A^\Lambda(0) - \intzwei{k} \bigl(3 M^\Lambda(p_0, \boldsymbol k) - C^\Lambda(p_0, \boldsymbol k)\bigr)\Bigr]^2.
\end{gather}
It can be solved easily in case the frequency and scale dependence of the effective interactions in the particle-hole channel are also neglected. These approximations seem to be well justified for a regular Fermi surface in the limit $U\rightarrow 0$. 
Sloppily removing the regulator from the non-singular particle-hole fluctuations then amounts to treating them perturbatively. The RG is subsequently applied to the singular particle-particle channel. A similar argumentation was applied in~\cite{Raghu2010} for the repulsive Hubbard model in the weak-coupling limit. Eventually, the RG equations reduce to
\begin{gather}
	\partial_\Lambda \Delta^\Lambda = - (U_\text{eff} + A^\Lambda(0)) \intdrei{p} S^\Lambda_{+-}(p) \\
	\partial_\Lambda A^\Lambda(0) = (U_\text{eff} + A^\Lambda(0))^2 \intdrei{p} \partial_\Lambda L^\Lambda_{11}(p,0)
\end{gather}
where
\begin{gather}
	U_\text{eff} = U - \intzwei{k} \bigl(3 M(p_0, \boldsymbol k) - C(p_0, \boldsymbol k)\bigr) \approx U - U^2 \intzwei{k} L_\text{PH}^\Lambda(0,\boldsymbol k).
\end{gather}
These equations are solved by the BCS gap equation for the anomalous self-energy and a Bethe-Salpeter like integral equation for the effective interaction in the Cooper channel with the bare interaction $U$ being replaced by $U_\text{eff}$, yielding
\begin{gather}
	\Delta = - U_\text{eff} \intdrei{p} F(p)
\end{gather}
for the gap at scale zero. This is exactly the weak-coupling result as obtained from equation~\eqref{eq:selfconU2} after replacing $a(\omega)$ by $a(0)$.

The above discussion illustrates that the computation of the superfluid gap from a renormalized gap equation is possible if scale-separation holds approximately: In that case, the renormalization of the effective interaction in the Cooper channel is due to high-energy particle-hole excitations, while pairing occurs for fermions near the Fermi surface. 

\chapter{Implementation of coordinate projection for enforcing Ward identities}
\label{sec:Appendix:CPWI}
In chapters~\ref{chap:AttractiveHubbard} and~\ref{chap:RepulsiveHubbard}, a coordinate projection scheme is used to enforce the Ward identity for global charge conservation connecting the anomalous self-energy with the vertex in the numerical solution of the flow equations. Schemes for imposing invariants on the numerical solution of ordinary differential equations were proposed for example in~\cite{Shampine1986,Gear1989,Ascher1994,Shampine1999}. In this section, the implementation of such a scheme is discussed for the attractive Hubbard model following Ascher~\etal~\cite{Ascher1994}. The application of the projection scheme to the repulsive Hubbard model is very similar.

The flow equations can be schematically written in the form
\begin{equation}
	\partial_\Lambda y^\Lambda = f^\Lambda(y^\Lambda)
	\label{eq:Appendix:FlowEqSchematic}
\end{equation}
where $y^\Lambda$ is a vector that contains all flowing couplings and $f^\Lambda(y^\Lambda)$ represents the right hand side of the RG equations. The Ward identity connecting the anomalous self-energy and the vertex reads
\begin{equation}
	h^\Lambda(y^\Lambda) = 0.
\end{equation}
where
\begin{equation}
	\begin{split}
		h^\Lambda(y^\Lambda) &=  \Delta_{(0)} + \Delta_{(0)} \intdrei{p} L^\Lambda_{22}(p,0)\Bigl[U + \Phi^\Lambda(0) g^{\Phi,\Lambda}(p_0)\\
	&\qquad + \tfrac{1}{2} \bigl(A^\Lambda(p-k) g^{A,\Lambda}(\tfrac{p_0}{2})^2 - \Phi^\Lambda(p-k) g^{\Phi,\Lambda}(\tfrac{p_0}{2})^2\bigr)\\
	&\qquad + C^\Lambda(p-k) g^{C,\Lambda}(\tfrac{p_0}{2})^2 - 3 M^\Lambda(p-k) g^{M,\Lambda}(\tfrac{p_0}{2})^2 \Bigr]_{k_0 = 0} - \Delta^\Lambda(0)
	\end{split}
\end{equation}
is the Ward identity evaluated at $k_0 = 0$. The equations for $k_0 \neq 0$ could be used as additional constraints, but this was not implemented for simplicity. As described already in chapter~\ref{chap:AttractiveHubbard}, $h^\Lambda(y^\Lambda)$ can be added to~\eqref{eq:Appendix:FlowEqSchematic} without changing the solution in case the Ward identity is fulfilled,
\begin{equation}
	\partial_\Lambda y^\Lambda = f^\Lambda(y^\Lambda) - (H^\Lambda \cdot H^\Lambda)^{-1} H^\Lambda h^\Lambda(y^\Lambda)
\end{equation}
where $H^\Lambda = \vec \partial_{y^\Lambda} h^\Lambda$ is the gradient of the Ward identity with respect to the couplings. If the solution does not fulfil the Ward identity, the correction pushes it back to the manifold that is spanned by $h^\Lambda(y^\Lambda) = 0$. Ascher~\etal~\cite{Ascher1994} noted that it is possible to do the projection after every step of the ODE solver algorithm. After obtaining the solution $y^\Lambda_\text{RK}$ from for example a Runge-Kutta routine, the couplings are projected in order to obtain the new couplings $y^\Lambda$,
\begin{equation}
	y^\Lambda = y^\Lambda_\text{RK} - \bigl[(H^\Lambda \cdot H^\Lambda)^{-1} H^\Lambda h^\Lambda(y^\Lambda)\bigr]_{y^\Lambda = y^\Lambda_\text{RK}}
\end{equation}
that are used to evaluate the next step of the ODE solver.

For simplicity, the projection was done only for the gap at vanishing fermionic frequency $\Delta^\Lambda(k_0 = 0)$, the mass of the phase mode $m^\Lambda_\Phi = -1/\Phi^\Lambda(0)$ and of the amplitude mode $m^\Lambda_A = -1/ A^\Lambda(0)$. The values of these quantities at finite frequencies or momenta were projected in a way that preserves the shape of the momentum and frequency dependences. For the computation of the gradient, the gap was written as $\Delta^\Lambda(k_0) = \Delta^\Lambda + \delta\Delta^\Lambda(k_0)$ where $\Delta^\Lambda = \Delta^\Lambda(k_0 = 0)$ and $\delta\Delta^\Lambda(k_0) = \Delta^\Lambda(k_0) - \Delta^\Lambda$. The frequency dependent masses were decomposed similarly in $m^\Lambda_i(q_0) = m^\Lambda_i + \delta m^\Lambda_i(q_0)$ where $\delta m^\Lambda_i(q_0) = m^\Lambda_i(q_0) - m^\Lambda_i$ for $i \in\{A, \Phi\}$. 
In the gradient, the derivatives of the Ward identity were approximated by those with respect to the frequency independent quantities $\Delta^\Lambda$, $m^\Lambda_\Phi$ and $m^\Lambda_A$, yielding
\begin{gather}
	\begin{split}
		\frac{\partial h^\Lambda(y^\Lambda)}{\partial \Delta^\Lambda} &=  \Delta_{(0)} \intdrei{p} \frac{\partial L^\Lambda_{22}}{\partial \Delta^\Lambda}(p,0)\Bigl[U + \Phi^\Lambda(0) g^{\Phi,\Lambda}(p_0)\\
	&\quad + \tfrac{1}{2} \bigl(A^\Lambda(p-k) g^{A,\Lambda}(\tfrac{p_0}{2})^2 - \Phi^\Lambda(p-k) g^{\Phi,\Lambda}(\tfrac{p_0}{2})^2\bigr)\\
	&\quad + C^\Lambda(p-k) g^{C,\Lambda}(\tfrac{p_0}{2})^2 - 3 M^\Lambda(p-k) g^{M,\Lambda}(\tfrac{p_0}{2})^2 \Bigr]_{k_0 = 0} - 1
	\end{split}\\
	\frac{\partial h^\Lambda(y^\Lambda)}{\partial m^\Lambda_\Phi} = \Delta_{(0)} \intdrei{p} L^\Lambda_{22}(p,0) \Bigl[(\Phi^\Lambda(0))^2 g^{\Phi,\Lambda}(p_0) - \tfrac{1}{2} (\Phi^\Lambda(p-k))^2 g^{\Phi,\Lambda}(\tfrac{p_0}{2})^2\Bigr]\\
	\frac{\partial h^\Lambda(y^\Lambda)}{\partial m^\Lambda_A} = \frac{\Delta_{(0)}}{2} \intdrei{p} L^\Lambda_{22}(p,0) (A^\Lambda(p-k))^2 g^{A,\Lambda}(\tfrac{p_0}{2})^2\Bigr].
\end{gather}
The square of the norm of $H^\Lambda$ in this approximation reads
\begin{equation}
	H^\Lambda \cdot H^\Lambda = \Bigl(\frac{\partial h^\Lambda(y^\Lambda)}{\partial \Delta^\Lambda}\Bigr)^2 + \Bigl(\frac{\partial h^\Lambda(y^\Lambda)}{\partial m^\Lambda_\Phi}\Bigr)^2 + \Bigl(\frac{\partial h^\Lambda(y^\Lambda)}{\partial m^\Lambda_A}\Bigr)^2.
\end{equation}
The projected values for the frequency dependent gap were obtained as
\begin{equation}
	\Delta^\Lambda(k_0) = \Delta^\Lambda_\text{RK}(k_0) - \Bigl[(H^\Lambda \cdot H^\Lambda)^{-1} \frac{\partial h^\Lambda(y^\Lambda)}{\partial \Delta^\Lambda} h^\Lambda(y^\Lambda)\Bigr]_{y^\Lambda = y^\Lambda_\text{RK}},
\end{equation}
while the momentum and frequency dependence of the amplitude and phase mode of the gap were projected as
\begin{gather}
	A^\Lambda(q) = -\Bigl[- \frac{1}{A^\Lambda(q)} - (H^\Lambda \cdot H^\Lambda)^{-1} \frac{\partial h^\Lambda(y^\Lambda)}{\partial m^\Lambda_A} h^\Lambda(y^\Lambda)\Bigr]^{-1}_{y^\Lambda = y^\Lambda_\text{RK}}\\
	\Phi^\Lambda(q) = -\Bigl[- \frac{1}{\Phi^\Lambda(q)} - (H^\Lambda \cdot H^\Lambda)^{-1} \frac{\partial h^\Lambda(y^\Lambda)}{\partial m^\Lambda_\Phi} h^\Lambda(y^\Lambda)\Bigr]^{-1}_{y^\Lambda = y^\Lambda_\text{RK}}.
\end{gather}
In the numerical solution of the flow equations in chapters~\ref{chap:AttractiveHubbard} and~\ref{chap:RepulsiveHubbard}, it turned out that the derivative of the Ward identity with respect to $m^\Lambda_A$ was much smaller than the other two derivatives. It is expected that other derivatives for example with respect to couplings in the particle-hole channel are even smaller, so that they can be neglected to a good approximation. This suggests that the above scheme captures the most important contributions to the gradient of the Ward identity, which is well approximated for the gap for small $k_0$ and for the amplitude and phase mode for small $q$.

This simple projection scheme yielded satisfactory results and significantly improved the fulfilment of the Ward identity in the numerical solution of the flow equations. Nevertheless, further improvements may be obtained by projecting not only after the steps of the ODE solver, but before every evaluation of the right hand side of the flow equations. Furthermore, a better approximation for the frequency dependence of the gap might be obtained by projecting as described above and subsequent self-consistent determination of $\Delta^\Lambda(k_0)$ from the Ward identity with fixed vertex and normal self-energy.

\chapter{Reduced \texorpdfstring{$s$}{s}- and \texorpdfstring{$d$}{d}-wave pairing and forward scattering model}
\label{sec:Appendix:SDRPFM}
In this appendix, the exact solution of a reduced model with interactions in the $s$- and $d$-wave pairing and forward scattering channels is presented. It yields information on the couplings that have to be taken into account in order to capture the exact solution within the functional RG, which is particularly useful with regard to the anomalous effective interactions. Such a model would appear within the one-particle irreducible framework of RG in case fluctuations are taken into account for the high-energy degrees of freedom while they are neglected at low scales, yielding a renormalized mean-field flow. In the following, it is assumed that the pairing interaction is attractive in the $d$-wave channel and that no instability occurs in the other channels (\ie\ no instability occurs in the particle-hole channel and the $s$-wave pairing interaction is repulsive).

The microscopic action of the model is the same as in equations~\eqref{eq:RPFM:Action},~\eqref{eq:RPFM:KinEnergy} and~\eqref{eq:RPFM:V0SC} to~\eqref{eq:RPFM:F0S}. The momentum dependence of the interaction is assumed to be separable with the same form factors in all channels for simplicity,
\begin{align}
	V^{(0)}(k,k') &= g_s^{(0)} + g_d^{(0)} f_d(\boldsymbol k) f_d(\boldsymbol k')\\
	F^{(0)}_c(k,k') &= g_{c,s}^{(0)} + g_{c,d}^{(0)} f_d(\boldsymbol k) f_d(\boldsymbol k')\\
	F^{(0)}_m(k,k') &= g_{m,s}^{(0)} + g_{m,d}^{(0)} f_d(\boldsymbol k) f_d(\boldsymbol k'),
\end{align}
where the subscripts ``$s$'' and ``$d$'' label couplings in the the $s$- and $d_{x^2-y^2}$-wave channels, respectively, and $f_s(\boldsymbol k) = 1$ and $f_d(\boldsymbol k) = \cos k_x - \cos k_y$ are lattice form factors. The Bethe-Salpeter like integral equation for the couplings~\eqref{eq:RPFM:IntegralEquation} then reads
\begin{equation}
	\begin{split}
	\sum_{\alpha,\beta} f_\alpha(\boldsymbol k) &\hat V^{\text{PH},\Lambda}_{\alpha\beta} f_\beta(\boldsymbol k') = \sum_{\alpha,\beta} f_\alpha(\boldsymbol k) \hat V^{\text{PH},\Lambda_0}_{\alpha\beta} f_\beta(\boldsymbol k') \\
	&\ + \sum_{p,\alpha,\beta,\gamma,\delta} f_\alpha(\boldsymbol k) \hat V^{\text{PH},\Lambda_0}_{\alpha\gamma} f_\gamma(\boldsymbol p) (2 \hat L^\Lambda_f(p)) f_\delta(\boldsymbol p)  \hat V^{\text{PH},\Lambda}_{\delta\beta} f_\beta(\boldsymbol k')
	\end{split}
\end{equation}
where the Greek indices sum over the $s$- and $d$-wave channels. Defining
\begin{equation}
	\hat L^\Lambda_{f,\gamma\delta} = \sum_p f_\gamma(\boldsymbol p) \hat L^\Lambda(p) f_\delta(\boldsymbol p),
\end{equation}
the matrices with regards to the form factor index read
\begin{align}
	(\hat V^{\text{PH},\Lambda})_{\alpha\beta} &= \begin{pmatrix}
	                                     	\hat V_{ss}^{\text{PH},\Lambda}		&		\hat V_{sd}^{\text{PH},\Lambda}\\
	                                     	\hat V_{ds}^{\text{PH},\Lambda}		&		\hat V_{dd}^{\text{PH},\Lambda}
	                                     \end{pmatrix}_{\alpha\beta}		&		
	(\hat L^\Lambda_f)_{\alpha\beta} = \begin{pmatrix}
	                                     	\hat L_{f,ss}^\Lambda		&		\hat L_{f,sd}^\Lambda\\
	                                     	\hat L_{f,ds}^\Lambda		&		\hat L_{f,dd}^\Lambda
	                                     \end{pmatrix}_{\alpha\beta}.
\label{eq:Appendix:FormFactorMatrix}
\end{align}
The sub-matrices have a Nambu structure similar to those that appeared in the explicit solution of the reduced model in chapter~\ref{chap:RPFM}, albeit with additional indices for the form factors. Eliminating the form factors with arguments $\boldsymbol k$ and $\boldsymbol k'$ by orthogonal projection, the integral equation reduces to
\begin{equation}
	\hat V^{\text{PH},\Lambda}_{\alpha\beta} = \hat V^{\text{PH},\Lambda_0}_{\alpha\beta}	+ \sum_{p,\gamma,\delta} \hat V^{\text{PH},\Lambda_0}_{\alpha\gamma} (2 \hat L^\Lambda_{f,\gamma\delta}) \hat V^{\text{PH},\Lambda}_{\delta\beta}.
	\label{eq:Appendix:BetheSalpeterMatrix}
\end{equation}
Note that this is a matrix equation with respect to Nambu and form factor indices. The matrix containing the fermionic loop integrands reads
\begin{equation}
	\hat L^\Lambda_{f,\gamma\delta} = \begin{pmatrix}
	                           	l^\Lambda_{m,\gamma\delta}		&		0		&		0		&		0\\
	                           	0		&		l^\Lambda_{a,\gamma\delta}		&		0		&		l^\Lambda_{x,\delta\gamma}\\
	                           	0		&		0		&		l^\Lambda_{\phi,\gamma\delta}		&		0		\\
	                           	0		& l^\Lambda_{x,\gamma\delta}		&		0		&		l^\Lambda_{c,\gamma\delta}
	                           \end{pmatrix}
\label{eq:Appendix:LMatrix}
\end{equation}
with matrix elements
\begin{equation}
	\begin{split}
	l^\Lambda_{m,\gamma\delta} = \sum_p f_\gamma(\boldsymbol p) L^\Lambda_{00}(p) f_\delta(\boldsymbol p)\\
	l^\Lambda_{a,\gamma\delta} = \sum_p f_\gamma(\boldsymbol p) L^\Lambda_{11}(p) f_\delta(\boldsymbol p)\\
	l^\Lambda_{\phi,\gamma\delta} = \sum_p f_\gamma(\boldsymbol p) L^\Lambda_{22}(p) f_\delta(\boldsymbol p)\\
	l^\Lambda_{c,\gamma\delta} = \sum_p f_\gamma(\boldsymbol p) L^\Lambda_{33}(p) f_\delta(\boldsymbol p)\\
	l^\Lambda_{x,\gamma\delta} = \sum_p f_\gamma(\boldsymbol p) L^\Lambda_{13}(p) f_\delta(\boldsymbol p).
	\end{split}
\end{equation}
The unbroken fourfold symmetry of the lattice leads to simplifications of the matrices: All loops that appear on the diagonal of~\eqref{eq:Appendix:LMatrix} are non-zero for $\gamma=\delta\in\{s,d\}$ because an even number of $d$-wave form factors appears in the integrands in this case. Most off-diagonal entries vanish due to the appearance of an odd number of $d$-wave form factors in the integrands that change sign under rotations of the loop momentum by $\frac{\pi}{2}$. The situation is different for the off-diagonal anomalous (3+1)-loops because these are linear in the anomalous propagator and the contained $d$-wave form factor needs one additional factor of $f_d$ to make the integral finite. The resulting matrix thus reads\footnote{For matrix elements that are diagonal with respect to the form factor index, one index is suppressed for convenience.}
\begin{equation}
	\begin{pmatrix}
	                                     	\hat L_{f,ss}^\Lambda		&		\hat L_{f,sd}^\Lambda\\
	                                     	\hat L_{f,ds}^\Lambda		&		\hat L_{f,dd}^\Lambda
	                                     \end{pmatrix} =
	\begin{pmatrix}
		l^\Lambda_{m,s}	&		0		&		0		&		0		&			0	&		0		&		0		&			0	\\
		0		&		l^\Lambda_{a,s}	&		0		&		0		&		0		&		0		&		0		&		l^\Lambda_{x,ds}\\
		0		&		0		&	l^\Lambda_{\phi,s}	&		0		&		0		&		0		&		0		&		0		\\
		0		&		0		&		0		&	l^\Lambda_{c,s}		&		0		&	l^\Lambda_{x,sd}		&		0		&		0\\
		0		&		0		&		0		&		0		&		l^\Lambda_{m,d}	&		0		&		0		&		0\\
		0		&		0		&		0		&	l^\Lambda_{x,sd}		&		0		&		l^\Lambda_{a,d}	&		0		&		0\\
		0		&		0		&		0		&		0		&		0		&		0		&		l^\Lambda_{\phi,d}		&		0\\
		0		&	l^\Lambda_{x,ds}	&		0		&		0		&		0		&		0		&		0		&		l^\Lambda_{c,d}
	\end{pmatrix}
\end{equation}
The matrix containing the couplings has the same structure and reads
\begin{equation}
\begin{pmatrix}
	                                     	\hat V_{ss}^{PH,\Lambda}		&		\hat V_{sd}^{PH,\Lambda}\\
	                                     	\hat V_{ds}^{PH,\Lambda}		&		\hat V_{dd}^{PH,\Lambda}
	                                     \end{pmatrix} =
\begin{pmatrix}
		g^\Lambda_{m,s}	&		0		&		0		&		0		&			0	&		0		&		0		&			0	\\
		0		&		\tfrac{1}{2} g^\Lambda_{a,s}	&		0		&		0		&		0		&		0		&		0		&		g^\Lambda_{x,ds}\\
		0		&		0		&	\tfrac{1}{2} g^\Lambda_{\phi,s}	&		0		&		0		&		0		&		0		&		0		\\
		0		&		0		&		0		&	g^\Lambda_{c,s}		&		0		&	g^\Lambda_{x,sd}		&		0		&		0\\
		0		&		0		&		0		&		0		&		g^\Lambda_{m,d}	&		0		&		0		&		0\\
		0		&		0		&		0		&	g^\Lambda_{x,sd}		&		0		&		\tfrac{1}{2} g^\Lambda_{a,d}	&		0		&		0\\
		0		&		0		&		0		&		0		&		0		&		0		&		\tfrac{1}{2} g^\Lambda_{\phi,d}		&		0\\
		0		&	g^\Lambda_{x,ds}	&		0		&		0		&		0		&		0		&		0		&		g^\Lambda_{c,d}.
	\end{pmatrix}
\end{equation}
At the scale $\Lambda_0$, the couplings $g$ have to be replaced by the corresponding $g^{(0)}$. 
The flowing couplings are obtained by solving equation~\eqref{eq:Appendix:BetheSalpeterMatrix} for $\hat V^{\text{PH},\Lambda}_{\alpha\beta}$, which is simplified by the block-diagonal structure of the above matrices. For the magnetic channel and the phase component of the vertex, this yields
\begin{align}
	g^\Lambda_{m,i} &= \Bigl( (g^{(0)}_{m,i})^{-1} - 2 l^\Lambda_{m,i}\Bigr)^{-1}\\
	g^\Lambda_{\phi,i} &= \Bigl( (g^{(0)}_{i})^{-1} - l^\Lambda_{\phi,i}\Bigr)^{-1}
\end{align}
for $i\in \{s,d\}$. The couplings for the density, amplitude and anomalous (3+1)-channels read
\begin{align}
	g^\Lambda_{a,s} &= \frac{\bigl(2 g^{(0)}_{c,d}\bigr)^{-1} - l^\Lambda_{c,d}}{d^\Lambda_{ds}}		&		g^\Lambda_{a,d} &= \frac{\bigl(2 g^{(0)}_{c,s}\bigr)^{-1} - l^\Lambda_{c,s}}{d^\Lambda_{sd}}\\
	g^\Lambda_{x,ds} &= \frac{l^\Lambda_{x,ds}}{2d^\Lambda_{ds}}		&		g^\Lambda_{x,sd} &= \frac{l^\Lambda_{x,sd}}{2d^\Lambda_{sd}}\\
	g^\Lambda_{c,d} &= \frac{\bigl(g^{(0)}_s\bigr)^{-1} - l^\Lambda_{a,s}}{2d^\Lambda_{ds}}		&		g^\Lambda_{c,s} &= \frac{\bigl(g^{(0)}_d\bigr)^{-1} - l^\Lambda_{a,d}}{2d^\Lambda_{sd}}
\end{align}
where
\begin{align}
	d^\Lambda_{ds} &= \Bigl(\bigl(g^{(0)}_s\bigr)^{-1} - l^\Lambda_{a,s}\Bigr) \Bigl(\bigl(2g^{(0)}_{c,d}\bigr)^{-1} - l^\Lambda_{c,d}\Bigr) - \bigl(l^\Lambda_{x,ds}\bigr)^2\\
	d^\Lambda_{sd} &= \Bigl(\bigl(g^{(0)}_d\bigr)^{-1} - l^\Lambda_{a,d}\Bigr) \Bigl(\bigl(2g^{(0)}_{c,s}\bigr)^{-1} - l^\Lambda_{c,s}\Bigr) - \bigl(l^\Lambda_{x,sd}\bigr)^2.
\end{align}
This result shows that the anomalous (3+1)-effective interactions couple the $s$-wave forward scattering channel to the $d$-wave pairing channel and vice versa. No anomalous (3+1)-coupling between the $s$-wave forward scattering and pairing channels appears in case the fourfold symmetry of the lattice is unbroken. The same is also true for the $d$-wave channel. Note that the $s$-wave pairing channel is described by two couplings, which were labelled ``amplitude'' and ``phase'' mode in analogy to the $d$-wave channel despite the fact that they remain non-singular.

The critical scale $\Lambda_c$ for $d$-wave pairing is determined from the condition
\begin{equation}
	g^{(0)}_{i} = l^{\Lambda_c}_{\phi,d} = l^{\Lambda_c}_{a,d}.
\end{equation}
At this scale, the $d$-wave pairing interaction would diverge in an RG flow without external pairing field. From the results in chapter~\ref{chap:RPFM}, it follows that the couplings $g^\Lambda_{\phi,d}$, $g^\Lambda_{a,d}$ as well as $g^\Lambda_{x,sd}$ are singular at the critical scale in case the external pairing field is sent to zero due to the appearance of $(g^{(0)}_2)^{-1} - l^\Lambda_{\phi,2}$ or $(g^{(0)}_2\bigr)^{-1} - l^\Lambda_{a,2}$ in the denominators. $g^\Lambda_{\phi,d}$ remains singular below the critical scale, while $g^\Lambda_{a,d}$ and $g^\Lambda_{x,sd}$ are regularized by the generated $d$-wave superfluid gap. All other couplings remain non-singular.

\chapter{Expansion of products of form factors}
\label{sec:Appendix:FormFactorExpansion}
In the flow equations for the self-energy and the vertex in the repulsive Hubbard model in chapter~\ref{chap:RepulsiveHubbard}, Brillouin zone integrals over products of form factors and exchange propagators of the form
\begin{equation}
	\sum_{\beta,\gamma} \intzwei{k} f_\alpha(\boldsymbol k + \boldsymbol p) f_\beta(\boldsymbol p + \tfrac{\boldsymbol k - \boldsymbol q}{2}) f_\gamma(\boldsymbol p + \tfrac{\boldsymbol k + \boldsymbol q}{2}) D^\Lambda_{\beta\gamma}(p_0, \boldsymbol k)
\end{equation}
appear, where $D^\Lambda_{\beta\gamma}$ is some combination of exchange propagators from~\eqref{eq:RH:Propagators}. For an efficient numerical evaluation of the flow equations, it is convenient to analytically decompose the form factor products in parts that depend on $\boldsymbol k$ and parts that do not. The $\boldsymbol k$-integrations are then performed before evaluating the right hand side of the flow equations at every scale. In this appendix, the frequency dependence of the exchange propagators is included as it would appear in the $C^\Lambda$-functions in an improved approximation where the frequency dependence of the exchange propagators but not of the fermion-boson vertices is kept. In the following, the fourfold symmetry of the lattice is assumed to be unbroken and is exploited for simplifying the resulting expressions.

The computation of the fluctuation contributions involving the diagonal propagators requires decompositions of the following products of form factors:
\begin{equation}
		\intzwei{k} f_{\alpha = s}(\boldsymbol k + \boldsymbol p) f_{\beta = s}(\boldsymbol p + \tfrac{\boldsymbol k + \boldsymbol q}{2}) f_{\beta = s}(\boldsymbol p + \tfrac{\boldsymbol k - \boldsymbol q}{2}) D^\Lambda_s(p_0, \boldsymbol k) = \intzwei{k} D^\Lambda_s(p_0, \boldsymbol k)
\end{equation}
\begin{equation}
	\begin{split}
		\intzwei{k} &f_{\alpha = s}(\boldsymbol k + \boldsymbol p) f_{\beta = d}(\boldsymbol p + \tfrac{\boldsymbol k + \boldsymbol q}{2}) f_{\beta = d}(\boldsymbol p + \tfrac{\boldsymbol k - \boldsymbol q}{2}) D^\Lambda_d(p_0, \boldsymbol k) =\\
		&= \tfrac{1}{2} (\cos q_x + \cos q_y) \intzwei{k} D^\Lambda_d(p_0, \boldsymbol k) \\
		&\quad + \tfrac{1}{2} \bigl(\cos(2p_x) + \cos(2p_y)\bigr) \intzwei{k} \cos k_x D^\Lambda_d(p_0, \boldsymbol k)\\
		&\quad -\bigl(\cos(p_x + \tfrac{q_x}{2}) \cos(p_y - \tfrac{q_y}{2}) + \cos(p_x - \tfrac{q_x}{2}) \cos(p_y + \tfrac{q_y}{2})\bigr)\times\\
		&\quad\quad\times\intzwei{k} \cos(\tfrac{k_x}{2}) \cos(\tfrac{k_y}{2}) D^\Lambda_d(p_0, \boldsymbol k)
	\end{split}
\end{equation}
\begin{equation}
	\begin{split}
	\intzwei{k} &f_{\alpha = d}(\boldsymbol k + \boldsymbol p) f_{\beta = s}(\boldsymbol p + \tfrac{\boldsymbol k + \boldsymbol q}{2}) f_{\beta = s}(\boldsymbol p + \tfrac{\boldsymbol k - \boldsymbol q}{2}) D^\Lambda_s(p_0, \boldsymbol k) =\\
	&=(\cos p_x - \cos p_y) \intzwei{k} \cos k_x D^\Lambda_s(p_0, \boldsymbol k)
	\end{split}
\end{equation}
\begin{equation}
	\begin{split}
	&\intzwei{k} f_{\alpha = d}(\boldsymbol k + \boldsymbol p) f_{\beta = d}(\boldsymbol p + \tfrac{\boldsymbol k + \boldsymbol q}{2}) f_{\beta = d}(\boldsymbol p + \tfrac{\boldsymbol k - \boldsymbol q}{2}) D^\Lambda_d(p_0, \boldsymbol k) =\\
	&=\tfrac{1}{4}(\cos p_x - \cos p_y) \intzwei{k} D^\Lambda_d(p_0, \boldsymbol k)\\
	&\quad + \tfrac{1}{2}(\cos p_x - \cos p_y) (\cos q_x + \cos q_y) \intzwei{k} \cos k_x D^\Lambda_d(p_0, \boldsymbol k)\\
	&\quad + \tfrac{1}{4}\bigl(\cos(3p_x) - \cos(3p_y)\bigr) \intzwei{k} \cos(2k_x) D^\Lambda_d(p_0, \boldsymbol k)\\
	&\quad + \tfrac{1}{2}\bigl(\cos p_x \cos(2p_y) - \cos(2p_x) \cos p_y\bigr) \intzwei{k} \cos k_x \cos k_y D^\Lambda_d(p_0, \boldsymbol k)\\
	&\quad + 2(\cos p_x - \cos p_y) \cos(\tfrac{q_x}{2}) \cos(\tfrac{q_y}{2}) \intzwei{k} \sin(\tfrac{k_x}{2}) \sin k_x \cos(\tfrac{k_y}{2}) D^\Lambda_d(p_0, \boldsymbol k)\\
	&\quad-(\cos p_x - \cos p_y) \bigl(\cos(p_x + \tfrac{q_x}{2}) \cos(p_y - \tfrac{q_y}{2}) + \cos(p_x - \tfrac{q_x}{2}) \cos(p_y + \tfrac{q_y}{2})\bigr)\times\\
	&\quad\quad \times \intzwei{k} \cos(\tfrac{3k_x}{2}) \cos(\tfrac{k_y}{2}) D^\Lambda_d(p_0,\boldsymbol k)
	\end{split}
\end{equation}

The products of form factors that appear in the fluctuation contributions from off-diagonal propagators have a somewhat different form. For 
\begin{align*}
	D^\Lambda_+(p_0, \boldsymbol k) &= X^\Lambda_{ds}(p_0, \boldsymbol k) + X^\Lambda_{sd}(p_0, \boldsymbol k)	&	D^\Lambda_-(p_0, \boldsymbol k) &= X^\Lambda_{ds}(p_0, \boldsymbol k) - X^\Lambda_{sd}(p_0, \boldsymbol k)
\end{align*}
and similarly with $X^\Lambda$ replaced by $\tilde X$, they read
\begin{equation}
	\begin{split}
		\intzwei{k}& f_{\alpha = s}(\boldsymbol k + \boldsymbol p) (f_{\beta=d}(\boldsymbol p + \tfrac{\boldsymbol k + \boldsymbol q}{2}) + f_{\beta=d}(\boldsymbol p + \tfrac{\boldsymbol k - \boldsymbol q}{2})) D^\Lambda_+(p_0, \boldsymbol k) =\\
	&\quad = 2 \bigl(\cos p_x \cos(\tfrac{q_x}{2}) - \cos p_y \cos(\tfrac{q_y}{2})\bigr) \intzwei{k} \cos(\tfrac{k_x}{2}) D^\Lambda_+(p_0, \boldsymbol k)
	\end{split}
\end{equation}
\begin{equation}
	\begin{split}
		\intzwei{k}& f_{\alpha = d}(\boldsymbol k + \boldsymbol p) \bigl(f_{\beta=d}(\boldsymbol p + \tfrac{\boldsymbol k + \boldsymbol q}{2}) + f_{\beta=d}(\boldsymbol p + \tfrac{\boldsymbol k - \boldsymbol q}{2})\bigr) D^\Lambda_+(p_0, \boldsymbol k) =\\
	&\quad = \bigl(\cos(\tfrac{q_x}{2}) + \cos(\tfrac{q_y}{2})\bigr) \intzwei{k} \cos(\tfrac{k_x}{2}) D^\Lambda_+(p_0, \boldsymbol k)\\
	&\quad\quad - 2\cos p_x \cos p_y \bigl(\cos(\tfrac{q_x}{2}) + \cos(\tfrac{q_y}{2})\bigr) \intzwei{k} \cos k_x \cos(\tfrac{k_y}{2}) D^\Lambda_+(p_0, \boldsymbol k)\\
	&\quad\quad + \bigl(\cos(2p_x) \cos(\tfrac{q_x}{2}) + \cos(2p_y) \cos(\tfrac{q_y}{2})\bigr) \intzwei{k} \cos(\tfrac{3k_x}{2}) D^\Lambda_+(p_0, \boldsymbol k)
	\end{split}
\end{equation}
\begin{equation}
	\begin{split}
		\intzwei{k}& f_{\alpha = s}(\boldsymbol k + \boldsymbol p) \bigl(f_{\beta=d}(\boldsymbol p + \tfrac{\boldsymbol k - \boldsymbol q}{2}) - f_{\beta=d}(\boldsymbol p + \tfrac{\boldsymbol k + \boldsymbol q}{2})\bigr) D^\Lambda_-(p_0, \boldsymbol k) =\\
	&\quad = -2\bigl(\sin p_x \sin(\tfrac{q_x}{2}) - \sin p_y \sin(\tfrac{q_y}{2})\bigr) \intzwei{k} \cos(\tfrac{k_x}{2}) D^\Lambda_-(p_0, \boldsymbol k)
	\end{split}
\end{equation}
\begin{equation}
	\begin{split}
		&\intzwei{k} f_{\alpha = d}(\boldsymbol k + \boldsymbol p) \bigl(f_{\beta=d}(\boldsymbol p + \tfrac{\boldsymbol k - \boldsymbol q}{2}) - f_{\beta=d}(\boldsymbol p + \tfrac{\boldsymbol k + \boldsymbol q}{2})\bigr) D^\Lambda_-(p_0, \boldsymbol k) =\\
	&\quad = 2\bigl(\sin p_x \cos p_y \sin(\tfrac{q_x}{2}) + \cos p_x \sin p_y \sin(\tfrac{q_y}{2})\bigr) \intzwei{k} \cos k_x \cos(\tfrac{k_y}{2}) D^\Lambda_-(p_0, \boldsymbol k)\\
	&\quad\quad-2 \bigl(\cos p_x \sin p_x \sin(\tfrac{q_x}{2}) + \cos p_y \sin p_y \sin(\tfrac{q_y}{2})\bigr) \intzwei{k} \cos(\tfrac{3k_x}{2}) D^\Lambda_-(p_0,\boldsymbol k).
	\end{split}
\end{equation}

%% file: Thesis.bbl
%

%% file: Thesis_Backmatter.tex
\addchap{Acknowledgements}

This thesis would not have been possible without the support of many people. First of all, I am indebted to Walter Metzner for giving me the opportunity to work on this thesis in his group in Stuttgart. I would like to thank him for proposing the fascinating and challenging topic, great supervision, inspiring discussions, helpful advice, encouraging me in all circumstances and keeping me believing that what I was trying was feasible. Alejandro Muramatsu and Manfred Salmhofer are thanked for co-refereeing this thesis.

The members of the department Metzner provided a friendly and stimulating environment. Intense discussions with Christoph Husemann, Tobias Holder and Benjamin Obert inspired my work and shaped some of the ideas in this thesis. Fruitful conversations with Johannes Bauer, Kay-Uwe Giering, Nils Hasselmann, Peter Horsch, Stefan Maier, Dirk Manske, Hiroyuki Yamase and Roland Zeyer provided me with some insights into the physics of correlated fermions and the Hubbard model. Jeanette Schüller-Knapp supported me a lot on administrative issues. I am grateful to Nils Hasselmann and Tobias Holder for careful proofreading of the manuscript of this thesis.

The IT department of the institute provided well functioning computing facilities and, in particular Armin Schuhmacher, helped in solving any kind of problems.

The International Max Planck Research School for Advanced Materials is acknowledged for financial support. Its coordinator Hans-Georg Libuda's constant help and advice throughout the PhD period are appreciated.

For making the time at the Max Planck Institute more pleasant, for example with interesting discussions during coffee breaks or by organizing common sports activities, I am grateful to many colleagues.
 
On the personal side, I am indepted to my parents Elisabeth and Wolfgang Eberlein for their help and advice over all the years. My sister Maria, my brother Matthias and many friends are thanked for various kinds of support and for having refreshing down-to-earth views on many issues.

\addchap{Academic curriculum vitae}

\newcommand{\BreitelinkeSpalte}{4cm}
\newcommand{\BreitemittelSpalte}{1.7cm}
\newcommand{\BreitekleinemittelSpalte}{1.3cm}
\newcommand{\PlatznachUeberschrift}{1.1mm}
\newcommand{\Platzlinks}{1.5em}
\newcommand{\Platzrechts}{1.5em}

\vspace*{\parskip}
\vspace*{1mm}

\begin{tabularx}{\linewidth}{@{}p{\BreitelinkeSpalte}X}
	Name & Andreas Eberlein\\
	Date of Birth & March 7th, 1984\\
	Place of Birth	&	Lichtenfels\\
\vspace*{0.1mm}

	\textbf{Education}		&\\[\PlatznachUeberschrift]
	01/2009 - today		&		Ph.D. student of Prof. Dr. W. Metzner, Max Planck
Institute for Solid State Research, Stuttgart\\[2mm]
10/2003 - 12/2008		& Studies of physics, Dresden University of
Technology\\[2mm]
	&		12/2008 Diploma in physics\\[2mm]
06/2003 & Abitur, Dientzenhofer Gymnasium, Bamberg\\[5mm]

\textbf{Scholarships}		&\\[\PlatznachUeberschrift]
	2009 - today	&	Fellow of International Max Planck Research School for
Advanced Materials\\[2mm]
	2003 - 2008	&	Studienstiftung des deutschen Volkes\\[10mm]
\end{tabularx}

\noindent
Stuttgart, January 2013